\documentclass[12pt,twoside,a4paper]{cernyrep} 
\usepackage[paperheight=29.5cm,paperwidth=21.cm,margin=0.6in,heightrounded]{geometry}
\usepackage{texnames}
\usepackage[T1]{fontenc}
\usepackage{hyperref}
\hypersetup{
linktocpage=true,
colorlinks=true,
linkcolor=blue,          
citecolor=mygreen,        
filecolor=red,      
urlcolor=red           
}  
\usepackage{url}
\usepackage{graphicx}
\usepackage{caption}
\usepackage{subcaption}
\usepackage{hepnames}
\usepackage{color}
\definecolor{mygreen}{rgb}{0.0,0.75,0.0}
\usepackage{timestamp}
\usepackage{times}
\usepackage{comment}
\usepackage{tikz}
\usetikzlibrary{arrows,decorations.pathmorphing,backgrounds,positioning,fit,petri}
\usepackage{mathtools}
\usepackage{enumitem}
\usepackage{fancyhdr} 
\usepackage{etoolbox}

\definecolor{amber}{rgb}{1.0, 0.75, 0.0}
\usepackage{pagecolor}
\usepackage{afterpage}  

\renewcommand{\headrulewidth}{0.4pt}   
\newlength\FHoffset
\fancyheadoffset{\FHoffset}
\newlength\FHleft
\newlength\FHright
\newbox\FHline
\setbox\FHline=\hbox{\hsize=\paperwidth%
  \hspace*{\FHleft}%
  \rule{\dimexpr\headwidth-\FHleft-\FHright\relax}{\headrulewidth}\hspace*{\FHright}%
}

\def\mathswitch#1{\relax\ifmmode#1\else$#1$\fi}
\def\mathswitchr#1{\relax\ifmmode{\mathrm{#1}}\else$\mathrm{#1}$\fi}

\newcommand{\Pt}{\mathswitchr t}
\def \sd {\texttt{SD}{}}
\def\order#1{\ensuremath{{\cal O}(#1)}}

\newcommand{\bhlumi}{{\tt BHLUMI}}

\newcommand{\Mmf}{\mathfrak{M}}

\newcommand{\MW}{\mathswitch {M_{\PW}}}
\newcommand{\MZ}{\mathswitch {M_{\PZ}}}
\newcommand{\GZ}{\mathswitch {\Gamma_{\PZ}}}
\newcommand{\MH}{\mathswitch {M_{\PH}}}

\newcommand{\mb}{\mathswitch {m_{\Pb}}}
\newcommand{\mt}{\mathswitch {m_{\Pt}}}

\newcommand{\scrs}{\scriptscriptstyle}
\newcommand{\sw}{\mathswitch {s_{\scrs\PW}}}
\newcommand{\cw}{\mathswitch {c_{\scrs\PW}}}

\newcommand{\as}{\alpha_{\mathrm s}}
\newcommand{\at}{\alpha_\Pt}
\newcommand{\seff}[1]{\sin^2\theta_{\rm eff}^{\rm #1}}

\newcommand{\gev}{\,\, \mathrm{GeV}}

\newcommand{\re}{\Re  \,}
\newcommand{\im}{\Im  \,}

\newcommand{\OO}{{\mathcal O}}

\newcommand{\bracket}[1]{{\left\langle #1 \right\rangle}}

\DeclareMathOperator{\Pexp}{Pexp}
\DeclareMathOperator{\Ima}{Im}
\DeclareMathOperator{\coIma}{coIm}
\DeclareMathOperator{\rank}{rank}

\newcommand{\nn}{\nonumber}

\newcommand*{\CO}{\mathcal{O}}
\newcommand*{\CI}{\mathcal{I}}  
\newcommand*{\R}{\mathbb{R}}      
\newcommand*{\C}{\mathbb{C}}      
\newcommand*{\CU}{\mathcal{U}}
\newcommand*{\CF}{\mathcal{F}}

\def \mbnum      {\texttt{MBnumerics}{}}

\def \ar {\texttt{AMBRE}{}}
\def \la {\texttt{LA}{}}
\def \ga {\texttt{GA}{}}
\def \mb {\texttt{MB}{}}
\def \pltest      {\texttt{PlanarityTest}{}}

\def \mbm  {\texttt{MB.m}{}}
\def \mbr  {\texttt{MB}{}}

\newcommand{\bea}{\begin{eqnarray}}
\newcommand{\eea}{\end{eqnarray}}

\newcommand{\Eqn}[1]{Eq.~(\ref{#1})}
\newcommand{\Fign}[1]{Fig.~\ref{#1}}

\newcommand\Vol{\mathop{\mathrm{Vol}}}
\newcommand\op[1]{\mathbf{#1}}
\newcommand\opI{\op{I}}
\newcommand\opQ{\op{C}}
\newcommand\opM{\op{M}}
\newcommand\opN{\op{N}}
\newcommand\rd{\mathrm{d}}

\newcommand\ri{\mathrm{i}}
\newcommand\Itot{I_{\mathrm{tot}}}
\newcommand\Etot{E_{\mathrm{tot}}}
\newcommand\epsrel{\varepsilon_{\mathrm{rel}}}
\newcommand\epsabs{\varepsilon_{\mathrm{abs}}}
\newcommand\ord{\mathcal{O}}
\newcommand\Code[1]{\ensuremath{\texttt{#1}}}
\newcommand\Var[1]{\ensuremath{\mathit{#1}}}
\newcommand\accel{_{\mathrm{accel}}}
\newcommand\cores{_{\mathrm{cores}}}

\newcommand{\babayagaNLO}{\texttt{{BabaYaga@NLO}}}
\newcommand{\babayaga}{\texttt{{BabaYaga}}}

\newcommand{\sss}[1]{\scriptscriptstyle{#1}}
\newcommand{\ds }{\displaystyle}

\newcommand{\vma}[2]{\delta_{#1}^{#2}}
\newcommand{\nll}{\nonumber\\}

\renewcommand{\thechapter}{\Alph{chapter}}

\usepackage{varwidth}

\usepackage{xcolor}

\begin{document}

\pagenumbering{roman}  
\pagestyle{empty}

\renewcommand{\chaptermark}[1]{\markboth{ #1}{}}
\renewcommand{\sectionmark}[1]{\markright{\thesection.\ #1}}

\thispagestyle{empty}
\setlength{\unitlength}{1mm}
\begin{picture}(0.001,0.001)
\put(-8,8){\large CERN Yellow Reports: Monographs}
\put(120,8){\large CERN-2019-003}

\put(-5,-60){\LARGE\bfseries
                                              Standard Model Theory for the FCC-ee \mbox{Tera-Z} stage}
\put(-5,-68){\small Report on the Mini Workshop}
\put(-5,-73){\small Precision EW and QCD Calculations for the FCC Studies: Methods and Tools}
\put(-5,-78){\small 12--13 January 2018, CERN, Geneva}

\put(-5,-94){\Large Editors:}

\put(0,-105){\Large A. Blondel}
\put(5,-109){\small DPNC University of Geneva, Switzerland}

\put(0,-120){\Large J. Gluza}
\put(5,-124){\small Institute  of Physics, University of Silesia, 40-007 Katowice, Poland}

\put(0,-135){\Large S. Jadach}
\put(5,-139){\small Institute of Nuclear Physics, PAN, 31-342 Krak\'ow, Poland}

\put(0,-150){\Large P. Janot}
\put(5,-154){\small CERN, CH-1211 Geneva 23, Switzerland}

\put(0,-165){\Large T. Riemann}
\put(5,-169){\small Institute  of Physics, University of Silesia, 40-007 Katowice, Poland}
\put(5,-173){\small Deutsches Elektronen-Synchrotron, DESY, 15738 Zeuthen, Germany}


\end{picture}
\newpage

\thispagestyle{empty}
\mbox{}
\vfill

\begin{flushleft}
CERN Yellow Reports: Monographs\\
Published by CERN, CH-1211 Geneva 23, Switzerland\\[3mm]

\begin{tabular}{@{}l@{~}l}
  ISBN & 978-92-9083-541-7 (paperback) \\
  ISBN & 978-92-9083-542-4 (PDF) \\
  ISSN & 2519-8068 (Print)\\ 
  ISSN & 2519-8076 (Online)\\ 
  DOI & \url{http://dx.doi.org/10.23731/CYRM-2019-003}\\
\end{tabular}\\[3mm]
Accepted for publication by the CERN Report Editorial Board (CREB) on 8 September 2019\\[1mm]
Available online at \url{http://publishing.cern.ch/} and \url{http://cds.cern.ch/}\\[3mm]

Copyright \copyright{} CERN, 2019\\[1mm]
Creative Commons Attribution 4.0\\[1mm]
Knowledge transfer is an integral part of CERN's mission.\\[1mm]
CERN publishes this volume Open Access under the Creative Commons Attribution 4.0 license\\
(\url{http://creativecommons.org/licenses/by/4.0/}) in order to permit its wide dissemination and use.\\
The submission of a contribution to a CERN Yellow Report series shall be deemed to constitute the contributor's agreement to this copyright and license statement. Contributors are requested to obtain any clearances that may be necessary for this purpose.\\[5mm]

This volume is indexed in: CERN Document Server (CDS)\\[5mm]

This volume should be cited as:\\[1mm]

Standard Model Theory for the FCC-ee Tera-Z stage\\Report on the mini workshop "Precision EW and QCD Calculations for the FCC Studies: Methods and Tools", 12--13 January 2018, CERN, Geneva\\ 
Eds. A.~Blondel, J.~Gluza, S.~Jadach, P.~Janot and T.~Riemann\\  CERN Yellow Reports: Monographs, CERN-2019-003 (CERN, Geneva, 2019), \url{http://dx.doi.org/10.23731/CYRM-2019-003}\\[3mm]

\end{flushleft}

\clearpage
%
%
\pagecolor{white}


\vphantom{tord}


{\bf\LARGE
\noindent Standard Model Theory for the FCC-ee Tera-Z stage}
\\[4mm] 
\small {Report on the Mini Workshop\\ 
Precision EW and QCD Calculations for the FCC Studies: Methods and Tools}\footnote{Workshop web site and presentations: \url{https://indico.cern.ch/event/669224/}}\\
\small {12--13 January 2018, CERN, Geneva 
 }
 
\vspace*{10mm}   

\noindent {\bf A.~Blondel$^{1}$,~ J.~Gluza$^{*,2}$,~  S.~Jadach$^3$,~ P.~Janot$^{4}$,~ T.~Riemann$^{2,5}$ (editors)}\\
A.~Akhundov$^{6,7}$, ~ 
A.~Arbuzov$^{8}$, ~ 
R.~Boels$^9$, ~ 
S.~Bondarenko$^{8}$, ~ 
S.~Borowka$^{4}$, ~ 
C.M.~Carloni~Calame$^{10}$, ~ 
I.~Dubovyk$^{5,9}$, ~ 
Y.~Dydyshka$^{11}$, ~ 
W.~Flieger$^2$, ~ 
A.~Freitas$^{12}$, ~  
K.~Grzanka$^2$, ~ 
T.~Hahn$^{13}$, ~ 
T.~Huber$^{14}$, ~ 
L.~Kalinovskaya$^{11}$, ~ 
R.~Lee$^{15}$, ~ 
P.~Marquard$^{5}$, ~ 
G.~Montagna$^{16}$, ~ 
O.~Nicrosini$^{10}$, ~ 
C.G.~Papadopoulos$^{17}$, ~ 
F.~Piccinini$^{10}$, ~ 
R.~Pittau$^{18}$, ~ 
W.~P\l{}aczek$^{19}$, ~ 
M.~Prausa$^{20}$, ~ 
S.~Riemann$^{5}$, ~ 
G.~Rodrigo$^{21}$, ~ 
R.~Sadykov$^{11}$, ~ 
M.~Skrzypek$^{3}$, ~ 
D.~St{\"o}ckinger$^{22}$, ~ 
J.~Usovitsch$^{23}$, ~ 
B.F.L.~Ward$^{24,12}$, ~ 
S.~Weinzierl$^{25}$, ~ 
G.~Yang$^{26}$, ~ 
S.A.~Yost$^{27}$



\begin{flushleft}
{\em\small
{$^{1}$DPNC University of Geneva, Switzerland}\\
{$^2$Institute  of Physics, University of Silesia, 40-007 Katowice, Poland}\\
{$^3$Institute of Nuclear Physics, PAN, 
31-342 Krak\'ow, Poland}\\
{$^{4}$CERN, CH-1211 Geneva 23, Switzerland}\\
{$^{5}$Deutsches Elektronen-Synchrotron, DESY, 15738 Zeuthen, Germany}\\
{$^{6}$Departamento de F\'{\i}sica Teorica, Universidad de Val\`{e}ncia, 46100 Val\`{e}ncia, Spain}\\ 
$^{7}${Azerbaijan National Academy of Sciences, ANAS,
Baku, 
Azerbaijan}\\
{$^{8}$Bogoliubov Laboratory of Theoretical Physics, JINR, Dubna, 141980 Russia}\\
{$^9$II. Institut f{\"u}r Theoretische Physik, Universit{\"a}t Hamburg,
22761 Hamburg,  Germany}\\
{$^{10}$Istituto Nazionale di Fisica Nucleare, Sezione di Pavia, Pavia, Italy}\\
{$^{11}$Dzhelepov Laboratory of Nuclear Problems, JINR, Dubna, 141980 Russia}\\
{$^{12}$Pittsburgh Particle Physics, Astrophysics \& Cosmology Center
(PITT PACC) and Department of Physics \& Astronomy, University of Pittsburgh, Pittsburgh, PA 15260, USA}\\
{$^{13}$Max-Planck-Institut f{\"u}r  Physik, 
80805 M{\"u}nchen, Germany}\\
{$^{14}$Naturwissenschaftlich-Technische Fakult\"at, Universit\"at Siegen, 
57068 Siegen, Germany}\\
{$^{15}$The Budker Institute of Nuclear Physics, 630090, Novosibirsk, Russia}\\
{$^{16}$Dipartimento di Fisica, Universit\`a di Pavia, Pavia, Italy}\\
{$^{17}$Institute of Nuclear and Particle Physics, NCSR Demokritos, 
15310, Greece}\\
{$^{18}$Dep. 
   de F\'{i}sica Te\'orica y del Cosmos and CAFPE, Universidad de Granada, 
E-18071 Granada, Spain}\\
{$^{19}$Marian Smoluchowski Institute of Physics, Jagiellonian University, 
30-348 Krak\'ow, Poland}\\
{$^{20}$Albert-Ludwigs-Universit\"at, Physikalisches Institut, Freiburg, Germany}\\
{$^{21}$Instituto de F\'{\i}sica Corpuscular, Universitat de Val\`{e}ncia -- 
CSIC, 46980 Paterna, Val\`{e}ncia, Spain}\\
{$^{22}$Institut f\"ur Kern- und Teilchenphysik, TU Dresden, 01069 Dresden, Germany}\\
{$^{23}$Trinity College Dublin -- School of Mathematics, Dublin 2, Ireland}\\
{$^{24}$Baylor University, Waco, TX, USA}\\
{$^{25}$PRISMA Cluster of Excellence, Inst. f{\"u}r Physik, Johannes Gutenberg-Universit{\"a}t, 55099 Mainz, Germany}\\
{$^{26}$CAS Key Laboratory of Theoretical Physics, 
Chinese Academy of Sciences, Beijing 100190, China}\\
{$^{27}$The Citadel, Charleston, SC, USA}
}
\end{flushleft}
\kern -3pt
\hrule width 2in
\kern 2.6pt
$^{*}$ Corresponding editor, email: janusz.gluza@cern.ch.

\cleardoublepage
\pagestyle{fancy}
\fancyhead[LO]{}
\fancyhead[RO]{}
\fancyhead[CO]{\thechapter ~ \leftmark}
\fancyhead[LE]{}
\fancyhead[CE]{}
\fancyhead[RE]{} 
\pagestyle{plain}
\pagenumbering{roman}
\setcounter{page}{5}
\section*{Abstract  \label{ch:abstract}}
\phantomsection   
\addcontentsline{toc}{chapter}{Abstract}



\noindent
{\small
The proposed $100\Ukm$ circular collider FCC at CERN is planned to operate in one of its modes as an electron-positron FCC-ee machine. We give an overview, comparing the theoretical status 
of Z boson resonance energy physics
with the experimental demands of one of four foreseen FCC-ee operating stages, called the FCC-ee Tera-Z stage. 
The FCC-ee Tera-Z will deliver the highest integrated luminosities, as well as very small systematic errors for a study of the Standard Model with unprecedented precision. In fact, the FCC-ee Tera-Z stage will allow the study of at least one more perturbative order in quantum field theory, compared with the precision
obtained using the LEP/SLC. This is an important new feature in itself, independent of specific `new physics' searches. 
Currently, the precision of theoretical calculations of the various Standard Model observables does not match that of the anticipated experimental measurements. 
The obstacles to overcoming this situation are identified. In particular, the issues of precise QED unfolding and the correct calculation of Standard Model pseudo-observables are critically reviewed. In an  {\it executive summary}, we specify which basic theoretical calculations are needed to meet the strong experimental expectations at the FCC-ee Tera-Z.  Finally, several methods, techniques, and tools needed for higher-order multiloop calculations are presented. By inspection of the Z boson partial and total decay width analyses, 
we argue that, until
the beginning of operation of the FCC-ee Tera-Z, the theoretical predictions will be precise enough
not to limit the physical interpretation of the measurements. This statement is based on 
the completion this year of two-loop electroweak radiative corrections to the Standard Model pseudo-observables 
and on 
anticipated progress in analytical and numerical calculations of multiloop and multiscale Feynman integrals.  
However,  on a time perspective over one or two decades, a highly dedicated and focused investment is needed by the community, to bring the state-of-the-art theory to the necessary level.

 }

\clearpage
\pagestyle{empty}
\cleardoublepage 

\pagestyle{plain}
\pagenumbering{roman}
\setcounter{page}{7}
\fancyhead[LO]{}
\fancyhead[RO]{}
\fancyhead[CO]{Foreword}
\lfoot[]{}
\cfoot[\thepage]{  \thepage \hspace*{0.075mm} }
\rfoot[]{}
\fancyhead[LE]{}
\fancyhead[CE]{A. Blondel, P. Janot}
\fancyhead[RE]{} 
\lfoot[]{}
\cfoot[\thepage]{  \thepage \hspace*{0.075mm} }
\rfoot[]{}

\chapter*{Foreword 
\label{foreword}}
\phantomsection     
\addcontentsline{toc}{chapter}
{Foreword 
        \\
        {\it A. Blondel, P. Janot} }

\pagestyle{fancy}
\fancyhead[LO]{}
\fancyhead[RO]{}
\fancyhead[CO]{Foreword}
\fancyhead[LE]{}
\fancyhead[CE]{A. Blondel, P. Janot}
\fancyhead[RE]{} 
%

\noindent
{\bf Authors:} Alain Blondel {[Alain.Blondel@cern.ch]} and Patrick Janot {[Patrick.Janot@cern.ch]}. \\Physics co-coordinators of the FCC-ee design study and members of the FCC coordination group.
\vspace*{.5cm}

\section*{Precision measurements at the FCC-ee and the wish-list to theory}

Particle physics has arrived at an important moment of its history. The discovery of the Higgs boson at the LHC, with a mass of $125\UGeV$, completes the matrix of particles and interactions that has now constituted the `Standard Model' for  several decades. The Standard Model is a very consistent and predictive theory, which has proven extraordinarily successful in describing the vast majority of phenomena accessible to experiment. The observed masses of the top quark and the Higgs boson are found to agree well with the values that could be predicted, before their direct observation, from a wealth of precision measurements collected at the LEP and SLC e$^+$e$^-$ colliders, at the Tevatron and from other precise low-energy experimental input. Given the top quark
 and Higgs masses, the Standard Model can even be extrapolated to the Planck scale without encountering a breakdown of the stability of the Universe. 

At the same time, we know that the story is not over. Several experimental facts require extension of the Standard Model, in particular: (i) in the composition of the observable Universe, matter largely dominates antimatter; (ii) the well-known evidence for dark matter from astronomical and cosmological observations; and (iii) more closely to particle physics, not only do neutrinos have masses, but these masses are about $10^{-7}$ times smaller than that of the electron. To these experimental facts can be added a number of theoretical issues of the Standard Model, including the hierarchy problem, the neutrality of the Universe and the stability of the Higgs boson mass on radiative corrections,
and the strong CP problem, to name a few. The problem faced by particle physics is that the possible solutions to these questions seem to require the existence of new particles or phenomena that can occur over an immense range of mass scales and coupling strengths. To make things more challenging, it is worth recalling that the predictions of the top quark and Higgs boson masses from precision measurements were made within the Standard Model framework, assuming that no other new physics exists, which would modify the loop corrections on which the predictions were made. 

The observation of new particles or phenomena may happen by luck, by increasing energy. The past has shown, however, that, for example, the existence of the W and Z bosons, of the top quark, and of the Higgs boson, as well as their properties, were predicted before their actual observations, from a long history of experiments and theoretical maturation. 

In this context, a decisive improvement in precision measurements of electroweak pseudo-observables (EWPOs) could play a crucial role, by integrating sensitivity to a large range of new physics possibilities. The observation of a significant deviation from the Standard Model predictions will definitely be a discovery. It will require not only a considerable improvement in precision, but also a large set of measured observables, in order to (i) eliminate spurious deviations and (ii) possibly reveal a pattern of deviations, enabling the guidance of theoretical interpretation. \textit{Improved precision equates discovery potential.}  

For these quantum effects to be measurable, however, the precision of theoretical calculations of the various observables within the Standard Model will have to match that of the experiment, \ie to improve by up to two orders of magnitude with respect to current achievements. This tour de force will require complete two- and three-loop corrections to be calculated. Probably, this will lead to the development of breakthrough computation techniques to keep the time needed for these numerical calculations within reasonable limits.

The 2013 European Strategy for Particle Physics, ESPP~\cite{strat}, states, ``To stay at the forefront of particle physics, Europe needs to be in a position to propose an ambitious post-LHC accelerator project at CERN by the time of the next Strategy update, when physics results from the LHC running at $14\UTeV$ will be available. CERN should undertake design studies for accelerator projects in a global context, with emphasis on proton--proton and electron--positron high-energy frontier machines.'' 

The importance of precision was not forgotten by the ESPP, however, which
goes on to state,  ``There is a strong scientific case for an electron--positron collider, complementary to the LHC, that can study properties of the Higgs boson and other particles with unprecedented precision and whose energy can be upgraded.''

The FCC international collaboration~\cite{fcc} has thus undertaken the study of a future $100\Ukm$ circular infrastructure, designed with the capability  to host, as its ultimate goal, a $100\UTeV$  pp collider (FCC-hh). Within the study, a considerable effort is going into the design of a high-luminosity, high-precision e$^+$e$^-$ collider, FCC-ee, which would serve as a first step, in a way similar to the LEP/LHC success story. The study established that FCC-ee is feasible with good expected performance, has a strong physics case \cite{tlep} in its own right and could technically be built within a time-scale so as to start seamlessly at the end of the HL-LHC programme. Thus, with a combination of synergy and complementarity, both in the infrastructure and for the physics, the FCC programme fulfils both recommendations of the ESPP. 

The FCC-ee is designed to  deliver e$^+$e$^-$ collisions to study the Z, W, and Higgs bosons and the top quark, as well as the bottom and charm quarks and the tau lepton. The run plan, spanning 15\,years, including commissioning, is shown in Table~\ref{tab:FCC-ee-runplan}. The number of Z bosons planned to be produced by the FCC-ee (up to $5 \times 10^{12}$), for example, is  more than five orders of magnitude larger than the number of Z bosons collected at the LEP ($2 \times 10^7$), and three orders of magnitude larger than that envisioned with a linear collider (${\sim}10^9$). Furthermore,  exquisite determination of the centre-of-mass energy by resonant depolarization available in the storage rings will allow measurements of the W and Z masses and widths with a precision of a few hundred kiloelectronvolts. The high-precision measurements and the observation of rare processes that will be  made possible by these large data samples will open opportunities for new physics discoveries, from very weakly coupled light particles that could explain the yet-unobserved dark matter or neutrino masses, to quantum effects of weakly coupled new particles up to masses up to the better part of $100\UTeV$. 

\begin{table}
\centering
\caption{Run plan for FCC-ee in its baseline configuration with two experiments. The WW event numbers are given for the entirety of the FCC-ee running at and above the WW threshold.
\label{tab:FCC-ee-runplan}}
\begin{tabular}{l l l l l } \hline \hline 
Phase & Run duration  & Centre-of-mass  &  Integrated   & Event   \\ 
& (years)  & energies & luminosity &  statistics \\ 
&   & (GeV) & (ab$^{-1}$) &   \\ 
\hline 
FCC-ee-Z & 4  & \phantom{3}88--95   & 150   & $3 \times 10^{12}$ visible Z decays   \\ 
FCC-ee-W & 2  & 158--162 &  \phantom{3}12   & $10^8$ WW events               \\ 
FCC-ee-H & 3  & 240     &  \phantom{3}5    & $10^6$ ZH events               \\ 
FCC-ee-tt & 5  & 345--365 &  \phantom{3}1.5 & $10^6$ $\mathrm{t}\bar{\mathrm{t}}$ events       \\
 \hline \hline
\end{tabular} 
\end{table}

Apart from the FCC-ee, other options are being considered internationally for future electron colliders. The International Linear Collider  (ILC) 
\cite{tlep,Moortgat-Picka:2015yla} 
and Compact Linear Collider (CLIC) \cite{clic} offer high-energy reach and are, to a large extent, complementary to the FCC-ee.   
The ILC proposal is presently in the final stage of negotiations in Japan.  
It is planned with a first step at a centre-of-mass of $250\UGeV$, and could be extended to  $500\UGeV$. While the present plan does not foresee intense running at the Z boson resonance energy, a `Giga-Z' run has been discussed. The CLIC,  built at CERN and based on  a high-gradient room-temperature acceleration system, would cover  energies between $0.5\UTeV$ and $3\UTeV$. 
Finally, the Circular Electron--Positron Collider  (CEPC) \cite{cepc} in China, 
similar to the FCC-ee, is designed for collisions from the Z to the ZH production maximum at $250\UGeV$. 
Among these projects, FCC-ee is the most ambitious for precision measurements;  we will concentrate on this project here. Precision calculations suitable for FCC-ee will, by definition, suit the other projects.

Table~\ref{tab:FCC-observables} summarizes  some of the most significant FCC-ee experimental accuracies and compares them with those of current measurements. 

\renewcommand{\arraystretch}{1.07}   
\begin{table} 
\centering
\caption{Measurement of electroweak quantities  at the FCC-ee, compared with current precisions
  \label{tab:FCC-observables}}
\resizebox{\textwidth}{!}{%
\begin{tabular}{lllllll}
\hline \hline
Observable  & Present &  &          &  FCC-ee  &  FCC-ee  &  
Source and   \\ 
            & value  &$\pm$& error  &  (statistical) &   (systematic)     &  dominant experimental error \\ 
\hline 
$ \mathrm{ m_Z  ~(keV/c^2) } $  &  91\,186\,700   & $\pm$ &  2200    & 5  & 100  & 
Z line shape scan  \\ 
$  $  &  & &    &   &  &  Beam energy calibration  \\
$ \mathrm{  \Gamma_Z  ~(keV) } $  & 2\,495\,200   & $\pm$ &  2300    & 8  & 100  & 
Z line shape scan  \\
$  $  &  & &    &   &   &  Beam energy calibration  \\
$ \mathrm{  R_{\ell}^{Z}} ~(\times 10^3) $  & 20\,767 & $\pm$ &  25   & 0.06   & 1   &  Ratio of hadrons to leptons \\
$  $  &  & &    &   &   &  Acceptance for leptons  \\
$ \mathrm{ \alpha_{s} (m_Z) } ~(\times 10^4) $  & 
 1196 & $\pm$ &  30  &  0.1  &  1.6  &   
$\mathrm{  R_{\ell}^{Z}}$ above\\
$ \mathrm{  R_b} ~(\times 10^6) $  & 216\,290 & $\pm$ &  660   & 0.3   &  <60  &  Ratio of  
$\mathrm {b\bar{b}}$ to hadrons  \\
$  $  &  & &    &   &   &  Stat. extrapol. from SLD 
~\cite{Abe:2005nqa}\\
$ \sigma_\mathrm{had}^0 ~(\times 10^3)$ (nb) & 41\,541 & $\pm$ &  37   & 0.1  &  4  &  Peak hadronic cross-section  \\
$  $  &  & &    &   &   &  Luminosity measurement  \\
$ \mathrm{  N_{\nu}}  (\times 10^3) $  & 2991  & $\pm$ & 7   & 0.005   &  1  &  Z peak cross-sections \\
$  $  &  & &    &   &  &   Luminosity measurement \\
$ \mathrm{ sin^2{\theta_{W}^{\rm eff}}} (\times 10^6) $  & 231\,480   & $\pm$ &  160   & 3   &  2--5  &   
$ \mathrm{ A_{FB}^{{\mu} {\mu}}}$  at Z peak\\
$  $  &  & &    &   &   &  Beam energy calibration  \\
$ \mathrm{1/\alpha_{QED} (m_Z) } (\times10^3) $  & 128\,952 
  & $\pm$ &  14   & 4   &  Small  &   
$ \mathrm{ A_{FB}^{{\mu} {\mu}}}$ off peak\\
$ \mathrm{A_{FB}^{b,0}} ~(\times 10^4) $  & 992 & $\pm$ &  16   & 0.02   &  <1  &  b quark asymmetry at Z pole  \\
$  $  &  & &    &   &   &  
Jet charge \\
$ \mathrm{A_{FB}^{pol,\tau} ~(\times 10^4)} $  & 1498 & $\pm$ &  49   & 0.15   &  <2  &  $\tau$ polar. and charge asymm.  \\
$  $  &  & &    &   &   &  $\tau$ decay physics \\

$ \mathrm{ m_W  ~(keV/c^2) } $  &  803\,500   & $\pm$ &  15\,000    & 600  & 300  &
WW threshold scan \\ 
$  $  &  & &    &   &  &  Beam energy calibration  \\
$ \mathrm{  \Gamma_W  ~(keV) } $  & 208\,500   & $\pm$ &  42\,000    & 1500  & 300  & 
WW threshold scan \\
$  $  &  & &    &   &   &  Beam energy calibration  \\
$ \mathrm{ \alpha_{s} (m_W) }  (\times 10^4)$  & 
 1170 & $\pm$ & 420    &  3  & Small  &   
  $ \mathrm{R_{\ell}^{W} }$\\
$ \mathrm{  N_{\nu}}  (\times 10^3) $  & 2920 & $\pm$ &  50   & 0.8   & Small   &   Ratio of invis. to leptonic \\
$  $  &  & &    &   &  & in radiative Z returns  \\
$ \mathrm{ m_{top}  ~(MeV/c^2) } $  &  172\,740   & $\pm$ &  500    & 20  & Small  & 
$\mathrm {t\bar{t}}$ threshold scan \\ 
$  $  &  & &    &   &  &  QCD errors dominate  \\
$ \mathrm{ \Gamma_{top}  ~(MeV/c^2) } $  &  1410   & $\pm$ &  190    & 40  & Small  & 
$\mathrm {t\bar{t}}$ threshold scan \\ 
$  $  &  & &    &   &  &  QCD errors dominate \\
$ \mathrm{ \lambda_{top}/\lambda_{top}^{SM}   } $  &  $m$ = 1.2 
   & $\pm$ &  0.3    & 0.08  & Small  & 
   $\mathrm {t\bar{t}}$ threshold scan \\ 
$  $  &  & &    &   &  &  QCD errors dominate \\
$ \mathrm{ t{\bar{t}}Z ~couplings   } $  &   
   & $\pm$ &  30\%   & <2\%  & Small  & 
   $ \mathrm{E_{CM}=365\UGeV}$ run  \\ 
\hline \hline
\end{tabular}
} 
\end{table}
\renewcommand{\arraystretch}{1} 

Some important comments are in order. 
\begin{itemize}
\item
FCC-ee will provide a set of ground breaking measurements of a large number of new-physics sensitive observables, with improvement with respect to the present status by a factor of 20--50 or even more; moreover, it will improve input parameters, $ \mathrm{ m_Z } $ of course, but also $ \mathrm{ m_{top} }$,
$ \mathrm{ \alpha_{s} (m_Z) }$ and, for the first time, a direct and precise measurement of  $ \mathrm{ \alpha_{QED} (m_Z) }$, with a precision that will reduce considerably the parametric uncertainties in the electroweak predictions. 
\item 
Table~\ref{tab:FCC-observables} is only a first sample of the accessible observables. Work on future projections of experimental and theory requirements is subject to intensive studies within the FCC-ee design study groups. Important contributions are expected from bottom, charm, and tau physics at the Z pole, such as forward--backward and polarization asymmetries. Also, the tau lepton branching fraction and lifetime measurements, especially if a more precise tau mass becomes available, will provide another dimension of precision measurements.   
\item
While the statistical precisions follow straightforwardly from the integrated luminosities, the systematic uncertainties do not. It is quite clear that the centre-of-mass energy uncertainty will dominate  for the Z and W mass and width, and that  the luminosity measurement error will dominate for the total cross-sections (and thus the number of neutrino determination). These have been the subject of considerable work already. However, there is no obvious limit to the experimental precision reachable for such observables as $ \mathrm{  R_{\ell}^{Z}}$ or  $ \mathrm{  R_b} $ or the top quark pair cross-section measurements.  
\item
While we have indicated a possible experimental error level for $ \mathrm{  R_{\ell}^{Z}}$ or  $ \mathrm{  R_b} $, these should be considered  \textit{indicative} and might improve with closer investigation. It is likely, however, that the interpretation of these measurements in terms of, \eg the bottom weak couplings, the strong coupling constant, or the top mass, width, and  weak and Yukawa couplings, will be limited by questions related to the precise definition of these quantities, or to issues such as, `What is the bottom
quark mass?' 
\end{itemize}

Table~\ref{tab:FCC-observables} clearly sets the requirements for theoretical work: the aim should be either to provide the tools to compare experiment and theory at a level of precision better than the experimental errors or
to identify which additional calculation or experimental input would be required to achieve it. Another precious line of research to be followed jointly by theorists and experimenters should be to try to find observables or ratios of observables for which theoretical uncertainties are reduced. 

The theoretical work that experiment requires from the theoretical community can be separated into a few classes. 
\begin{itemize}
\item
QED (mostly) and QCD corrections to cross-sections and angular distributions that are needed to convert experimentally measured cross-sections back to `pseudo-observables': couplings, masses, partial widths, asymmetries, etc., that are close to the experimental measurement (\ie the relation between measurements and these quantities does not alter the possible `new physics' content). Appropriate event generators are essential for the implementation of these effects in the experimental procedures. 
\item
Calculation of the pseudo-observables with the precision required in the framework of the Standard Model  so as to take full advantage of the experimental precision. 
\item 
Identification of the limiting issues and, in particular, the questions related to the definition of parameters, in particular, the treatment of quark masses and, more generally, QCD objects. 
\item 
An investigation of the sensitivity of the proposed experimental observables (or new ones) to the effect of new physics in a number of important scenarios. This is an essential work to be done early, before the project is fully designed,  since it potentially affects the detector design and  the running plan.
\end{itemize}

The workshop at CERN {\em Precision EW and QCD calculations for the FCC studies: methods and tools} was the start of a process that will be both exciting and challenging.  The precision calculations might look like a high mountain to climb but may contain the gold nugget: the discovery of the signals pointing the particle physics community towards the solution of some of our deep questions about the Universe. 

It is a pleasure to thank Janusz Gluza, Staszek Jadach and Tord Riemann for their competence and enthusiasm in organizing the workshop and the write-up, and all the participants for their contributions. We look forward to this adventure together.  

\clearpage
\pagestyle{empty}
\cleardoublepage
 
\chapter*{Editors' note} \label{ch:exsum2}
\vspace*{1.cm}
\addcontentsline{toc}{chapter}{Editor's note}

\chead[]{}
\lhead[\thechapter.~~ \leftmark]{\thechapter.~~ \leftmark}
\rhead[\rightmark]{\rightmark}
\lfoot[]{}
\cfoot[\thepage]{\thepage}
\rfoot[]{}

 
\noindent
We should think big. 
And the FCC-ee project is a big chance to step up in our case for understanding the physical world at the smallest scales. 
We were 
\textcolor{black}{impressed}
by the enthusiasm of the participants of the workshop on the theoretical backing of the FCC-ee Tera-Z stage and by the subsequent 
commitments. 
The present report documents this.
The contributions by the participants of the workshop, with several additional invited submissions and accomplished with approximately 800
detailed references, prove that the theory community needs and enjoys such demanding long-term projects as the FCC, including the FCC-ee mode running at the Z peak.
They push particle physics research into the most advanced challenges, which stand  ahead of twenty-first- century science. 

  A quote from the report's body:
{\it
\begin{quote}
Such huge improvements will allow the FCC-ee Tera-Z stage to test the Standard Model at an unprecedented precision level. 
The increment in precision corresponds to the increment that was represented by the LEP/SLC 
in their time; they tested the Standard Model at a  precision that 
needed `complete' one-loop corrections, plus leading higher-order terms.
The FCC-ee Tera-Z stage will need `complete' two-loop corrections, plus leading higher-order terms.
Even without 
{
an explicit} reference to new physics, the  
FCC-ee Tera-Z stage lets us expect exciting, qualitatively new results.
\end{quote}
}


On the eve of  LEP1 in 1989, the year-long workshop on Z physics at LEP1 summarized the world knowledge on the Z resonance physics of that time. The results were made available as CERN Report CERN 89-08 
\cite{Altarelli:1989YRFred, Altarelli:1989YRTom, Altarelli:1989YRDick} 
with a total of about 1000 pages, covering  the `standard physics' in  vol. I, of 465 pages.
In 1995, the CERN Report 95-03 
\cite{Bardin:1995XX} with `Reports of the Working Group on Precision Calculations for the Z resonance' summarized, in parts I to III, over 410 pages, the state of the art of the radiative corrections for the Z resonance.
Its Part I.2 contains the `Electroweak Working Group Report' 
\cite{Bardin:1997xq}, 
with a careful analysis of the influence of complete one-loop and leading higher-order electroweak and QCD corrections.
Now, in 2018, 23 years on, we see again the need for collective tackling of perturbative contributions to the Z resonance.     
To control the theoretical predictions with an accuracy of up to about $10^{-6}$, as the FCC-ee Tera-Z stage project deserves, will necessitate years of dedicated work.
The start-up studies are presented here.

We hope that the workshop was a good starting point for further regular workshops on FCC-ee physics,  as well as future international collaboration. Innovations, endorsement, and stability on a long-term scale give us a unique chance to accomplish the big FCC-ee vision.

We thank all authors for their excellent work.

A. Blondel, J. Gluza, S. Jadach, P. Janot, T. Riemann

\clearpage
\cleardoublepage

\pagestyle{plain}
\fancyhead[LO]{}
\fancyhead[RO]{}
\fancyhead[CO]{\thechapter ~ \leftmark}
\fancyhead[LE]{}
\fancyhead[CE]{}
\fancyhead[RE]{} 

\tableofcontents
\clearpage
\cleardoublepage

\pagestyle{plain}
\pagenumbering{arabic} 
\pagestyle{empty}
\lfoot[\thepage]{-  \thepage \hspace*{0.075mm} -}
\cfoot{-  \thepage \hspace*{0.075mm} -}
\rfoot[\thepage]{-  \thepage \hspace*{0.075mm} -}
\chapter*{Executive summary} \label{ch:exsum}
\addcontentsline{toc}{chapter}{Executive summary}

\chead[]{}
\lhead[\thechapter.~~ \leftmark]{\thechapter.~~ \leftmark}
\rhead[\rightmark]{\rightmark}
\lfoot[]{}
\cfoot[\thepage]{\thepage}
\rfoot[]{}


\noindent
The message of this report may be summarized as follows.
\begin{enumerate}
\item 
One of the highlights of the FCC-ee  scientific programme is a comprehensive campaign of measurements of Standard Model  precision observables,
spanning the Z pole, the W pair threshold, Higgs production, and the top quark threshold. The statistics available, up to  ${\sim}10^{12}$ visible Z decays (Tera-Z) and ${\sim}10^8$ W pairs, leads to an experimental precision improved by one to two orders of magnitude compared with the state of the art. This increased precision opens a broad reach for discovery but puts strong demands on the precise calculations of Standard Model 
and 
QED corrections, especially at the Z pole,
on which this first report concentrates. This will involve  controlling perturbation theory at more than two loop levels,
and constitutes in itself a novel technical and mathematical scientific undertaking. 
This is discussed in the foreword
 and {in Chapters~\ref{ch-s1},
 \ref{sthstatus}, \ref{sunfold}, and  \ref{chmt}}.
\item 
To meet the experimental precision of the FCC-ee Tera-Z stage for electroweak pseudo-observables (EWPOs), even three-loop calculations of the $\mathrm{Z}\mathrm{f}{\bar {\mathrm{f}}}$ vertex will be needed,
comprising the loop orders 
${\cal{O}}(\alpha \alpha_\mathrm{s}^2)$, 
${\cal{O}}(N_\mathrm{f} \alpha_{}^2 \alpha_\mathrm{s})$,
${\cal{O}}(N_\mathrm{f}^2 \alpha_{}^3)$ 
and corresponding QCD four-loop terms. This is a key problem and is discussed in Chapters~\ref{sthstatus} and \ref{ch-sff}.
\item 
Real-photon emission is the other key problem, of a complexity that is  comparable to the loop calculations.
{A joint, concise  treatment of electroweak and QCD loop corrections with the real-photon corrections, and their interplay, must be worked out.}
In practice, 
{a variety of peculiarities,  as well as a huge complexity of expressions, will result.} 
This is discussed in Chapter~\ref{sthstatus}.
\item
The $\mathrm{Z}\mathrm{f}{\bar {\mathrm{f}}}$-vertex corrections are embedded in a structure describing the hard scattering process $\mathrm{e}^+\mathrm{e}^- \to \mathrm{f}  {\bar{\mathrm{f}}}$, based on matrix elements in the form of a Laurent series around the $\mathrm{Z}$ pole. Here, additional non-trivial contributions, such as two-loop weak box diagrams, show up. 
This is discussed in Chapter~\ref{sunfold}.
\item 
Full two-loop corrections to the $\mathrm{Z}\mathrm{f}{\bar {\mathrm{f}}}$-vertex have recently been completed.
We estimate that future calculations of the aforementioned higher-order terms would meet the experimental FCC-ee-Z  demands if they are performed with a 10\% accuracy, corresponding to two significant figures.
A specific issue is the treatment of the electroweak $\gamma_5$ problem.
This is discussed in Chapters~\ref{sthstatus} and~\ref{ch-sff}.
\item 
The central techniques for the electroweak loop calculations will be numerical. This is due to the large number of scales involved. 
To achieve the accuracy goals, we have identified and discussed these and several additional exploratory strategies, methods, and tools in Chapters~\ref{sunfold}, \ref{ch-sff}, and \ref{chmt}.
\item
The treatment of four-fermion processes will be required for the W mass and width measurements and will be addressed with as much synergy as possible in future work. Special treatment will be required for the Higgs and top physics, but the experimental precision is less demanding. The high-precision Z pole work will provide a strong basis for these further studies.
\end{enumerate}
The techniques for higher-order Standard Model corrections are basically understood, but not easily worked out or extended. 
We are confident that the community knows how to tackle the described problems. 
We anticipate that, at the beginning of the FCC-ee campaign of precision measurements,  
the theory will be precise enough not to limit their physics interpretation.  This statement is, however, conditional to sufficiently strong support by  the physics community and the funding agencies, including strong training programmes.

Cooperation of experimentalists and theorists will be highly beneficial, as well as the blend of experienced and younger colleagues; the workshop gained considerably from joining the experience from the LEP/SLC with the most advanced theoretical developments.
We hope that the meeting was a starting event for forming an active
SM@FCC-ee  
community.
 
\clearpage \pagestyle{empty} 
\cleardoublepage



\chapter
[Introduction to basic theoretical problems connected with precision
calculations for the FCC-ee
\\
{\it J. Gluza, S. Jadach, T. Riemann}]
{Introduction to basic theoretical problems connected with precision
        calculations for the FCC-ee}
        \label{ch-s1}

\pagestyle{fancy}
\fancyhead[CO]{\thechapter \hspace{1mm} Introduction}
\fancyhead[RO]{}
\fancyhead[LO]{}
\fancyhead[LE]{}
\fancyhead[CE]{}
\fancyhead[RE]{}
\fancyhead[CE]{J. Gluza, S. Jadach, T. Riemann}
\lfoot[]{}
\cfoot{  \thepage \hspace*{0.075mm} }
\rfoot[]{}
      
\noindent
{\bf Authors: Janusz Gluza, Stanis\l aw Jadach,  Tord Riemann}
\\ 
Corresponding author: 
Stanis\l aw~Jadach [stanislaw.jadach@cern.ch]
\vspace*{.5cm}

\section[Introduction]{Introduction}
This report includes a collection of material devoted to a discussion of the status of theoretical efforts 
towards the calculation of higher-order Standard Model  corrections 
needed for the FCC-ee. 
It originates from presentations at the FCC-ee mini workshop 
{\it Precision EW and QCD Calculations for the FCC Studies: Methods and Tools},
12--13 January 2018, CERN, Geneva, Switzerland \cite{mini}. Both at the workshop and  in this survey we have deliberately focused on the FCC-ee Tera-Z mode, see Table \ref{tab:FCC-ee-runplan} in the foreword. It will be the first operational stage of the FCC-ee.
The mini workshop was intended to initiate a discussion on several topics.
\begin{enumerate}
\item What are the necessary precision improvements of theoretical calculations
 in the Standard Model, such that
 they match the needs of the experiments 
 at the planned FCC-ee collider in the Z peak mode?
\item A focus is on the calculation of Feynman diagrams in terms of Feynman integrals.
Which calculational techniques are available or must be developed 
in order to attain the necessary precision level?
\item
Besides the high-loop vertex corrections, the realistic QED contributions are highly non-trivial.
\item
What else has to be calculated and put together 
for a data analysis? Respecting thereby gauge-invariance, analyticity, and unitarity.
\end{enumerate}
Although the theoretical calculations are universal 
and will be crucial for the success of any future high-luminosity collider, 
we will focus on the FCC-ee Tera-Z  study as the most precise facility.
Its precision would be useless without the corresponding higher-order Standard Model 
predictions. 
 
To obtain an understanding of the unprecedented accuracy of the FCC-ee Tera-Z project, 
let us look at the electron asymmetry parameter $A_\mathrm{e}$.
The most precise theoretical prediction in the Standard Model is given in Refs. \cite{Awramik:2006ar,Awramik:2006uz}.
The actual LEP/SLC-based value is 
$\sin^2\theta_{\rm eff}^{\rm lept} = 0.23152 \pm 16 \times 10^{-5}$ \cite{ALEPH:2005ab}; 
see also Table~\ref{tab:FCC-observables} in the foreword.
Interestingly,  the LHC can also play a role for electroweak precision measurements.  
The currently claimed ATLAS measurement is 
$\sin^2\theta_{\rm eff}^{\rm lept} = 0.23140 \pm 36 \times 
10^{-5}$ \cite{ATLAS-CONF-2018-037}. It is planned to  improve on this. 
The FCC-ee Tera-Z stage will measure the leptonic effective weak mixing angle with highest precision from the muon charge 
asymmetry, namely
$\delta \sin^2\theta_{\rm eff}^{\rm lept} = \pm 0.3 \times 10^{-5}$, see Table~\ref{tab:FCC-observables}.
This may be compared with the so-called  Giga-Z option of the ILC  \cite{Baer:2013cma}, which might have a certain degree 
of polarization.
Currently, the option is not in the priority list of the ILC.
For the Giga-Z stage, a relative uncertainty of
$\delta \sin^2\theta_{\rm eff}^{\rm lept} = \pm 1.3 \times 10^{-5}$ is expected \cite{Erler:2000jg}.
In conclusion, the FCC-ee Tera-Z stage will improve the $A_\mathrm{e}$ measurement by at least one order of magnitude.

Similarly, the values of mass and width of the Z boson will be improved by factors of 20. 

The meeting and the report presented here are based on two complementary 
sources of knowledge.
First, the knowledge base accumulated by physicists who have worked for many years 
at the LEP/SLC. 
Their expertise
is accelerating the FCC-ee studies at every stage.
A second, equally important, source is the huge progress made in 
the last two decades
in the area of analytical and numerical methods and practical tools for
multiloop calculations in perturbative field theory.
These two strong pillars are manifest in this document;
see Chapters \ref{sunfold} and \ref{sthstatus} for the first 
and Chapters \ref{ch-sff} and \ref{chmt} for the second.
The report also reflects another component
 for the success of such a complicated long-term project --
the many contributions by young,
talented, mathematically oriented colleagues, who are
contributing bold and new ideas to this study.
Further, a certain degree of coherence of the community is needed and this workshop 
has shown that we may reach it.
 
As stated already in the foreword, 
the FCC-ee claims a dramatically improved experimental accuracy
compared with that of the LEP/SLC for practically all electroweak measurements.
 
In the following sections, we introduce the issues of importance from the perspective 
of  FCC-ee Tera-Z theoretical studies.

\section{Electroweak pseudo-observables (EWPOs) \label{ss-int-ewpos}}

The so-called electroweak pseudo-observables (EWPOs),
are quantities like the Z mass and width, the various  $2\to 2$ Z peak cross-sections,
and all kinds of  $2\to 2$ charge and spin asymmetries at the Z peak;  one may also add the equivalent effective electroweak mixing angles.
These are derived directly from experimental data, 
such that QED contributions and kinematic cut-off effects are removed.
For a more detailed definition of EWPOs, 
see Section~\ref{sunfold}.\ref{s-qed}.

For the Z width, the experimental error will go down to about $\pm 0.1\UMeV$, 
which is about 1/20 of the LEP/SLC accuracy. 
For the measurement of the effective electroweak mixing angles from asymmetries,
an improvement by a factor of up to about 50 is envisaged, 
see Table~\ref{tab:FCC-observables} in the foreword.

{\it
Such huge improvements will allow the FCC-ee Tera-Z stage to test the Standard Model 
at an unprecedented precision level.
The increment in precision corresponds to the increment that was represented by the LEP/SLC 
in their time; they tested the Standard Model at a  precision that 
needed `complete' one-loop corrections, plus leading higher-order terms.
The FCC-ee Tera-Z 
{\color{black} stage will need} `complete' two-loop corrections, plus leading higher-order terms.
Even without 
{\color{black} 
an explicit} reference to new physics, the  
FCC-ee Tera-Z stage lets us expect exciting, qualitatively new results.}

Consequently, the tremendous precision of the FCC-ee
will require serious efforts in the area of three-loop electroweak calculations.
We would like to mention that the two-loop  electroweak  calculations 
for Z physics were completed only recently
\cite{Dubovyk:2016aqv,Dubovyk:2018rlg}.
In the QCD sector of the Standard Model,
some additional four-loop calculations seem to be necessary as well.
Discussing the status and prospects of three-loop weak and four-loop QCD calculations 
is a main subject of Chapter \ref{sthstatus}, which elaborates and summarizes
why and how Standard Model calculations must improve; this subject was also discussed  recently on several 
occasions~\cite{Gluza:Jan2018,Jadach:April2018,Gluza:April2018}.

Very briefly, for Standard Model calculations of the Z boson width,
the complete EW two-loop and some leading partial QCD or mixed three-loop 
terms are known.
The current so-called `intrinsic' theoretical  error due to
uncontrolled higher-order terms is estimated to be
$\delta \Gamma_\mathrm{Z} \sim \pm 0.4$ MeV in \cite{Dubovyk:2018rlg}; 
see also Table \ref{tab2} in Chapter~\ref{sthstatus}. 
This value is below the experimental accuracy of the LEP, 
but it is larger than the anticipated accuracy of the FCC-ee Tera-Z stage. 
Here, a new round of calculations is indispensable.
For other quantities, the situation is similar, see Table {\ref{tabfin}}. 

The essential questions are: 
`How difficult is the calculation of EW three-loop and QCD four-loop contributions?' and
`Do we know how to do this?'
This issue is addressed in more details in Chapters \ref{sthstatus} and \ref{ch-sff}.

Let us also note the following aspect of higher-order Standard Model corrections for the FCC-ee.
At the LEP, it was a standard procedure that QED was extracted such that only the first and higher EW effects  remained 
in EWPOs. 
The elimination of QED from EWPOs was a natural task at LEP
because, since the 1970s, QED theory and higher-order techniques
were already fully established.
Hence, in the LEP data analysis physicists were interested only 
in  exploration of the QCD and EW effects. 
Thanks to the LHC, QCD is presently treated in high-energy studies similarly to QED.
It is quite likely that, in future FCC-ee Tera-Z searches for new physics, the
EW higher-order effects will be treated the same way,
namely as known, calculable effects to be removed from data.
But at least at the two-loop level!
Notably, particles like W, Z, and H bosons, as well as the top quark,
are considered to be heavy from the current perspective.
In future, they will be regarded as light particles compared with
the 20--$50\UTeV$ mass scales of the effective theories 
used to analyse FCC data.
Such a change of perspective  poses a non-trivial practical question 
for the future strategy of the data analysis:
How are we to treat EW higher-order corrections?
Should we keep them in the EWPOs or extract them, like QED effects? 
This kind of question is natural in the context of  Chapter \ref{sunfold}.
Generally, we propose to start from what was done at the LEP,
but keeping an eye on potential modifications, in particular on
possible new  definitions of EWPOs. 
Accordingly, in Chapter~\ref{sunfold}, the use of Laurent series around the Z pole and an S-matrix-inspired framework for $2\to 2$ scattering are discussed in more detail. It might also happen that a consistent description of real processes and the extraction of effective couplings at the FCC-ee Tera-Z stage will necessitate a change from analysing differential squared amplitudes to analysing spin amplitudes, merged properly with a Monte Carlo analysis. In this respect, a possible modification of the language with EWPOs into a language with EWPPs (EW `pseudo-parameters') is discussed in Section \ref{sunfold}.\ref{s-qed}.
See also further remarks connected with these issues in the QED section,
Section A.\ref{ss-int-qed}.

Another non-trivial issue is how to test  non-standard (BSM) physics. 
As discussed in Ref. \cite{Czakon:1999ha}, the structure of
higher-order corrections can be quite different when comparing the Standard Model with  extensions.
This is actually the case for all models with 
$\rho_{\rm tree} \equiv M_\mathrm{W}^2/(M_\mathrm{Z}^2\cos^2{\theta_\mathrm{W}}) \neq 1$.
In the Standard Model effective field theory (SMEFT) approach, EWPOs with dimension-6 operators are considered in the  `Warsaw' basis \cite{Grzadkowski:2010es} or in the `SILH' basis \cite{Giudice:2007fh,Contino:2013kra}. For some recent one-loop SMEFT analysis, see Ref.\ \cite{Bakshi:2018ics}. In Ref. \cite{Dawson:2018jlg},
 SMEFT corrections to Z boson decays are considered. For a current SMEFT global analysis, see, \eg Ref. \cite{deBlas:2017wmn}. It seems that the SMEFT framework is also the most practical method for parametrizing new physics at other FCC-ee stages, namely the FCC-ee-W, FCC-ee-H, and FCC-ee-tt stages (Table~\ref{tab:FCC-ee-runplan} in the foreword.) 
 
\section{QED issues \label{ss-int-qed}}

The development of a better control of the sizeable QED effects at the FCC-ee Tera-Z stage is vital.
The theoretical precision of QED calculations has to be better by 
a factor of 20--100 in comparison with the LEP era.
This is more than one perturbative order; hence,
not trivial. There are subtle problems, owing to high-order infrared singularities, 
many-particle final states in the real cross-sections, and higher-$n$-point functions.  
We need a theoretically well-justified, clear, and clean recipe
for disentangling the QED component of the Standard Model from the 
electroweak QCD part working at the two- to four-loop level.
This includes, for instance, massless and massive double-box diagrams with internal photons, as well as  
initial--final-state interference radiative effects; as is well-known, both are related.
The general answer is known in principle. 
But one has to define and implement an efficient methodology
of subtracting and resumming QED corrections due to universal, process-independent
soft and collinear parts of the perturbative series, known to infinite order,
while process-dependent  small non-soft and non-collinear remnants 
may remain together with the `pure' EW corrections.
Once this problem is fixed, 
practical methods of removing QED effects from data, the so-called `QED deconvolution',
at a much higher precision level
than at the LEP have also to be elaborated, especially for those observables measured near the Z peak,
where the boost of experimental precision will be biggest.

There are three groups of observables near the Z peak 
where the QED issues look different.
\begin{enumerate}
\item
Observables related to resonance phenomena: the Z line shape as a function of $s$, 
\ie the various total cross-sections at  the peak, 
as well as  the Z mass and total and partial decay widths and branching ratios.
\item
Charge and spin asymmetries  at the Z peak, related to angular distributions
of the final fermion pairs, including  wide-angle Bhabha scattering.
\item
Small-angle Bhabha scattering, photon pair production, 
the $Z$ radiative return above the Z peak,
WW, ZH, tt production and other processes with multiparticle final states over some range in $s$.
\end{enumerate}

For the first group, the so-called flux function approach was used in the  LEP analyses 
\cite{Bardin:1999yd}.
There is a chance that it may still be sufficiently precise at the FCC-ee.
In this approach, the integrated
cross-sections could be formulated with sufficient accuracy by using a
one-dimensional residual integration, describing the folding of 
the  hard scattering kernels due to weak interactions (or effective Born terms) with flux functions representing the
loss of centre-of-mass energy due to single initial- 
(or final-) state photon emission plus exponentiated soft photon emission.
The Z resonance is, mathematically seen, 
a Laurent series in the centre-of-mass energy  squared $s$.
This fact could be sufficiently well described as a Breit--Wigner resonance, 
interfering with the `background' of photon exchange. 
It is known how to include weak loop effects in this approach.
Accordingly,  EWPOs
were defined in the LEP data analysis and related to the hard kernels  due to weak interactions
(\ie the effective Born cross-sections with effective coupling constants)
with relatively simple relations,
practically neglecting any factorization problems or imaginary parts.
This methodology worked well, owing to the limited precision of LEP data. 
The non-factorizable QED corrections (initial--final-state interferences) 
and other effects, such as those due to imaginary parts of the  effective  $Z$ couplings
could be neglected. 
This was numerically controlled with tools like ZFITTER and TOPAZ0.
It is not proved yet that a similar approach will work at FCC-ee Tera-Z precision
for the Z line-shape-related EWPOs, but a chance exists and should be explored.
The gain would be a relatively simple and fast analysis methodology. 

For the second group of observables, such as the leptonic charge asymmetry
and the tau spin asymmetry, the QED issue is much more serious.
For instance, the muon charge asymmetry will be measured 
${\sim}50$ times more precisely than at the LEP, 
see Table \ref{tab:FCC-observables} in the foreword.
The non-factorizable initial--final-state QED interference (IFI),
which was about 0.1\% and could simply be  neglected at the LEP,
will have to be calculated at the FCC-ee Tera-Z stage with a two-digit precision 
and then explicitly removed from the data.
Moreover, there is, at LEP accuracy, a set of simple formulae for the IFI 
with a one-dimensional convolution over some flux functions
and hard effective Born terms at the next-to-leading-order. 
This is implemented in ZFITTER \cite{Burgers:1985qg}.
There is no such simple formula beyond the next-to-leading-order.
The well-known formula of this kind for the charge asymmetry, 
based on soft photon approximation, involves a four-dimensional convolution. 
In addition, IFI-type corrections become mixed up with electroweak corrections 
for the FCC-ee Tera-Z stage at the three- or four-loop level.
Fortunately, the methodology of disentangling pure QED corrections from
pure electroweak corrections at the amplitude level, summing up soft photon
effects to infinite order (exponentiation) and adding QED collinear non-soft
corrections order by order (independently of the EW part) is well-known and
numerically implemented in the {\tt KKMC} program \cite{Jadach:1999vf}.
This program includes, so far, QED non-soft corrections to second-order and pure EW corrections
up to first-order (with some second-order EW improvements, QCD,
etc.) 
using the weak library DIZET \cite{Bardin:1989tq} of ZFITTER \cite{Bardin:1999yd}.
The calculational scheme of {\tt KKMC} can be extended to two to four loops in a natural way.
The interrelations of hard kernels  due to weak interactions and QED folding has long been understood
at the level of sophistication  needed for FCC-ee studies.
A correct treatment of the Z resonance as a Laurent series in the hard kernel,
namely, using the S-matrix approach \cite{Leike:1991pq,Riemann:1992gv},
which was formulated  in the 1990s,
fits very well in this scheme.
It is basically clear how to do this in principle and in practice,
also going beyond the flux function approach.
All of this is described in several sections  in Chapter~\ref{sunfold}.

All this describes a scenario in which the QED and EW parts are separated
in a systematic, clean manner at the amplitude level, and where the hard kernels  due to weak interactions
 encapsulate all two- to four-loop EW or QCD corrections.
However, in the construction of EWPOs at the LEP, the  hard kernel was replaced
by effective Born cross-sections (\ie squared amplitudes) with effective couplings, which, on the one hand, were fit to the data
and, on the other hand, could be compared with the best knowledge of Standard Model predictions. The muon charge asymmetry  without QED effects was, with sufficient precision, simply a one-to-one representation
of the effective couplings of the Z boson,
neglecting $s$-channel photon exchange, a non-factorizing component, and all QED effects.
It is not to be excluded that a similar method might work at FCC-ee precision.
The important difference is that, instead of the primitive flux method, a 
Monte Carlo approach with sophisticated matrix elements
would take care of all QED effects, including non-factorizable parts.

Note that the inclusion of loop corrections to the hard kernel 
and its formal simplification in terms of an effective Born
cross-section will generate a need for the calculation of additional, 
more complicated contributions, such as additional massive two-loop  box terms. 
Some general formulations in this direction at the amplitude level 
are given in Chapter~\ref{sunfold} for the effective Born
cross-section. 
This approach is elaborated in Section~\ref{sunfold}.\ref{s-qed}; 
see also Section~\ref{sunfold}.\ref{s-smatrix}.

Finally, a few remarks on the QED effects in the third group of observables,
which include luminosity measurements
using small-angle Bhabha scattering,  photon pair production,  and, for neutrino counting, the radiative return above the Z peak. 
They are so strongly dependent on the experimental event selection and the cut-offs
that the only way to take them into account 
is the direct comparison of experimental data
with the results of the Monte Carlo programs with sophisticated QED matrix elements.
These QED matrix elements must also  cover the relevant weak effects.

Having all this in mind, we try to describe how a theoretical treatment of 
the measurements of the Z peak parameters might be  formulated for the 
FCC-ee Tera-Z 
stage or for similar projects.
We will not answer all immediate questions and  will not work out all ideas completely, 
nor will we be able to perform the  necessary detailed numerical calculations.
Here, the QED expertise of the Krak\'ow group 
and the formalism of the SMATRIX language worked out in Zeuthen with the support of other groups should come 
together in the exploration of unsolved problems. 

These issues are defined in Chapter~\ref{sunfold}. 
 If needed, they may be treated in more detail.
However, owing to the amount of necessary resources and research work, 
such a project definition would not be 
unconditional, concerning any kind of support. 

The JINR/Dubna SANC/ZFITTER group has expressed interest 
in cooperating to create a new tool, like SANC/ZFITTER/SMATASY. 
Together with the \babayaga\ and \bhlumi\ groups, they fittingly close  Chapter 
\ref{sunfold}, mainly treating  luminosity problems in future colliders.

Certainly, over a larger time-scale, 
it would be a highly welcome situation
if other independent groups would form, in order to start  work on these issues. 

\section{Methods and tools}

Our general conclusion from the discussions during and after the workshop
is that the techniques and software available today 
would not be sufficient for an appropriate FCC-ee Tera-Z data analysis.
The issues of methods, techniques, and tools for the calculation of Feynman integrals 
and higher-order loop effects are discussed in Chapter~\ref{chmt}.
Moreover, it is quite probable that approaches that were developed for higher-loop effects 
in other areas of research, and  are not discussed here, 
can  also be used in future for the calculation of EWPOs of the FCC-ee Tera-Z
stage. 
Let us mention only the case of the decay $B \to X_s 
\gamma$ \cite{Misiak:2010sk,Misiak:2017woa}, 
where the Z propagator with unitary cut at the four-loop level is 
equivalent to three-loop EWPOs of the Z boson decay studies. 

Chapter \ref{chmt}, on methods and tools, includes descriptions of both analytical 
and numerical methods for the calculation of higher-order corrections. 
Five contributions -- Sections 
\ref{chmt}.\ref{subs:mbn}, \ref{chmt}.\ref{contr:mbambre}, \ref{chmt}.\ref{contr:mprausa}, 
\ref{chmt}.\ref{sec-1loop},
and \ref{chmt}.\ref{sec:mbth} -- 
deal completely or partly with the Mellin--Barnes (\mbr) method. 
In one contribution, Section \ref{chmt}.\ref{sec:secdec}, 
the purely numerical sector decomposition (\sd) method is described. 
Both methods are used heavily in current studies and are  thought to be crucial for FCC-ee Tera-Z calculations.

The reasons for a preference of the two numerical methods mentioned, {\mbr}
 and \sd, are twofold. First,
integrals depend on $M_\mathrm{Z}, M_\mathrm{W}, M_\mathrm{H}, m_\mathrm{t}$ plus $s$ for vertices -- and, for box integrals, additionally on $t$ , \ie 
on up to four or five dimensionless ratios of the parameters.
We have no analytical tools to cover that.
Further, the integrals contain infrared singularities.
The {\mbr} and the {\sd} methods are the only known numerical methods with algorithms to deal with these singularities
systematically at all loop orders. 

\textit{There is a consensus in the community that, to achieve the goals of precision,
it will be most crucial to have the numerical integrations efficiently implemented
for Feynman integrals in  Minkowskian kinematics.}
 
This is discussed in Sections \ref{chmt}.\ref{subs:mbn} and \ref{chmt}.\ref{sec:secdec}.

Sections~\ref{chmt}.\ref{sec:rlee} and ~\ref{chmt}.\ref{sec:costas} deal 
with different approaches to differential equations, 
including a discussion of cut Feynman integrals. In Section \ref{chmt}.\ref{sec:sweinz},
first steps are discussed towards solutions for multiscale, multiloop Feynman integrals. 
Special functions are introduced, which go into the topics of elliptic functions.
Some sections deal with  still-exploratory  ideas. 
For example, {\mbr} thimbles are discussed  in Section \ref{chmt}.\ref{sec:mbth}.
A new, low-dimensional, numerically efficient approach to {\mbr} representations at the one-loop level 
 is introduced in Section \ref{chmt}.\ref{sec-1loop}, 
which might also be generalized to multiloop cases. 
In Sections \ref{chmt}.\ref{sec:rp} and \ref{chmt}.\ref{sec:gr},  direct numerical 
calculations of Feynman integrals in $d=4$ are explored.  
Further, to achieve the goals of precision, 
we are also interested  in methods and tools used to calculate extensions of the Standard Model. 
In fact, extensions of the Standard Model 
are, in general, more complex in structure. 
A representative example is studied in Section~\ref{chmt}.\ref{contr:rboels}.
We cannot exclude the possibility that some of the methods covered here will become   
standard or complementary tools in precision calculations in future.  

Finally, in Section~\ref{chmt}.\ref{sec:hcuba}, 
the \textsc{Cuba} library for numerical integrations is described, exhibiting some features 
of importance for multidimensional integral calculations.


\clearpage  \pagestyle{empty} 
\cleardoublepage
\chapter
[Theory status of Z boson physics
\\
{\it I. Dubovyk, A.M. Freitas, J. Gluza, K. Grzanka, S. Jadach, T. Riemann, J. Usovitsch}]
{Theory status of Z boson physics}
\label{sthstatus}


\pagestyle{fancy}
\fancyhead[LO]{}
\fancyhead[CO]{\thechapter
        \hspace{1mm}
        Theory status  of Z boson physics
}

\fancyhead[CE]{I. Dubovyk, A.M. Freitas, J. Gluza, K. Grzanka, S. Jadach, T. Riemann, J. Usovitsch}
\fancyhead[RE]{} 

\noindent
{\bf Authors: 
Ievgen Dubovyk, Ayres M. Freitas, Janusz Gluza, Krzysztof Grzanka, Stanis\l aw Jadach, Tord Riemann, Johann Usovitsch}  
\\
Corresponding author: Ayres M. Freitas [afreitas@pitt.edu]
\vspace*{.5cm}

\noindent The number of Z bosons collected at the LEP, approximately 17 million in total,
made it possible to determine a large amount of electroweak observables with very 
high precision through measurements of the Z line shape and of cross-section asymmetries, combined with high-precision parity-violating asymmetries measured at 
the SLC  \cite{ALEPH:2005ab}.
 
These measurements are typically expressed through the  cross-section $\mathrm{e}^+\mathrm{e}^- \to \mathrm{f}\bar{\mathrm{f}}$ at the
Z pole, $\sigma^0_\mathrm{f} \equiv \sigma_\mathrm{f} (s= M_\mathrm{Z}^2)$, for different final states $\mathrm{f}\bar{\mathrm{f}}$, the total width of the Z boson, $\GZ$, determined from the shape of $\sigma_\mathrm{f}(s)$, and branching ratios of various final states:
\begin{align}
\sigma^0_{\rm had} &= \sigma[\mathrm{e}^+\mathrm{e}^- \to 
\text{hadrons}]_{s=M_\mathrm{Z}^2};  \label{eqsig}\\
\Gamma_\mathrm{Z} &= \sum_\mathrm{f} \Gamma[\mathrm{Z} \to \mathrm{f}\bar{\mathrm{f}}]; \\
R_\ell &= \frac{\Gamma[\mathrm{Z} \to \text{hadrons}]}{\Gamma[\mathrm{Z} \to \ell^+\ell^-]}, \quad
 \ell = \mathrm{e},\mu,\tau;\\
R_\mathrm{q} &= \frac{\Gamma[\mathrm{Z} \to \mathrm{q}\bar{\mathrm{q}}]}{\Gamma[\mathrm{Z} \to \text{hadrons}]},\quad
 \mathrm{q} = \mathrm{u},\mathrm{d},\mathrm{s},\mathrm{c},\mathrm{b}. \label{ewpos}
\end{align}
In the definition of these quantities, contributions from $s$-channel photon exchange, virtual box 
contributions, and 
initial-state as well as initial--final-state interference QED radiation are understood to be already 
subtracted; see, \eg 
Refs.~\cite{ALEPH:2005ab,Riemann:Jan2018}.

{\it The precise calculation of the terms to be subtracted, at variable centre-of-mass energy $\sqrt{s}$ around the \textup{Z} peak, will be a substantial part of the theoretical analysis for the FCC-ee Tera-\textup{Z} stage.
Further, for a determination of $M_\mathrm{Z}$ and $\Gamma_\mathrm{Z}$, we will have to confront  cross-section data and predictions around the Z peak position as part of the analysis.
}

Correspondingly,
Chapter \ref{sunfold} of this report contains  
an updated discussion of QED unfolding in the context of the demanding FCC-ee needs. 
To clarify this fact, the parameters of Eqs.\ (\ref{eqsig})--(\ref{ewpos}) 
have become known as the so-called electroweak pseudo-observables (EWPOs), rather than true observables.
However,  Eqs.\  (\ref{eqsig})--(\ref{ewpos}) still include the effect of final-state QED and QCD radiation. Fortunately, the final-state radiation effects factorize from the massive electroweak corrections almost perfectly;  
see, \eg Refs.~\cite{Arbuzov:2005ma,Chetyrkin:1994jsFred, Chetyrkin:1994jsTom}. Therefore, it is possible to compute the latter, as well as 
potential contributions from new physics, without worrying about 
effects from soft and collinear real radiation. 

The remaining basic pseudo-observables are cross-section asymmetries, 
measured at the Z pole. The forward--backward
asymmetry is defined as
\begin{equation}
A^\mathrm{f}_{\rm FB} = \frac{\sigma_\mathrm{f} \left[\theta<\frac{\pi}{2}\right]-
 \sigma_\mathrm{f} \left[ \theta>\frac{\pi}{2} \right]}{\sigma_\mathrm{f} \left[ \theta<\frac{\pi}{2} \right] +
 \sigma_\mathrm{f} \left[ \theta>\frac{\pi}{2} \right] }, \label{eq:afb}
\end{equation}
where $\theta$ is the scattering angle between the incoming e$^-$ and the
outgoing f. It can be  approximately written as a product of two terms (for more precise discussion, see Section \ref{sunfold}.\ref{s-smat-22MEasy}):
\begin{equation}
A^\mathrm{f}_{\rm FB} = \frac{3}{4}{\cal A}_\mathrm{e}{\cal A}_\mathrm{f}, 
\end{equation}
with
\begin{equation}
{\cal A}_\mathrm{f} = \frac{1-4|Q_\mathrm{f}|\seff{f}}{1-4|Q_\mathrm{f}|\seff{f}+8 (Q_\mathrm{f}\seff{f})^2}.
\end{equation}
The $\seff{f}$ is called the effective weak mixing angle, which contains the 
net contributions from all the radiative corrections. The most precise measurements of $A^\mathrm{f}_{\rm FB}$ have been obtained for leptonic and bottom quark final states ($\rm f=\ell,b$).
In the presence of
polarized electron beams, one can also measure the parity-violating left--right asymmetry:
\begin{equation}
A^\mathrm{f}_{\rm LR} = \frac{\sigma_\mathrm{f} \left[ P_\mathrm{e}<0 \right]-
 \sigma_\mathrm{f} \left[ P_\mathrm{e}>0 \right] }{\sigma_\mathrm{f} \left[ P_\mathrm{e}<0 \right] +
 \sigma_\mathrm{f} \left[ P_\mathrm{e}>0 \right] }={\cal A}_\mathrm{e}|P_\mathrm{e}|.
\end{equation}
Here, $P_\mathrm{e}$ denotes the polarization degree of the incident electrons, where $P_\mathrm{e}<0$ and $P_\mathrm{e}>0$
refer to left-handed and right-handed polarizations, respectively. Since 
$A^\mathrm{f}_{\rm FB}$  and $A^\mathrm{f}_{\rm LR}$ are defined as normalized asymmetries, they do not depend on (parity-conserving) 
initial- and final-state QED and QCD radiation 
effects.\footnote{Here, it is assumed that any issues related to the determination of the experimental acceptance have 
been evaluated and unfolded using Monte Carlo methods.} 

The present and predicted future experimental values for the most relevant EWPOs are given in  \Tref{tab:FCC-ee-runplan} in the foreword. In the following, we will compare these numbers with the current theoretical situation and with estimates for future precision calculations. In this context, a discussion of theoretical errors connected with these calculations is crucial.

Table \ref{THtab1} shows the FCC-ee experimental goals for the basic EWPOs. As is evident from the table, the theoretical intrinsic uncertainties of the current results (TH1) are safely below the current experimental errors (EXP1). However, they are not sufficiently small, in view of the FCC-ee experimental precision targets (EXP2).

\begin{table}
\renewcommand{\arraystretch}{1.2}
\caption{
Current total experimental errors (EXP1) and, estimated in 2014 
\cite{Awramik:2006ar,Awramik:2008gi,Freitas:2014hra}, theoretical intrinsic errors (TH1) for selected EW observables, as well as   corresponding error estimates for the FCC-ee Z resonance mode  
(EXP2), see foreword.}
\label{THtab1}
\centering
\begin{tabular}{lllllll}
\hline \hline
 & $\delta \Gamma_\mathrm{Z}\;(\rm MeV)$ &  $\delta R_\ell \; (10^{-4})$ & $\delta R_\mathrm{b}\; (10^{-5})$ & 
 $\delta \seff{ell}\; (10^{-6})$ & 
 $\delta \seff{b}\; (10^{-5})$  
 \\ \hline
 \multicolumn{6}{l}{Current EWPO errors} 
\\
EXP1 \cite{ALEPH:2005ab}& 2.3 & $250$ & $66$ & $160$ 
& 1600 
\\
 TH1 \cite{Awramik:2006ar,Awramik:2008gi,Freitas:2014hra} 
  & 0.5  &  \phantom{1}50   & $15$ & $\phantom{1}45$ & $\phantom{160}5$ 
  \\  
\multicolumn{6}{l}{FCC-ee-Z EWPO error estimates}  
\\
EXP2 \cite{Alain:Jan2018} \& Table \ref{tab:FCC-observables}   & 0.1 & \phantom{1}10 &  $2\div 6$  & $\phantom{61}6$ & \phantom{16}70  
\\
\hline \hline
\end{tabular}
\end{table}

This situation, as seen from the perspective of 2014, underlines the goals and strategic plan for improvements in the theoretical calculation of radiative Standard Model corrections defined here. Historically, the complete one-loop corrections to the Z pole EWPOs were reported for the first time in 
Ref.~\cite{Akhundov:1985fc}. Over the next 32 years, many groups, using many methods, determined partial 
two- and three-loop corrections to EWPOs. A more detailed list of the relevant types of radiative corrections will be given later.

In the last 2\,years, as discussed in Ref.~\cite{Gluza:Jan2018}, substantial progress in numerical calculations of multiloop and multiscale Feynman integrals was made and the calculation of the last set of two-loop corrections, of order ${\cal{O}}(\alpha_\mathrm{bos}^2)$, to all Z pole EWPOs \cite{Dubovyk:2016aqv,Dubovyk:2018rlg} became possible. Here `bos' denotes diagrams without closed fermion loops. 

All the numerical results discussed next are based on the input parameters gathered in Table \ref{THtab:input}.

\begin{table}
\renewcommand{\arraystretch}{1.2}
\caption{
Input parameters used in the numerical analysis \cite{Patrignani:2016xqp,Steinhauser:1998rq-new,Davier:2010nc}
\label{THtab:input}}
\centering
\begin{tabular}{ll}
\hline \hline
Parameter & Value   \\
\hline
$\MZ$ & 91.1876 GeV  \\
$\Gamma_\mathrm{Z}$ & 2.4952 GeV  \\
$\MW$ & 80.385 GeV \\
$\Gamma_\mathrm{W}$ & 2.085 GeV  \\
$\MH$ & 125.1 GeV  \\
$\mt$ & 173.2 GeV  \\
 $m_{\rm b}^{\overline{\rm MS}}$ & 4.20 GeV \\
 $m_{\rm c}^{\overline{\rm MS}}$ & 1.275 GeV \\
 $m_\tau$ & 1.777 GeV \\
  $m_\mathrm{e},m_\mu,m_\mathrm{u},m_\mathrm{d},m_\mathrm{s}$ & 0 \\
 $\Delta\alpha$ & 0.05900 \\
 $\as(\MZ)$ & 0.1184 \\
 $G_\mu$ & $1.16638 \times 10^{-5}\UGeV^{-2}$ \\
\hline \hline
\end{tabular}
\end{table}

As a concrete example, let us discuss the different higher-order contributions to the Standard Model prediction for the bottom quark effective weak mixing in more detail. It can be written as
\begin{equation}
\seff{b} = \left (1-\frac{{\MW}^2}{{\MZ}^2} \right )(1+\Delta\kappa_{\Pb}),
\end{equation}
where $\Delta\kappa_{\Pb}$ contains the contributions from radiative corrections.
Numerical results from loop corrections of different orders are shown in
  \Tref{THtab:orders}. 
Altogether, the corrections included in \Tref{THtab:orders} are: electroweak $\OO(\alpha)$ \cite{Akhundov:1985fc} and 
  fermionic $\alpha^2_{\rm ferm}$ 
\cite{Barbieri:1992nz,Barbieri:1992dq,Fleischer:1993ub,Fleischer:1994cb,Awramik:2008gi} and bosonic $\alpha^2_{\rm bos}$ 
\cite{Dubovyk:2016aqv} EW $\OO(\alpha^2)$  contributions; $\OO(\alpha\as)$ corrections to internal gauge boson self-energies
\cite{Djouadi:1987gn,Djouadi:1987di,Kniehl:1989yc,Kniehl:1991gu,Djouadi:1993ss};
 leading three- and four-loop corrections in the large-$\mt$ limit, of
orders $\OO(\alpha_\Pt\as^2)$ \cite{Avdeev:1994db,Chetyrkin:1995ix}, $\OO(\alpha_\Pt^2\as)$, $\OO(\alpha_\Pt^3)$ \cite{vanderBij:2000cg,Faisst:2003px},
and $\OO(\alpha_\Pt\as^3)$ \cite{Schroder:2005db,Chetyrkin:2006bj,Boughezal:2006xk},
where $\alpha_\Pt \equiv \alpha {(\mt^2)}$; and non-factorizable vertex contributions $\OO(\alpha\as)$ \cite{Czarnecki:1996ei,Harlander:1997zb,Fleischer:1992fq,Buchalla:1992zm,Degrassi:1993ij,Chetyrkin:1993jp},
which account for the fact that the factorization between 
virtual EW corrections  and final-state radiation effects  is not exact.

\begin{table}
\renewcommand{\arraystretch}{1.2}
\caption{Comparison of different kinds of radiative correction to
$\Delta\kappa_{\Pb}$ \cite{Dubovyk:2016aqv}, using the input parameters in Table~\ref{THtab:input}. Here, $\at = y_\Pt^2/(4\pi)$, where $y_\Pt$ is the top Yukawa coupling.
\label{THtab:orders}}
\centering
\begin{tabular}{ll}
\hline \hline
Order & Value ($10^{-4}$) \\
\hline
$\alpha$ & \phantom{$-$}468.945 \\
$\alpha\as$ & \phantom{$4$}$-42.655$ \\
$\at\as^2$ & \phantom{$44$}$-7.074$ \\
$\at\as^3$ & \phantom{$44$}$-1.196$ \\
$\at^2\as$ & \phantom{$-44$}1.362 \\
$\at^3$ & \phantom{$-44$}0.123 \\
$\alpha^2_{\rm ferm}$ & \phantom{$-44$}3.866 \\
$\alpha^2_{\rm bos}$ & \phantom{$44$}$-0.986$ \\
\hline \hline
\end{tabular}
\end{table}

The most recently determined correction, the ${\cal O}(\alpha^2_{\rm bos})$ electroweak 
two-loop correction, amounts to 
$
\Delta\kappa_{\Pb}^{(\alpha^2,\rm bos)} = -0.9855 \times 10^{-4},
$
which is comparable in magnitude to the fermionic corrections. 
Taking into account this new result, an updated error estimate due to missing higher-order terms will 
be discussed later on, see \Tref{THtab:orders}. 

Table \ref{THtab:res1} summarizes the known contributions to Z boson production and decay vertices, order by order. The technically challenging  \emph{bosonic} two-loop calculation was completed very recently~\cite{Dubovyk:2018rlg}.
This result has been achieved   through a combination of different methods: (a) numerical integration of Mellin--Barnes 
(\mbr) representations with contour rotations and contour shifts, for a substantial improvement of the convergence; (b) 
sector decomposition (\sd) with
numerical integration over Feynman parameters; and (c) dispersion relations for subloop insertions. The \mbr{} and \sd{} 
methods were discussed intensively at the workshop \cite{Dubovyk:Jan2018,Borowka:Jan2018}; see Chapter~\ref{chmt} for details.

\begin{table}
\renewcommand{\arraystretch}{1.2}
\caption{Loop contributions to the partial and total Z widths with
fixed $\MW$ as input parameter. Here $N_\mathrm{f}$ and $N_\mathrm{f}^2$ refer to corrections with 
one and two closed fermion loops, respectively, whereas $\alpha^2_{\rm bos}$
denotes contributions without closed fermion loops. Furthermore, $\at = y_\Pt^2/(4\pi)$, where $y_\Pt$ is the top Yukawa coupling. Table taken from Ref.~\cite{Dubovyk:2018rlg} (Creative Commons Attribution Licence, CC BY).
\label{THtab:res1}}
\centering
\begin{tabular}{lllllll}
\hline \hline
\multicolumn{1}{l}{$\Gamma_i$ (MeV)} & $\Gamma_\mathrm{e}\;\;$ & $\Gamma_\nu\;\;$ & $\Gamma_\mathrm{d}\;\;$ & $\Gamma_\mathrm{u}\;\;$ & 
 $\Gamma_\mathrm{b}\;\;$ & $\Gamma_\mathrm{Z}\;\;$ \\
\hline 
$\OO(\alpha)$ & 2.273 & 6.174 & 9.717 & 5.799 & 3.857 & 60.22 \\
$\OO(\alpha\as)$ & 0.288 & 0.458 & 1.276 & 1.156 & 2.006 & \phantom{6}9.11 \\
$\OO(\at\as^2,\,\at\as^3,\,\at^2\as,\,\at^3)$ &
 0.038 & 0.059 & 0.191 & 0.170 & 0.190 & \phantom{6}1.20 \\
$\OO(N_\mathrm{f}^2\alpha^2)$ & 0.244 & 0.416 & 0.698 & 0.528 & 0.694 & \phantom{6}5.13 \\
$\OO(N_\mathrm{f}\alpha^2)$ & 0.120 & 0.185 & 0.493 & 0.494 & 0.144 & \phantom{6}3.04 \\
$\OO(\alpha^2_{\rm bos})$ & 
                    0.017 & 0.019 & 0.059 & 0.058 & 0.167 & \phantom{6}0.51 \\
\hline \hline
\end{tabular}
\end{table}

As is evident from \Tref{THtab:res1}, the two-loop electroweak corrections to the Z boson partial decay widths are sizeable, of the same order as the $\OO(\alpha\as)$ terms. The bosonic corrections $\OO(\alpha^2_{\rm bos})$ are smaller than the fermionic ones, but larger than previously estimated \cite{Freitas:2014hra}. This demonstrates that theoretical error evaluations are always to be taken with a grain of salt. 

For the total width $\Gamma_{\PZ}$, the corrections are also significantly larger than the projected future experimental error (EXP2) given in \Tref{THtab1}.

These numerical examples demonstrate that radiative electroweak corrections beyond the two-loop level must be calculated for future high-luminosity $\mathrm{e}^+\mathrm{e}^-$ experiments. 
In Table \ref{THtab:res1}, corrections are calculated using $M_\mathrm{W}$ as an 
input. By calculating $M_\mathrm{W}$ obtained from $G_\mu$, we get a value of $0.34\UMeV$ for $\OO(\alpha^2_{\rm bos})$  instead of $0.51\UMeV$  \cite{Dubovyk:2018rlg}.

Let us discuss the impact of radiative corrections in more detail by estimating their potential values. 

On the one hand, a source of uncertainty for the Standard Model prediction for any EWPO is the dependence on input parameters, as listed in \Tref{THtab:input}. The impact of input parameters is best evaluated through a global fit, as shown, \eg in 
Refs.~\cite{Patrignani:2016xqp,deBlas:2017wmn}. On the other hand, a separate source of uncertainty 
is the missing knowledge of theoretical higher-order corrections.

To estimate
the latter, one can take different approaches, each of which has its own advantages and 
disadvantages \cite{Freitas:2016sty}.
\begin{enumerate}
\item 
Determination of relevant prefactors of a class of higher-order corrections, such as 
couplings, group factors, particle multiplicities,
mass ratios, \etc, and assuming the remainder of the loop
amplitude to be order \order{1}.  
\item
Extrapolation under the assumption that higher-order radiative corrections can be approximated by a geometric series.
\item
Testing the scale-dependence of a given fixed-order result obtained using the  $\overline{\text{MS}}$ renormalization scheme, in order to estimate the size of the missing higher orders; this is used more often in QCD. 
\item
Comparing results obtained using the on-shell
and $\overline{\text{MS}}$ schemes, where the differences are of the next order in the perturbative expansion.
\end{enumerate}
In \Tref{tab2}, the intrinsic errors are shown for the Z boson decay width. Numerical estimates 
that are mainly based on the geometric series extrapolation, but corroborated by some of the other methods, are denoted TH1. In Ref. 
\cite{Dubovyk:2018rlg}  the $ {\alpha_\mathrm{bos}^2}$ contribution is given as $+0.505\UMeV$ with a net numerical precision of about four digits, which eliminates the uncertainty associated with that term completely.   It also shifts some of the geometric series extrapolations, such as
\begin{eqnarray}
\OO(\alpha^3)-\OO(\at^3) &\sim &
 \frac{\OO(\alpha^2)-\OO(\at^2)}{\OO(\alpha)} 
 \OO(\alpha^2) \sim  0.2\; \rm{MeV},  \label{est1}
\end{eqnarray}
where the full $\order{\alpha^2}$ term was previously not available.
The new error estimate, TH1-new, is $\pm0.4\UMeV$. As we can see, 
the estimated theoretical error is still much larger than that needed for the 
projected EXP2 goals in Table~\ref{THtab1}, which is for the Z boson decay width  $\lesssim 
\pm 0.1\UMeV$. The dominant remaining uncertainty stems from unknown three-loop contributions with 
either QCD loops, $\order{\alpha\as^2}$ and $\order{\alpha^2\as}$, or { electroweak} fermionic 
loops, $\order{N_\mathrm{f}^2\alpha^3}$, where $N_\mathrm{f}^2$ refers to diagrams with at least two closed fermion loops.

\begin{table}
\renewcommand{\arraystretch}{1.2}
\caption{Intrinsic theoretical error estimates (TH1) for $\Gamma_\mathrm{Z}$
 \cite{Freitas:2014hra,Freitas:2016sty}, updates taking into account the newly completed $\order{\alpha^2_{\rm bos})}$ corrections ({TH1-new}) \cite{Dubovyk:2018rlg}
and a projection into the future, assuming $\delta_{2,3}$ and the fermionic parts of $\delta_1$ to be known (TH2).
\label{tab2}}
\centering
\begin{tabular}{lllllll}
\hline \hline 
 & $\delta_1 $& $\delta_2\; $ & $\delta_3\; $ &
$\delta_4\;  $ &
$\delta_5\;  $ & $\delta \Gamma_Z$ (MeV)  \\
 & $ {\cal{O}}(\alpha^3)$& $ {\cal{O}}(\alpha^2 \as)$ & $  {\cal{O}}(\alpha \as^2)$ &
$ {\cal{O}}(\alpha \as^3)$ &
$ {\cal{O}}(\alpha_\mathrm{bos}^2)$ & $  =\sqrt{\sum_{i=1}^5 \delta_i^2}$ \\
\hline
 \multicolumn{7}{l}{TH1 (estimated error limits from geometric series of perturbation)}\\
&  0.26 & 0.3 & 0.23 & 0.035 & 0.1 & 0.5 \\
 \multicolumn{7}{l}{TH1-new (estimated error limits from geometric series of perturbation)}\\
& 0.2 & 0.21 & 0.23 & 0.035 & $<10^{-4}$ & 0.4 \\
\hline \hline
& $\delta'_1 $& $\delta'_2\; $ & $\delta'_3\;$ & $\delta_4\;$ & & $\delta \Gamma_\mathrm{Z}$ (MeV) \\
& $ {\cal{O}}(N_\mathrm{f}^{\leq 1}\alpha^3)$& ${\cal{O}}(\alpha^3 \as)$ & ${\cal{O}}(\alpha^2 \as^2)$ & ${\cal{O}}(\alpha \as^3)$ & & $= \sqrt{\delta_1^{\prime 2}+\delta_2^{\prime 2}+\delta_2^{\prime 3}+\delta_4^2}$ \\
 \hline
  \multicolumn{7}{l}{TH2 (extrapolation through prefactor scaling)}\\
 
&  0.04 & 0.1 & 0.1 & 0.035 & $10^{-4}$ & 0.15 \\
  \hline \hline
\end{tabular}
\end{table}

Once these corrections become available, with a robust intrinsic numerical precision of at least two digits, the remaining theoretical error will become dominated by missing four-loop terms. Estimating these future errors is rather unreliable at this time using geometric series of perturbation, since two orders of extrapolation are required. Nevertheless, a rough guess can be obtained by using the following experience-based scaling relations: each order of $N_\mathrm{f}\alpha$ and $\alpha_{\rm bos}$ generate corrections of about 0.1 and 0.01, respectively, and
$n$ orders of $\as$ produce a correction of roughly $n! \times (0.1)^n$, where the $n!$ factor accounts for the combinatorics of the SU(3) algebra.
In this fashion, one arrives at the TH2 scenario in \Tref{tab2}.\footnote{Accounting for `everything else' besides the specific orders listed in \Tref{tab2}, one may assign a more 
conservative future theoretical error estimate of $\delta\Gamma_\mathrm{Z} \sim 0.2\UMeV$; see 
also Ref.~\cite{Freitas:2016sty}.}

For a safe interpretation of FCC-ee-Z measurements,
the theoretical error must be subdominant relative to the experimental 
uncertainties. Comparing the TH2 scenario with the EXP2 numbers, one can see that it does not yet fit this bill. 
This 
implies that calculation of four-loop corrections, or at least the leading parts thereof, will be necessary to fully match 
the planned precision of the FCC-ee experiments. Since estimates of future theoretical errors are highly uncertain, and 
four-loop contributions are two orders beyond the current state of the art, we do not attempt to make a quantitative 
estimate of the achievable precision, but it seems plausible that the remaining uncertainty will be well below the 
EXP2 targets.
 
Let us now come back to
the prospects for computing the missing three-loop contributions. Two 
basic factors  play a role: the number of Feynman diagrams (or, correspondingly, the number of Feynman integrals) 
and the precision with which single Feynman integrals can be calculated. Some basic bookkeeping concerning the number 
of diagram topologies and different types of diagrams is given in \Tref{tabKG}. First, let us compare the known 
number of diagram topologies and individual diagrams at two  and 
 three loops. Comparing the genuine three-loop fermionic diagrams, which are 
simpler than the bosonic ones, with the already known two-loop bosonic diagrams, there is about an order of magnitude 
difference in their number: 17\,580 diagrams for $\mathrm{Z} \to \mathrm{bb}$ (and 13\,104 diagrams for $\mathrm{Z} \to \mathrm{e}^+\mathrm{e}^-$) at 
$\order{\alpha^3_{\rm 
ferm}}$ versus 964 (and 766) diagrams at $\order{\alpha^2_{\rm bos}}$. In general, however, the number 
of diagrams is, of course, not equivalent to the number of integrals to be calculated. At $\order{\alpha^3_{\rm ferm}}$, 
we expect $\order{10^3} - \order{10^4}$ distinct three-loop Feynman integrals before a reduction to a basis, because 
different classes of diagrams often share parts of their integral bases.

\begingroup
\renewcommand{\arraystretch}{1.1}
\begin{table}
\caption{Number of topologies and diagrams  for $\mathrm{Z} \to \mathrm{f} \bar{\mathrm{f}}$ decays in the Feynman gauge. Statistics for planarity, QCD, and EW-type diagrams are also given.  
Label `A' denotes statistics after elimination of tadpoles and wavefunction corrections, and label `B' denotes 
statistics after elimination of  topological
symmetries of diagrams.} 
\label{tabKG}
\centering
\begin{tabular}[c]{llll}
\hline \hline
\multicolumn{1}{l}{{{ $\mathrm{Z} \rightarrow \mathrm{b} \bar{\mathrm{b}}$}}} & 1 loop  & 2 loops\hspace{0.9cm} &  3 loops  
\\
\hline
Number of topologies
& 1 & 14 $\stackrel{(\mathrm{A})}{\rightarrow}7\stackrel{(\mathrm{B})}{\rightarrow}5$   & 211 $\stackrel{(\mathrm{A})}{\rightarrow}84\stackrel{(\mathrm{B})}{\rightarrow}51$\\
{Number of diagrams}& 15 & 2383 $\stackrel{(\mathrm{A},\mathrm{B})}{\rightarrow}1074$ & 490\,387 $\stackrel{(\mathrm{A},\mathrm{B})}{\rightarrow}120\,472$\\
{Fermionic loops} &0 & ${150}$ & $17\,580$\\
{Bosonic loops} &15 & ${924}$ & ${102\,892}$\\
{Planar / non-planar} &15 / 0 & ${981 / 133}$ & ${84\,059 / 36\,413}$\\
{QCD / EW} &1 / 14&{98 / 1016}& ${10\,386 / 110\,086}$\\
\multicolumn{4}{l}{\textbf{{$\mathrm{Z} \rightarrow \mathrm{e}^{+} \mathrm{e}^{-}, \dots$}}} \\
{Number of topologies} 
& 1 & 14 $\stackrel{(\mathrm{A})}{\rightarrow}7\stackrel{(\mathrm{B})}{\rightarrow}5$   & 211 $\stackrel{(\mathrm{A})}{\rightarrow}84\stackrel{(\mathrm{B})}{\rightarrow}51$\\
{Number of diagrams}& 14 & 2012 $\stackrel{(\mathrm{A},\mathrm{B})}{\rightarrow}880$ & 397\,690 $\stackrel{(\mathrm{A},\mathrm{B})}{\rightarrow}91\,472$\\
{Fermionic loops} &0 & ${114}$ & $13104$\\
{Bosonic loops} &14 & ${766}$ & ${78\,368}$\\
{Planar / non-planar} &14 / 0 & ${782 / 98}$ & ${65\,487 / 25\,985}$\\
{QCD / EW} &0 / 14&{0 / 880}& ${144 / 91\,328}$\\
\cline{2-4}
\hline \hline
\end{tabular}
\end{table}
\endgroup

Second, the accuracy with which three-loop diagrams can be calculated must be estimated. 
For two-loop bosonic vertex integrals, results have been obtained with a high level of accuracy; eight digits 
in most cases and at least six digits for the few worst integrals, with some room for improvement.
The final accuracy of the complete results for the bosonic two-loop corrections to the EWPOs was at the level of 
at least four digits \cite{Dubovyk:2016aqv,Dubovyk:2018rlg}.
To achieve this goal, the Feynman integrals have been calculated numerically, directly in the Minkowskian region, using 
two main 
approaches: (i) \sd{}, as implemented in the  packages {\tt FIESTA\,3} \cite{Smirnov:2013eza} and {\tt
SecDec\,3} \cite{Borowka:2015mxa},
and (ii) \mbr{} integrals, as implemented in the package {\tt MBsuite} 
\cite{Gluza:2007rt,Gluza:2010rn,Bielas:2013rja,Dubovyk:2015yba,Dubovyk:2016ocz,Dubovyk:2017cqw}.
Because fermionic three-loop diagrams are technically not much more complicated than two-loop bosonic 
integrals (\eg in the case of self-energy insertions, the dimensionality of {\mbr} integrals increases by only one), an overall two-digit precision for the final phenomenological results appears to be within reach. This estimate is based on current knowledge and available methods and tools.
 
Two further remarks are in order. First, the previously estimated value of the bosonic two-loop 
correction  to $\Gamma_\mathrm{Z}$ 
based on the geometric series (TH1) was at the level of $0.1\UMeV$, which is much smaller than its actual calculated value \cite{Freitas:2016sty,Dubovyk:2018rlg}. This is partly based on the fact that all 
final-state flavours sum up because they contribute to $\Gamma_\mathrm{Z}(\alpha^2_{\rm bos})$ with the same 
sign, which was not foreseen in the previous estimate. Thus, care should be taken in interpreting any theoretical error 
estimates. Nonetheless, owing to the lack of a better strategy, we \emph{assume} that the values TH1-new in \Tref{tab2} 
are representative of the actual size of the currently unknown three-loop corrections. 
Second, the achievement of at 
least two digits intrinsic net 
numerical precision for the 
three-loop electroweak corrections 
will probably require the evaluation of
single Feynman integrals with much greater precision than in the two-loop case, since the larger 
number of diagrams leads to more numerical cancellations, and each new diagram topology poses new 
challenges for the numerical convergence. 
 
Thus, besides straightforward improvements in numerical calculations based on \sd{} and \mbr{} methods, 
work on new innovative numerical and analytical techniques (and combinations thereof) should continue and may lead to 
accelerated progress.
There are many other places for future improvements,
\eg optimizations at the three- and four-loop levels of the minimal number of 
{\mbr} integral dimensions (see Section E.\ref{sec-borowka} in this report), 
integration-by-parts (IBP) reductions to master integrals, or  reliable practical prescriptions for the $\gamma_5$ issue at three loops and beyond. 
The numerical methods will certainly be complemented by progress in analytical and 
semi-analytical approaches (both in methods and tools), to which Chapter~\ref{chmt} is devoted.
Similarly, other EWPOs can be discussed. 
Table~\ref{tabfin} collects all present and expected theoretical intrinsic error estimates (see, \eg 
Ref.~\cite{Freitas:2016sty}).

\begin{table}
\caption{Comparison of experimental FCC-ee precision goals for selected EWPOs (EXP2, from Table~\ref{THtab1}) with various scenarios for theoretical error estimations. TH1-new, current theoretical error based on extrapolations through geometric series; TH2, estimated  theoretical error (using prefactor scalings), assuming that electroweak three-loop corrections are known; TH3, a scenario where the dominant four-loop corrections are also available. Since  reliable quantitative estimates of TH3 are not possible at this point, only conservative upper bounds of the theoretical error are given.
\label{tabfin}}
\centering
\begin{tabular}{lllll}
\hline \hline
& \multicolumn{4}{l}{FCC-ee-Z EWPO error estimates}  \\
 & $\delta \Gamma_\mathrm{Z} \;(\rm MeV)$ & $\delta R_\ell\; (10^{-4})$ & $\delta R_\mathrm{b}\; (10^
{-5})$ &$\delta \seff{\ell}\; (10^{-6})$
 \\ 
 \hline
EXP2 \cite{Alain:Jan2018}   & 0.1 & 10 &  $2\div 6$  & \phantom{4}$6$ 
\\
TH1-new   & 0.4 & 60 & 10  &  45 \\
TH2   & 0.15 & 15 & \phantom{4}5 & 15 \\
TH3   & ${<}0.07$ & ${<}7$ & ${<}3$ & ${<}7$ \\
\hline \hline
\end{tabular}
\end{table}

To summarize, FCC-ee-Z imposes very strong demands on future
theoretical calculations of currently unknown higher-order quantum EW and QCD corrections. 
As shown here, different estimates lead to predictions for EWPO error bands 
that are at the level of or of the order of future experimental demands. 
Then actual calculations may shift the values and diminish the errors of EWPOs substantially, as 
has been shown recently in the case of the Z boson decay width \cite{Dubovyk:2018rlg}. Here, the result for the bosonic two-loop corrections was found to be greater than the previous estimate by a factor of 3--5, depending on the chosen input parametrization. One of the most promising avenues for addressing the challenges of these future calculations is the use of numerical integration methods. These are more flexible than analytical techniques, but are limited by the achievable numerical precision.
Our estimates bring us to the conclusion that an accuracy of at least two digits  in future three- and four-loop calculations of EWPOs is needed. Therefore, dedicated and increased efforts by the theory community will be important to meet the experimental demands of 
the FCC-ee-Z or other lepton collider projects in the Z line shape mode without
limiting the physical interpretation of the corresponding precision measurements.

Let us stress that, apart from the problems mentioned here, there is also the  
issue of extracting EWPOs from real processes, including  QED unfolding. This is the subject of 
Chapter~\ref{sunfold}; see also Refs.~\cite{Riemann:Jan2018, Jadach:Jan2018}.

\clearpage \pagestyle{empty} 
\cleardoublepage 
\chapter{Theory meets experiment}
\label{sunfold}


\section
[Cross-sections and electroweak pseudo-observables (EWPOs)
\\ {\it J. Gluza, S. Jadach, T. Riemann}]
{Cross-sections and electroweak pseudo-observables (EWPOs) \label{sec:unfoldintro}}

\pagestyle{fancy}
\fancyhead[RO]{}
\fancyhead[CO]{}
\fancyhead[LO]{}
\fancyhead[CO]{\thechapter.\thesection  
\hspace{1mm} Cross-sections and electroweak pseudo-observables (EWPOs)}
\fancyhead[LE]{}
\fancyhead[CE]{J. Gluza, S. Jadach, T. Riemann}
\fancyhead[RE]{} 

\noindent
{\bf Authors: Janusz Gluza, Stanis\l aw Jadach, Tord Riemann}
\\
Corresponding author: Tord Riemann [Tord.Riemann@cern.ch]
\vspace*{.5cm}
 
\noindent The interpretation of real cross-sections at the Z peak is a delicate problem for the FCC-ee, owing to its 
incredible precision. 
We consider here exclusively fermion pair production.
The real cross-section describes the reaction
\begin{equation}\label{sigmareal}
 \mathrm{e}^+ \mathrm{e}^- \to \mathrm{f}^+ \mathrm{f}^-  + {\mathrm{invisible}}~ (n~\gamma + {\mathrm{e^+e^- pairs}} + \cdots),
\end{equation}
\ie fermion pair production including those additional final-state configurations that stay invisible in the detector.
It is well-known that one may describe such a reaction with 
multidimensional generic ansatzes, \eg
\begin{equation}\label{structurefunctionansatz}
 \sigma^{\mathrm{e}^+ \mathrm{e}^- \to \mathrm{f}^+ \mathrm{f}^- +  \cdots}(s) = \int \mathrm{d}x_1 \mathrm{d}x_2 ~ g(x_1)  ~ g(x_2)  ~ \sigma^{\mathrm{e}^+ \mathrm{e}^- \to \mathrm{f}^+ 
\mathrm{f}^-}(s') ~ \delta(s'-x_1x_2 s).
\end{equation}
In the one-loop approximation with soft photon exponentiation, or the flux function approach,   $x_2=1-x_1$, 
resulting in the generic ansatz
\begin{equation}
 \sigma^{\mathrm{e}^+ \mathrm{e}^- \to \mathrm{f}^+ \mathrm{f}^- +  \cdots}(s) = \int \mathrm{d}x  ~ f(x) ~  \sigma^{\mathrm{e}^+ \mathrm{e}^- \to \mathrm{f}^+
\mathrm{f}^-}(s') ~ \delta(x-s'/s).
\end{equation}
The $\sigma^{\mathrm{e}^+ \mathrm{e}^- \to \mathrm{f}^+ \mathrm{f}^-}$ is called the underlying hard scattering cross-section or the effective Born cross-section.
The kernel functions $g(x)$ and $f(x)$ depend on the process, the observable to be described,  and  experimental 
conditions, such as the choice of variables 
and cuts.
Further, if initial--final-state radiation interferences are considered, combined with box diagram contributions,  
the hard scattering basic Born function in the flux function approach has a more general structure
\cite{Bardin:1988xt,Bardin:1989cw,Bardin:1989di,Bardin:1989tq,Bardin:1990fu,Bardin:1990de,Bardin:1992jc,Bardin:1999yd,
Christova:1999cc}: %
\begin{equation}
\label{sigma-ini-fin}
 \sigma^{\mathrm{e}^+ \mathrm{e}^- \to \mathrm{f}^+ \mathrm{f}^-}(s') \to  \sigma^{\mathrm{e}^+ \mathrm{e}^- \to \mathrm{f}^+ \mathrm{f}^-}(s,s').
\end{equation}
An example from Ref.  \cite{Christova:1999cc} is reproduced in \Eref{sigmainifin}. 

Concerning the extraction of physical parameters from real cross-sections, one may follow two different 
strategies.
\begin{enumerate}
 \item 
 Direct fits of $\sigma^\mathrm{real}$ in terms of such quantities as $M_\mathrm{Z}, \Gamma_\mathrm{Z}$ and other parameters. 
 The other parameters are called {\it electroweak pseudo-observables} (EWPOs).
 \item 
 Extraction of the various hard $2 \to 2$ scattering cross-sections $\sigma_\mathrm{tot,FB,\dots}^{(0)}$ from the real cross-sections $\sigma^\mathrm{real}$ and  a subsequent 
analysis of the hard cross-sections in terms of such quantities as $M_\mathrm{Z}, \Gamma_\mathrm{Z}$ and other parameters, such as $A_\mathrm{f}$.
 \end{enumerate}
In practice, at the LEP, the second approach was chosen by all experimental collaborations \cite{ALEPH:2005ab}.

For a Z line shape analysis, the structure functions or flux functions are assumed to be known from theoretical 
calculations with sufficient accuracy to match the experimental demands.
Before the unfolding, data have to be prepared using Monte Carlo programs, \eg {\tt KKMC} \cite{Jadach:1999vf},
 to match the 
simplified unfolding conditions of  analysis programs, \eg ZFITTER 
\cite{Bardin:1992jc,Bardin:1999yd,Arbuzov:2005ma,Akhundov:2013ons,web-sanc.zfitter:2016}.

\textit{To determine the structure function or flux function kernels for data preparation or for unfolding is one of 
the 
challenges of FCC-ee-Z physics.}

At the LEP, the Z line shape analysis was performed using the ZFITTER package.
ZFITTER relies completely on the flux function approach, which is sufficiently accurate, if the photonic next-to-leading-order (NLO) corrections plus soft photon exponentiation dominate the invisible terms in \Eref{sigmareal}.
This is in accordance with the condition $x_2=1-x_1$. 
ZFITTER contains a variety of  flux functions $f(x)$, which have been determined in a series of theoretical papers; as explained in the following.
Details may also be found in the ZFITTER descriptions cited previously.

The crucial point in the unfolding procedure is that
the result of the unfolding depends on the ansatz chosen.
This sounds trivial.
But the statement reflects the need of knowing sufficiently many details of the analytical structure of the hard scattering process, as a function of the chosen physical parameters.

Be it a so-called model-independent approach or the Standard Model ansatz, parameters like $M_\mathrm{Z}$ and 
$\Gamma_\mathrm{Z}$ must be introduced in a proper way, respecting, \eg their universality (channel independence, \etc), as well as the quantum field theoretical structure of the underlying theory; see Ref. \cite{Bardin:1988xt} for a discussion.

\textit{To determine the correct hard scattering ansatz in a model-independent-based or Standard Model-based ansatz is another challenge of FCC-ee-Z 
physics.
}

At the LEP, it was possible to determine the mass and width of the Z boson with experimental errors of $2\UMeV$ each 
\cite{ALEPH:2005ab,Schael:2013ita}. 
The  experimental challenges are manyfold, including high event statistics, good 
apparatus systematics, and good knowledge of the beam energy. 
In all these respects, FCC-ee claims to be much better, resulting in unprecedented error estimates.
Several of the anticipated experimental errors are reproduced in the FCC-ee-Z wish-list, given in the foreword.
We mention here as challenging experimental aims:
\begin{align}
\frac{\delta M_\mathrm{Z}}{M_\mathrm{Z}} &\approx \frac{0.1 \UMeV}{93 \UGeV} \approx 10^{-6},
\\
\frac{\delta \Gamma_\mathrm{Z}}{\Gamma_\mathrm{Z}} &\approx \frac{0.1 \UMeV}{2 \UGeV} \approx 5 \times 10^{-5}.
\end{align}
Certain asymmetry errors are also anticipated to be smaller than at the LEP by orders of magnitude, \eg
\begin{eqnarray}
 \delta A_\mathrm{FB}^{{\bar {\mathrm{b}}}\mathrm{b}} \approx 10^{\color{black}{-5} },
\end{eqnarray}
which is about two orders of magnitude better than at the LEP \cite{Patrignani:2016xqp,Tenchini:April2018},
$A_\mathrm{FB}^{{\bar{ \mathrm{b}}}\mathrm{b}} = 0.0992 \pm  160 \times 10^{-5}$, corresponding to 
$A_\mathrm{b} = 0.923 \pm  200 \times 10^{-4}$.
Similarly, the planned improvements of net experimental accuracies compared with those obtained at the LEP amount to a factor of 20 for $M_\mathrm{Z}$
and $\Gamma_\mathrm{Z}$ and are so high that the unfolding of real observables has to be re-analysed 
compared with LEP physics.
Such a demand was first described in some detail in an article  predicting the complete  leptonic weak mixing in the 
Standard Model at two loops \cite{Awramik:2006uz}. It has been argued there that the structure of the ansatz in ZFITTER 
contradicts, beginning at  two-loop accuracy, the structure predicted by perturbative quantum field theory around a resonance like the Z boson. 

In the following two sections, we  will describe the current state of the art and the  progress needed in order 
to meet the demands from FCC-ee measurements on the issues of unfolding in a model-independent-based or the Standard Model-based approaches.
For the proper unfolding ansatzes of hard scattering, both in some model-independent- and the Standard Model ansatz, see Section 
C.\ref{s-smatrix}.  
For the determination of the structure functions at three-loop orders combined with appropriate 
exponentiations of larger higher-order terms see Section C.\ref{s-qed}.

{\it To summarize, packages like {\tt KKMC} and ZFITTER must be improved in two qualitatively different respects, namely: 
(i) 
considering the necessary higher orders in perturbation theory; and (ii) respecting thereby the correct structure of the hard 
scattering ansatz.   What this means in detail will be the subject of the next two sections.}


The relevant $2\to 2$ hard scattering matrix elements to be used in Monte Carlo programs will be discussed at length in the 
next section.
The $2\to 2$ total cross-section 
$\sigma_\mathrm{tot}^{(0)}(s)$
and the various asymmetries, based on these matrix elements, are defined as follows:
\begin{align}
\label{e-2to2tot}
\sigma_\mathrm{tot}^{(0)}(s) &= \int_{-1}^{+1} \mathrm{d}\cos \theta \frac{\mathrm{d} \sigma^{(0)}}{d\cos\theta},
\\
A_\mathrm{FB}^{(0)}(s) &= \frac{\sigma_\mathrm{FB}^{(0)}(s)}{\sigma_\mathrm{tot}^{(0)}(s)} 
= \frac{1}{\sigma_\mathrm{tot}^{(0)}(s)} 
 \left( \int_{0}^{+1} - \int_{-1}^{0}\right) \mathrm{d}\cos \theta \frac{\mathrm{d} \sigma^{(0)}}{\mathrm{d}\cos\theta},
 \\
 A_\mathrm{LR}^{(0)}(s) &= \frac{\sigma_\mathrm{LR}^{(0)}(s)}{\sigma_\mathrm{tot}^{(0)}(s)} 
= 
\frac{1}{\sigma_\mathrm{tot}^{(0)}(s)} 
 \int_{-1}^{+1} \mathrm{d}\cos \theta \left( \frac{\mathrm{d} \sigma^{(0)}_{\mathrm{e}_\mathrm{L}^-}}{\mathrm{d}\cos\theta} - \frac{\mathrm{d} 
\sigma^{(0)}_{\mathrm{e}_\mathrm{R}^-}}{\mathrm{d}\cos\theta}\right),
  \\
 A_\mathrm{pol}^{(0)}(s) &= \frac{\sigma_\mathrm{pol}^{(0)}(s)}{\sigma_\mathrm{tot}^{(0)}(s)} 
= \frac{1}{\sigma_\mathrm{tot}^{(0)}(s)} 
 \int_{-1}^{+1} \mathrm{d}\cos \theta \left( \frac{\mathrm{d} \sigma^{(0)}_{\mathrm{f}_\mathrm{R}^-}}{\mathrm{d}\cos\theta} - \frac{\mathrm{d} 
\sigma^{(0)}_{\mathrm{f}_\mathrm{L}^-}}{\mathrm{d}\cos\theta}\right),
  \\
 A_\mathrm{LR,pol}^{(0)}(s) &= \frac{\sigma_\mathrm{LR,pol}^{(0)}(s)}{\sigma_\mathrm{tot}^{(0)}(s)} \nonumber \\
 &= \frac{1}{\sigma_\mathrm{tot}^{(0)}(s)} 
 \int_{-1}^{+1} \mathrm{d}\cos \theta 
 \left( \frac{\mathrm{d} \sigma^{(0)}_{\mathrm{e}_\mathrm{L}^-,\mathrm{f}_\mathrm{R}^-}}{\mathrm{d}\cos\theta} - \frac{\mathrm{d} 
\sigma^{(0)}_{\mathrm{e}_\mathrm{L}^-,\mathrm{f}_\mathrm{L}^-}}{\mathrm{d}\cos\theta}
- \frac{\mathrm{d} \sigma^{(0)}_{\mathrm{e}_\mathrm{R}^-,\mathrm{f}_\mathrm{R}^-}}{\mathrm{d}\cos\theta}
+\frac{\mathrm{d} \sigma^{(0)}_{\mathrm{e}_\mathrm{R}^-,\mathrm{f}_\mathrm{L}^-}}{\mathrm{d}\cos\theta}
\right),
\\
\label{e-2to2LRFB}
 A_\mathrm{LR,FB}^{(0)}(s) &= \frac{\sigma_\mathrm{LR,FB}^{(0)}(s)}{\sigma_\mathrm{tot}^{(0)}(s)} \nonumber \\ 
&= \frac{1}{\sigma_\mathrm{tot}^{(0)}(s)} 
 \left( \int_{0}^{+1} - \int_{-1}^{0}\right) \mathrm{d}\cos \theta  
\left( 
 \frac{\mathrm{d} \sigma^{(0)}_{\mathrm{e}_\mathrm{L}^-}}{\mathrm{d}\cos\theta} - \frac{\mathrm{d} \sigma^{(0)}_{\mathrm{e}_\mathrm{R}^-}}{\mathrm{d}\cos\theta}
 \right)
 ,
 \\
 \label{e-2to2fbpol}
 A_\mathrm{pol,FB}^{(0)}(s) &= \frac{\sigma_\mathrm{FB,pol}^{(0)}(s)}{\sigma_\mathrm{tot}^{(0)}(s)} 
= \frac{1}{\sigma_\mathrm{tot}^{(0)}(s)} 
 \left( \int_{0}^{+1} - \int_{-1}^{0}\right) \mathrm{d}\cos \theta  
 \left( 
 \frac{\mathrm{d} \sigma^{(0)}_{\mathrm{f}_\mathrm{R}^-}}{\mathrm{d}\cos\theta} - \frac{\mathrm{d} \sigma^{(0)}_{\mathrm{f}_\mathrm{L}^-}}{\mathrm{d}\cos\theta}
 \right)
 \\ 
 A_\mathrm{LR,pol,FB}^{(0)}(s) 
 &= \frac{\sigma_\mathrm{LR,pol,FB}^{(0)}(s)}{\sigma_\mathrm{tot}^{(0)}(s)}
\nonumber \\ 
 &= \frac{1}{\sigma_\mathrm{tot}^{(0)}(s)} 
 \left( \int_{0}^{+1} - \int_{-1}^{0}\right) \mathrm{d}\cos \theta  
  \left( \frac{\mathrm{d} \sigma^{(0)}_{\mathrm{e}_\mathrm{L}^-,\mathrm{f}_\mathrm{R}^-}}{\mathrm{d}\cos\theta} - \frac{\mathrm{d} 
\sigma^{(0)}_{\mathrm{e}_\mathrm{L}^-,\mathrm{f}_\mathrm{L}^-}}{\mathrm{d}\cos\theta}
- \frac{\mathrm{d} \sigma^{(0)}_{\mathrm{e}_\mathrm{R}^-,\mathrm{f}_\mathrm{R}^-}}{\mathrm{d}\cos\theta}
+\frac{\mathrm{d} \sigma^{(0)}_{\mathrm{e}_\mathrm{R}^-,\mathrm{f}_\mathrm{L}^-}}{\mathrm{d}\cos\theta}
\right)
 . \label{e-2to2lrpolfb}
\end{align}
Here, `L' and `R' are helicities of the massless external particles. 
For unpolarized scattering, initial-state helicities are assumed to be 
averaged, and final helicities combine incoherently to the final-state polarization $A_\mathrm{pol}^{(0)}(s)$.

Let us assume that an unbiased unfolding is possible from real cross-sections to the $2\to 2$ processes.
Here, we understand by real cross-sections those cross-sections and also cross-section asymmetries that may be 
measured in the detector.
The experiments determine numbers with dimensions of, \eg square centimetres or nanobarns. Theory, however, works with either matrix elements or integrated squared matrix elements.

To be definite and to shorten the notation, let us work first with squared matrix elements and use the flux 
function approximation.
This is how the analysis tool ZFITTER is composed.
We will have to improve that considerably for the FCC-ee applications.

The total cross-section and asymmetries are defined generically through
\begin{align}
\label{e-totreal-0}
\sigma_\mathrm{A}^\mathrm{real}(s) &= \int \frac{\mathrm{d}s'}{s} ~  \rho_\mathrm{tot}(s'/s) ~ \sigma_\mathrm{A}^{(0)}(s'),
~~~~ \mathrm{A} = \mathrm{tot, LR, pol, LRpol}, 
  \\\label{e-totreal-f}
  \sigma_\mathrm{A}^\mathrm{real}(s) &= \int \frac{\mathrm{d}s'}{s} ~  \rho_\mathrm{FB}(s'/s) ~ \sigma_\mathrm{A}^{(0)}(s'),
  ~~~~ \mathrm{A} = \mathrm{FB, LRFB, polFB, LRpolFB}.
  \end{align}
  using the notation
  \begin{align}
   s' \equiv s_{\mathrm{f}^+\mathrm{f}^-} &= s\left(1 - \frac{2 E_{\gamma}}{\sqrt{s}}\right),
   \\
   R \equiv 1-v &= \frac{s'}{s}
   .
  \end{align}
Here, $s$ is the centre-of-mass energy squared, $s'$ the invariant mass squared of the final fermion pair, and
$E_{\gamma}$ the photon energy.
As indicated in these equations, the flux functions $\rho_\mathrm{T,FB}$ for  $\cos\theta$-even and for $\cos\theta$-odd integrals differ.
They also depend  on the other experimental cuts.
Only four of the seven observables shown are independent because the $2\to 2$ scattering of (practically) massless 
external spin-1/2 particles has only four helicity degrees of freedom.

When taking the complete photonic $O(\alpha)$ 
corrections into account, including initial- and final-state radiations and 
their interferences, the cross-section foldings have the following general structure:
\begin{align} 
\sigma_\mathrm{A}^\mathrm{real}(s) &= \sigma_\mathrm{A}^{(0)}(s)  + \sigma_\mathrm{A}^\mathrm{real,ini}(s) +
\sigma_\mathrm{A}^\mathrm{real,fin}(s) + \sigma_\mathrm{A}^\mathrm{real,int}(s)
 \nonumber \\ \nonumber 
&=
 \sigma_\mathrm{A}^{(0)}(s)
+ 
\int \mathrm{d}R \sigma_\mathrm{A}^{(0)}(s') \rho_\mathrm{A}^\mathrm{ini}(R) 
+
\sigma_\mathrm{A}^{(0)}(s) \int \mathrm{d}R \rho_\mathrm{A}^\mathrm{fin}(R)
\nonumber 
\\ \label{sigmainifin} 
& \quad 
+ \int \mathrm{d}R \sum_{V_i,V_j=\gamma, \mathrm{Z}} \sigma_{\bar{\mathrm{A}}}^{(0)}(s,s',i,j) \rho_\mathrm{A}^\mathrm{int}(R,i,j)
.
\end{align}
In the initial--final-state interferences, the effective Born cross-sections depend on both $s$ and $s'$,
as well as on the type of exchanged vector particles $V_i$ (\eg photon or Z).
Additionally, one has to be aware in the interference that for $\mathrm{A}
= \mathrm{tot,LR,pol}$ one needs ${\bar {\mathrm{A}}}= \mathrm{FB}$ and for $\mathrm{A} = \mathrm{FB,LRFB}$ 
one needs ${\bar {\mathrm{A}}}= \mathrm{tot}$.
The compositions of real cross-sections are modified when contributions are exponentiated, \eg for
 initial-state soft photon exponentiation of $\sigma_\mathrm{tot}$ \cite{Greco:1975rm,Bardin:1989cw}:
\begin{align} 
 \sigma_\mathrm{tot}^{(0)}(s)  + \sigma_\mathrm{tot}^\mathrm{real,ini}(s) &\to 
\int \mathrm{d}R~ \sigma_\mathrm{tot}^{(0)}(s')~ \rho_\mathrm{tot}^\mathrm{ini}(R),  
\\ \label{eq:3}
  \rho_\mathrm{tot}^\mathrm{ini}(R) &= \left(1+{\bar S}\right)\beta (1-R)^{\beta-1} 
   + {\bar H}_\mathrm{tot}^\mathrm{ini}(R),
 \\   \label{eq:4}
         {\bar S} &= \frac{3}{4}\beta +\frac{\alpha}{\pi}Q_\mathrm{e}^2
\left(\frac{\pi^2}{3} - \frac{1}{2}\right) + \mathrm{h.o.},
\\
  \beta &= \frac{2\alpha}{\pi} Q_\mathrm{e}^2 L_\mathrm{e},
  \\
  L_\mathrm{e} &= \left( \ln \frac{s}{m_\mathrm{e}^2}-1\right) ,
\\  \label{eq:3a}
 {\bar H}_\mathrm{T}^\mathrm{ini}(R)&=\left[ H_\mathrm{BM}(R)-\frac{\beta}{1-R}\right] + \mathrm{h.o.} ,
\end{align}
where `h.o.' stands for higher orders, and $H_\mathrm{BM}(R)$ is the Bonneau--Martin kernel
\cite{Bonneau:1971mk}:
\begin{eqnarray}
  \label{eq:bm}
  H_\mathrm{BM}(R) = \frac{1}{2}~\frac{1+R^2}{1-R}~\beta.
\end{eqnarray}
The radiator function for the forward--backward antisymmetric cross-section differs, owing to the different integration 
over the scattering angle. 
To show the simplest term, we reproduce here the ${\cal O}(\alpha)$ approximated initial-state radiation hard 
scattering part for 
$\sigma_\mathrm{FB}$ \cite{Bardin:1989cw}:
\begin{equation}
\label{eq-rhoFB}
 \rho_\mathrm{FB}^\mathrm{ini}\left(\frac{s'}{s}\right) \sim h_\mathrm{e}(v) 
= Q_\mathrm{e}^2 \frac{\alpha}{\pi} \left( L_\mathrm{e}-1 
 -\ln \frac{1-v}{ \left(1-\frac{v}{2}\right)^2}
 \right)
 \frac{1+(1-v)^2}{v} \frac{1-v}{\left(1-\frac{v}{2}\right)^2}.
\end{equation}
The $v$ vanishes in the soft photon limit with $s'\to s$,  and then $\rho_\mathrm{FB}$ approaches $\rho_\mathrm{tot}$.

{\it In these formulae, we assume that we can consider $\gamma$ exchange and Z exchange as independently defined.
In Section C.\ref{s-smat-gammaZ}, we will discuss how such an assumption is the result of careful considerations, 
starting with one and only one Z amplitude as a Laurent series in $s$. }

The unfolding of realistic observables 
can be performed with the analysis tools TOPAZ0 
\cite{Montagna:1993ai,Montagna:1995ja,Montagna:1998kp} and  ZFITTER. The latter  relies on 
the 
work quoted  here for $\rho_\mathrm{tot}$ and $\rho_\mathrm{FB}$.
Evidently, the result of unfolding depends on the model chosen for the hard process $\sigma^0_\mathrm{tot}(s')$ or 
$\sigma^0_\mathrm{FB}(s')$.
This fact is reflected by the various model-dependent so-called {\it interfaces} of ZFITTER.

The parameter $s'$ is distinguished when soft photon emission is 
exponentiated, which runs in this variable.

The complications in the derivation of flux functions arise from (i) higher-order corrections, (ii) more 
sophisticated cuts, and (iii) the resonance 
character of the Z peak.

In the simplest case, with no cuts at all, and to order $\cal O(\alpha)$, one has extremely simple expressions for the 
photonic corrections, which may be completely integrated out, see Ref. \cite{Bardin:1989cw}.
With soft photon exponentiation, this becomes modified, as discussed in Refs. \cite{Greco:1975rm,Bardin:1989cw}.
Flux functions, being differential in the scattering angle, with and without soft photon exponentiation, are derived in Refs. 
\cite{Bardin:1990fu,Bardin:1990de}.
Additional common exponentiation of initial- and final-state radiation may be found in Refs. 
\cite{Bardin:1992jc,Bardin:1999yd,Arbuzov:2005ma}, where also many properties of two-particle scattering at the Z 
peak 
are discussed.
The introduction of an additional acollinearity cut between the two final-state fermions is treated in Refs. 
\cite{Bilenky:1989zg,Bardin:1989cw,Christova:1999cc,Jack:2000as}.
Whenever any unfolding of real cross-sections is discussed, one has to have in mind that a real cross-section is, seen 
by theory, a relatively complicated aggregation of terms.

For massive final states (\eg top-pair production, opening kinematically much above the Z peak region), one has two 
additional degrees of freedom \cite{Fleischer:2003kk,Bardin:2000kn}.

The real asymmetries from the experiment are pure numbers with errors, typically defined  as follows:
\begin{align}
 A_\mathrm{FB}^\mathrm{real}(s) &= \frac{\sigma_\mathrm{FB}^\mathrm{real}(s)}{\sigma_\mathrm{tot}^\mathrm{real}(s)},
 \\
  A_\mathrm{LR}^\mathrm{real}(s) &= \frac{\sigma_\mathrm{LR}^\mathrm{real}(s)}{\sigma_\mathrm{tot}^\mathrm{real}(s)},
  \\
   A_\mathrm{pol}^\mathrm{real}(s) &= \frac{\sigma_\mathrm{pol}^\mathrm{real}(s)}{\sigma_\mathrm{tot}^\mathrm{real}(s)}
.
\end{align}
Here, the indices `tot', `FB', `LR', and `pol' refer to the definitions introduced in Eqs. 
\eqref{e-2to2tot} to \eqref{e-totreal-f}. 
As mentioned, only three of the asymmetries are independent quantities besides $\sigma_\mathrm{tot}^\mathrm{real}$.

At this stage, there is no difference between a model-independent- or Standard-Model-based data analysis.
In the next section, we will introduce explicit expressions for the effective Born cross-sections and discuss their 
relevance for an exact analysis of the data.
The notion of pseudo-observables, or electroweak pseudo-observables (EWPOs), will be introduced there. They are all  defined at $s=M_\mathrm{Z}^2$.
It will also be shown how the notation used for the analysis of cross-section formulae can be 
taken over 
to the modelling with Monte Carlo (MC) programs, which rely not on cross-sections but on effective Born matrix 
elements.

\clearpage \pagestyle{empty} 
\cleardoublepage
%


\section
[Higher-order radiative corrections, matrix elements, EWPOs
\\ \noindent {\it A. Freitas, J. Gluza, S. Jadach, T. Riemann}]
{\label{s-smatrix}
Higher-order radiative corrections, matrix elements, EWPOs
}

\pagestyle{fancy}
\fancyhead[LO]{}
\fancyhead[RO]{}
\fancyhead[CO]{\thechapter.\thesection 
\hspace{2mm}
Higher-order radiative corrections, matrix elements, EWPOs}
\fancyhead[LE]{}
\fancyhead[CE]{A. Freitas, J. Gluza, S. Jadach, T. Riemann}
\fancyhead[RE]{}

\noindent
{\bf{Authors:}  Ayres Freitas, Janusz Gluza, Stanis\l aw Jadach, Tord Riemann}
\\
Corresponding author: Tord Riemann [Tord.Riemann@cern.ch]
\vspace*{.5cm}

\newcommand{\Oaa}{\mathswitch{{\cal{O}}(\alpha^2)}}
\newcommand{\GW}{\mathswitch {\Gamma_W}} 
\newcommand{\GF}{\mathswitch {G_\mu}}
\newcommand{\SinEff}{\mathswitch {\sin^2\theta^{\mbox{\footnotesize lept}}_{\mbox{\footnotesize eff}}}}
\newcommand{\sineff}{\sin^2\theta^{\mbox{\scriptsize lept}}_{\mbox{\scriptsize eff}}}
\newcommand{\SinEfff}{$\sin^2\theta^\sinefff{f}_{\mbox{\footnotesize eff}}$}
\newcommand{\sinefff}{ \sin^2\theta^\mathrm{f}_{\mbox{\scriptsize eff}}}

\noindent In this section, we will consider the 2$\to$2 reaction, 
\begin{eqnarray}\label{e-smat-eeff}
 \mathrm{e}^+ \mathrm{e}^- \to \mathrm{f} {\bar {\mathrm{f}}}
 .
\end{eqnarray}
At the Z peak, we may neglect the initial- and final-state fermion masses at many places.
Assuming the planned FCC-ee accuracy, one may expect that, at least for $\mathrm{b}{\bar {\mathrm{b}}}$ production, the Born and one-loop 
contributions have to be described with full account of the b quark mass.
This part of the predictions is available in several software packages, see, \eg\ Refs.
\cite{Fleischer:2002nn,FRW:2002sw,Fleischer:2003aa,%
Bardin:2000kn,Hahn:2003ab,%
Gluza:2004tq,Lorca:2004dk,%
Lorca:2004fg,Fleischer:2006ht} and will
not be discussed here in more detail. 

People often analyse a set of so-called pseudo-observables in a chosen theory frame, \eg a `measured' 
forward--backward asymmetry.
The exact expression will be given in Eq. \eqref{e-afbcomplete0}. Usually, the following approximated ansatz has 
been sufficiently correct:
\begin{eqnarray}\label{e-smat-poth1}
 A_\mathrm{FB}^{\mathrm{meas}} &=& \frac{3}{4} A_\mathrm{e}^{\mathrm{th}} A_\mathrm{f}^{\mathrm{th}},
\end{eqnarray}
where $ A_\mathrm{e}, A_\mathrm{f}$ are simple expressions in terms of the effective weak mixing angle.
In fact, $A_\mathrm{FB}^{\mathrm{meas}}$ has to be extracted with the anticipated accuracy from an experimentally 
measured cross-section. 
The $A_\mathrm{f}^{\mathrm{th}}$ has a simple definition:
\begin{align}
\label{e-smat-poth2}
 A_\mathrm{f}^{\mathrm{th}} &= 2\frac{\frac{v_\mathrm{f}^2}{a_\mathrm{f}^2}}{1+\frac{v_\mathrm{f}^2}{a_\mathrm{f}^2}},
\\
\frac{v_\mathrm{f}}{a_\mathrm{f}}&= 1-4|Q_\mathrm{f}|\sin^2\theta_{\mathrm{eff}}^\mathrm{f} ~~\equiv~~ 
 1-4|Q_f|  \Re  (\kappa_\mathrm{f}) ~   \sin^2\theta_\mathrm{W}
.
\end{align}
The right-hand side has to be calculated with the necessary precision from some theory, here the Standard Model at some 
loop order.
For massless final-state fermions, there are only two form factors, the vector and the axial-vector 
couplings, taken at $q^2=M_\mathrm{Z}^2$.
At the FCC-ee Tera-Z stage, the precisions are  typically $10^{-5}$ or better, so one should bear in mind that Eq. 
\eqref{e-smat-poth1} is approximate and its validity must be checked before use.
{
Further, as we will discuss later, in addition to the weak mixing angle $\sin^2\theta_{\mathrm{eff}}^\mathrm{f}$, which is defined in terms of the vector and axial-vector couplings of the $\mathrm{Z}\mathrm{f}\bar {\mathrm{f}}$ vertex, there are two other often-used definitions of the weak mixing angle. The various definitions are not identical but agree approximately.}

In this section we  explain in detail the definitions of and the relations between the following quantities:
\begin{enumerate}
 \item
the matrix element for the vertex $\mathrm{Z} \to \mathrm{f}{\bar {\mathrm{f}}}$;
\item 
the exact matrix element for 2$\to$2 scattering with loop corrections;
\item 
the 2$\to$2 effective Born cross-sections.
\end{enumerate}
The relations 
of 2$\to$2 Born cross-sections (iii)  with real observables was 
discussed in Section C.\ref{sec:unfoldintro}.

{\it Here, in this section, we introduce the most general massless 2$\to$2 matrix element, which is characterized by 
four 
form factors.}

These form factors have to be determined by theory, be it the Standard Model with $n$ loops, supersymmetry in 
the Born approximation, or something else.
In principle, these four form factors would be what should be measured in an experiment.\footnote{We mention that, for 
the production of massive top quarks, six form factors are needed for a general ansatz 
\cite{Fleischer:2003kk,Bardin:2000kn}.}
In practice, it is more comfortable to measure some derived quantities, called electroweak pseudo-observables (EWPOs). 
A better name would SMPOs, for Standard Model pseudo-observables, thus including the QCD part in the abbreviation.

At the FCC-ee, it might be even better to measure some other building blocks instead of the EWPOs, but for our didactic 
purposes, it is fine to deal here with EWPOs. 
Some ideas on alternative strategies are introduced in Section 
C.\ref{s-qed}. 

We try to describe the Z resonance near its pole position with the greatest numerical precision.
Consequently, special attention is given to a correct treatment of the 2$\to$2 matrix elements in several respects.
\begin{enumerate}
 \item
 Treat the 2$\to$2 matrix element strictly in its most general form 
 with four complex kinematics-dependent form factors 
 in order to cover, in principle, all 
aspects of physics and in all needed orders of perturbation theory. The form factors may be chosen to be, 
\eg $\rho_\mathrm{ef}, \kappa_\mathrm{e}, \kappa_\mathrm{f}, \kappa_\mathrm{ef}$ or, alternatively, $\rho_\mathrm{ef}, v_\mathrm{e}, v_\mathrm{f}, v_\mathrm{ef}$, while the axial 
couplings are fixed to some real number, \eg $a_\mathrm{f} = I_{3,L}(f) = \pm \frac{1}{2}$.
\item 
Respect unitarity, analyticity, and gauge-invariance for quantum field theoretic perturbative corrections close to the 
resonance peak;
this is fulfilled by the so-called pole renormalization scheme, combined with the use of a Laurent series representation for 
the matrix element (\ie for the four form factors), with one single first-order pole in the $s$-plane, located at $s_0 
= 
M_\mathrm{Z}^2 -\mathrm{i} M_\mathrm{Z} \Gamma_\mathrm{Z}$.  
\item 
Besides the Z boson exchange, there are substantial contributions from photon exchange.
A correct treatment of the coexistence of Z boson and $\gamma$ exchange, and of their  interferences, deserves a 
careful discussion. 
\item
The theoretical treatment of real observables has to respect items (i) to (iii), and a careful re-analysis of its 
structural aspects is needed.
\end{enumerate}
People who were active in LEP analyses and who have some knowledge in using the ZFITTER package will notice that many 
features of the ZFITTER approach might be taken over from the 1980s and 1990s.
Several conceptual changes and a questioning of approximations are necessary when going from 
LEP accuracy to FCC-ee accuracy.
In the Standard Model, ZFITTER covers 2$\to$2 scattering with an accuracy corresponding to one electroweak loop (and 
two loops with QCD), plus certain leading two-loop terms (and three-loop terms with QCD). Let us call this 1+1/2 loops.
Real scattering has to scope with that;  flux functions basically do this job.
For the FCC-ee, the theory must include 2+1/2 loops, in the sense that complete electroweak two-loop calculations are 
needed (and exist by today), accompanied by selected three- and four-loop terms, which are yet to be determined.
This concerns vertex corrections. 
Weak box diagrams are needed at one-loop order less than vertices, \ie at two-loop order.
Here, we should mention that the two-loop weak box terms are unknown so far.

We try to cover the whole calculational scheme in a language that allows experimentalists to understand much 
of the  conceptual detail.
Evidently, we can work out here the principal ideas, but an analysis of their numerical relevance for real 
observations 
at the FCC-ee-Z has to be left for future studies.

\subsection{\label{s-smat-}Renormalization in a nutshell} 
We want to give a short introduction to few of the most important definitions used.

The weak mixing angle $s^2_\mathrm{W} \equiv \sin^2\theta_\mathrm{W}$ has three potential different meanings or functions in the 
model-building.
\begin{enumerate}
 \item  It describes the ratio of the two gauge couplings, 
 \begin{equation}\label{sw2gc}
 g'/g = c_\mathrm{W}/s_\mathrm{W},
 \end{equation} 
 usually in the $\overline{\text{MS}}$  scheme.
 \item It describes the ratio of two gauge boson (on-shell) masses, 
 \begin{equation}\label{sw2fix}
 s_\mathrm{W}^2=1-\frac{M_\mathrm{W}^2}{M_\mathrm{W}^2}.
 \end{equation}
\item It describes the ratio of the vector and axial-vector couplings of an (on-shell) Z boson to fermions,
\begin{equation}\label{sw2va}
\frac{v_\mathrm{f}}{a_\mathrm{f}} = 1-4|Q_\mathrm{f}| s_\mathrm{W}^2.
\end{equation}
This definition is called the \emph{effective weak mixing angle}, henceforth denoted  $\sin^2\theta_\mathrm{W}^{\mathrm{f},\mathrm{eff}}$.
\end{enumerate}
At the Born level of the Standard Model, all these definitions agree, with $a_\mathrm{f}^\mathrm{B} = \pm 1/2$.
\footnote{In ZFITTER and related literature,  $a_\mathrm{f} = 1$ is used universally: this is convenient 
but contradicts the by-now generally accepted conventions.}

At orders in perturbation theory, one has to define a renormalization scheme. In the on-shell scheme of
Alberto Sirlin \cite{Sirlin:1980nh}, one uses the following renormalization conditions:
all particle masses are fixed, and additionally one coupling constant $\alpha$: $e^2 = 4\pi \alpha$. 
The weak mixing angle is not an independent quantity in this scheme.

There are several variations of this renormalization scheme. For example, one may use a renormalization condition for the Fermi constant $G_\mathrm{F}$ instead of $M_\mathrm{W}$. 
$G_\mathrm{F}$ is related to other Standard Model parameters via
\begin{equation}
\label{e-smat-ewdefs2}
\frac{G_\mathrm{F}}{\sqrt{2}} =
 \frac{\pi\alpha_\mathrm{em}} {2\sin^2\theta_\mathrm{W} \cos^2\theta_\mathrm{W}\MZ^2}(1+\Delta r),
\end{equation}
where $\Delta r$ contains the contributions from radiative corrections.

These choices are fully equivalent and are mostly a matter of convenience. Some renormalization schemes lead to smaller one-loop corrections, but arguments like these become more complicated beyond the one-loop level.

Owing  to certain approximations, which we describe in this section, the EWPOs are related to the
effective vector and axial-vector couplings of the Z boson to a fermion type f, $v_\mathrm{f}$ and $a_\mathrm{f}$.\footnote{
It is necessary to checked whether these approximations are still valid at FCC-ee-Z accuracy: this is relatively `easy' but 
necessary and deserves a dedicated study \`a la Ref. \cite{Bardin:1999gt}.} These, in turn, are related to the weak mixing angle according to
\begin{eqnarray}\label{e-sw2Zffkappa}
\frac{v_\mathrm{f}}{a_\mathrm{f}}=1-4|Q_\mathrm{f}| \sin^2\theta_\mathrm{W}^{\mathrm{f},\mathrm{eff}} \equiv 1-4|Q_\mathrm{f}| s_\mathrm{W}^2 \,  \kappa_\mathrm{f}^\mathrm{Z}
,
\end{eqnarray}
where, from now on, Eq. \eqref{sw2fix} holds. The $s_\mathrm{W}^2=1-M_\mathrm{W}^2/M_\mathrm{Z}^2$ is the on-shell weak mixing angle, and $\kappa_\mathrm{f}^\mathrm{Z}$ 
captures the effect of radiative corrections (in the Born approximation, $\kappa_\mathrm{f}^\mathrm{B} = 1$).
The form factor $\kappa_\mathrm{f}^\mathrm{Z}$ was introduced by A. Sirlin \cite{Sirlin:1980nh,Marciano:1980pb}.
Although $s_\mathrm{W}^2$ is not a (pseudo-)observable in itself, it is a useful quantity for discussions of the EWPOs with 
higher-order corrections.

\subsection{\label{s-smat-2to2me}The 2$\to$2 matrix elements}
We now consider the definitions of the hard scattering kernels used in Eqs. \eqref{e-2to2tot}--\eqref{e-2to2fbpol}.
All momenta are assumed to be incoming. 
The scattering of massless fermions,
\begin{eqnarray}\label{e-process}
\mathrm{e}^-(p_1)+\mathrm{e}^+(p_2)\to \mathrm{f}^-(p_3)+\mathrm{f}^+(p_4),
\end{eqnarray}
depends on two invariants, chosen here to 
be the squared centre-of-mass energy 
$s$ and the four-momentum transfer $t$:
\begin{align}
\label{e-kin}
 s &= (p_1+p_2)^2 = 4E_\mathrm{beam}^2,
 \\ \label{e-kint}
 T = -t &= -(p_2+p_3)^2 = \frac{s}{2}\left( 1 - \cos\theta \right),
\end{align}
where $\cos\theta$ is the scattering angle in the centre-of-mass system. We will further  use
\begin{equation}
\label{e-kinu}
  U = -u = -(p_2+p_4)^2 = \frac{s}{2}\left( 1 + \cos\theta \right).
\end{equation}
Often, the matrix element is split into contributions from $s$-channel 
{
photon exchange and the remaining contributions, which include $s$-channel Z boson exchange}:
\begin{equation}
\label{e-gzborn}
 {\cal M}^{(0)}(\mathrm{e}^-\mathrm{e}^+\to \mathrm{f}^-\mathrm{f}^+) 
 =  
  {\cal M}_{\gamma}^{(0)}(\mathrm{e}^-\mathrm{e}^+\to \mathrm{f}^-\mathrm{f}^+) +  {\cal M}_\mathrm{Z}^{(0)}(\mathrm{e}^-\mathrm{e}^+\to \mathrm{f}^-\mathrm{f}^+)
.
\end{equation}
At the Born level, one has
\begin{align}
 \label{e-gzbornz}
  {\cal M}_\mathrm{Z}^{(0,\mathrm{B})}(\mathrm{e}^-\mathrm{e}^+\to \mathrm{f}^-\mathrm{f}^+)
  &= 4\mathrm{i}e^2\frac{\chi_\mathrm{Z}(s)}{s} \,
  \left(  v_\mathrm{e}^\mathrm{B}-a_\mathrm{e}^\mathrm{B} \gamma_5 \right)\gamma_{\alpha} \otimes 
 \left(  v_\mathrm{f}^\mathrm{B}-a_\mathrm{f}^\mathrm{B} \gamma_5 \right)\gamma^{\alpha},
 \\
 \label{e-gzborng}
   {\cal M}_{\gamma}^{(0,\mathrm{B})}(\mathrm{e}^-\mathrm{e}^+\to \mathrm{f}^-\mathrm{f}^+) &= \frac{\mathrm{i}e^2}{s}\,Q_\mathrm{e} Q_\mathrm{f}\, \gamma_\alpha \otimes \gamma^\alpha, 
    \end{align}
where we have used the shorthand notation
\begin{equation}
\Gamma_1\otimes\Gamma_2 \equiv
(\bar{v}_\mathrm{e}\Gamma_1u_\mathrm{e})(\bar{u}_\mathrm{f}\Gamma_2v_\mathrm{f}).
\end{equation}
Furthermore,
\begin{equation}
\label{e-smat-gam1}
 \chi_\mathrm{Z}(s) = \frac{G_\mathrm{F}M_\mathrm{Z}^2}{\sqrt{2}~8\pi\alpha_\mathrm{em}}~K_\mathrm{Z}(s) ~ \approx ~ 
0.37393 \,K_\mathrm{Z}(s).
\end{equation}
In the \emph{pole scheme}, where $\bar{M}_\mathrm{Z}$ is defined as the real part of the pole of the S-matrix, one has
\begin{equation}
\label{e-kzpole}
K_\mathrm{Z}(s) = (1+\mathrm{i}\bar\Gamma_\mathrm{Z}/\bar M_\mathrm{Z})^{-1}\,\frac{s}{s-\bar{M}_\mathrm{Z}^2+\mathrm{i}\bar{M}_\mathrm{Z}\bar\Gamma_\mathrm{Z}}
.
\end{equation}
In many analyses of the LEP era, it was assumed that the Z width is $s$-dependent, because it is a result of 
renormalizing the Z boson self-energy, which depends on $s$. 
Owing to the absence of production thresholds around the Z peak, one finds, to a very good level of precision:
\begin{equation}\label{e-smat-gam2}
K_\mathrm{Z}(s) = \frac{s}{s-M_\mathrm{Z}^2+\mathrm{i}M_\mathrm{Z}\Gamma_\mathrm{Z}(s)}, \qquad
\Gamma_\mathrm{Z}(s) =  \frac{s}{M_\mathrm{Z}^2} ~ \Gamma_\mathrm{Z}.
\end{equation}
Both definitions are related \cite{Bardin:1988xt}:
\begin{align}
\label{e-smat-gam3}
 {\bar M}_\mathrm{Z} & \approx M_\mathrm{Z} - \frac{1}{2} \frac{\Gamma_\mathrm{Z}^2}{M_\mathrm{Z}}   \approx M_\mathrm{Z} - 34 \UMeV, \\
 {\bar \Gamma}_\mathrm{Z} & \approx \Gamma_\mathrm{Z} - \frac{1}{2} \frac{\Gamma_\mathrm{Z}^3}{M_\mathrm{Z}^2}  \approx \Gamma_\mathrm{Z} - 0.9 \UMeV.
\end{align}
 The Born couplings are:
 \begin{align}
  Q_\mathrm{e} &= -1, 
  \\[2mm]
  a_\mathrm{f}^\mathrm{B} &= \pm \tfrac{1}{2},
  \label{e-aB}
  \\[2mm]
  v_\mathrm{f}^\mathrm{B} &= a_\mathrm{f}^\mathrm{B} \left( 1- 4|Q_\mathrm{f}| \sin^2\theta_\mathrm{W}\right),
  \label{e-vB}
\intertext{with}
  \sin^2\theta_\mathrm{W} &= 1-\frac{M_\mathrm{W}^2}{M_\mathrm{Z}^2}.
\nonumber
  \end{align}

Beyond the Born level, one can write
 \begin{align}
 \label{e-qedR}
{\cal M}^{(0)}_\gamma(\mathrm{e}^-\mathrm{e}^+\to \mathrm{f}^-\mathrm{f}^+) &=
 \frac{4\pi \mathrm{i} \alpha_\mathrm{em}(s)}{s} Q_\mathrm{e} Q_\mathrm{f}
  \gamma_\alpha \otimes \gamma^\alpha, \\
\label{e-genmatrix}
 {\cal M}^{(0)}_\mathrm{Z}(\mathrm{e}^-\mathrm{e}^+\to \mathrm{f}^-\mathrm{f}^+)
 &=  
4\mathrm{i}e^2\frac{\chi_\mathrm{Z}(s)}{s}
\left[
M_{vv}^\mathrm{ef}          
 \gamma_{\alpha}  \otimes \gamma^{\alpha}
- M_{av}^\mathrm{ef}
  \gamma_{\alpha}\gamma_5  \otimes \gamma^{\alpha}   
 - M_{va}^\mathrm{ef}   
 \gamma_{\alpha}  \times \gamma^{\alpha}\gamma_5 
 +  M_{aa}^\mathrm{ef}   
   \gamma_{\alpha}\gamma_5 \otimes \gamma^{\alpha}\gamma_5
 \right]
 .
  \end{align}
Here, $\alpha_\mathrm{em}(s)$ is the running electromagnetic coupling in the Standard Model, whose value is extracted from 
$\mathrm{e}^+\mathrm{e}^-$ data and 
theory {{\cite{Steinhauser:1998rq-new, Teubner:2012qb,Davier:2017zfy,Jegerlehner:2017zsb,Keshavarzi:2018mgv}}}.
The complex coefficients $M_{vv}$, $M_{va}$, $M_{av}$, and $M_{aa}$ are, by construction, gauge-invariant and contain all contributions of a given perturbative order. It is important to note that they contain not only Z boson vertex and self-energy corrections, but also corrections to the $s$-channel photon-exchange contribution (beyond the running of $\alpha$) and box contributions. Therefore, the general matrix element 
(Eq. \eqref{e-genmatrix}) does not factorize into initial- and final-state form factors of the Z boson.

  From a comparison with the Born matrix element ${\cal M}_\mathrm{Z}^{(0,\mathrm{B})}$ (\Eref{e-gzbornz}), we see the correspondences:
  \begin{eqnarray}\label{e-borndefs}
   M_{vv}^\mathrm{ef,B} = v_\mathrm{e}^\mathrm{B}v_\mathrm{f}^\mathrm{B},
\qquad
   M_{va}^\mathrm{ef,B} = v_\mathrm{e}^\mathrm{B}a_\mathrm{f}^\mathrm{B},
\qquad
   M_{av}^\mathrm{ef,B} = a_\mathrm{e}^\mathrm{B}v_\mathrm{f}^\mathrm{B},
\qquad
   M_{aa}^\mathrm{ef,B} = a_\mathrm{e}^\mathrm{B}a_\mathrm{f}^\mathrm{B}.
     \end{eqnarray}

Beyond the Born level, there exists, in general, no set of couplings $v_e, a_\mathrm{e}, v_\mathrm{f}, a_\mathrm{f}$ 
allowing one to write the matrix element as in 
\Eref{e-borndefs}.
Of course, if the non-factorizing 
part of radiative corrections in the Standard Model, \eg from  weak insertions to the photon self-energy or 
from one-loop box diagrams at the Z peak,  or in new physics models is small, 
then factorization is approximately fulfilled. 

An alternative language, in terms of the weak vertex form factors $\rho_\mathrm{f}$ and 
$\kappa_\mathrm{f}$, was first introduced in Refs. \cite{Sirlin:1980nh,Marciano:1980pb} for Z boson decay.
In Refs. \cite{Bardin:1988by,Bardin:1989tq,Bardin:1992jc}, based on Ref. \cite{Bardin:1982sv}, the concept was generalized to 
2$\to$2 scattering by splitting the effective 
weak mixing angle into three of them.  
This goes as follows:
  \begin{align}
  \label{e-smat-rhokappa}
  M_{aa}^\mathrm{ef} &= I_\mathrm{e}I_\mathrm{f}  \rho_\mathrm{Z}  
  {
  = \pm\tfrac{1}{4}\rho_\mathrm{Z} a_\mathrm{e}^\mathrm{ZF} a_v^\mathrm{ZF}}
,
   \\\label{e-rhokappa2}
\frac{M_{av}^\mathrm{ef}}{M_{aa}^\mathrm{ef}}
&\equiv  1-4 |Q_\mathrm{f}| \kappa_\mathrm{f} \sin^2\theta_\mathrm{W}
 {{ =  \frac{v_\mathrm{f}^\mathrm{ZF}}{ a_\mathrm{f}^\mathrm{ZF}} }}
,
   \\
\frac{M_{va}^\mathrm{ef}}{M_{aa}^\mathrm{ef}}
&\equiv  1-4 |Q_\mathrm{e}| \kappa_\mathrm{e} \sin^2\theta_\mathrm{W}
 {{ =  \frac{v_\mathrm{e}^\mathrm{ZF}}{ a_\mathrm{e}^\mathrm{ZF}} }}
,
   \\\label{e-rhokappa3}
\frac{M_{vv}^\mathrm{ef}}{M_{aa}^\mathrm{ef}}
&\equiv   
 1-4(|Q_\mathrm{e}|\kappa_\mathrm{e}+|Q_\mathrm{f}|\kappa_\mathrm{f})\sin^2\theta_\mathrm{W}
 + 
16 |Q_\mathrm{e}Q_\mathrm{f}|^2 \sin^4\theta_\mathrm{W} \kappa_\mathrm{ef}
 {{ =  \frac{v_\mathrm{ef}^\mathrm{ZF}}{ a_\mathrm{e}^\mathrm{ZF} a_\mathrm{f}^\mathrm{ZF}} }}
,
\end{align}
where $I_\mathrm{f}=\pm\frac{1}{2}$ is the weak isospin of fermion f.
We indicate here the relation to ZFITTER notions by introducing the effective couplings $v_\mathrm{e}^\mathrm{ZF},v_\mathrm{f}^\mathrm{ZF},v_\mathrm{ef}^\mathrm{ZF}$, while $a_\mathrm{e}^\mathrm{ZF} = a_\mathrm{f}^\mathrm{ZF} = 1$. {
Note the different normalization compared with the symbols without superscript `ZF'.}
Further, 
\begin{align}
\label{e-factoriz}
v_\mathrm{ef}^\mathrm{ZF} &= v_\mathrm{e}^\mathrm{ZF} v_\mathrm{f}^\mathrm{ZF} + \Delta_\mathrm{ef},
\\ 
\label{e-Deltaef}
\Delta_\mathrm{ef} &= 16|Q_\mathrm{e}Q_\mathrm{f}| \sin^4\theta_\mathrm{W} (\kappa_\mathrm{ef} -\kappa_\mathrm{e}\kappa_\mathrm{f})
.
\end{align}
In terms of $\rho_\mathrm{Z}$ and $\kappa_i$, the Z exchange matrix element may be again  rewritten into a simple form.
We 
quote 
from Eq. (3.3.1) in Ref. \cite{Bardin:1999yd}:
\begin{align}
 {\cal M}^{(0)}_\mathrm{Z}(s,t)
\sim 
        4 \mathrm{i} e^2   \frac{\chi_\mathrm{Z}(s)}{s}
       ~ I_\mathrm{e}I_\mathrm{f} \rho_\mathrm{Z}(s,t) ~
         &\bigl\{
        \textcolor{black}{  
        \gamma_{\alpha}( 1-\gamma_5 ) \otimes \gamma^{\alpha}( 1-\gamma_5 )
        } \nonumber   
              \\
               \nonumber 
               & \quad
-4 |Q_\mathrm{e}| \sin^2\theta_\mathrm{W} {\kappa_\mathrm{e}(s,t)}
        \textcolor{black}
        {
\gamma_{\alpha} \otimes \gamma^{\alpha}( 1-\gamma_5 )
        }
\\
 \nonumber 
  & \quad
       -4 |Q_\mathrm{f}| \sin^2 \theta_\mathrm{W} {\kappa_\mathrm{f}(s,t)}
        \textcolor{black}{
 \gamma_{\alpha}( 1-\gamma_5 ) \otimes \gamma^{\alpha}
        }                             
             \\
                              & \quad
+
16 |Q_\mathrm{e} Q_\mathrm{f}| \sin^4 \theta_\mathrm{W}  \kappa_\mathrm{ef}(s,t)
        \textcolor{black}{
\gamma_{\alpha}\otimes \gamma^{\alpha}
        } \bigr\}.
\label{e-smat-processrhokappadef}
\end{align}
Here, the matrix element is not written in terms of vector and axial-vector components, as in \Eref{e-genmatrix}, but in 
terms of left-handed and vector components.

As examples, 
Eqs.~(4.12) and (4.21) in Ref. \cite{Bardin:1980fe} give the contributions to the 
general 2$\to$2 matrix element in the unitary gauge. They correspond to a generic massive crossed one-loop box diagram of the 
ZZ and WW box type (Fig.~4.11 in Ref. \cite{Bardin:1980fe}) and to the two-photon box diagrams (Fig.~4.20 in Ref. \cite{Bardin:1980fe}). 
The  two-photon box diagrams must be combined with initial--final-state interference soft photon bremsstrahlung in order to get a finite result.
   Reference \cite{Bardin:1980fe} demonstrates how the box diagrams contribute to the form factors introduced in \Eref{e-genmatrix}.
For detailed definitions, we refer to the original.
External fermion masses are taken into account only in the pole terms,  to allow the exact 
cancellation of these pole terms in observable quantities. 
From the QED boxes, there are divergent contributions only to the structure $\gamma_\alpha \otimes \gamma^\alpha$, 
while finite terms contribute to the two combinations 
$\gamma_\alpha \otimes \gamma^\alpha \pm \gamma_\alpha\gamma_5 \otimes \gamma^\alpha\gamma_5$.
Thus, the photon box terms go exclusively into $M_{a_\mathrm{e}a_\mathrm{f}} = a_\mathrm{e}a_\mathrm{f} \rho_\mathrm{Z} $ and  $M_{v_\mathrm{ef}} = \rho_\mathrm{Z} v_\mathrm{ef}$, 
while the ZZ and WW boxes contribute to the various form factors quite differently. 
Evidently, all box diagrams violate the factorization of the Born matrix element (\Eref{e-gzbornz}).
From a closer inspection of the analytical representations, one may easily read off the contributions to residues 
$R$ and background terms $B^{(n)}$ of the corresponding Laurent series, introduced in the next subsection.



\subsection{\label{s-smat-laurent}The Z resonance as a Laurent series}
So far, we have kept the c.m.\ energy arbitrary. However, the most sensitive $\mathrm{e}^+\mathrm{e}^-$ measurements have been performed near the Z pole, so that we aim for a precise description for $s \approx M_\mathrm{Z}^2$;
let us mention Ref. \cite{Borrelli:1989bd} for {\textcolor{black}{a}} model-independent approach and Refs. \cite{Stuart:1991xk,Stuart:1991cc} for 
a discussion using the Standard Model.
The completely model-independent S-matrix approach, ignoring the notion of loop corrections, was advocated in Refs. 
\cite{Leike:1991pq,Riemann:1992gv,Kirsch:1994cf,Riemann:2015wpn}.
Near a resonance, the perturbative series 
{{can be considered as a meromorphic function in $s$ with a simple pole; close to the pole, it 
may be represented by a Laurent series.
This introduces the pole position $s_0=s-\bar M_\mathrm{Z}^2+\mathrm{i} \bar M_\mathrm{Z} \bar \Gamma_\mathrm{Z}$ as}}
an additional parameter, and when $s-\bar M_\mathrm{Z}^2$ becomes small, the Z 
boson width, which itself is a higher-order term in perturbative quantum
field theory (QFT),  becomes important.
The width appears in the Breit--Wigner resonance term $1/(s-\bar M_\mathrm{Z}^2 + \mathrm{i} \bar M_\mathrm{Z} \bar\Gamma_\mathrm{Z})$, see \Eref{e-kzpole}.
Further, one has to respect gauge-invariance and unitarity.
These conditions are fulfilled by the `pole scheme', where the matrix element is constructed as a Laurent expansion in 
the $s$-plane with a single simple pole and a Taylor series called `background' 
\cite{Willenbrock:1991hu,Sirlin:1991fd,Stuart:1991xk,Leike:1991pq,Stuart:1991cc,Veltman:1992tm,Passera:1998uj,%
Gambino:1999ai,%
Bohm:2000jw}:
 \begin{equation}\label{JGTR-lau}
{\cal M} = \frac{R}{s-s_0} + \sum_{n=0}^{\infty} (s-s_0)^n ~ B^{(n)}
,
\end{equation}
where
\begin{equation}\label{JGTR-lau2}
s_0 = {\bar M}_\mathrm{Z}^2 + \mathrm{i} {\bar M}_\mathrm{Z}  {\bar \Gamma}_\mathrm{Z}
.
\end{equation}
{{The derivation of a scheme for realistic analyses from 
this ansatz is called the S-matrix approach 
\cite{Leike:1991pq,Riemann:1992gv,Kirsch:1994cf,gruenewald-smatasy:2005}.
A mini-review with a  collection of experimental applications is given in Ref. \cite{Riemann:2015wpn}.}}

The residue $R$ and the coefficients $B^{(n)}$ are complex numbers characteristic of the process, and ${\bar M}_\mathrm{Z}$ and 
${\bar \Gamma}_\mathrm{Z}$ are the universal mass and width of the Z particle. 
The full process can be represented 
by introducing the four independent helicity matrix elements that 
describe the 2$\to$2 scattering of massless fermions.
We follow Ref. \cite{Cahn:1986qf},
and we will use the notation 
\begin{eqnarray}\label{e-matrix4}
 i = [1,2,3,4] = \mathrm{[(LR)(LR), (LR)(RL),(RL)(LR),(RL)(RL)]}.
\end{eqnarray}
The matrix element with photon exchange is helicity blind: 
\begin{align}
\label{e-matrix4cahn5-1}
{\cal M}_{1,\gamma} &= {\cal M}_{4,\gamma} =
  4\pi\alpha_\mathrm{em}(s)  Q_\mathrm{e} Q_\mathrm{f}  \frac{U}{s},
  \\
\label{e-matrix4cahn5-2}
{\cal M}_{2,\gamma} &= {\cal M}_{3,\gamma} 
=
  4\pi\alpha_\mathrm{em}(s)  Q_\mathrm{e} Q_\mathrm{f}  \frac{T}{s}.
\end{align}
We remind the reader that the definitions of $U$ and $T$ are given in Eqs.  \eqref{e-kint} and \eqref{e-kinu}.
For the weak amplitudes, ${\cal M}_{i,\mathrm{Z}}$, one finds 
\begin{align}
\label{e-smat-matrix1} 
 {\cal M}_{1,\mathrm{Z}} &=  {\cal M}_\mathrm{Z} (\mathrm{e}^-_\mathrm{L}\mathrm{e}_\mathrm{R}^+\to \mathrm{f}_\mathrm{L}^-\mathrm{f}_\mathrm{R}^+)
\,=\, C
\left[
M_{aa}^\mathrm{ef} + M_{av}^\mathrm{ef} + M_{va}^\mathrm{ef} + M_{vv}^\mathrm{ef}
 \right]
  \frac{U}{s-s_0}
 ,
 \\  \label{e-smat-matrix2}
 {\cal M}_{2,\mathrm{Z}} &= 
 {\cal M}_\mathrm{Z} (\mathrm{e}^-_\mathrm{L}\mathrm{e}_\mathrm{R}^+\to \mathrm{f}_\mathrm{R}^-\mathrm{f}_\mathrm{L}^+)
\,=\, C
\left[
-M_{aa}^\mathrm{ef} + M_{av}^\mathrm{ef} - M_{va}^\mathrm{ef} + M_{vv}^\mathrm{ef}
 \right]
  \frac{T}{s-s_0},
 \\\label{e-smat-matrix3}
 {\cal M}_{3,\mathrm{Z}} &= 
  {\cal M}_\mathrm{Z} (\mathrm{e}^-_\mathrm{R}\mathrm{e}_\mathrm{L}^+\to \mathrm{f}_\mathrm{L}^-\mathrm{f}_\mathrm{R}^+)
\,=\, C
\left[
-M_{aa}^\mathrm{ef} - M_{av}^\mathrm{ef} + M_{va}^\mathrm{ef} + M_{vv}^\mathrm{ef}
 \right]
  \frac{T}{s-s_0},
 \\\label{e-smat-matrix4}
 {\cal M}_{4,\mathrm{Z}} &=
{\cal M}_\mathrm{Z} (\mathrm{e}^-_\mathrm{R}\mathrm{e}_\mathrm{L}^+\to \mathrm{f}_\mathrm{R}^-\mathrm{f}_\mathrm{L}^+)
\,=\, C
\left[
M_{aa}^\mathrm{ef} - M_{av}^\mathrm{ef} - M_{va}^\mathrm{ef} + M_{vv}^\mathrm{ef}
 \right]
  \frac{U}{s-s_0}
  , \\
C &= 2\frac{G_\mathrm{F}}{\sqrt{2}}\;\frac{\bar M_\mathrm{Z}^2}{1+\mathrm{i}\bar \Gamma_\mathrm{Z}/\bar M_\mathrm{Z}}.
   \end{align}
   In the Born approximation, 
using  \Eref{e-borndefs} and $\bar\Gamma_\mathrm{Z} \to 0$, we recover
the helicity 
amplitudes of Ref. \cite{Cahn:1986qf}. ({Here, one must observe a sign convention difference for the axial 
couplings. In Ref. \cite{Cahn:1986qf}, it is $Q_\mathrm{L} \sim Q_V-Q_A$. Intuitively, in a left-handed or $(V-A)$ theory, 
there is only a left-handed coupling, and the vector and axial couplings are equal,  $V=+A$, and so it would be 
$Q_\mathrm{L} = V-A=0$.  Owing to a sign change of $Q_A$ in Ref. \cite{Cahn:1986qf}, the combination $V+A$ becomes $Q_V-Q_A$, \etc})

We mention here that 
{{the introduction of a photon-exchange amplitude may be formally avoided by putting the corresponding terms  into the 
Z amplitude with the replacement}}
\begin{eqnarray}
 \label{e-smat-ffzg}
M_{vv}^\mathrm{ef} \to M_{vv}^\mathrm{ef}
 + \frac{s-s_0}{s} ~ Q_\mathrm{e}Q_\mathrm{f} \frac{4\pi\alpha}{C}.
\end{eqnarray}
This is a consequence of the previous definitions, where both $v_\mathrm{ef}$ and the $\gamma$ exchange go together with the 
matrix element structure  $\gamma\otimes\gamma$.
See also Subsection C.\ref{s-smat-gammaZ}.

Each of the helicity matrix elements (Eqs. \eqref{e-smat-matrix1}--\eqref{e-smat-matrix4}) may then be expanded in a Laurent series according to \Eref{JGTR-lau}.
   As a result, 
the four helicity matrix elements
\begin{equation}
\label{e-matrixiifull}
 {\cal M}_i(\mathrm{e}^+\mathrm{e}^-\to \mathrm{f}{\bar {\mathrm{f}}}) = {\cal M}_{i,\gamma}^\mathrm{f} + {\cal M}_{i,\mathrm{Z}}^\mathrm{f}
\end{equation}
have  the generic form
\begin{equation}
\label{e-matrix} 
 {\cal M}_{i}(\mathrm{e}^+\mathrm{e}^-\to \mathrm{f}{\bar {\mathrm{f}}})
 = 
 \frac{s(1 \pm \cos\theta)}{2} ~
 \left(
 Q_\mathrm{e}Q_\mathrm{f}\frac{4\pi\alpha_\mathrm{em}(s)}{s} + \frac{R_\mathrm{f}^{(i)}}{s-s_0} + \sum_{n=0}^{\infty} B_{\mathrm{f},n}^{(i)} (s-s_0)^n
 \right)
 .
 \\
\end{equation}
This last equation
is the central element of a phenomenological analysis.
The form results from the demand of a correct determination of the loop corrections in accordance with 
unitarity, analyticity, and gauge-invariance. 
A phenomenological ansatz, which is expected not to contradict perturbation theory, should also respect this form.

We allow for an explicit account of $s$-channel photon exchange besides the Z, as previously discussed.
Because a Laurent series has only one pole term, the ansatz (\Eref{e-matrixiifull}) seems to contradict a consistent 
mathematical structure by also introducing the photon pole $1/s$.
In Section C.\ref{s-smat-gammaZ}, we will show that, in fact, this treatment can be put on a solid theoretical footing \cite{Riemann:2015wpn}. 

The  residues $R_\mathrm{f}^{(i)}$, the universal pole location $s_0$, and the coefficients $B_{\mathrm{f},n}^{(i)}$ 
can be computed in the Standard Model or some other model, or in a model-independent parametrization of new physics.
They are the basic building blocks for any experimental analysis. For the phenomenologically most important residue 
terms, one gets, for the four helicity matrix elements, 
\begin{equation}
\label{e-matrix-02}
R_\mathrm{f}^{(i)} = C \left [\pm M_{aa}^\mathrm{ef} \pm M_{av}^\mathrm{ef} \pm M_{va}^\mathrm{ef} + M_{vv}^\mathrm{ef} \right ]_{s=s_0}
.
\end{equation}
The sign conventions are introduced in \Eref{e-matrix4}.
In the Born approximation, the background terms vanish. They arise only from radiative corrections or from new physics; if 
not, the photon is formally made part of the background, see Section C.\ref{s-smat-gammaZ}.

In Monte Carlo approaches, which rely on the use of helicity matrix elements instead of  squared matrix elements, 
these considerations may be used to construct the  radiatively corrected matrix elements that must replace 
the Born matrix elements in the codes.

In Monte Carlo programs that already contain matrix elements with weak corrections,
such as {\tt KKMC} \cite{Jadach:1999vf},
where the original weak library DIZET \cite{Bardin:1989tq} of the ZFITTER project is implemented, one has to check whether 
the higher-order contributions in these matrix elements respect the origin from the pole renormalization scheme with Laurent 
series. 

\subsection{\label{s-smat-22MEasy}Electroweak pseudo-observables}
Let us now calculate the basic 2$\to$2 observables (Eqs. \eqref{e-2to2tot}--\eqref{e-2to2fbpol}) without specific 
assumptions 
about 
the underlying matrix elements.
The starting point are the differential cross-sections, composed of the incoherent sums of squared helicity 
matrix 
elements (Eqs. \eqref{e-smat-matrix1}--\eqref{e-smat-matrix4}):
\begin{align}
\label{eq-smat-sigma}
 \frac{\mathrm{d} \sigma_\mathrm{A}^{(0)}}{\mathrm{d} \cos\theta} 
 &=
 \frac{\pi \alpha^2}{2s} \left |\chi_\mathrm{Z}(s) \right |^2
 \left[
\left (1+\cos^2 \theta \right )  \textcolor{black}{k_\mathrm{A}} + (2 \cos\theta)  \textcolor{black}{k_{\mathrm{A},\mathrm{FB}}}
 \right] 
  \nonumber \\
 &= 
\frac{N_c}{32\pi s} \left(
 c_{\mathrm{A},1} |{\cal M}_1|^2 + c_{\mathrm{A},2}|{\cal M}_2|^2 + c_{\mathrm{A},3}|{\cal M}_3|^2 + c_{\mathrm{A},4}|{\cal M}_4|^2
\right )
 ,\quad  \mathrm{A}  = \mathrm{T,LR,pol,LRpol}
 .
\end{align}
The subscripts stand for: `T', total cross-section; `LR', electron's left--right asymmetry; `pol',  final-state 
polarization; `LRpol', the combination  of the last two asymmetries.
The coefficients are:
\begin{align}
\label{eq-smat-sigma5}
c_{\mathrm{T},i} &= [+1,+1,+1,+1],
\\
c_{\mathrm{LR},i} &= [+1,+1,-1,-1],
\\
c_{\mathrm{pol},i} &= [-1,+1,-1,+1],
\\
c_{\mathrm{LRpol},i} &= [-1,+1,+1,-1]
.
\end{align} 
The  terms in \Eref{eq-smat-sigma} with proportionality to $(1+\cos^2\theta)$ will contribute to the total cross-sections:
\begin{align}
\label{e-sigmatot22}
%
  \sigma_\mathrm{T}^{(0)}(s)
  &= 
\int_{-1}^{+1} \mathrm{d} \cos\theta \;  \frac{\mathrm{d} \sigma_\mathrm{T}^{(0)}}{\mathrm{d} \cos\theta} , \nonumber
  \\ 
  \nonumber 
  &=
  \frac{64\pi \alpha^2N_c}{3s} |\chi_\mathrm{Z}|^2 
  \bigl ( |M_{aa}^\mathrm{ef}|^2 + |M_{va}^\mathrm{ef}|^2 + |M_{av}^\mathrm{ef}|^2 + |M_{vv}^\mathrm{ef}|^2 \bigr ) 
 \\ \nonumber &=
{{
  \frac{4\pi \alpha^2}{3s} |\chi_\mathrm{Z}|^2 ~ 
 \left[ |a_\mathrm{e}^\mathrm{ZF}a_\mathrm{f}^\mathrm{ZF}|^2 + |v_\mathrm{e}^\mathrm{ZF}a_\mathrm{f}^\mathrm{ZF}|^2 + |a_\mathrm{e}^\mathrm{ZF}v_\mathrm{f}^\mathrm{ZF}|^2  + |v_\mathrm{ef}^\mathrm{ZF}|^2 \right] }}
  \\ 
  &\stackrel{\!\!\! \!\!\! \text{Z only} \!\!\! \!\!\!}{=} ~~   
  \frac{64\pi \alpha^2N_c}{3s} |\chi_\mathrm{Z}|^2 
 \left(|a_\mathrm{e}|^2 + |v_\mathrm{e}|^2 \right) \left(|a_\mathrm{f}|^2 + |v_\mathrm{f}|^2 \right)
 ,
 \end{align}
Here, and in the following, we do not show the photon-exchange terms explicitly.
Further, the notion that `Z only' restricts kinematics to $s=M_\mathrm{Z}^2$, and the non-factorizing parts are neglected. 
The $\chi_\mathrm{Z}(s)$ is defined in \Eref{e-smat-gam1}.
 In the foregoing, the terms $M_{aa}^\mathrm{ef}$, \etc, are assumed to cover both photon and Z exchange contributions. 
The approximated last line indicates the result after removal of the photon-exchange and box 
contributions, 
where the remaining terms $a_\mathrm{f}$ and $v_\mathrm{f}$ are the higher-loop generalizations of the Zf$^+$f$^-$ vertex couplings 
introduced in  Eqs. \eqref{e-aB} and \eqref{e-vB}.
They are related to the vertex form factor $\kappa_\mathrm{f}^\mathrm{Z}$ introduced in \Eref{e-sw2Zffkappa}.

The forward--backward asymmetric cross-section stems from the $(2\cos\theta)$ term:
 \begin{align}
 \label{e-sigmatot23}
 \sigma_\mathrm{T,FB}^{(0)} (s)
   &= 
\left[\int_{0}^{+1} - \int_{-1}^{0} \right] \mathrm{d} \cos\theta    \frac{\mathrm{d} \sigma_\mathrm{T}^{(0)}}{\mathrm{d} \cos\theta} ,
  \\ \nonumber 
&= \frac{16\pi\alpha^2}{s} |\chi_\mathrm{Z}|^2 \,
2\Re  \left \{M_{av}^\mathrm{ef} \left(M_{va}^\mathrm{ef} \right )^* + M_{aa}^\mathrm{ef}
\left (M_{vv}^\mathrm{ef} \right )^* \right \}
\nonumber \\ 
&= 
{{
\frac{16\pi \alpha^2}{s} |\chi_\mathrm{Z}|^2 {
|\rho_\mathrm{Z}|^2 ~ 
 2\Re  \left( a_\mathrm{e}^\mathrm{ZF*} a_\mathrm{f}^\mathrm{ZF*}  v_\mathrm{ef}^\mathrm{ZF} +  a_\mathrm{e}^\mathrm{ZF*} a_\mathrm{f}^\mathrm{ZF} v_\mathrm{e}^\mathrm{ZF} v_\mathrm{f}^\mathrm{ZF*}\right)}
 }}
    \nonumber \\
 &\stackrel{\!\!\! \!\!\! \text{Z only} \!\!\! \!\!\!}{=} ~~   
  \frac{16\pi \alpha^2}{s} |\chi_\mathrm{Z}|^2
\, \Re \{
\left(a_\mathrm{e} v_\mathrm{e}^* + v_\mathrm{e} a_\mathrm{e}^* \right) \left(a_\mathrm{f} v_\mathrm{f}^* + v_\mathrm{f} a_\mathrm{f}^* \right)\}
 .
  \end{align}
The forward--backward asymmetry can be evaluated as follows:
\begin{align}
\label{e-afbcomplete0}
 A_\mathrm{T,FB}^{(0)} (s) &= 
 \frac{\sigma_\mathrm{T,FB}^{(0)} (s)}{\sigma_\mathrm{T}^{(0)} (s)}
\nonumber  \\
  \nonumber
&= 
\frac{\left[\int_{0}^{+1} - \int_{-1}^{0} \right] \mathrm{d}\cos\theta \;
(|{\cal M}_1|^2 + |{\cal M}_2|^2  + |{\cal M}_3|^2 + |{\cal M}_4|^2)}
    {\int_{-1}^{+1} \mathrm{d} \cos\theta\;
    (|{\cal M}_1|^2 + |{\cal M}_2|^2  + |{\cal M}_3|^2 + |{\cal M}_4|^2)}
  \\ 
  \nonumber
 &= 
 \frac{3}{4}\,
\frac{2\,\Re \{M_{av}^\mathrm{ef}(M_{va}^\mathrm{ef})^* + M_{aa}^\mathrm{ef}(M_{vv}^\mathrm{ef})^* \}}{|M_{aa}^\mathrm{ef}|^2 + |M_{va}^\mathrm{ef}|^2 
+ |M_{av}^\mathrm{ef}|^2 + |M_{vv}^\mathrm{ef}|^2}
\nonumber \\ 
&=
\frac{3}{4}
\frac{ 2 \Re  \{
a_\mathrm{e}^\mathrm{ZF} v_\mathrm{e}^\mathrm{ZF*} a_\mathrm{f}^\mathrm{ZF*} v_\mathrm{f}^\mathrm{ZF} + a_\mathrm{e}^\mathrm{ZF}a_\mathrm{f}^\mathrm{ZF} v_\mathrm{ef}^\mathrm{ZF*}
\}
}  
{ |a_\mathrm{e}^\mathrm{ZF}a_\mathrm{f}^\mathrm{ZF}|^2 + |v_\mathrm{e}^\mathrm{ZF}a_\mathrm{f}^\mathrm{ZF}|^2 + |a_\mathrm{e}^\mathrm{ZF}v_\mathrm{f}^\mathrm{ZF}|^2  + \color{black}{|v_\mathrm{ef}^\mathrm{ZF}|^2 } }
\nonumber \\
 \nonumber
&=
\frac{3}{4}\,
\frac{
(a_\mathrm{e} v_\mathrm{e}^* + v_\mathrm{e} a_\mathrm{e}^* ) (a_\mathrm{f} v_\mathrm{f}^* + v_\mathrm{f} a_\mathrm{f}^* )
+ \Delta_\mathrm{FB}
}{\left( |a_\mathrm{e}|^2 + |v_\mathrm{e}|^2 \right) \left( |a_\mathrm{f}|^2 + |v_\mathrm{f}|^2 \right) + \Delta_\mathrm{T}
}
\\
 \nonumber
&\simeq 
\frac{3}{4}
\Biggl[
\frac{(a_\mathrm{e} v_\mathrm{e}^* + v_\mathrm{e} a_\mathrm{e}^* ) (a_\mathrm{f} v_\mathrm{f}^* + v_\mathrm{f} a_\mathrm{f}^* )}
{
 \left( |a_\mathrm{e}|^2 + |v_\mathrm{e}|^2 \right) 
 \left( |a_\mathrm{f}|^2 + |v_\mathrm{f}|^2 \right) }
 +
 \frac{\Delta_\mathrm{FB}}{\left( |a_\mathrm{e}|^2 + |v_\mathrm{e}|^2 \right) 
 \left( |a_\mathrm{f}|^2 + |v_\mathrm{f}|^2 \right)} 
\nonumber \\
 \nonumber
& \qquad \quad  -  
 \frac{ (2\,\Re \{a_\mathrm{e} v_\mathrm{e}\})  \, (2\, \Re  \{a_\mathrm{f} v_\mathrm{f}\})  
}
{
\left[ \left( |a_\mathrm{e}|^2 + |v_\mathrm{e}|^2 \right) 
 \left( |a_\mathrm{f}|^2 + |v_\mathrm{f}|^2 \right)\right]^2 }
 \Delta_\mathrm{T} 
 \Biggr]
 \nonumber \\
   &\equiv  \frac{3}{4} A_\mathrm{e} A_\mathrm{f} + c_\mathrm{FB} \Delta_\mathrm{FB} - c_\mathrm{T} \Delta_\mathrm{T}
\end{align}
Here, we use the abbreviations:
\begin{align}
\Delta_\mathrm{T}
&= 
|\Delta v_\mathrm{ef}|^2 + 2 \Re  \left( \frac{v_\mathrm{e}}{a_\mathrm{e}} \frac{v_\mathrm{f}}{a_\mathrm{f}} 
\Delta v_\mathrm{ef}^*\right),
\\
\Delta_\mathrm{FB} &= 2 \Re  \Delta v_\mathrm{ef} .
\end{align}
The $\Delta v_\mathrm{ef}$ is defined in \Eref{e-Deltaef}.
 Assuming factorization, 
 {
 which is the case for the $s$-channel Z boson contribution,} one has $\Delta_\mathrm{T} = \Delta_\mathrm{FB} =0$.
The $\Delta_\mathrm{T}$ and $\Delta_\mathrm{FB}$ capture subleading 
contributions from photon-exchange and box contributions that are suppressed by $(s-s_0)/s$ (these include the 
non-factorizing term $\Delta M_{vv}$, introduced in the previous subsection). 
 
In the last line of \Eref{e-afbcomplete0}, we identify the asymmetry parameters $A_\mathrm{e}, A_\mathrm{f}$:
 \begin{equation}
 A_\mathrm{f} = \frac{2\,\Re  \{a_\mathrm{f}v_\mathrm{f}\}}{a_\mathrm{f}^2+|v_\mathrm{f}|^2}
 \equiv  
 \frac{1-4|Q_\mathrm{f}|\sin^2\theta_\mathrm{f}^{\mathrm{eff}}}{1 - 4 |Q_\mathrm{f}|\sin^2\theta_\mathrm{f}^{\mathrm{eff}}
 + 8 |Q_\mathrm{f}|^2 \sin^4 \theta_\mathrm{f}^{\mathrm{eff}}     }
.
\end{equation}
The effective weak mixing angle is defined as
\begin{equation}
\label{e-smat-ewdef}
 \sin^2\theta_\mathrm{f}^{\mathrm{eff}} \equiv \Re  \{\kappa_\mathrm{Z}^\mathrm{f}\}  \sin^2\theta_\mathrm{W}
 ,
 \end{equation}
where $\kappa_\mathrm{Z}^\mathrm{f}$ is the Z exchange contribution of the vertex form factor $\kappa_\mathrm{f}$, see Eq. \eqref{e-rhokappa2}.

The $A_\mathrm{f}$ has a very weak dependence on 
the c.m.s. energy $s$, resulting from the loop corrections in $\kappa_\mathrm{Z}^\mathrm{f}$ or, equivalently, in $v_\mathrm{f}/a_\mathrm{f}$. 
$\Delta_\mathrm{T}$ and $\Delta_\mathrm{FB}$ depend more significantly on $s$.

The assumption that the form factors are independent of the scattering angle is correct for self-energies and vertices, 
but not for box diagrams, see Eqs.~(4.12) and (4.21) in Ref. \cite{Bardin:1980fe} as examples.  
There, the scalar one-loop functions, such as 
$A(q^2,q^2-S,M_1^2,M_2^2)$, $B(q^2,q^2-S,M_1^2,M_2^2)$, $C(q^2,q^2-S,M_1^2,M_2^2)$,
and $D(q^2-S,S)$, of the spin 
structures 
$\gamma_\alpha \otimes 
\gamma^\alpha$, \etc,  in the ZZ and WW box functions have additional angular dependences; they depend on both $S=s$ and  
$q^2=t$, where the latter contains the scattering angle.  The case is similar for the other box 
diagrams.

In a precision study, one has to determine by explicit calculations whether the angular dependences of the four-point functions 
may be neglected or not when studying angular-integrated observables.  

The remaining asymmetries are:
\begin{align}
 A_\mathrm{LR}^{(0)}(s) &= \frac{\sigma_{\mathrm{e}_\mathrm{L}} - \sigma_{\mathrm{e}_\mathrm{R}}}{\sigma_{\mathrm{e}_\mathrm{L}} + \sigma_{\mathrm{e}_\mathrm{R}}}
\nonumber \\
&=
\frac{ \int_{-1}^1 \mathrm{d} \cos\theta (
|{\cal M}_1|^2 + |{\cal M}_2|^2  - |{\cal M}_3|^2 -  |{\cal M}_4|^2)}
{
 \int_{-1}^1 \mathrm{d} \cos\theta (|{\cal M}_1|^2 + |{\cal M}_2|^2 + |{\cal M}_3|^2 +  |{\cal M}_4|^2)} 
\nonumber \\
&=
\frac{2\,\Re \left \{M_{aa}^\mathrm{ef} \left (M_{va}^\mathrm{ef} \right
)^* + M_{av}^\mathrm{ef} \left (M_{vv}^\mathrm{ef} \right )^* \right \}}{|M_{aa}^\mathrm{ef}|^2 + |M_{va}^\mathrm{ef}|^2 + |M_{av}^\mathrm{ef}|^2 + |M_{vv}^\mathrm{ef}|^2}
\nonumber \\  
 &\stackrel{\!\!\! \!\!\! \text{Z only} \!\!\! \!\!\!}{=} ~~ A_\mathrm{e} 
 ,
  \\   \label{e-2to2pol-2}
 \lambda_\mathrm{f} \equiv A_\mathrm{pol}^{(0)}(s) 
   &= \frac{\sigma_{\mathrm{f}_\mathrm{R}} - \sigma_{\mathrm{f}_\mathrm{L}}}{\sigma_{\mathrm{f}_\mathrm{R}} + \sigma_{\mathrm{f}_\mathrm{L}}}
\nonumber \\
&=
\frac{ \int_{-1}^1  \mathrm{d}\cos\theta (
-|{\cal M}_1|^2 + |{\cal M}_2|^2  - |{\cal M}_3|^2 +  |{\cal M}_4|^2)}
{
 \int_{-1}^1 \mathrm{d}\cos\theta (|{\cal M}_1|^2 + |{\cal M}_2|^2 + |{\cal M}_3|^2 +  |{\cal M}_4|^2)} 
   \nonumber \\
 &=
-\frac{2\,\Re \left \{M_{aa}^\mathrm{ef} \left (M_{av}^\mathrm{ef} \right)^* + M_{va}^\mathrm{ef} \left (M_{vv}^\mathrm{ef} \right)^* \right \}}{|M_{aa}^\mathrm{ef}|^2 + |M_{va}^\mathrm{ef}|^2 + |M_{av}^\mathrm{ef}|^2 + |M_{vv}^\mathrm{ef}|^2}
  \nonumber \\ 
&\stackrel{\!\!\! \!\!\! \text{Z only} \!\!\! \!\!\!}{=} ~~ 
 - A_\mathrm{f},   
   \\
     \label{e-2to2lrpol-2}
 A_\mathrm{LRpol}^{(0)}(s) &= 
 \frac{
 \int_{-1}^1 \mathrm{d} \cos\theta (
 -|{\cal M}_1|^2 + |{\cal M}_2|^2  + |{\cal M}_3|^2 - |{\cal M}_4|^2)}
{
 \int_{-1}^1 \mathrm{d} \cos\theta (|{\cal M}_1|^2 + |{\cal M}_2|^2 + |{\cal M}_3|^2 +  |{\cal M}_4|^2)} 
  \nonumber \\
 &=
-\frac{2\,\Re \left \{M_{av}^\mathrm{ef}\left (M_{va}^\mathrm{ef} \right)^* + M_{aa}^\mathrm{ef} \left (M_{vv}^\mathrm{ef} \right )^* \right \}}{|M_{aa}^\mathrm{ef}|^2 + |M_{va}^\mathrm{ef}|^2 + |M_{av}^\mathrm{ef}|^2 + |M_{vv}^\mathrm{ef}|^2}
 \nonumber \\
&\stackrel{\!\!\! \!\!\! \text{Z only} \!\!\! \!\!\!}{=} ~~ 
- A_\mathrm{e} A_\mathrm{f}  
,
 \\
 \label{e-2to2fbpol-2}
 A_\mathrm{polFB}^{(0)}(s) &= 
\frac{  
\left[\int_{0}^{+1} - \int_{-1}^{0} \right] \mathrm{d} \cos\theta
(
-|{\cal M}_1|^2 + |{\cal M}_2|^2  - |{\cal M}_3|^2 + |{\cal M}_4|^2 )}
{\int_{-1}^1 \mathrm{d} \cos\theta (|{\cal M}_1|^2 + |{\cal M}_2|^2 + |{\cal M}_3|^2 +  |{\cal M}_4|^2)}
\nonumber   \\ 
  \nonumber
  &=
  -\frac{3}{4}\;
\frac{2\,\Re \left \{M_{aa}^\mathrm{ef} \left (M_{va}^\mathrm{ef} \right
)^* + M_{av}^\mathrm{ef} \left (M_{vv}^\mathrm{ef} \right )^* \right \}}{|M_{aa}^\mathrm{ef}|^2 + |M_{va}^\mathrm{ef}|^2 + |M_{av}^\mathrm{ef}|^2 + |M_{vv}^\mathrm{ef}|^2}
  \\  
&\stackrel{\!\!\! \!\!\! \text{Z only} \!\!\! \!\!\!}{=} ~~ 
  - \frac{3}{4} A_\mathrm{e}                                    
 \\[2mm]
 \label{e-2to2lrfb-2}
  A_\mathrm{LRFB}^{(0)}(s)
 &= \frac{3}{4} 
\frac{2\,\Re \left \{M_{aa}^\mathrm{ef} \left (M_{av}^\mathrm{ef} \right
)^* + M_{va}^\mathrm{ef} \left (M_{vv}^\mathrm{ef} \right)^* \right \}}{|M_{aa}^\mathrm{ef}|^2 + |M_{va}^\mathrm{ef}|^2 + |M_{av}^\mathrm{ef}|^2 + |M_{vv}^\mathrm{ef}|^2}
    \nonumber \\
  &\stackrel{\!\!\! \!\!\! \text{Z only} \!\!\! \!\!\!}{=} ~~ \frac{3}{4} A_\mathrm{f } 
  ,
  \\
   A_\mathrm{LRpolFB,Z}^{(0)}(s) 
  &=  
- \frac{3}{4} \,
{
      \frac{ \left (a_\mathrm{e}^\mathrm{ZF} \right )^2 \left (a_\mathrm{f}^\mathrm{ZF}
      \right )^2 + \left  (a_\mathrm{e}^\mathrm{ZF} \right )^2 |v_\mathrm{f}^\mathrm{ZF}|^2+ |v_\mathrm{e}^\mathrm{ZF}|^2 \left (a_\mathrm{f}^\mathrm{ZF} \right)^2 + |v_\mathrm{ef}^\mathrm{ZF}|^2 }
    {\left ( |a_\mathrm{e}^\mathrm{ZF}a_\mathrm{f}^\mathrm{ZF}|^2 + |v_\mathrm{e}^\mathrm{ZF}a_\mathrm{f}^\mathrm{ZF}|^2 + |a_\mathrm{e}^\mathrm{ZF}v_\mathrm{f}^\mathrm{ZF}|^2  + |v_\mathrm{ef}^\mathrm{ZF}|^2 \right )}}
         \nonumber \\
  &= - \frac{3}{4}
.
  \end{align}
  
At the Z peak, $s=M_\mathrm{Z}^2$, the photon and box terms become suppressed by their proportionality to $(s-s_0)/s \to 
\mathrm{i}\Gamma_\mathrm{Z}/M_\mathrm{Z}$.
The photon term and the $\gamma \mathrm{Z}$ interference are  generally  numerically  important for experimental analyses.

For definition and measurement of $A_\mathrm{polFB,Z}^{(0)}$ and other asymmetries, see Ref. \cite{Alexander:1996ha}.
We follow the definitions used in Ref. \cite{Riemann:2010zz}, where most of the asymmetries are defined and calculated 
in the Born approximation.
Further, 
$\lambda_\mathrm{f}$ is the polarization of 
$\mathrm{f}^-$.

All these considerations were independent of assuming the Standard Model or any other underlying model.
They are introduced at this length in order to show that the hard scattering process around the Z peak contains exactly four form 
factors per final-state fermion f, which suffice to describe any observable:
\begin{equation}
\label{e-2to24a}
M_{aa}^\mathrm{ef}=a_\mathrm{e}a_\mathrm{f},
\qquad 
M_{av}^\mathrm{ef}=a_\mathrm{e}v_\mathrm{f}, 
\qquad 
M_{va}^\mathrm{ef}=v_\mathrm{e}a_\mathrm{f}, 
\qquad
M_{vv}^\mathrm{ef}=v_\mathrm{ef}, 
\end{equation}
or 
\begin{eqnarray}\label{e-2to24b}
\rho_\mathrm{Z}, \kappa_\mathrm{e}, \kappa_\mathrm{f}, \kappa_\mathrm{ef},
\end{eqnarray}
or 
\begin{eqnarray}\label{e-2to24b-02}
\rho_\mathrm{Z}, \sin^2\theta_\mathrm{W}^{\mathrm{e},\mathrm{eff}} , \sin^2\theta_\mathrm{W}^{\mathrm{f},\mathrm{eff}}, \sin^2\theta_\mathrm{W}^{\mathrm{ef},\mathrm{eff}}
.
\end{eqnarray}
If we further restrict ourselves to the Z exchange amplitude only and neglect non-factorizing terms,  
these reduce to three independent quantities, $M_{aa}^\mathrm{ef,Z}, M_{av}^\mathrm{ef,Z}, M_{va}^\mathrm{ef,Z}$, since  $M_{vv}^\mathrm{ef,Z} = M_{av}^\mathrm{ef,Z}M_{va}^\mathrm{ef,Z}/M_{aa}^\mathrm{ef,Z}$ and 
$\kappa_\mathrm{Z}^\mathrm{ef}=\kappa_\mathrm{Z}^\mathrm{e} \kappa_\mathrm{Z}^\mathrm{f}$. Another choice for these three quantities is, in this restricted sense:
\begin{equation}
\label{e-2to24b-0}
\rho_\mathrm{Z},  \sin^2\theta_\mathrm{W}^{\mathrm{e},\mathrm{eff}}, \sin^2\theta_\mathrm{W}^{\mathrm{f},\mathrm{eff}}
.
\end{equation}
Similarly, quantities such as $A_\mathrm{f}$ are only defined for the Z exchange amplitude; thus, additional steps are needed to extract them from the observables.
The art of a Z line shape analysis relies on the ability to reduce the many degrees of freedom from experiment to 
a sufficiently small set of intermediate variables, which are 
easily described by theory. 
With one-loop accuracy (and a little beyond), this was prepared in the ZFITTER package 
\cite{Bardin:1992jc,Bardin:1999yd,Arbuzov:2005ma,Akhundov:2013ons} and studied on many occasions, most notably, as reported in Refs. 
\cite{Bardin:1999gt,ALEPH:2005ab,Schael:2013ita} and references therein. 
\subsection{Loops in the 2$\to$2 matrix element
\label{s-loopstoEWPOs}}

We now address the question of how to compute the 2$\to$2 pseudo-observables in the Standard Model, or any extended models, up to some desired loop order. For concreteness, we show explicit expressions for the example of an evaluation to two-loop order in this subsection, but they can be straightforwardly extended to higher-loop orders.
This was reported notably in Refs. \cite{Stuart:1991xk,Stuart:1991cc,Veltman:1992tm} and in  Ref.
\cite{Awramik:2006uz} and the many 
references therein.

To proceed, one has to expand the four independent quantities, \eg $M_{aa}^\mathrm{ef}, M_{av}^\mathrm{ef}, 
M_{va}^\mathrm{ef}, M_{vv}^\mathrm{ef}$, in a Laurent series about the complex pole $s_0 = \bar{M}_\mathrm{Z}^2 + \mathrm{i}\bar{M}_\mathrm{Z}\bar{\Gamma}_\mathrm{Z}$. The most relevant pseudo-observables are defined for $s=\bar{M}_\mathrm{Z}^2$, so that the expansion parameter becomes $s-s_0 = -\mathrm{i}\bar{M}_\mathrm{Z}\bar{\Gamma}_\mathrm{Z}$. Since $\bar{\Gamma}_\mathrm{Z}/\bar{M}_\mathrm{Z} \sim {\cal O}(g^2)$, where $g$ is the weak coupling, one has to perform the expansion \emph{simultaneously} in $s-s_0$ and in the loop order of the form factors.
For example, if the residue term $R$ in \Eref{JGTR-lau} is expanded to two-loop order, then $B^{(0)}$ and $B^{(1)}$ should be expanded to one-loop and tree-level order, respectively, while all higher $B^{(n)}$ terms can be neglected. Specifically, the residue contribution of $M_{vv}^\mathrm{ef}$ at two-loop order is given by Refs. \cite{Veltman:1992tm,Awramik:2006uz}:
\begin{align}
\label{e-residueR}
R_{vv} &=
 v_\mathrm{e}^{(0)}  R_\mathrm{ZZ}  v_\mathrm{f}^{(0)} + \left[ v_\mathrm{e}^{(1)}
 \left(M_\mathrm{Z}^2 \right)\, v_\mathrm{f}^{(0)} +
   v_\mathrm{e}^{(0)}  v_\mathrm{f}^{(1)} \left(M_\mathrm{Z}^2 \right) \right] 
   \left[ 1+ {\Sigma_\mathrm{Z Z}^{(1)}}' \left(M_\mathrm{Z}^2 \right) \right] 
  \nonumber  \\
    \nonumber
 & \quad +  v_\mathrm{e}^{(2)} \left (M_\mathrm{Z}^2 \right )  v_\mathrm{f}^{(0)} 
+
   { v_\mathrm{e}^{(0)}  v_\mathrm{f}^{(2)}\left (M_\mathrm{Z}^2 \right )}
   + v_\mathrm{e}^{(1)}\left (M_\mathrm{Z}^2 \right)  v_\mathrm{f}^{(1)}
   \left (M_\mathrm{Z}^2 \right )
   \\
    & \quad - \mathrm{i} M_\mathrm{Z} \Gamma_\mathrm{Z} \left[ \mbox{$v_\mathrm{e}^{(1)}$}'
    \left (M_\mathrm{Z}^2 \right ) v_\mathrm{f}^{(0)} +
   v_\mathrm{e}^{(0)}  \mbox{$v_\mathrm{f}^{(1)}$}' \left (M_\mathrm{Z}^2
   \right ) \right],
\\
 R_\mathrm{ZZ} &= 
 1 - {\Sigma_\mathrm{Z Z}^{(1)}}' \left (M_\mathrm{Z}^2 \right) 
\nonumber \\
  \nonumber
 & \quad  - {\Sigma_\mathrm{Z Z}^{(2)}}' \left (M_\mathrm{Z}^2 \right) + \left( {\Sigma_\mathrm{Z Z}^{(1)}}' \left (M_\mathrm{Z}^2 \right) \right)^2
   + \mathrm{i} M_\mathrm{Z} \Gamma_\mathrm{Z} {\Sigma_\mathrm{Z Z}^{(1)}}''
   \left (M_\mathrm{Z}^2 \right) 
   \\
     & \quad - \frac{1}{M_\mathrm{Z}^4} \left( {\Sigma_{\gamma \mathrm{Z}}^{(1)}}
     \left (M_\mathrm{Z}^2 \right) \right)^2
   + \frac{2}{M_\mathrm{Z}^2}  {\Sigma_{\gamma \mathrm{Z}}^{(1)}}(M_\mathrm{Z}^2)      {\Sigma_{\gamma \mathrm{Z}}^{(1)}}' \left (M_\mathrm{Z}^2 \right).
\end{align}
Here, the superscript $(n)$ indicates the loop order and a prime denotes the derivative with respect to $s$.
The $\Sigma_{V_1V_2}, V_i= \mathrm{Z},\gamma$ stand for transverse gauge
boson self-energies, whereas $v_\mathrm{f}(s)$ is the vector form factor used in \Eref{e-sigmatot22}~ff. Note that
$v_\mathrm{f}^{(n)}(s)$ and $a_\mathrm{f}^{(n)}(s)$ are understood to include the effects of $\gamma$--Z mixing; see Ref. \cite{Awramik:2006uz} for more details.
Similar expressions for $M_{va}^\mathrm{ef},M_{av}^\mathrm{ef},M_{aa}^\mathrm{ef}$ can be obtained in an obvious way. 

The residue $R$ is essentially factorized into the product of the two renormalized vertex functions
$v_\mathrm{e}(M_\mathrm{Z}^2)$ and $v_\mathrm{f}(M_\mathrm{Z}^2)$, sandwiching $R_\mathrm{ZZ}$.
The $R_\mathrm{ZZ}$ subsumes the self-energy contributions.
The subsequent terms, $B_vv^{(n)}$ in the Laurent expansion contain contributions from the Z exchange amplitude, as well as the photon-exchange and box amplitudes \cite{Veltman:1992tm}.

Using these expressions, one can derive explicit formulae for the pertinent pseudo-observables. Since the latter are defined solely for the Z exchange, we assume that the photon-exchange and box contributions have been removed. Thus, the following expressions  are given solely in terms of Z amplitude form factors.

When computing the asymmetries, one can explicitly verify that terms involving self-energies and derivatives cancel, leading to the naively expected result \cite{Awramik:2006uz}
\begin{align}
\sinefff 
&\equiv 
\left(1- \frac{\bar{M}_\mathrm{W}^2}{\bar{M}_\mathrm{Z}^2} \right) 
        \re \left\{ 1 + \Delta\overline{\kappa}_\mathrm{Z}^\mathrm{f} \left
        (M_\mathrm{Z}^2 \right ) \right\} 
        \\ \nonumber
 &= \left(1- \frac{\bar{M}_\mathrm{W}^2}{\bar{M}_\mathrm{Z}^2} \right) \re
 \Biggl\{ 1+
   \frac{a_\mathrm{f}^{(1)}  v_\mathrm{f}^{(0)} - 
        v_\mathrm{f}^{(1)}  a_\mathrm{f}^{(0)}}{a_\mathrm{f}^{(0)} \left
        (a_\mathrm{f}^{(0)}-v_\mathrm{f}^{(0)}
        \right )}
        \Biggr|_{k^2 = M_\mathrm{Z}^2} 
        \\ \nonumber
 &
  \qquad  \qquad \qquad \qquad \quad   +   \frac{a_\mathrm{f}^{(2)}  v_\mathrm{f}^{(0)}  a_\mathrm{f}^{(0)} - 
        v_\mathrm{f}^{(2)} \left (a_\mathrm{f}^{(0)} \right )^2 -
        \left (a_\mathrm{f}^{(1)} \right )^2  v_\mathrm{f}^{(0)} +
        a_\mathrm{f}^{(1)}  v_\mathrm{f}^{(1)}  a_\mathrm{f}^{(0)}}{\left
        (a_\mathrm{f}^{(0)}
        \right )^2
       \left (a_\mathrm{f}^{(0)}-v_\mathrm{f}^{(0)} \right )} \Biggr|_{s = M_\mathrm{Z}^2}
   \Biggr\}, 
\label{eq:sin2}
\end{align}
which is compatible with
\begin{equation}
 {\bar \kappa}_\mathrm{Z}^\mathrm{f}(s) = \frac{1}{4|Q_\mathrm{f}|\sin^2\theta_\mathrm{W}} \left( 1 - \frac{v_\mathrm{f}(s)}{a_\mathrm{f}(s)} \right),  
\end{equation}
As before, the barred quantities refer to the pole scheme, see Subsection~C.\ref{s-smat-2to2me}.
At two-loop accuracy,  it is not necessary to
distinguish between barred and unbarred masses in the radiative corrections,
since $\overline{M}_\mathrm{Z}^2 - M_\mathrm{Z}^2 = \Oaa$.
For further details of notation, we refer to Ref. \cite{Awramik:2006uz}.

For the cross-section, one also needs the form factor $\bar\rho_\mathrm{ef}$ in addition to $\bar\kappa_\mathrm{f}$, see \Eref{e-smat-processrhokappadef}. Again, assuming that photon-exchange and box contributions have been removed, the result for the Z exchange contribution can be written as \cite{Freitas:2013dpa,Freitas:2014hra}
\begin{align}
\bar\rho_\mathrm{Z}^\mathrm{ef} &= F_A^\mathrm{e} F_A^\mathrm{f} (1+\delta X), \\
F_A^\mathrm{f} &= \Biggl[ \left (a_\mathrm{f}^{(0)} \right )^2 \left[ 1-\re {\Sigma_\mathrm{Z Z}^{(1)}}' - \re{\Sigma_\mathrm{Z Z}^{(2)}}' + \left (\re{\Sigma_\mathrm{Z Z}^{(1)}}' \right )^2 \right] + 2 \re \left \{a_\mathrm{f}^{(0)}a_\mathrm{f}^{(1)}
\right \}
\left[1-\re{\Sigma_\mathrm{Z Z}^{(1)}}'\right] \notag \\
&\qquad + 2 \re \left \{a_\mathrm{f}^{(0)}a_\mathrm{f}^{(2)} \right \} +
|a_\mathrm{f}^{(1)}|^2 - \tfrac{1}{2}M_\mathrm{Z}\Gamma_\mathrm{Z} (a_\mathrm{f}^{(0)})^2
\,\im {\Sigma_\mathrm{Z Z}^{(1)}}''\Biggr]_{s = M_\mathrm{Z}^2}.
\end{align}
Note that $F_A^\mathrm{f}$ is also the contribution from the axial-vector form factor to the partial decay width $\Gamma(\mathrm{Z}\to \mathrm{f}\bar{\mathrm{f}})$. There is an additional  factor $\delta X$, which stems from Z propagator correction terms that are present in the cross-section for $\mathrm{e}^+\mathrm{e}^- \to \mathrm{f}\bar{\mathrm{f}}$, but not in the partial widths \cite{Freitas:2013dpa,Freitas:2014hra}. It first occurs at two-loop order, where it is given by
\begin{equation}
\delta X^{(2)} = - \left (\im {\Sigma_\mathrm{Z Z}^{(1)}}'' \right)^2 - 2M_\mathrm{Z}\Gamma_\mathrm{Z}\im {\Sigma_\mathrm{Z Z}^{(1)}}''.
\end{equation}

The effective vector and axial-vector Z vertex couplings, $v_\mathrm{f}$ and $a_\mathrm{f}$ have been computed at one loop in Refs. 
\cite{Akhundov:1985fc,Jegerlehner:1988ak,Bernabeu:1987me,Beenakker:1988pv} and at two loops in Refs. 
\cite{Djouadi:1987di,Kniehl:1989yc,Kniehl:1991gu,Djouadi:1993ss,Czarnecki:1996ei,Harlander:1997zb,
Awramik:2004ge,Awramik:2006ar,Awramik:2006uz, 
Awramik:2008gi,Freitas:2012sy,Freitas:2013dpa,Freitas:2014hra,Dubovyk:2016aqv,Dubovyk:2018rlg}. In addition, some leading higher-order corrections at three- and four-loop levels are known \cite{Avdeev:1994db,Chetyrkin:1995ix,vanderBij:2000cg,Faisst:2003px,Schroder:2005db,Chetyrkin:2006bj,Boughezal:2006xk}.

The quantity ${\bar \kappa}_\mathrm{ef}$ has a non-factorizing part from photon exchange and from  
(one-loop) ZZ and WW boxes, as expected from the general considerations of Subsection 
C.\ref{s-smat-gammaZ}.
We also refer  to Figs.~4.11 and ~4.20 in Ref. \cite{Bardin:1980fe}, where explicit expressions for ZZ and WW boxes are reproduced 
from Ref. 
\cite{Bardin:1980fe}. But one should be aware that Ref. \cite{Bardin:1980fe} 
and the ZFITTER project use the unitary gauge, while Ref. 
\cite{Awramik:2006uz} uses the 't~Hooft-Feynman gauge, and only the complete form factors (and residues) are gauge-invariant. 

When focusing on Z exchange amplitudes only, however, factorization is recovered, and one can write $\bar\kappa_\mathrm{Z}^\mathrm{ef} = \bar\kappa_\mathrm{Z}^\mathrm{e} \bar\kappa_\mathrm{Z}^\mathrm{f}$.

From $\bar\rho_\mathrm{Z}^\mathrm{ef}$, $\bar\kappa_\mathrm{Z}^\mathrm{e}$ and $\bar\kappa_\mathrm{Z}^\mathrm{f}$ one can straightforwardly obtain the hadronic peak cross-section, $\sigma^0_{\rm had}$, while the (partial) $\mathrm{Z} \to \mathrm{f}\bar{\mathrm{f}}$ width is a function of $F_A^\mathrm{f}$ and $\bar\kappa_Z^\mathrm{f}$.

\subsection{\label{s-smat-gammaZ}A coexistence of photon and Z exchange}
In this overview, it is not intended to cover the common treatment of Z boson and $\gamma$ exchange in detail.
One may consult many documents for that, \eg Refs. \cite{Schael:2013ita,Bardin:1992jc,Bardin:1989tq,Leike:1991pq} and Section \ref{sunfold}.\ref{pathbhlumi}.
In this subsection, we concentrate  on the interplay of the loop-corrected propagators under the assumption that they were, after renormalization, finite and well-defined objects.

There are two sets of integrating notation.
They may be applied if one does not want to use photonic and weak Born amplitudes in parallel, or 
if one wants to use a factorizing weak Born amplitude.
To construct the latter, one rewrites the two matrix elements with the structure $\gamma_\alpha \otimes 
\gamma_\alpha$ as follows:
\begin{multline}
\label{e-factweak}
\frac{M_\gamma}{s} \gamma_\alpha \otimes \gamma^\alpha +
{{
\frac{M_{vv}}{s-s_0}}} \gamma_\alpha \otimes \gamma^\alpha
\\ 
 \to
  \left[ \frac{M_\gamma}{s}   + \frac{1}{s-s_0}\left( M_{vv} -  \frac{M_{va}   M_{av}}{M_{aa}}\right) \right] \gamma_\alpha \otimes \gamma^\alpha 
  + \frac{1}{s-s_0}\;\frac{M_{va}   M_{av}}{M_{aa}} \gamma_\alpha \otimes
\gamma^\alpha.
\end{multline}
This construction results in a photon term that is corrected by the non-factorizing weak part, plus a weak structure that has a Born-like factorized form (although the terms $M_{va,av,aa}$ receive contributions from initial- and final-state vertex corrections and box corrections).

Introducing the measure for the degree of non-factorization 
$\Delta M_{vv} = M_{vv} -  M_{va}   M_{av}/M_{aa}$,
one may rewrite the sum of the general matrix elements for Z and $\gamma$ exchange 
(Eqs. \eqref{e-gzbornz}--\eqref{e-gzborng}) as follows:
\begin{align}
\label{e-gzbornz2}
  {\cal M}^{(0)} \left (\mathrm{e}^-\mathrm{e}^+\to \mathrm{f}^-\mathrm{f}^+
  \right )
&= {\cal M}_\mathrm{Z}^{(0,\mathrm{mod})} \left (\mathrm{e}^-\mathrm{e}^+\to \mathrm{f}^-\mathrm{f}^+ \right ) +  {\cal M}_{\gamma}^{ (0,\mathrm{mod})}
\left (\mathrm{e}^-\mathrm{e}^+\to \mathrm{f}^-\mathrm{f}^+ \right ),
\\ 
{\cal M}_\mathrm{Z}^{(0,\mathrm{mod})} \left (\mathrm{e}^-\mathrm{e}^+\to \mathrm{f}^-\mathrm{f}^+ \right )
  &\sim
\frac{M_{aa}^\mathrm{ef}}{s-s_0}
\bigl[-M_{v/a}^\mathrm{e} \gamma_\alpha + \gamma_{\alpha}\gamma_5 \bigr]
\otimes 
\bigl[-M_{v/a}^\mathrm{f} \gamma_\alpha + \gamma_{\alpha}\gamma_5 \bigr],
  \\
  \label{e-gzborng2}
   {\cal M}_{\gamma}^{(0,\mathrm{mod})} \left (\mathrm{e}^-\mathrm{e}^+\to \mathrm{f}^-\mathrm{f}^+ \right ) &\sim 
   \left(Q_\mathrm{e} Q_\mathrm{f} + c_\mathrm{fact} ~ \Delta M_{vv} \right)
   \gamma_{\alpha}  \otimes \gamma^{\alpha}
   .
\end{align}
Here
\begin{align}
M_{v/a}^\mathrm{f} &= M_{av}^\mathrm{ef}/M_{aa}^\mathrm{ef}, \qquad
M_{v/a}^\mathrm{e} = M_{va}^\mathrm{ef}/M_{aa}^\mathrm{ef}, \\
\label{e-cfact}
 c_\mathrm{fact} &= \frac{s}{s-s_0} \times \frac{C}{8\pi\alpha}
 .
\end{align}

This ansatz looks trivial, but it has a similar {\em factorized} form for the Z exchange amplitude as the Born matrix 
element -- 
without any loss of generality. The price is that the photon amplitude undergoes a finite 
renormalization.
Such an ansatz is a useful starting point for writing Monte Carlo programs based on non-squared matrix 
elements. 
Note, however, that $M_{v/a}^\mathrm{f}$ and $M_{v/a}^\mathrm{e}$ are not identical to the loop-corrected Z couplings, 
since they still include contributions from photon vertex corrections and box diagrams.

Alternatively, the general amplitude ${\cal M}(\mathrm{e}^+\mathrm{e}^- \to \mathrm{f}^+\mathrm{f}^-)$ may also be written as {\em one} amplitude, by using the replacement in \Eref{e-smat-ffzg}.
We see in that approach that the photon will become inevitably a part of the Z resonance background.

This ansatz is natural when describing the Z resonance matrix element as a Laurent series in the 
complex $s$-plane with a single pole at $s=s_0$.
The photon-exchange contribution, as well as its radiative corrections, to the cross-section at the Z peak are non-negligible. However, assuming four helicity matrix elements of the form
\begin{equation} \label{smatrix2}
{\bf \cal M}^i (s) = \frac {R^i_{\gamma}}{s} + \frac {R_\mathrm{Z}^i}{s-s_{0}} + F^i(s), \qquad i=1,\ldots 4,
\end{equation}
with {\it two poles}, would be mathematically inconsistent with the very idea of a Laurent series; this 
is already discussed  in, \eg\ Refs. 
\cite{Stuart:1991xk,Bohm:2004zi}.
The criticism is well-founded, but  the entire amplitudes (\Eref{smatrix2}) are not yet Laurent series.
They are physical ansatzes where the Z parts  already have the correct form.
When expanding them around the pole at $s_0$, the 
photon contribution looks as follows \cite{Riemann:2015wpn}:  
\begin{align}
 \frac {R^i_{\gamma}(s)}{s} &= \frac{\sum_{n=0}^{\infty}B_n^i (s-s_0)^n}{s} 
= \frac{\sum_{n=0}^{\infty}B_n^i (s-s_0)^n}{s_0-(s_0-s)} 
={\sum_{n=0}^{\infty}B_n^i (s-s_0)^n} \frac{1}{s_0} \frac{1}{1-\frac{s_0-s}{s_0}} 
\nonumber \\ &= 
{\sum_{n=0}^{\infty}B_n^i (s-s_0)^n} \frac{1}{s_0}  \left[ 1+ \frac{s_0-s}{s_0} + 
\left(\frac{s_0-s}{s_0}\right)^2 
\cdots\right] .
\label{e-Rg-exp}
\end{align}
The $R^i_{\gamma}(s)$ are defined in \Eref{e-qedR}.
The message of \Eref{e-Rg-exp} is that the photon-exchange term $R^i_{\gamma}(s)/s$ may be understood as part of the the background term $B(s)$ of ${\cal 
M}_\mathrm{Z}$. It depends on the phenomenological application whether \Eref{e-gzbornz2} or \Eref{e-Rg-exp} is the more 
appropriate description.

We see that there are several opportunities to include photon exchange in a formally correct Laurent series ansatz. 
Either 
treating the photon subseries of background terms additively as a separate matrix element interfering with the Z 
exchange amplitude, leading to the well-known form
\begin{eqnarray}\label{JGTR-lau9}
 M_\gamma \sim \frac{\alpha_\mathrm{em}(s)}{s},
\end{eqnarray}
or as a part of the $v_\mathrm{ef}$ term of the Z amplitude itself, as indicated in \Eref{e-smat-ffzg}.
Both approaches are exactly equivalent.
Normally, in Z peak phenomenology one prefers the intuitive first approach of the two.

The considerations of this and the following
subsections apply in all cases of finite matrix elements, after renormalization.
This is assumed here.
Unfortunately, nature is somewhat complex and, normally, 
the matrix elements are {\it not finite} after renormalization.
This is due to the infrared (IR) singularities arising from soft 
and collinear situations with massless particles.
Elimination of the IR singularities is a complex procedure 
and needs the combination of several matrix elements of different nature, 
\eg $2\to 2$ matrix elements with~$2 \to 3$ matrix elements of some (different) loop order.
Further, one has to regularize the singularities, usually by dimensional regularization.
At one loop, the IR problem is solved; see, \eg the formulae in Section C.\ref{ss-smat-zf}.
An isolation is difficult in multiloop situations. 
Analytical solutions are preferred,
but in the most difficult cases one has to use numerical tools, such as sector decomposition 
and Mellin--Barnes representations, for the IR-divergent Feynman integrals. 
For short introductions to these methods, see Sections E.\ref{sec:secdec} 
and E.\ref{contr:mbambre} of this report and the references cited therein.
The systematic treatment of IR singularities in the presence of a resonance 
has not been solved in full generality%
\footnote{Systematic treatment of IR divergences in QED to infinite order
 for charged narrow resonances ($\mathrm{W}^\pm$) is still missing in the literature.}
and with the accuracy needed for the FCC-ee Tera-Z stage.  
However, there are several approaches in specific cases.
One  is the exact two-loop renormalization of small-angle QED Bhabha 
scattering for small electron mass \cite{Penin:2005kf,Penin:2005eh,Kuhn:2001hz}. 
A systematic approach to the solution of the QED infrared problem 
in $\mathrm{e}^+\mathrm{e}^-$ annihilation including resummation and proper treatment
of the narrow neutral resonances, like the Z peak, was deduced by the 
Krak\'ow group~\cite{Jadach:1998jb,Jadach:2000ir,Jadach:1999vf},
and is briefly introduced in the following subsection.

The understanding and safe numerical handling of the higher-order 
IR structure of cross-sections around the Z peak is, of course, an old topic of research,
see Refs.~\cite{Greco:1975ke,Greco:1975wq,Greco:1980mh,%
Consoli:1982ib,Greco:1986dc,Nicrosini:1986sm,Nicrosini:1987sw,%
Aversa:1991rw,Fadin:1992ue,Fadin:1994ny,%
Jadach:1998jb,Jadach:2000ir,Jadach:1999vf}.
The IR problem is certainly one of the most demanding 
theoretical issues of future FCC-ee Tera-Z studies.

\subsection
[Electroweak and QED corrections in the CEEX scheme of {\tt KKMC}]
{\label{s-smat-CEEX} Electroweak and QED corrections in the CEEX scheme of {\tt KKMC}}

\def\talpha{\tilde{\alpha}}
\def\tbeta{\tilde{\beta}}
\def\bbeta{\bar{\beta}}
\def\hbeta{\hat{\beta}}
\newcommand{\Meu}{{\cal M}}
\newcommand{\sfac}{\mathfrak{s}}

Let us explain briefly in the following short overview how the EW parts of the Standard Model corrections 
to fermion pair production in electron--positron annihilation are actually embedded
in the most sophisticated scheme CEEX%
\footnote{CEEX stands for {\em coherent exclusive exponentiation}.}
of the QED calculations with soft photon resummation 
of Refs.~\cite{Jadach:1998jb,Jadach:2000ir},
as implemented in the {\tt KKMC} Monte Carlo event generator~\cite{Jadach:1999vf}.
We are going to follow the notation of Ref.~\cite{Jadach:2000ir},
suppressing spin or spinor indices for simplicity.
It will  also be shown that it is rather easy to modify the existing 
implementation of the EW part in CEEX of {\tt KKMC}, such that
it precisely follows  the S-matrix approach advocated in this section, 
\ie following in practice what is described around \Eref{e-cfact}.

In the CEEX {\em factorization scheme}, the cross-section for the process
\[
 \mathrm{e}^{-}(p_a)+\mathrm{e}^{+}(p_b)\to \mathrm{f}(p_c)+\bar{\mathrm{f}}(p_d)+\gamma(k_1),\dots,\gamma(k_n)
\]
with complete perturbative corrections up to \order{\alpha^r} and soft photon resummation
reads as follows:
\begin{equation}
  \label{eq:master-bis}
  \sigma^{(r)} = 
  \sum_{n=0}^\infty {1\over n!}
  \int \mathrm{d}\tau_{n} ( p_1+p_2 ;\; p_3,p_4,\; k_1,\dots,k_n)\;
  \mathrm{e}^{2\alpha\Re B_4(p_a,\dots,p_d)}
  {1\over 4}\sum_{\rm spin} \left| \Mmf^{(r)}_n \left(p, k_1, k_2, \dots k_n \right) \right|^2
  ,
\end{equation}
where the virtual form factor $B_4$ is factorized (exponentiated)
and real emission factors $\sfac$ are also factorized out:%
\footnote{Momenta of all fermions $p_a,p_b,p_c,p_d$ are denoted collectively as $p$.}
%
\begin{equation}
  \label{eq:beta-trunc}
  \begin{split}
  & \Mmf_n^{(r)}(p,k_1,k_2,k_3,\dots ,k_n) = 
  \prod_{s=1}^n \sfac(k_s) \left \{ \hbeta^{(r)}_0(p)
    +\sum_{j=1}^n      {\hbeta^{(r)}_1(p,k_j) \over \sfac(k_j) }
    +\sum_{j_1<j_2}    {\hbeta^{(r)}_2(p,k_{j_1},k_{j_2}) \over \sfac(k_{j_1})\sfac(k_{j_2}) }
    + \cdots
   \right \},
  \end{split}
\end{equation}
such that the subtracted amplitudes $\hbeta^{(r)}_j$ are IR-finite.
Resummation, that is spin summing or averaging of the squared amplitudes
and the phase space integration $\int \mathrm{d}\tau_n$, is performed numerically in a separate
Monte Carlo module of the {\tt KKMC}, independent from the other part of
the {\tt KKMC}
where spin amplitudes $\Mmf_n^{(r)}(p,k_1,k_2,k_3, \allowbreak \dots ,k_n)$ are constructed and evaluated.
The S-matrix methodology of Eqs. \eqref{e-factweak}--\eqref{e-cfact} is relevant
for the $2\to{2}$ Born-like object $\hbeta^{(r)}_0$.
In the \order{\alpha^2} ($r=2$) implementation of {\tt KKMC}, this object reads:
\begin{equation}
  \label{eq:virtual-subtraction}
  \hbeta^{(2)}_0(p)= \Mmf^{(2)}_0(p) =
  \left[ \mathrm{e}^{-\alpha B_4(p)} \Meu^{(2)}_0(p) \right]\Big|_{{\cal O}(\alpha^2) },
\end{equation}
where $\Meu^{(2)}_0(p)$ represents Born spin amplitudes 
corrected up to two loops, derived directly from Feynman diagrams.
In practice, the non-soft parts of the QED corrections 
are complete in $\hbeta^{(2)}_0(p)$
up to two loops, while the EW corrections are taken from DIZET 6.21~\cite{Bardin:1999yd}
(\ie they are at 1+1/2 loops), exactly according to the prescription shown 
in \Eref{e-from0608099-sig}; see also Eqs.~(21)--(24) in Ref.~\cite{Jadach:1999vf}.
This implementation of the EW corrections in {\tt KKMC} can  easily
be modified
to be compatible with the S-matrix approach, following the prescription of
Eqs.~\eqref{eq:c126}--\eqref{eq:c130}).

Concerning the EW corrections to the $2\to 3$ process, they would enter into
\begin{equation}
\label{eq:BetRecursive2}
\begin{split}
&  \hbeta^{(2)}_1(p,k_1)     
    =  \Mmf^{(2)}_1(p,k_1) -\hbeta^{(1)}_0(p)  \sfac(p,k_1), \qquad
\Mmf^{(2)}_1(p,k_1) = 
  \left[ \mathrm{e}^{-\alpha B_4(p)} \Meu^{(2)}_1(p,k_1) \right]\Big|_{{\cal O}(\alpha^2) }.
\end{split}
\end{equation}
In the {\tt KKMC} implementation, one-loop complete QED corrections are
included in $\hbeta^{(2)}_1$, 
mandatory for the completeness of the \order{\alpha^2} QED,%
\footnote{Except of 
numerically small five-point graphs in Fig. 5 of Ref.~\cite{Jadach:1999vf}.}
neglecting the one-loop EW part.
For the future FCC-ee applications, it will also be necessary to include 
one-loop (five-point) EW corrections, see the discussion in Subsection C.\ref{ss-smat-5pt}.
However, it should be stressed that, in order to be useful in the CEEX scheme,
they will have to be properly subtracted at the amplitude level,
instead of combined with real emission for the differential cross-sections
\`a la Bloch--Nordsieck.

In Subsection C.\ref{sec:EW-QED-separation},
we  also discuss two examples of proper treatment of EW
and non-soft QED corrections in the CEEX scheme at the amplitude level for three-loop
corrections in $\hbeta^{(3)}_0(p)$ and two-loop corrections in $\hbeta^{(2)}_1(p,k_1)$.

Summarizing,
it is straightforward to insert arbitrary EW and QED loop corrections into an existing, 
sophisticated QED calculation based on the 
 CEEX factorization scheme, following the S-matrix approach.
For test purposes, it is mandatory to look at the numerical effects, 
if not even for realistic calculations for the  FCC-ee Tera-Z stage.

\subsection{Radiative loops: five-point functions \label{ss-smat-5pt}}
An isolated topic is that of the contributions from so-called radiative loop diagrams.
The simplest cases are five-point functions, which result from one-loop box diagrams with an additional emission of a 
(soft) photon.
This is a 2$\to$3 process and will interfere with other 2$\to$3 processes; thus, it is a  next-to-next-to-leading-order (NNLO) contribution to 2$\to$2 
scattering. 
In principle, it is well-known how to calculate such processes and there are several one-loop packages that can deal with 
tensor five-point functions, for instance 
{\rm FeynArts/FormCalc}~\cite{Hahn:2000kx,Nejad:2013ina}, 
{\rm CutTools}~\cite{Ossola:2007ax}, 
{\rm    Blackhat}~\cite{Berger:2008sj}, 
{\rm   Helac-1loop}~\cite{Kanaki:2000ey,vanHameren:2009dr}, 
{\rm Samurai}~\cite{Mastrolia:2010nb},
 {\rm Madloop}~\cite{Hirschi:2011pa}, 
 {\rm GoSam}~\cite{Cullen:2011ac}, 
 {\rm OpenLoops}~\cite{Cascioli:2011va},
 {\rm RECOLA}~\cite{Actis:2016mpe},
  and
PJFry~\cite{Fleischer:2007ph,Diakonidis:2008ij,Fleischer:2010sq,Fleischer:2012et,Fleischer:2011zz,pjfry-project},
as well as 
{\rm COLLIER}~\cite{Denner:2016kdg}.
 In precision analysis, the numerically stable treatment of 
tensor reduction is very  important, and we  especially advocate the recent developments of the latter two approaches. 
 See also the 
mini-review of Ref. \cite{AlcarazMaestre:2012vp}, but we are not aware of dedicated studies for $\mathrm{e}^+\mathrm{e}^-$-scattering, with the 
notable exclusion of PJFry \cite{Actis:2010gg,CarloniCalame:2011zq,CarloniCalame:2011aa,Campanario:2013uea}.
The numerical impact of these 2$\to$3 processes, including fermion pair emission, is very small at 
meson factories, with their typical precision 
\cite{Actis:2010gg,Gluza:2012yz,CarloniCalame:2011aa,CarloniCalame:2011zq,Campanario:2013uea}.
In view of the extraordinary precision of the FCC-ee-Z, one has to reconsider corresponding  studies.

\subsection{%
Bhabha scattering: massive loops at NNLO \label{ss-smat-bhmf}}
Bhabha scattering is a basic process for theoretical luminosity precision determination. 
It will be discussed in more detail in 
Section \ref{sunfold}.\ref{pathbhlumi}, where
the present status of \bhlumi{} is given with a path to the better precision needed at the FCC-ee compared with that at the LEP. Section \ref{sunfold}.\ref{pathbhlumi} also discusses the uncertainty due to photonic higher-order corrections, the hadronic vacuum polarization, and light 
fermion pair emissions. These issues at the level of accuracy needed at the FCC-ee should be confronted in 
future Monte Carlo FCC-ee studies with fixed two-loop corrections, including QED insertions with massive fermions.
These issues are also discussed in Section {\textcolor{black}{\ref{sunfold}.\ref{egbabayaga}}}.

For Bhabha scattering, the results for two-loop Feynman diagrams 
have been determined by at least two independent groups, relying on different methods.
\begin{enumerate}
 \item \emph{Photonic corrections:} computed in Refs. \cite{Penin:2005kf,Penin:2005eh} and
recalculated in Ref. \cite{Becher:2007cu}.
\item \emph{Electron $N_\mathrm{f}=1$ corrections:} computed in Ref. \cite{Bonciani:2004gi} and cross-checked
in Ref. \cite{Actis:2007gi} (with full $m_\mathrm{e}$ dependence) and  Ref. \cite{Becher:2007cu} (small electron mass limit).
\item \emph{Heavy-fermion $N_\mathrm{f}=2$ contributions:} determined with two independent 
methods in the limit $m_\mathrm{f}^2 \ll s,t$ in Refs. \cite{Actis:2007gi,Becher:2007cu} and 
for any mass $m_\mathrm{f}$  in Refs. \cite{Actis:2007pn,Actis:2007fs} (dispersive approach) and
Refs. \cite{Bonciani:2008zz,Bonciani:2008ep} (analytical result).
\item  \emph{Virtual hadronic NNLO contributions}, including both reducible self-energy insertions and irreducible 
vertex and box corrections \cite{Actis:2007fs,Actis:2008br,Kuhn:2008zs}.
\end{enumerate}

\section*{Acknowledgements}
We thank 
A. Blondel,
S. Heinemeyer,
P. Janot,
F. Piccinini,
R. Tenchini,
S. Riemann,
M. Zralek
and many other colleagues for encouragement, fruitful discussions, and support.

\clearpage \pagestyle{empty} \cleardoublepage

\cleardoublepage

\pagestyle{fancy}
\fancyhead[RO]{}
\fancyhead[LO]{}
\fancyhead[CO]{\thechapter.\thesection \hspace{1mm} 
QED deconvolution and pseudo-observables at FCC-ee precision}
\fancyhead[LE]{}
\fancyhead[CE]{A. Freitas, J. Gluza, S. Jadach}
\fancyhead[RE]{}

\section
[QED deconvolution and pseudo-observables at FCC-ee precision 
\\ {\it A. Freitas, J. Gluza, S. Jadach}]
{QED deconvolution and pseudo-observables at FCC-ee precision}
\label{s-qed}
\noindent
{\bf Authors: Ayres Freitas, Janusz Gluza, Stanis\l aw Jadach}
\\
Corresponding author: Stanis\l aw Jadach [Stanislaw.Jadach@cern.ch]
                       
\vspace*{.5cm}

\noindent The concept of the electroweak {\em pseudo-observables}
was essential in the final analysis of the LEP1 data of Ref.~\cite {ALEPH:2005ab}.
Electroweak pseudo-observables (EWPOs) were instrumental in 
(a) combining data from four LEP collaboration and SLD experiments and 
(b)  conveniently organizing the procedure
of fitting the Standard Model to experimental data.
The EWPOs used in the final analysis of the LEP data~\cite {ALEPH:2005ab}
near Z resonance were defined and thoroughly tested in Ref.~\cite{Bardin:1999gt}.
Both works  exploited ZFITTER~\cite{Arbuzov:2005ma} 
and TOPAZ0~\cite{Montagna:1993ai,Montagna:1995ja} programs.

The effects of QED on data, even if large, are, in principle,
perfectly calculable with arbitrary precision.
Once they are removed, the remaining EWPOs of the LEP include
smaller pure electroweak corrections and
possibly signals of a new physics beyond the Standard Model (BSM).
The procedure of removing deformation of the data by QED effects,
commonly referred to as {\em QED deconvolution},%
\footnote{The term `deconvolution' is not quite adequate -- operationally
    it is just a fitting procedure -- but  is kept for historical reasons.}
is an essential part of the definition or construction of the EWPOs.
Separating the QED part from the higher-order EW part
consistently and systematically is an important and delicate
issue, especially at higher orders, and we shall come back to it later on.

Note that, for some processes of the low-angle Bhabha process used
for the measurement of luminosity, Z$\gamma$ production 
for $s^{1/2}>M_\mathrm{Z}$ (radiative return) above the Z peak or production of W pairs,
the technique of EWPOs including QED deconvolution could not be used 
and data were compared directly using  Monte Carlo programs 
(\bhlumi{}, KORALW, \etc), mainly because of
the more complicated dependence of QED effects 
on the event selection criteria (experimental cut-offs).

Before addressing the challenge of constructing EWPOs at the very high
precision level of the FCC-ee, we are going to summarize, briefly, 
the definition and  use of the EWPOs in LEP data analysis.

\subsection{EWPOs in the LEP era}

\begin{figure}
\centering
  \includegraphics[width=0.8\textwidth]{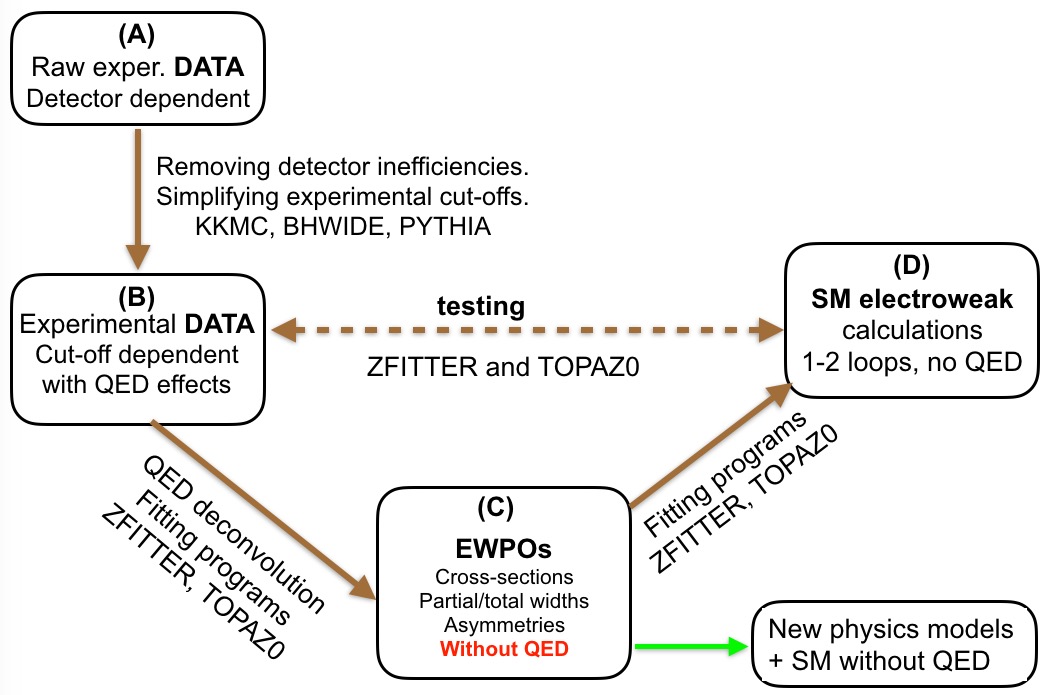} 
\caption{Construction of  EWPOs in data analysis of the LEP}
\label{fig:EWPO-LEP}
\end{figure}

The entire procedure of constructing, testing, and using
EWPOs in the final analysis of the LEP/SLC data~\cite {ALEPH:2005ab}
is schematically illustrated in \Fref{fig:EWPO-LEP}.

In the first step, from (A) to (B) in \Fref{fig:EWPO-LEP},
raw data are transformed; they are corrected for
inefficiencies of the detector
and kinematic cut-offs are rounded up to a simpler shape, 
which can be dealt with using semi-analytical non-MC `fitter'
programs based on ZFITTER and TOPAZ0.
The transition from (A) to (B) was achieved using sophisticated 
Monte Carlo event generators, such as KORALZ, {\tt KKMC}, {\tt BHWIDE}, and PYTHIA.
Data for stage (B) were obtained separately for each LEP collaboration.
The important point is that the choice of the type or shape 
of simplified cut-offs
was dictated by the limited capabilities of the fitter programs in dealing with realistic experimental cuts%
.\footnote{For the same types of experimental cut, the values of the 
  cut-off parameters could be different between experiments.}

The aim of the next most important transformation of data, 
from (B) to (C) in Fig.~\ref{fig:EWPO-LEP},
using fitter programs based on ZFITTER and TOPAZ0, 
was to remove QED effects and cut-off dependence,
such that the resulting `pseudo-data', that is EWPOs, 
do not depend on specific details of the individual experiments 
and are not `polluted' by QED.
It is important to keep in mind that this step  introduces
certain well-known losses in data precision; this was tolerable at the LEP precision,
but may be not tolerable at the FCC-ee.
This procedure introduces some small, hopefully negligible, dependence
on the Standard Model parameters and the details of the QED calculations used.

The fitter programs based on ZFITTER and TOPAZ0 were able to calculate cross-sections and asymmetries for some classes of simplified kinematic cut-offs 
(minimum mass of fermion pair, acollinearity,
angular range of one of fermions),
 combining the inclusive initial-state radiation (ISR) electron structure function 
of Ref.~\cite{Jadach:1992aa},
${\cal O}(\alpha^1)$ QED analytical calculations,
and the effective Born amplitudes of the EWPO scheme.
As already noticed and strongly emphasized in Ref.~\cite{Bardin:1999gt},
the sticking point was that these scenarios could be invalidated
by the initial--final-state interference (IFI) contributions,
for various reasons.
For instance, the convolution of the ISR structure function involves integration
over the effective mass $\sqrt{s'}$ 
after ISR and before final-state radiation (FSR).
If IFI is switched on, this variable loses its physical meaning.
The solution was to introduce an acollinearity cut, 
which  approximately limited $s'$, accompanied with a cut-off
of the angle of one of the final fermions, 
leaving the angle of the other one uncontrolled.

In the  (B)$\to$(C) transition in Fig.~\ref{fig:EWPO-LEP},
an effective Born term is used 
in the fitter programs instead  of complete EW corrections.
The differential distribution of the effective Born term is obtained from spin
amplitudes of the $\mathrm{e}^-\mathrm{e}^+\to \mathrm{f}\bar{\mathrm{f}}$ process, 
with the carefully defined (real) effective coupling constants 
of $\gamma$ and Z bosons to electrons and other fermions 
$\mathrm{f}=\mathrm{e},\mu,\tau,\mathrm{u},\mathrm{d},\mathrm{s},\mathrm{c},\mathrm{b}$.
In fact, the differential distribution of the effective Born term
in Eq.~(1.34) of Ref.~\cite {ALEPH:2005ab} 
is in one-to-one correspondence with the spin amplitudes of
\Eref{e-gzborn}, or the Born version of 
Eqs.~\eqref{e-matrix4cahn5-1}--\eqref{e-matrix4cahn5-2} 
and \eqref{e-smat-matrix1}--\eqref{e-smat-matrix4},
with adjustable parameters being 
$M_\mathrm{Z}$, $\Gamma_\mathrm{Z}$, $\alpha_\mathrm{em}(M_\mathrm{Z})$ 
and Z couplings for each fermion type, $a_\mathrm{f}$ and $v_\mathrm{f}$.

This one-to-one correspondence of
the parameters of the effective Born term at the amplitude level,
that is, four couplings per  fermion, and the mass and width of the Z boson
--
which will be referred to as EW `pseudo-parameters' (EWPPs)\footnote{The prefix `pseudo-' emphasizes
  the fact that these parameters are different from the Standard Model Lagrangian parameters.} --
means, in practice, that from their values
one easily obtains  partial widths proportional to $a_\mathrm{f}^2+v_\mathrm{f}^2$,
hadronic peak cross-sections,
and all possible charge and spin asymmetries, 
being simple functions of $v_\mathrm{f}/a_\mathrm{f}$
(Eqs.~(1.37), (1.45), and (1.51)-(1.54) in Ref.~\cite {ALEPH:2005ab}),
either during the data fitting procedure
or when  obtaining final or fitted EWPOs for each experiment.

The list of EWPOs in Ref.~\cite {ALEPH:2005ab}
representing LEP/SLC data consists of
$M_\mathrm{Z}$, $\Gamma_\mathrm{Z}$, $\sigma^{(0)}_\mathrm{had}$, 
$R^{(0)}_\mathrm{f} $, $A^{(0),\mathrm{f}}_\mathrm{FB}$, $\mathrm{f} = \mathrm{e},\mu,\tau,
\mathrm{c} ,\mathrm{b}$
(see Tables~2.5, 2.13, and 5.10 therein).
The EWPOs created at stage (C) separately for each LEP and SLD collaboration
were then combined into common EWPOs, 
with the experimental error reduced by roughly a factor of two.\footnote{%
 In principle, EWPPs can be re-derived from EWPOs after combining over experiments.
}
The number of the combined EWPOs 
was still much greater than the number
of independent parameters of the Standard Model Lagrangian.
For instance, there were several values of $\sin\theta^2_\mathrm{eff}$, derived
from various types of asymmetry for different species of final fermions.

The peculiarity of the LEP EWPOs is that final-state radiation (FSR)
effects are included in the partial Z widths,\footnote{This is to ensure that 
  partial widths sum up to the total Z width.}
see Section 3.5.3 in Ref.~\cite {ALEPH:2005ab},
including the non-factorizable \order{\alpha\alpha_s} 
FSR corrections of Ref.~\cite{Czarnecki:1996ei}. This  means that these corrections are removed from the data when fitting EWPPs
and later on reinstalled in the EWPOs related to partial widths.

The EWPOs at stage (C) in Fig.~\ref{fig:EWPO-LEP}
are, in principle, independent of
parameters in the Standard Model Lagrangian.%
\footnote{In practice, a residual dependence
 of EWPOs on EW or QED details remains in fitted EWPOs.}
The fitting procedure of the Standard Model Lagrangian parameters to EWPOs,
transition (C)$\to$(D) in Fig.~\ref{fig:EWPO-LEP},
is achieved with QED effects switched off.
 
For a given LEP experiment, 
one could fit the Standard Model directly to data from stage (B),
as illustrated in transition (B)$\to$(D) in Fig.~\ref{fig:EWPO-LEP}.
In such a fit, without the use of EWPOs,
one avoids the introduction of an additional bias
present in the two-step fitting of  (A)$\to$(C)$\to$(D).
This is an important cross-check on the size of this kind of bias.
The bias was previously estimated to be smaller than
LEP experimental errors \cite{Bardin:1999gt}.
Such a cross-check was also made using  data of each LEP collaboration separately,
as reported in Section 2.5.4 
of Ref.~\cite {ALEPH:2005ab}; see also Table~2.12 therein.

The disadvantage of the direct fitting (C)$\to$(D) is that
it consumes more CPU time (owing  to the QED component) and
makes combining data from several experiments more difficult.
Note that the effective Born term in the aforementioned LEP construction of EWPOs
is chosen such that it encapsulates one-loop and higher-order-loop pure EW corrections in  the most effective way, thus minimizing
the bias introduced in the two-step variant of fitting Standard Model parameters
using EWPOs of the intermediate step (C),
but there is no guarantee that such a bias is below the required FCC-ee precision level.

In this procedure, QED effects are taken into account
in the (A)$\to$(B) step, thanks to the sophistication of the MC programs,
in a most complete form,
up to \order{\alpha^2} with soft photon resummation,
and also for non-factorizable IFI,
and collinear logs up to \order{\alpha^3}.
Fitter programs used in the (B)$\to$(C) step use a simplified representation of QED, 
typically with ISR integrated over photons to the level
of the so-called radiator (flux) function convoluted in a single dimension
with the effective Born term. 
Other non-ISR QED effects (including IFI) were taken in the \order{\alpha^1}
and combined with ISR additively.
The aforementioned treatment of QED in the  (B)$\to$(C) step was 
carefully proven ~\cite{Bardin:1999gt}
to be precise enough for EWPOs near the Z peak,
as compared with the precision of the LEP data, 
but most probably requires much improvement to be compatible
with the FCC-ee precision, as discussed next.

\subsection{Potential problems with LEP deconvolution at FCC-ee precision}
Generally, one may still hope that, for
the line-shape-related group of EWPOs, such as the mass of the Z boson,
the peak cross-section, and the total and partial Z widths, one can still employ
in FCC-ee data analysis the same LEP scheme using QED-independent
(experiment-independent) EWPOs in the intermediate step on the way between
experimental data and the Standard Model or SM+BSM, as depicted in Fig.~\ref{fig:EWPO-LEP}.

In the LEP EWPOs scheme,
there were some solvable problems, compared with LEP data precision. 
However, these may get much worse, owing to the higher precision of the FCC-ee.
\begin{enumerate}
\item
The errors introduced in the transition (A)$\to$(B),
adjusting true experimental cut-off to unrealistic cut-offs 
required by the non-MC fitter programs, may be too big and not tolerable.
It will be necessary to check these points for future data analysis.
\item
The non-factorizable IFI contribution, which could be neglected at the LEP
near the Z resonance, will be too difficult to control at
a factor of ${\sim}50$ better precision level.
In particular, a reliable extension of the simple one-dimensional
convolution formula used for ISR,  also incorporating the IFI contribution, does not exist.%
\footnote{It is possible to integrate analytically over
  soft photon energies and angles, keeping IFI in the game,
  even for energies comparable with the resonance width $\Gamma_\mathrm{Z}$,
  as  shown in Refs.~\cite{Greco:1975rm,Greco:1975ke}
  and more recently developed further in Ref.~\cite{Jadach:2018lwm}:
  however, the remaining four-dimensional convolution integral 
  over photon energies cannot be reduced to one dimension.
}
\item
The methodology of the separation of QED and EW parts 
in perturbative Standard Model calculations,
which was conceptually and practically relatively simple in the LEP era,
owing to the \order{\alpha^1} restriction, will require an update in order
to work for two- and three-loop EW corrections decorated with extra photon insertions.
\end{enumerate}

The issue of item (1) is discussed later on in the following.
Let us only indicate that the restriction in stage (B) to cut-offs
that are manageable by the fitter programs will be removed.

Concerning IFI contribution to the line shape and all related EWPOs,
it is rather small, owing to suppression by the $\Gamma_\mathrm{Z}/M_\mathrm{Z}$ factor
-- as reported in Ref.~\cite{Jadach:1999gz}, it was estimated near the Z resonance to be
at the level of $\delta\sigma/\sigma =10^{-4}$ for the integrated cross-section.
Line shape experimental precision at the FCC-ee will improve by a factor
of ${\sim}7$ for $\sigma_\mathrm{tot}(M_\mathrm{Z})$
and ${\sim}20$ for $M_\mathrm{Z}$ and $\Gamma_\mathrm{Z}$, see Table~2 in the foreword;
hence, the LEP scheme with ISR radiator function neglecting IFI
may still work, at least for these observables,
although its precision will have to be reexamined,
especially for energies up to $\pm 3.5\UGeV$ away from $M_\mathrm{Z}$,
where IFI suppression deteriorates significantly, 
see for instance Ref.~\cite{Jadach:2018lwm}.

The situation will be worse for the angular-distribution-related EWPOs,
where experimental precision at the FCC-ee will improve 
for charge asymmetry by a factor of ${\sim}50$.
Again, the IFI effect near Z resonance for cut-offs 
on photon energies not too strong is suppressed by the $\Gamma_\mathrm{Z}/M_\mathrm{Z}$ factor; in the LEP experiment, the IFI effect was compatible with the LEP experimental error, 
which for muon charge asymmetry 
was $\delta A_\mathrm{FB} \sim 0.1\%$ near Z resonance 
and $\delta A_\mathrm{FB} \sim 1\%$ further away.
In the rare cases when IFI approached the level of the LEP experimental precision,
it was taken into account (subtracted form data) in the construction of EWPOs
in the ${\cal O}(\alpha^1)$ approximation, additively, without any resummation,
using LEP-era ZFITTER and TOPAZ0,
in the (A)$\to$(B) step using Monte Carlo programs, such as KORALZ.

This LEP-era approach to IFI problem is inherently limited
to a rather loose experimental cut-off of the total photon energy;
hence, the treatment of IFI in the LEP-era fitter programs
is definitely inadequate for  FCC-ee purposes,
as $A_\mathrm{FB}$ will be measured at the FCC-ee down to 
$\delta A_\mathrm{FB} \sim 10^{-5}$ precision,
which means that an IFI contribution of ${\sim}0.1\%$
will have to be controlled with two-digit precision or better.
Note that it will also be  possible to eliminate IFI in the earlier step,
in the transition (A)$\to$(B),
with the help of sophisticated MC programs
of the {\tt KKMC} class~\cite{Jadach:2000ir,Jadach:1999vf}.

Summarizing  item (2),
the methodology of the QED deconvolution of the LEP,
and the resulting construction of EWPOs, see Fig.~\ref{fig:EWPO-LEP},
might not work for important observables
at  much higher experimental precision of FCC-ee,
especially for asymmetries.
This is why we are already motivated to think  about
alternative solutions,
which do not suffer from these problems
and, in addition, have several advantages.
This is what we are going to do in the following.

Before we present new alternative scenarios of
removing cut-off dependent QED effects from data,
let us first comment on the fundamental question of the
existence of a theoretically clean methodology of separating, 
 beyond first-order,
QED effects from the remaining pure EW part in the perturbative calculations,
as indicated in item (3).
In fact, the methodology of disentangling pure QED corrections from
pure EW corrections {\em at the amplitude level}, 
resumming  soft photon effects to infinite order (exponentiation) 
and adding QED collinear non-soft corrections order by order,
independently of the EW part, is well-known.
Multiphoton spin amplitudes of the {\tt KKMC} program~\cite{Jadach:2000ir,Jadach:1999vf}
are constructed according to such a scheme,
the coherent exclusive exponentiation (CEEX) scheme
given in Ref.~\cite{Jadach:1998jb}.

It should be stressed immediately,
that this CEEX factorization of the scattering matrix
element into a universal part, encapsulating all leading QED corrections,
and a remaining part 
(EW connections and IR-finite QED remnants, without collinear mass logs)
works at any order and for arbitrary precision.

The CEEX scheme is a complete scheme for 
QED infrared factorization or resummation
and matching, consistently, finite non-IR contribution with the resummed parts,
also working  for narrow resonances, such as  Z resonance.
The role of the Monte Carlo  method is merely
to square the CEEX matrix element, sum up over spins of photons and fermions,
integrate over the soft and hard photon phase space
and sum up over photon multiplicities,
numerically and without  approximation.
All this was implemented for the $\mathrm{e}^-\mathrm{e}^+\to \mathrm{f}\bar{\mathrm{f}}$ process
in the {\tt KKMC} event generator~\cite{Jadach:1999vf} 
but the CEEX technique is universal and can be used for any other process.
The IFI contributions due to real photons  simply result
from squaring and spin summing
of the scattering matrix element; hence, are automatically
and fully taken into account (virtual contributions must be calculated separately).

More precisely, spin amplitudes in {\tt KKMC} include QED non-soft corrections 
to second-order and pure EW correction
up to first-order (with some second-order EW improvements, QCD, etc.)
using the DIZET library~\cite{Bardin:1989tq}.
The CEEX calculation scheme of {\tt KKMC} can be extended in a natural way
to higher orders, including EW corrections up to two and three loops.
More details on the CEEX scheme will be given in  
Section~C.\ref{sec:EW-QED-separation}.
 
\subsection{Electroweak pseudo-observables at the FCC-ee}

\begin{figure}
\centering
  \includegraphics[width=0.8\textwidth]{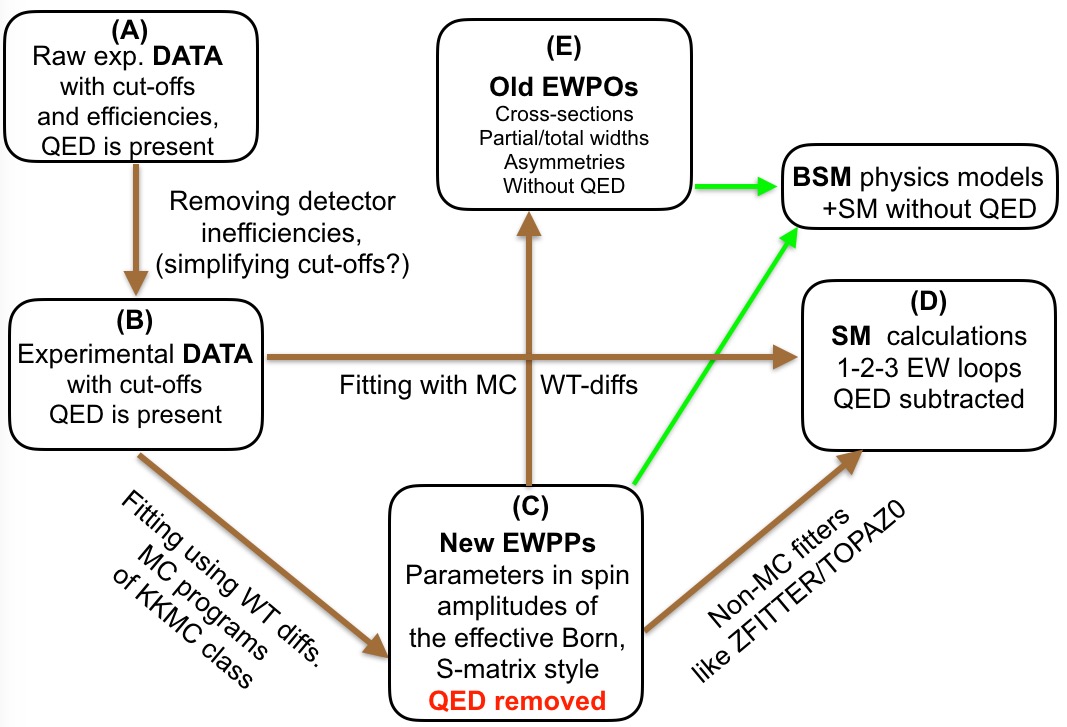} 
\caption{Possible scheme of construction of  EWPPs in data analysis of the
FCC-ee}
\label{fig:EWPO-FCC}
\end{figure}

With all these introductory remarks in mind, let us present an alternative
scheme of QED deconvolution, which should work at the FCC-ee precision
and is free of the indicated problems of the LEP EWPO scheme.
This new scheme is illustrated in Fig.~\ref{fig:EWPO-FCC}.

In the first step, (A)$\to$(B), detector inefficiencies are removed.
Kinematic boundaries of the detectors can  also be replaced
by simpler phase space boundaries in terms of some kinematic cuts,
without any loss in precision, 
using MC event generators with sophisticated QED matrix element
and full phase space coverage, interfaced with  detector simulation programs.
Contrary to LEP procedure,  we are not restricted here 
to the limited choice of the semi-realistic cut-offs of the non-MC programs,
which may be far away from the true experimental cut-offs,
owing to greater use of the Monte Carlo programs in the following steps.

In the new scheme of \Fref{fig:EWPO-FCC}, the role of the direct
fitting of the Standard Model internal parameters in the single step (B)$\to$(D) will increase.
By contrast with the former LEP scenario, this step is now implemented using
a sophisticated Monte Carlo program, because only this kind of tool
is capable of calculating QED effects for arbitrary cut-offs
and properly combining IR-resummed QED and two- and three-loop EW  corrections
with arbitrary precision.
The  MC programs cannot be used in the fitting in a straightforward
way, owing to slowness of the MC event generators 
(even without detector simulation).
However, according to  more detailed discussion in the following,
it will be possible using weight difference methodology (WT-diff).

Of course, the two-step scenario (B)$\to$(C)$\to$(D) in Fig.~\ref{fig:EWPO-FCC}
will be preferred for some types of EW pseudo-observables at the intermediate stage (C).
However,  as in the older LEP scheme, 
the single step (B)$\to$(C) will be more precise
and will be used to cross-check biases introduced
in the two-step scenario (B)$\to$(C)$\to$(D).
If these biases are acceptable, in view of the FCC-ee experimental precision,
then the two-step scenario (B)$\to$(C)$\to$(D) will be preferred,
otherwise the (B)$\to$(D) scenario will be the principal one,
and intermediate stage (C)
will lose much of its attractiveness.

Let us now concentrate on the transformation of
data in transition (B)$\to$(C) of Fig.~\ref{fig:EWPO-FCC}.
The principal aim of removing large QED effects
and the dependence on the kinematic cut-off specific to a given detector
remains the same, 
and the use MC programs (for arbitrary experimental cut-offs)
instead of non-MC fitters is again envisaged.

{\em 
The key strategic point is now whether, in the transition {\rm (B)$\to$(C)},
we intend to remove only all QED effects, as in the LEP, 
or  we decide courageously
to remove, in addition to the QED
part, most of the pure EW higher-order corrections as well?
}
In any case, all of the available EW+QED corrections will be included
in the two steps (B)$\to$(C)$\to$(D), from data to Standard Model (or to SM+BSM).
The question is only whether pure EW effects are located in
(B)$\to$(C) or in (C)$\to$(D), or distributed among them in a clever way, 
such that most of the convenient features of the LEP EWPOs are preserved.

Operationally, in step (B)$\to$(C), we would use the same MC tool as in step
(B)$\to$(D),
but with the hard process part of the multiphoton 
spin amplitudes in the MC program
simplified to an effective Born term with the mass and width of the Z boson
and some couplings or form factors,
possibly for each fermion type, fitted to data of  stage (B).
All higher-order QED corrections, including IFI, will be kept, of course.
Non-factorizable QED corrections to one-loop EW corrections
(two-loop in the Standard Model sense) will be neglected and will have to be restored
in the (C)$\to$(D) step towards the Standard Model or SM+BSM.
If the validation using direct fitting in step (B)$\to$(D) proves
that the additional bias in the two-step scenario is below  FCC-ee precision,
then this kind of LEP-like solution
will be considered  acceptable.
Similarly to the LEP case, EWPPs will be $M_\mathrm{Z}$, $\Gamma_\mathrm{Z}$
and two Z couplings for each fermion type.
Derived EWPOs following step (C)$\to$(E) will  again be peak cross-section
partial widths and various asymmetries (calculated without QED).
For instance, charge asymmetry can be calculated from EWPPs
using Eq.~(\ref{e-afbcomplete0}).

A new interesting option to be considered is that of including,
in the effective Born spin amplitudes
(to be decorated according to the CEEX scheme with multiphoton
QED real and virtual (resummed) contributions),
not only Born-type couplings of Z, 
but also some additional factors representing higher-order EW corrections,
 for instance, the amplitudes 
of Eqs.~\eqref{e-matrix4cahn5-1}--\eqref{e-matrix4cahn5-2} 
and \eqref{e-smat-matrix1}--\eqref{e-smat-matrix4},
following the S-matrix approach.
As advocated in Section~\ref{sunfold}.\ref{s-loopstoEWPOs}, 
this kind of spin amplitude efficiently encapsulates two- and three-loop EW corrections,
and does not  contradict gauge-invariance, unitarity, \etc
In such an approach, two well-defined additional parameters,
for instance $\rho_\mathrm{Z}$ and $\kappa_\mathrm{ef}$ for each final state of \Eref{e-2to24b}, will be calculated in the Standard Model
and will depend slightly on the Standard Model input parameters.
This dependence will have to be well controlled and recorded,
along with the remaining coupling constants $a_\mathrm{f}$ and $v_\mathrm{f}$.
Altogether, $M_\mathrm{Z}$, $\Gamma_\mathrm{Z}$, $a_\mathrm{f}$, and $v_\mathrm{f}$ 
will again form EW pseudo-parameters, EWPPs, as in the case
of the LEP, and will be fitted to data using $\rho_\mathrm{Z}$ and $\kappa_\mathrm{ef}$
calculated from the Standard Model for some well-known Standard Model input parameters.
The advantage will be that in the next step of fitting the Standard Model,
 step (C)$\to$(D), only very subtle missing two- and three-loop corrections
will have to be included.
The same is true for the case of fitting to the SM+BSM.

Can we include FSR corrections to partial widths  in our new scheme, in
the same way as in the LEP scenario?
For hadronic final states, owing to almost perfect 100\% acceptance,
we may proceed in the same way as in the LEP; 
also for \order{\alpha\alpha_s} 
corrections~\cite{Czarnecki:1996ei,Freitas:2014hra},
\ie using a multiplicative overall factor.
For leptonic final states, one may worry about dependence on the cut-off%
\footnote{Lepton separation from beams indirectly cuts on FSR.}
of the real photons emitted from the EW one-loop vertex.
Since we envisage, in the (B)$\to$(C) construction of EWPPs, the use
of MC with sophisticated CEEX-type spin amplitudes,
it will also be possible to include at this step two-loop
and one-real-one-virtual non-factorizable EW+QED corrections
without any approximation.%
\footnote{%
   This would also streamline the use of fitted EWPPs, in
   comparison with the SM+BSM predictions.
   }
For more details, see the next section.

Finally, let us come back to the problem that MC programs
were always regarded as too slow for the fitting procedure,
if used in a straightforward way.
For instance, recent calculations of the cross-sections and $A_\mathrm{FB}$
with five-digit precision, using {\tt KKMC} \cite{Jadach:2018lwm},
have required the generation of ${\sim}10^{10}$ events,
which  took 3\,days of CPU time on the 100-processor farm.
The use of the MC in the fitting procedure is, however, not so hopeless, 
because such a long MC run to get high precision is needed only once, 
for central values of a few parameters in the effective Born term
of the EWPP construction.
Small deviations, ${\sim}0.1\%$
away from the central value, can be calculated very quickly with ${<}10^{-4}$
precision using small differences of the MC weights corresponding 
to EWPP deviations from the central values,
needed for the fitting procedure.
Moreover, the dependence of the realistic observable on these
small variations of EWPPs is ideally linear; hence, linear extrapolation
using only two--three points in each variable will work with sufficient precision
over the entire interesting range of all EWPPs 
in the matrix element of the effective Born term required in the fits.
In addition, modern methods of calculating two- and three-loop EW corrections numerically,
see Chapter~\ref{chmt}, also consume  a great deal of CPU time 
per phase space point; then look-up tables and interpolation over space
of input parameters and phase space of external legs will be mandatory anyway.
This kind of pre-calculated EW higher-order correction will also
considerably speed up  fitting procedures using MC programs.

Summarizing,  Monte Carlo integration or simulation will always be
the ultimate tool for implementing perfect factorization
into QED and pure EW parts at any order, at any needed precision level.
In principle, it could be used to extract or fit EWPPs
corresponding to a well-defined effective Born matrix element
(before squaring it), for any kind of experimental cut-off,
without any problems due to the presence of  non-factorizable QED
corrections, such as IFI.
This would come at the price of some effort.
In particular, the simplified effective Born matrix element
will have to be implemented in the MC, optionally,
in parallel with the full Standard Model matrix element,
and subprograms providing weight differences due to variation
of the EWPPs have to be added.
For implementing scenario (B)$\to$(D), better two- and three-loop corrections
(with subtracted QED and interpolation over input parameters)
have to be added in  the MC event generator of the {\tt KKMC} class,
along with the most sophisticated QED matrix element.

\subsection{More on QED and EW separation beyond first-order}
\label{sec:EW-QED-separation}
The essential feature of the LEP EWPOs was that they were defined
using the $2\to 2$ process; it would be desirable to keep this
feature in the proposed FCC-ee extension.
For LEP EWPOs, construction was based essentially on \order{\alpha^1}
Standard Model calculations (including QED), so it was easy to justify such a restriction,
because  EW and QED loops combine additively 
and a single real photon of \order{\alpha^1} was attached 
to the tree-level $2\to 2$ process.
In the FCC-ee environment, where two- and three-loop EW calculation will be  standard, 
one has to consider two-loops with mixed EW and QED content
and EW corrections to the $2\to 3$ process
with extra real photons attached to EW loops.
In the following, we shall discuss how to separate QED and EW parts
in a consistent and practical way in the calculations beyond first-order
and how this should be done in the scheme of \Fref{fig:EWPO-FCC}.

The methodology formulated in Refs.~\cite{Jadach:1998jb,Jadach:2000ir}
of separating the (resummed) EQD component from Standard Model (or SM+BSM) amplitudes
will be illustrated with
the help of a representative example of the WW box diagram
and of FSR \order{\alpha^2} `non-factorizable' EW+QED contributions.
This technique is referred to as the coherent exclusive exponentiation
(CEEX) factorization or resummation technique.

Before coming to the more complicated two-loop case, let us recall briefly
how the $\gamma$Z box is treated in the  CEEX scheme.
In this scheme, the $\gamma$Z box is split at the amplitude level to form an IR part, 
which is combined later on with the corresponding real emission part,
in the resummation over photons in the soft approximation to infinite order.
(MC does all the numerics of that, also executing IR cancellations).
The remaining non-soft, non-collinear, pure \order{\alpha^1} part
is not included in the resummation and
is treated the same way as all other EW corrections, 
\ie is included in the Born-like spin amplitudes of the $2\to 2$ process.
Note that, in CEEX implementation, one never deals analytically 
with the differential cross-section;
all that is done numerically: squaring, spin summing,
integrating over phase space. 
In addition, the specific part of the $\gamma$Z box involved in the IFI suppression
near the Z resonance of the class $\ln(\Gamma_\mathrm{Z}/M_\mathrm{Z})$ is isolated
and also properly resummed to infinite order.

Let us now elaborate on the two-loop example of the WW box diagram,
with  $\mathrm{W}^+$ and $\mathrm{W}^-$ lines connecting the incoming electron line 
and the outgoing muon line,
and with insertion of an additional photon line 
attached to all six internal and external lines of charged fermions and bosons,
forming a gauge-invariant (in QED) two-loop spin amplitude.
According to the well-known analysis of 
Yennie, Frautschi, and Suura~\cite{Yennie:1961ad} (YFS),
infrared (IR) divergences reside in $6+4=10$ diagrams,
where one or both photons are attached to external legs.
The actual decomposition of the IR terms among these diagrams depends
on the choice of the gauge, but the sum is unique and gauge-independent.

One is, of course, tempted to apply
the classic method of dealing with IR cancellations using the Bloch--Nordsieck method,
that is, to take the interference of the previous  amplitude with the Born 2$\to$2 process
and combine it with the squared amplitude, where one real photon is attached to
all charged external and internal particles in the basic one-loop WW diagram, 
integrated up to some photon energy limit.
{\em  This is, however, the wrong way to proceed, 
in case the experimental acceptance affects the real-photon phase space; the only practical method to apply is that using the CEEX scheme.}
The results of the Bloch--Nordsieck method are completely useless for construction
of the CEEX matrix element implemented in   MC programs, \eg\ {\tt KKMC},
simply because the MC program is managing all IR cancellations by itself,
so using the Bloch--Nordsieck method is counterproductive.
Instead, the construction of the QED$\times$EW matrix element in CEEX
requires subtraction of the IR part from the previous two-loop matrix element,
using the well-known four-point YFS virtual form factor.
Similarly, one should subtract from the corresponding 
one-real-photon spin amplitude the well-known soft factor%
\footnote{%
    A kind of square root of the eikonal factor.}
multiplied by the WW box amplitude.
These two IR-finite objects are then used as building blocks
in the construction of the CEEX matrix element. (They will still include non-soft collinear mass logs, which
  would also be subtracted in the case when collinear resummation
  was done in the basic QED matrix element in the {\tt KKMC}.
  So far, collinear resummation is not done in the CEEX matrix element
  of {\tt KKMC}, but it could be done.)

Another group of diagrams with mixed QED$\times$EW content can be obtained
by means of inserting a virtual- or real-photon line attached to
all internal and external lines of the first-order diagram with final or initial
EW vertex corrections.
These corrections are non-factorizable in the sense of Ref.~\cite{Czarnecki:1996ei},
simply because of non-planar diagrams.%
\footnote{For FSR, the relevant diagrams are the same as in
  Ref.~\cite{Czarnecki:1996ei}, except for replacement of gluons by photons.}
As already stated, within FCC-ee precision, one may be concerned over these corrections
about the dependence on cut-off of real photons, for leptonic final states.
Like the previous WW box case, the CEEX scheme allows one to treat these issues without any approximation.
First of all, in the CEEX scheme, as implemented in {\tt KKMC}, the 
\order{\alpha\alpha_\mathrm{em}} class effects are already taken into account
in the soft photon approximation. 
What remains is to calculate  
non-soft IR-finite parts and add them to spin amplitudes.
The  
non-soft IR-finite parts are obtained by means of IR subtraction from spin amplitudes
of the YFS form factor from two-loop diagrams 
and subtraction of the soft factor
(electromagnetic current) from the diagrams with real-photon insertion in diagrams with EW loops.
All additional comments made in the case of WW box diagram apply here as well.

Similarly, one may also insert one photon to any other two-loop pure EW diagram,
creating a family of three-loop diagrams, 
for which the same procedure of CEEX subtractions will work --
one-real-photon insertion to a two-loop diagram would then enter  the game.

This CEEX subtraction scheme of IR singularities at the amplitude level
extends to higher orders, with an arbitrary number of photons inserted
into the basic diagram with pure EW content.
For instance, insertion of two photons in the one-loop 
WW box diagram creates a family of three-loop diagrams.

In other words, in all two- and three-loop corrections in the Standard Model, 
one can isolate groups of gauge-invariant subsets, as in these examples,  obtained by means of one or two photon insertions 
to basic one- or two-loop pure EW diagrams.
For each group, CEEX subtraction at the amplitude level can be done independently.
Also, for EW diagrams with one or two real-photon insertions,
a similar CEEX subtraction procedure at the amplitude level
is followed {\em independently} in parallel.

Infrared regulation for the real-photon subtracted object 
is, obviously, not necessary.
Conversely,
since the YSF form factor has a simple integral representation, in principle,
for each gauge-invariant group of diagrams 
obtained with the virtual photon insertions,
it is also possible to do subtraction before the loop virtual phase space integration, 
thus avoiding the need of IR regulation whatsoever.
In practice, it may be more convenient to use some IR regulator 
in the intermediate steps.
Contrary to the Bloch--Nordsieck approach, this regulator may be chosen
differently for each group of IR-entangled diagrams and YFS form factors.

Summarizing, the CEEX scheme briefly described here, which has  
already been tested in practice {\tt KKMC},
provides for a well-defined, clear, and clean methodology of the separation of QED
and EW parts in the one- to three-loop perturbative calculation.
This scheme also covers one or two real-photon emissions attached to diagrams with EW loops.
More details of the CEEX subtraction or resummation scheme
may be found in Refs.~\cite{Jadach:1998jb,Jadach:2000ir,Jadach:1999vf}.

In this way, one of the main obstacles towards  establishing a new scheme
for obtaining  QED-independent EWPOs, or rather EWPPs, compatible with the high precision
of the FCC-ee can be viewed as solved.

In the practice of the scheme illustrated in Fig.~\ref{fig:EWPO-FCC},
this means that the IR-subtracted
remnants of the diagrams with EW loops decorated with photon insertions
will be exactly taken into account  in the MC programs used
in the data fitting at steps  (B)$\to$(C) or (B)$\to$(D).
These contributions are expected to be small; by means of switching
them on or off, one will be able to test how strongly they influence
the bias of the two-step (B)$\to$(C)$\to$(D) procedure with respect
to the single step path (B)$\to$(D).

\section*{Acknowledgements}
 We thank Tord Riemann for useful feedback concerning issues discussed in this section.


\clearpage \pagestyle{empty} 
\cleardoublepage


\section
[The ZFITTER project 
\\
{\it A. Akhundov, A. Arbuzov, L. Kalinovskaya, S. Riemann, T. Riemann}]
{The ZFITTER project}
\label{ss-smat-zf}
\pagestyle{fancy}
\fancyhead[LO]{}
\fancyhead[RO]{}
\fancyhead[CO]{\thechapter.\thesection  \hspace{1mm} The ZFITTER project}
\fancyhead[LE]{}
\fancyhead[CE]{A. Akhundov, A. Arbuzov, L. Kalinovskaya, S. Riemann, T. Riemann}
\fancyhead[RE]{}

\noindent
{\bf{Authors:}  Arif Akhundov, Andrej Arbuzov, Lida Kalinovskaya, Sabine Riemann, Tord Riemann}
\\
Corresponding author: Tord Riemann [tordriemann@gmail.com]
\vspace*{1cm}

\begin{minipage}{8cm}
{\it Dedicated to Dima Bardin \\
~~~
\small{19.4.1945--30.6.2017}
}
\end{minipage}
\begin{minipage}{7.5cm}
\includegraphics[width=0.32\textwidth]{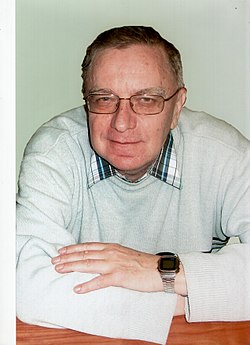}
~
 \href{https://commons.wikimedia.org/wiki/File:DYuBardin-2.jpg}{Dima Bardin}
\\
\copyright Anna Kalinovskaya, Moscow \\ (CC-BY-SA-4.0)
\end{minipage}
 

\vspace*{.5cm}

\subsection{Introduction\label{zfsma-intro}}
In the foregoing sections, it was mentioned on several occasions that the ZFITTER project was the preferred theoretical 
basis for  
physics analyses at LEP1/SLC.
The ZFITTER project has the following features.
\begin{itemize}
 \item 
  ZFITTER 
  calculates cross-sections for the reaction $\mathrm{e}^+\mathrm{e}^-\to \mathrm{f} \bar {\mathrm{ f}}$ at energies around the Z peak.
 It is not a fitting program, and it is no Monte Carlo program.
 The efficient modelling of Z peak phenomena, combining exact weak loops and one-dimensional representations of
the photonic corrections, is the basis for a preferred use of ZFITTER in data fitting. 
The latest  version released is ZFITTER 6.43.\footnote{The latest documented version is ZFITTER 6.42 \cite{Arbuzov:2005ma}; version 6.43 contains additional higher-order QCD corrections to the Z width of no numerical relevance for LEP1/SLC physics, see Ref. \cite{Akhundov:2013ons}.} 
 ZFITTER contains two related parts. One is the Standard Model library (also called the weak library), whose 
central part is DIZET.
The other part consists of QED corrections from (mainly) additional photon emission.
 \item 
 The Standard Model library DIZET of ZFITTER was, in its first-published  version in 1989, accurate at one loop, and later at 
1+1/2~loops.  
 This means that the one-loop corrections are complete and supplemented by selected higher-order-loop terms.  
 For EWPOs, higher orders are implemented \`a la Ref. \cite{Awramik:2006uz}, which is, by far, sufficient for 
LEP1/SLC physics accuracy.
  \item 
  The QED corrections in ZFITTER are available in several calculational `chains'.
  They are the result of analytical phase space integrations, so that only a few simple cuts on the final-state kinematics are possible. All versions include complete QED NLO corrections plus soft photon exponentiation, plus a few higher-order terms. 
  Roughly 
speaking, they also comprise 1+1/2 perturbative orders.  
\item 
The user can select quite different `interfaces' and flags, allowing  ZFITTER to be adapted to the description of 
certain EWPOs, or just to some Z line shape ansatz, be it defined in the Standard Model or  model-independent.
The strict modularity of ZFITTER was an important feature for this, but became unnecessarily restricted, owing to later 
implementations of higher-order corrections. 
\item 
ZFITTER is open-source from the beginning. A careful user support is being delivered permanently, supplemented by more 
than 300 pages of program descriptions. Until 2011, many program versions were available for anonymous download from a webpage and a CERN AFS account.
\end{itemize}
All this is described in much detail in the published program versions
\cite{Bardin:1989tq,Bardin:1992jc,Bardin:1999yd,Arbuzov:2005ma}
and in the theoretical papers quoted therein.
See also the 2013 ZFITTER review \cite{Akhundov:2013ons} with plots from version 6.43, the ZFITTER webpages 
\cite{web-sanc.zfitter:2016,web-zfitter.com:2015,web-zfitter.education:2015},
and the talks on the {\it Dima Day} at the conference CALC2018 
\cite{CALC2018:2018}.  

{\it An interesting question is that on the future of the ZFITTER package.}

There was a prediction by Lew Okun in 2000 \cite{RiemannTord:CALC2018}. 
He wrote the following in his review on the ZFITTER project for the JINR scientific prize 2000:
{\it \small
\begin{quote}
In the long term, with the advent of more precise experiments, ZFITTER will allow to take into account
all two-loop electroweak corrections. 
\\
$\dots$
\\
The series of theoretical articles on precision tests of the Standard Model at electron--positron colliders
certainly deserves the award of the JINR prize 2000. 
\end{quote}
}

In the following subsections, we will present our vision of the future package for the analysis of FCC-ee Tera-Z precision 
data.
We will show that there are many promising features of ZFITTER.
But in a perspective of about 20\,years from now, it is evident that a new generation of physicists should create their 
own, modern, tool.

\subsection{The form factors $\rho, v_\mathrm{e}, v_\mathrm{f}, v_\mathrm{ef}$}
The effective Born approximation of ZFITTER, \ie its $2\to2$ cross-section ansatz is:
\begin{multline}
\label{e-from0608099-sig}
{\cal A}[\mathrm{e}^+\mathrm{e}^- \to \mathrm{f} \bar{\mathrm{f}}] 
= 4\pi \mathrm{i} \alpha_\mathrm{em}(s)   \frac{Q_\mathrm{e} Q_\mathrm{f}}{s}
 \gamma_\mu \otimes \gamma^\mu 
+  \mathrm{i} \frac{\sqrt{2} G_\mu M_\mathrm{Z}^2}{1 + \mathrm{i} \Gamma_\mathrm{Z}/M_\mathrm{Z}}  I_\mathrm{e}^{(3)}  I_\mathrm{f}^{(3)} 
  \frac{1}  {s - \overline{M}_\mathrm{Z}^2 + \mathrm{i} \overline{M_\mathrm{Z}} \overline{\Gamma}_\mathrm{Z}   } 
  \\
  \times \rho_\mathrm{ef}  \Bigl[
        \gamma_\mu(1+\gamma_5) \otimes \gamma^\mu(1+\gamma_5) 
        -  4 |Q_\mathrm{e}| \sw^2  \kappa_\mathrm{e}  \gamma_\mu \otimes
  \gamma^\mu(1+\gamma_5) 
  \\  
-  4 |Q_\mathrm{f}| \sw^2  \kappa_\mathrm{f}  \gamma_\mu(1+\gamma_5) \otimes
 \gamma^\mu 
+ 16 |Q_\mathrm{e} Q_\mathrm{f}| \sw^4  \kappa_\mathrm{ef}  \gamma_\mu \otimes
  \gamma^\mu \Bigr]
 .
\end{multline}
This implementation depends on the form factors $\alpha_\mathrm{em}(s)$ and $\rho_\mathrm{ef}(s,t), \kappa_\mathrm{e}(s,t), \kappa_\mathrm{f}(s,t), 
\kappa_\mathrm{ef}(s,t)$.
These parametrizations were invented for Z peak physics  \cite{Bardin:1989di} in 1989, following an implementation  
\cite{Bardin:1988by} in which deep inelastic scattering was investigated.
The original calculations \cite{Bardin:1980fe,Bardin:1982sv} used form factors $F_{Z,i}$ and were defined close to $M_{vv}^\mathrm{ef}$ to $M_{aa}^\mathrm{ef}$, see \Eref{e-genmatrix}. 
The introduction of $\rho_\mathrm{Z}$ and $\kappa_\mathrm{e}, \kappa_\mathrm{f}, \kappa_\mathrm{ef}$ were intended as an intuitive generalization of the form 
factors invented by A. Sirlin for the Z vertex \cite{Sirlin:1980nh}.
They correspond to a renormalization of $G_\mu$ by $\rho_\mathrm{Z}$ and three generalized weak vector couplings $v_\mathrm{e}, v_\mathrm{f}, 
v_\mathrm{ef}$.
The relations between the various parameters are given in Eqs. \eqref{e-smat-rhokappa}--\eqref{e-rhokappa3}.

Having these remarks in mind, the weak library of ZFITTER is also operable  in future.
Present restrictions  arise from broken modularity, owing to too-pragmatic implementations of higher-order corrections.
Further, ZFITTER does not rely on Laurent expansions of the loop corrections, as introduced in Subsection C.\ref{s-smat-laurent}.
This leads to small deviations of derived quantities.
In Ref. \cite{Awramik:2006uz}, it is shown that, \eg the weak mixing angle of a strict pole renormalization differs from the ZFITTER default.
The difference in $\sin^2\theta^\mathrm{f}_\mathrm{eff}$ 
between {ZFITTER} and the pole 
scheme is found to be, expressed in terms of the notation of Ref. \cite{Awramik:2006uz}:
\begin{align}
\nonumber
 {\sin^2\theta^\mathrm{f}_{\mbox{\scriptsize eff}}}_{\mbox{, \scriptsize\sc ZFITTER}} \hfill
&= 
\sw^2  
        \re \left\{ \kappa_\mathrm{Z}^\mathrm{f} \left (M_\mathrm{Z}^2 \right) \right\}
        ,
        \\
         \nonumber
{\sin^2\theta^\mathrm{f}_{\mbox{\scriptsize eff}}}_{\mathrm{,pole}}
&= 
\overline{s}_{\scrs\PW}^2 
        \re \left\{ \overline{\kappa}_\mathrm{Z}^\mathrm{f} \left (M_\mathrm{Z}^2
        \right ) \right\}
        \\
         \nonumber
        &=  {\sin^2\theta^\mathrm{f}_{\mbox{\scriptsize eff}}}_{\mbox{, \scriptsize\sc ,ZFITTER}} 
    - \frac{\Gamma_\mathrm{Z}}{M_\mathrm{Z}}  \frac{q^{(0)}_\mathrm{f}}{a^{(0)}_\mathrm{e}
        \left (a^{(0)}_\mathrm{f} - v^{(0)}_\mathrm{f} \right)}  \im \left\{ p^{(1)}_\mathrm{e} \right\} 
        ,
  \label{eq:shift}
\end{align}
with the component
\begin{equation}\nonumber 
 \overline{s}_{\scrs\PW}^2 = \left(1- \frac{\overline{M}_\mathrm{W}^2}{\overline{M}_\mathrm{Z}^2} \right) =
  \sw^2 \left[1+ \frac{\cw^2}{\sw^2}
  \left(\frac{\Gamma_\mathrm{W}^2}{\MW^2} - \frac{\Gamma_\mathrm{Z}^2}{M_\mathrm{Z}^2} \right)\right]^{-1}
  . 
\end{equation}
 As a result, the ZFITTER value of $ {\sin^2\theta^\mathrm{f}_\mathrm{eff,  ZFITTER}} $
has to be corrected by a shift:
\begin{equation}
\label{eq:c126}
 s_\mathrm{W}^2 \delta\kappa_\mathrm{f} = 
  - \frac{\Gamma_\mathrm{Z}}{M_\mathrm{Z}}   \frac{q^{(0)}_\mathrm{f}}{a^{(0)}_\mathrm{e}
       \left  (a^{(0)}_\mathrm{f} - v^{(0)}_\mathrm{f} \right )}   \im   \left\{ p^{(1)}_\mathrm{e}\right\}
        \approx
        1.5 \times 10^{-6}
\end{equation}
to be compared with the actual accuracy of not better than  
$170 \times 10^{-6}$ \cite{ALEPH:2005ab}, or the planned accuracy at the FCC-ee  Tera-Z stage of $3 \times 10^{-6}$. Here, this difference will become relevant.

An additional difference between the ZFITTER renormalization and the pole scheme arises from 
the Laurent expansions of
weak loop contributions.
Look at the ZZ and WW box diagrams, which are included in {ZFITTER}
at the one-loop level, being sufficient for the next-to-next-to-leading-order
calculation in the pole scheme.
Look at the $v_\mathrm{e}$ term, which is related to the asymmetry parameter $A_\mathrm{e}$.
The  $v_\mathrm{e}$ arises from the ratio of two Laurent series, where we introduce here in the background series only the box contributions, 
$b_{a_\mathrm{e} a_\mathrm{f}}, b_{v_\mathrm{e} a_\mathrm{f}}$:
\begin{align} 
M^\mathrm{ef}_{a_\mathrm{e} a_\mathrm{f}} &= \frac{r_{a_\mathrm{e}a_\mathrm{f}}}{s - s_0} + \epsilon ~ b_{a_\mathrm{e}a_\mathrm{f}},
\\ 
M^\mathrm{ef}_{v_\mathrm{e}a_\mathrm{f}}  &=  \frac{r_{v_\mathrm{e}a_\mathrm{f}}}{s - s_0} + \epsilon ~ b_{v_\mathrm{e}a_\mathrm{f}},
\\ 
v_\mathrm{e}  &=  \frac{M^\mathrm{ef}_{v_\mathrm{e}a_\mathrm{f}}}{M^\mathrm{ef}_{a_\mathrm{e}a_\mathrm{f}}},
\\ \label{eq:c130}
\kappa_\mathrm{e}  
&=  - \frac{v_\mathrm{e}/a_\mathrm{e} - 1}{4 |Q_\mathrm{e}| \sin^2\theta_\mathrm{W}}
  \nonumber \\
&= 
\frac{r_{a_\mathrm{e} a_\mathrm{f}} - r_{v_\mathrm{e} a_\mathrm{f}}}{4 r_{a_\mathrm{e} a_\mathrm{f}} \sin^2\theta_W} 
- \epsilon ~ (s - s_0) 
\frac{b_{v_\mathrm{e}a_\mathrm{f}} r_{a_\mathrm{e}a_\mathrm{f}} - b_{a_\mathrm{e}a_\mathrm{f}} r_{v_\mathrm{e}a_\mathrm{f}}}{4 r_{a_\mathrm{e}a_\mathrm{f}}^2 \sin^2\theta_\mathrm{W}}
+ {\cal O}[(s-s_0),\epsilon^2]
.
\end{align}
Setting, for simplicity (as is done in ZFITTER), $a_\mathrm{e}=a_\mathrm{f}=1$, and setting $\epsilon$  a coupling constant, typically $\alpha_\mathrm{em}/(4\pi)$,
we see that the one-loop box terms $b_{a_\mathrm{e}a_\mathrm{f}}$ and $b_{v_\mathrm{e}a_\mathrm{f}}$ are suppressed compared with the pole term, as expected.
In $\kappa_\mathrm{e}$, an extra
term  arises from the box contributions, which is proportional to
$\mathrm{i}M_\mathrm{Z}\Gamma_\mathrm{Z}$. 
The term is due to the difference of $(s-s_0)$ (pole scheme) compared with $(s-M_\mathrm{Z}^2)$ (ZFITTER scheme). 
However, it does not 
contribute to the squared matrix element because weak  box diagrams are real at $s=M_\mathrm{Z}^2$; they are here below the production threshold 
of Z or W pairs and thus have no
absorptive part.\footnote{A special case is Bhabha 
scattering, $\mathrm{f}=\mathrm{e}$, where
additional box and
   $t$-channel diagrams contribute.}

To conclude,
the Standard Model loop library of ZFITTER should undergo several adaptations and improvements in order to make it fit for FCC-ee applications at the $Z$ peak.
Its general structure would allow this, although completely new programming seems to be preferable.

\subsection{Fortran versus {\tt C++}: modernity and modularity}
ZFITTER is written in Fortran. 
At the time, this was the only natural choice. The history of ZFITTER began with the central subroutine 
ZRATE, then ZWRATE (calculating $\Gamma_\mathrm{W}$ and $\Gamma_\mathrm{Z}$, based on Appendix F of Ref. \cite{Bardin:1982sv} with $m_\mathrm{t}=0$ and 
on Refs. \cite{Akhundov:1985fc,Bardin:1986fi} with the exact top mass dependences) in about 1981--1986.

Whether Fortran is a sufficiently `modern' programming language is an open question.
Maybe no more in 20\,years' time, when FCC will come into operation!
To beat Fortran in calculational speed with {\tt C++} is possible only with a very good optimization of {\tt C++} 
programming.\footnote{V. Yundin, private communication to T.R.} 
Conversely, young colleagues no longer learn Fortran programming.
And it is easier to perform structured, modular programming with an object-oriented language like {\tt C++}.
So the answer might be: for the user, the choice plays no role, as long as compilers exist, but for the source 
programmers, {\tt C++} will be the choice. However, programming in {\tt C++} has its own essential drawbacks;  maybe the language C, or some other alternative, is a better choice?\footnote{J. Vermaseren, private communication to T.R.}

\noindent
The authors' team, H. Fl\"acher, M. Goebel, J. Haller, A. Hoecker, K. M{\"o}nig \textit{et al.,} of the early Gfitter project (see Ref. \cite{Fred} 
 and references therein) summarizes in the introduction of Ref. \cite{Tom}:
{\it \small
\begin{quote}
Several theoretical libraries within and beyond the SM have been developed in the past, which, tied to a multiparameter minimization program, allowed
one to constrain the unbound parameters of the SM [\dots].
However, most of these programs are relatively old, were implemented in outdated programming languages,
and are difficult to maintain in [\ldots ] progress. It is unsatisfactory to rely on them during the forthcoming era
\dots \\
$\dots$
\\
None of the previous programs were modular enough to [\ldots ] allow [\ldots ] to be extended  [\ldots]  beyond the SM, 
and
they are usually tied to a particular minimization package.
These considerations led to the development of the generic fitting package Gfitter [\dots], designed to provide a
framework for model testing in high-energy physics. Gfitter is implemented in {\tt C++} \ldots
\\

\end{quote}
}

To comment on this briefly: all these statements on ZFITTER, TOPAZ0, and  DAPP are wrong, with the notable exclusion that the 
programs are really `old'.

Ironically, all early versions of the Gfitter code are, to this day, non-declared one-to-one copies of large parts of 
ZFITTER 6.42, translated from Fortran to a very simple, Fortran-like, {\tt C++} 
\cite{web-desy-zfitter-gfitter:2014-c}. 
A motivation for the copy--paste actions was, according to the Gfitter authors, to retain ZFITTER for the future.

In any case, 
we need to build a 
suitable theory framework. ZFITTER/DIZET will not be a useful basis for the
FCC-ee, since it is structured to achieve 
consistent (1+1/2)-loop precision, but not beyond. No Laurent-series approach is foreseen in the kernel ZFITTER; but see Subsection 
C.\ref{smat-ZF-smata} on the SMATASY project and its applications to data.
Further, later versions of the code lost modularity, owing to too-lazy additions concerning this item.
We will have to begin developing a 
new 
program framework -- probably object-oriented, \eg {\tt C++} -- that is general enough to be extended to any loop order and 
to
different assumptions about QED and inputs. 
All the future calculations, covering up to weak three loops and QCD four
loops 
should be performed to fit into this new framework.

\subsection{Prospects of QED flux functions
\label{smat-ZF-flux}}
We discussed several reasons to plan  a newly written package, substituting ZFITTER.
In view of the high-precision demands of the FCC-ee Tera-Z stage on the QED descriptions, one might assume that the flux function 
approach of ZFITTER is outdated; see  Appendix \ref{smat-ZF-app} for some introduction to that approach. 
In fact, this is not clear.

It is true that the flux function approach has a limited accuracy.
Conversely, it is relatively close to the real data, depending on the experimental situation, with a precision 
of  ${\approx}0.5\%$.
Additionally, the account of experimental cuts is limited.
Already at the LEP, ZFITTER's QED part was not precise enough for fits in accordance with the experimental accuracy.
However, ZFITTER is fast and flexible in many other respects.
Thus, experimentalists prepared data such that they would correspond to ZFITTER's layout, using a Monte Carlo program, 
preferably the {\tt KKMC} \cite{Bardin:1999yd}.  
It is of some importance here that {\tt KKMC} uses the weak library of ZFITTER
 for the data simulation, so that, in 
this respect, there is sufficient coherence of the tools.\footnote{In Subsection C.\ref{smat-ZF-smata}, we demonstrate that 
the data simulation may be performed without the use of a weak library. This is then a strictly model-independent approach.}

As a result of combining {\tt KKMC} and ZFITTER, the limited accuracy of ZFITTER's QED ansatz and its limited cut structure 
are {\it not limiting} the accuracy of the net fitting environment.
Of course, the high {\it technical} precision of both packages is understood on default.

Let us now look at the FCC-ee Tera-Z data.
They demand a higher accuracy than LEP data.
Let us assume that the higher-loop terms are implemented properly.
Applying again the combination {\tt KKMC}/ZFITTER, the reduction of data to the ZFITTER model can proceed in exactly the same 
way as at the  LEP.
The crucial difference from LEP analyses would be that {\tt KKMC} has to model the additionally needed higher-order QED 
corrections properly before that reduction.
However, the procedure could basically remain unchanged.

{\it The somewhat unexpected conclusion is that a ZFITTER successor package might well be suited to FCC-ee Tera-Z demands, 
concerning the QED part.}

There are subtleties to be considered carefully.
To mention one example, the higher-order terms of the initial--final-state interferences have a specific dependence on 
the weak form factors (in the Born approximation: the weak couplings). 
Another example is the potential importance of the usually neglected dependence of the weak couplings on $s$ and $\cos 
\theta$.
The influence of all that has to be studied. 

Another interesting future application of ZFITTER might happen at meson factories.\footnote{This was 
pointed out by Torben Ferber, private communication.}
The Belle II physics programme foresees measurement of the $A_\mathrm{FB}(\mathrm{e}^+\mathrm{e}^-\to\mu^+\mu^-)$ at $\sqrt{s}=10.58\UGeV$.
See Section 5.14 of Ref. \cite{Aushev:2010bq}.
The authors estimate: ``With a statistical error of $\sigma(A_\mathrm{FB}) = \pm 1 \times 10^{-5}$ on the charge 
asymmetry, the corresponding error on $\sin^2\theta_\mathrm{W}$ is $\sigma(\sin^2\theta_\mathrm{W}) = 5 \times 10^{-4}$.''
The expected systematic errors are not discussed there, for that see Refs. \cite{Ferber:dfg2015,vanderBij:2014mxa}.

 The point of interest here is the statement in Ref. \cite{Aushev:2010bq} that
Belle II will detect about { $10^9$ $\mu^+\mu^-$ pairs} at {$\sqrt{s}=10.58\UGeV$}, with a need of theoretical 
precision of about $10^{-3}$ or even better.
Evidently, one has to check whether the account of the final-state muon mass is needed, ${m_{\mu}^2}/{s}\approx 10^{-4}$. 

{\it The physical analysis of the high statistics of $\mu$-pair events at BELLE II with such a high precision seems to 
be feasible only with one combination of tools: {\tt KKMC}+ZFITTER.\footnote{Torben Ferber, private communication.}
ZFITTER 6.42 would deserve a test as to whether the $\mu$ mass has to be considered in the weak library \`a la Refs.
\cite{Lorca:2004fg,Fleischer:2003kk}, while for {\tt KKMC} one would have to check whether the necessary higher orders in QED are 
available.
}

In fact, from the formulae of Section C.\ref{smat-ZF-app} (see also Ref. \cite{Fedorenko:1986hw}), it is evident which 
contributions dominate the analysis, from the point of view of weak physics: it is the $\gamma$Z interference.
The forward--backward asymmetry is, in the notation of Ref. \cite{Aushev:2010bq}:
\begin{align}
\label{smat-afbbelle}
 A_\mathrm{FB} 
 &\approx 
 \frac{3c_2}{8c_1} 
 = \frac{3}{8} \frac{-4Q_\mathrm{e}Q_\mu a_\mathrm{e}a_\mu \chi + 8 a_\mathrm{e}a_\mu v_\mathrm{e} v_\mu \chi^2}
 {Q_\mathrm{e}^2Q_\mu^2 + 2 Q_\mathrm{e}Q_\mu v_\mathrm{e} v_\mu + (a_\mathrm{e}^2+v_\mathrm{e}^2)(a_\mu^2+v_\mu^2)\chi^2  }  
 \nonumber \\
&\approx -\frac{3}{2}  a_ea_\mu \chi,
 \\
 \label{smat-chiz}
 \chi|_{s=10\UGeV} &= 
 \frac{G_\mathrm{F} M_\mathrm{Z}^2} {\sqrt{2}~8\pi\alpha_\mathrm{em}}~K(s)|_{s=10\UGeV}  \approx  - 6.49 \times 10^{-3}
 ,
\end{align}
with $a_\mathrm{e}=a_\mu=-1$ and $v_\mathrm{e}=v_\mu= -1+4s_W^2$, and
\begin{equation}\label{smat-ks}
K(s) = \frac{s}{s-M_\mathrm{Z}^2+\mathrm{i} M_\mathrm{Z} \Gamma_\mathrm{Z}}
.
\end{equation}
Combining Eqs. \eqref{smat-afbbelle} and \eqref{smat-chiz}, we see that the weak library will be needed 
for $\Delta A_\mathrm{FB} \sim 10^{-5}$ 
at only one-loop accuracy, since the factor $\chi \sim 10^{-3}$ suppresses its numerical influence.

The definitions are very close to those in \Eref{e-smat-gam1}.
An alternative notion of $\chi(s)$ is $\chi_2(s)$:
\begin{equation}
  \label{smat-chi2}
 \chi_2(s)  = 
 \frac{1}{16s_\mathrm{W}^2 c_\mathrm{W}^2} K(s)
 .
\end{equation}
The difference of the two definitions of $\chi$ arises from renormalization; see Subsection C.\ref{s-smat-}.
In Ref. \cite{Aushev:2010bq}, a factor $\rho$ is introduced and set at $\rho=1$ for the Standard Model.
In fact, with radiative corrections, the only change of \Eref{smat-afbbelle} is the replacement $ a_\mathrm{e}a_\mu \to 
\rho_\mathrm{Z} a_\mathrm{e}a_\mu$, as discussed in detail in this section.
Evidently,  measurement of the running of the leptonic weak mixing angle, as it arises from an $s$-dependence of 
$v_\mathrm{e}(s)$ is not possible from \Eref{smat-afbbelle} because $A_\mathrm{FB}$ is independent of the vector couplings.

{\it By contrast, a measurement of $A_\mathrm{FB}$ at $\sqrt{s}=10.56\UGeV$ allows a measurement of the radiative corrections to 
the Fermi constant $G_\mu$. It will be a true single parameter measurement with a high precision.}

If one uses instead $A_\mathrm{FB}$ via the definition of \Eref{smat-chi2} for a determination of $s_\mathrm{W}^2$, the meaning of that 
is not clear, at least not in the context presented in Ref. \cite{Aushev:2010bq}. It is impossible to interpret fine-tuned 
measurements without an account of the weak corrections and without using a well-defined renormalization scheme.

\subsection{The SMATASY interface \label{smat-ZF-smata}}
The correct quantum field theoretical treatment of higher-order loop contributions to a resonance like the Z boson 
in $\mathrm{e}^+\mathrm{e}^- \to \mathrm{f} \bar{\mathrm{f}}$, \ie in $2 \to 2$ scattering, 
has been  studied since 1991 by several authors, see Refs. \cite{Stuart:1991xk,Stuart:1991cc,Stuart:1991rv,Stuart:1992jf,Stuart:1995zr} and \cite{Veltman:1992tm}.
The essential finding was that one has to interpret the loop corrections as a Laurent series around a single pole at $s_0=\bar M_\mathrm{Z}^2-\mathrm{i} \bar M_\mathrm{Z} \bar \Gamma_\mathrm{Z}$ in the complex $s$-plane.
This corresponds to the understanding of the weak amplitudes as meromorphic functions with simple poles originating from particle resonances.
Another name for this approach is the S-matrix approach.

This approach led to the idea of trying to fit  the Z resonance by Laurent series, where only the residues and the pole position are fitted.
In that way, the underlying theory gets no interpretation, while the mass and width of the Z boson are measured \cite{Leike:1991pq}.
This worked quite well. 
Of course, one has to adopt the effects from bremsstrahlung, a 2-to-3 or even 2-to-$n$ reactions, and also the interference of the Z exchange amplitude with the photon-exchange amplitude. This goes beyond the S-matrix model, so that one should more accurately call the resulting formalism `S-matrix inspired'. 

An unpublished {\it stand-alone} software ZPOLE was used first, with a simplified treatment of the QED corrections.
An observed systematic shift of the central value of the S-matrix Z mass determination, compared with ZFITTER fits, was not understood (and not documented), see Ref. \cite{Leike:1991pq}.
To clarify this, an SMATRIX {\it interface} for ZFITTER was written, so that the S-matrix fits have the same QED environment as conventional ZFITTER fits.\footnote{This was initiated by S. Riemann, unpublished.} 
A careful investigation showed that the observed systematic shift of the S-matrix Z mass was due to an internal (undocumented) fixing of the hadronic $\gamma$Z interference term in ZFITTER with some input for the quark couplings.
When this was done, the numerical effect on fits was considered to be negligible, but in the meantime the accuracy was improved. Further, the measured values for the  quark couplings were shifted. In sum, an artificial Z mass shift in ZFITTER resulted from that programming.
After repairing, the S-matrix Z mass determination agreed with its central value with the conventional ZFITTER fit, as it should be. A new, corrected ZFITTER version was released. 

Why is this episode mentioned? It demonstrates how subtleties of the fitting libraries may influence the fitting in a completely unexpected manner.

The S-matrix (inspired) approach was extended to the fit of asymmetries, taking account of QED corrections, in Ref. \cite{Riemann:1992gv}, and finally the interface package SMATASY (1991--2005) to ZFITTER 6.42 was written \cite{Kirsch:1994cf,gruenewald-smatasy:2005}. 
Its release allowed  experimental groups to perform S-matrix (inspired) fits to their data from LEP and TRISTAN.
The first application of the S-matrix (inspired) approach to LEP data was made by the L3 collaboration \cite{Adriani:1993sx}.
Later, ALEPH used the approach to study the $\gamma$Z interference at the Z peak \cite{Buskulic:1996ua}.
Further experimental analyses followed \cite{Abbiendi:2000hu,Sachs:2003ja,Holt:2014moa}, and at TRISTAN 
\cite{Yusa:1999dx,Miyabayashi:1994ej}.
A  mini-review of the SMATASY approach and various experimental 
applications may be found in Ref. 
\cite{Riemann:2015wpn}. See also the recent discussion related to FCC-ee physics in Ref. \cite{Tenchini:April2018}.

Today, when facing the immense accuracy of the FCC-ee Tera-Z data, it seems natural to combine the potential of ZFITTER for describing QED contributions with SMATASY as an interface to the language of the Laurent series and to introduce into the latter an interface to a weak library with two- to four-loop accuracy.
Such an approach would constitute a prototype of a fit library for the EWPOs with FCC-ee Tera-Z accuracy.

Instead of describing in detail how SMATASY works, let us only quote a few important facts. We will reproduce a short introduction to the essentials of the SMASY approach, and start from the generic matrix elements (\Eref{e-matrix}).
Calculating from these the integrated cross-sections, 
we arrive at generic forms, which  are now understood also to include  the contributions from photon exchange and from the $\gamma$Z interference~\cite{Leike:1991pq}.
The self-explanatory notation of Ref.  \cite{Kirsch:1994cf} is used:


\newcommand{\imag}{\mbox{$\Im$}}
\newcommand{\real}{\mbox{$\Re$}}

\def\gam{{\gamma}}%
\def\Zo{{\mathrm {Z}}}
\def\ovMZ{{{\overline m}_\mathrm{Z}}}
\def\ovGZ{{{\overline \Gamma}_\mathrm{Z}}}
\newcommand{\rtlep} {r^{\mathrm{lep}}_{\mathrm{tot}}}
\newcommand{\jtlep} {j^{\mathrm{lep}}_{\mathrm{tot}}}
\newcommand{\rfblep}{r^{\mathrm{lep}}_{\mathrm{fb}}}
\newcommand{\jfblep}{j^{\mathrm{lep}}_{\mathrm{fb}}}
\def\rtl{{r^\ell_{\mathrm{tot}}}}
\def\jtl{{j^\ell_{\mathrm{tot}}}}
\def\jtf{{j^f_{\mathrm{tot}}}}
\def\rtf{{r^f_{\mathrm{tot}}}}
\def\jaf{{j^f_{\mathrm{A}}}}
\def\raf{{r^f_{\mathrm{A}}}}
\def\rfbl{{r^\ell_{\mathrm{fb}}}}
\def\jfbl{{j^\ell_{\mathrm{fb}}}}
\def\rfbf{{r^f_{\mathrm{fb}}}}
\def\jfbf{{j^f_{\mathrm{fb}}}}
\def\rpolf{{r^f_{\mathrm{pol}}}}
\def\jpolf{{j^f_{\mathrm{pol}}}}
\def\rfbpolf{{r^f_{\mathrm{fb-pol}}}}
\def\jfbpolf{{j^f_{\mathrm{fb-pol}}}}
\def\rthad{{r^{\mathrm{had}}_{\mathrm{tot}}}}
\def\jthad{{j^{\mathrm{had}}_{\mathrm{tot}}}}
\def\Rgf{{\mathrm {R}^\mathrm{f}_\gam}}
\def\Rg{{\mathrm {R}_\gam}}
\def\RZfi{{\mathrm {R_\mathrm{Z}}^{\mathrm{f}i}}}
\def\RZ{{\mathrm {R_\mathrm{Z}}}}
\def\Ffin{{\mathrm {F}_n^{fi}}}
\def\RZli{{\mathrm {R_\mathrm{Z}}^{\ell i}}}
\def\RZti{{\mathrm {R_\mathrm{Z}}^{\tau i}}}
\def\RZfn{{\mathrm {R_\mathrm{Z}}^{f0}}}
\def\RZfe{{\mathrm {R_\mathrm{Z}}^{f1}}}
\def\RZfz{{\mathrm {R_\mathrm{Z}}^{f2}}}
\def\RZfd{{\mathrm {R_\mathrm{Z}}^{f3}}}
\def\RZtn{{\mathrm {R_\mathrm{Z}}^{\tau 0}}}
\def\RZte{{\mathrm {R_\mathrm{Z}}^{\tau 1}}}
\def\RZtz{{\mathrm {R_\mathrm{Z}}^{\tau 2}}}
\def\RZtd{{\mathrm {R_\mathrm{Z}}^{\tau 3}}}

\begin{align}
\renewcommand{\arraystretch}{2.2}
\label{eqn:smxs}
\nonumber
\sigma_\mathrm{A}^0(s)
&=
\displaystyle{\frac{4}{3} \pi \alpha^2
\left[ 
\frac{r^{\gamma \mathrm{f}}_\mathrm{A}}{s} +
\frac {s r^\mathrm{f}_\mathrm{A} + (s - \ovMZ^2) j^\mathrm{f}_\mathrm{A}} {(s-\ovMZ^2)^2 + \ovMZ^2 \ovGZ^2} 
+ \frac{r_\mathrm{A}^{\mathrm{f}0}}{\ovMZ^2} + {\mathrm{further ~ background}}
\right]  },
\\
\nonumber
&\to
\displaystyle{\frac{4}{3} \pi \alpha^2
\left[ \frac{r^{\gamma \mathrm{f}}_\mathrm{A}}{s} +
\frac {s r^\mathrm{f}_\mathrm{A} + (s - \ovMZ^2) j^\mathrm{f}_\mathrm{A}} {(s-\ovMZ^2)^2 + \ovMZ^2 \ovGZ^2} 
+ \frac{r_\mathrm{A}^{\mathrm{f}0}}{\ovMZ^2}\right]},
\qquad A = \mbox{ T, FB, pol, polFB,} \ldots \\ 
&\approx
\displaystyle{\frac{4}{3} \pi \alpha^2
\left[ \frac{r^{\gamma \mathrm{f}}_\mathrm{A}}{s} +
\frac {s r^\mathrm{f}_\mathrm{A} + (s - \ovMZ^2) j^\mathrm{f}_\mathrm{A}} {(s-\ovMZ^2)^2 + \ovMZ^2 \ovGZ^2}
\right]}
.
\renewcommand{\arraystretch}{1.}
\end{align}
The $r^{\gamma \mathrm{f}}_\mathrm{A}$ is the photon-exchange term,
\begin{eqnarray}
\label{qedr}
r^{\gamma \mathrm{f}}_\mathrm{A} =
\displaystyle{\frac{1}{4}  c_\mathrm{f} \sum_{i=1}^4 \{\pm 1\}
\left| \Rgf \right|^2 } R_\mathrm{QCD}.
\end{eqnarray}
For  better precision, one may use $\alpha_\mathrm{em}(s)$ instead of the constant parameter $r^{\gamma \mathrm{f}}_\mathrm{A}$; but assuming a resummation of background parts, as  discussed in Subsection C.\ref{s-smat-gammaZ}. 
This term vanishes for all asymmetric cross-section combinations, \ie for $\mathrm{A} \neq \mathrm{T}$.
It is $c_\mathrm{f}=1,3$ for leptons and quarks, respectively.
The QCD corrections for quarks are taken into account by the factor $R_\mathrm{QCD}$ of Ref.
\cite{Bardin:1999yd}. 
The Z exchange residues $r^\mathrm{f}_\mathrm{A}$ and the $\gamma \mathrm{Z}$-interference terms
$j^\mathrm{f}_\mathrm{A}$ are:
\begin{align}
r^\mathrm{f}_\mathrm{A} & =  \displaystyle{c_\mathrm{f} \left\{\frac{1}{4}\sum_{i=1}^4 \{\pm 1\}
\left| \RZfi \right|^2 + 2 \frac{\ovGZ}{\ovMZ} \imag C^\mathrm{f}_\mathrm{A} \right\}} R_\mathrm{QCD}, \nonumber
\\ 
\label{eqn:rrjj}
j^\mathrm{f}_\mathrm{A} & =  c_\mathrm{f}\left\{2 \real C^\mathrm{f}_A - 2 \displaystyle{\frac{\ovGZ}{\ovMZ}}
\imag C^\mathrm{f}_\mathrm{A}\right\} R_\mathrm{QCD},
\end{align}
and the $C^\mathrm{f}_\mathrm{A}$ are the complex-valued interferences of the residues $R_\mathrm{f}^{(i)}$ of \Eref{e-matrix} with $\alpha_\mathrm{em}(s)$,
\begin{equation}
\label{eqn:cfa}
C^\mathrm{f}_\mathrm{A}  = \left (R^\mathrm{f}_\gamma \right )^* \left(\displaystyle{\frac{1}{4}\sum_{i=1}^4
\{\pm 1\} \RZfi}\right) .
\end{equation}
The factors $\{\pm 1\}$ in Eqs.~(\ref{qedr}) and~(\ref{eqn:rrjj})
indicate that the signs of $\left| R_{\gamma}^\mathrm{f} \right|$, 
$\left| \RZfi \right| ^2$,
and  $\RZfi$ correspond to the signs of $\sigma_i$ in Eqs.
\eqref{e-smat-matrix1}--\eqref{e-smat-matrix4}.
Explicit expressions for the background contributions may well become of numerical importance in the FCC-ee Tera-Z era.
We refer, for explicit expressions, to the literature quoted.

When measuring the Z line shape, the independent parameters of a pseudo-observable are $M_\mathrm{Z}$ and $\Gamma_\mathrm{Z}$, as well as the residue term $r_\mathrm{A}$ and the $\gamma \mathrm{Z}$ interference term $j_\mathrm{A}$, and potentially 
more parameters arising from the background.
So each of the observables depends on a least four parameters.

{\it 
For an adequate fit of the Z line shape, one needs, consequently, at least five independent data points taken at different values of $s$. When also
fitting the background, one needs more data points.}

From the correlation tables, not shown here, one derives that $M_\mathrm{Z}$ and $j_\mathrm{T}$ are strongly correlated, as well as  $\Gamma_\mathrm{Z}$ and 
$r_\mathrm{T}$.
In fact, there is a simple, but very accurate relation between an uncertainty (or shift) of the Z peak position $s_{\max}$ (the centre of gravity of the Z peak is defined by the Z mass) and the $\gamma$ Z term 
$j_\mathrm{T}$ \cite{Leike:1992uf}: 
\begin{eqnarray}\label{e-correlMj}
\Delta M_\mathrm{Z} \sim 
 \Delta(\sqrt{s_{\max}}) = \frac{1}{4}  \frac{\Gamma_\mathrm{Z}^2}{M_\mathrm{Z}} \times \frac{j_\mathrm{T}}{r_\mathrm{T}} \approx 17\UMeV \times 
\frac{j_\mathrm{T}}{r_\mathrm{T}}
.
\end{eqnarray}
The mass and width of the Z boson are best determined from the hadronic cross-section 
$\sigma_\mathrm{had}$, so that we may 
set: $j_\mathrm{T} \to j_\mathrm{T, had}$ and  $R_\mathrm{T} \to R_\mathrm{T, had}$.
A conclusion for fitting is as follows.

{\it 
For an adequate fit of $M_\mathrm{Z}$, one must control the correlation with the $\gamma \mathrm{Z}$ interference term $j_\mathrm{T,had}$. Fixing the latter produces a smaller error of $M_\mathrm{Z}$, but in a model-independent understanding, this is not correct.
For the FCC-ee Tera-Z applications, one must study the interplay of all the fitting components  carefully.
Similar statements apply to the correlation of $\Gamma_\mathrm{Z}$ and $r_\mathrm{T, had}$.
}

A further remark concerns the asymmetries \cite{Riemann:1992gv}.
They  are all ratios of Laurent series, thus becoming themselves Taylor expansions in $s$ around $s_0$.
If there were  no contributions from photon exchange, the asymmetries would be, at first approximation, independent of $s$, if 
there were no other considerable background contributions.

How do such observations change if we do not consider the 2$\to$2 cross-sections and observables, but the real cross-sections?
This has been studied, assuming the applicability of a flux function approach with sufficient accuracy 
\cite{Leike:1991pq,Riemann:2015wpn} and can be further studied using ZFITTER/SMATASY 
\cite{Kirsch:1994cf,gruenewald-smatasy:2005}.
If QED is described by some Mont Carlo code, all the conclusions remain unchanged because the formulae are truly generic concerning QED.

The main results on the influence of QED corrections are as follows.

{\it 
One must distinguish helicity  asymmetries and angular asymmetries due to their different QED corrections.
}

The former undergo the same photonic corrections as the total cross-section, and the latter go with the forward--backward (FB) antisymmetric cross-sections. 
Both differ from each other, and they differ more, if more hard radiation happens.
See Section  \ref{smat-ZF-app} for details on that issue.

In fact, one may derive very elegant and universal formulae for the real asymmetries in the S-matrix (inspired) approach. This was done 
in Ref. \cite{Riemann:1992gv}, in which  the accessible EWPOs are explicitly exhibited, but  in the SMATRIX (inspired) language.

The $2\to2$ forward--backward asymmetry is 
\begin{eqnarray}
 A_{0,\mathrm{FB}}(s) = A_0 + A_1 (s-s_0) + A_2 (s-s_0)^2 + \cdots  
\end{eqnarray}
Here, 
\begin{align}
A_0 &= \frac{r_\mathrm{FB}}{r_\mathrm{T}}
,
\\ \label{sma-a1}
A_1 &= \frac{j_\mathrm{FB}}{r_\mathrm{FB}} - \frac{j_\mathrm{T}}{r_\mathrm{T}},
\end{align}
etc.
One may show that the QED-corrected asymmetries are quite similar to the Born asymmetries. They are modified by certain factors $c_n$, which are smooth functions of the kinematics.
The real forward--backward asymmetry is
\begin{eqnarray}
 A_\mathrm{FB}(s) = c_{0,\mathrm{FB}}(s) A_0 +c_{1,\mathrm{FB}}(s) A_1 (s-s_0) + c_{2,\mathrm{FB}}(s) A_2 (s-s_0)^2 + \cdots  
\end{eqnarray}
The QED functions $c_{i,\mathrm{FB}}(s)$ are independent of the underlying model.
While
\begin{eqnarray}
 c_{0,\mathrm{T}}(s) =1,
\end{eqnarray}
it is 
\begin{align} \label{sma-c0}
 c_{0,\mathrm{FB}}(s)&= \frac{C^R_\mathrm{FB}(s)}{C^R_\mathrm{T}(s)},
\\
 c_{1,\mathrm{FB}}(s)&= c_{0,\mathrm{FB}}(s) ~ \frac{C^J_\mathrm{T}(s)}{C^R_\mathrm{T}(s)}
 ,
 \end{align}
etc. 
The functions $c_{i,\mathrm{A}}(s)$ depend on the QED treatment. 
For the account of initial-state radiation \`a la ZFITTER:
\begin{align}
 C^R_\mathrm{T,FB} &= \int \mathrm{d} \frac{s'}{s}  \rho^\mathrm{ini}_\mathrm{T,FB}(s')
~ \frac{s'}{s}
~ \frac{|s-s_0|^2}{|s'-s_0|^2},
\\
 C^J_\mathrm{T,FB} &= \int \mathrm{d} \frac{s'}{s}  \rho^\mathrm{ini}_\mathrm{T,FB}(s')
~ \frac{s'-M_\mathrm{Z}^2 }{s-M_\mathrm{Z}^2}
~ \frac{|s-s_0|^2}{|s'-s_0|^2},
\end{align}
etc.
Evidently,  $C^{R,J}_\mathrm{T}$ and  $C^{R,J}_\mathrm{FB}$ are smooth in $s$ and independent of the weak interactions model. 

In the case of another treatment of the QED corrections, the QED factors will become modified by a necessary replacement of the flux functions $\rho^\mathrm{ini}_\mathrm{T,FB}(s')$, while the QED `kernels' 
${s'}/{s}~ {|s-s_0|^2}/{|s'-s_0|^2}$  and 
${s'-M_\mathrm{Z}^2}/{s-M_\mathrm{Z}^2}~ {|s-s_0|^2}/{|s'-s_0|^2}$, \etc,
will stay unchanged.

\begin{figure}
\centering
  \includegraphics[width=0.6\textwidth]{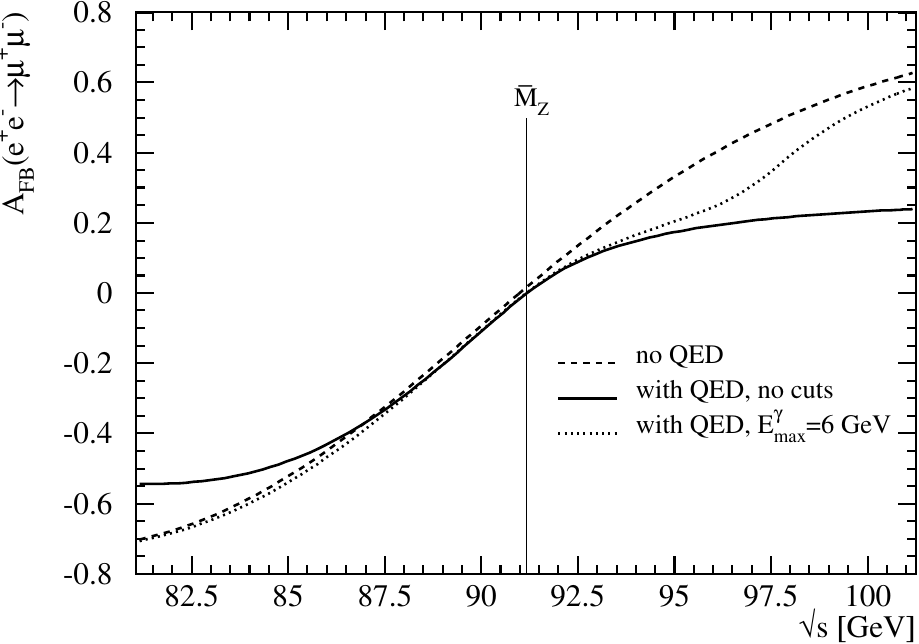} 
         \caption{Forward--backward asymmetry for the process $\mathrm{e}^+ \mathrm{e}^- \rightarrow \mu^+ \mu^-$ near the Z peak. Reproduced from Ref. \cite{Kirsch:1994cf}, with permission from Elsevier, licence number 4551260314704 (17 March  2019).
  \label{asyfromCPC}}
\end{figure}

Figure \ref{asyfromCPC} shows a typical forward--backward asymmetry around the Z peak, here for muon pair production.
The non-vanishing $\gamma$ Z interference close to the Z peak prevents the asymmetry from staying constant, a 2to2 effect. 
Because the photonic corrections differ for $\sigma_\mathrm{T}$ and $\sigma_\mathrm{FB}$, the real asymmetry does not agree with the Born asymmetry at $s=M_\mathrm{Z}^2$; see \Eref{sma-c0}.
If  no hard photons were emitted, the real asymmetry would be equal to the Born approximation because, in that case, the QED corrections would cancel out, see Section C.\ref{smat-ZF-app}; this is very well fulfilled at $s=M_\mathrm{Z}^2$, where hard photon emission is suppressed. 
The increase in $A_\mathrm{FB}$  around the Z peak is due to the $\gamma$Z interference.
Further, the behaviour right of the peak is markedly due to the radiative corrections.
The curve with QED and no cuts gets strongly damped at larger $s$ because the $\sigma_T$ develops a resonating behaviour 
(resonance tail due to radiative return), but the $\sigma_\mathrm{FB}$ does not; see the formulae and  comments in Section C.\ref{smat-ZF-app}.
If one cuts the maximal allowed energy of the photon emission, the radiative return is 
suppressed and we see a QED-corrected line shape close to the Born shape -- as  appears left of the Z peak.

\subsection{Conclusions}
ZFITTER is a mighty tool for all its applications with precision goals like those at LEP1/SLC.
For true precision, the Monte Carlo tool used for preparing the measurements  (\eg\ {\tt KKMC}) should rely on the same weak library as ZFITTER. 
The combination ZFITTER/SMATASY is a role model for applying the S-matrix (inspired) ansatz, using Laurent series.
If interpreted in the Standard Model, it would be fit for multiloop FCC-ee Tera-Z applications.
Preferable is the development of a new package.
The SANC collaboration has already made  steps in this direction, see Section C.\ref{contr:sanc}.
It is fair to say that a completely numerical Monte Carlo approach might also be appropriate for FCC-ee Tera-Z applications. One has to study the pros and the cons of the semi-analytical and the pure numerical approaches.
In any case, the weak library of the unfolding and that of the fitting procedures have to be the same, in order to be consistent and accurate in a controlled way. 

A completely model-independent fit is possible, when based on the residues and background coefficients of the Laurent series of the matrix elements, not interpreting them in terms of the Standard Model.
The correlations of fit parameters must be very carefully studied.

\subsection{Appendix: QED flux functions \label{smat-ZF-app}}
The origin of ZFITTER is twofold.
First, the seminal articles with the one-loop weak form factors, together with the Fortran code BCF (in 
Euclidean metrics),  became the basis of the weak library DIZET of ZFITTER \cite{Bardin:1980fe,Bardin:1982sv}, after 
translation to the Minkowski metrics.
The other basis was the analytical calculation of the integrated complete $\cal O(\alpha)$ photonic 
corrections to $\mathrm{e}^+\mathrm{e}^-\to \mu^+\mu^-$ scattering around the Z peak  without a cut.
The original work remained unpublished, because the referee of \textit{Nuclear Physics B} estimated the contents as too far from true physics 
-- because no cuts were included.
As a compensation, the preprint appeared twice \cite{Bardin:1987hv,Bardin:1988ze}.
For didactic purposes, we reproduce the main results.
The formulae are exact, have no kinematic cuts, and can be useful for basic tests of MC programs.
They are extremely compact.
The corresponding Fortran program is presumably lost. It had been very carefully checked in 1988 against the 1982 Monte Carlo code 
MUSTRAAL \cite{Berends:1982ie,Berends:1983mi}, which made basically the same calculation, but numerically.
The explicit expressions are:
\begin{align}
\sigma_\mathrm{T} 
&= 
\sigma_0 \Bigl \{ Q_\mathrm{e}^2Q_\mathrm{f}^2 
\left[ 1+\frac{\alpha}{\pi} \left (Q_\mathrm{e}^2 F_0^\mathrm{T}+Q_\mathrm{f}^2F_2^\mathrm{T}
\right ) \right] + \frac{\alpha}{\pi}Q_{\mathrm{e},a}Q_{\mathrm{f},a}Q_\mathrm{e}^2Q_\mathrm{f}^2 \Re 
\left (F_4^\mathrm{T} \right )
\nonumber 
 \\
& \qquad \quad  + 2 |Q_\mathrm{e}Q_\mathrm{f}| v_\mathrm{e} v_\mathrm{f} \Re \left[ \chi + \frac{\alpha}{\pi} \chi \left (Q_\mathrm{e}^2G_0^\mathrm{T}+Q_\mathrm{f}^2 G_2^\mathrm{T} \right ) \right]
+2 |Q_\mathrm{e}Q_\mathrm{f}| a_\mathrm{e} a_\mathrm{f} 
\frac{\alpha}{\pi} Q_\mathrm{e}Q_\mathrm{f} \Re \left  (\chi G_4^\mathrm{T} \right
) \nonumber 
\\
& \qquad \quad + (v_\mathrm{e}^2+a_\mathrm{e}^2)(v_\mathrm{f}^2+a_\mathrm{f}^2)|\chi|^2 \left[ 1+ \frac{\alpha}{\pi}\Re \left (Q_\mathrm{e}^2H_0^\mathrm{T}+Q_\mathrm{f}^2 H_2^\mathrm{T} \right ) \right] \nonumber
\\
 \label{f-324-d-A}
& \qquad \quad
+ 4 v_\mathrm{e} a_\mathrm{e} v_\mathrm{f} a_\mathrm{f} |\chi|^2 \frac{\alpha}{\pi} Q_\mathrm{e} Q_\mathrm{f} \Re \left [H_4^\mathrm{T} \right ] 
 \Bigr \}
,
\\ 
A_\mathrm{FB} 
&= 
\frac{\sigma_0}{\sigma_\mathrm{T}} 
\Biggl\{    Q_\mathrm{e}Q_\mathrm{f}                        
\frac{\alpha}{\pi} 
\left[  Q_\mathrm{e}^2Q_\mathrm{f}^2 {\Re } \left(F_1^\mathrm{T} \right)+ 4Q_{\mathrm{e},a}Q_{\mathrm{f},a}Q_\mathrm{e}Q_\mathrm{f}{\Re } \left (Q_\mathrm{e}^2 F_3^\mathrm{T}  + Q_\mathrm{f}^2 F_5^\mathrm{T} \right )\right] 
\nonumber \\ 
\nonumber
& \quad + 2 |Q_\mathrm{e}Q_\mathrm{f}|v_\mathrm{e} v_\mathrm{f} \frac{\alpha}{\pi} Q_\mathrm{e}Q_\mathrm{f} {\Re } \left (\chi G_1^\mathrm{T} \right )
+ 2 |Q_\mathrm{e}Q_\mathrm{f}| a_\mathrm{e} a_\mathrm{f} {\Re } \left [\frac 34 \chi + \frac{\alpha}{\pi} \chi \left  (Q_\mathrm{e}^2 G_3^\mathrm{T}+Q_\mathrm{f}^2 G_5^\mathrm{T} \right ) \right ]  
\\ 
& \quad 
+ \left (v_\mathrm{e}^2+a_\mathrm{e}^2 \right ) \left (v_\mathrm{f}^2+a_\mathrm{f}^2
\right) |\chi|^2 
  \frac{\alpha}{\pi}  Q_\mathrm{e}Q_\mathrm{f} {\Re }(H_1^\mathrm{T}) \nonumber 
\\ 
& \quad
+ 4 v_\mathrm{e} a_\mathrm{e} v_\mathrm{f} a_\mathrm{f} |\chi|^2 
\left[ \frac 34 + \frac{\alpha}{\pi}  {\Re } \left [Q_\mathrm{e}^2 H_3^\mathrm{T}+Q_\mathrm{f}^2 H_5^\mathrm{T} \right ] \right] 
\Biggr\}  
, \label{smat-afb666}
\end{align} 
with
\begin{eqnarray}
 \chi = \frac{G_\mu M_\mathrm{Z}^2} {\sqrt{2}8\pi\alpha_\mathrm{em}} \frac{s}{s-M^2}.
\end{eqnarray}
In the following, the suffix `br' stands for `bremsstrahlung' and the suffix
`box' stands for one-loop QED $\gamma \gamma$ and $\gamma$Z 
box terms.

The initial-state one-loop contributions to $\sigma_\mathrm{T}$ are:
\begin{align}
 F_0^\mathrm{T} &= A+B \left[ L_\mathrm{f}-\frac 76 \right]
 ,
 \\
 G_0^\mathrm{T} &= A+B \left[ R+\frac 12 + \left (1+R^2 \right)L_R  \right]
 ,
 \\
 H_0^\mathrm{T} &= A+B \left[ 2R+\frac 12 -|R|^2 + \frac {\mathrm{i}}{g} (1-R^*)R
 \left (1+R^2
 \right )L_R  \right]
 ,
 \end{align}
with 
\begin{align}
 A&=\frac{\pi^2}{3}-\frac 12
 ,
 \\
 B&=L_\mathrm{e}-1
 .
\end{align}
The well-known global enhancement $L_\mathrm{e}=\ln(s/m_\mathrm{e}^2)$ appears only in the initial-state contributions.
The enhancement $L_\mathrm{f}=\ln(s/m_\mathrm{f}^2)$ in the pure photonic part disappears with cuts.
A logarithmic term is related to the Z boson:
\begin{align}\label{smat-lr}
 L_R &= \ln\left(1-\frac 1R \right),
 \\
 R&=\frac{M^2}{s},
 \\
 M^2 &= M_\mathrm{Z}^2 - {\mathrm{i}} M_\mathrm{Z} \Gamma_\mathrm{Z}
 .
\end{align}
A trace of the Z propagator treatment, enabling the analytical integrations,  
\begin{eqnarray}
\frac{1}{ \left |s-M^2 \right |^2} = \frac{\mathrm{i}}{2M_\mathrm{Z}\Gamma_\mathrm{Z}} \left(\frac{1}{s-M^2} - \frac{1}{s-M^{*2}}\right), 
\end{eqnarray}
is seen in $H_0^\mathrm{T}$, and only there: 
the pure Z exchange initial-state radiation term develops at energies above $s=M_\mathrm{Z}^2$, the so-called radiative tail.
This is a huge enhancement, proportional to $M_\mathrm{Z}/\Gamma_\mathrm{Z}$, of the cross-section,
\begin{align}
\Re   \left( \frac {\mathrm{i}}g L_R\right) &= \Re  \left[ \frac{\mathrm{i}}{\Gamma_\mathrm{Z}} \frac{s}{M_\mathrm{Z}} \ln\left(1-\frac{s}{M^2}  \right) 
 \right]
\nonumber \\ 
\nonumber
&\equiv  \frac{M_\mathrm{Z}}{\Gamma_\mathrm{Z}} \frac{s}{M_\mathrm{Z}^2} ~ \Re  \left[ \mathrm{i} \times \ln(1-s/M_\mathrm{Z}^2 - \mathrm{i} s\Gamma_\mathrm{Z}/M_\mathrm{Z}^3) \right]
\\
&= 0 \qquad \mathrm{if~} s< M_\mathrm{Z}^2, \qquad \mathrm{(the~} \ln~ \mathrm{ function~~is ~real ~there)}
.
\end{align}
The enhancement is possible due to the prefactor  ${\mathrm{i}} /g$.
If $M_\mathrm{Z}/\Gamma_\mathrm{Z}$ is small, then
\begin{align}
 \nonumber
\Re   \left( \frac {\mathrm{i}}g L_R\right)
&= \frac{M_\mathrm{Z}}{\Gamma_\mathrm{Z}} \frac{s}{M_\mathrm{Z}^2}
\Re  \left[\mathrm{i} \left[\ln(1-s/M_\mathrm{Z}^2) - \mathrm{i} 
\pi \right] \right]
\\
&\sim 
-\frac{s}{M_\mathrm{Z}^2}  \frac{M_\mathrm{Z}}{\Gamma_Z} 
\qquad \mathrm{if~} s> M_\mathrm{Z}^2
.
\end{align}
The same tail function as in $H_0^\mathrm{T}$ will also appear  in the Z squared initial-state part $H_3^\mathrm{T}$ of $\sigma_\mathrm{FB}$.
 
The final-state photonic corrections are known to be small and universal:
\begin{eqnarray}
 F_2^\mathrm{T} =  G_2^\mathrm{T} =  H_2^\mathrm{T} = \frac{3}{4}. 
\end{eqnarray}

Finally, we come to the initial--final-state interferences, combined with box contributions. 
The initial--final-state interferences fulfil the relation \cite{Riemann:1988gy}:
\begin{eqnarray}
 G_i = \frac 12 (F_i + H_i),
\end{eqnarray}
so that only the pure photonic and the pure Z exchange functions are to be determined:
\begin{align}
{F_4^\mathrm{T}}_\mathrm{br} &= 6 \bar{P}_\mathrm{IR} - 9
,
\\
{H_4^\mathrm{T}}_\mathrm{br} &= 6 \bar{P}_\mathrm{IR}-9+3 R[1+(1+R)L_R] ,  \label{f-324-g} 
\\
{F_4^\mathrm{T}}_\mathrm{box} &= -6 \bar{P}_\mathrm{IR}+\frac{9}{2},
  \label{f-324-a} 
  \\
  {H_4^\mathrm{T}}_\mathrm{box} &= -6 \bar{P}_\mathrm{IR}+9-3 R[1+(1+R)L_R]-3 L_\mathrm{Z} ,  \label{f-324-b}
\end{align} 
    with
    \begin{align}
  \label{f-324-c0}
  L_\mathrm{Z}    &= - \ln R, 
  \\
\bar{P}_\mathrm{IR} &= {P}_\mathrm{IR} + \frac 12 \ln \frac{s}{M_\mathrm{W}^2} = \ln \frac{2E}{\lambda},
\\
{P}_\mathrm{IR} &= \frac{1}{n-4}+\frac 12 \gamma_E -\ln (2 \sqrt{\pi}) = \ln 
\frac{M_\mathrm{W}}{\lambda}. 
    \end{align} 
The $\lambda$ is a small regularizing photon mass.
As in $H_0^\mathrm{T}$ and  $H_3^\mathrm{T}$, as well as in $H_4^\mathrm{T}$, we meet the $L_R$ of \Eref{smat-lr}, but here it is not in connection with a factor 
$\mathrm{i}/\Gamma_\mathrm{Z}$, so no tail property will arise. 
The total interference contributions to $\sigma_\mathrm{T}$ are:
\begin{align}
 F_4^\mathrm{T} &= -\frac 92,
 \\
  G_4^\mathrm{T} &= \frac 12 \left (F_4^\mathrm{T}+H_4^\mathrm{T} \right ),
  \\
  H_4^\mathrm{T} &= -3 L_\mathrm{Z}.
\end{align}
The function $F_4^\mathrm{T}$ does not appear in $\sigma_\mathrm{T}$, because the axial couplings of the photon vanish, see 
\Eref{f-324-d-A}.
It will appear when sufficiently many of the scattering particles are polarized; see the modifications of the coupling 
constants due to spin effects as given, \eg in Eq. (19) of Ref. \cite{Bardin:1988ze}.
For a polarized electron, among others, the changes $v_\mathrm{e}v_\mathrm{f}\to (v_\mathrm{e}-\lambda a_\mathrm{e})v_\mathrm{f}$ and $a_\mathrm{e}a_\mathrm{f}\to (a_\mathrm{e}-\lambda 
v_\mathrm{e})a_\mathrm{f}$ appear. For the electric charges (vector couplings), this means additional mixing terms. If  both beam particles are now 
polarized,  the combination $Q_{\mathrm{e},a}Q_{\mathrm{f},a}$  produces vector couplings in the cross-section expressions and so  
contributions of $F_1^\mathrm{T}$ and $F_4^\mathrm{T}$ will appear.
For a coexistence not only of photon and Z boson, but also a second $\mathrm{Z}'$ boson, the terms $F_3$ and $F_4$ are also  needed 
 in unpolarized scattering \cite{Leike:1989ah}.

We come now to the contributions to the forward--backward asymmetry $A_\mathrm{FB}$,
initial-state corrections $F_3^\mathrm{T}, G_3^\mathrm{T}, H_3^\mathrm{T}$, initial--final interferences: $F_1^\mathrm{T}, G_1^\mathrm{T}, H_1^\mathrm{T}$, and final-state 
corrections: 
\begin{eqnarray}
F_5^\mathrm{T} = G_5^\mathrm{T} = H_5^\mathrm{T} = 0.
\end{eqnarray}
The initial-state contributions are:
\begin{align}
\frac 13 G_3^\mathrm{T} &=
-\frac 18 + \frac{(1-R)}{(1+R)} (1-\ln 2) + (1-R) \ln^2 2 +\frac 14 (1+2R) \mathrm{Li}_2(1)
\nonumber \\
 & \quad
+ \frac 12 \frac{(1-R)}{(1+R)} \left[ -\frac{(1+3R)}{(1+R)}\ln 2 +\frac{(7-R)}{4(1-R)} \right]\left(L_\mathrm{e}-1 \right)
+R\frac{ \left (1+R^2 \right )}{(1+R)^2} D_3,
\\
\frac 13 H_3^\mathrm{T} &=
-\frac{2R}{|1+R|^2} - \left| \frac{(1-R)}{(1+R)}\right|^2\ln 2 + |1-R|^2\ln^2 2 + \frac 14 \left (1+4R-2|R|^2 \right )\mathrm{Li}_2(1)
\nonumber 
\\ & \quad 
+ \left(\frac 78 - 2 \frac{R}{|1+R|^2} \right) L_\mathrm{e} +\left(-\frac 52 +2\frac{(6R-1)}{|1+R|^2} 
+ 4 \frac{(1-R^2)}{|1+R|^4}\right)\left( L_\mathrm{e}-1\right)\ln 2
\nonumber
\\ & \quad
+ R^2  \frac{(1+R^2)}{(1+R)^2}\left[ 2+\frac {\mathrm{i}}{g} (1-R) \right] D_3
.
\end{align}
The QED function $F_3^\mathrm{T}$ was implemented in ZFITTER later, so we refer to the code for this term. 
With $F_4^\mathrm{T}$, the situation is simpler because it may be reconstructed from $G_4^\mathrm{T}$ and $H_4^\mathrm{T}$, which were determined 
explicitly.
The function $D_3$ introduces through its $L_R$ the term $\frac {\mathrm{i}}{g} \ln(1-1/R)$, also producing the radiative tail of $H_0^\mathrm{T}$ 
 in $H_3^\mathrm{T}$. 

The photonic part  $F_1^\mathrm{T}$ from the initial--final-state interference is a constant:
\begin{align}
{F_1^\mathrm{T}}_\mathrm{br} &=  (1+8 \ln 2)\bar{P}_\mathrm{IR} +\frac 34 [1-16 \ln 2 -\mathrm{Li}_2(1)]  , \label{f-324-...A} \\
{F_1^\mathrm{T}}_\mathrm{box} &= - (1+8 \ln 2)\bar{P}_\mathrm{IR} +\frac 34 (1+6 \ln 2 +\ln^2 2)-\frac{\mathrm{i} \pi}{2}(2-5 \ln 2)   
\label{f-324-...}
. 
\end{align}
The sum of the two $F_1^\mathrm{T}$ in \Eref{smat-afb666} contributes to $A_\mathrm{FB}$:
\begin{eqnarray}
{\Re } \left(F_1^\mathrm{T} \right) = -\frac 34 \mathrm{Li}_2(1)+\frac 34 \ln^2 2-\frac{15}{2} \ln 2+\frac 32 = -4.572
.
\end{eqnarray}
This value agrees with Eq. (25) of Ref. \cite{Fedorenko:1986hw}.

The initial--final-state interference function  due to pure Z boson exchange and the two $\gamma$Z box diagrams 
is $H_1^\mathrm{T}$:
\begin{align}
{H_1^\mathrm{T}}_\mathrm{br} &= (1+8 \ln 2)\bar{P}_\mathrm{IR} 
- \frac 14 (5 R-3) - \frac 34 \left (1+R-2R^2 \right ) \mathrm{Li}_2(1)
\nonumber \\
& \quad + \frac{1}{1+R} \left (-12+3R+8R^2+5R^3 \right )\ln 2 
\nonumber \\
& \quad + \frac 12 (1-R) \left (5-R+R^2 \right )\mathrm{Li}_2 \left (\frac 1R \right
)- \frac{R}{1+R} \left (1-4R+R^2 \right )L_R
+\frac 12 \left  (5-3R+6R^2 \right )D_1 
\nonumber \\
& \quad +  \frac R2 \left (6-3R+5R^2 \right )D_2 \label{eq412}
,
\\
{H_1^\mathrm{T}}_\mathrm{box} &= -(1+8 \ln 2)\bar{P}_\mathrm{IR} 
+ \frac 32-R+ \left (9-4R-4R^2 \right ) \ln 2 +2 \ln^2 2+
\frac 12 (-5+4R) L_\mathrm{Z}  
\nonumber \\
& \quad
+ 4R^3 \left[ \mathrm{Li}_2(1)-\mathrm{Li}_2 \left (1-\frac 1R \right ) \right] 
\nonumber \\
& \quad + \frac 12 \left[ 4-9R+3R^2+2 \left (-5+3R-6R^2 \right ) \ln 2 \right] L_R   
\nonumber \\
& \quad + (1-3R+6R^2-8R^3) \left[\mathrm{Li}_2 \left (1-\frac{1}{2R} \right )-\mathrm{Li}_2 \left (1-\frac 1R \right )\right]
\label{eq413}
.
\end{align}
It is 
\begin{align}
 D_{0,1} &= \mathrm{Li}\left(\pm \frac{1+R}{1-R} \right) + \frac 12 L_R^2,
 \\
 D_2 &= D_0 + \mathrm{Li}_2\left(-\frac{1}{R}\right) -\mathrm{Li}_2(1) + 
\ln\left(\frac{R+1}{R-1}\right)\ln\left(-\frac{1}{R}\right), 
\\
D_3 &= D_1 + D_2 +\ln^2(2) +\left(L_\mathrm{e}-1-2\ln(2)\right) L_R
\end{align}
The initial--final interference, $\gamma$Z term in the asymmetry $A_\mathrm{FB}$ contains the sum of two, 
\begin{eqnarray}
G_1^\mathrm{T}={F_1^\mathrm{T}}+{H_1^\mathrm{T}}.
\end{eqnarray}

The phase space to be integrated was the following:
\begin{eqnarray}
 \int \mathrm{d} \Gamma = \frac{\pi^2}{4s} ~ \int_{-1}^1 \mathrm{d} \cos \theta \int_0^1 x \mathrm{d}x \frac{1-x}{1-x+m_\mathrm{f}^2/s}
 ~ \frac{1}{4\pi}\int_{0}^{2\pi} \mathrm{d}\phi_\gamma~\int_{-1}^{1} \mathrm{d} \cos\theta_\gamma
 ,
\end{eqnarray}
where $x=2E_{\mathrm{f}^+}/\sqrt{s}$ and the photon variables are defined in the rest system of fermion+photon 
\cite{Passarino:1982zp}.
The problematic feature of these results is the lack of variable $s'$, the invariant mass of the final-state 
fermion pair, related the photon energy by 
$s'/s=1-2E_\gamma/\sqrt{s}$.
Because of that lack, soft photon exponentiation may not be implemented, and it was decided to change to other variables.

As a consequence, a new parametrization of QED corrections was published in a series of three papers from May 1989 to October 1990  
\cite{Bardin:1989cw,Bardin:1990fu,Bardin:1990de} and implemented in ZFITTER:
integrated cross-sections with no cut, differential in the scattering-angle $\cos\theta$ cross-sections, and cross-sections integrated with a cut on $\cos\theta$. Additionally, cuts on $s'$ are possible.
Later, this was accomplished with the account of an alternative $\mathrm{f}^+\mathrm{f}^-$-collinearity cut $\zeta$  
\cite{Bilenky:1989zg,Christova:1999cc,Jack:2000as}
using techniques developed in Refs. \cite{Bardin:1976qa,Akhundov:1984mm,Bardin:1987ht}.

The phase space was now chosen to allow for  a cut on the photon energy:
\begin{equation}
   \int \mathrm{d} \Gamma = \frac{\pi^{2}}{4 s}
   \int_{-c_{m}}^{c_{m}} \mathrm{d}\cos\theta
   \int_{0}^{v_{m}} \mathrm{d} v
  \int_{v_{2,\min}}^{v_{2,\max}} \mathrm{d} v_{2}\frac{1}{2\pi}
   \int_{0}^{2\pi} \mathrm{d} \varphi_{\gamma},
\label{eq:iii}
\end{equation}
with
\begin{align}
  v_{2,\max(\min)} &= \frac{1}{2} v \left[1\pm v_{m}(s')^{1/2}
   \right],
\\
  v_{m}(s) &= 1 - 4m_\mathrm{f}^{2}/s.
\label{eq:v23}
\end{align}
In \Eref{eq:iii}, $\varphi_{\gamma}$ is the azimuthal
angle of the photon in the c.m.s. and
\begin{equation}
  v_{2} = - \frac{2}{s} p_{\mathrm{f}^+} p_{\gamma} = 1 - 2 E_{\mathrm{f}^{-}}/\sqrt{s}.
\end{equation}The integration boundaries for $\cos \theta$ are chosen:
\begin{equation}
  c_{m} = 1 - 2 m_\mathrm{e}^{2}/s.
\label{eq:ccc}
\end{equation}
Finally, $x$ is the
momentum  fraction  of  $\mathrm{f}^{+}$ in the c.m.s. in units  of  the  beam
energy:
\begin{equation}
x = -2 p_{\mathrm{f}^{+}} ( p_{\mathrm{e}^{+}} + p_{\mathrm{e}^{-}})/s = 2 E_{\mathrm{f}^{+}}/\sqrt{s}.
   \end{equation}

The corresponding formulae are published, so we need not  reproduce them here.
For didactic reasons, we show some initial--final interference formulae for the totally integrated cross-sections with 
numerical integration over 
$s'/s = 1 - 2E_{\gamma}/\sqrt{s}$, taken from Ref.
\cite{Bardin:1988ze}. They  are 
reproduced in Eqs. \eqref{f-324-c1}--\eqref{f-324-c}.
It is
\begin{equation}
  \label{f-324-c1}
A_\mathrm{FB}(s) = \frac{\sigma_\mathrm{FB}}{\sigma_\mathrm{T}},
\end{equation}
 and the corresponding cross-sections are 
 \begin{equation}
\sigma_\mathrm{T} = \sum\limits_{\begin{array}{l}\mathrm{a}=\mathrm{e,i,f} \\k,l=1,2\end{array}}^{}  \Re \int\limits_{0}^{1}\mathrm{d}v 
\sigma_\mathrm{T}^{\mathrm{a},0}(s,s';B_k,B_l) r_\mathrm{T,a}(v;B_k,B_l) \label{f-324-c2-A}
,
 \end{equation}
 and
 \begin{equation}
\sigma_\mathrm{FB} = \sum\limits_{\begin{array}{l}\mathrm{a}=\mathrm{e,i,f} \\k,l=1,2\end{array}}^{}  \Re \int\limits_{0}^{1}\mathrm{d}v 
\sigma_\mathrm{FB}^{\mathrm{a},0}(s,s';B_k,B_l) r_\mathrm{FB,a}(v;B_k,B_l) \label{f-324-c2-B}
,
 \end{equation}
 with $B_1= \mathrm{Z},B_2=\gamma$ and
   \begin{align}
\sigma_\mathrm{FB}^{i,0}(s,s';B_k,B_l) & =\frac 34 \sigma_\mathrm{T}^{0}(s,s';B_k,B_l),   \label{f-324-c3} \\
\sigma_\mathrm{A}^{0}(s,s';B_k,B_l) &=\frac{4\pi \alpha^2}{3s'}C_\mathrm{A}(B_k,B_l) \times \frac 12 \left[ 
\kappa_k(s')\kappa_l^\ast(s)+ 
\kappa_k(s) \kappa_l^\ast(s')\right], \qquad \mathrm{A}=\mathrm{T,FB}   \label{f-324-c4} 
\\
 \kappa_l(s)&=\frac{s}{s-m_l^2}; \\
 m_l^2 &= M_l^2-\mathrm{i} M_l \Gamma_l(s),\\
 \Gamma_l(s)&= \frac{s}{M_l^2} \Gamma_l,
  \label{f-324-c5}
  \end{align}
 where $C_\mathrm{A}(B_k,B_l)$ are corresponding coupling constant combinations of the kind used in the one-loop formulae; see 
any ZFITTER reference.

The relevant points here are the convolution weights for $\sigma_\mathrm{T}$ and $A_\mathrm{FB}$:
\begin{equation}
r_\mathrm{A,a}(v;B_k,B_l)=\delta(v) s_a(B_k,B_l)+\theta(v-\epsilon)H_\mathrm{A,a}(v), \qquad \mathrm{A}=\mathrm{T,FB}, \qquad \mathrm{a}=\mathrm{e,i,f},
\end{equation}
where
\begin{equation}
B(\mathrm{Z,Z})=H_{1\mathrm{box}}^\mathrm{T},\qquad B(\gamma,\gamma)=F^\mathrm{T}_{1\mathrm{box}},\qquad
B(\mathrm{Z},\gamma)=\frac 12 \left[ B^\ast (\gamma,\gamma)+B(\mathrm{Z,Z}) 
\right]. \label{f-324-c}
\end{equation} 
 The soft functions are:
 \begin{align} 
 S_\mathrm{a} &= s_\mathrm{a}, \qquad \mathrm{a}=\mathrm{e,f},
 \\
 s_\mathrm{a} &=  Q_\mathrm{a}^2 \frac{\alpha}{\pi} \left[(L_\mathrm{a}-1)
 \left [2\ln(\epsilon)+\frac{3}{2}
 \right ] + \frac{\pi^2}{3}-\frac{1}{2} \right], 
\qquad \mathrm{a}=\mathrm{e,f},
 \\
 S_i(B_k,B_l)&= s_i(B_k,B_l),
 \\
s_i(B_k,B_l)&= Q_\mathrm{e}Q_\mathrm{f} \frac \alpha \pi \left(-(1+8 \ln 2)\ln \frac{2 \epsilon}{\lambda}+4 \ln^22+\ln 2+\frac 12 +\frac 
13 \pi^2-B(B_k,B_l) \right)
. 
 \end{align}
 The hard functions for $\sigma_\mathrm{T}$ are $H_\mathrm{T,a}(v)\equiv H_\mathrm{a}(v)$:
 \begin{align}\label{smathh1}
 H_\mathrm{e}(v) &= Q_\mathrm{e}^2 \frac{\alpha}{\pi} \left(L_\mathrm{e}-1\right)\frac{1+(1-v)^2}{v}
  ,
  \\ \label{smathh2} 
  H_\mathrm{i}(v) &= Q_\mathrm{e} Q_\mathrm{f} \frac{\alpha}{\pi} \frac{3}{v}(1-v)(v-2)
  ,\\ \label{smathh3}
  H_\mathrm{f}(v) &=Q_\mathrm{f}^2 \frac{\alpha}{\pi}  \frac{1}{v} \left [1+(1-v)^2 \right
  ] \left
  [(L_\mathrm{e}-1)+\ln(1-v)
  \right ]
  .
 \end{align}
The hard FB functions are $H_\mathrm{FB,a}(v)\equiv h_\mathrm{a}(v)$:
 \begin{align}
 \label{smathh4}
  h_\mathrm{e}(v) &=  
  Q_\mathrm{e}^2 \frac{\alpha}{\pi} \frac{(1+(1-v)^2)}{v} \frac{(1-v)}{(1-v/2)^2}
  \left[L_\mathrm{e}-1-\ln\frac{(1-v)}{(1-v/2)^2} \right]
 ,
 \\  \label{smathh}
 \nonumber 
 h_\mathrm{i}(v) 
 &=  Q_\mathrm{e}Q_\mathrm{f} \frac{\alpha}{\pi}\frac{2}{3v}
 \Bigl[  
 2(1-v) \left (v^2+2v-2 \right ) + (1-v) \left (5v^2-10v+8 \right )\ln(1-v) 
 \\ 
 & \qquad \qquad \qquad 
 + \left  (5v^3-18v^2+24v-16 \right)\ln(2-v)
 \Bigr],
 \\ \label{smathh6}
 h_\mathrm{f}(v) 
 &=  Q_\mathrm{f}^2 \frac{\alpha}{\pi} \frac{2}{v}\left[ (1-v)(L_\mathrm{f}-1) +\ln(1-v) +\frac{1}{2} v^2 L_\mathrm{f}  \right]
 .
 \end{align}
 
  Soft photon exponentiation is described by the following replacements for the initial-state terms:
 \begin{align}\label{smat-ss1} 
 \bar s_\mathrm{e} = \bar S_\mathrm{e} &= Q_\mathrm{e}^2 \frac{\alpha}{\pi} \left[ \frac 32 (L_\mathrm{e}-1) + \frac 13 \pi^2-\frac 12\right],
  \\\label{smat-ss2}
  \bar h_\mathrm{e}(v) &=  h_\mathrm{e}(v) -\frac{\beta_\mathrm{e}}{v},
  \\\label{smat-ss3}
   \bar H_\mathrm{e}(v) &=  H_\mathrm{e}(v) -\frac{\beta_\mathrm{e}}{v}
  .
  \end{align}
  The soft cross-sections become modified:
    \begin{equation}
  \int_0^1 \mathrm{d}v \sigma_\mathrm{T,FB}(s') \left [\delta(v)(1+S_\mathrm{e}) + \theta(v-\epsilon) \frac{\beta_\mathrm{e}}{v} \right ]
  \to
  (1+\bar S_\mathrm{e}) \int_0^1 \mathrm{d}v \sigma_\mathrm{T,FB}(s')
 \left  [\beta_\mathrm{e} v^{\beta_\mathrm{e}-1} \right ]
.
  \end{equation}
The term $\theta(v-\epsilon) {\beta_\mathrm{e}} / {v}$ comes from the hard parts, so that these have to be modified as in Eqs.
\eqref{smat-ss2} and \eqref{smat-ss3}.

Here follows an important phenomenological remark.
It is known that, at the Z peak, soft photon emission dominates hard photon emission.
This is because the emission of hard photons moves the $\sigma(s')$ away from the peak, thus making it  rapidly smaller.
Further, for soft emission, $v$ is  small.
For small $v$, the hard initial-state kernels $H_\mathrm{e}(v)$ and $h_\mathrm{e}(v)$ approach each other, see Eqs. \eqref{smathh1} and 
\eqref{smathh4}.
The consequence is that, at the Z peak, the radiation due to $\sigma_\mathrm{T}$ and due to $\sigma_\mathrm{FB}$ are much alike.
This explains that the $A_\mathrm{FB}$ has small radiative corrections at the Z peak; a well-know phenomenon.
Further, helicity asymmetries have same-type QED corrections in both numerator and denominator, and thus get practically no photonic corrections; this is
also a 
well-known fact.

\subsection*{Acknowledgement}
We would like to thank Janusz Gluza for his contribution in an early stage of the write-up.

\clearpage \pagestyle{empty} 
\cleardoublepage


\pagestyle{fancy}
\fancyhead[LO]{}
\fancyhead[RO]{}
\fancyhead[CO]{\thechapter.\thesection \hspace{1mm} 
The event generator \babayaga}
\fancyhead[LE]{}
\fancyhead[CE]{C.M. Carloni Calame, G. Montagna, O. Nicrosini, F. Piccinini}
\fancyhead[RE]{} 

\section
[The event generator \babayaga \\ {\it C.M.~Carloni~Calame, G.~Montagna, O.~Nicrosini, F.~Piccinini}]
{The event generator \babayaga} 
\label{egbabayaga}
\noindent
{\bf Authors: C.M.~Carloni~Calame, G.~Montagna, O.~Nicrosini, F.~Piccinini }
\\
Corresponding author: F.~Piccinini {[fulvio.piccinini@pv.infn.it]}  
\vspace*{.5cm}

\noindent The Monte Carlo event generator \babayaga\ has been developed for high-precision simulation of QED processes ($\mathrm{e}^+\mathrm{e}^-\to
\mathrm{e}^+\mathrm{e}^-$, $\mathrm{e}^+\mathrm{e}^-\to \mu^+\mu^-$ and $\mathrm{e}^+\mathrm{e}^-\to
\gamma\gamma$) at flavour factories, chiefly for luminometry purposes,
with an estimated theoretical accuracy at the 0.1\% level or
better. QED radiative corrections are included by means
of a parton shower in QED matched with exact next-to-leading-order
corrections to reach the required accuracy. The latter is assessed
by means of consistent comparisons with independent calculations and
an estimate of the size of missing higher-order corrections. The main theoretical
framework is overviewed and the status of the generator is
summarized. Some possible developments of the generator for FCC-ee
physics are addressed and considered.

\subsection{Introduction}
\label{intro}
\noindent
Knowledge of the luminosity $\mathcal L$ is an important ingredient
for any measurement at $\mathrm{e}^+\mathrm{e}^-$ machines. A
common strategy is to calculate it through the relation ${\mathcal L} = N_\mathrm{obs}/\sigma_\mathrm{th}$,
where $\sigma_\mathrm{th}$ is the theoretical cross-section of a
QED process, namely $\mathrm{e}^+\mathrm{e}^-\to \mathrm{e}^+\mathrm{e}^-$ (Bhabha),
$\mathrm{e}^+\mathrm{e}^-\to \mu^+\mu^-$, or $\mathrm{e}^+\mathrm{e}^-\to\gamma\gamma$,
and $N_\mathrm{obs}$ is the number of observed events. QED
processes are the best choice because of their clean signal and low background, and the
possibility of pushing the theoretical accuracy up to the $0.1\%$
level or better. The latter requires the inclusion of the relevant radiative
corrections  in the cross-section calculation and their implementation in Monte Carlo (MC) event
generators  in order to easily account for realistic event selection criteria.

Modern event generators used for luminometry simulate QED processes by including the
exact next-to-leading-order (NLO) QED corrections or a
leading-logarithmic (LL) approximation of
higher-order (h.o.)
effects~\cite{CarloniCalame:2000pz,CarloniCalame:2001ny,CarloniCalame:2003yt,Balossini:2006wc,Balossini:2008xr,Jadach:1995nk,Arbuzov:2005pt,Berends:1983fs,Berends:1980px}. 
The consistent inclusion and matching of NLO and h.o. LL contributions is mandatory, in
view of the required theoretical accuracy. In the following, it is
discussed how this is achieved and implemented in the \babayaga\ event generator.

\subsection{The event generator \babayaga\ and \babayagaNLO}
\label{babayaga}
\noindent
The  \babayaga\ event generator was originally developed for the precise simulation
of large-angle Bhabha scattering at low-energy $\mathrm{e}^+\mathrm{e}^-$ colliders, with centre-of-mass energies up to $10\UGeV$. It was later extended to simulate 
$\mu^+\mu^-$ and $\gamma\gamma$ final states in the same energy
regime. For the sake of clarity, in this section we focus on Bhabha
scattering as
a reference process to discuss the theoretical framework of the generator.

In its first version \cite{CarloniCalame:2000pz,CarloniCalame:2001ny}, the generator relied on a QED
parton shower (PS) to account for the LL photonic corrections, resummed up to all
orders in perturbation theory. The PS is a MC
algorithm that gives an exact iterative solution of the Dokshitzer--Gribov--Lipatov--Altarelli--Parisi
(DGLAP) equation in QED for the non-singlet QED structure function $D(x,Q^2)$, which reads
\begin{align}
 Q^2\frac{\partial}{\partial Q^2}D(x,Q^2) & =\frac{\alpha}{2\pi}
\int_x^{1}\frac{\mathrm{d}y}{y} P_+(y) D \left (\frac{x}{y},Q^2 \right )\nonumber \\
 P_+(x) & =\frac{\left ( 1+x^2 \right )}{\left (1-x \right )}-\delta(1-x)\int_0^1 \mathrm{d}t P(t),
\label{qeddglap}
\end{align}
where $P_+(x)$ is the regularized Altarelli--Parisi vertex, $x$ is the
fraction of energy lost because of radiation, and $Q^2$ is the scale of
the process.
The QED structure functions account for photon radiation emitted
by both initial- and final-state fermions and
allow the inclusion of universal virtual
and real photonic corrections, resummed up to all orders of perturbation
theory. The advantage of the PS solution is that the kinematics of the
emitted photons can be recovered (within some approximation) and hence
an exclusive event generation can be performed, \ie all the momenta of
the final-state particles (fermions and an indefinite number of
photons) can be reconstructed.

Within the structure function approach, the corrected
cross-section can be written as 
\begin{equation}
\sigma(s)=\int \mathrm{d} x_1\mathrm{d} x_2\mathrm{d} y_1\mathrm{d} y_2\int \mathrm{d}\Omega\times
D \left (x_1,Q^2 \right )D \left (x_2,Q^2 \right )D \left (y_1,Q^2 \right
)D \left (y_2,Q^2 \right )\times\frac{\mathrm{d}\sigma_0(x_1x_2s)}{d\Omega}\Theta(\mathrm{cuts}).
\label{sezfs}
\end{equation}

Despite its advantages, the PS is intrinsically accurate at the LL
level and a precision better than $0.5$--$1\%$ can not be expected in the
calculation of the cross-section~(\Eref{sezfs}). To improve the accuracy, 
matching with the exact NLO radiative
corrections is mandatory, in such a way that the
features of the PS are preserved (\ie exclusive event
generation and
resummation of LL corrections up to all orders), while avoiding the
double counting of the ${\cal O}(\alpha)$ LL corrections, present both in the
PS approach and in the NLO calculation. 

An original matching algorithm has been implemented in the latest
version of \babayaga\ \cite{CarloniCalame:2003yt,Balossini:2006wc,Balossini:2008xr} (\babayagaNLO), which includes
exact NLO corrections in a PS framework and achieves an accuracy at
the $0.1\%$ level in the cross-section calculation. 

Without spelling out the details, the fully differential PS cross-section
implicit in Eq.~(\ref{sezfs})
can be recast in the form
\begin{equation}
\mathrm{d}\sigma^{\infty}_\mathrm{PS}=
{\Pi} \left (Q^2,\varepsilon \right )~
\sum_{n=0}^\infty \frac{1}{n!}~|{\cal M}_{n,\mathrm{PS}}|^2~\mathrm{d}\Phi_n,
\label{generalPS}
\end{equation}
where ${\Pi}(Q^2,\varepsilon)$ is the Sudakov form factor, accounting
for virtual and soft (up to $x=\varepsilon$) radiation, ${\cal M}_{n,\mathrm{PS}}$
is the amplitude for the emission of $n$ real photons in the PS
approximation and $\mathrm{d}\Phi_n$ is the exact phase space for the emission
of $n$ real photons (with $x\geq\varepsilon$), divided by the incoming
flux factor. Equation (\ref{generalPS})
can be improved to include the missing NLO contributions by defining the
 correction factors
\begin{align}
F_\mathrm{SV}&= 1+\frac{\mathrm{d}\sigma^\mathrm{NLO}_\mathrm{SV}-\mathrm{d}\sigma^{[\alpha,\mathrm{PS}]}_\mathrm{SV}}{\mathrm{d}\sigma_0}
\nonumber\\
F_{i,{H}}&= 1+\frac{\mathrm{d}\sigma^\mathrm{NLO}_{i,H}-\mathrm{d}\sigma^{[\alpha,\mathrm{PS}]}_{i,H}}
{\mathrm{d}\sigma^{[\alpha,\mathrm{PS}]}_{i,H}},
\label{ffactors}
\end{align}
where SV stands for soft and virtual photon corrections, $H$ for non-soft
real-photon corrections, $[\alpha,\mathrm{PS}]$ stands for the ${\cal O}(\alpha)$ expansion of
the PS contribution, $i$ runs over the emitted photons, and $\mathrm{d}\sigma_0$
is the lowest-order differential cross-section. With these
definitions, the matched differential cross-section can be written in
the form
\begin{equation}
\mathrm{d}\sigma^{\infty}_\mathrm{matched}=
{\Pi}(Q^2,\varepsilon)  F_\mathrm{SV} 
\sum_{n=0}^\infty \frac{1}{n!}|{\cal M}_{n,\mathrm{PS}}|^2\;F_{n,H} \mathrm{d}\Phi_n,
\label{matched}
\end{equation}
which is the master formula according to which event generation and
cross-section calculation are performed in \babayagaNLO.

A few comments are in order with respect to the master
formula (\Eref{matched}).
\begin{enumerate}
\item By construction, the factors in Eq.~(\ref{ffactors})
are infrared and collinear safe quantities and they let the ${\cal O}(\alpha)$
expansion of $\mathrm{d}\sigma^{\infty}_\mathrm{matched}$ in Eq.~(\ref{matched}) coincide with the
exact NLO result.
\item The resummation of h.o. LL corrections is preserved.\item The correction factors (\Eref{ffactors}) tend to be larger in those phase space regions
where the PS is more unreliable, typically in phase space regions where
the photon is hard and not collinear to one of
the charged particles.
\item Equation~(\ref{matched}) is cast
in a completely differential form, so that events
can be generated exclusively, as in the pure PS
approach.
\item The theoretical error is shifted to corrections of order
  $\alpha^2$, \ie at the next-to-next-to-leading-order
  (NNLO) level.
\end{enumerate}

Equation (\ref{matched}) is used in \babayagaNLO\ to generate events for Bhabha,
$\mathrm{e}^+\mathrm{e}^-\to\mu^+\mu^-$ and $\mathrm{e}^+\mathrm{e}^-\to\gamma\gamma$~\cite{Balossini:2008xr} processes, with an
indefinite number of extra photons to account for NLO and h.o. QED radiative
corrections.
Some numerical results and an estimate of the theoretical accuracy of the approach
are sketched in the next sections.

\subsection{Results and estimate of the theoretical error at
  flavour factories}
\label{accuracyff}
\noindent
\Tref{table:sigmas} shows the impact of different classes of
QED radiative
corrections in the determination of large-angle Bhabha cross-sections
within typical event selection criteria used for luminometry at flavour
factories. Set-ups {\it a} and {\it b} correspond to $\sqrt{s}=1.02\UGeV$ and set-ups
{\it c} and {\it d} to $\sqrt{s}=10\UGeV$, for large ({\it a} and {\it c}) and narrower
({\it b} and {\it d}) angular acceptance cuts. The detailed description of the set-ups, which tend to select elastic events, is
reported in 
Ref.~\cite{Balossini:2006wc}.

\begin{table}
\caption{Bhabha cross-section (in nanobarns) according to different precision
  levels. The table is reproduced from
  Ref.~\cite{Balossini:2006wc} (where the experimental set-ups are also detailed), with permission from Elsevier, licence number 4545601325609 (10 March 2019).}
\label{table:sigmas}
\centering
\begin{tabular}{lllll}
      \hline \hline
      Set-up               &     {\it a}   &
      {\it b}   &      {\it c}   &     {\it d} \\
          \hline
      $\sigma_{0}   $      &
      $6855.74$      &     $529.463$  &
      $71.333$   &     $5.502$    \\
      $\sigma_{0,\mathrm{VP}}$      &
      $6976.5$   &     $542.66$  &
      $74.763$   &     $5.8552$    \\
      $\sigma_\mathrm{NLO}$     &
      $6060.1$ &     $451.523$  &
      $59.90$  &     $4.425$    \\
      $\sigma_\mathrm{PS}^\alpha$      &
      $6083.6$ &     $454.50$  &
      $60.14$  &     $4.456$    \\
      $\sigma_\mathrm{PS}^\infty$      &
      $6107.6$   &     $458.44$  &
      $60.62$   &     $4.530$    \\
      $\sigma_\mathrm{matched}^{\infty}$      &
      $6086.7$   &     $455.86$  &
      $60.42$   &     $4.504$    \\
      \hline \hline
\end{tabular}
\end{table}

The first row of \Tref{table:sigmas} is the lowest-order cross-section, the second includes the effects of vacuum polarization (VP), the
third  is the exact NLO result, the fourth is the
${\cal  O}(\alpha)$ expansion of Eq.~(\ref{generalPS}) (\ie
the PS approximation of the NLO result), the fifth is
Eq.~(\ref{generalPS}) and the last  corresponds to the most
accurate matched formula of Eq.~(\ref{matched}). By comparing the cross-sections
calculated at different theoretical accuracies, it can be inferred
that VP affects the cross-section at some level, fixed
${\cal O}(\alpha)$ QED corrections at the $10$--$20\%$ level and
h.o. QED corrections have an impact of $0.5$--$1.5\%$: all these classes of
correction are important to achieve the theoretical precision
required by  experiment and are consistently included and simulated in
\babayagaNLO\ by means of the master formula (\Eref{matched}).

To assess the theoretical accuracy of
the approach implemented in \babayagaNLO,
an important step is the performance of tuned comparisons with independent MC event generators.

\begin{figure}
\centering
\includegraphics[width=0.6\textwidth]{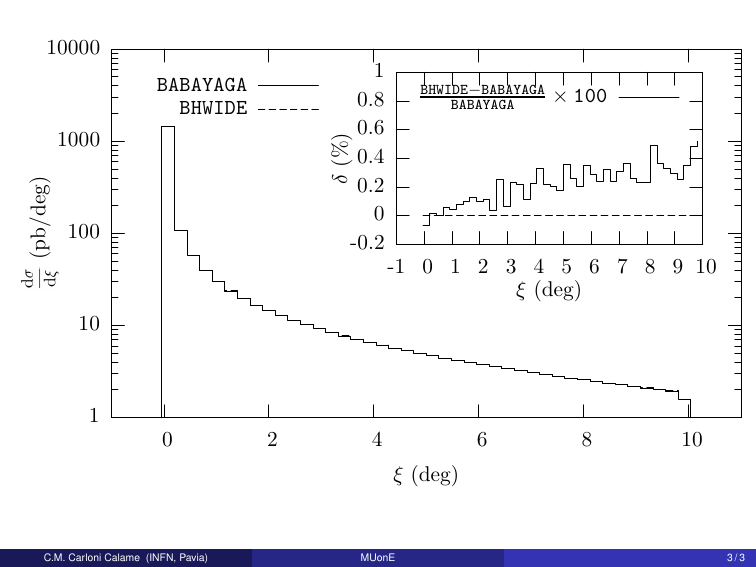}
\caption{Comparison between \babayagaNLO\ and {\tt BHWIDE} for the
  acollinearity distribution}
\label{vsbhwideacoll}       
\end{figure}

\begin{figure}
\centering
\includegraphics[width=0.6\textwidth]{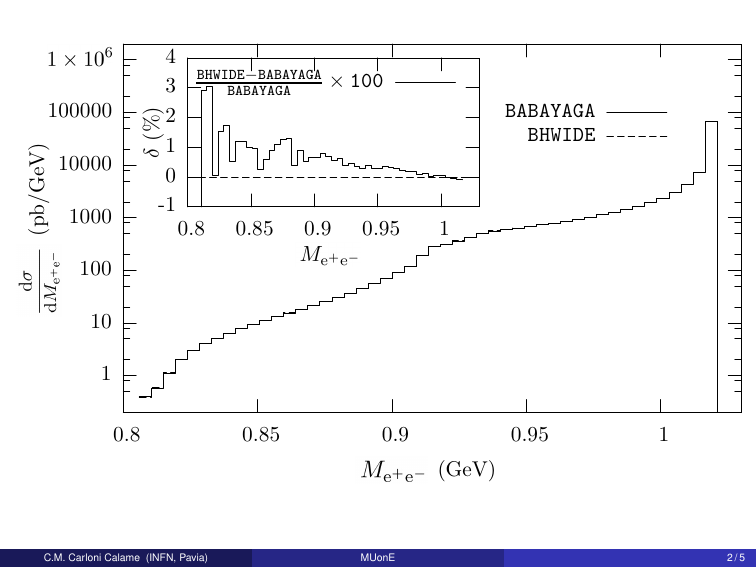}
\caption{Comparison between \babayagaNLO\ and {\tt BHWIDE} for the
  final-state invariant mass distribution.}
\label{vsbhwideinv}       
\end{figure}

As an example, in Figs. \ref{vsbhwideacoll} and \ref{vsbhwideinv},
the acollinearity and final-state invariant mass distributions
obtained by \babayagaNLO\ and the independent event generator
{\tt BHWIDE} \cite{Jadach:1995nk} are compared for set-up {\it a}. While the difference in
the integrated cross-sections is below the $0.01\%$ level, some
differences at the $1\%$ level appear (see the insets in the figures) in the differential
distributions and are due to the different theoretical frameworks on
which 
the two event generators are based. However, it is worth noticing that such differences appear
where the differential cross-section drops by several orders of
magnitude with respect to the elastic region and where the different treatment of hard
radiation beyond NLO can have an impact.

A further step in estimating the theoretical accuracy is to compare with
existing calculations of the NNLO corrections to Bhabha scattering,
which have been published over the years~\cite{Bonciani:2004gi,Bonciani:2004qt,Penin:2005kf,Penin:2005eh,Bonciani:2003cj,Bonciani:2003ai,Bonciani:2003te,Bonciani:2005im,Czakon:2006pa,Czakon:2004wm} (see also
Ref.~\cite{Actis:2010gg} and references therein). The dominant part of
such corrections are already included in the master
formula (\Eref{matched}), which can be expanded up to NNLO and the NNLO
terms of which can be
unambiguously compared with  exact analytical results: any difference
should be considered  a theoretical error in the
formulation of \babayagaNLO. A
detailed and non-trivial comparison has been reported in Ref.~\cite{Balossini:2006wc}, finding that the NNLO
corrections not included in Eq.~(\ref{matched}) impact the cross-section
at the level of a few units in $10^{-4}$, when typical event selection
criteria are taken into account.

In the luminosity section of Ref.~\cite{Actis:2010gg},  other sources of
theoretical uncertainties were also considered, such as the uncertainty in
the hadronic contribution to VP and extra light-pair emission (for
which, see also Refs.~\cite{CarloniCalame:2011aa,CarloniCalame:2011zq}). The main conclusion of that
section is that a sound estimate of the current theoretical error on the
luminosity determination via Bhabha scattering at flavour factories
lies in the region of $0.1\%$ or slightly below, which is enough for
the present experimental requirements. Any further improvement would
require the inclusion of the full exact NNLO radiative
corrections in MC event generators
for Bhabha scattering,
matched with LL h.o. contributions. This should be considered a feasible, although not trivial, task.

In a similar way, the matching algorithm of Eq.~(\ref{matched}) has
been  been applied to the $\mathrm{e}^+\mathrm{e}^-\to\gamma\gamma$ process and a
phenomenological study at flavour factories has been presented in Ref.~\cite{Balossini:2008xr}.

As an example, in \Tref{table:gg} the impact of different classes of
radiative corrections for typical event selection criteria at flavour
factories is shown, in analogy with \Tref{table:sigmas} for Bhabha scattering. In
the $\mathrm{e}^+\mathrm{e}^-\to\gamma\gamma$ case, the impact of NLO radiative
correction is in the
$5$--$10\%$ range, while h.o. corrections change the cross-sections by
$0.1$--$0.5\%$: both effects must be accounted for to reach a theoretical accuracy
at the $0.1\%$ level. To the best of our
knowledge, \babayagaNLO\ is the only MC event generator implementing a matching of NLO and
h.o. corrections in a PS approach for the $\gamma\gamma$ final state
that is to reach such a theoretical accuracy.

\begin{table}
\caption{Photon pair production cross-sections (in nanobarns) to different 
accuracy levels (from Ref.~\cite{Balossini:2008xr}; see there for details).}
\label{table:gg}
\centering
\begin{tabular}{llll}
\hline \hline
$\sqrt{s}\ \left(\textrm{GeV}\right)$ & $1$ & $3$ & $10$ \\
\hline
$\sigma_0$ & $137.53$ & $15.281$ & $1.3753$ \\
$\sigma^{\alpha}_\mathrm{PS}$ & $128.55$ & $14.111$ & $1.2529$ \\
$\sigma_\mathrm{NLO}$ & $129.45$ & $14.211$ & $1.2620$ \\
$\sigma^{\infty}_\mathrm{PS}$ & $128.92$ & $14.169$ & $1.2597$ \\
$\sigma_\mathrm{matched}$ & $129.77$ & $14.263$ & $1.2685$ \\
\hline \hline
\end{tabular}
\end{table}

It is also worth mentioning that an
extension of \babayaga\ has been developed for dark photon searches at
low energy via the radiative return method~\cite{Barze:2010pf} and has been
used extensively  by the KLOE-2 collaboration to set limits on the
dark photon couplings and mass~\cite{Anastasi:2015qla,Babusci:2014sta}.

Furthermore, it is planned to exploit the same theoretical framework
described here for the development of a high-precision event generator for the QED
process $\mu \mathrm{e}\to\mu \mathrm{e}$, which is needed for the measurement of the hadronic
contribution to the running of $\alpha_\mathrm{QED}$ in the space-like region
and, in turn, for an independent determination of the leading-order hadronic corrections to the muon
anomalous magnetic moment. The proposed experiment ($\mu$-on-e) and the underlying
ideas are discussed in Refs.~\cite{Abbiendi:2016xup,Calame:2015fva}.

\subsection{Exploratory results at the FCC-ee}
\label{accuracyfcc}
\noindent
In this section, we present some exploratory and preliminary results
obtained with \babayagaNLO\ at the FCC-ee, considering
different scenarios for its centre-of-mass energy, namely the Z peak ($\sqrt{s} = 91\gev$),
WW threshold ($\sqrt{s} = 160\gev$), above ZH threshold ($\sqrt{s}= 240\gev$)
and $\mathrm{t}\bar{\mathrm{t}}$ threshold ($\sqrt{s} = 350\gev$). The aim is to show
some phenomenological results in view of a luminosity monitoring with
Bhabha scattering at small angles or with the $\mathrm{e}^+\mathrm{e}^-\to\gamma\gamma$
process at large angles. At this stage, we refrain from firmly assessing the achievable
theoretical precision in the calculation of the
cross-section. From past experience, it
is beyond dispute that going below the 0.1--0.05\% accuracy will require the
inclusion of the complete set of NNLO QED corrections, together with the
resummation of higher orders.

\Tref{table:fccbhabha} reports the Bhabha cross-section (in nanobarns)  for different energies, calculated by requiring that the final $\mathrm{e}^-$ state
lies between $3^\circ$ and $6^\circ$, the final $\mathrm{e}^+$ state  lies between
$174^\circ$ and $176^\circ$, and the product of their energies
$E^+E^-$ is larger than $0.2\times s$. The first row is the QED
tree-level cross-section (improved with VP effects), the
second row shows the impact of the inclusion of the Z exchange
tree-level diagrams, the third row shows the impact of QED NLO radiative
corrections and the
fourth row represents the higher-order QED corrections, as implemented in
\babayagaNLO. It can be seen that Z contributions are at  the  ${\sim}0.01\%$
level, that NLO QED radiative
corrections change the cross-section by 15--20\%, and
that the size of higher orders is at the 0.5--1\% level. The bottom two rows of the table show the effect of VP (which first enters at NLO
in the Bhabha process) and
the uncertainty induced in the integrated cross-section by the error on the hadronic
contribution to VP.\footnote{We used the routine (2012 version) developed by Professor Fred
  Jegerlehner to evaluate the hadronic contribution to VP and its
  errors~\cite{Eidelman:1995ny,Jegerlehner:2008rs}.}
We stress that the error in the hadronic part of VP induces, by itself, an
uncertainty at the  ${\sim}10^{-4}$ level, which can become a limiting
factor in the knowledge of the theoretical cross-section, in view of a
very precise luminosity determination. Since the hadronic contribution
on VP relies, through dispersion relations and the optical theorem, on
the measurement of $\mathrm{e}^+\mathrm{e}^-\to\mathrm{ hadrons}$ cross-sections at low
energies, any future improvement in this respect is of critical importance.

\begin{table}
\caption{Small-angle Bhabha cross-section at FCC-ee for different centre-of-mass energies} 
\label{table:fccbhabha}
\centering
\begin{tabular}{lllll}
\hline \hline
$\sqrt{s}$ (GeV) & {$91$}     & {$160$}    & {$240$}      & {$350$}\\
\hline
$\sigma^\mathrm{VP}_0$ (nb)& $36.0030$  & $11.8062$  & $5.28998$  & $2.50709$ \\
+ Z              & \phantom{1}$+0.064\%$ & \phantom{1}$-0.062\%$ & \phantom{1}$-0.044\%$ & \phantom{1}$-0.030\%$\\
+ QED NLO          & $-17.41\%$ & $-18.73\%$ & $-19.57\%$ & $-20.35\%$ \\
+ QED h.o.         & \phantom{1}$+0.54\%$  & \phantom{1}$+0.66\%$  & \phantom{1}$+0.71\%$  & \phantom{1}$+0.77\%$\\
VP                        & \phantom{1}$+5.17\%$     & \phantom{1}$+6.27\%$     & \phantom{1}$+7.14\%$    & \phantom{1}$+7.99\%$ \\
$\pm\delta\Delta\alpha_\mathrm{h}(q^2)$ & \phantom{1}$\pm 0.021\%$ & \phantom{1}$\pm 0.027\%$ & \phantom{1}$\pm 0.030\%$ & \phantom{1}$\pm 0.032\%$ \\
\hline \hline
\end{tabular}
\end{table}

Alternatively, the luminosity could be monitored with the
$\mathrm{e}^+\mathrm{e}^-\to\gamma\gamma$ process, which is not affected, at least up to
NNLO, by VP hadronic errors (nor by Z `contamination') and is simulated in \babayagaNLO\   with
the same theoretical accuracy as the Bhabha scattering, as far as
photonic radiative
corrections are concerned \cite{Balossini:2008xr}.

In \Tref{table:fccgg}, the cross-section for the
$\mathrm{e}^+\mathrm{e}^-\to\gamma\gamma$ process at large angles is shown as a function of the collider
energy, where we consider events with at least two photons between
$10^\circ$ and $170^\circ$ and requiring the product of the energies of the
two most energetic ones (falling in the angular range), $E_1E_2$, to be
larger than $0.2\times s$. We notice that the cross-section is a
factor of ${\sim}500$ smaller than the small-angle Bhabha process considered
previously and that the impact of different classes of radiative corrections
is roughly a factor of two smaller than in the previous case. From
these preliminary considerations on the $\gamma\gamma$ final state, we
remark that it is a promising and complementary alternative to Bhabha
scattering as a luminosity monitor and is worthy of more detailed study \cite{lumigg1,lumigg2}.

\begin{table}
\caption{Large-angle $\mathrm{e}^+\mathrm{e}^-\to\gamma\gamma$ cross-section at FCC-ee for different centre-of-mass energies} 
\label{table:fccgg}
\centering
\begin{tabular}{lllll}
\hline \hline
$\sqrt{s}$ (GeV) & {$91$}     & {$160$}    & {$240$}      & {$350$}\\
\hline
$\sigma_0$ ({pb}) & $60.962$  & $19.785$    & $8.793$    & $4.135$ \\
+ QED NLO              & $-8.61\%$ & $-9.06\%$   & $-9.40\%$  & $-9.71\%$ \\
+ QED h.o.             & $-0.37\%$  & $-0.41\%$  & $-0.42\%$  & $-0.44\%$\\
\hline \hline
\end{tabular}
\end{table}

\subsection{Conclusions}
\label{concl}
\noindent
The theoretical framework of the \babayaga\ and \babayagaNLO\ event generators has been over\-viewed and
summarized. \babayagaNLO\ is a Monte Carlo event generator developed for high-precision simulation
of QED processes (Bhabha scattering, $\mathrm{e}^+\mathrm{e}^-\to\mu^+\mu^-$ and
$\mathrm{e}^+\mathrm{e}^-\to\gamma\gamma$) at flavour factories, mainly for luminometry
purposes, and  is based on an original algorithm to match exact NLO
with higher-order QED radiative corrections in a parton shower
approach. Within typical event selection criteria for luminometry at
flavour factories, the cross-sections can be calculated with an overall theoretical error at the level
of $0.1\%$ or better. The assessment and  scrutiny of the theoretical error are
based on consistent and detailed comparisons with independent
generators and existing exact NNLO calculations of the relevant
radiative corrections.

The generator can also be used  to study QED processes at the FCC-ee, and
we showed some exploratory results for small-angle Bhabha scattering
and large-angle $\mathrm{e}^+\mathrm{e}^-\to\gamma\gamma$ process. We notice that the
error in the hadronic part of VP, as known in the current parametrizations,
could become a limiting factor of the precise knowledge of the Bhabha
cross-section with an accuracy at the $10^{-4}$ level. We thus advocate the
study of the large-angle $\mathrm{e}^+\mathrm{e}^-\to\gamma\gamma$ process as an alternative way to
monitor the FCC-ee collider luminosity.

An assessment of the theoretical accuracy of \babayaga/\babayagaNLO\ for FCC-ee
predictions at the sub-per-mille level would require  inclusion in the
code of pure weak and QED NNLO corrections.

\clearpage \pagestyle{empty} 
\cleardoublepage


\pagestyle{fancy}
\fancyhead[LO]{}
\fancyhead[RO]{}
\fancyhead[CO]{}
\fancyhead[CO]{\thechapter.\thesection \hspace{1mm} 
\bhlumi{}: the path to $0.01\%$ theoretical luminosity precision}
\fancyhead[LE]{}
\fancyhead[RE]{} 
\fancyhead[CE]{} 
\fancyhead[CE]{S. Jadach,  W. Placzek, M. Skrzypek, B.F.L. Ward, S.A. Yost}

\section
[\bhlumi{}: the path to $0.01\%$ theoretical luminosity precision
\\ {\it 
S. Jadach,  W. Placzek, M. Skrzypek, B.F.L. Ward, S.A. Yost}
]
{\bhlumi{}: the path to $0.01\%$ theoretical luminosity precision} 
\label{pathbhlumi}

\vspace{2mm}
\noindent
{\bf Authors: S.\ Jadach,  W.\ Placzek, M. Skrzypek,\ B.F.L. Ward, S.A. Yost}
\\
Corresponding author: Stanis\l aw~Jadach {[Stanislaw.Jadach@cern.ch]} 


\vspace*{.5cm}

\noindent The state of the art of  theoretical calculations 
for the low-angle Bhabha (LABH) process $\mathrm{e}^-\mathrm{e}^+ \to \mathrm{e}^-\mathrm{e}^+$
is critically examined and it is indicated how to attain
a level comparable to the FCC-ee experimental precision
of $0.01\%$, see Ref. \cite{Gomez-Ceballos:2013zzn}.
The analysis starts from the context of the LEP experiments, 
through their current updates, up to prospects 
of their improvements for the sake of the FCC-ee.
Possible upgrades of the Monte Carlo event generator \bhlumi\ 
and other similar MC programs for
the LABH process are also discussed.

In the following, we are going to recall the main aspects of
the theoretical precision in the LEP luminosity measurement
and current important components of the corresponding error budget.
Next, we will present improvements
on some of these components that are already feasible.
Finally, we will discuss, in detail, prospects for
reaching the $0.01\%$ theoretical precision for the FCC-ee
luminometry and outline ways of upgrading the main Monte 
Carlo program for this purpose, \bhlumi, in this respect. 

Luminosity measurements of all four LEP collaborations at CERN, 
and also of the SLD at the SLAC,
relied on theoretical predictions for the low-angle
Bhabha process obtained using the \bhlumi\ Monte Carlo multiphoton event generator,
featuring a sophisticated QED matrix element with soft photon resummation.
Version 2.01 of this event generator was published in 1992 (see Ref.~\cite{Jadach:1991by})
and the upgraded version, 4.04, is published in Ref~\cite{Jadach:1996is}.

The theoretical uncertainty of the \bhlumi\ Bhabha prediction,
initially rated at 0.25\% \cite{Jadach:1991cg}, 
was re-evaluated in 1996 after extensive tests and debugging 
to be 0.16\%~\cite{Jadach:1995pd}.
From that time, the code of \bhlumi\ version 4.04
used by all LEP collaborations in their data analysis remains frozen.
All  following re-evaluation of its precision came from investigations
using external calculations outside the \bhlumi\ code.
For instance, the 0.11\% estimate of Ref.~\cite{Arbuzov:1996eq},
was based on better estimates of the QED corrections missing in \bhlumi\ 
and on improved knowledge of the vacuum polarization contribution.
The detailed composition of the final estimate of the theoretical uncertainty
$\delta \sigma / \sigma \simeq 0.061\%$
of the \bhlumi~4.04 prediction in 1999, based on published work, 
is shown in the second column of Table~\ref{tab:lep-update},
where we use $L_\mathrm{e}=\ln(|t|/m_\mathrm{e}^2)$,
following Ref.~\cite{Ward:1998ht}.
This value was used in the final LEP1 data analysis
in Ref.~\cite{ALEPH:2005ab}.
Conversely, at LEP2, the experimental error was substantially
greater than the QED uncertainty of the Bhabha process;
see Ref.~\cite{Ward:1998ht} for more details.

All four LEP collaborations  quoted  experimental luminosity errors  below 0.05\% 
for LEP1 data; this is less than the theoretical error.
The best experimental luminosity error, 0.034\%,
was quoted by the OPAL collaboration%
\footnote{The OPAL collaboration has found all their experimental
   distributions for low-angle Bhabha data to be in striking agreement 
   with the \bhlumi\ Monte Carlo simulation~\cite{Abbiendi:1999zx}.}
-- they also quoted a slightly smaller theory error, 0.054\%, 
using the improved light fermion pair calculations of
Refs.~\cite{Montagna:1998vb,Montagna:1999eu};
see also the review article~\cite{CarloniCalame:2015zev}
and workshop presentations~\cite{jadach:2006fcal,CarloniPisa:2015}.

From the end of the LEP until the present time, there has been limited progress
on practical calculations for low-angle Bhabha
scattering at energies around and above the Z resonance.%
\footnote{This is in spite of considerable effort on the
 ${\cal O}(\alpha^2)$ so-called `fixed-order' 
 (without resummation) QED calculations for the Bhabha process; 
 see further for more discussion.}
A new Monte Carlo generator, \babayaga, based on the parton shower
algorithm, was developed%
\cite{CarloniCalame:2000pz,CarloniCalame:2001ny,Balossini:2006wc,CarloniCalame:2015zev}. 
It was intended mainly for low-energy electron--positron colliders 
with $\sqrt{s} \leqslant 10\UGeV$, claiming precision at $0.1\%$,
but was not validated for energies near the Z peak.

There was, however, a steady improvement in the precision 
of the vacuum polarization in the $t$-channel photon propagator;
see the recent review in the FCC-ee workshop~\cite{JegerlehnerCERN:2016}.
Using the uncertainty $\delta\Delta^{(5)}_\mathrm{had.}=0.63 \times 10^{-4}$
at $\sqrt{-t}=2\UGeV$ quoted in Ref.~\cite{Jegerlehner:2017zsb}, one obtains 
$\delta \sigma/\sigma =1.3 \times 10^{-4}$.
This is shown in the third column
in Table~\ref{tab:lep-update}, marked `Update 2018'.
The improvement of the light-pair corrections of
Refs.~\cite{Montagna:1998vb,Montagna:1999eu}
is also taken into account there.

\begin{table}
\caption{
Summary of  total (physical+technical) theoretical uncertainty for  typical
calorimetric LEP luminosity detector within a generic angular range
of $18$--$52$\,mrad.
Total error is summed in quadrature.
}
\label{tab:lep-update}
\centering
\begin{tabular}{l l l l}
\hline \hline
Type of correction~or~error
    &  1999
        & Update 2018
\\ \hline 
(a) Photonic ${\cal O}(L_\mathrm{e}\alpha^2 )$
    & 0.027\% ~\cite{Jadach:1999pf}
        & \phantom{(}0.027\%
\\ 
(b) Photonic ${\cal O}(L_\mathrm{e}^3\alpha^3)$
    & 0.015\%~\cite{Jadach:1996ir}
        & \phantom{(}0.015\%
\\ 
(c) Vacuum polarization
    &0.040\%~\cite{Burkhardt:1995tt,Eidelman:1995ny} 
        & \phantom{(}0.013\%~\cite{JegerlehnerCERN:2016}
\\ 
(d) Light pairs
    & 0.030\%~\cite{Jadach:1996ca}
        & \phantom{(}0.010\%~\cite{Montagna:1998vb,Montagna:1999eu}
\\ 
(e) Z and $s$-channel $\gamma$ exchange
    &0.015\%~\cite{Jadach:1995hv,Arbuzov:1996eq}
        & \phantom{(}0.015\%
\\ 
(f) Up--down interference
    &0.0014\%~\cite{Jadach:1990zf}
        & \phantom{(}0.0014\%
\\ 
(f) Technical precision& -- & (0.027)\%
\\   
Total
    & 0.061\%~\cite{Ward:1998ht}
        & \phantom{(}0.038\%
\\ \hline \hline  
\end{tabular}
\end{table}

It should be emphasized that the technical precision,
which is marked in parentheses as $0.027\%$ in \Tref{tab:lep-update}, 
is not included in the total sum
because, according to Ref.~\cite{Arbuzov:1996eq}, it is
included in the uncertainty of the photonic corrections.
Future reduction of the photonic correction error will require a
clear separation of the technical precision, which may turn out to be a dominant
error, from other uncertainties.

Let us now address the following important question: 
{\em what steps are needed on
the path to the ${\leq}0.01\%$ precision required
for the low-angle Bhabha (LABH) luminometry at the FCC-ee experiments?}
The last column in Table~\ref{tab:lep2fcc} summarizes 
this goal component-by-component
in the precision forecast for the FCC-ee luminometry.
All necessary improvements
in the next version or \bhlumi, which could bring us to 
the FCC-ee precision level, will  also be specified.

\begin{table}
\caption{
Anticipated total (physical + technical) theoretical uncertainty 
for a FCC-ee luminosity calorimetric detector with
an angular range of $64$--$86\,$mrad (narrow) near the Z peak.
Description  in brackets of photonic contributions is related to the 3rd column.
Total error is summed in quadrature.
}
\label{tab:lep2fcc}
\centering
\begin{tabular}{l l l l}
\hline \hline
Type of correction~or~error
        & Update 2018
                &  FCC-ee forecast
\\ \hline 
(a) Photonic   $[{\cal O}(L_e\alpha^2 )]$ ${\cal O}(L_e^2\alpha^3 )$
        & \phantom{(}0.027\%
                &  $ 0.1 \times 10^{-4} $
\\ 
(b) Photonic   $[{\cal O}(L_e^3\alpha^3)]$ ${\cal O}(L_e^4\alpha^4)$
        & \phantom{(}0.015\%
                & $ 0.6 \times 10^{-5} $
\\
(c) Vacuum polarization
        & \phantom{(}0.014\%~\cite{JegerlehnerCERN:2016}
                & $ 0.6 \times 10^{-4} $
\\
(d) Light pairs
        & \phantom{(}0.010\%~\cite{Montagna:1998vb,Montagna:1999eu}
                & $ 0.5 \times 10^{-4} $
\\
(e) Z and $s$-channel $\gamma$ exchange
        & \phantom{(}0.090\%~\cite{Jadach:1995hv}
                & $ 0.1 \times 10^{-4} $
\\ 
(f) Up--down interference
    &\phantom{(}0.009\%~\cite{Jadach:1990zf}
        & $ 0.1 \times 10^{-4} $
\\
(f) Technical precision & (0.027)\% 
                & $ 0.1 \times 10^{-4} $
\\  
Total
        & \phantom{(}0.097\%
                & $ 1.0 \times 10^{-4} $
\\ \hline \hline 
\end{tabular}
\end{table}

In the following discussion, we must keep in mind certain basic
features of the QED perturbative calculations for FCC-ee and LABH detectors.
First of all, the largest photonic QED effects due to multiple real
and virtual photon emission are strongly cut-off dependent.
Monte Carlo implementation of QED perturbative results is mandatory,
because the event acceptance of the LABH luminometer is quite complicated,
and cannot be dealt with analytically.
The LABH detector at the FCC-ee will be similar to that of the LEP, 
with calorimetric detection of electrons and photons (not distinguishing them)
within the angular range $(\theta_{\min},\theta_{\max})$ on opposite
sides of the collision point~\cite{lumigg2}.
The detection rings are divided into small cells; the angular range
on both sides is slightly different, in order to minimize QED effects.
The angular range at the FCC-ee is planned 
to be $64$--$86\,$mrads (narrow)~\cite{lumigg2},
while at the LEP it was typically $28$--$50\,$mrads
(narrow range, ALEPH/OPAL silicon detector);
see Fig.~2 in Ref.~\cite{Jadach:1995pd}
(also Fig.~16 in Ref.~\cite{Jadach:1996gu}) 
for an idealized detection algorithm of the generic LEP silicon detector.
The average $t$-channel transfer near the Z resonance will be 
$|\bar{t}|^{1/2} = \langle |t| \rangle^{1/2} \simeq 3.25$\,GeV at the FCC-ee,
instead of 1.75\,GeV at the LEP.%
\footnote{At $350\UGeV$, the FCC-ee luminometer will have $|\bar{t}| = 12.5\UGeV$.}
The important scale factor controlling photonic QED effects,
$\gamma=\frac{\alpha}{\pi}\ln\frac{|\bar{t}|}{m_\mathrm{e}^2}=0.042$
for the FCC-ee, is only slightly greater than $0.039$ for the LEP.
Conversely, the factor $x=|t|/s$ suppressing
$s$-channel contributions will be $1.27\times 10^{-3}$,
significantly larger than $0.37 \times 10^{-3}$ for the LEP.

The {\em photonic higher-order and subleading corrections} 
components in Tables~\ref{tab:lep-update} and \ref{tab:lep2fcc}
are large, but this is mainly due to collinear and soft singularities.
Fortunately, they are known in QED at any perturbative order,
hence can be resummed to infinite order.
Why is the cross-section of the  LABH luminometer highly sensitive
to the emission of real soft and collinear photons?
This is because the emission of even very soft or collinear photons 
in the initial state (ISR)
may pull final electrons outside the acceptance angular range,
while final-state photons can easily change the shape of the final-state `calorimetric cluster' in the detector.
This is why resummation of the multiphoton effects 
is of paramount importance, and must be implemented in an exclusive way, using the method of
exclusive exponentiation (EEX), as in \bhlumi~\cite{Jadach:1996is},
or  the parton shower (PS) method, 
as in \babayaga~\cite{CarloniCalame:2000pz}.
It is well-known, see, for instance, Ref.~\cite{slac-talk},
that the so-called `fixed-order'
\order{\alpha^2} calculations without resummation%
\footnote{For instance, the calculations of 
  Refs.~\cite{Penin:2005eh,Czakon:2005gi}.}
are completely inadequate for the LABH luminometry.

Photonic uncertainty of \order{L_\mathrm{e} \alpha^2} 
in item (a) in Table~\ref{tab:lep-update},
neglected in the matrix element in \bhlumi\ version 4.04, 
scales as $L_\mathrm{e}=\ln(|t|/m_\mathrm{e}^2)$,
where $t$ is the relevant squared momentum transfer.
Assuming that the technical precision is dealt with separately
(see the later discussion),
this item will disappear from the error budget completely
once the EEX matrix element of \bhlumi\ is upgraded to include missing
\order{L_\mathrm{e}\alpha^2} contributions, which are already known and published.
In fact, these \order{L_\mathrm{e} \alpha^2} corrections 
consist of two real-photon contributions,
one-loop corrections to one-real emission, and two-loop corrections.

Efficient numerical and analytic methods of calculating 
the exact \order{\alpha^2} matrix element (spin amplitudes)
for two real photons, keeping fermion masses, have been known for decades; 
see Refs.~\cite{Kleiss:1985yh,Berends:1984qf}.
In Ref.~\cite{Jadach:1992tf}, exact two-photon amplitudes were compared
with the matrix element of \bhlumi.
The pioneering work on \order{L_\mathrm{e} \alpha^2, L_\mathrm{e}^0 \alpha^2} 
virtual corrections to one-photon distributions was reported in Ref.~\cite{Jadach:1995hy}.
These were calculated 
by neglecting interference terms between $\mathrm{e}^+$ and $\mathrm{e}^-$ lines, 
which, near the Z peak, are of the order of
$\big(\frac{\alpha}{\pi}\big)^2  \frac{|t|}{s} L_\mathrm{e} \sim 10^{-7}$
multiplied by  some logarithm of the cut-off.
Note that, from the $s$-channel analogue in Ref.  \cite{Jadach:2006fx},
we know that the pure \order{L_\mathrm{e} \alpha^2} 
correction of this class (neglecting the \order{L_\mathrm{e}^0 \alpha^2} term)
is amazingly compact --
it consists of merely a three-line formula at the amplitude level.%
\footnote{Let us add for completeness that this correction
was also calculated numerically~\cite{Actis:2009uq}.}
Finally, in Ref.\cite{Ward:1998ht}, the
two-loop \order{L_\mathrm{e} \alpha^2} $t$-channel photon form factor relevant
for the LABH process (keeping in mind $|t|/s$ suppression) was continued
analytically from the known $s$-channel result of Ref.~\cite{Berends:1987ab},
thus completing the entire photonic \order{L_\mathrm{e} \alpha^2} matrix element.
This is known but not yet included in the MC \bhlumi\ version 4.04.
Once the  well-known photonic \order{L_\mathrm{e} \alpha^2} 
part is added in the future upgrade of the EEX matrix element in \bhlumi, 
the corresponding item will disappear from the list of  projected FCC-ee luminometry uncertainties,
as in \Tref{tab:lep2fcc}.

In view of this discussion, it is clear that the major effort
of calculating the complete \order{\alpha^2} QED correction to
low and wide-angle Bhabha processes in 
Refs.~\cite{Czakon:2005gi, Penin:2005eh, Bern:2000ie,Bonciani:2005im},
see also Refs. \cite{Actis:2007gi,Actis:2008br,Bonciani:2007eh,Kuhn:2008zs},
is of rather limited practical importance for the LABH luminometry at the
FCC-ee.
It is more relevant for the wide-angle Bhabha processes, 
provided they are included in the MC with soft photon resummation.
However, this is rather problematic, because, in all these works,
soft real-photon contributions are added to loop corrections
 \`a la Bloch--Nordsieck, instead of subtracting the well-known
virtual form factor from virtual loop results already
at the amplitude level, before squaring them.
All these works essentially add previously unknown 
\order{L_\mathrm{e}^0 \alpha^2} corrections,
which are of the order of ${\sim}10^{-5}$.

Another important photonic correction, listed
as item (b) in Table~\ref{tab:lep-update}
as an uncertainty of \bhlumi, is the
\order{\alpha^3 L_\mathrm{e}^3} correction (third-order LO).
It is already known from Ref.~\cite{Jadach:1996bx,Jadach:1996ir} 
and is currently omitted in the EEX matrix element of version 4.04 of \bhlumi,
although it is already included in the LUMLOG subgenerator of \bhlumi\
based on collinear kinematics.
It would be relatively simple to add
this class of corrections to the main EEX matrix element of \bhlumi,
such that it would disappear from the uncertainty list.
Once this is done, the uncertainty due to 
\order{\alpha^4 L_\mathrm{e}^4} and \order{\alpha^3 L_\mathrm{e}^2}
should be estimated and included in the list of photonic
uncertainties of the updated \bhlumi.
Using scaling rules of thumb indicated in the previous discussion,
one may estimate an error due to missing \order{\alpha^4 L_\mathrm{e}^4}
as $0.015\% \times \gamma= 0.6\times 10^{-5}$ near the Z peak
and that of the missing \order{\alpha^3 L_\mathrm{e}^2} 
should  also be of a similar order, $\gamma^2 \alpha/\pi \simeq 10^{-5}$,
although such estimates are rather uncertain.

The so-called up--down interference between photon emission from $\mathrm{e}^+$ 
and $\mathrm{e}^-$ lines was  calculated in Ref.~\cite{Jadach:1990zf}
at \order{\alpha^1} to be roughly $\delta\sigma/\sigma \simeq 0.07\; |t|/s$.
At LEP1, this contribution was rated as negligible, see Table~\ref{tab:lep-update},
but at the FCC-ee luminometer it will be a factor of ten larger,
owing to the wider angular range of the FCC-ee detector, and must be included
in the matrix element of the upgraded \bhlumi.
Once this is done, its uncertainty will be negligible,
see Table~\ref{tab:lep2fcc}, 
where we used $ 2\gamma \times 0.07\; |t|/s $,
as a crude estimator.

\bhlumi\ multiphoton distributions obey a clear separation
into an exact Lorentz invariant phase space and a squared matrix element.
The matrix element is an independent part of the program and is currently
built according to exclusive exponentiation (EEX) based on the
Yennie--Frautshi--Suura~\cite{Yennie:1961ad} (YFS)
soft photon factorization and resummation
performed on the spin-summed squared amplitude.
It includes complete \order{\alpha^1}
and \order{L_\mathrm{e}^2 \alpha^2} corrections,
neglecting interference terms between electron and positron lines,
suppressed by a $|t|/s$ factor.
It has not been changed in the upgrades 
since version 2.01~\cite{Jadach:1991by}.
It would be desirable to introduce the results from Refs.~\cite{Jadach:1995hy,Jadach:1996bx,Jadach:1996ir,Ward:1998ht}
into the EEX matrix element, that is \order{\alpha^2 L_\mathrm{e}}
and \order{\alpha^3 L_\mathrm{e}^3}, again neglecting  some ${\sim}|t|/s$ terms.

Conversely, a preferable solution would be to 
keep the same underlying multiphoton phase space
MC generator of \bhlumi\ and exploit 
the results from Refs.~\cite{Jadach:1995hy,Jadach:1996bx,Jadach:1996ir,Ward:1998ht} 
to implement a more sophisticated matrix element of
the CEEX~\cite{Jadach:2000ir} type,
where CEEX stands for coherent exclusive exponentiation.
In the CEEX resummation methodology, soft photon factors
are factorized at the amplitude level and  matching
with fixed-order results is also done
at the amplitude level (before squaring and spin summing).
The big advantage of CEEX over EEX is that the separation of the infrared (IR)
parts and matching with the fixed-order result are much simpler and 
more transparent -- all IR cancellations for
complicated interferences are managed automatically and numerically.
The inclusion of the $s$-channel Z and photon exchange and $t$-channel Z
exchange including \order{\alpha} corrections, soft photon
interference between electron and positron lines, 
and so on, would be much easier to take into account 
for CEEX than in the case of EEX.
However, the inclusion of \order{\alpha^3 L_\mathrm{e}^3} in CEEX will have to
be worked out and implemented.

The uncertainty of the low-angle Bhabha cross-section
due to {\em imprecise knowledge of the QED running coupling constant}
of the $t$-channel photon exchange is simply
${\delta_\mathrm{VP}\sigma} /{\sigma}= 2 {\delta\alpha_\mathrm{eff}(\bar{t})}
/ {\alpha_\mathrm{eff}(\bar{t})}$,
where $\bar{t}$ is the average transfer of the $t$-channel photon.
For the FCC-ee luminometer, it will be $|\bar{t}|^{1/2} \simeq 3.5\UGeV$
near the Z peak and $|\bar{t}|^{1/2} \simeq  13\UGeV$ at $350\UGeV$.
The uncertainty of $\alpha_{\rm eff}(t)$ is mainly due to the use of
the experimental cross-section $\sigma_{\rm had}$
for $\mathrm{e}^-\mathrm{e}^+\to \mathrm{hadrons}$ below $10\UGeV$
as an input to the (subtracted) dispersion relations.
A comprehensive review of the corresponding methodology
and the latest update of the results can be found
in Refs.~\cite{Jegerlehner:2006ju,Jegerlehner:2017gek}; 
see also the FCC-ee workshop presentation~\cite{JegerlehnerCERN:2016}.
The hadronic contribution to $\alpha_{\rm eff}$
from the dispersion relation is encapsulated in 
$\Delta \alpha^{(5)}(-s_0)$, where $2\UGeV \leq s_0^{1/2} \leq 10\UGeV$
in order to minimize the dependence on $\sigma_\mathrm{had}(s)$,
such that the main contribution comes from $s^{1/2}\leq 2\UGeV$.
\footnote{The main contribution to the error in the measurement 
of the muon $g-2$ is from the same cross-section range~\cite{Jegerlehner:2006ju}.}

The parameter range $2\UGeV \leq s_0^{1/2} \leq 10\UGeV$,
which is incidentally of paramount interest for  FCC-ee luminometry,
is part of a wider strategy in
Refs.~\cite{Jegerlehner:2006ju,Jegerlehner:2017gek}
of obtaining $\alpha_{\rm eff}(M_\mathrm{Z}^2)$ in two steps.
First, $\Delta \alpha^{(5)}(-s_0)$ is obtained from dispersion relations
and next the difference $\Delta \alpha^{(5)}(M_\mathrm{Z}^2) - \Delta \alpha^{(5)}(-s_0)$
is calculated using the perturbative QCD technique of 
the Adler function~\cite{Eidelman:1998vc}.
Taking $s_0^{1/2} = 2.0\UGeV$ and the value 
$\Delta \alpha^{(5)} (-s_0) = (64.09 \pm 0.63) \times 10^{-4}\; $
of Ref.~\cite{Jegerlehner:2017zsb} as a benchmark,
in Table~\ref{tab:lep-update} we quote 
$(\delta_{\rm VP} \sigma)/\sigma = 1.3 \times 10^{-4}$.
With the anticipated improvements of data for 
$\sigma_{\rm had}(s)$, $s^{1/2}\leq 2.5\UGeV$,
one may expect an improvement of a factor of two by the time of the FCC-ee experiments,
resulting in $\delta_{\rm VP} \sigma/\sigma = 0.65 \times 10^{-4}$ 
near the Z peak, see Table~\ref{tab:lep2fcc}.
At the high-energy end of the FCC-ee, $350\UGeV$,
owing to the increase in the average transfer $|\bar{t}|=12.5\UGeV$,
one currently obtains, from the dispersion relation,
$\delta \alpha_{\rm eff}/\alpha_{\rm eff}=1.190\times 10^{-4}$
and $\delta_{\rm VP} \sigma/\sigma \simeq 2.4 \times 10^{-4}$,
and again with the possible improvement of a factor of two, 
so that the FCC-ee expectation%
\footnote{We thank F.\ Jegerlehner for elucidating
         private communications on these predictions.}
is $(\delta_\mathrm{VP} \sigma)/\sigma \simeq 1.2 \times 10^{-4}$.

There are also alternative proposals for 
the measurement of $\alpha_{\rm eff}(t)$,
not relying (or relying less) on dispersion relations --
in Ref.~\cite{Abbiendi:2016xup}, it was proposed
to measure  $\alpha_{\rm eff}(t)$ directly, for $t\sim -1\UGeV^2$,
from the elastic scattering of energetic muons on atomic
electrons; this sounds interesting, but requires more studies.

The {\em light fermion pair effect} is not currently included in \bhlumi;
hence, all of it is accounted for in the error budget of Table~\ref{tab:lep2fcc}
(current state of the art).

Let us start by summarizing  the LEP-era state of the art.
The largest correction to the light fermion pair
due to additional electron pair production
is given in Ref.~\cite{Montagna:1998vb},
where the process $\mathrm{e}^+\mathrm{e}^-\to \mathrm{e}^+\mathrm{e}^-\mathrm{e}^+\mathrm{e}^-$ 
is calculated using the numerical approach of Ref.~\cite{Caravaglios:1995cd},
combined with virtual or soft corrections of
Refs.~\cite{Barbieri:1972as,Barbieri:1972hn,Burgers:1985qg},
resulting in a theoretical error of 0.01\%.\footnote{The emission of a $\mu$-pair is also discussed in Ref.~\cite{Montagna:1998vb}.}
In Refs.~\cite{Arbuzov:1995qd, Arbuzov:1995cn},
a semi-analytical method of obtaining NLO accuracy is exploited, 
omitting non-logarithmic corrections
and taking virtual corrections from Refs. \cite{Barbieri:1972as,Barbieri:1972hn}. 
The third-order LO corrections due to simultaneous emission of
the additional $\mathrm{e}^+\mathrm{e}^-$ pair (non-singlet and singlet) and additional photons
are also evaluated there.
The overall precision of the results in
Refs.~\cite{Arbuzov:1995qd, Arbuzov:1995cn} is estimated to be $0.006\%$,
mainly owing to the omission of the heavier lepton pairs 
($\mu^+\mu^-$, $\tau^+\tau^-$) and quark pairs ($0.005\%$). 
In  Ref.~\cite{Jadach:1992nk}, the complete LO semi-analytical 
calculations of additional pair production based on the electron structure functions technique
are given up to the third-order for the non-singlet%
\footnote{This is contrary to the incorrect statement 
  in Ref.~\cite{Arbuzov:1995qd}: third-order non-singlet $\mathrm{e}^+\mathrm{e}^-\gamma$ corrections are realized in Ref.~\cite{Jadach:1992nk} 
  by second-order structure functions with  running coupling.}
and singlet structure functions. 
Contrary to Ref.~\cite{Arbuzov:1995qd}, 
results are also provided  for the asymmetric acceptances.
In the approach of Ref. \cite{Jadach:1996ca},
based on the extension of the YFS~\cite{Yennie:1961ad},
soft $\mathrm{e}^+\mathrm{e}^-$ pair emission are resummed similarly as soft photons 
(omitting up--down interference, multiperipheral graphs, \etc)
-- with relevant real and virtual soft ingredients calculated in Ref. \cite{Jadach:1993wk}.
Numerical implementation exists in the unpublished \bhlumi{} version 2.30 MC code. 
The accuracy of the results was estimated  to be 0.02\% 
for the asymmetric angular acceptance, 
\ie $3.3^\circ$--$6.3^\circ$ and $2.7^\circ$--$7.0^\circ$,  
with the energy cut $1-s'/s <z_\mathrm{cut} =0.5$.
Reference~\cite{Montagna:1998vb} claims that
this precision is even better, $6\times 10^{-5}$ for $z_\mathrm{cut} \leq 0.5$, while
for hard emission, $z_{\rm cut}> 0.5$,
owing to significant multiperipheral components,
it deteriorates to 0.01\%.

{\em What should be done  to consolidate or extend
these, mostly LEP-era, calculations of the fermion pair contribution
and to reach an even better precision level needed for the FCC-ee?}

As in Ref.~\cite{Montagna:1998vb}, for the additional real 
$\mathrm{e}^+\mathrm{e}^-$ pair radiation, the complete matrix element should be used,
because non-photonic graphs can 
contribute as much as 0.01\% for the cut-off $z_\mathrm{cut}\sim 0.7$. 
A number of MC generators for the $\mathrm{e}^+\mathrm{e}^-\to 4\mathrm{f}$ process, 
developed for the LEP2 physics, are available to be exploited for that purpose.
To improve on the 0.005\% uncertainty of Ref.~\cite{Arbuzov:1995qd},
owing to the emission of the $\mu^+\mu^-$, $\tau^+\tau^-$, and quark pairs,
one may use LO  calculation of Ref.~\cite{Jadach:1992nk},
incorporating lepton pair contributions by means of
the modification of the running coupling.
A naive rescaling of the electron logarithm
(owing to the mass of the muon)
$\ln^2\frac{|t|}{m_\mu^2}/\ln^2\frac{|t|}{m_\mathrm{e}^2}=0.16$,
applied to the  $\mathrm{e}^+\mathrm{e}^-$ pair contribution of 0.05\% provides
an estimate of the muon pair contribution of 0.008\%.\footnote{This is less optimistic than the estimate in Ref.~\cite{Arbuzov:1995qd}.
}
A similar estimate for tau lepton pairs shows that
this contribution can be neglected.
Note that adding $\mu^+\mu^-$ pairs to the \bhlumi{} version\ 2.30 code
of Ref.~\cite{Jadach:1993wk} would be straightforward. 
Moreover, in the  approach of Refs.~\cite{Montagna:1998vb,Montagna:1999eu} 
this should be possible.%
\footnote{The other option is to use the previously described 
  general-purpose LEP2 4f codes, including also
  the earlier-discussed corresponding virtual corrections.}
The contribution of light quark pairs ($\pi$ pairs, etc.) 
can be roughly estimated using the quantity 
R$_\mathrm{had} = \sigma_\mathrm{had}/\sigma_{\mu}\simeq 3$
for the effective hadronic production threshold of the order of $1\UGeV$. 
One obtains 
R$_\mathrm{had}\ln^2\frac{|t|}{0.5^2\UGeV^2}/\ln^2\frac{|t|}{m_\mu^2}= 0.9$, 
\ie this contribution is of the size of the muon pair contribution,
that is,  of the order of 0.008\%. (This is less optimistic than the estimate in Ref.~\cite{Arbuzov:1995qd}.
Adding  quadrature errors due to muon and light quark pairs, one obtains 
0.011\%, rather than the 0.006\% of Ref.~\cite{Arbuzov:1995qd}. 
However, it is
consistent with the estimate of Ref.~\cite{Montagna:1998vb}.)

Another important group of corrections to light fermion pair emission
are the higher-order terms. 
The emission of two (or more) electron pairs is suppressed by another 
factor $(\frac{\alpha} {\pi} \ln\frac{|t|}{m_\mathrm{e}^2})^2 \sim 10^{-3}$ 
and is negligible. 
The additional $\mathrm{e}^+\mathrm{e}^- +n\gamma$ correction is non-negligible. 
Its evaluation was based 
either on LO structure functions (Table 1 in Ref. \cite{Arbuzov:1995qd}, Fig. 8 in Ref. 
\cite{Montagna:1998vb}, \cite{Jadach:1992nk}) 
or on the YFS~\cite{Yennie:1961ad}
soft approximation (Fig. 4 in Ref. \cite{Jadach:1996ca}),
resulting in quite different results,
and their comparison is rather inconclusive.
They are, at most, of the order of 0.5--0.75 of the additional $\mathrm{e}^+\mathrm{e}^-$ correction
(without $\gamma$). 
The remaining non-leading, non-soft additional $\mathrm{e}^+\mathrm{e}^-+n\gamma$ corrections 
are suppressed by another $1/ \ln\frac{|t|}{m_\mathrm{e}^2} \sim 0.06$ 
and should be negligible (${\sim}0.003\%$). 
It would also be  possible 
to calculate the additional $\mathrm{e}^+\mathrm{e}^- +\gamma$ real emission in a way similar to 
the existing code for LEP2 physics \cite{Denner:2002cg}.

We summarize as follows on uncertainties due to light fermion pair emissions:
(1) the contribution of light quark pairs must be calculated with an accuracy 
of 25\%, \ie 0.0027\%; 
(2) the contribution of the muon pairs will be known to 10\%,
\ie to 0.0008\%; 
(3) the non-leading, non-soft additional $\mathrm{e}^+\mathrm{e}^-+n\gamma$ corrections
will be treated as an error of 0.003\%. 
Adding points (1)--(3) in quadrature, we obtain 0.004\%.
Applying a safety factor of 1.25, we end up with 0.005\% for the possible pair production 
uncertainty forecast for the FCC-ee, quoted in Table~\ref{tab:lep2fcc}.
These improvements can be implemented either 
directly in the upgraded \bhlumi\
or using a separate calculation, 
such as \bhlumi~version 2.30~\cite{Jadach:1996ca} code,
or external MC programs, like 
those of Refs.~\cite{Montagna:1998vb,Montagna:1999eu}.

In addition to $\gamma$ exchange in the
$t$-channel $\gamma_t$, there are also, in the Bhabha process,
contributions from $\gamma$ exchange in the $s$-channel $\gamma_s$ and Z exchange in both $t$- and $s$-channels, ${Z}_t$ and ${Z}_s$. 
Once they are added at the amplitude level (relevant Feynman diagrams)
and then squared to obtain the differential cross-section,
$|\gamma_t+ {Z}_t + \gamma_s+ {Z}_s |^2$
gives rise to six interference and three squared contributions.
Apart from the pure $t$-channel $\gamma$ exchange, $\gamma_t\otimes \gamma_t$,
numerically the most important
are interferences of other contributions with the $\gamma_t$ amplitude,
owing to the enhancement factor ${\sim}s/|t|$.
Near the Z peak, among them, the most sizeable 
is the interference $\gamma_t \otimes {Z}_s$.
This is discussed in  detail in
Ref.~\cite{Jadach:1995hv} for two types of detector: 
SICAL, with an angular coverage of ${\sim}1.5^{\circ}$--$3^{\circ}$, 
and LCAL, with an angular coverage of ${\sim}3^{\circ}$--$6^{\circ}$. 
The implementation of the $\gamma_t \otimes {Z}_s$ contribution 
in \bhlumi\ version 4.02 and the assessment of its theoretical precision for the LEP
luminosity measurement is based on this work.
The following estimated theoretical errors of all other contributions 
beyond the dominant $\gamma_t\otimes \gamma_t$
will be also based on the study of Ref.~\cite{Jadach:1995hv}.
Since the angular coverage of the planned FCC-ee 
luminometer~\cite{lumigg2} 
is close to the LCAL one,
we shall exploit the corresponding results of Ref.~\cite{Jadach:1995hv}.

The Born level $\gamma_t \otimes Z_s$ contribution is
up to ${\sim}1\%$ and changes from being positive below the Z peak to negative above it, reaching the maximal absolute value at about $\pm 1\UGeV$ from the peak. 
Corrections due to additional photon emission are sizeable, up to ${\sim}0.5\%$.
\bhlumi\ includes the QED corrections and running-coupling effects 
for this contribution within the ${\cal O}(\alpha)$ YFS exclusive exponentiation.
The theoretical uncertainty for $\gamma_t \otimes Z_s$ 
was estimated at $0.090\%$ for the LCAL detector
and is used as an initial estimate of the theoretical error 
in Table~\ref{tab:lep2fcc}.

The other contributions will be estimated by means
of relating them to the $\gamma_t \otimes Z_s$ or $\gamma_t \otimes \gamma_t$,
using rescaling factors,
$|t|/s \approx 1.3\times 10^{-3}$ and
$\tilde{\gamma}_\mathrm{Z} = \Gamma_\mathrm{Z}/M_\mathrm{Z} \approx 2.7\times 10^{-2}$.
The interference $\gamma_t\otimes\gamma_s$ (included in \bhlumi),
the next-most-sizeable contribution near the Z peak,
is smaller at the Born level than the 
$\gamma_t \otimes Z_s$ contribution by the factor%
\footnote{The factor of four comes from the ratio of 
   the corresponding coupling constants.}
${\sim}4\,\tilde{\gamma}_\mathrm{Z} \approx 0.1$.
Taking ${\sim}1\%$ for the Born level $\gamma_t \otimes Z_s$,
we get ${\sim}0.1\%$ for $\gamma_t\otimes\gamma_s$.
For not-too-tight cuts on radiative photons,
the missing photonic QED corrections 
are smooth near the Z peak and should stay within $10\%$,
leading to an estimate of
the theoretical precision of the $\gamma_t\otimes\gamma_s$ contribution
in \bhlumi\ for the FCC-ee luminometry of ${\sim}0.01\%$.
At the Born level, the resonant $Z_s \otimes Z_s$,
is smaller than the $\gamma_t \otimes Z_s$ term by the factor 
${\sim}|t|/s\times 1/(4\,\tilde{\gamma}_\mathrm{Z}) \approx 1.3\times 10^{-2}$;
thus, its size is ${\sim}0.01\%$.
This enters into the theoretical error as a whole,
as it is omitted in the current version of \bhlumi.
However, it can be included rather easily,
such that, finally, only missing radiative corrections will matter.
They can reach ${\sim}50\%$ of the Born level contribution, owing to the
presence of the Z resonance;
hence, the corresponding theoretical error estimate is ${\sim}0.005\%$.
The less important $t$-channel interference $\gamma_t \otimes Z_t$ 
is estimated by multiplying the $\gamma_t \otimes Z_s$ contribution by
the ${\sim}|t|/s \times \tilde{\gamma}_\mathrm{Z} \approx 3.5\times 10^{-5}$ factor.
It can be easily implemented in \bhlumi,
by leaving out only photonic corrections below $10^{-5}$.
The non-resonant $s$-channel $\gamma_s\otimes\gamma_s$ contribution 
is suppressed by the factor
${\sim}(4\,\tilde{\gamma}_\mathrm{Z})^2 \approx 0.01$ with respect to $Z_s \otimes Z_s$
(which is worth ${\sim}0.01\%$),
so is rated at an order of $10^{-6}$.
Finally, the $Z_t \otimes Z_t$ contribution is smaller than the dominant 
$\gamma_t \otimes \gamma_t$ one by the factor ${\sim}(|t|/s/4)^2 < 10^{-6}$;
thus, it is completely negligible.

Combining all these theoretical errors in the quadrature,
the total uncertainty (contributions omitted in \bhlumi)
due to the Z exchanges and $\gamma_s$ exchange 
for the FCC-ee luminometer near the Z peak 
is estimated at the level of $0.090\%$ 
and quoted as the current state of the art
(for the FCC-ee luminometer) in Table~\ref{tab:lep2fcc}.

It is worth emphasizing that
this uncertainty is completely dominated by the uncertainty of the
$\gamma_t \otimes Z_s$ contribution, which comes from a rather conservative
estimate in Ref.~\cite{Jadach:1995hv} 
based on comparisons of \bhlumi\ with the MC generator 
{\tt BABAMC}\cite{Berends:1987jm} and the semi-analytical program
{\tt ALIBABA} \cite{Beenakker:1990mb,Beenakker:1990es},
the latter including higher-order leading-log QED effects.
Later on, the new MC event generator {\tt BHWIDE} \cite{Jadach:1995nk} 
was developed for  wide-angle Bhabha scattering,
including all Born level contributions for the Bhabha process and 
${\cal O}(\alpha)$ YFS exponentiated EW radiative corrections.
A comparison of \bhlumi\ with {\tt BHWIDE} for the FCC-ee luminometer
would help to reduce all these theoretical errors.
In principle, not only the Born level but also the ${\cal O}(\alpha)$ QED
matrix elements of {\tt BHWIDE}, could be implemented in
the EEX-style matrix elements of \bhlumi.
This would reduce the theoretical error 
for this group of contributions below $0.01\%$,
as indicated in Table~\ref{tab:lep2fcc}.
However, the most efficient and elegant
way to reduce the uncertainty of these contributions
practically to zero would be to include these Z exchanges 
and $s$-channel photon exchange
into the CEEX matrix element at \order{\alpha^1} in \bhlumi.
Concerning  EW corrections,
it would most probably be sufficient to add them
in the form of effective couplings in the Born amplitudes.
Conversely, such a CEEX matrix element 
with the Z exchanges
in \bhlumi\ would serve as a starting point for a better
wide-angle Bhabha MC generator, much as \bhlumi~version 4.04 served
as a starting point for {\tt BHWIDE} \cite{Jadach:1995nk}.

\label{sec:tech-prec}
The question of the {\em technical precision}
due to programming bugs, numerical instabilities, technical cut-off parameters, \etc
is quite non-trivial and most difficult.
The evaluation of the technical precision of \bhlumi\ version 4.04
with YFS soft photon resummation and complete \order{\alpha^1}
relies on two pillars: the comparison with semi-analytical
calculations given in Ref.~\cite{Jadach:1996bx} and comparisons
with two hybrid MC programs {\tt LUMLOG+OLDBIS} and {\tt SABSPV},
reported in Ref.~\cite{Jadach:1996gu}.
This precision was established to be 0.027\%
(together with missing photonic corrections).
This was not an ideal solution, because the
two hybrid MCs did not feature complete soft photon 
resummation and disagreed with \bhlumi\ by more than $0.17\%$ 
for sharp cut-offs on the total photon energy.

Conversely, after the LEP era, another MC program, \babayaga%
~\cite{CarloniCalame:2000pz,CarloniCalame:2001ny,Balossini:2006wc},
with soft photon resummation, has been developed using a parton shower (PS) technique, 
and could, in principle, also  be used for a better validation
of the technical precision of both \bhlumi\ and \babayaga.
In fact, such a comparison with {\tt BHWIDE} MC~\cite{Jadach:1995nk}
was made for $s^{1/2} \leq 10\UGeV$
and  $0.1\%$ agreement was found. It is quite likely
that such an agreement persists near $s^{1/2}\simeq M_\mathrm{Z}$.
Note that the complete \order{\alpha^1}
was included in \babayaga\  before the three technologies of matching
fixed-order NLO calculations with a parton shower (PS) algorithm were unambiguously established:
{\tt MC@NLO}~\cite{Frixione:2002ik},
{\tt POWHEG}~\cite{Nason:2004rx}, and {\tt KrkNLO}~\cite{Jadach:2015mza}.
The algorithm of NLO matching in \babayaga\ 
is quite similar to that of {\tt KrkNLO}. (Single MC weight  introduces NLO correction in both methods,
but in {\tt KrkNLO} it sums over real photons,
while in \babayaga\ it takes the product over them. 
However, it is the same when truncated to \order{\alpha^1}.
We are grateful to the authors of \babayaga\ for clarification on this point.)

Ideally, in the future validation of the upgraded \bhlumi,
 to establish its {\em technical precision}
at a  level of $10^{-5}$ for the total cross-section and of $10^{-4}$
for single differential distributions, one would need to compare
it with another MC program developed independently,
which properly implements the soft photon resummation,
LO corrections up to \order{\alpha^3 L_\mathrm{e}^3}, and
 second-order corrections with the complete $\order{\alpha^2 L_\mathrm{e}}$.
In principle, an extension of a program like \babayaga\
to the level of NNLO for the hard process,
keeping the correct soft photon resummation,
would be the best partner for the upgraded \bhlumi,\ to establish the
technical precision of both programs at the $10^{-5}$
precision level.%
\footnote{The upgrade of the \bhlumi\ distributions will be
  relatively straightforward because its multiphoton 
  phase space is exact~\cite{Jadach:1999vf} for any number of photons.}
In the meantime, a comparison between the upgraded
\bhlumi\ with EEX and CEEX matrix elements would also offer
a very good test of its technical precision,
since the basic multiphoton phase space integration module of \bhlumi\
was already well tested~\cite{Jadach:1996bx}
and such a test can be repeated at an even higher precision level.

Summarizing, we conclude that an upgraded new version of \bhlumi\ 
with the error budget of 0.01\% shown in Table~\ref{tab:lep2fcc}
is perfectly feasible.
With appropriate resources, such a version of \bhlumi\
with the \order{\alpha^2} CEEX matrix element and
a precision tag of $0.01\%$, necessary for  FCC-ee physics,
could be realized.
Keeping in mind that the best experimental  luminosity error achieved at
the LEP was 0.034\% \cite{Abbiendi:1999zx},
it would be interesting to study whether the systematic error
of the designed FCC-ee luminosity detector \cite{lumigg1}
can match this anticipated theory precision.

Let us also remark that the process $\mathrm{e}^+\mathrm{e}^-\to 2\gamma$
is also considered for FCC-ee luminometry, 
see Refs.~\cite{CarloniCalame:2015zev,Balossini:2008xr} 
for further discussion on the QED radiative corrections to this process.

\clearpage \pagestyle{empty} 
\cleardoublepage


\clearpage
\pagestyle{empty}

\section
[The SANC project
\\
{\it A. Arbuzov, S. Bondarenko, Y. Dydyshka, L. Kalinovskaya, R. Sadykov}]
{The SANC project} \label{contr:sanc}

\pagestyle{fancy}
\fancyhead[LO]{}
\fancyhead[CO]{\thechapter.\thesection \hspace{1mm} 
The SANC project}
\fancyhead[RO]{}
\fancyhead[LE]{}
\fancyhead[CE]{A. Arbuzov, S. Bondarenko, Y. Dydyshka, L. Kalinovskaya, R. Sadykov}
\fancyhead[RE]{} 

\noindent
{\bf Authors: A.~Arbuzov, S.~Bondarenko, Y.~Dydyshka, L.~Kalinovskaya, R.~Sadykov}
\\
Corresponding author: Lidia~Kalinovskaya {[lidia.kalinovskaya@cern.ch]}
\vspace*{.5cm}

\noindent In this section, we present a plan of studies that will be
performed in the framework of the new stage of the SANC project 
aimed at the preparation of the electron--positron collider  research 
programme on precision tests of the Standard Model.
The accuracy of the corresponding experimental studies improves continuously 
with increased luminosity, novel detector technologies, elaboration of new 
analysis techniques, etc. All this leads to new requirements on the accuracy 
of theoretical predictions challenged by the experimental data.

The particular goal of our project is the development of 
a Monte Carlo event generator for processes that will be studied 
at electron--positron colliders. QED, QCD, and electroweak (EW) effects 
will be treated in the complete one-loop accuracy level within the Standard Model.
Numerically relevant higher-order radiative corrections will be implemented in 
the generator on availability. 
In particular, we will include the known higher-order corrections 
through the $\Delta \rho$ parameter
(all calculations for the EW and QCD sectors that exist today).
Possible polarization of electron and positron beams will be allowed.
In the creation of the new Monte Carlo event generator, we will use the experience
of our group, accumulated in similar projects such as 
(Pol)HECTOR~\cite{Arbuzov:1995id},
ZFITTER~\cite{Bardin:1999yd,Arbuzov:2005ma}, 
and SANC~\cite{Andonov:2004hi,Arbuzov:2015yja}.

We will concentrate on the processes that are most relevant for the verification 
of the electroweak sector of the Standard Model, including W and Z gauge bosons, 
the Higgs boson, and the top quark. In particular, future high-energy $\mathrm{e}^+\mathrm{e}^-$ 
colliders will provide an experimental environment to study, with excellent 
precision, the following processes and ingredients of the Standard Model:
\begin{itemize}
\item the Bhabha scattering process $\mathrm{e}^+ \mathrm{e}^- \to \mathrm{e}^+ \mathrm{e}^- $,
since it will be used for luminosity measurements, detector calibration,
\etc;

\item the top quark mass, width, and coupling constants in
$\mathrm{e}^+ \mathrm{e}^- \to \mathrm{t}\bar {\mathrm{t}}$;

\item the sine of the effective weak mixing angle, $\sin\vartheta_\mathrm{W}^\mathrm{eff}$;

\item the Higgs boson properties via
the Higgsstrahlung process $\mathrm{e}^+ \mathrm{e}^- \to \mathrm{ZH}$;

\item the WW-fusion process $\mathrm{e}^+ \mathrm{e}^-\to \mathrm{H} \nu_\mathrm{e} \bar{\nu_{\mathrm{e}}}$.
\end{itemize}

Special attention will be devoted to the processes of electron--positron
annihilation into a fermion pair at the Z resonance centre-of-mass energy.   
We will describe these processes by means of a
Monte Carlo generator, which will include the complete
set of one-loop corrections supplemented with the most important
higher-order contributions. A new feature of our generator will
be inclusion of the effects due to beam polarization.

The general plan of the project is as follows.

\begin{itemize}

\item{Development of a Monte Carlo (MC) event generator
at the level of the complete one-loop and leading multiloop
radiative corrections, taking into account longitudinal or transverse beam polarization
for the processes $\mathrm{e}^+\mathrm{e}^- \rightarrow \mathrm{e}^+\mathrm{e}^-$ $(\mu^+\mu^-$, $\tau^+ \tau^-$, 
$\mathrm{t}\bar{{\mathrm{t}}}$, HZ, $\mathrm{H} \gamma$, $\mathrm{Z} \gamma$, ZZ, $\mathrm{H} \nu \bar\nu$, $\mathrm{H} \mu^+\mu^-$, 
$\mathrm{f}\bar{{\mathrm{f}}} \gamma$, $\gamma \gamma$).}

\item{Creation of an interface to supplement electroweak radiative
correction to PYTHIA.}
In the possible implementation of `the best we have',
the first stage (PYTHIA + PHOTOS) is to account for the lowest-order 
plus QCD or QED parton showers (LO+PS) in the final state and
the second stage {MCSANC} is to account for the NLO EW and higher-order EW and QCD 
corrections via the $\Delta\rho$ parameter.
The possibility of constructing such an interface has been 
demonstrated in Ref.~\cite{Richardson:2010gz}, where events generated by 
our MCSANC code were transferred to (PYTHIA + PHOTOS) 
or (HERWIG + PHOTOS) to simulate QCD and QED parton showers.

\item{Implementation of the single-resonance approach to describe 
complex cascade processes.}

\item{Elaboration of the standard SANC procedure for calculating
helicity amplitudes for processes $ 2 \rightarrow 3, 4$.}

\item{Creation of additional SANC building blocks to include the complete
  weak two-loop (EW) and three-loop (QCD) calculations, and the leading three-loop (EW) 
  and four-loop (QCD) ones.}

\end{itemize}

The helicity amplitude formalism will be applied. It allows one to obtain
a cross-section by summing up the squares of helicity amplitudes
(instead of squaring the sum of amplitudes in the conventional approach),
see, \eg\ Ref. \cite{MoortgatPick:2005cw}.
Therefore, we have a basis for implementation of both transverse and 
longitudinal beam polarizations.

The active environment of the existing SANC system allows one to obtain analytical 
results for scalar form factors, helicity amplitudes, 
and Bremsstrahlung contributions. To calculate helicity amplitudes, 
the Vega--Wudka~\cite{Vega:1995cc} and  Kleiss--Stirling~\cite{Kleiss:1985yh} 
methods are used. 
All calculations of helicity amplitudes are implemented using single-thread computing.
The advantage of the calculation of one-loop corrections using the helicity 
amplitude method is the possibility of taking 
polarization effects into account in a simple way.

The SANC system contains programs in the language of symbolic computation 
FORM~\cite{Ruijl:2017dtg} to calculate Standard Model processes of the following 
types: $4\mathrm{f} \to 0$, $4\mathrm{b} \to 0$ and $2\mathrm{f}2\mathrm{b} \to 0$. 
These programs compute one-loop ultraviolet-finite scalar form factors 
and amplitudes. 
To verify the correctness of the results, first,  we check 
the Ward identities and the independence of scalar form factors 
on the gauge parameters and, second, 
we perform extensive numerical comparisons of our results with independent 
calculations known in the literature: 
 ZFITTER~\cite{Bardin:1999yd,Arbuzov:2005ma}, 
FeynArts~\cite{Hahn:1999mt}, 
the GRACE system~\cite{Belanger:2003sd}, 
the topfit program~\cite{Fleischer:2002nn}, etc.

The first steps for the basis for the future MC generator are:
(a) the library for the parameter $\Delta \rho$ (in preparation) and
(b)  recent experience with the description of polarized Bhabha scattering 
at the one-loop level~\cite{Bardin:2017mdd}.

\subsection{Higher-order radiative corrections for massless 
four-fermion processes}
  
A large group of the dominant radiative corrections can be absorbed into 
the shift of the $\rho$ parameter from its lowest-order value $\rho_\mathrm{Born} = 1$.
These groups of radiative corrections are:
\begin{equation}
  \Delta\rho =
    \Delta\rho_{X_t}                          
  + \Delta\rho_{\alpha\alpha_s}     
  + \Delta\rho_{X_t\alpha_s^2} +
  + \Delta\rho_{(X_t^3 + X_t^2\alpha_s)}
  + \Delta\rho_{X_t\alpha_s^2}
  + \Delta\rho_{\alpha_t^2}
  + \Delta\rho_{X_t \alpha_s^3}
  + \Delta\rho_{X_t^2(\mathrm{bos})}
  +\Delta\rho_{X_t^3}.\nonumber \\
\end{equation}

\subsection{Bhabha scattering}

A theoretical description of Bhabha scattering with radiative corrections
taken into account is crucial for high-precision measurements of this process 
and thus for luminosity monitoring at future $\mathrm{e}^+\mathrm{e}^-$ colliders.
The Bhabha scattering cross-section with the one-loop QED contribution, 
including transverse and longitudinal polarizations of the incoming beams,
is presented in Refs.~\cite{Hollik:1981bu,Hollik:1982wr}.

Our first attempt to estimate the theoretical uncertainty
for this process at the one-loop level by MC is described in Ref.~\cite{Andonov:2004hi}.
We describe this process by means of a Monte Carlo generator that now includes 
 the complete set of one-loop corrections~\cite{Bardin:2017mdd}. 
The most relevant higher-order contributions will be supplemented. 
A new feature of our results is that they take  beam polarization
into account.

We have shown that the complete ${\mathcal{O}}(\alpha)$ electroweak 
radiative corrections provide a considerable impact on the differential 
cross-section and the left--right asymmetry. Moreover, the corrections 
themselves are rather sensitive to polarization degrees of the initial beams.

Next we present the analytical expressions for the helicity amplitudes of Bhabha
scattering. There are six non-zero helicity amplitudes, however, since for Bhabha scattering 
${\cal F}^{\mathrm{Z}}_{LQ}={\cal F}^{\mathrm{Z}}_{\sss QL}$, 
the number of independent helicity amplitudes is actually reduced to four.

We obtained the compact expression for the Born level
(${\cal F}_{{\sss QL},{\sss LL},{\sss QQ}} =1$) 
and the virtual (loop) contributions to the cross-section in the helicity amplitude approach:
\begin{align}
\label{VIRT_HA}
{\cal H}_{++++} &= {\cal H}_{----} =   
-2 e^2\,\frac{s}{t} \Bigl[ {\cal F}^{(\gamma,\mathrm{Z})}_{\sss QQ}(t,s,u) 
              - {\chi_\mathrm{Z}}(t)\vma{\mathrm{e}}{} {\cal F}^\mathrm{Z}_{\sss QL}(t,s,u)\Bigr],
\\
    {\cal H}_{+-+-} &= {\cal H}_{-+-+}=  
-2 e^2\,\frac{t}{s}
\Bigl[ {\cal F}^{(\gamma,\mathrm{Z})}_{\sss QQ}(s,t,u)-{\chi_\mathrm{Z}}(s)\vma{\mathrm{e}}{}{\cal F}^\mathrm{Z}_{LQ}(s,t,u)\Bigr],
\\
    {\cal H}_{+--+} &=
    \phantom{-}2 e^2\,\frac{u}{s}
\left( \left[ {\cal F}^{(\gamma,\mathrm{Z})}_{\sss QQ}(s,t,u)
+{\chi_{\mathrm{Z}}}(s)\left( {\cal F}^{\mathrm{Z}}_{\sss LL}(s,t,u) 
-2\vma{\mathrm{e}}{} {\cal F}^\mathrm{Z}_{ LQ}(s,t,u) \right) \right]+\frac{\ds s}{\ds t} \left[ s\leftrightarrow t  \right]
\right),
 \\
    {\cal H}_{-++-} &=
    \phantom{-}2 e^2\,\frac{u}{s}
\left (\left[{\cal F}^{(\gamma,\mathrm{Z})}_{\sss QQ}(s,t,u) \right]
 +\frac{\ds s}{\ds t} \left[  s\leftrightarrow t  \right] \right ),
\end{align}
where
\begin{equation} {\cal F}^{(\gamma,\mathrm{Z})}_{ QQ}(a,b,c)={\cal F}^{(\gamma)}_{QQ}(a,b,c)
+{\chi_{ Z}}(a)\vma{\mathrm{e}}{2}{\cal F}^{( \mathrm{Z})}_{ QQ}(a,b,c) \,
.
\end{equation}

To study the case of  longitudinal polarization, we produce the helicity amplitudes
and make a formal application of Eq.~(1.15) from Ref.~\cite{MoortgatPick:2005cw}:
\begin{align}
\frac{\mathrm{d}\sigma(P_{\mathrm{e}^-},P_{\mathrm{e}^+})}{\mathrm{d}\cos\vartheta}
&=\frac{1}{128\pi s}  \Bigl[
    (1+P_{\mathrm{e}^-})(1+P_{\mathrm{e}^+}) \sum_{ij}\mid{\cal H}_{++ij}\mid^2
  +(1+P_{\mathrm{e}^+})(1-P_{\mathrm{e}^-}) \sum_{ij}\mid{\cal H}_{+-ij}\mid^2
\nll
&\quad
   +(1-P_{\mathrm{e}^+}) (1+P_{\mathrm{e}^-})\sum_{ij}\mid{\cal H}_{-+ij}\mid^2
   +(1-P_{\mathrm{e}^+})(1-P_{\mathrm{e}^-}) \sum_{ij}\mid{\cal H}_{--ij}\mid^2\Bigr].
\label{PolXSec}
\end{align}
For the cross-check, we obtain, analytically, zero for the difference 
between the square of the covariant amplitude (we introduced  
the spin density matrix in our procedures) and Eq.~(\ref{PolXSec}).

The left--right asymmetry $A_\mathrm{LR}$ and the relative correction $\delta$ 
are defined as
\begin{align}
A_\mathrm{LR} &=
      \frac{\mathrm{d}\sigma(-1,1)-\mathrm{d}\sigma(1,-1)}{\mathrm{d}\sigma(-1,1)+\mathrm{d}\sigma(1,-1)},\qquad
 \nonumber \\
\delta & =
       \frac{\mathrm{d}\sigma^{\text{1-loop}}(P_{\mathrm{e}^-},P_{\mathrm{e}^+})}{\mathrm{d}\sigma^{\text{Born}}(P_{\mathrm{e}^-},P_{\mathrm{e}^+})}-1,
\end{align}
where we omitted $\mathrm{d}\cos\vartheta$ for brevity.

All numerical results are obtained for the set of energies 
$E_\mathrm{cm}=250$, $500$, and $1000\UGeV$
for the following magnitudes of the electron $(P_{\mathrm{e}^-})$ and positron $(P_{\mathrm{e}^+})$
beam polarizations: $(0,0)$, $(-0.8,0)$, $(-0.8,-0.6)$, and $(-0.8,0.6)$.

\begin{figure*}
\[
\begin{array}{ccc}
  \includegraphics[width=80mm]{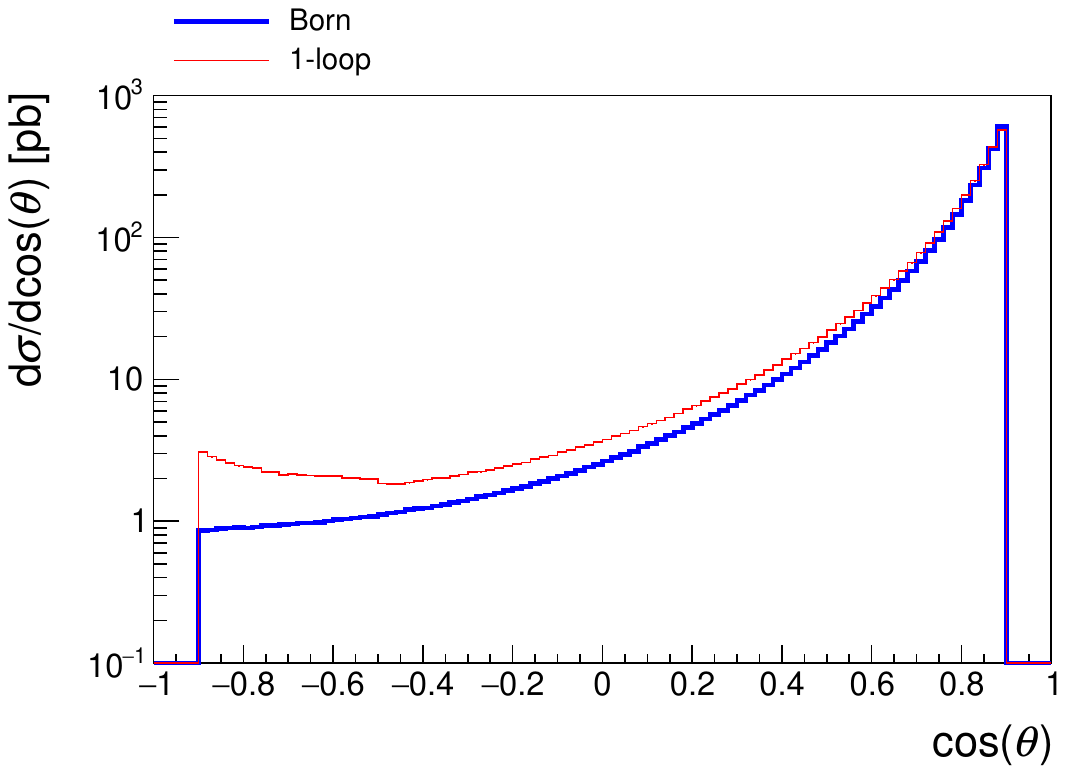}
  &&
  \includegraphics[width=80mm]{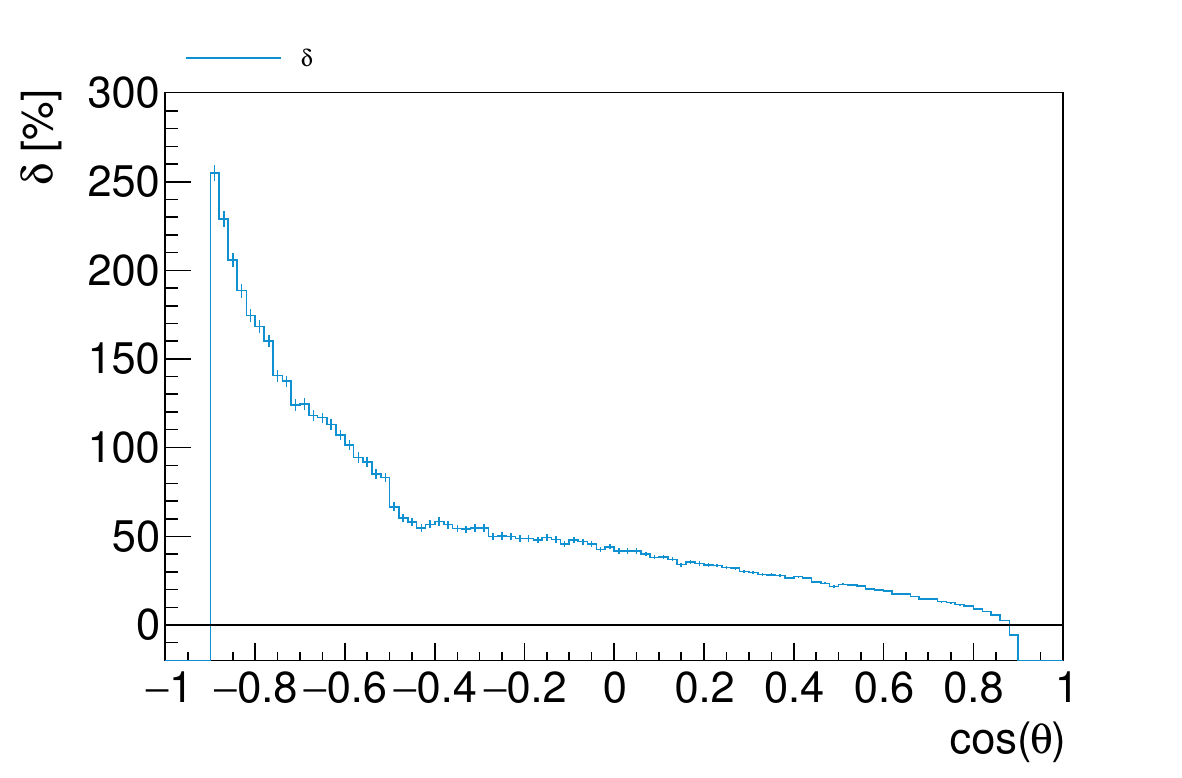}
  \end{array}
\]
\caption{Differential cross-section (left) and relative correction $\delta$ (right) as a function of the cosine 
of the electron scattering angle for $\sqrt{s}=250\UGeV$.}
\label{el_delta_costh_250}
\end{figure*}

\begin{figure*}
\[
\begin{array}{ccc}
  \includegraphics[width=80mm]{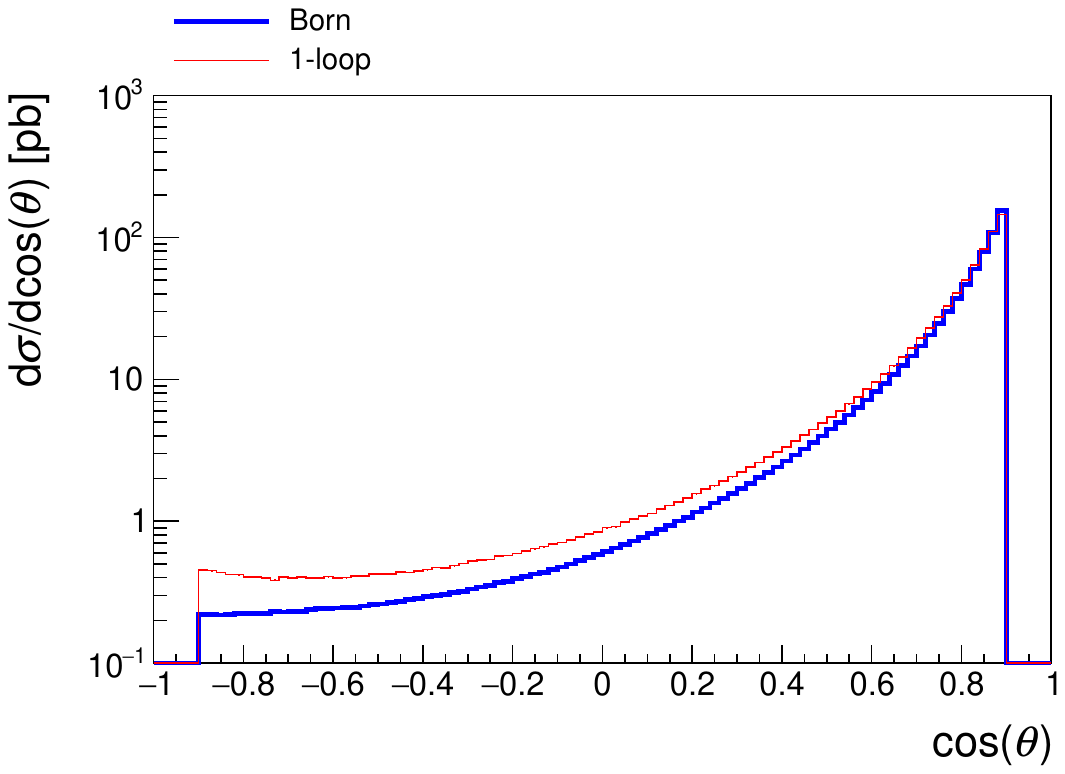}
  &&
  \includegraphics[width=80mm]{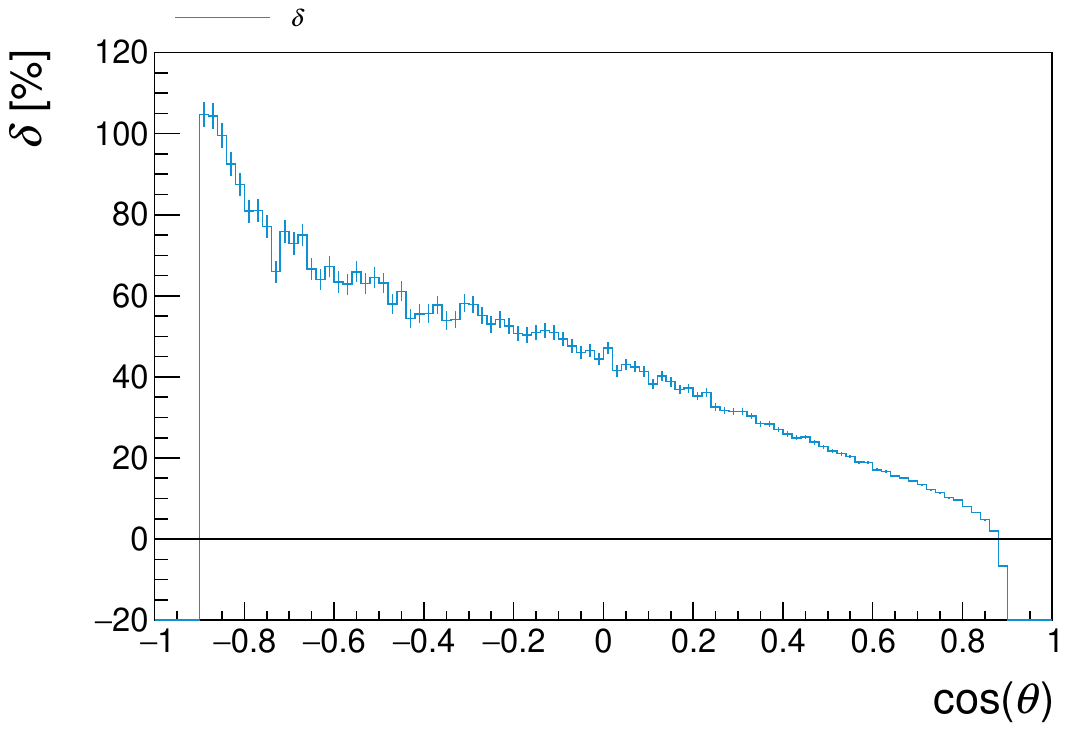}
  \end{array}
\]
\caption{Differential cross-section (left) and 
relative correction $\delta$ (right) as a function of the cosine 
of the electron scattering angle for $\sqrt{s}=500\UGeV$.}
\label{el_delta_costh_500}
\end{figure*}

\begin{figure*}[!h]
\[
\begin{array}{ccc}
  \includegraphics[width=80mm]{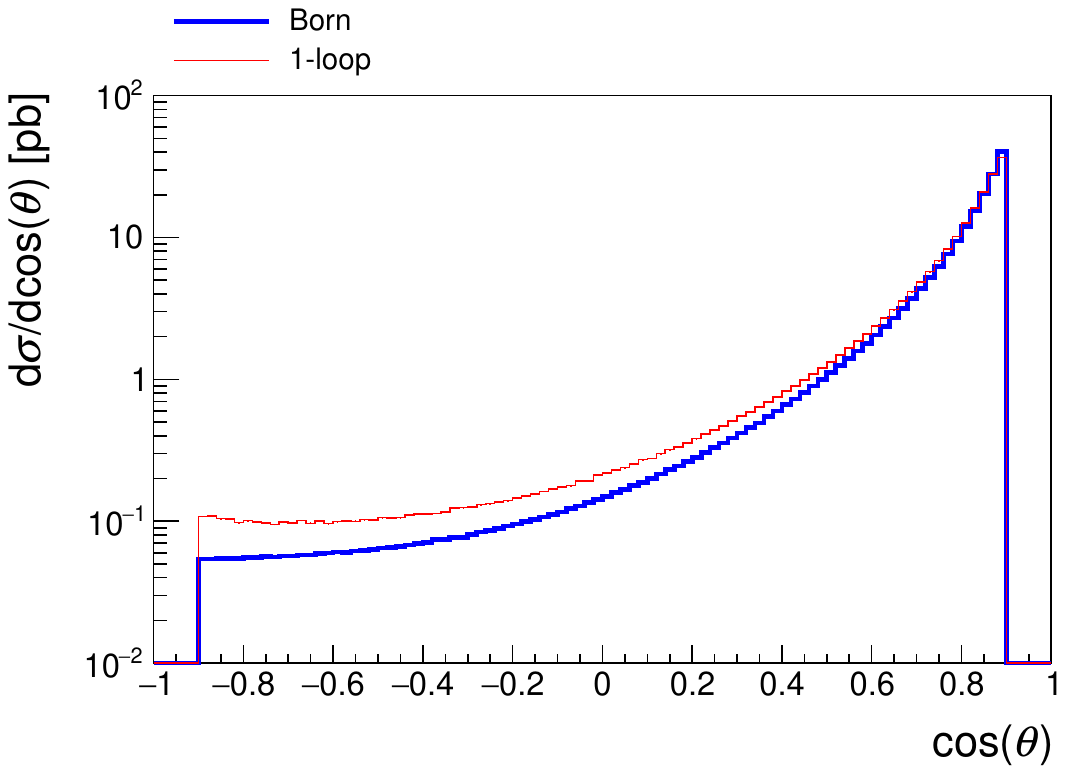}
  &&
  \includegraphics[width=80mm]{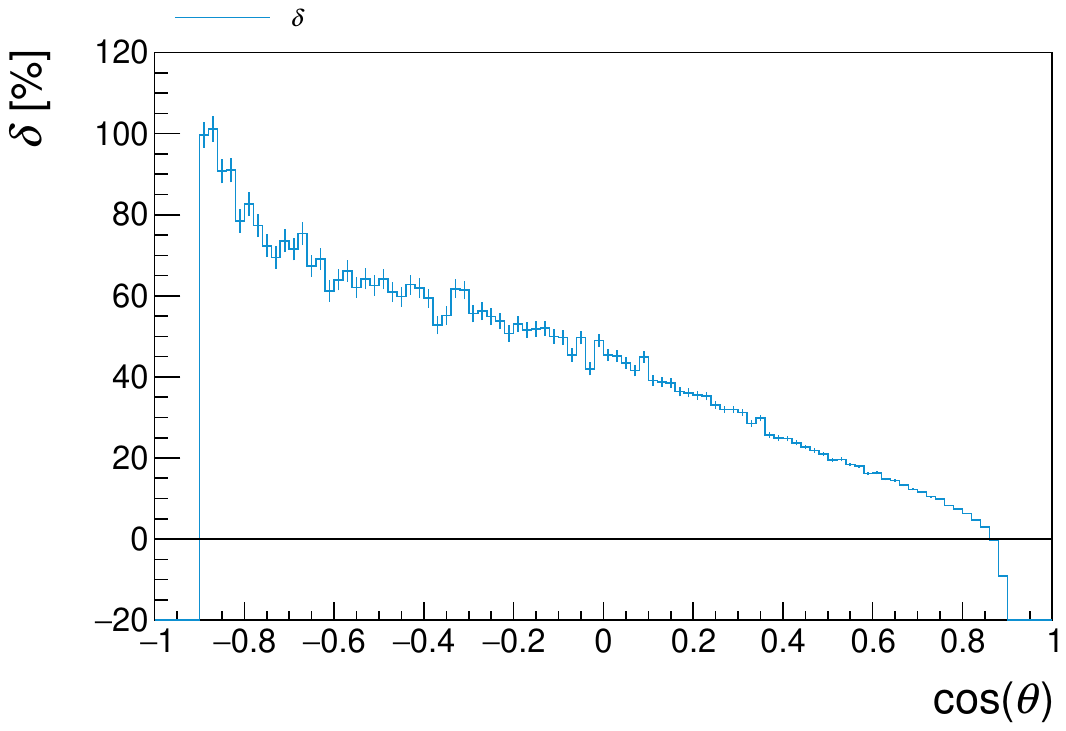}
  \end{array}
\]
\caption{Differential cross-section (left)  and 
 relative correction $\delta$ (right) as a function of the cosine 
of the electron scattering angle for $\sqrt{s}=1000\UGeV$.}
\label{el_delta_costh_1000}
\end{figure*}

The unpolarized differential cross-section of Bhabha scattering
and the corresponding relative ${\mathcal{O}}(\alpha)$ 
correction $\delta$ (as a percentage), as a function
of the electron scattering angle, are shown in Figs.~\ref{el_delta_costh_250}--\ref{el_delta_costh_1000} for $|\cos\theta|<0.9$
and different centre-of-mass energies.
The huge relative radiative corrections for the
backward scattering angles are due to the smallness of the Born cross-section 
in this domain; this does not mean any problem with the perturbation theory.

The integrated cross-section of the Bhabha scattering and the relative
correction $\delta$ are given in Table~\ref{Table:sanc_delta} for
various energies and beam polarization degrees.

\begin{table}
\caption{Born and one-loop cross-sections of Bhabha scattering and corresponding 
relative corrections $\delta$ for $\sqrt{s} = 250$, $500$, and $1000\UGeV$.}
\label{Table:sanc_delta}
 \centering
  \begin{tabular}{lllll}
  \hline \hline
$P_{\mathrm{e}^-}$, $P_{\mathrm{e}^+}$ & 0, 0 & $-$0.8, 0 & $-$0.8, $-$0.6 & $-$0.8, 0.6\\    
\hline
\multicolumn{5}{l}{$\sqrt{s} = 250\UGeV$}\\
$\sigma_{\mathrm{e}^+\mathrm{e}^-}^{\text{Born}}$ (pb) & 56.6763(1) & 57.7738(1) & 56.2725(4) & 59.2753(5)\\
$\sigma_{\mathrm{e}^+\mathrm{e}^-}^{\text{1-loop}}$ (pb) & 61.731(6) & 62.587(6) & 61.878(6) & 63.287(7)\\
$\delta$ (\%) & 8.92(1) & 8.33(1) & 9.96(1) & \phantom{$-$}6.77(1) \\
\multicolumn{5}{l}{$\sqrt{s} = 500\UGeV$}\\
$\sigma_{\mathrm{e}^+\mathrm{e}^-}^{\text{Born}}$ (pb) & 14.3789(1) & 15.0305(1) & 12.7061(1) & 17.3550(2)\\
$\sigma_{\mathrm{e}^+\mathrm{e}^-}^{\text{1-loop}}$, pb & 15.465(2) & 15.870(2) & 13.861(1) & 17.884(2)\\
$\delta$ (\%) & 7.56(1) & 5.59(1) & 9.09(1) & \phantom{$-$}3.05(1)\\
\multicolumn{5}{l}{$\sqrt{s} = 1000\UGeV$}\\
$\sigma_{\mathrm{e}^+\mathrm{e}^-}^{\text{Born}}$ (pb) & \phantom{1}3.67921(1) & \phantom{1}3.90568(1) & \phantom{1}3.03577(3) &\phantom{1}4.77562(5)\\
$\sigma_{\mathrm{e}^+\mathrm{e}^-}^{\text{1-loop}}$ (pb) & \phantom{1}3.8637(4) & \phantom{1}3.9445(4) & \phantom{1}3.2332(3) & \phantom{1}4.6542(7)\\
$\delta$ (\%) & 5.02(1) & 0.99(1) & 6.50(1) & $-$2.54(1)\\
\hline \hline
\end{tabular}
\end{table}

\begin{figure*}
\[
\begin{array}{ccc}
  \includegraphics[width=80mm]{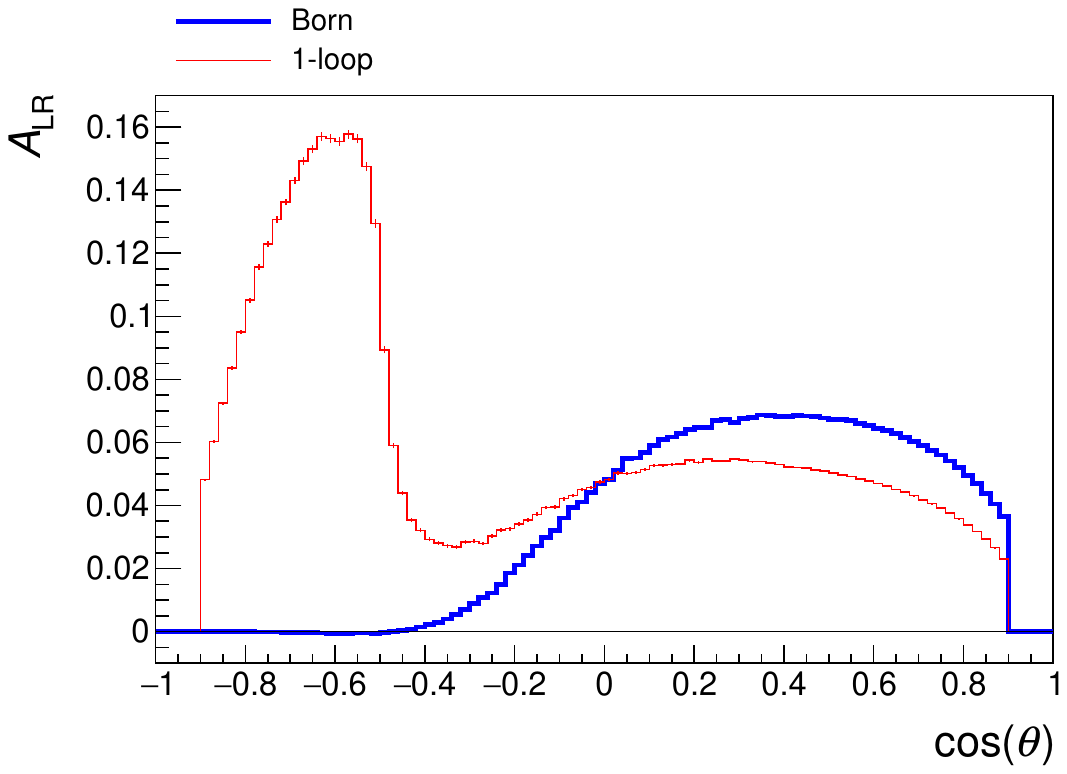}
  &&  
  \includegraphics[width=80mm]{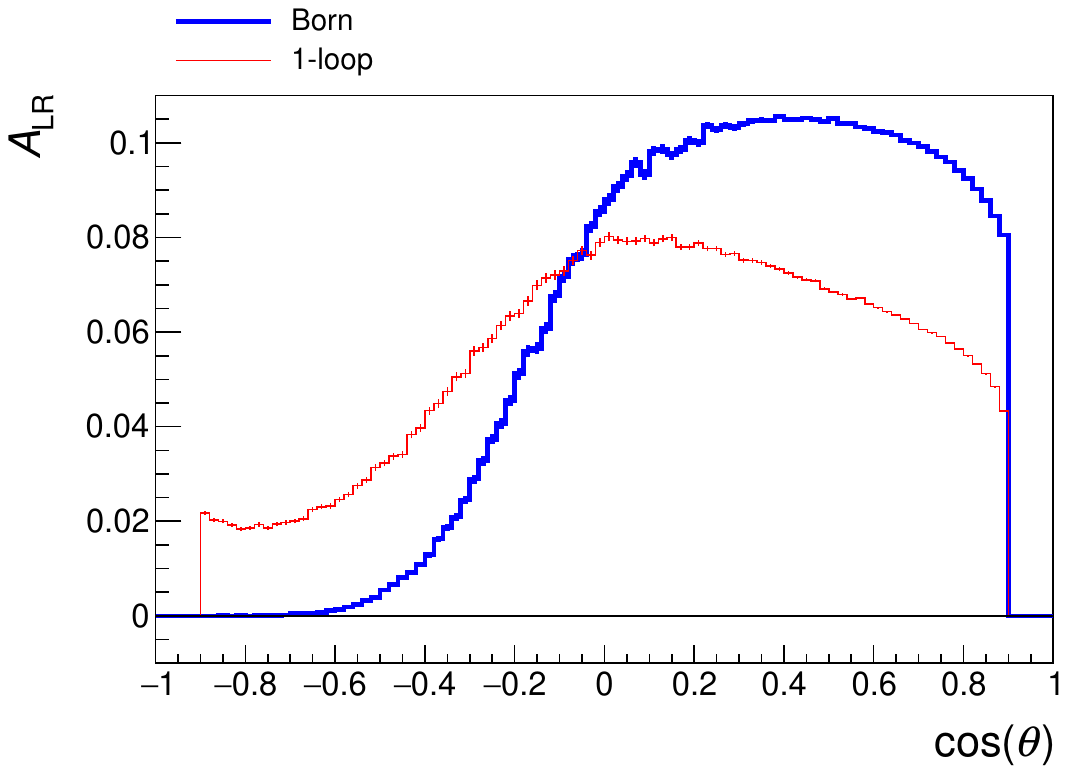}
  \end{array}
\]
  \caption{Left--right asymmetry $A_\mathrm{LR}$ as a function of the cosine of the electron 
  scattering angle at $\sqrt{s}=250\UGeV$ (left) and $\sqrt{s}=500\UGeV$ (right).}
\label{a_el_costh-250}
\end{figure*}

\begin{figure*}
\centering
  \includegraphics[width=80mm]{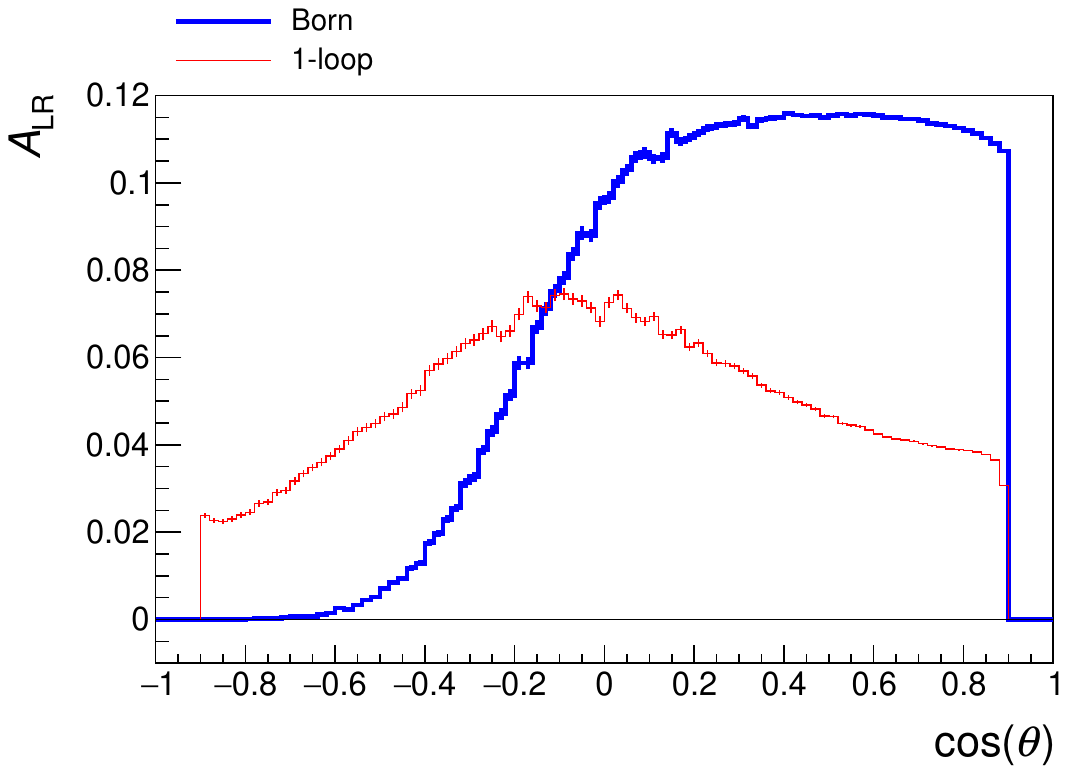}  
\caption{Left--right asymmetry $A_\mathrm{LR}$ as a function of the cosine of the electron scattering angle at $\sqrt{s}=1000\UGeV$.}
\label{a_el_costh-1000}
\end{figure*}

The $A_\mathrm{LR}$ asymmetry at $\sqrt{s}=250$, $500$, and $1000\UGeV$
is shown in Figs.~\ref{a_el_costh-250} and \ref{a_el_costh-1000}. 
One can see that the EW radiative corrections affect the asymmetry very strongly.

One can see that the one-loop EW radiative corrections are quite large and
should  certainly be taken into account. However, to satisfy the precision
requirements of  future experiments, it will be necessary to include higher-order effects. Only a part of the latter effects can be introduced as 
oblique or factorizable corrections, while many other corrections are process-
and kinematics-dependent. This is why the creation of a Monte Carlo generator
with explicit treatment of various contributions is of great importance.


\clearpage
\pagestyle{empty}
\chapter
{Towards three- and four-loop form factors} 
\label{ch-sff} 

\pagestyle{fancy}
\fancyhead[LO]{}
\fancyhead[RO]{}
\fancyhead[CO]{}
\fancyhead[CO]{\thechapter.\thesection
\hspace{1mm} 
Form factors and $\gamma_5$
}
\fancyhead[LE]{}
\fancyhead[CE]{}
\fancyhead[RE]{} 
\fancyhead[CE]{P. Marquard, D. St\"ockinger}

\section
[Form factors and $\gamma_5$ \\ {\it P. Marquard, D. St\"ockinger}]
{Form factors and $\gamma_5$}
\label{sec-ffms}

\noindent
{\bf Authors: Peter Marquard, Dominik St\"ockinger}
\\
Corresponding authors: Peter Marquard {[peter.marquard@desy.de]},
\\
\hspace*{3.65cm} 
Dominik St\"ockinger  {[Dominik.Stoeckinger@tu-dresden.de]}
\vspace*{.5cm}

\subsection{Introduction}

As discussed in Chapter \ref{sunfold}, the key objects in the deconvolution of observables in the FCC-ee-Z are $2 \to 2$ processes.   
The leading residuum terms are, besides self-energies, products of vertices.  
Thus, vertices are central objects that can be calculated systematically order by order in radiative corrections. 
They constitute,  in the  form of form factors, the hard, pure,
virtual part of the calculation and allow for the study of the
point-like nature of the involved particles. Their poles correspond to
universal anomalous dimensions, which govern the infrared behaviour
of QCD.
In this chapter, we explore vertices in detail by discussing the current status and perspective as, see Chapter \ref{sthstatus}, three- and four-loop vertex corrections to the Z boson decay will be needed. Some important technicalities,
such as the  $\gamma_5$ issue or frameworks beyond the Standard Model  are also discussed, see Section~\ref{ch-sff}.\ref{contr:rboels}.

In general, we must distinguish the two cases of heavy, massive, or
light, massless, quarks or gluons. We need to keep the mass of the quark
if the centre-of-mass energy is comparable to the mass of the quark in
question. If the centre-of-mass energy is well above the production
threshold, we can safely neglect the quark mass in higher-order
calculations. Thus, at a future electron--positron collider,  quarks
up to the bottom quark will, in general, be considered massless, while
the mass of the top quark is retained. 

We will summarize the status and prospects for the
two cases in the following sections.

\subsection{Massive form factor}
\label{sec:massive-form-factor}

In the case of the massive-quark form factor, we have to consider couplings
of a pair of heavy quarks to an external current through a scalar, pseudo-scalar, vector, or
axial-vector coupling. All of these coupling, except the pseudo-scalar
one,  are realized within the Standard Model through the couplings of
quarks to the gauge bosons and the Higgs boson. Thus, we will consider
the following general coupling structure of a pair of massive quarks
with momenta $q_1$ and $q_2$:
\begin{align}
 \Gamma^{\mu} &= \Gamma_\mathrm{V}^{\mu} + \Gamma_\mathrm{A}^{\mu}
 \nonumber\\
 &= -\mathrm{i} \left[ v_\mathrm{Q} \left( \gamma^{\mu}~F_{\mathrm{V},1} + \frac{\mathrm{i}}{2 m_\mathrm{Q}} \sigma^{\mu\nu} q_{\nu} ~F_{\mathrm{V},2} \right)
+ a_\mathrm{Q} \left( \gamma^{\mu} \gamma_5~F_{\mathrm{A},1}
         + \frac{1}{2 m_\mathrm{Q}} q^{\mu} \gamma_5 ~ F_{\mathrm{A},2}  \right) \right],
\end{align}
where $\sigma^{\mu\nu} = {\mathrm{i}}/{2} [\gamma^{\mu},\gamma^{\nu}]$,
$q=q_1+q_2$, and $v_\mathrm{Q}$ and $a_\mathrm{Q}$ are the Standard Model vector and axial-vector
coupling constants, as defined by
\begin{equation}
  v_\mathrm{Q} = \frac{e}{\sin \theta_\mathrm{w} \cos \theta_\mathrm{w}} \left ( \frac{T_3^Q}{2} - \sin^2 \theta_\mathrm{w} Q_\mathrm{Q} \right ) \,,
  \qquad
  a_\mathrm{Q} = - \frac{e}{\sin \theta_\mathrm{w} \cos \theta_\mathrm{w}} \frac{T_3^Q}{2} \,, 
\end{equation}
where $e$ denotes the elementary charge, $\theta_\mathrm{w}$ the weak mixing angle,
$T_3^Q$ the third component of the weak isospin, and $Q_\mathrm{Q}$ the charge
of the heavy quark. Here, $F_{\mathrm{V},1}$ and $F_{\mathrm{V},2}$ correspond to the
electric and magnetic form factor, respectively.

In the case of  (pseudo-)scalar coupling, we have
\begin{equation}
 \Gamma = \Gamma_\mathrm{S} +  \Gamma_\mathrm{P}
 = - \frac{m}{v}  \left[ s_\mathrm{Q}  F_\mathrm{S} + \mathrm{i} p_\mathrm{Q} \gamma_5  F_\mathrm{P} \right] \,,
\end{equation}
where $s_\mathrm{Q}$ and $p_\mathrm{Q}$ are the scalar and pseudo-scalar coupling,
respectively.

Higher-order QCD corrections to all form factors corresponding to the different currents have been
studied extensively in the past decades.  The results are, in general,
too lengthy to reproduce here; we refer to the literature for the
explicit expressions. The form factors have first been studied at
one-loop order  \cite{Arbuzov:1991pr, Djouadi:1994wt}. At two loops, they have, together
with the required master integrals, first been calculated
in Refs.~\cite{Bernreuther:2004th,Bernreuther:2005rw,Bernreuther:2005gw,Bernreuther:2004ih,Bernreuther:2005gq}.
These calculation have later been extended to include more orders in
$\epsilon$ for the vector current in Ref.~\cite{Gluza:2009yy} and for all other
currents in Ref.~\cite{Ablinger:2017hst}. At three-loop order,  at present, no complete calculation is available. Only partial results, the
colour-planar, \ie leading colour, contributions and the full
light-fermionic contributions
exist~\cite{Lee:2018nxa,Lee:2018rgs,Henn:2016tyf,Ablinger:2018yae}. The
corresponding master integrals have been calculated, as given in Ref. 
\cite{Henn:2016kjz} and, independently, in Ref.~\cite{Ablinger:2017hst}.  For a review of general strategies for
the calculation of master integrals see,
\eg\ Ref. \cite{Blumlein:2018cms}. At three loops, these calculations
successfully reproduce the universal cusp anomalous dimensions
given in Refs. \cite{Grozin:2015kna,Grozin:2014hna}. The universal infrared
structure, both in general, and, in particular, for the heavy quark form factors, has
been studied\cite{Mitov:2006xs,Becher:2009kw}.

The massive form factors depend on the ratio $q^2/m_\mathrm{Q}^2$; it has proved
favourable to express them through the variable $x$, defined via
$q^2 / m_\mathrm{Q}^2 = - (1-x)^2 / x$. Expressed through the variable $x$, the
available results for the massive form factors at three-loop order
only contain normal harmonic polylogarithms, with 
$1/x$, $1/(1-x)$, $1/(1+x)$, and cyclotomic harmonic polylogarithms, with
$1/(1-x+x^2)$, $x/(1-x+x^2)$, or their corresponding linearized
sixth-root-of-unity counterparts. Beyond the colour-planar
contributions, elliptic structures are expected to appear, a topic
that currently attracts a lot of attention.

The axial and pseudo-scalar form factors suffer from the general
problem of an ill-defined $\gamma_5$ in dimensional
regularization. While we can safely
use a naive anticommuting $\gamma_5$  for the non-singlet contributions, this is no longer possible for
the singlet contributions. We discuss general aspects of this issue
and possible solutions in Section D.\ref{sec:remarks-regul-gamm}. 

\subsection{Massless form factor}
\label{sec:massless-form-factor}

The form factors where all quarks are considered massless also play an
important role. In the context of massless form factors, the scalar
$\mathrm{h}\to gg$ and vector $\gamma \to q\bar q $ form factor, but also the
axial-vector form factor, contributing to $\mathrm{Z} \to q\bar q $, have
received the most attention.  The scalar form factor constitutes the
virtual correction for Higgs production through gluon fusion in the
Higgs effective theory. $\mathrm{e}^+ \mathrm{e}^- \to \mbox{2 jets}$ is governed purely
by the vector form factor, while for Z boson production and decay we need both
vector and axial-vector form factors. If only QCD corrections are
considered, the vector and axial-vector form factors agree. 

Considering only pure QCD corrections, a plethora of results for the
form factors is available.  They have been calculated at the two-loop
level 
\cite{Kramer:1986sg,Matsuura:1988sm,Matsuura:1987wt,Harlander:2000mg,Ravindran:2004mb,Gehrmann:2005pd}
and obtained at
three-loop order 
\cite{Baikov:2009bg,Gehrmann:2010ue}. These results have later been extended to higher
orders in the dimensional regulator 
\cite{Gehrmann:2010ue,Gehrmann:2010tu}.  The
relevant master integrals have been calculated
\cite{Heinrich:2009be,Gehrmann:2006wg,Heinrich:2007at}. The infrared structure of massless
form factors is again universal and has been studied
\cite{Catani:1998bh,Sterman:2002qn,Becher:2009cu,Gardi:2009qi,Ravindran:2004mb}.

At four-loop order, again only partial results exist. For the quark form
factor, the colour-planar, \ie leading colour, contributions have been
obtained first for  fermionic contributions  \cite{Henn:2016men}
and later for  purely bosonic ones  \cite{Lee:2016ixa}. The
complete result for contributions containing at least two fermion loops,
\ie $n_\mathrm{f}^2$, has been calculated  \cite{Lee:2017mip}. For the gluon
form factor, only the $n_\mathrm{f}^3$ contributions are available 
\cite{vonManteuffel:2016xki}. While, in the massive case, the main
difficulty lies in  calculation of the master integrals, in the
massless one, even the reduction to master integrals poses a major
challenge.

The
massless form factor has also been considered in $N=4$ super Yang--Mills (SYM) theory. For recent progress, see Section~\ref{ch-sff}.\ref{contr:rboels}.

\subsection{Remarks on regularization and $\gamma_5$}
\label{sec:remarks-regul-gamm}

One technical step in quantum field theoretical higher-order
calculations is the regularization of loop and phase space
integrals. Here, we point out that the regularization of multiloop
electroweak calculations is particularly non-trivial and progress
needs to be made for the desired computations.

Generally, the topic of regularization has already received increasing
attention in recent years. On the one hand, existing schemes, such as
dimensional regularization, dimensional reduction, or the
four-dimensional helicity scheme, have been investigated in more detail
and relations between them have been better understood. On the other
hand, novel reformulations of dimensional schemes, as well as purely
four-dimensional schemes, have been developed. These developments have
been summarized in a recent review \cite{Gnendiger:2017pys}. A major
driving force of those developments was the need for highly
non-trivial higher-order calculations for the LHC -- clearly
further progress will have to be made for higher-order calculations
for the physics at the FCC-ee.

An issue of particular importance for electroweak calculations is the
regularization of $\gamma_5$, which is notoriously problematic in
dimensional schemes. In four dimensions, three properties hold:
\begin{align}
  \{\gamma_5,\gamma^\mu\}&=0,\\
  \label{traceformula}
  \text{Tr}(\gamma_5\gamma^\mu\gamma^\nu\gamma^\rho\gamma^\sigma)&=4\mathrm{i}\epsilon^{\mu\nu\rho\sigma},\\
  \text{Tr}(\Gamma_1\Gamma_2)&=\text{Tr}(\Gamma_2\Gamma_1)\,.
\end{align}
The last equality means that traces are cyclic. In $D=4-2\epsilon$
dimensions, it is inconsistent to require these properties
simultaneously, and one has to give  one of them up. As a result, there
is a plethora of proposals for how to treat $\gamma_5$. The standard one,
which is known to be mathematically well-defined and consistent, is
the so-called HVBM scheme \cite{tHooft:1972tcz,Breitenlohner:1977hr}. This scheme gives up the
anticommutation property of $\gamma_5$; it is consistent in the sense
that it is compatible with unitarity and causality of quantum field
theory, but it does not manifestly lead to the correct
conservation or non-conservation properties of currents -- these must
be ensured by hand.

A variant of the HVBM scheme has been developed by Larin
\cite{Larin:1993tq}; Larin's scheme has been applied regularly in
multiloop QCD calculations involving the axial current. In this
context, the vector current is manifestly preserved. The
anomalous non-conservation of the axial current is restored by adding
a certain finite counterterm. The determination of this counterterm is
difficult. Reference\ \cite{Larin:1993tq} uses an anomalous Ward identity
to determine the two-loop counterterm via a three-loop
calculation. Published recently, Ref.\ \cite{Gnendiger:2017rfh} suggests a
simplification that could enable determination of the $n$-loop
counterterm via only an $n$-loop calculation.

It is important to realize that applying the HVBM scheme to
electroweak physics is significantly more complicated. In this case,
neither the vector- nor the axial-vector current is conserved, but
the left-handed SU(2) gauge current must be conserved  to
preserve gauge-invariance of the theory. This is not manifestly  the
case. Hence, the HVBM scheme breaks electroweak gauge-invariance,
\ie the regularized Green functions do not satisfy the
Ward or Slavnov--Taylor identities for the electroweak Standard Model and
gauge-invariance-restoring counterterms are required, which do not
arise from the usual renormalization transformation. The form of these
counterterms is unknown at the ${\ge}2$-loop level.

As an alternative to HVBM, recipes have been proposed that promise to
preserve electroweak gauge-invariance, either completely or to a larger
extent than HVBM. To exemplify the set of problems, we describe here
three existing electroweak multiloop calculations in which the
$\gamma_5$-problem was relevant.
\begin{itemize}
\item
  In the evaluation of the two-loop electroweak contributions to $g-2$
  of the muon \cite{Heinemeyer:2004yq}, a simplified treatment was
  possible where $\gamma_5$ was treated as anticommuting but the trace
  formula 
  (\Eref{traceformula}) 
  was used in its four-dimensional form by
  hand. It was shown that, because of the particular structure of
  diagrams, the inconsistency did not lead to incorrect results;
  rather, gauge-invariance was manifestly preserved, simplifying the
  computation.
\item
  In the evaluation of two-loop electroweak contributions to muon
  decay \cite{Freitas:2002ja}, the fermionic one-loop subdiagrams were
  first evaluated in the HVBM scheme and gauge-invariance-restoring
  counterterms were added. Then the result was inserted into the
  second loop, which could be evaluated in four dimensions.
\item
  In Refs.\ \cite{Bednyakov:2015ooa,Zoller:2015tha}, the four-loop
  $\beta$-function for $\alpha_s$ in the full Standard Model has been
  computed using various prescriptions involving anticommuting
  $\gamma_5$ and the reading-point prescription of
  Ref.\ \cite{Korner:1991sx} was used, with conflicting results.
\end{itemize}
These examples show that no established practical procedure exists
that could be applied to, \eg three-loop calculations in the electroweak
Standard Model, but progress can be made in several directions.

One avenue for improvement is to study further the recipes used in
Refs.\ \cite{Heinemeyer:2004yq,Bednyakov:2015ooa,Zoller:2015tha} and
to clarify when simplified prescriptions are
possible. References\ \cite{Jegerlehner:2000dz,Korner:1991sx} have
provided arguments that such simpler prescriptions, based, \eg on
reading points for $\gamma_5$ and traces, should exist.

Apart from dimensional regularization schemes, a variety of non-dimensional
regularization schemes, which remain in the physical four-dimensional
space, have been developed in recent years, see Ref. \cite{Gnendiger:2017pys}. 
These schemes might offer practical advantages with respect to the
treatment of $\gamma_5$. Reference\ \cite{Bruque:2018bmy}  considers a wide class of
four-dimensional regularization schemes, which have the important
property of momentum-routing invariance and do not break gauge-invariance as immediately as, \eg the Pauli--Villars scheme. This
reference shows clearly that all these schemes have very similar
problems for $\gamma_5$ as dimensional schemes. The reason is that in
those schemes the regularization is essentially achieved by replacement
rules, and those replacement rules do not necessarily commute with
applying, \eg cyclicity of traces. Nevertheless, these four-dimensional
schemes offer promising alternative avenues for future progress in
developing practical treatments of $\gamma_5$.

Within dimensional schemes, Ref.\ \cite{Gnendiger:2017rfh} has
recently shown that one particular recent reformulation, the so-called
four-dimensional formulation  approach \cite{Fazio:2014xea} offers a simpler way to treat
$\gamma_5$. So far, this approach exists only at the one-loop
level; a study of its extension to the
multiloop level is clearly motivated.

Finally, it remains, of course, a viable possibility to use the HVBM
scheme. In this case, it is important to determine, step by step, the required gauge-invariance-restoring  at the two- and three-loop levels. A
systematic procedure to achieve this has been put forward in
Ref.\ \cite{Martin:1999cc}, based on the quantum action principle, and
Ref.\ \cite{Grassi:2001zz}, based on the quantum action principle and
the background field method. These methods should be determined and
applied to the multiloop level.

\clearpage \pagestyle{empty} 
\cleardoublepage

\pagestyle{fancy}
\fancyhead[LO]{}
\fancyhead[RO]{}
\fancyhead[CO]{\thechapter.\thesection
\hspace{1mm} 
Four-loop form factor in $N = 4$ super Yang--Mills theory
}
\fancyhead[LE]{}
\fancyhead[CE]{R.H.  Boels, T. Huber, G. Yang}
\fancyhead[RE]{} 

\section
[Four-loop form factor in $N = 4$ super Yang--Mills theory \\ {\it R.H. Boels, T. Huber, G. Yang}]
{Four-loop form factor in $N = 4$ super Yang--Mills theory
\label{contr:rboels}}
\noindent
{\bf Authors: Rutger H.  Boels, Tobias Huber, Gang Yang} \\
Corresponding author: Tobias Huber {[huber@physik.uni-siegen.de]}
\vspace*{.5cm}

\subsection{General motivation: $N = 4$ super Yang--Mills theory as a toy model}
In preparing for an order of magnitude increase in computational capabilities, it is important to track progress at the cutting edge of what is currently possible. A common approach in physics is to first trial tools and techniques in toy models, which are still general enough to capture all essential details of more physical computations. The most common approach to compute quantities such as form factors and scattering amplitudes in high-energy physics beyond the trivial tree-level, for instance, involves a number of steps.
\begin{enumerate}
\item Generate an integrand for the quantity under study, for instance through Feynman graphs or by employing unitarity methods. This integrand still has to be integrated over unobserved loop momenta.
\item Simplify the integrand by solving integration-by-parts identities. The result is now given in terms of a basis of remaining so-called master integrals.
\item Compute the master integrals analytically or numerically, typically as an expansion in terms of the dimensional regularization parameter $\epsilon$.
\end{enumerate}
The interest in this report is, ultimately, collider physics, so the goal is to implement these steps in QCD or even the fully fledged Standard Model. It is known, however, that this is, algebraically, a complicated problem, even at step one, while the output generated is also complex. For any new tools and techniques, it is, therefore, a good idea to study a less physical but more symmetric theory, where the initial inputs are simpler and there is more control over the expected output. In the context of gauge theories, a prime candidate toy model is the $\mathcal{N}=4$ maximally supersymmetric gauge theory in dimensional regularization, \ie in $4-2\epsilon$ dimensions. 

The  $\mathcal{N}=4$ super Yang--Mills (SYM) theory plays a prime role in many branches of formal high-energy physics, for instance in the AdS/CFT correspondence \cite{Maldacena:1997re}. For the purposes of this report, it serves two purposes. First, it leads to integrands that are much simpler, technically simpler. Second, it is known through many examples that results in this theory typically obey a special number-theory property known as the maximal transcendentality principle. Very roughly speaking, the latter means that the hardest part of the integration of the master integrals appearing in physical theories such as QCD is the same as that in $\mathcal{N}=4$, making the latter an ideal toy model to explore concrete simplification and integration techniques. This subsection gives an application of this philosophy. In general, we expect that $\mathcal{N}=4$ super Yang--Mills theory can play a lighthouse role in the push for the precision required for the FCC-ee-Z experimental programme. For this to work, it is important that the focus remains squarely on tools and techniques that, at least in principle, apply directly to physical theories. This is the route followed here. The results listed next include work first reported in Refs. \cite{Boels:2017skl, Boels:2017ftb}. 

\subsection{Concrete motivation: the non-planar cusp anomalous dimension at four loops}
In this subsection, we study the Sudakov form factor of $\mathcal{N}=4$ SYM, which consists of one off-shell member of the stress-tensor multiplet and two on-shell supersymmetric massless gluon multiplets, in dimensional regularization in the four-dimensional helicity scheme in momentum space. Supersymmetry determines the form factor to be proportional to the tree form factor. Kinematically, this is a single-scale problem, which can be expressed at the $l$-loop order as  
\begin{equation}
{\cal F}^{(l)} = {\cal F}^{\textrm{tree}} g_{\textrm{ym}}^{2 {l}} (-q^2  )^{-l \epsilon } F^{(l)}(C_i) \,,
\end{equation}
where $q^2= (p_1 + p_2)^2$ with $p_1^2=p_2^2=0$ the on-shell gluon multiplet momenta, $C_i$ stands for all possible Casimir invariants of the underlying gauge group, and $g_{\textrm{ym}}$ is the Yang--Mills coupling constant. Up to three loops, only the quadratic Casimir invariant $C_A$ --~defined via $f^{acd} f^{bcd} = C_{A} \delta^{ab}$~-- appears, raised to the $l$th power. The four-loop order is the first place in which a new, quartic, Casimir invariant can appear, see \eg Ref. \cite{Boels:2012ew}, defined through $d_{44} = d_A^{abcd}d_A^{abcd}/N_A$ with
\begin{equation}
d_A^{abcd} = \frac{1}{ 6} \left [ f^{\alpha a}{}_{ \beta} f^{\beta b}{}_{ \gamma} f^{\gamma c}{}_{ \delta} f^{\delta d}{}_{ \alpha} + {\text{perms.}(b,c,d)} \right ] \, . 
\end{equation}
For gauge group $\mathrm{SU}(N_c)$, the values of the Casimir invariants are $N_A = N_c^2 -1$, $C_A = N_c$, and $d_{44} = N_c^2/24 \, (N_c^2+36)$. Hence, four loops is the first loop order for which the form factor acquires a colour subleading, non-planar, correction in 't Hooft's large $N_c$ limit \cite{tHooft:1973alw}. Since $\mathcal{N}=4$ SYM famously has no ultraviolet divergences, the form of the form factor is determined by powerful infrared exponentiation theorems
\cite{Bern:2005iz,Mueller:1979ih,Collins:1980ih,Sen:1981sd,Magnea:1990zb}, which relate it to two universal functions, the cusp ($\gamma_{\textrm{cusp}}^{(l)}$) and collinear (${\cal G}_{\textrm{coll}}^{({l})}$) anomalous dimension at $l$ loops, through 
\begin{equation}
(\log F)^{(l)} = -  \bigg[ \frac{\gamma_{\textrm{cusp}}^{({l})} }{(2 {l} \epsilon)^2} + \frac{{\cal G}_{\textrm{coll}}^{({l})} }{2 {l} \epsilon} + {\rm Fin}^{(l)} \bigg] + {\mathcal O}\left(\epsilon\right) \, .
\end{equation}
These two functions are common to many situations that involve soft or collinear divergences and are fixed by the theory under study. Their universal nature leads to a plethora of possible calculational approaches and potential applications, see, \eg\ Refs. \cite{Korchemsky:1988si, Kotikov:2000pm, Gubser:2002tv}. In maximal super Yang--Mills theory, the leading term in the large $N_c$ limit, referred to as the planar contribution, of the cusp anomalous dimension
is known to be captured by a universal equation at any value of the coupling~\cite{Beisert:2010jr}. Direct computations of the four-loop planar cusp anomalous dimension can be found in Refs. ~\cite{Bern:2006ew, Cachazo:2006az, Henn:2013wfa}.  The first numerical result for the first non-planar correction to the cusp anomalous dimension at four loops in any theory appeared in Ref. \cite{Boels:2017skl}, see also Refs.  \cite{Grozin:2017css, Moch:2017uml, Moch:2018wjh, Bruser:2018aud, Vogt:2018miu}. Previous work on the Sudakov form factor integrals includes calculations at two loops \cite{vanNeerven:1985ja}  and at three loops \cite{Baikov:2009bg,Lee:2010cga,Gehrmann:2010ue,Gehrmann:2010tu,vonManteuffel:2015gxa}, see also Refs.  ~\cite{Henn:2016men,Lee:2016ixa,Ahmed:2017gyt,vonManteuffel:2016xki, Davies:2016jie, Lee:2017mip, Dixon:2017nat, Grozin:2018vdn} for progress on four-loop integrals. 

Apart from the general high interest in high-loop computations, as evidenced in the many contributions to this document, a particular motivation to compute the non-planar correction to the cusp anomalous dimension was a conjecture based on naive extrapolation of results through three loops, given in Ref. \cite{Becher:2009qa}, that, quite generically, the cusp anomalous dimension in perturbation theory was proportional to an appropriate power of the quadratic Casimir invariant only, see also Ref.  \cite{Korchemsky:1988si}. This would require the non-planar correction to the cusp anomalous dimension at four loops to vanish, in any theory. By the results in Ref. \cite{Boels:2017skl}, this conjecture is now known to be false, see also Refs.  \cite{Anzai:2010td, Grozin:2017css, Moch:2017uml, Boels:2017ftb, Moch:2018wjh, Bruser:2018aud, Vogt:2018miu}. The activity this conjecture has sparked has, however, yielded a veritable treasure trove of insight into gauge theory and its IR divergences, for instance Refs. \cite{Gardi:2009qi, Dixon:2009gx, Becher:2009qa, Becher:2009kw, Dixon:2009ur, Ahrens:2012qz, Korchemsky:2017ttd}.

\subsection{Constructing the integrand, briefly}
In principle, the integrand for the problem at hand can be constructed using textbook Feynman graphs. This leads to two problems: one is the prohibitive intermediate expression swell that is to be expected in a Feynman graph computation of this size. The second problem is less immediate but more serious: there is no off-shell, Lorentz-covariant formulation of a theory with maximal supersymmetry. Hence, the resulting expression will have either apparent UV singularities or unphysical poles, such as those that arise in the light-cone gauge. Bad apparent UV behaviour translates directly into integrals that are much harder to reduce to an integral basis by integration-by-parts identities,
owing to the high numerator count.

To circumvent these problems, we have applied  \cite{Boels:2012ew} a conjectural idea for the form of the integrand: that of colour-kinematic duality  \cite{Bern:2008qj,Bern:2010ue}, see, \eg Ref.~\cite{Carrasco:2015iwa} for a pedagogical introduction. In essence, this conjecture of colour-kinematic duality poses that the integrand of quite general observables in gauge theories can be written in a form where not only do the colour factors obey a colour-Jacobi identity, but the kinematic, loop-dependent parts do so as well. Although there is no known way, in general, to generate such a representation in field theory, even the suspicion of this duality is sufficient as an ansatz generator. With the ansatz in hand, one fixes coefficients from unitarity cuts. In fact, up until the findings
given in Ref.~\cite{Boels:2012ew},  it was thought that this approach only applied to scattering amplitudes; we extended it to a variety of form factors ~\cite{Boels:2012ew}, while in subsequent work, it was extended even to the five-loop level~\cite{Yang:2016ear}. The appearing integrals can be written generically as
\begin{equation}\label{eq:deffeynmanint}
I = (-q^2)^{2+4\epsilon} e^{4\epsilon\gamma_E} \int {\mathrm{d}^D l_1 \over \mathrm{i}\pi^{D/2}} \cdots {\mathrm{d}^D l_4 \over \mathrm{i}\pi^{D/2}} \; \frac{{N}(l_i, p_j)}{ \prod_{k=1}^{12} D_k } \,,
\end{equation}
where $D_i$ are 12 propagators and ${N}(l_i, p_j)$ are dimension-four numerators in terms of inner products of the four independent loop and two independent external on-shell momenta. 

Having obtained the integrand for the quantity under study (it can be found in Ref.~\cite{Boels:2012ew}), the next step, as outlined previously, is integration-by-parts reduction. Integration-by-parts (IBP) identities \cite{Chetyrkin:1981qh, Tkachov:1981wb} follow from 
\begin{equation}\label{eq:IBPidsDeriv}
\int \mathrm{d}^D l_1 \ldots \mathrm{d}^Dl_L   \frac{\partial}{\partial l_i^{\mu}}  ({\rm integrand}) = 0 \,,
\end{equation}
which is the observation that the value of the integrals over loop momenta are invariant under linear transformations of the momenta. For the case at hand, a reduction of the integrals obtained in Ref. ~\cite{Boels:2012ew} was eventually obtained in Ref.~\cite{Boels:2015yna} using the {\tt Reduze} code~\cite{vonManteuffel:2012np}, whose output is  expressions in terms of a relatively small basis of master integrals.\footnote{There are various private and public implementations of IBP reduction, such as {\tt AIR} \cite{Anastasiou:2004vj}, {\tt FIRE} \cite{Smirnov:2008iw, Smirnov:2013dia, Smirnov:2014hma}, and {\tt Reduze} \cite{vonManteuffel:2012np,2010CoPhC.181.1293S}, which are all variants based on the Laporta algorithm \cite{Laporta:2001dd};  {\tt LiteRed} \cite{Lee:2012cn, Lee:2013mka} implements a somewhat different approach to IBP reduction.}   The problem with this is that one can {\emph{choose}} a set of basis integrals. Since the basis integrals obtained in Refs. ~\cite{Boels:2015yna} evaded direct integration methods (see next), a new criterion was needed to select a natural integral basis. For this, we turn to number theory. 

\subsection{Maximally transcendental integrals from a conjecture}
In general, the Feynman integrals that appear in high-energy physics, including those in  \Eref{eq:deffeynmanint}, have a special property when computed as an expansion in the dimensional regularization parameter: there is a notion of transcendental weight, which functions as the number-theory version of mass-dimension. Assigning weight $-1$ to $\epsilon$ at each order, only a sum of rational multiples of certain constants appears, where the constants all have positive, integer-valued weight. Generally, at fixed-order, all terms have transcendental weight less than  or equal to a maximum. These constants are mostly multiple zeta values, which also obey several algebraic relations. A basis for these constants for low weights is given by (see, \eg Ref. \cite{Blumlein:2009cf})
\begin{equation}\label{eq:firstMZVs}
\{1\}_0, \{\phantom{v}\}_1, \{\pi^2\}_2, \{\zeta_3\}_3,\{\pi^4\}_4, \{\pi^2 \zeta_3, \zeta_5\}_5, \ldots
\end{equation}
with increasing weight denoted by the subscripts. Although more general numbers, such as Euler sums (which also have well-defined transcendental weight), could appear, it is known through three loops that, for the Sudakov form factor, only multiple zeta values appear in the component integrals. 

The notion of transcendental weight appears in a natural mathematical question: in a given class of integrals, such as those in \Eref{eq:deffeynmanint} for a given integral topology, can one find special linear combinations such that the $\epsilon$ expansion is maximally transcendental? That is, the expansion at each order only contains maximal weight terms, not the subleading ones. These integrals will be called uniformly transcendental (UT). This notion is important, since it is known in examples that, for the maximally supersymmetric Yang--Mills theory, such observables as the planar cusp anomalous dimension are typically maximally transcendental. Even better,  a general conjecture known as the `maximal transcendentality principle' \cite{Kotikov:2002ab,Kotikov:2004er} states that the maximal transcendental terms appearing in QCD are directly related to the $\mathcal{N}=4$ SYM. On this basis alone, it is already interesting to explore the transcendentality properties of Feynman integrals, especially for observables in maximally supersymmetric Yang--Mills theory. Even better, the three-loop form factor in $\mathcal{N}=4$ SYM is written in terms of UT integrals in Ref. \cite{Gehrmann:2011xn}. Hence, a natural expectation for the four-loop form factor is that its expansion is maximally transcendental. This property would be guaranteed if it can be expressed in terms of UT master integrals with rational number coefficients.

This leads to the question of how UT integrals may be identified other than by  prohibitively complicated direct integration. There are three known criteria.
\begin{itemize}
\item 
If an integral can be written in so-called d\,log form, it is UT \cite{Arkani-Hamed:2014via, Bern:2014kca}.
\item
A full set of UT integrals will allow differential equations to be written in so-called $\epsilon$-form \cite{Henn:2013pwa}.
\item
Conjecturally, the residues at all possible poles of a UT integral's integrand (also known as the leading singularities) must always be a constant \cite{Bern:2014kca,Bern:2015ple,Henn:2016men}. 
\end{itemize}
The differential equations in Mandelstam invariants mentioned in the second point are not directly applicable to the Sudakov form factor at hand, since it is a single-scale problem. See, however, Refs. \cite{Henn:2013nsa, Henn:2016men} for a way around this problem, which introduces a more complicated class of integrals with an additional off-shell leg in a first step, and later takes the limit of this leg going on-shell. The first criterion requires one to find a special and typically highly non-linear transformation of variables. Although we succeeded in deriving a d\,log form for some integrals, this currently falls more into the art category. The last criterion, however, enables the construction of an algorithm.

To compute the residues, one first parametrizes all loop momenta in a four-dimensional basis of momenta,
\begin{equation}
l_i = \sum_{j=1}^4 \alpha_{i,j} v^j
\end{equation}
In the case at hand, there is a preferential choice for the $v^{j}$ vectors,
owing to the special kinematics: the set $\{p_1, p_2, e_1, e_2\}$. Here, $e_i$ are massless momenta spanning the plane orthogonal to that spanned by $p_1$ and $p_2$. As a result, the only inner products that do not vanish are $p_1 \cdot p_2$ and $e_1 \cdot e_2$, which can be taken to be equal. Then, by mass-dimension homogeneity, the remaining scale factors out of all integrands, and we are left with a pure function of the 16 $\alpha_{i,j}$ parameters. The criterion introduced previously states that an integral is UT if and only if all of the possible poles taken consecutively are single. In principle, this leads to more than $16!$ possible orders of taking single poles to check --~an unachievable task in practice. In our experience, however, if an integral is not UT, it will fail the criterion rather quickly, typically within  100 attempts with randomly chosen pole orders. Hence, given an integral, one can relatively quickly check if it is potentially a UT integral. 

Given standard, mass-dimension-two propagators in a topology with $12$ different propagators in the denominator, the criterion requires a generic mass-dimension-four numerator. For a fixed integral topology, this class of numerators can be parametrized as a linear combination of $191$ basis elements. Given the simple-pole criterion, one can now derive linear constraints on this set by requiring
the absence of higher-order poles. Solving these constraints then leads to a smaller ansatz. This process can be iterated until no further constraints are found in a large number of random residue checks (typically several thousand). Then the remaining class of integrals is a set of candidate UT integrals.

For the next step, one should realize that most numerical integration algorithms greatly prefer to have input numerators given as products of propagators (mass-dimension-two Lorentz invariants involving loop momenta). Hence, one should find linear combinations of the candidate UT integrals that obey this criterion. Given the form of the integrand as derived previously, we have performed this task by a combination of brute-force equation solving and inspired guesswork. In several cases that involved the full 12-line integrals, it turned out to be impossible to find good product-form UT candidates. In these cases, a linear combination was obtained with one or more ten-line integrals added which consequently possess unit numerator. As an example, consider 
\begin{multline}
I_{6}^{(26)}  =  \hskip -.3cm \begin{tabular}{c}{\includegraphics[height=1.6cm]{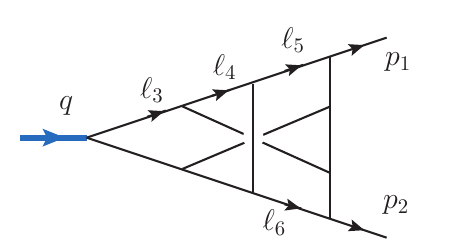}}\end{tabular}   \hskip -.5cm
 \times  \Big\{ [(\ell_3-\ell_4-\ell_5)^2-(\ell_3-\ell_4-p_1)^2-(\ell_6-p_2)^2-\ell_5^2]  \nonumber  \\
\shoveright{ \times [\ell_5^2-\ell_4^2-\ell_6^2+(\ell_4-\ell_6)^2]}
  \nonumber \\
 +4\, \ell_5^2  (\ell_6-p_2)^2 + (\ell_4-\ell_5)^2  (\ell_3-\ell_4+\ell_6-p_2)^2 \Big\},  
\end{multline}
where the entries on the last line are ten-line remnants, see Ref. \cite{Boels:2017ftb} (where the same picture appears) for more details.

\subsection{Putting the pieces together: results}
For all $34$ different integral topologies of the Sudakov form factor~\cite{Boels:2012ew}, a list of UT candidate integrals in product form (sometimes with ten-line corrections) was obtained. For both planar and non-planar form factors, we have related the result in Ref.~\cite{Boels:2012ew} to a rational sum over UT candidate integrals. This is non-trivial, since a subset of the general IBP identities, which can be derived from \Eref{eq:IBPidsDeriv}, consists of IBP relations without dependence on $\epsilon$. This subset will be referred to as rational IBP identities; see Ref. \cite{Boels:2016bdu} for an explanation of how to derive them from a full reduction. The rational IBP identities induce relations between UT candidate integrals. Solving them requires a basis choice. We have found that a choice of $32$ and $23$ UT integrals,   for planar and non-planar cases, respectively, works consistently in minimizing the number of different UT integrals in the problem. The fact that the Sudakov form factor found in Ref.~\cite{Boels:2012ew} can be expressed in terms of UT candidate integrals with only rational coefficients is quite non-trivial and requires intricate cancellations between the different integral topologies. Having settled on a list of UT candidate integrals, we have performed more than 10\,000 random pole checks to make sure that they are UT. This provides strong evidence that the four-loop form factor in $\mathcal{N}=4$ super Yang--Mills theory, including the fact that its non-planar correction is, indeed, maximally transcendental. A d\,log form was found for several
integrals. 

Considering the physical motivations, the next step is integration of the integrals, especially in the non-planar sector. Owing to the conjecture of  Ref.\cite{Becher:2009qa}, the physically relevant question is to what extent the result is in tension with a vanishing cusp  and collinear anomalous dimensions. The problem is that, while the form factor in total in the non-planar sector diverges as ${1} / {\epsilon^2}$, the individual integrals diverge as  ${1}
/ {\epsilon^8}$. This requires the integrals to be expanded up to six and seven orders in the epsilon expansion  to obtain information on the cusp and collinear anomalous dimensions, respectively. Checking vanishing of the sums up to and including ${1} / {\epsilon^3}$ provides a useful sanity check on the performance of any integration method. 

For the problem at hand, we have employed a mix of two integration methods: Mellin--Barnes represen\-tations ~\cite{Smirnov:1999gc,Tausk:1999vh,Anastasiou:2005cb}, as well as sector decomposition  \cite{Binoth:2000ps,Heinrich:2008si}. Both methods are discussed in more detail in other contributions to this report, so the focus here will be on the salient details of our applications. Public packages written in several different languages exist for either method, \eg  \ FIESTA  \cite{Smirnov:2008py, Smirnov:2009pb, Smirnov:2013eza, Smirnov:2015mct} and SecDec \cite{Carter:2010hi, Borowka:2012yc, Borowka:2015mxa} for sector decomposition, and other packages  for {\mbr} representations ~\cite{Smirnov:2004ym,Smirnov:2006ry,Anastasiou:2005cb,Czakon:2005rk,Smirnov:2009up}. If possible, {\mbr} representations tend to yield faster and more accurate results. The problem is that efficient {\mbr} represen\-tations that are valid (integrable) for non-planar integrals are hard to derive automatically. Hence, for the cases where we have succeeded, these representations were obtained by a mix of methods, while we  still lack a straightforward strategy to obtain low-dimensional {\mbr} representations for non-planar integral topologies. This is an obvious area of potential improvement; see other contributions to this report. For the remaining integrals, we have mainly used FIESTA, with cross-checks for simpler integrals using SecDec. To speed up and simplify FIESTA, it pays to implement  sector symmetries as much as possible. These are graph symmetries of the parent topology. If the integrand has a manifest graph symmetry,  one can reduce the number of independent sectors. What remains is, however, still a very hard computation, which we have eventually solved using brute force.

Since numerical methods are employed, a thorough discussion of numerical integration errors is called for. The precision on {\mbr} integrals is many orders of magnitude higher than sector decomposition; therefore, {\mbr} integrals can safely be ignored for this discussion. Where possible, we  cross-checked FIESTA and {\mbr} results. FIESTA employs the CUBA integration library \cite{Hahn:2004fe}, used here in VEGAS mode. VEGAS reports an error, which, given enough evaluation points,  asymptotically approaches the standard deviation of a Gaussian error. There have been reports in the literature, \eg in Ref. \cite{Marquard:2016dcn}, that this reported error underestimates the real error. To guard for this, we have, for each integral, carefully evaluated the consistency of the central value and error while increasing the number of evaluation points, finding no significant instabilities.

\begin{figure}
  \centering
    \label{fig:subfig:a} 
    \includegraphics[width=7.2cm]{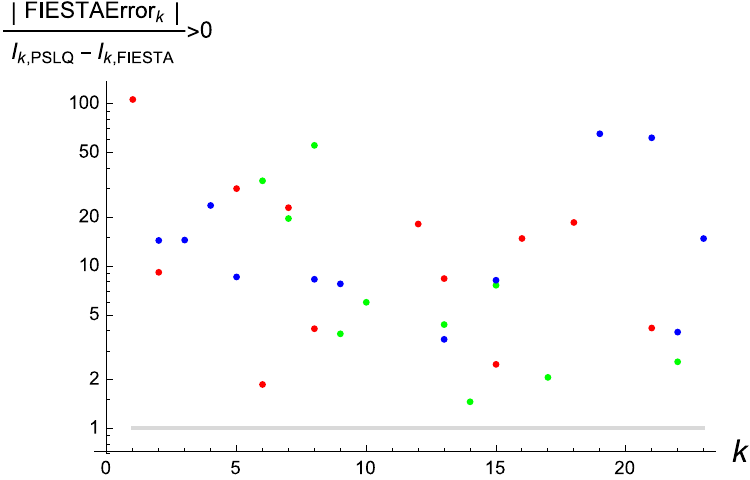}
    \label{fig:subfig:b} 
    \includegraphics[width=8.2cm]{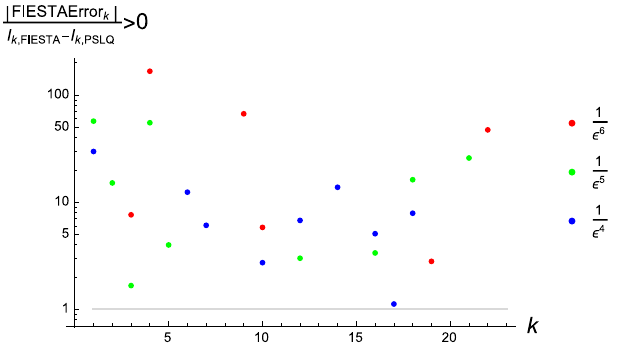}
  \caption{Relative error of FIESTA results compared with PSLQ results for the $\epsilon^{\{-6,-5,-4\}}$ coefficients. Left: plot of cases $({\textrm{FIESTA error}) /( I_{\rm PSLQ} - I_{\rm FIESTA}}) > 0$. Right: plot of cases $({\textrm{FIESTA error}) / (I_{\rm FIESTA} - I_{\rm PSLQ}}) > 0$. A logarithmic scale appears on the vertical axis, and all ratios larger than $200$ are not shown in the figures. All ratios are larger than unity; this clearly indicates that the FIESTA errors are conservative estimates. Deviations of FIESTA results from PSLQ results show no tilt to the positive or negative, which supports the conclusion that there is no source of systematic errors.}
  \label{fig:PSLQcheck} 
\end{figure}

Since the integrals are expected to be UT, their epsilon expansion is governed by multiple zeta values. From the list in \Eref{eq:firstMZVs}, one can see that, for the first five orders of expansion, one would expect the coefficients to be a single multiple zeta value multiplied by a rational. Given a numerical result with a precision of at least a few digits, one can guess which rational number this is through the PSLQ algorithm ~\cite{Ferguson:1999:API:307090.307114}, which is even implemented  in \textsc{Mathematica} ({\tt FindIntegerNullVector}). If we take this to be the exact answer,  this gives a platform to relate the true error to the FIESTA-reported error. Hence, we can use the number-theoretic properties of integrals to evaluate a numerical integration algorithm. The outcome of this is summarized in  \Fref{fig:PSLQcheck}, which is taken from
Ref. \cite{Boels:2017ftb} (distribution under the terms of the Creative Commons Attribution 4.0 International Licence). Basically, there is ample evidence that the FIESTA-reported error overestimates the true error by a considerable margin. 

Pushing through the computation then gives the following results for the non-planar, light-like cusp and collinear anomalous dimensions in $\mathcal{N}=4$ super Yang--Mills theory:
\begin{equation}
\label{eq:results}
\gamma_{\textrm{cusp, NP}}^{(4)}  = -3072\times( 1.60 \pm 0.19 ) \frac{1}{N_c^2} \qquad \textrm{and} \qquad {\cal G}_{\textrm{coll, NP}}^{(4)}  = -384\times( -17.98 \pm 3.25 ) \frac{1}{N_c^2} \, .
\end{equation}
These numbers emerge from the $1/\epsilon^2$ and $1/\epsilon$-pole of the non-planar part of the four-loop form factor, respectively. They are both  statistically significantly different from zero, while all the higher-order poles from $1/\epsilon^8$ to $1/\epsilon^3$ vanish within error bars.

\subsection{Discussion and conclusion}

This contribution has discussed the steps needed to compute, for the first time, the non-planar part of the Sudakov form factor in $\mathcal{N}=4$ SYM, which first appears  at four loops. The result shows that, contrary to a conjecture, the cusp and collinear anomalous dimensions do not vanish in the non-planar sector, for this particular theory. This implies that the structure of IR divergences for general gauge theories is more complicated than (perhaps naively) expected. The computation at various places involves educated guesswork and manual labour. In the context of the FCC, this should be taken as direct motivation for the development of automated tools and techniques. The UT finding algorithm described here is especially ripe for much more general applications. There has also been some quite encouraging development of automated tools for {\mbr} representations, some of which was reported at the workshop. The integrals treated in this contribution are, however, beyond the current round of development and will need further improvement. As always, having explicit results for specific integrals greatly benefits the development of code of this type. 

One interesting direction is to investigate two-point form factors like the Sudakov one beyond the confines of $\mathcal{N}=4$ SYM, probably first in massless QCD. This will pave the way to more intricate computations in  the electroweak sector. The ease of integration for UT integrals that we have encountered here is, in itself, sufficient motivation. Furthermore, in massless QCD for the four-loop Sudakov form factor, for instance, one could seek for an expansion in terms of UT integrals 
that have 
a genuine Taylor expansion around $\epsilon=0$; 
see Ref.~\cite{Schabinger:2018dyi} for progress in this direction. 

Perhaps the most immediately important observation made in the execution of the project is that UT integrals are much simpler to integrate using sector decomposition than generic integrals in the same class. This applies both to  integration timing as well as to accuracy. In fact, we have observed that the expansion coefficients of UT integrals typically grow by one decade per expansion order, whereas, for generic integrals, these can grow by several decades per expansion order. Whereas the first behaviour leads to a numerical precision that is just enough to obtain the physical result, with the latter it would have been out of the question. This offers an immediate advantage to be exploited for explicit computations beyond the scope of the present project, for instance, for self-energy computations at high-loop orders. For integrals with more complicated kinematics, it would be interesting to explore UT finding methods, perhaps to help in finding so-called $\epsilon$-form differential equations; see the talk of Johannes Henn \cite{Henn:Jan2018} and
Section~\ref{chmt}.\ref{sec:rlee}.

 \label{sec-rbff}
\clearpage
\pagestyle{empty}
\cleardoublepage
\pagestyle{empty}
\chapter
{Methods and tools} \label{chmt}

\section[
Introduction to methods and tools in multiloop calculations
\\ {\it J. Gluza, S. Jadach, T. Riemann}
]
{Introduction to methods and tools in multiloop calculations}

\pagestyle{fancy}
\fancyhead[CO]{\thechapter.\thesection \hspace{1mm} Introduction to methods and tools in multiloop calculations}
\fancyhead[RO]{}
\fancyhead[LO]{}
\fancyhead[CE]{J. Gluza, S. Jadach, T. Riemann}
\fancyhead[RE]{}
\fancyhead[LE]{}

\noindent
{\bf Authors: Janusz Gluza, Stanis\l aw Jadach, Tord Riemann} 
\vspace*{.5cm}
\label{sec:mtintro}
  
\noindent
There are two frontiers of theoretical research.
One is the quest for new basic concepts. 
The other  is the quest for greater precision. 
Both together have been the  fundamental of physics for centuries, 
and they will be so  in future. Both need special methods and tools to explore, though this is especially important in the case of precision studies and theoretical calculations, where frontier knowledge from mathematics and computational physics must be used. Studies there very often have an interrogative character, going into so-called exploratory mathematics \cite{Borwein:2004,math3020337}.

In this chapter, we try to cover all  available methods and tools, relevant from today's perspective,
which are needed in higher-order radiative correction calculations, specifically at the FCC-ee Tera-Z stage. 
Among these, there are already well-established methods, such as the  differential equation method, 
sector decomposition numerical method, and Mellin--Barnes analytical and numerical method. These are already widely in use in particle physics.  
There are also some emerging methods, which we will discuss. 
Usually, methods must be accompanied by suitable tools; here this  means that specific software must be developed. This is also discussed in this chapter. 
  The calculations that lie ahead, as seen from the FCC-ee-Z perspective, are very  challenging and must be more precise than the experimental accuracy to be reached. In many sections of this chapter, accuracy, limitations, and ways  to reach the goal  ahead of us are discussed.
  
Of  course, though basically all relevant methods and tools are represented in this chapter, it is not possible to explore them on the same footing,   concerning the computational complexity and the different 
theoretical, mathematical, and computer algebraic details.

Here, we explore especially the sector decomposition and Mellin--Barnes methods, as they have recently brought  us  to completion of the two-loop weak corrections to the Z boson decay \cite{Dubovyk:2016ocz,Dubovyk:2018rlg}, as EWPOs are actually the main issues of this report, explored in previous chapters. Our estimate is that these methods can be used for the next step   needed, namely three-loop electroweak calculations. However, there are also  many places for improvement in existing packages. 

Some more materials related to the issue of loop calculations for the FCC-ee
Tera-Z stage can be found in talks \cite{mini} that are not included in this report. Namely, for the unitarity approach, see the talk by Harald~Ita \cite{Ita:Jan2018}, with first two-loop numerical evaluation \cite{Page:2018flm,Abreu:2018gii}. For bootstrapping at NNLO, see the talk by Johannes~Henn \cite{Henn:Jan2018}. For IBP methods and difference equations, see the talk by Vladimir~Smirnov \cite{Smirnov:Jan2018}. Finally, for tools on the canonical form of differential equations, see the talk by Oleksandr Gituliar \cite{Gituliar:Jan2018}. Other useful talks can be found in web pages of recent conferences: `Loops and Legs in QFT' \cite{LL2018} and `LoopFest' \cite{LoopFest2018}. 
In addition to the advanced material that we are presenting here, we recommend some excellent works and textbooks.  For analytical methods of multiloop calculations, see Refs.
\cite{Smirnov:2004ym,Smirnov:2006ry,Bogner:2010kv}; for numerical and general computation methods, see Ref. \cite{Freitas:2016sty}. 
  
 \label{sec-mtintro}
\clearpage \pagestyle{empty}  
\cleardoublepage

\newpage
\section 
[The MBnumerics project 
\\ {\it J. Usovitsch, I. Dubovyk, T. Riemann}]
{The MBnumerics project \label{subs:mbn}}

\pagestyle{fancy}
\fancyhead[LO]{}
\fancyhead[RO]{}
\fancyhead[CO]{}
\fancyhead[CO]{\thechapter.\thesection \hspace{1mm} The MBnumerics project}
\fancyhead[LE]{}
\fancyhead[CE]{}
\fancyhead[RE]{} 
\fancyhead[CE]{J. Usovitsch, I. Dubovyk, T. Riemann}

\noindent
{\bf Authors: Johann~Usovitsch, Ievgen~Dubovyk, Tord~Riemann}
\\
Corresponding author: Johann~Usovitsch {[jusovitsch@googlemail.com]}
\vspace*{.5cm}

\newcommand{\ddD}{\mathrm{d}}
\subsection{Introduction \label{mbn-s-1}}
In 2012, we decided to launch the calculation of the last unknown, most complicated, electroweak Z boson parameters: the {\em bosonic corrections} to the decay
\begin{equation} \label{mbn-e-1}
 \mathrm{Z}\to \mathrm{b}{\bar {\mathrm{b}}}
 .
\end{equation}
The corresponding weak mixing angle was determined in 2016 \cite{Dubovyk:2016aqv}.
{\textcolor{black}{
The remaining  {\em bosonic corrections} to the $\mathrm{Z}\to \mathrm{b}{\bar {\mathrm{b}}}$ parameters were determined in 2018 \cite{Dubovyk:2018rlg}.
The corresponding {\em fermionic corrections} have been known since 2008
or 2014
\cite{Awramik:2008gi,Freitas:2012sy,Freitas:2013dpa,Freitas:2014hra}.
For the most accurately known {\em leptonic weak mixing angle}, we may refer to  Ref.
\cite{Awramik:2004ge} for the fermionic and to \cite{Awramik:2006ar}  for the bosonic corrections.} Besides analytical integrations, differential equations, Mellin--Barnes representations, sector decomposition, and integral reductions to a basis, three advanced methods are introduced in Ref. \cite{Awramik:2008gi}:
(i) dispersion relations,
(ii) asymptotic expansions for a large top quark mass $m_\mathrm{t}$, and (iii)
 a semi-numerical method based on the Bernstein--Tkachov (BT) algorithm.

The calculation of the Feynman integrals was the bottleneck for the determination of the bosonic two-loop corrections to $\mathrm{Z}\to \mathrm{b}{\bar {\mathrm{b}}}$ because questions of diagram calculations and  renormalization could be considered to be solved.
One faces two-loop vertex integrals at $m_\mathrm{b}^2=0$, at the kinematic point $s=M_\mathrm{Z}^2+\mathrm{i}\delta$, with dependences on up to {\textcolor{black}{ four dimensionless scales, built from the ratios of $s, M_\mathrm{Z}^2, M_\mathrm{W}^2, M_\mathrm{H}^2, m_\mathrm{t}^2$.}}
To limit the complexity of the study of \Eref{mbn-e-1}, it was decided to apply pure methods.
Two of them are known to enable a stringent treatment of infrared problems:
(i) sector decompositions ({\sd}) and (ii) Mellin--Barnes representations ({\mbr}).
Versions of {\sd} software for arbitrary kinematics were created at the time of the project start with SecDec 2.0 
\cite{Borowka:2012yc} and, shortly after, also with FIESTA 3.0 \cite{Smirnov:2013eza}.
The \ar{} project still resided with its numerics on the built-in Fortran module MBintegrate of the {\mbr} package \cite{Czakon:2005rk}, which was intended for Euclidean numerical tests only.
In fact, solving arbitrary kinematics algorithmically with Mellin--Barnes integrals  was considered to be desirable, but non-trivial \cite{Czakon:2005rk,Anastasiou:2005cb}.

The project, now known as MBnumerics, solves the problem of arbitrary kinematics for Mellin--Barnes representations of multiloop Feynman integrals; see Refs. 
\cite{Dubovyk:2016ocz,Usovitsch:2018shx} and references therein.
The most crucial limitation is of purely technical origin: the number of {\mbr}
dimensions should not be too large; typically, four of them may be treated with high accuracy.
To be definite,
 to calculate a two-loop correction with a three-loop accuracy in the Standard Model, one must control approximately six digits, as compared with a Born+one-loop number.
This may be estimated from a simple-minded scale argument by counting powers of the coupling:
another power of $\alpha_\mathrm{em}/(4\pi) \approx 10^{-3}$ for two loops, and its square 
$(\alpha_\mathrm{em}/(4\pi))^2 \approx 10^{-6}$ for three loops.
The FCC-ee project estimates the experimental accuracies of the  FCC-ee Tera-Z
stage option to reach  about $10^{-6}$, needing leading electroweak three-loop and QCD four-loop terms, see Chapter~\ref{sthstatus}. 
Consequently, two-loop terms should be known at that accuracy too, constituting a need for about four significant digits.

To obtain about four significant digits needs, in a problem with thousands of individual terms, an individual accuracy per Feynman integral of about, say, eight digits.
This was the envisaged aim  when the MBnumerics project started.

At our project start,  two {\sd} packages were already
available.
It later emerged that both these packages have certain limitations in Minkowski kinematics, concentrated on cases with a small number of scales and, at the same time, with a more pronounced infrared singular behaviour, independently of the topological structure of the diagram.
This was slightly astonishing, since the {\sd} approach is expected to solve infrared problems safely.
In any case, it is these integrals where the Mellin--Barnes method works at its best.
Summarizing this short introduction, we may conclude that there are now two complementary numerical approaches that, in  combination, enable a stable evaluation of all the electroweak or QCD vertices needed in Z resonance physics and beyond.

After an orientational phase of algorithmic developments, the further MBnumerics 
project
is being developed under the responsibility
 of one of the authors, JU; see Ref.  \cite{Usovitsch:PhD2018} and references therein.

\subsection{Notation and representations \label{mbn-s-2}}
Let us start with a scalar Feynman integral:
\begin{equation}
 G_{L}=\int\prod_{j=1}^L \frac{{\mathrm d}^{D} k_{j}}{\mathrm{i}\pi^{D/2}}\;
 \frac{
 1
 }
 {P_{1}^{\nu_{1}}\dots P_{N}^{\nu_{N}}}
 ,
 \label{eq:tensorIntegral}
\end{equation}
where the $N$ functions $P_{i}$ are composed of the $L$ loop-momenta $k_{l}$
and the $E$ linearly independent external momenta $p_{e}$:
\begin{equation}
P_{i}
=\left(\sum\limits_{l=1}^{L}a_{il}k_{l}+\sum\limits_{e=1}^{E}b_{ie}p_{e}\right)^{2}-m_{i}^{2}+\mathrm{i}\delta,
\mathrm{~~with~}
\;a_{il},\;b_{ie}\in\{-1,0,1\}
.
\label{eq:propagator}
\end{equation}
The propagator exponents $\nu_{i}$ are complex variables if not stated otherwise.
In dimensional regularization, $D=4-2\epsilon$ denotes the dimension of space-time. 
To evaluate these integrals, one may introduce Feynman parameters:
\begin{equation}
 \frac{(-1)^{\nu}}{\prod\limits_{j=1}^{N}(-P_{j})^{\nu_{j}}}=\frac{(-1)^{\nu}\Gamma(\nu)
 \left(\prod\limits_{j=1}^{N}\tilde n_{j}\right)
 \delta \left (1-{\sum\limits_{j=1}^{N_{G}}}x_{j} \right)}{( -k_{l}^{\mu}M_{ll'}k_{l'\mu}+2k_{l}^{\mu}Q_{l\mu}+J-\mathrm{i}\delta)^{\nu}},
 \mathrm{~~with~}
 \;\nu=\sum\limits_{j=1}^{N}\nu_{j},
 \label{eq:FeynmanIntroduction}
\end{equation}
where
\begin{equation}
M_{ll'}=\sum\limits_{j=1}^{N}a_{jl}a_{jl'}x_{j}
\label{eq:matrixM}
\end{equation}
is an $L\times L$ symmetric matrix, 
\begin{equation}
Q_{l}^{\nu}=-\sum\limits_{j=1}^{N}x_{j}a_{jl}\sum\limits_{e=1}^{E}b_{je}p_{e}^{\nu}
\label{eq:vectorQ}
\end{equation}
is a vector with $L$ components, and 
\begin{equation}
J=-\sum\limits_{j=1}^{N}x_{j}\left(\sum\limits_{e=1}^{E}b_{je}p_{e}^{\mu}\sum\limits_{e'=1}^{E}p_{e'}^{\nu}b_{je'}g_{\mu\nu}-m_{j}^{2}
\right ),
\label{eq:scalarJ}
\end{equation}
and the  $x_{j}$ are the Feynman parameters.
The variables $\tilde n_{j}$ in \Eref{eq:FeynmanIntroduction} are defined in Ref. 
\cite{Usovitsch:2018shx}.
Finally, the Feynman integral (\Eref{eq:tensorIntegral}) can be written as follows:
\begin{equation}
G_{L}=(-1)^{\nu}\Gamma(\nu-LD/2)
\left(\prod\limits_{j=1}^{N}\tilde n_{j}\right)
\delta \left(1-\sum\limits_{j=1}^{N_{G}}x_{j}\right)\frac{\CU(x)^{\nu-(L+1)D/2}}{\CF(x)^{\nu-LD/2}},
\label{eq:finalFeynman}
\end{equation}
where
\begin{eqnarray}
 \CU(x)&=&\det M
  ,
 \label{eq:Upolynom}
 \\
 \CF(x)&=&\CU(x)(Q_{l}^{\mu}M_{ll'}^{-1}Q_{l'\mu}+J-\mathrm{i}\delta)
{\textcolor{black}{ ~=~ \sum_{n_i+n_j+\cdots+ n_l=L+1} 
f_{ij\cdots l}~x_i^{n_i} ~ x_j^{n_j} ~ \cdots x_l^{n_l}}}
.
 \label{eq:Fpolynom}
\end{eqnarray}
The functions $\CF(x)$ and $\CU(x)$ are homogeneous in the Feynman parameters $x_{i}$. The function 
$\CU(x)$ has degree $L$, while  
$\CF(x)$ depends on the invariants composed of the external momenta and on the propagator masses. It is of degree $L+1$. 
The $\CF(x)$ and $\CU(x)$ are known as Symanzik polynomials.

\subsubsection{Feynman integrals represented as Mellin--Barnes integrals \label{mbn-3}}
Feynman integrals may have ultraviolet and infrared divergences. 
The latter pose subtle problems.  
Two methods are known for the treatment of the divergences in a consistent and automated way: the Mellin--Barnes integral approach \cite{Smirnov:1999gc,Tausk:1999vh,Heinrich:2004iq,%
Czakon:2004wm,Czakon:2005rk,Anastasiou:2005cb,Gluza:2007rt,Gluza:2010rn,%
Dubovyk:2015yba,Dubovyk:2016ocz,Prausa:2017frh} and the sector decomposition approach \cite{Hepp:1966eg,Binoth:2000ps,Binoth:2003ak,Binoth:2004jv,Denner:2004iz,Heinrich:2008si}.

Systematic derivations of Mellin--Barnes representations for Feynman integrals can be made with the \ar{} package \cite{%
ambrewww,%
Gluza:200704v1, Gluza:2007rt,Gluza:200704v12, Gluza:2009mj,Gluza:200704v13, Gluza:2010rn,Gluza:2010v20,Gluza:2010v21,Gluza:2010v22,%
Dubovyk:201509v30x,  
Gluza:2010mz,%
Dubovyk:2015yba}, together with auxiliary packages of the {\mbr} suite 
\cite{mbtoolsMBsuite}.
One may try either the loop-by-loop approach \cite{Gluza:2007rt} or the global approach 
\cite{Dubovyk:2015yba}. 
Both techniques apply the Mellin--Barnes integral master formula to the $\CF(x)$ and $\CU(x)$ functions in \Eref{eq:finalFeynman}:
\begin{equation}
\frac{1}{(a+b)^\nu}=\int\limits_{-\mathrm{i}\infty}^{\mathrm{i}\infty} \frac{ {\mathrm d} z}{2\pi \mathrm{i}}\,  \frac{a^{z}b^{-z-\nu}\Gamma(-z)\Gamma(\nu+z)}{\Gamma(\nu)},\quad |\arg a-\arg b| < \pi
.
\end{equation}
The integration path has to separate the poles of $\Gamma(-z)$ and $\Gamma(\nu+z)$.
Iterations are performed until all the integrations over Feynman parameters are made by Euler's beta function:
\begin{equation}
B(\xi,\chi)=\int\limits_{0}^{\infty}\frac{x^{\xi-1}}{(1+x)^{\xi+\chi}}{\mathrm d} x=\frac{\Gamma(\xi)\Gamma(\chi)}{\Gamma(\xi+\chi)},\quad \re \xi>0,\;\re \chi>0.
\end{equation}
The method leads to a number of Mellin--Barnes integrands, depending on ratios of Euler's gamma functions 
$\Gamma$, which, by themselves, depend on the {\mbr} integration variables $z_{i}$.
Further, the kinematics are contained in the factors 
$f_{ij\cdots l}$ of \Eref{eq:Fpolynom}, raised to powers with $z_{i}$-dependences.
The number of {\mbr} integrations scales with the number of terms in the  Symanzik polynomials.
It will rise quickly for arbitrary kinematic situations with, \eg $N$ internal masses and $E$ external momenta. 

\subsubsection{A sample two-loop vertex integral \label{mbn-s-3}}

As an example we study the Feynman integral of \Fref{fig:0h0w33r}, 
\begin{equation}
I_{\text{0h0w33r}}= \int \frac{{\mathrm d}^{D} k_{1}}{\mathrm{i}\pi^{D/2}} \frac{{\mathrm d}^{D} k_{2}}{\mathrm{i}\pi^{D/2}}\,\frac{\exp(2\epsilon \gamma_{E})}{(k_{1}^{2}-M_\mathrm{Z}^{2})(k_{1}-k_{2})^{2}k_{2}^{2}(k_{2}-p_{1})^{2}((k_{2}+p_{2})^{2}-M_\mathrm{Z}^{2})}
 .
\end{equation}
The Mellin--Barnes integral representation is:
\begin{align} 
I_{\text{0h0w33r}}=&
-(-s-\mathrm{i}\delta)^{-5 - 2 \epsilon} ~ s^{4} ~ \Gamma(1 - \epsilon)\notag\\
&\times\int\limits_{-\frac{2}{3}-\mathrm{i}\infty}^{-\frac{2}{3}+\mathrm{i}\infty}\frac{{\mathrm d} z_{1}}{2\pi \mathrm{i}}\int\limits_{-\frac{1}{3}-\mathrm{i}\infty}^{-\frac{1}{3}+\mathrm{i}\infty}\frac{{\mathrm d} z_{2}}{2\pi \mathrm{i}}\,
\frac{\left (-\frac{M_\mathrm{Z}^{2}}{\textcolor{black}{s+\mathrm{i}\delta}}\right )^{z_{1}}
\Gamma(-2 \epsilon - z_{1}) \Gamma(1 + 2 \epsilon + z_{1}) \Gamma(-\epsilon - z_{3}) }
{\Gamma(1 - 3 \epsilon - z_{1}) \Gamma(1 - 2 \epsilon - z_{3}) \Gamma(1 - 2 \epsilon - z_{1} + z_{3})} \notag\\
&\times \Gamma(-2 \epsilon - z_{1} + z_{3}) \Gamma(-\epsilon - z_{1} + z_{3})\Gamma(-z_{1} + z_{3})\Gamma(-z_{3})\Gamma(1 + z_{1} - z_{3}).
\label{eq:MBexample}
\end{align}
As mentioned, it is $\epsilon=(4-D)/2$.

\begin{figure}
   \centering
\includegraphics[width=0.48\linewidth]{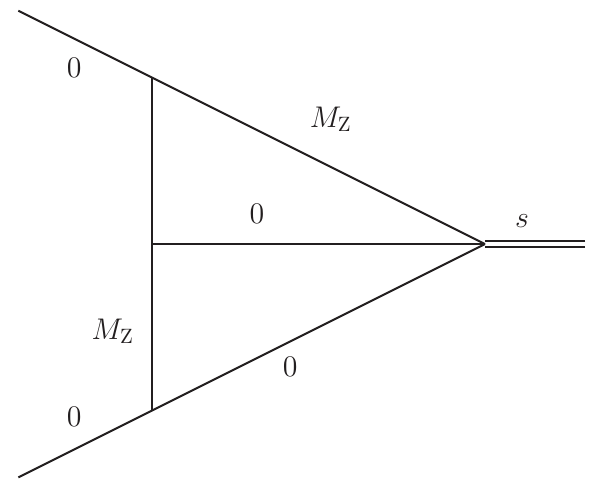}
\caption[]{
One-parameter two-loop vertex Feynman integral with two internal massive lines  and invariants $p^{2}_{1,2}=0$, $2p_{1}p_{2}=s$.
\label{fig:0h0w33r}}
\end{figure}

Straight integration contours 
$z_{1}=-2/3+\mathrm{i} t_{1}$ and $z_{2}=-1/3+\mathrm{i} t_{1}$ with $t_{1,2}\in [-\infty,\infty]$,
parallel to the imaginary axes, separate the poles as needed, 
see \Fref{fig:contour}.
The integral is finite, and an analytical continuation to small $\epsilon$ with the package {\mbr} is not needed here.
The expansion around $\epsilon=0$ in \Eref{eq:MBexample} leads to one finite {\mbr} integral for the lowest-order term in $\epsilon$:
{\color{black}
\begin{align} 
I_{\text{0h0w33r}}=&
\int\limits_{-\frac{2}{3}-\mathrm{i}\infty}^{-\frac{2}{3}+\mathrm{i}\infty}\frac{{\mathrm d} z_{1}}{2\pi \mathrm{i}}\int\limits_{-\frac{1}{3}-\mathrm{i}\infty}^{-\frac{1}{3}+\mathrm{i}\infty}\frac{{\mathrm d} z_{2}}{2\pi \mathrm{i}}\,
\frac{\left (-\frac{M_\mathrm{Z}^2}{s+i\delta} \right )^{z_{1}} \Gamma(-z_{1}) \Gamma(1 + z_{1}) \Gamma(1 + z_{1} - z_{2}) \Gamma(-z_{2})^{2} \Gamma(-z_{1} + z_{2})^{3}}{s \Gamma(1 - z_{1}) \Gamma(1 - z_{2}) \Gamma(1 - z_{1} + z_{2})}.
   \label{eq:MBexampleseries}
\end{align}
}

\begin{figure}
\centering
   \includegraphics[width=0.48\linewidth]{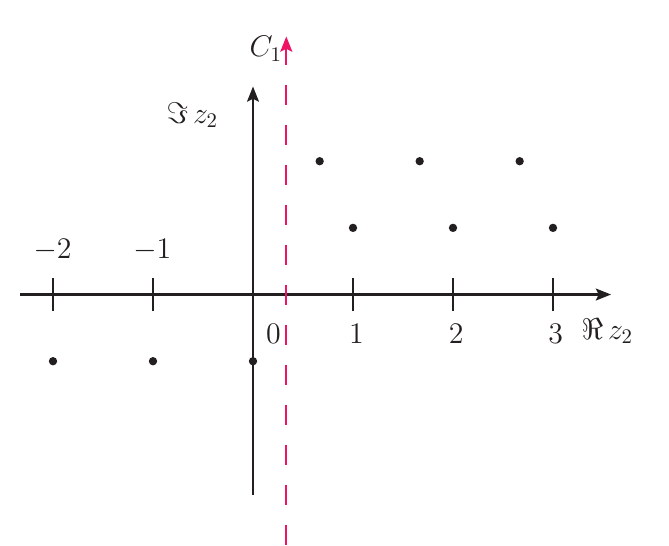}
  \caption[]{
  The black dots are the poles of the integrand in \Eref{eq:MB0h0w33finitelinearTransformation} in the $z_{2}$ complex plane. The dashed line is the integration contour parallel to the imaginary axis.
\label{fig:contour}}
\end{figure}

\subsubsection{Specifics of Minkowskian kinematics}

To illustrate potential convergence problems of the Mellin--Barnes integrals, we apply the  Stirling approximation formula,
\begin{equation}
\Gamma(z)\underset{|z|\to\infty}{\approx}z^{z-1/2}\mathrm{e}^{-z}\sqrt{2\pi},\quad|\arg{z}|<\pi
,
\label{eq:stirling}
\end{equation}
to the integrand in \Eref{eq:MBexampleseries} and examine the asymptotic behaviour when setting 
{\color{black}$z_{1}=- {2}/{3}+\mathrm{i}t_{1}$, and $z_{2}=- {1}/{3}+\mathrm{i}t_{2}$, $t_{1}\to-t$, and $t_{2}\to -t$:
\begin{equation}
 \CI_{\text{0h0w33r}}\underset{t\to\infty}{\approx} t^{-2+x_{1}-x_{2}}|_{x_{1}=-2/3,\,x_{2}=-1/3}.
\end{equation}
}

For the numerical treatment of Mellin--Barnes integrals, this polynomial asymptotic behaviour leads to numerous numerical instabilities, including
the following.
\begin{itemize}
 \item The integrand oscillations are less damped compared with the Euclidean case.
 \item Integrals may be not absolutely convergent if the asymptotic behaviour is worse than $1/t^{a}$, with $a<2$.
 \item At any level of accuracy, one has to evaluate the integrands for larger absolute values $t_{i}$ than in the case of Euclidean kinematics.
 \item In particular, because we are interested in highly accurate results, we have to evaluate 
 $\Gamma$ functions for very big arguments, again introducing numerical instabilities.
\end{itemize}

\subsection{Selected techniques for the improved treatment of Mellin--Barnes integrals in Minkowskian kinematics \label{mbn-s-4}}
\subsubsection{Contour deformations for one-dimensional {\mbr} integrals \label{mbn-s-5}}
Henceforth, it is assumed that the treatment of one-dimensional Mellin--Barnes integrals is solved, namely by means of contour deformation 
\cite{Freitas:2010nx,Gluza:2016fwh,Peng:2012zpa,Dubovyk:2016ocz}. 
MBnumerics includes a use of this technique. 
In the next subsections, we will describe additional techniques that may be applied to 
{\em multidimensional} Mellin--Barnes integrals. These techniques, as well as contour deformation, are automatized in the \textsc{Mathematica} package MBnumerics.
\subsubsection{Linear transformations of {\mbr} integration variables}
For Mellin--Barnes integrals, a linear transformation of integration variables may lead to improvements of the numerical integrations.
Applying a variable change {\color{black}{$z_{2}\to z_{2}+z_{1}$}} to the example integral in \Eref{eq:MBexampleseries},
one obtains the representation:
{\color{black}
\begin{align}
 I_{\text{0h0w33}}=&
 -\int\limits_{-\frac{2}{3}-\mathrm{i}\infty}^{-\frac{2}{3}+\mathrm{i}\infty}
 \frac{{\mathrm d} z_{1}}
 {2\pi \mathrm{i}}\int\limits_{\frac{1}{3}-\mathrm{i}\infty}^{\frac{1}{3}+\mathrm{i}\infty}
 \frac{{\mathrm d} z_{2}}
 {2\pi \mathrm{i}}\,
 \frac{\left(-\frac{M_\mathrm{Z}^{2}}{\textcolor{black}{s+\mathrm{i}\delta}} \right)^{z_{1}} \Gamma(-z_{1}) \Gamma(1 + z_{1}) \Gamma(1 - z_{2}) 
 \Gamma(-z_{1} - z_{2})^2 \Gamma(z_{2})^3}{s \Gamma(1 - z_{1}) \Gamma(1 - z_{1} - z_{2}) \Gamma(1 + z_{2})}.
      \label{eq:MB0h0w33finitelinearTransformation}
\end{align}
}%
Applying  the Stirling formula to the integrand in \Eref{eq:MB0h0w33finitelinearTransformation}, and studying the asymptotic behaviour for
 $z_{1}=- {2}/{3}+\mathrm{i}t_{1}$, $z_{2}= {1}/{3}+\mathrm{i}t_{2}$, $t_{1}\to-t$, and $t_{2}\to 0$, one finds
\begin{equation}
 \CI_{\text{0h0w33}}\underset{t\to\infty}{\approx} t^{-2-x_{2}}|_{x_{2}=1/3},
\end{equation}
\ie the polynomial asymptotic behaviour depends only on $x_{2}$. 
One concludes that linear integration variable transformations may be used for non-trivial cross-checks of the numerical evaluation of the Mellin--Barnes integrals since the integrands have different asymptotic behaviour before and after the linear transformation,
{\textcolor{black}{ although they do not improve the asymptotics in a crucial way.}}

\subsubsection{{\mbr} integrand mappings}
An obvious numerical improvement results from a cotangent mapping $t= {1}/{\tan(-\pi d)}$, 
which transforms the integration boundaries from $t\in[-\infty,\infty]$ to $d\in[0,1]$. Applying this mapping to a polynomial function gives
\begin{equation}
 \frac{1}{t^{a}}=\tan(-\pi d)^a
 .
 \label{eq:cotangMapping}
\end{equation}
The corresponding Jacobian is
\begin{equation}
J = \frac{\pi}{\sin(\pi d)^2}
.
\end{equation}
The integrand becomes, at the boundaries of the new integration domain,
\begin{equation}
  \lim\limits_{d\to0,d\to1}\frac{\pi\tan(-\pi d)^a}{\sin(\pi d)^2}= \begin{cases}
  \frac{1}{0}, & a<2, \\
  \pi, & a=2, \\
  0, & a>2.
  \end{cases}
\end{equation}
Compared with the cotangent mapping introduced here, the logarithmic mapping of the program MB.m \cite{Czakon:2005rk} always leads to infinities at the new  integration boundaries, which will  subsequently lead to numerical instabilities.
Once we use the cotangent mapping, it is mandatory to transform the integrand as follows:
\begin{equation}
 \prod_{i}\Gamma_{i}\to\exp\left(\sum_{i}\log\Gamma_{i}\right)
 .
 \label{eq:logGammaMapping}
\end{equation}
The underlying observation is that  $\log\Gamma(z_{i})$ functions grow more
slowly than  $\Gamma(z_{i})$ functions for large absolute values of their arguments $z_{i}$.
\subsubsection{Parallel shifts of {\mbr} integration paths}
If one shifts a Mellin--Barnes integration contour according to
\begin{equation}
 z_{i}=x_{i}+\mathrm{i}t_{i}+n_{i},\quad n_{i}\,\in \R
 ,
\end{equation}
the asymptotic behaviour of the  integrand will change due to these shifts $n_{i}$.
In the example,
{\color{black}
\begin{equation}
 \CI_{\text{0h0w33}}\underset{t\to\infty}{\approx} t^{-2-x_{2}-n_{2}}|_{x_{2}=1/3}.
\end{equation}
}%
As a result, it appears to be possible to improve the polynomial asymptotic behaviour by 
{\em systematic, heuristic}
tuning of the shifts $n_{i}$.
If, by changing the values of $n_{i}$, the integration contour crosses some poles of the Mellin--Barnes integrand, one has to collect their residues as additive terms of the net original {\mbr} integral.

Evidently, parallel shifts may be used as a method to improve the evaluation of Mellin--Barnes integrals in Minkowskian regions.
The integration value from summing over a shifted contour may be numerically smaller, by orders of magnitude, than the original integral. The potentially huge `rest' comes from the residues sum.
{\color{black} For example, the original integral in \Eref{eq:MB0h0w33finitelinearTransformation},}
{\color{black}
with $s=M_\mathrm{Z}^2 +\mathrm{i}\delta$ at $M_\mathrm{Z}=1$, evaluated along the contour $C_1$, see \Fref{fig:contourcolor},} {\color{black} gives $-3.6528409137$. 
The shifted integral with $n_{2}=2$, evaluated along the contour $C_2$, gives $0.060767748603 + 0.210051961900\,\mathrm{i}$. 
In addition, the following equation holds:
\begin{equation}
 \int 
 \frac{{\mathrm d}\,z_{1}}
 {2\pi \mathrm{i}}\int_{C_{1}}\frac{{\mathrm d}\,z_{2}}{2\pi \mathrm{i}}\CI_\text{0h0w33}  = \int
 \frac{{\mathrm d}\,z_{1}}
 {2\pi \mathrm{i}}\int_{C_{2}}
 \frac{{\mathrm d}\,z_{2}}
 {2\pi \mathrm{i}}\CI_\text{0h0w33} - \overset{\text{1-dim. integrals}}{\overbrace {\int 
 \frac{{\mathrm d}\,z_{1}}{2\pi \mathrm{i}}\left(
 \sum\limits_{z_{0}}\mathrm{Res}_{z_{0}}\CI_\text{0h0w33}\right)}}.
\label{eq:shiftequation}
\end{equation}
We must supplement the result of the integral along the shifted contour by calculating four residues, {corresponding} to the  poles enclosed by the contour $C_{3}$. On integrating them over $z_{1}$, their sum is  $-3.7136086623783
\allowbreak -  0.2100519619004\,\mathrm{i}$. 
 }
In general, shifting the contour of an $n$-fold Mellin--Barnes integral will yield residue terms, which will be $(n-1)$-fold Mellin--Barnes integrals and hence {\em }simpler and more accurate} to evaluate.

\begin{figure}
\centering
\includegraphics[width=0.48\linewidth]{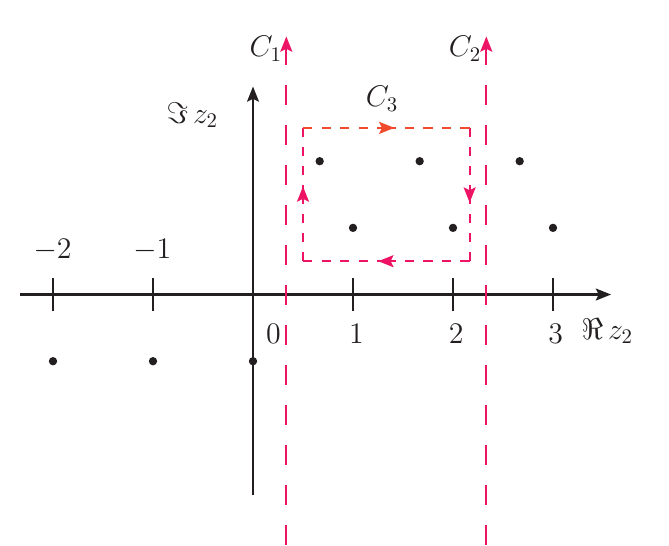}
  \caption[]{
  The original contour $C_{1}$ is shifted by {\color{black}{$n_{2}=+2$}} to a contour $C_{2}$. The third contour $C_{3}$ encircles the crossed poles in order to correct the change of the integral due to the shift. 
  \label{fig:contourcolor}}
\end{figure}

\subsection{Building a grid \label{mbn-s-grid}}

{\color{black}{
The Feynman integral
\begin{align} 
\label{eq:soft}
I_{\text{soft1}} &=
\\ \nonumber
&
\int
\frac{ {\mathrm d} ^{D}  k_{1}} {\mathrm{i}\pi^{D/2}}
\frac{ {\mathrm d} ^{D}  k_{2}} {\mathrm{i}\pi^{D/2}}
\,
\frac
{ \exp(2\epsilon \gamma_{E}) (M_\mathrm{Z}^{2})^{2+2\epsilon}}
{
(k_{1})^{2}((k_{1}-k_{2})^{2} - M_\mathrm{W}^2) ((k_{2})^{2} - m_{t}^2) ((k_{1}+p_{1})^{2})((k_{2}+p_{1})^{2} - M_\mathrm{W}^2)(k_{1}+p_{1}+p_{2})^{2}
}, 
\end{align} 
}}
is shown in \Fref{fig:soft}.
We evaluate this integral with the method of shifts as it is implemented in MBnumerics. 
The results are collected in \Tref{tab:MBnumerics}. Once MBnumerics terminates evaluating a Feynman integral, it saves 
to disk
all the Mellin--Barnes integrals that where found to suffice to reach the desired accuracy. 
All the  Mellin--Barnes integrals are improved in their convergence.
Thus, re-evaluating them to compute the Feynman integral for different kinematics is possible, see, as an example, \Fref{fig:grid}. 
{\color{black}The evaluation at the initial kinematic point  is shown in 
\Tref{tab:MBnumerics}. {\color{black} It took 8\,min. The remaining 120 points took 1\,min 18\,s on average each, which is an improvement by $84\%$.

\begin{figure}
   \centering
\includegraphics[width=0.48\linewidth]{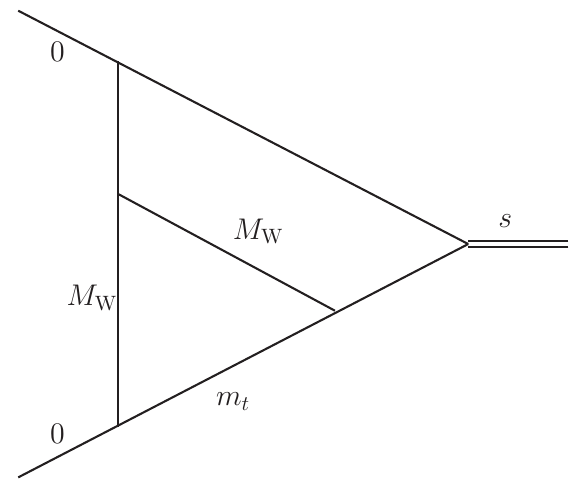}
\caption{
Feynman integral with three scales $s$, $M_\mathrm{W}^2$, and $m_{t}^2$
 \label{fig:soft}}
\end{figure}

\begin{table}
  \caption{Numerics for the Feynman integral (\Eref{eq:soft}). The numbers labelled `{\mbr}' are evaluated with MBnumerics. The numbers labelled 
`{\sd}' are evaluated with SecDec version 3 \cite{Borowka:2015mxa}.
The scales are: $ s= M_\mathrm{Z}^2+\mathrm{i}\delta$, $M_\mathrm{Z}=91.1876\UGeV$, $M_\mathrm{W}=80.385\UGeV$, and $m_{t}=173.2\UGeV$.
  \label{tab:MBnumerics}}
\renewcommand{\arraystretch}{1.0}
\centering
\begin{tabular}{lll}
\hline \hline
{\color{black}
Method} & Numerics& \\\hline
    {\mbr}          & $-0.3380005111031239\,\epsilon^{-2}$& \\
    {\sd} - 90 Mio & $-0.3380005111\,\epsilon^{-2}$& \\
    {\mbr}          & $(0.1232496327$ & $-1.0618599226\,\mathrm{i})\,\epsilon^{-1}$\\
    {\sd} - 90 Mio & $(0.1232503$  & $-1.0618602\,\mathrm{i})\,\epsilon^{-1}$\\
    {\mbr}          &$(1.54245262840$& $+0.24731366189\,\mathrm{i})+\CO(\epsilon)$\\
    {\sd} - 90 Mio &$(1.54219$ & $-0.24724\,\mathrm{i})+\CO(\epsilon)$\\
    \hline \hline
  \end{tabular}
\end{table}

\begin{figure}
\centering
\includegraphics[width=1\linewidth]{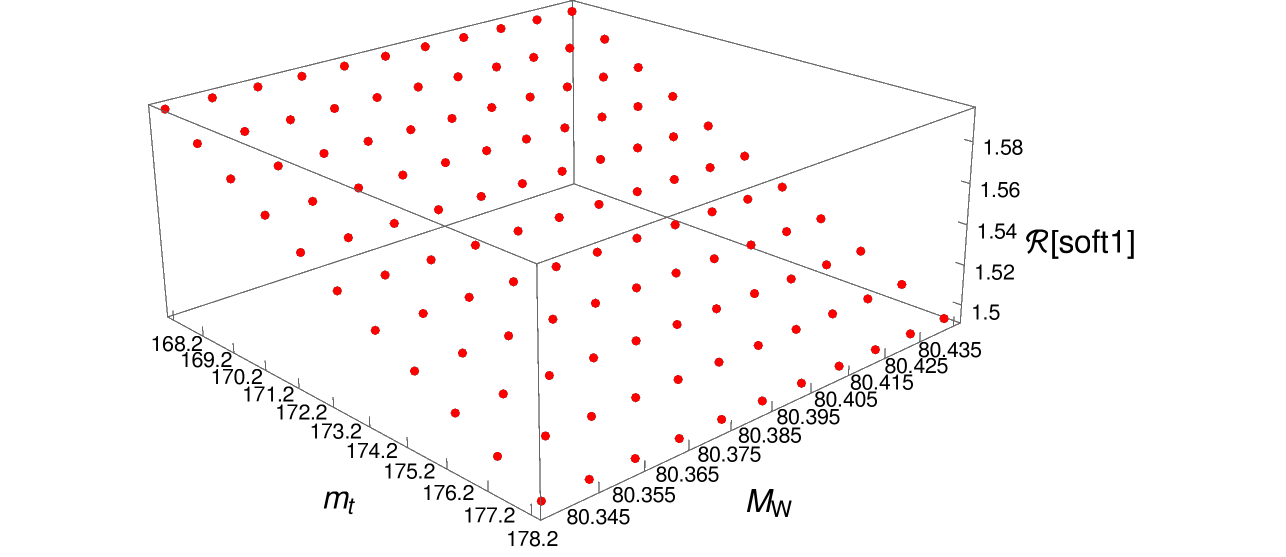}
  \caption{
  Grid for the numerical values of the Feynman integral $I_{\text{soft1}}$
of   \Eref{eq:soft}, varying the input values of  
  $M_\mathrm{W}$ and $m_{t}$ at fixed $s=M_\mathrm{Z}^2+\mathrm{i}\delta$. Only the real part is shown.}
  \label{fig:grid}
\end{figure}

In the example, the Mellin--Barnes integral representation is, at most, three-dimensional. 
With the sector decomposition approach, the Feynman integral (\Eref{eq:soft}) is five-dimensional. 
A general rule is the following.
If one can find a Mellin--Barnes integral representation whose dimension is smaller than or equal to that of the sector decomposition representation,
the method of shifts turns out to be very successful in computing the Feynman integral in Minkowskian regions.

{\color{black} As was mentioned, for a variety of Feynman integrals the SD-method is more advantageous.
An example is integral  {\tt xhxwxz63} of the $Zf{\bar f}$ project \cite{Dubovyk:2018rlg}:
\begin{align}
\int\frac{\mathrm{d}^{D} k_{1}}{i\pi^{D/2}}\frac{\mathrm{d}^{D} k_{2}}{i\pi^{D/2}}&\,\frac{\exp(2\epsilon \gamma_{E})(k_{1}p_{1})^{2}}{
(k_{1}^{2}-M_{Z}^{2})
((k_{1}-k_{2})^{2} - m_{t}^2) 
(k_{2}^{2} - m_{t}^2) 
((k_{1}-k_{2}+p_{1})^{2}-M_{W}^{2})   }\notag\\
&\times\frac{1}{((k_{2}+p_{2})^{2}-M_{W}^{2})
((k_{1}+p_{1}+p_{2})^{2}-M_{H}^{2})},
\end{align} 
shown in Fig.~\ref{mbn-fig:xhxwxz-NN} and reproduced in Tab.~\ref{tab:xhxwxz}.
Although the integral depends on as much as four dimensionless scales, the resulting higher-dimensional MB-method may serve here as an independent, efficient cross-check for the SD-calculation. }
\begin{table}[htpb]
  \caption{  The table reproduces the constant term of the $\epsilon$-expansion at 
  ${s}=M_{Z}^2+i\delta$,
  $M_{Z}=91.1876\; \mathrm{GeV}$,
  $M_{W}=80.385\; \mathrm{GeV}$,
  $M_H  =125.1\; \mathrm{GeV}$, 
    and $m_{t}=173.2\; \mathrm{GeV}$. Its MB-dimension is eight, while the SD-dimension is only five.}
\renewcommand{\arraystretch}{1.0}
\begin{center}
  \begin{tabular}[\linewidth]{ll}
    \hline
    MB  &$(0.0029\pm0.0008)+\CO(\epsilon)$ \\
    SD - 15 Mio &$0.00313383363+\CO(\epsilon) $\\
    \hline
  \end{tabular}
  \end{center}
  \label{tab:xhxwxz}
\end{table}

\begin{figure}[t]
   \begin{center}
\includegraphics[width=0.48\linewidth]{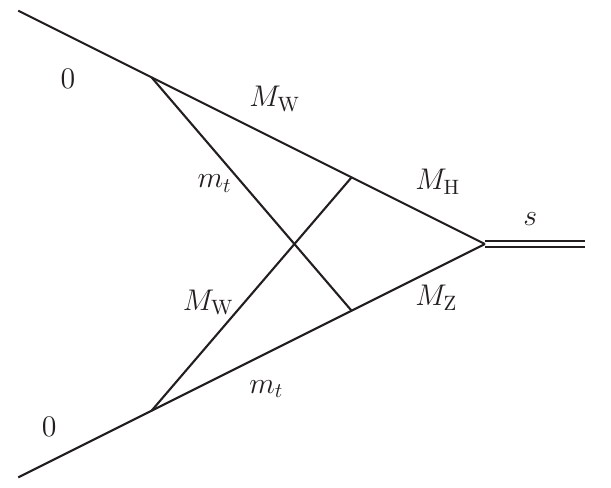}
   \end{center}
  \caption{
  \color{black}
  The integral  {\tt xhxwxz63} of the $Zf{\bar f}$ project \cite{Dubovyk:2018rlg}
  is real in Minkowskian kinematics and not divergent at $D=4$.
   \label{mbn-fig:xhxwxz-NN}
  }
\end{figure}

\subsection{The MBnumerics package: present and future \label{mbn-s-9}}
The MBnumerics package will be made available on request from the author after an improved automatization of its use  
\cite{Usovitsch:2018shx}, depending also on further checks and numerical improvements.
It consists of about 2000 lines of \textsc{Mathematica} code.
Details of the numerical implementations described here are given in Ref. \cite{Usovitsch:PhD2018}.
The parameter integrations of MBnumerics are performed using the CUHRE implementation of the CUBA library \cite{Hahn:2004fe}. 
 
From the sector decomposition approach, we know that, for $L$-loop integrals with $N$ internal lines, 
$(N-1)$-dimensional parameter integrals will result, independent of the specific topology and for any number of physical parameters.
In this respect, the {\sd} method is very robust.
An important issue for the {\mbr}-approach
is the search for lower-dimensional realizations of Feynman integrals in terms of Mellin--Barnes integrals in those cases where their dimensionality is high, owing to the many kinematic invariants.

Concerning the dimensionality of {\mbr} representations, 
we would like to draw  the reader's attention to recent progress for {\em one-loop} Feynman integrals.
In Ref. \cite{Bluemlein:2017rbi},
a recursion relation was derived yielding $(N-1)$-dimensional {\mbr} integrals for $N$-point functions, independent of the details of their kinematics.
This feature corresponds to the efficiency of the {\sd} approach.
In Ref. \cite{Usovitsch:Jan2018}, 
it is demonstrated that this representation allows for very stable, efficient numerics when using a modified MBnumerics package, even for small or vanishing Gram determinants.

For future projects related to the Z boson resonance, one has to envisage numerical solutions for electroweak three-loop vertex integrals with up to four dimensionless parameters.
Another issue, arising from unfolding the Z resonance curve, are electroweak two-loop box integrals.

%
\subsection*{Acknowledgements}
We would like to thank Ayres Freitas and Janusz Gluza for fruitful discussions. We enjoyed the opportunity to complete  with them the calculation of the electroweak two-loop
corrections to the Z boson resonance physics. 

 \label{sec-mbnum}  
\clearpage \pagestyle{empty} 
\cleardoublepage


\cleardoublepage
\section
[Mini-review on the pragmatic evaluation of multiloop multiscale 
integrals using Feynman parameters
\\ {\it S. Borowka}
]
{Mini-review on the pragmatic evaluation of multiloop multiscale 
integrals using Feynman parameters} \label{sec:secdec}

\pagestyle{fancy}
\fancyhead[LO]{}
\fancyhead[RO]{}
\fancyhead[CO]{}
\fancyhead[CO]{\thechapter.\thesection
\hspace{1mm} 
Pragmatic evaluation of multiloop multiscale 
integrals using Feynman parameters}
\fancyhead[LE]{}
\fancyhead[CE]{S. Borowka}
\fancyhead[RE]{} 

\noindent
{\bf Author: Sophia Borowka} 
{~~[sophia.borowka@cern.ch]}
\vspace*{.5cm}

\subsection{Introduction}
\label{sec:sb:intro}

The numerical evaluation of multiloop multiscale integrals has
become a vital pillar for phenomenological predictions at high
energies. In particular, at current and future colliders, like
the proposed FCC-ee, the masses of heavy quarks, vector
bosons, and the Higgs boson can be resolved. To arrive at the desired 
accuracy for predictions of exclusive processes, these need to be taken
into account exactly. A particularly prominent example is Higgs boson
pair production at the NLO in gluon fusion, where naive approximations assuming
an infinitely large top quark mass fail
dramatically~\cite{Borowka:2016ehy,Borowka:2016ypz}.\footnote{It should be
noted that an approximation supplemented by the appropriate threshold
expansion performs very well, see Ref. \cite{Grober:2017uho}.}
As a result of the additional non-negligible scales, the Feynman
integrals involved in the computation 
of perturbative predictions become harder to compute. In particular, the
virtual contributions beyond the
one-loop level contain massive propagators, which can lead to the
appearance of elliptic structures, see,
\eg\ Refs. \cite{SABRY1962401,Aglietti:2007as,vonManteuffel:2017hms}. However,
these can just as well appear in fully massless integrals, as  shown
in Ref.~\cite{CaronHuot:2012ab} for the two-loop hexagon integral in $N=4$
supersymmetric Yang--Mills theory. Since the analytical treatment of these
structures is still under way, see, \eg Ref. \cite{Broedel:2017kkb} and
references therein, other methods to compute these types of integral
become highly attractive. Two of these methods are reviewed in this
contribution, see Sections~E.\ref{sec:sb:pysecdec} and E.\ref{sec:sb:tayint}. The first approach 
\cite{Carter:2010hi,Borowka:2012yc,Borowka:2013cma,Borowka:2015mxa,Borowka:2017idc} 
is highly algorithmic and allows for a full numerical evaluation of 
multiloop multiscale integrals. While the second method 
\cite{Borowka:2018dsa} is based on Taylor-series expansions and is of 
similar algorithmic nature, 
the generation of a library for one 
integral result takes considerably more time. Yet once the library is 
generated, the results for arbitrary kinematic values can be evaluated 
instantaneously and to very high accuracy. 

Before considering explicit calculations of (multi)loop integrals,
it is advisable to check whether analytical results are already available in 
the literature. A growing subset of these is listed on the community-driven
database \textsc{Loopedia} \cite{Bogner:2017xhp}, introduced in Section~E.\ref{sec:sb:loopedia}.

%
%
\subsection{Preliminaries}
\label{sec:sb:prelim}

A general Feynman loop integral $G$ at $L$ loops with $N$ propagators, where 
the propagators $P_j$ can have, in principle, arbitrary powers $\nu_j$ and mass $m_j$,  
has the following representation in momentum space
\begin{gather}
G_{l_1 \dots l_R}^{\mu_1 \dots \mu_R} (\{p\},\{m\})  =
 \prod\limits_{l=1}^{L} \int \text{d}^D\kappa_l 
\frac{k_{l_1}^{\mu_1} \cdots k_{l_R}^{\mu_R}} {\prod\limits_{j=1}^{N} P_{j}^{\nu_j}
\left(\{k\},\{p\},m_j^2
\right )} \label{eq:genfeynintegral} \\
\text{d}^D\kappa_l = \frac{\mu_r^{4-D}}{\mathrm{i}\pi^{\frac{D}{2}}}\,\text{d}^D k_l\;, \qquad P_j \left(\{k\},\{p\},m_j^2 \right )=q_j^2-m_j^2+\text{i}\delta\;,
\label{eq:propagatordefinition}
\end{gather}
where the $q_j$ are linear combinations of external momenta $p_i$ 
and loop momenta $k_l$. The integral is of rank $R$; the 
indices $l_i$ indicate the loop momentum associated with Lorentz index $\mu_i$. 
In what follows, the renormalization scale $\mu_r$ is set to one by default.

After rewriting the integral in terms of Feynman parametrization and integration of the angular
component, a scalar integral reads
\begin{align}
G =&
\frac{(-1)^{N_{\nu}}}{\prod_{j=1}^{N}\Gamma(\nu_j)}
\,\prod_{j=1}^{N}\,  \int_{0}^{\infty}  
\mathrm{d}t_j\,\,t_j^{\nu_j-1}\,\delta \left (1-\sum_{l=1}^N t_l \right ) \; 
\frac{{\cal U}^{N_{\nu}-(L+1) D/2}}
{{\cal F}^{N_\nu-L D/2}} \text{ ,}
\label{eq:thefeynmanloopintegral}
\end{align}
where $\mathcal{U}$ and $\mathcal{F}$ denote the 
first and second Symanzik polynomials. They are
homogeneous in the Feynman parameters and of degree $L$ and $L+1$, respectively. 

An integral in Feynman parametrization can possess physical singularities  
in the ultraviolet or infrared limit that need to be regulated 
dimensionally. Their proper factorization can be performed using sector 
decomposition \cite{Hepp:1966eg,Roth:1996pd,Binoth:2000ps,Heinrich:2008si}. The
integration of 
the $\delta$-distribution in Eq.~(\ref{eq:thefeynmanloopintegral}) can 
 be done by splitting the integral into $N$
primary sectors before performing an iterated factorization of the poles. This
has the advantage of moving the integration boundaries of the remaining 
Feynman parameters to zero and one. Alternatively, the Cheng--Wu 
theorem \cite{Cheng:1987ga,Smirnov:2006ry} can be utilized. In this case, the resulting
integration boundaries of the remaining Feynman parameters can only be mapped to
zero and one, when the factorization of the poles is performed with a deterministic sector decomposition
strategy based on algebraic geometry, first introduced in 
Refs.~\cite{Kaneko:2009qx,Kaneko:2010kj}. The resulting decomposition strategy \cite{Borowka:2015mxa,Schlenk:2016a}
has proven most efficient in terms of the number of
generated sectors and the complexity of the resulting functions. For completeness,
other strategies are presented and compared in
Refs.~\cite{Bogner:2007cr,Bogner:2008ry,Smirnov:2008py,Smirnov:2008aw,Smirnov:2009pb}. 

\subsection{\textsc{Loopedia} -- a database for loop integrals}
\label{sec:sb:loopedia}
Before embarking on the computation
of (multi)loop integrals, it is preferable to have a database of loop integrals,
where  existing results are listed and linked to the associated literature
references. Until recently, such a database did not exist. \textsc{Loopedia}~\cite{Bogner:2017xhp}
attempts to fill this gap, though it is not limited to bibliographic information.
The description field of each record can hold any kind of textual information (\eg 
links to software). In addition, arbitrary files can be uploaded, for example, Fortran programs
or Maple worksheets. Not only can the database  be searched, but new entries can also
be submitted and are then reviewed by the \textsc{Loopedia} administrators.

Integrals in the database are generally associated with their underlying graphs.
In the case of non-trivial numerators, linear combinations, or a set of integrals in a 
topology, the reference is indexed by the scalar integral of the topology. In this
context, a topology is defined as a fixed set of propagators and external lines with
fixed mass assignments.
Tensor integrals are intentionally not covered. In legitimate cases, they can be added 
as an additional entry for the scalar integral of the same topology by providing
the additional information
in the description box of the submission. 

\textsc{Loopedia} is located at \texttt{loopedia.org}. Its landing page is depicted
in \Fref{fig:loopedia}. 

\begin{figure}
  \centering
      \includegraphics[width=0.8\textwidth]{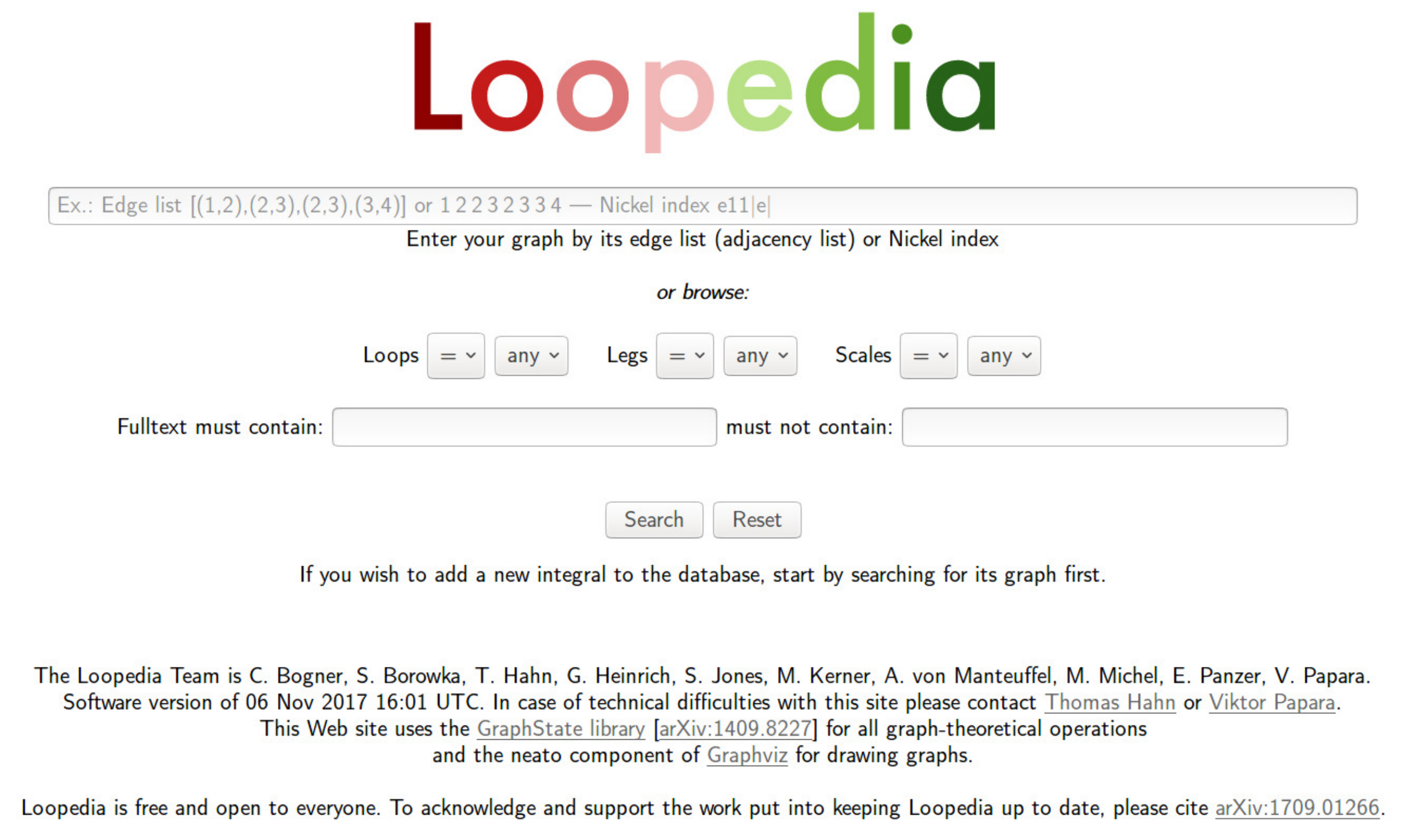}
      \caption{\textsc{Loopedia}'s landing page}
    \label{fig:loopedia}
\end{figure}

\subsection{Numerical evaluation with the program \textsc{pySecDec}}
\label{sec:sb:pysecdec}

For a highly automated and user-friendly calculation of multiloop and multiscale 
Feynman integrals the publicly available program 
\textsc{pySecDec} \cite{Borowka:2017idc} is introduced. 
\textsc{pySecDec} is the direct successor of the program
\textsc{SecDec} \cite{Carter:2010hi,Borowka:2012yc,Borowka:2013cma,Borowka:2015mxa}. 
It is based on the method of sector decomposition and is one  of several
publicly available programs \cite{Bogner:2007cr,Gluza:2010rn,Smirnov:2008py,Smirnov:2009pb,Smirnov:2013eza,Smirnov:2015mct}
for  the calculation of
multiloop integrals numerically. While \textsc{sector\_decomposition} \cite{Bogner:2007cr,Gluza:2010rn}
is restricted to the Euclidean region or single-scale integrals,
\textsc{(py)SecDec} and \textsc{FIESTA} \cite{Smirnov:2008py,Smirnov:2009pb,Smirnov:2013eza,Smirnov:2015mct}
allow for an evaluation in the physical region of, in principle, arbitrary
multiloop multiscale integrals. 

Unlike previous versions of \textsc{SecDec}, \textsc{pySecDec} relies on
open-source software only. It performs the automated factorization of
dimensionally regulated poles in Feynman and more general parametric
integrals and the subsequent numerical evaluation of the finite coefficients. 

The algebraic part of the program is written in the form of Python
modules, allowing for very flexible usage. An optimized {\tt C++} code is
generated using the program \textsc{FORM} \cite{Vermaseren:2000nd,Kuipers:2013pba},
leading to faster numerical convergence with
respect to previous versions of the program. To facilitate the integration itself, the
program is linked to the publicly available \textsc{Cuba} library
\cite{Hahn:2004fe,Agrawal:2011tm,Hahn:2014fua}. For
one-dimensional integrals, the integrator \textsc{CQuad} contained in the GNU
scientific library \cite{Gough:2009:GSL:1538674} can be used. 
An overview of the operational sequence is depicted in \Fref{fig:flowchart}.
The generated files can be used as an integral library, which can be interfaced
with user-specific codes, \eg for the evaluation of matrix elements.

\begin{figure}
  \centering
      \includegraphics[width=0.8\textwidth]{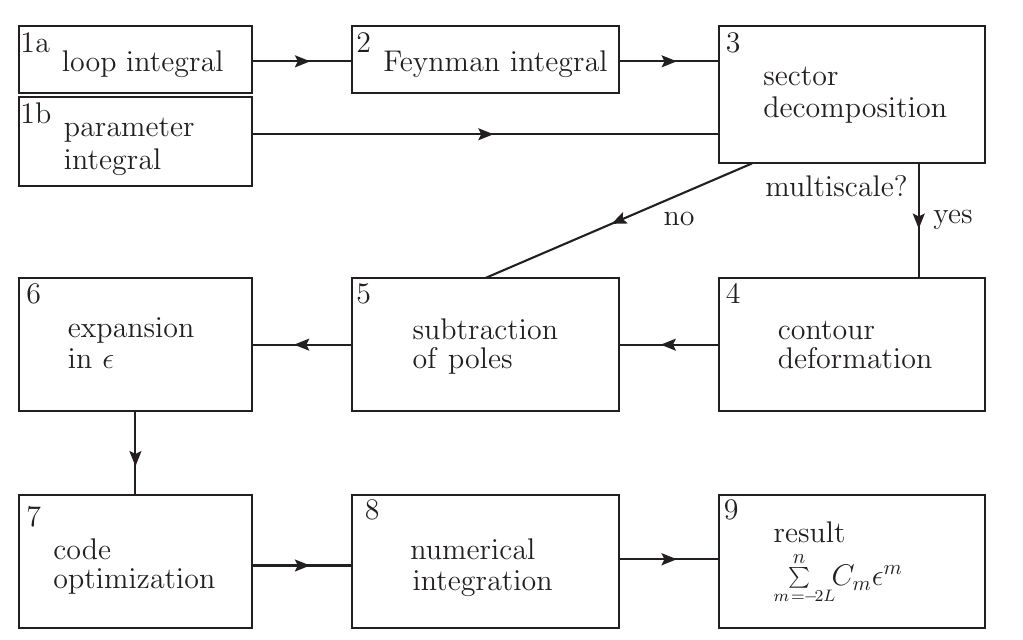}
      \caption{The main building blocks of \textsc{pySecDec}.
        Steps 1 to 6 are executed in Python. \textsc{FORM} is used in step 7 to
        produce optimized {\tt C++} code.}
    \label{fig:flowchart}
\end{figure}

\textsc{pySecDec} has further new features.
\begin{itemize}
\item The evaluation of multiple integrals or even amplitudes is now 
possible, using the generated {\tt C++} library.
\item The treatment of numerators of loop integrals is more flexible. Numerators 
can be defined in terms of contracted Lorentz vectors, inverse
propagators, or a combination of both.

\item The functions can have any number of regulators for endpoint singularities, 
not only the dimensional regulator $\epsilon$.
\item The inclusion of additional functions that do not enter the 
decomposition has been facilitated and extended.
\item A symmetry finder~\cite{2013arXiv1301.1493M} has been implemented, which can detect 
isomorphisms between sectors.
\item A procedure has been implemented to detect and remap singularities
at $x_i = 1$ that result from special kinematic configurations.
\item The  treatment  of  poles  that  are  higher  than  logarithmic  has  
been improved.
\item If the propagators are given in terms of an edge list, the resulting 
loop diagram can be drawn using neato~\cite{graphviz}. This functionality is optional and the program
runs normally if neato is not installed.
\item The  distinction  between `general  functions'  and  `loop  integrands'
is removed, in the sense that all features are available for both loop
integrals and general polynomial functions (as far as they make sense
outside the loop context).
\item Complete re-structuring and usage of open-source software only.
\end{itemize} 

More details on the usage are found in Refs.~\cite{Borowka:2017idc} and 
\cite{Borowka:2017esm}.

\subsection{Accurate approximation using the \textsc{TayInt} approach}
\label{sec:sb:tayint}

To shorten the evaluation times with respect to a fully numerical evaluation, the 
results must be algebraic in the kinematic parameters (Mandelstam invariants, 
external and internal particle 
masses). Then the evaluation at each kinematic 
point takes just as long as the time needed for the insertion of 
the numerical values. 
Algebraic results can be obtained if the integrands are Taylor expanded in the 
Feynman parameters. To ensure that the approximation 
is quickly converging in all regions of parameter space, each integrand must be 
manipulated so that it is 
in a form optimized for a Taylor expansion. This task and the expansion itself 
is performed by \textsc{TayInt} \cite{Borowka:2018dsa}, a program 
to analytically approximate loop integrals. Its algorithm can be summarized as follows.
\begin{enumerate}
\item Input an integral.
\item Reduce the integral to a quasi-finite basis, introduced in 
Refs.~\cite{Panzer:2014gra,vonManteuffel:2014qoa}, 
such that the divergences are in the coefficient of the simplest 
integrals. An automated script using the libraries of the publicly available 
program {\tt Reduze} \cite{vonManteuffel:2012np,vonManteuffel:2014qoa,Bauer:2000cp} 
performs this.
\item For those basis integrals that are not known in analytical form, carry out 
a decomposition into subsectors with smoother integrands. These are obtained using the publicly available program 
\textsc{SecDec 3} \cite{Carter:2010hi,Borowka:2012yc,Borowka:2015mxa,Borowka:2017idc}, 
without its contour deformation option. The subsector integrands are analytical 
within the integration region, but may contain integrable singularities 
over thresholds and at endpoints.  
\item Use a conformal mapping to move the singularities outside of the region 
of integration as far away as is possible. This is done using \textsc{Mathematica} \cite{Wolfram}. 
\item \begin{enumerate} 
\item To produce a result valid below the kinematic thresholds, the integrand 
is Taylor expanded, and integrated over the Feynman parameters. 
This is all done using FORM \cite{Vermaseren:2000nd,Kuipers:2013pba}.
\item To produce a result valid above thresholds, a separate algorithm is 
implemented in \textsc{Math\-ematica}. The subsectors are 
first mapped onto the complex half plane. The algorithm then determines which 
configuration to use for each sector, that is, which contour orientation to use 
for the multiple variable integration and how to partition the subsequent 
region into smaller pieces. The Taylor expansion and integration are then 
performed on the new integrands specified by \textsc{TayInt}. 
\end{enumerate}
\end{enumerate}

The method and its applications are described in comprehensive detail in Ref.~\cite{Borowka:2018dsa}. 
Among the applications is the prediction of results for the finite I39 integral, see \Fref{fig:I39}, 
which is, as yet, not available in the literature and contributes, \eg to Higgs$+$jet production 
at the NLO. Its coefficient at finite order is given in \Fref{fig:I39OTE0funcplot}. Results for 
higher orders in the dimensional regulator $\epsilon$ are found in Ref.~\cite{Borowka:2018dsa}.

\begin{figure}
\centering     
\includegraphics[height=30mm]{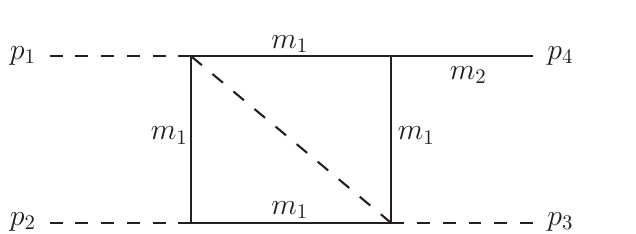}
\caption{The box type integral I39 is so far unknown analytically. Dashed 
lines indicate massless and solid lines massive propagators.}
\label{fig:I39}
\end{figure}

\begin{figure}
\centering
\includegraphics[width=1\textwidth]{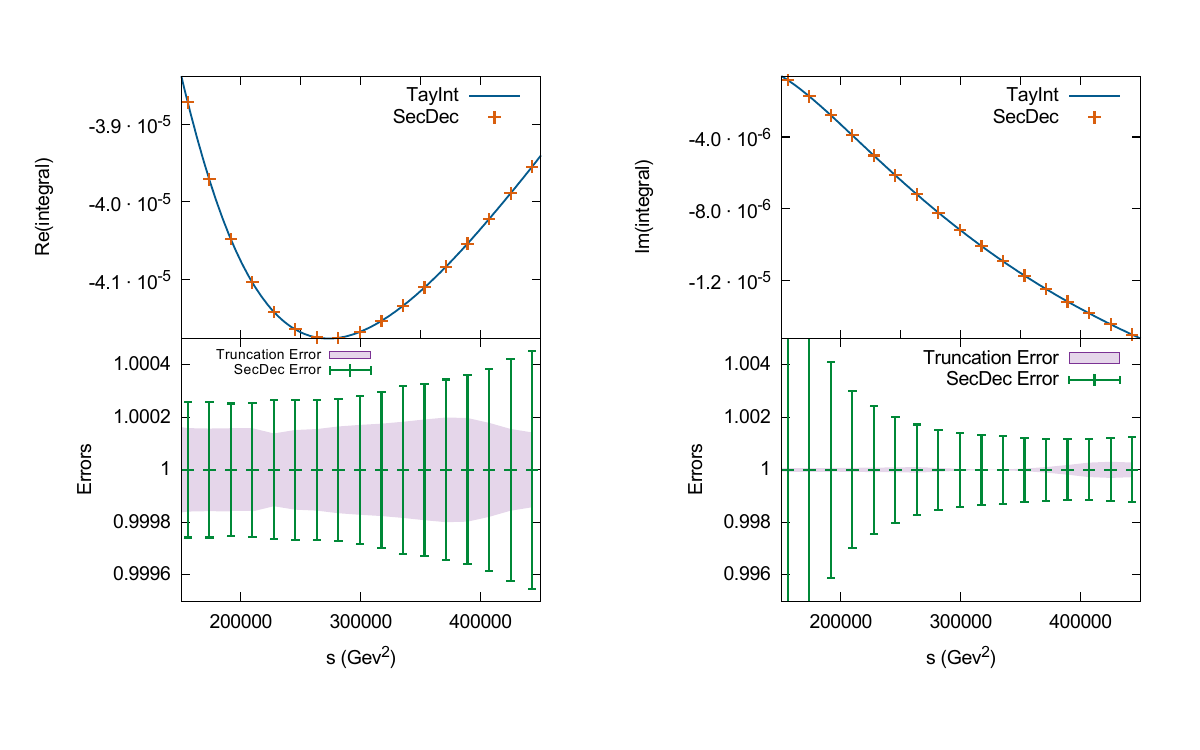}
\caption{The I39 integral calculated at $\mathcal{O}(\epsilon^0)$ with a fourth-order series expansion. The scale $s$ is over the $4m_1^2$ threshold, with $u=-59858 \UGeV^2$, 
$m_2= m_1/\sqrt{2}$ and $m_1=173\UGeV$. The lower plots show the relative \textsc{TayInt} and \textsc{SecDec} uncertainties, respectively.
}
\label{fig:I39OTE0funcplot}
\end{figure}


\subsection{Summary and outlook}
\label{sec:sb:summary}

First, the community-driven public online database for loop integrals, \textsc{Loopedia}, was 
introduced. 
Next, a mini-review on the pragmatic evaluation of multiloop multiscale integrals using 
Feynman parameters was presented, introducing two different methods that do not rely 
on the analytical evaluation when dealing with highly complicated integrals. 
As both approaches are highly algorithmic, the computation of multiloop multiscale integrals 
entering predictions for the FCC-ee can be further automated.

\section*{Acknowledgements}
I want to thank my collaborators, C. Bogner, T. Gehrmann, T. Hahn, G. Heinrich, D. Hulme, S. Jahn,
S.P. Jones, M. Kerner, A. von Manteuffel, M. Michel, E. Panzer, J. Schlenk,
and V. Papara, for the fruitful collaboration.

 \label{sec-borowka}
\clearpage
\pagestyle{empty}
\cleardoublepage

\cleardoublepage

\section
[\ar{} -- construction of Mellin--Barnes integrals for two- and three-loop Z boson vertices
\\ {\it I. Dubovyk, J. Gluza, T. Riemann}
]
{\ar{} -- construction of Mellin--Barnes integrals for two- and three-loop Z boson  decays
} 
\label{contr:mbambre}

\pagestyle{fancy}
\fancyhead[LO]{}
\fancyhead[RO]{}
\fancyhead[CO]{}
\fancyhead[CO]{\thechapter.\thesection 
\hspace{1mm} 
\ar{} -- construction of Mellin--Barnes integrals for two- and three-loop Z boson vertices}
\fancyhead[LE]{}
\fancyhead[CE]{I. Dubovyk, J. Gluza, T. Riemann}
\fancyhead[RE]{} 

\noindent
{\bf Authors: Ievgen Dubovyk, Janusz Gluza, Tord Riemann} 
\\ 
Corresponding author: Ievgen~Dubovyk {[e.a.dubovyk@gmail.com]}
\vspace*{.5cm}

\subsection{Introduction}

\ar{} is {a} project devoted to the representation of Feynman integrals by a finite number 
of Mellin--Barnes integrals ({\mbr} integrals) in $d=n-2\epsilon$ dimensions, $n \in \mathbb{Z}$. 
It includes (i) automatic construction of planar and  non-planar diagrams with up to three loops and 
(ii) treatment of tensor integrals.
   
{Descriptions} of the \ar{} project with details and examples can be found in Refs. \cite{ambrewww,
Gluza:2007rt,Gluza:2010rn,Gluza:2007bd,Gluza:2009mj,Gluza:2010mz,Dubovyk:2016zok,Dubovyk:2016ocz}. 
Two basic strategies are realized in {the} construction of \mbr; these are called loop-by-loop (\la) and global (\ga) approaches.

In practice,  a choice between \la{} and \ga{} strategies for a given integral aiming at the lowest 
\mb{} dimensionality of integrals seems not to be unique. For instance,  in {the case of a } massless non-planar two-loop {vertex}, a minimal fourfold \mb{} representation was derived, starting from the
global Feynman parameter representation {in Ref.} \cite{Tausk:1999vh}. Conversely, in the massive case, 
it appears that  \la{} is more efficient; {an}  eightfold \mb{} representation can be obtained \cite{Heinrich:2004iq}. 
In this case, {a most optimal}  tenfold \mb{} representation {can be obtained} if we start from global Feynman parameters \cite{Heinrich:2004iq}. As discussed there  and in Ref. \cite{Czakon:2007wk}, those representations may also differ 
 concerning the regularization of singularities. This is another subtlety that shows up when constructing \mb{} representations in real applications.  

Taking these examples into account, it is clear why it is not easy to obtain a general and efficient 
program for {the}  construction of a broad class of \mb{} representations. 
Next we give a short {overview on the} procedure. 

\subsection{\ar{} how-to: present status}

Figure \ref{scheme1} shows the flow of operations; {the} present versions of {the} corresponding software 
for {the} generation of \mb{} integrals and the numerical solutions are indicated.

The procedure starts from identification of the diagram's topology. Based on this, an appropriate version of the \ar{} program is used. 
Drawing and topology identification of the diagrams  is based on the knowledge of propagators (masses and momenta flow) and can be 
made using  \pltest{} \cite{Bielas:2013rja}. This seems to be trivial, but
 in the case of more loops it is sometimes  not 
easy to determine the planarity of the diagram by eye and an automation is desired.
An example is given in \Fref{figsm}. It is planar, though it was identified as non-planar in Ref. \cite{Smirnov:2015mct}. 

After \mb{} construction, an analytic $\epsilon$ continuation in $d=4-2\epsilon$ dimensions can be made 
using  \mbm{} or {\tt MBresolve.m}{} \cite{Czakon:2005rk,Smirnov:2009up}. Then a further optimization 
beyond what is already done by \ar{} can be made using the {\texttt{barnesroutines}{}} from the MBtools web 
page \cite{mbtools}. Finally, numerical calculations can be made using \mbm{} or \mbnum{}.

\begin{figure}
\centering
\begin{tikzpicture}
 \node[rectangle, inner sep=10pt, draw=black!50, fill=black!20] (A)
{
Input integral
};

\node[rectangle, inner sep=5pt, align=left, draw=black!50, fill=black!20] (B) [right=of A]
{
Diagram analysis:\\ 
PlanarityTest \cite{Bielas:2013rja}
};

\node[rectangle, inner sep=5pt, align=left, draw=black!50, fill=black!20] (C) [right=of B]
{
\mbr{} construction: \\
\ar{} software \\
\cite{Gluza:2010v22, Gluza:2010rn, Dubovyk:201509v30x,  Dubovyk:2015yba}
};

\node[rectangle, inner sep=5pt, align=left, draw=black!50, fill=black!20] (D) [below=of C]
{
$\epsilon$ continuation: \\
MB.m \cite{Czakon:2005rk} \\
MBresolve.m \cite{Smirnov:2009up}
};

\node[rectangle, inner sep=5pt, align=left, draw=black!50, fill=black!20] (E) [left=of D]
{
Optimization of output: \\
barnesroutines.m \cite{mbtools-kosower}
};

\node[rectangle, inner sep=5pt, align=left, draw=black!50, fill=black!20] (F) [left=of E]
{
Numerical integration: \\
MB.m \\
MBnumerics.m \cite{Usovitsch:2018shx}
};

\draw [->] (A) to (B);
\draw [->] (B) to (C);
\draw [->] (C) to (D);
\draw [->] (D) to (E);
\draw [->] (E) to (F);

\end{tikzpicture}

\vspace{3mm}
\caption{ \label{scheme1}Operational sequence of the \mbr{} suite. The flowchart shows the main steps and the corresponding software for 
 producing a Mellin--Barnes representation for a Feynman integral and performing its numerical {evaluation}.}
\end{figure}
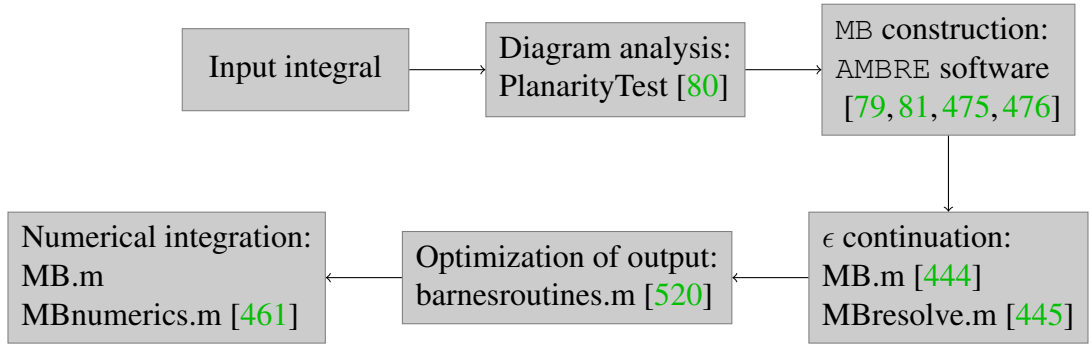
 
\begin{figure}
  \centering
  \includegraphics[scale=0.65]{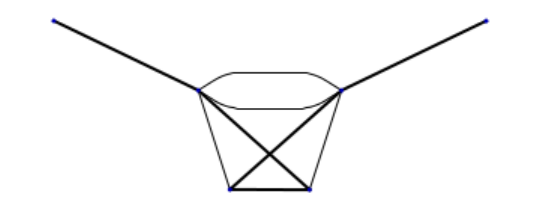}
\caption{Example of a planar four-loop diagram \label{figsm} }
\end{figure}
  
For planar diagrams, the  \la{} method is kept, independently of the number of loops involved.  
However, we would like to make clear that, in general, the construction of \mb{} integrals beyond 
three loops results in high-dimensional \mb{} integrals. This is a natural limitation 
of the \mb{} method, 
as it is used in the \ar{} project.\footnote{For one-loop integrals, a way to reduce the number of {\mbr} dimensions is given in Refs. \cite{Bluemlein:2017rbi,Riemann:April2018}, see  Section \ref{chmt}.\ref{sec-1loop}.}
On top of that, both methods are not necessarily optimal, if more masses and 
legs are involved. This is a bottleneck of the present automatic method. In special cases, the final choice 
may depend on the particular physical problem to be solved.  This is why the  user has, in addition, an opportunity 
to improve the factorization of the $U$ and $F$ polynomials manually and can continue either either \la{} (using {\texttt{AMBRE} version 1.3.1) or \ga{}.  
For two-loop non-planar instances we proceed in a fully automatic way  using  \ga{}. 
In this case, independent of the diagram types, the $U$ polynomials are unique. 
In three-loop cases, we divide the procedure, depending on the subgraphs involved.  

In summary, we  use the following methods and corresponding \ar{} software:
\begin{itemize}
 \item iteratively to each subloop -- \la{} 
       ({\texttt{AMBRE} version 1.3.1} and {\texttt{AMBRE} version 2.1.1});
 
 \item in one step to the complete $U$ and $F$ polynomials -- \ga{} 
       ({\texttt{AMBRE} version 3.1.1});

 \item a combination of these methods -- hybrid approach (under development -- {\texttt{AMBRE} version 4}). 
\end{itemize}

Examples, descriptions, and links to basic tools and literature can be found in Ref. \cite{ambrewww}.
Let us refer to other recent approaches,  presented in Section D.\ref{contr:rboels} and the method of brackets presented in Section E.\ref{contr:mprausa}.
 

We would like to mention some important facts, which were already described on a few occasions on the  \ar{} website \cite{ambrewww}, in
publications, and in a series of talks.
\begin{itemize}
\item For \la{}, the momentum flow {\color{black} really} matters. 
\item The first and second Barnes lemmas are key ingredients. For instance, 
  $-s x_1 x_2 - s x_1 x_4 + \dots = -s x_1 ( x_2 + x_4 ) + \dots 
  $ is equivalent to the first Barnes lemma \cite{Gluza:2007rt}.
\item \la{} may also work  for non-planar diagrams. For instance, for the diagram in \Fref{npex},  
  \ar{} version 2.1.1 gives a six-dimensional \mb{} integral, while \ar{}
  version 3.1.1  gives a 13-dimensional \mb{} integral. 
  However, the problem here is that, even in Euclidean kinematics, highly oscillatory factors may appear, \eg  $(-s)^z s^z$.
   
\item \ga{} works for both planar and non-planar diagrams. However,  less-dimensional \mb{} 
  integrals are often expected when using \ar{} packages tuned for \la{}.
\end{itemize}
  
  \begin{figure}
  \begin{center}
  \includegraphics[scale=0.45]{
  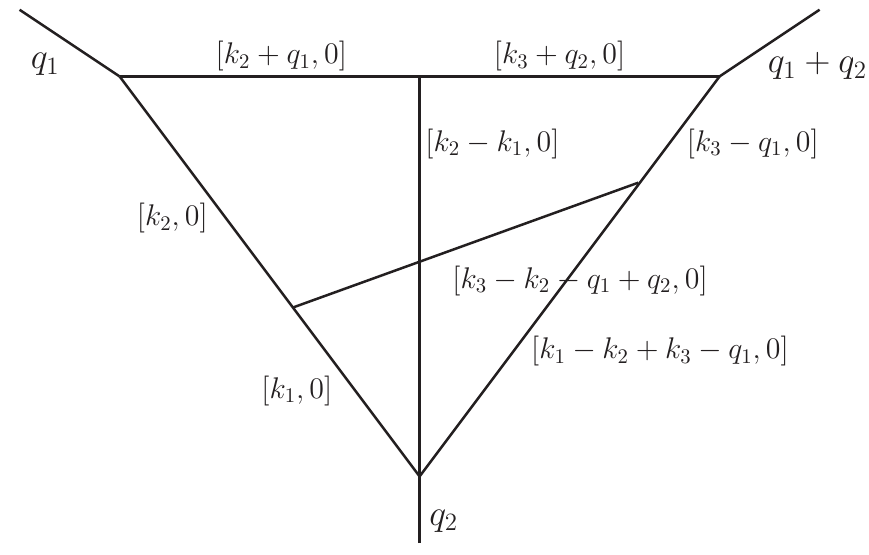}
  \end{center}
\caption{Example of a three-loop non-planar integral where \la{} gives a less-dimensional \mbr{} integral than \ga{}, 
though its numerical stability can be worse, owing to more complicated kinematic factors; even
in Euclidean kinematic regions. \label{npex} }
\end{figure}


\subsection{Three-loop \ar{} representations} 

To be able to use the \mb{} method efficiently for FCC-ee calculations of EWPOs at three loops,  further tuning 
of the packages is necessary. 
It appears that, at the three-loop level, there are only two basic topologies 
from which all physical diagrams can be obtained \cite{Cvitanovic:1974uf}. These 
are shown in \Fref{fig:skel}. 
To obtain any three-loop diagram, one must attach a corresponding amount of external
legs to lines or vertices of the skeleton diagram.

\begin{figure}
\centering
\includegraphics[scale=0.4]{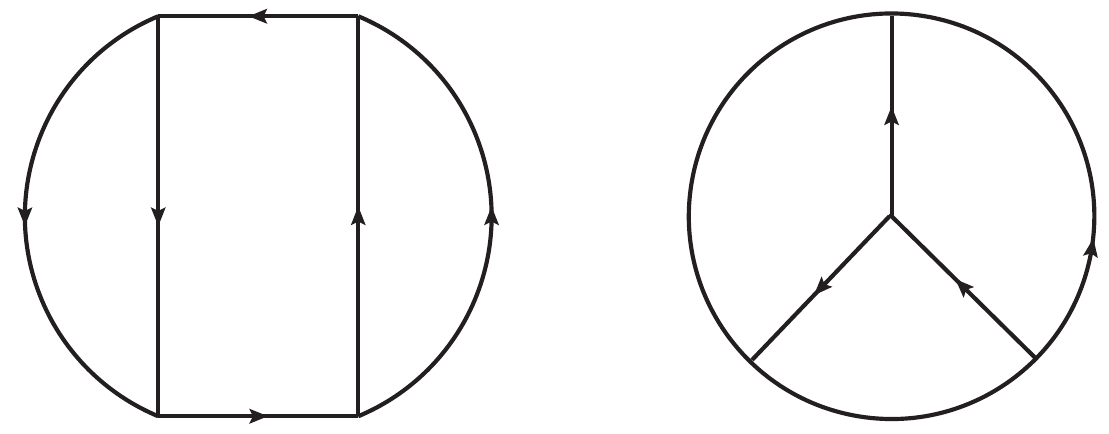}
\caption{Basic skeleton-generating diagrams for three-loop topologies \cite{Cvitanovic:1974uf} 
\label{fig:skel}}
\end{figure} 


What we need in the first instance, in the FCC-ee case, are three-loop vertex  diagrams.
The diagram on the left side in \Fref{fig:skel} and all its derivatives can be divided into two one-loop pieces, see \Fref{fig:case1}.
Vertex diagrams of this type have a planar subloop and can be treated using the hybrid approach, see the next section.

\begin{figure}
  \centering
  \includegraphics[scale=0.5]{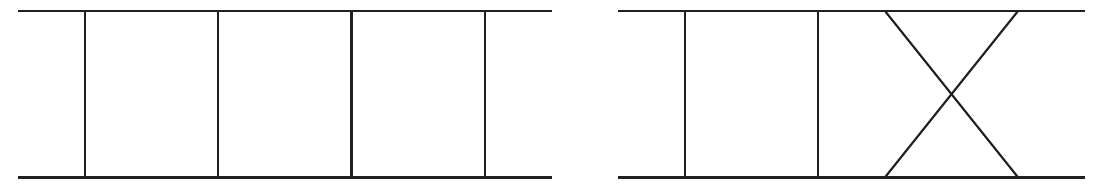}
\caption{Three-loop topologies generated from the left skeleton diagram in \Fref{fig:skel} \label{fig:case1} }
\end{figure} 

The skeleton diagram on the right side in \Fref{fig:skel} generates the most complicated non-planar topologies, 
see \Fref{fig:case2}. Here, the more advanced \ga{} is needed.

\begin{figure}
\centering
\includegraphics[scale=0.5]{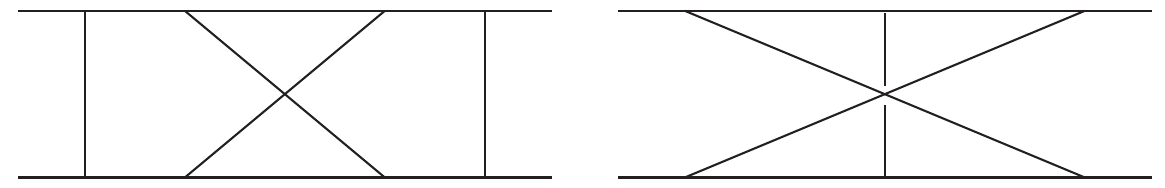}
\caption{Three-loop topologies generated from the right skeleton diagram in \Fref{fig:skel} \label{fig:case2} }
\end{figure}

As in the two-loop case, after a proper transformation of variables, all child diagrams have the same $U$ function, of the form
\begin{multline}
U = v_1 v_2 v_3+v_1 v_2 v_4+v_1 v_3 v_4+v_1 v_2 v_5+v_1 v_3 v_5+v_2 v_3 v_5+v_2 v_4 v_5+v_3 v_4 v_5  \\
    + v_1 v_2 v_6 +v_2 v_3 v_6+v_1 v_4 v_6+v_2 v_4 v_6+v_3 v_4 v_6+v_1 v_5 v_6+v_3 v_5 v_6+v_4 v_5 v_6. \label{eqcase2}
\end{multline}
In the next step, \Eref{eqcase2} can be simplified using the Cheng--Wu theorem \cite{Cheng:1987ga, Smirnov:2002,Blumlein:2014maa,Dubovyk:2015yba},
keeping three variables in the common $\delta$-function:
\begin{equation}
 \int^{\infty}_0 \mathrm{d} v_2 \mathrm{d} v_3 \mathrm{d} v_4 \int_0^1 \mathrm{d} v_1 \mathrm{d} v_\mathrm{d} d v_6 \delta (1 - v_1 - v_5 - v_6 ) \, ,
\end{equation}
which gives
\begin{equation}
 U_\mathrm{CW} = v_2 v_3 + v_2 v_4 + v_3 v_4 + v_1 v_2 v_5 + v_1 v_3 v_5 + 
           v_1 v_2 v_6 + v_1 v_4 v_6 + v_1 v_5 v_6 + v_3 v_5 v_6 + v_4 v_5 v_6
.
\end{equation} 
This form is not unique; the variables in the common $\delta$-function can be chosen in four different ways. 
One  possibility for performing an integration from $0$ to $\infty$ is to use the following factorization trick:
\begin{equation}
 U_\mathrm{CW} = v_2 (v_3 + v_4 + v_1 v_5) + v_3 (v_4 + v_1 v_5) + 
           v_1 v_6 (v_2 + v_5) + v_4 v_6 (v_1 + v_5) + v_3 v_5 v_6
           .
\label{u_factor}           
\end{equation}
There are six different ways to get four terms in $U$; altogether, we have 24 variants for
applying the Cheng--Wu theorem and  factorizing $U$. The final choice is based, first, on the minimization of the number of terms in $F$ and, second,
on the presence of  the same factorization patterns as in \Eref{u_factor} for $U$, as well as in $F$.
Finally, the $U$ polynomial can be reduced to four  \mb{} terms. 
As a rule of thumb, \ga{}   usually gives optimal representations if, from the beginning, $\mbox{length($U$)} \leq \mbox{length($F$)}$.   

\subsubsection{The mixed three-loop approach: an example}

At three loops, we can also approach propagators and subloops combining  \la{} and \ga{}. This depends on whether 
or not a given topology includes a planar subtopology that can be disconnected or not. 

In the first case, \Fref{fig:boxsub}, propagators connected with a planar subloop in the form 
of the one-loop box are  transformed into an \mb{} form in a first step, defining a new effective propagator. 
In this way, an effective two-loop diagram is created. In the second case, \Fref{fig:noboxsub}, a non-planar 
disconnected subgraph can be identified. In this case, the first step is to take five propagators 
of the two-loop connected propagators, leading in a straightforward  way to an effective one-loop topology.

\begin{figure}
\centering
\includegraphics[scale=0.55]{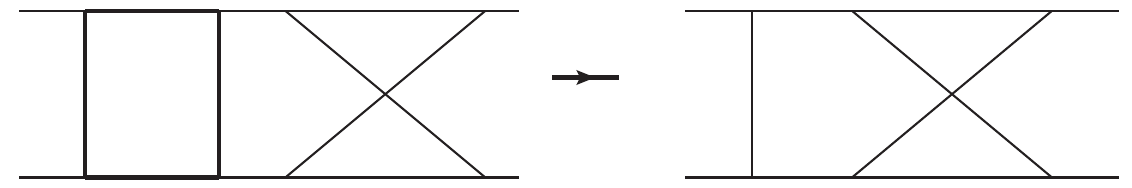}
\caption{Transforming propagators of the one-loop box at the left side  into an effective propagator that 
changes the whole diagram into a two-loop topology. \label{fig:boxsub} }
\end{figure} 

\begin{figure}
\centering
\includegraphics[scale=0.55]{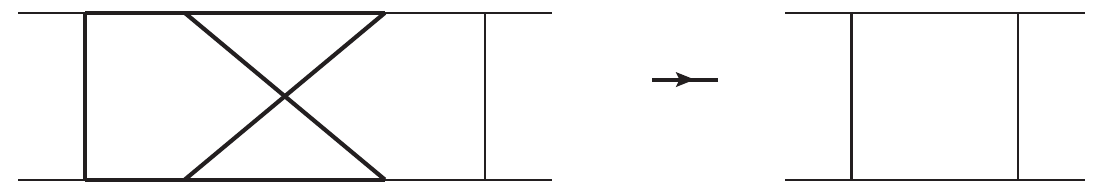}
\caption{Transforming five propagators of the chosen two-loop subdiagram into an
effective propagator leading directly to the one-loop topology. \label{fig:noboxsub} }
\end{figure}  

These strategies are realized in the new \ar{} version 4.0 package, which is under development. 

Next, as an example,  the  
 \ar{} output  for the momenta flow, as given 
in \Fref{fig:3loopex}, is shown. 
Some parts of the output are omitted for simplicity.
\begin{verbatim}
MBrepr[{1},{PR[k1, 0, n1]      PR[-k1 + k2, 0, n2] PR[k2, 0, n3]
            PR[k2 + q1, 0, n4] PR[k3 + q2, 0, n5]  PR[k3 - q1, 0, n6] 
            PR[-k2 + k3 - q1 + q2, 0, n7] PR[k1 - k2 + k3 - q1, 0, n8]},
                                {k1,{k2,k3}}];
\end{verbatim}
First, the integration is performed over the \texttt{k1} subloop. In \Fref{fig:3loopex}, this is shown 
by a thick line:
\begin{verbatim}
--iteration nr: 1 with momentum: k1
\end{verbatim}
\begin{verbatim}
--integral: PR[k1, 0, n1] PR[-k1 + k2, 0, n2] PR[k1 - k2 + k3 - q1, 0, n8]
\end{verbatim}
Coefficients in the $F$ polynomial can be represented in form of propagators, as in the case of  \la{}:
\begin{verbatim}
   F polynomial during this iteration 
-PR[k2,0] X[1] X[2]-PR[k2-k3+q1,0] X[1] X[3]-PR[k3-q1,0] X[2] X[3]
\end{verbatim}
In the next step, the  integration is performed over the remaining two-loop momenta \texttt{k2} and \texttt{k3},
using  \ga{}: 
\begin{verbatim}
--iteration nr: 2 with momentum: k2, k3
\end{verbatim}
\begin{verbatim}
--integral: PR[k2, 0, n3 + z1] 
            PR[k3 - q1, 0, n6 - z1 - z2 - n1 - n2- n8 + d/2] 
            PR[k2 + q1, 0, n4]            PR[k3 + q2, 0,n5] 
            PR[-k2 + k3 - q1 + q2, 0, n7] PR[k2 - k3 + q1, 0, z2]
\end{verbatim}
Propagators that remain after the first integration, together with those appearing in $F$, form the two-loop non-planar vertex.
Now it is easily seen that the final \mb{} representation is four-dimensional.

\begin{figure}
\centering
\includegraphics[scale=0.45]{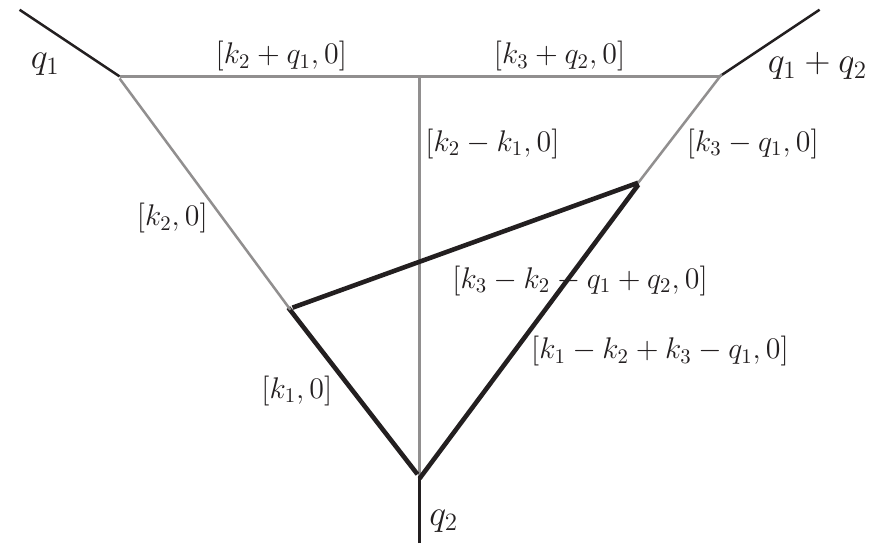}
\caption{First integration over momentum $k_1$ in the three-loop diagram. \label{fig:3loopex} }
\end{figure}

\subsubsection{Comparisons with the method of brackets}

Recently, a very interesting new approach to the construction of \mbr{} representations  was introduced in Ref. 
\cite{Prausa:2017frh}. Here, we compare some basic results from this method of brackets 
(see Section E.\ref{contr:mprausa} for details) with results by \ar{} version
4.0, in which the previously discussed 
 three-loop non-planar cases were optimized.
 

From \Fref{fig:ambre} and  \Tref{tab:comp}, it is clear that the new version of \ar{} is much better optimized towards 
three-loop Z boson studies than the previous public version {\texttt{AMBRE} version 3, which is discussed in Ref. \cite{Prausa:2017frh}.

\begin{figure}
      \centering
      \begin{subfigure}[b]{.19\textwidth}
        \centering
        \includegraphics[scale=.9]{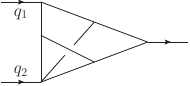}
        \caption{}
        \label{fig:int1}
      \end{subfigure}%
      \begin{subfigure}[b]{.19\textwidth}
        \centering
        \includegraphics[scale=.9]{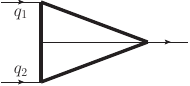}
        \caption{}
        \label{fig:int2}
      \end{subfigure}%
      \begin{subfigure}[b]{.19\textwidth}
        \centering
        \includegraphics[scale=.9]{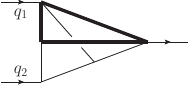}
        \caption{}
        \label{fig:int3}
      \end{subfigure}%
      \begin{subfigure}[b]{.19\textwidth}
        \centering
        \includegraphics[scale=.9]{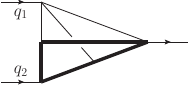}
    \caption{}
        \label{fig:int4}
      \end{subfigure}%
      \begin{subfigure}[b]{.19\textwidth}
        \centering
        \includegraphics[scale=.9]{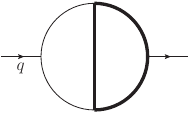}
        \caption{}
        \label{fig:int5}
      \end{subfigure}%
      \\
      \begin{subfigure}[b]{.19\textwidth}
        \centering
        \includegraphics[scale=.9]{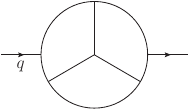}
        \caption{}
        \label{fig:int6}
      \end{subfigure}
      \begin{subfigure}[b]{.19\textwidth}
        \centering
        \includegraphics[scale=.9]{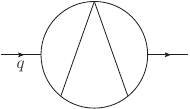}
        \caption{}
        \label{fig:int7}
      \end{subfigure}%
      \begin{subfigure}[b]{.19\textwidth}
    \centering
        \includegraphics[scale=.9]{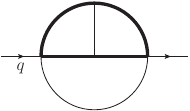}
        \caption{}
        \label{fig:int8}
      \end{subfigure}%
      \begin{subfigure}[b]{.19\textwidth}
        \centering
        \includegraphics[scale=.9]{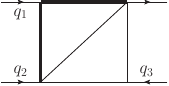}
        \caption{}
        \label{fig:int9}
      \end{subfigure}%
      \begin{subfigure}[b]{.19\textwidth}
        \centering
        \includegraphics[scale=.9]{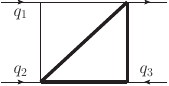}
        \caption{}
        \label{fig:int10}
      \end{subfigure} 
      \caption{ \label{fig:ambre}
        Examples of two- and three-loop diagrams. Bold and thin lines represent
massive and massless propagators, respectively. Taken from Ref.\cite{Prausa:2017frh} (Creative Commons Attribution 4.0) and
discussed in \Tref{tab:comp}.}
    \end{figure}

\begin{table}
\caption{Number of {\mbr} integrations of the representation constructed by the method of brackets, 
compared with  previous \ar{} versions ~\cite{Gluza:2007rt,Gluza:2010rn,Blumlein:2014maa,Dubovyk:2016ocz} 
and the latest version, \ar{} version  4.   \label{tab:comp}}
      \centering
      \begin{tabular}{lllll} \hline \hline
        Diagram & Method of brackets & \texttt{AMBRE} & Planarity & \ar{} 4$^*$ \\ \hline
        \Fref{fig:ambre}(a)  & \textbf{7} & 13  & Non-planar & \textbf{4} \\
        \Fref{fig:ambre}(b)  & \textbf{1} & \phantom{1}2 & Planar & \textbf{1} \\
        \Fref{fig:ambre}(c)  & \textbf{7} & \phantom{1}9  & Non-planar & \textbf{5} \\
        \Fref{fig:ambre}(d)  & \textbf{7} & \phantom{1}8  & Non-planar & 8 \\
        \Fref{fig:ambre}(e)  & 5 & \phantom{1}\textbf{3}  & Planar & \textbf{3} \\
        \Fref{fig:ambre}(f)  & 9 & \phantom{1}\textbf{4}  & Planar & \textbf{4} \\
        \Fref{fig:ambre}(g)  & 7 & \phantom{1}\textbf{4} & Planar & \textbf{4} \\
        \Fref{fig:ambre}(h)  & 5 & \phantom{1}\textbf{4} & Planar & \textbf{4}\\
        \Fref{fig:ambre}(i)  & \textbf{2} & \phantom{1}\textbf{2}  & Planar & \textbf{2} \\
        \Fref{fig:ambre}(j) & \textbf{2} & \phantom{1}\textbf{2}  & Planar & \textbf{2} \\
        \hline \hline
      \end{tabular}        
\end{table}

\subsection{Conclusions and outlook}
 
 \begin{enumerate}
  \item The dimensionality of {\mbr} representations depends strongly  on topology, number of legs and loops, and internal and external masses.
  \item The {\texttt{AMBRE}} software is based on two different approaches: \\
  \la{} -- general planar and some non-planar diagrams; \\
  \ga{} -- two-loop planar and non-planar, three-loop non-planar diagrams with massless external legs.
  \item The new \texttt{AMBRE} version 4, which is not yet public, combines all the advantages of these methods; 
  it has been constructed and  preliminary tests have been conducted. It is optimized for FCC-ee precision three-loop Z boson studies.
 \end{enumerate}

 \label{sec-mbambre}
\clearpage
\pagestyle{empty}
\cleardoublepage

\section
[Mellin--Barnes meets method of brackets \\ {\it M. Prausa}]
{Mellin--Barnes meets method of brackets} \label{contr:mprausa}

\pagestyle{fancy}
\fancyhead[RO]{}
\fancyhead[LO]{}
\fancyhead[CO]{\thechapter.\thesection 
\hspace{1mm}
Mellin--Barnes meets method of brackets
}
\fancyhead[LE]{}
\fancyhead[CE]{M. Prausa}
\fancyhead[RE]{} 

\noindent
{\bf Author: Mario Prausa}  {~~[prausa@physik.rwth-aachen.de]}
\vspace*{.5cm}

\subsection{Introduction}
The evaluation of Feynman integrals by means of Mellin--Barnes ({\mbr}) representation is one of the most successful techniques in the field of multiloop computation.
Over the years, a large collection of public tools was developed for the numerical as well as analytical calculation of {\mbr} representations~\cite{Czakon:2005rk,Smirnov:2009up,MBasymptotics,Ochman:2015fho}.
An important criterion for the applicability of all of these tools is a rather small dimensionality of the {\mbr} representation.

The construction of a {\mbr} representation for a Feynman integral is a highly non-trivial task.
The public tool \texttt{AMBRE}~\cite{Gluza:2007rt,Gluza:2010rn,Blumlein:2014maa,Dubovyk:2016ocz} offers two powerful approaches to addressing this problem.
The loop-by-loop approach~\cite{Gluza:2007rt,Gluza:2010rn} is the preferable method for planar Feynman integrals, which, in most cases, leads to a very small dimensionality of the {\mbr} representation.
Non-planar Feynman integrals can be treated using the global approach~\cite{Blumlein:2014maa,Dubovyk:2016ocz}.
A hybrid approach, applicable for non-planar Feynman integrals with planar subgraphs, combining the two previously mentioned methods is presented in Section ~\ref{chmt}.\ref{contr:mbambre}.

While the loop-by-loop approach is a well-developed method for purely planar Feynman integrals, the treatment of non-planar integrals might still have room for improvement.
In this contribution, a different method for the construction of {\mbr} representations of Feynman integrals, published in Ref.~\cite{Prausa:2017frh}, is explained with a non-planar two-loop example.
This technique is applicable for both planar and non-planar topologies and leads, in some cases (especially non-planar ones), to a smaller dimensionality than the \texttt{AMBRE} package.
Our approach is based on the `method of brackets' introduced in Ref.~\cite{Gonzalez:2007ry,Gonzalez:2010,Gonzalez:2010uz}, which is a technique for the construction of multifold sums, starting directly from a Schwinger-parametrized Feynman integral.

\subsection{The method of brackets}
\label{sect:mbbrackets:mob}
In this section, we introduce the method of brackets developed by Gonzalez
\textit{et al.} in Refs.~\cite{Gonzalez:2007ry,Gonzalez:2010,Gonzalez:2010uz} and motivate our modifications (see Ref.~\cite{Prausa:2017frh}) in order to obtain {\mbr} representations instead of multifold sums.

If a function $g(x)$ has a formal power series
\[
    g(x)
    =
    \sum\limits_{n=0}^\infty
    G(n)
    \frac{(-x)^n}{n!}
    \,,
\]
the integral
\[
    \int\limits_0^\infty \mathrm dx\;
    x^{\alpha-1}
    g(x)
    =
    \Gamma(\alpha)
    G(-\alpha)
\]
holds true.
This is known as Ramanujan's master theorem, which is the basis for the method of brackets formulated in Refs.~\cite{Gonzalez:2007ry,Gonzalez:2010,Gonzalez:2010uz}.
In these references, a special notation
\begin{equation} \label{eq:mbbrackets:bracket}
    \bracket{\alpha} 
    \equiv
    \int\limits_0^\infty \mathrm dx\;
    x^{\alpha-1}\,,
\end{equation}
denoted the `bracket', is introduced in order to write this theorem in a more compact way,
\begin{equation} \label{eq:mbbrackets:ramanujan}
    \sum\limits_{n=0}^\infty
    \frac{(-1)^n}{n!}
    G(n)
    \bracket{n+\alpha}
    =
    \Gamma(\alpha)
    G(-\alpha)
    \,.
\end{equation}
In particular, this notation is useful to express the generalization of Ramanujan's theorem to multifold sums
\[
    \sum\limits_{n_1=0}^\infty
    \hspace{-3pt}
    \cdots
    \hspace{-2pt}
    \sum\limits_{n_K=0}^\infty
    \frac{(-1)^{n_1+\cdots+n_K}}{n_1!\cdots n_K!}
    G(\vec n)\;
    \bracket{\beta_1 + \vec\alpha_1 \cdot \vec n}
    \cdots
    \bracket{\beta_K + \vec\alpha_K \cdot \vec n}
    =
    \frac1{|\det(A)|}
    \Gamma(-n_1^*)
    \cdots
    \Gamma(-n_K^*)
    G(\vec n^{\;*})
    \,,
\]
where $A=(\vec\alpha_1,\cdots,\vec\alpha_K)^\mathrm{T}$, $\vec n = (n_1,\cdots,n_K)^\mathrm{T}$, $\vec n^{\;*} = -A^{-1} \vec\beta$, and $\vec\beta = (\beta_1,\cdots,\beta_K)^\mathrm{T}$.
This is the master formula for the original method of brackets.

The idea behind the original method of brackets is that the integrand of a Schwinger-parametrized Feynman integral can be formally Taylor expanded in the Schwinger parameters.
The Schwinger parameter integrals can then be written as \emph{brackets}, which makes the master formula applicable to some of the sums that originate from the Taylor expansion.

The first step to modify the method of brackets so that it yields {\mbr} representations instead of multifold sums is to find a new master formula in which the sums are replaced by contour integrals.
The Mellin inversion theorem provides exactly what is needed for this step.
Using the bracket notation, the inversion theorem can be written as
\[
    \int\frac{\mathrm dz}{2\pi \mathrm i} F(z) \bracket{z+\alpha} = F(-\alpha)\,,
\]
which resembles Ramanujan's theorem (\Eref{eq:mbbrackets:ramanujan}).
The generalization to multidimensional {\mbr} integrals, which will be the master formula for our modified method of brackets, reads
\begin{equation} \label{eq:mbbrackets:master}
    \int\frac{\mathrm dz_1}{2\pi\mathrm i}
    \hspace{-1pt}
    \cdots
    \hspace{-2pt}
    \int\frac{\mathrm dz_K}{2\pi\mathrm i}
    F(\vec z)
    \;
    \bracket{\beta_1 + \vec\alpha_1 \cdot \vec z}
    \cdots
    \bracket{\beta_K + \vec\alpha_K \cdot \vec z}
    =
    \frac1{|\det A|}
    F(-A^{-1} \vec\beta)
    \,,
\end{equation}
where $\vec z = (z_1,\cdots,z_K)^\mathrm{T}$.

The method of brackets consists of a small set of simple rules on how to rewrite a Schwinger-parametrized integral in a form to which this master formula can be applied.
Instead of describing these rules in their most general form (as is done in Ref.~\cite{Prausa:2017frh}), we follow a more didactic way in this contribution, \ie  we will apply the method immediately to a non-trivial two-loop example.

\subsection{The example} \label{sect:mbbrackets:example}

Throughout this contribution, we consider the example in \Fref{fig:mbbrackets:example}, where the thin lines are massless propagators and the bold lines carry mass $m$.
The kinematics are given by
\[
    q_1^2 = q_2^2 = 0\,, \qquad
    (q_1+q_2)^2 = M^2
    \,.
\]
The Schwinger parametrization of this two-scale integral is given by
\begin{equation} \label{eq:mbbrackets:schwinger}
    \frac1{\Gamma(a_1)\cdots\Gamma(a_6)}
    \int\limits_0^\infty \mathrm dx_1\; x_1^{a_1-1}
    \cdots
    \int\limits_0^\infty \mathrm dx_6\; x_6^{a_6-1}
    \;
    \frac{\mathrm{e}^{-F/U-\sum_i x_i m_i^2}}{U^{d/2}}\,,
\end{equation}
where the Symanzik polynomials $U$ and $F$ and the sum over the squared internal masses are
    \begin{align*}
        U &= x_2 x_5 + x_1 x_2 + x_4 x_5 + x_2 x_4 + x_2 x_6 + x_3 x_6 + x_1 x_5 
          + x_1 x_6 + x_3 x_5 + x_1 x_3 + x_4 x_6 + x_3 x_4\,, \\[3pt]
        F &= -M^2 \left[ x_2 x_3 x_5 + x_1 x_3 x_6 + x_1 x_2 x_3 + x_2 x_4 x_5 + x_2 x_3 x_4 + x_2 x_3 x_6 \right]\,, \\[3pt]
        {\textstyle \sum_i m_i^2 x_i} &= m^2\left[x_1 + x_2 + x_3 + x_4\right]\,.
    \end{align*}

\begin{figure}
    \centering
    \includegraphics{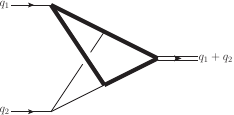}
    \caption{Two-loop non-planar vertex-type Feynman integral}
    \label{fig:mbbrackets:example}
\end{figure}

Our example has a $U$ polynomial with twelve terms, an $F$ polynomial with six terms, four massive lines, and six propagators.
A naive application of the method of brackets would lead to a $12+6+4-6-1 = 15$-dimensional {\mbr} representation.
This number can  easily be derived from the set of rules given in Ref.~\cite{Prausa:2017frh}.
Such a large number of dimensions is most certainly very inconvenient.

\subsection{Optimization} \label{sect:mbbrackets:optimization}
In the context of the original method of brackets, Ref.~\cite{Gonzalez:2007ry} suggests an optimization procedure to be applied to the graph polynomials, which drastically reduces the multiplicity of the sums in the final result.
The same procedure can be used to reduce the dimensionality of the final {\mbr} representations in the modified method of brackets.

The optimization consists of \emph{recursively} identifying common subexpressions in $U$, $F$, and $\sum_i m_i^2 x_i$.
The common subexpressions are then replaced by auxiliary variables.
If a common subexpression with $N$ terms is found to appear $J$ times, this leads to a reduction of the dimensionality by $(J-1)(N-1)$.

In the example, such an optimization leads to
    \begin{align*}
        r_1 &= x_1 + x_4 \,, \\
        r_2 &= x_2 + x_3 \,, \\
        r_3 & = r_1 + r_2 \,, \\
        r_4 &= x_5 + x_6 \,, \\
        U & = r_4 r_3 + r_2 r_1 \,, \\
        F & =     -M^2 \left[r_4 x_2 x_3 + x_1 x_3 x_6 + r_1 x_2 x_3 + x_2 x_4 x_5 \right] \,, \\
        {\textstyle \sum_i m_i^2 x_i} & = m^2 r_3 \,,
    \end{align*}
where the auxiliary variables are denoted  $r_i$.
Here, the recursive nature of the factorization becomes clear, since the auxiliary variable $r_3$ depends on further auxiliary variables $r_1$ and $r_2$.
We will see later on that this optimized form leads to only four {\mbr} integrals in the final result.

\subsection{The rules} \label{sect:mbbrackets:rules}
In this section, we apply a set of four simple rules, derived in Ref.~\cite{Prausa:2017frh}, to the Schwinger-parametrized Feynman integral (\Eref{eq:mbbrackets:schwinger}), in order to rewrite it as a {\mbr} representation. 

\paragraph*{Rule A}
Inserting $F$ and $\sum_i m_i^2 x_i$ in optimized form into \Eref{eq:mbbrackets:schwinger} leads to
\[
    \frac1{\Gamma(a_1)\cdots\Gamma(a_6)}
    \int\limits_0^\infty \mathrm dx_1\; x_1^{a_1-1}
    \cdots
    \int\limits_0^\infty \mathrm dx_6\; x_6^{a_6-1}
    \;
    \frac{\mathrm{e}^{M^2 \left[r_4 x_2 x_3 + x_1 x_3 x_6 + r_1 x_2 x_3 + x_2 x_4 x_5 \right]/U-m^2 r_3}}{U^{d/2}}\,.
\]
The exponential function in the integrand  now has to be split into factors using $\mathrm{e}^{-\sum_i A_i} = \prod_i \mathrm{e}^{-A_i}$, so that $A_i$ consists only of a monomial or a monomial divided by $U$,
\[
    \frac1{\Gamma(a_1)\cdots\Gamma(a_6)}
    \int\limits_0^\infty \mathrm dx_1\; x_1^{a_1-1}
    \cdots
    \int\limits_0^\infty \mathrm dx_6\; x_6^{a_6-1}
    \;
    \frac{\mathrm{e}^{M^2 r_4 x_2 x_3/U} \mathrm{e}^{M^2 x_1 x_3 x_6/U} \mathrm{e}^{M^2 r_1 x_2 x_3/U} \mathrm{e}^{M^2 x_2 x_4 x_5/U} \mathrm{e}^{- m^2 r_3}}{U^{d/2}}\,.
\]
To each exponential function, the Cahen--Mellin formula
\[
    \mathrm{e}^{-A_i}
    =
    \int
    \frac{\mathrm dz_i}{2\pi \mathrm i}
    A_i^{z_i} \,
    \Gamma(-z_i)
\]
can be applied, which introduces a {\mbr} integral.
This yields
    \begin{multline*} 
            \int\limits_0^\infty \mathrm dx_1\; x_1^{a_1-1}
        \cdots
        \int\limits_0^\infty \mathrm dx_6\; x_6^{a_6-1}
        \int\frac{\mathrm dz_1}{2\pi\mathrm i}
        \cdots
        \int\frac{\mathrm dz_5}{2\pi\mathrm i}
      \left  (-M^2 \right )^{z_{1234}}
       \left (m^2 \right )^{z_5}
        \frac{
        \Gamma(-z_1) 
        \cdots
        \Gamma(-z_5)
        }{\Gamma(a_1)\cdots\Gamma(a_6)}
        \\ \times
        r_1^{z_3} 
        r_3^{z_5}
        r_4^{z_1} 
        U^{-z_{1234}-d/2}
        x_1^{z_2} 
        x_2^{z_{134}} 
        x_3^{z_{123}} 
        x_4^{z_4} 
        x_5^{z_4} 
        x_6^{z_2}\,, 
    \end{multline*}        
with the widely used index notation $z_{ijk\cdots} = z_i + z_j + z_k + \cdots$.
Note that all powers are arranged in such a way that every base appears only once, which is important for later steps.

\paragraph*{Rule B}
Now the Symanzik polynomial $U$, as well as the auxiliary variables $r_4$, $r_3$, $r_2$, and $r_1$, must be reinserted in that order.
This requires a rule to rewrite multinomials raised to some power in terms of {\mbr} integrals and brackets.
Such a rule is given by
\[
    (A_1 + \cdots + A_J)^\alpha
    =
    \frac1{\Gamma(-\alpha)}
    \int\frac{\mathrm dz_1}{2\pi \mathrm i}
    \cdots
    \int\frac{\mathrm dz_J}{2\pi \mathrm i}
    \,
    \bracket{z_1 + \cdots + z_J - \alpha} 
    A_1^{z_1}
    \cdots
    A_J^{z_J}
    \,
    \Gamma(-z_1)
    \cdots
    \Gamma(-z_J)\,.
\]
Inserting $U$ and applying the formula leads to
    \begin{multline*} 
        \int\limits_0^\infty \mathrm dx_1\; x_1^{a_1-1}
        \cdots
        \int\limits_0^\infty \mathrm dx_6\; x_6^{a_6-1}
        \int\frac{\mathrm dz_1}{2\pi\mathrm i}
        \cdots
        \int\frac{\mathrm dz_7}{2\pi\mathrm i}
        \left(-M^2 \right)^{z_{1234}}
       \left (m^2 \right )^{z_5}
        \bracket{z_{123467}+d/2}
        \\ 
        \times
        \frac{
        \Gamma(-z_1) 
        \cdots
        \Gamma(-z_7)
        }{\Gamma(a_1)\cdots\Gamma(a_6)\Gamma(z_{1234}+d/2)}
        r_1^{z_{37}} 
        r_2^{z_7}
        r_3^{z_{56}}
        r_4^{z_{16}} 
        x_1^{z_2} 
        x_2^{z_{134}} 
        x_3^{z_{123}} 
        x_4^{z_4} 
        x_5^{z_4} 
        x_6^{z_2}
        \,.
    \end{multline*}        
Note that the powers had to be arranged properly, in the same way as in Rule~A.

Repeating this step for $r_4$, $r_3$, $r_2$, and $r_1$ yields
 \scriptsize
    \begin{multline*} 
        \int\limits_0^\infty \mathrm dx_1
        \cdots
        \int\limits_0^\infty \mathrm dx_6
        \int\frac{\mathrm dz_1}{2\pi\mathrm i}
        \cdots
        \int\frac{\mathrm dz_f}{2\pi\mathrm i}
       \left(-M^2 \right)^{z_{1234}}
       \left (m^2 \right )^{z_5}
        \bracket{z_{123467}+d/2}
        \bracket{z_{89} - z_{16}}
        \bracket{z_{ab} - z_{56}}
        \bracket{z_{cd} - z_{7b}}
        \bracket{z_{ef} - z_{37a}}
        \\
         \times
        x_1^{a_1+z_{2e}-1} 
        x_2^{a_2+z_{134c}-1} 
        x_3^{a_3+z_{123d}-1} 
        x_4^{a_4+z_{4f}-1} 
        x_5^{a_5+z_{48}-1} 
        x_6^{a_6+z_{29}-1} 
        \\
      \times
        \frac{
        \Gamma(-z_1) 
        \cdots
        \Gamma(-z_f)
        }{\Gamma(a_1)\cdots\Gamma(a_6)\Gamma(z_{1234}+d/2)\Gamma(-z_{16})\Gamma(-z_{56})\Gamma(-z_{7b})\Gamma(-z_{37a})}
        \,,
    \end{multline*}        
\normalsize%
where the integration variables are counted alphabetically after $z_9$.

\paragraph*{Rule C}
Since, at this point, all auxiliary variables are eliminated, the $x_i$-integrals are all of the form of the right-hand side of \Eref{eq:mbbrackets:bracket}, such that they can be rewritten in terms of brackets.
This yields
\begin{multline} \label{eq:mbbrackets:presolution}
          \int\frac{\mathrm dz_1}{2\pi\mathrm i}
        \cdots
        \int\frac{\mathrm dz_f}{2\pi\mathrm i}
       \left (-M^2\right )^{z_{1234}}
       \left (m^2 \right)^{z_5}
        \bracket{z_{123467}+d/2}
        \bracket{z_{89} - z_{16}}
        \bracket{z_{ab} - z_{56}}
        \bracket{z_{cd} - z_{7b}}
        \bracket{z_{ef} - z_{37a}}
        \\
        \times
        \bracket{a_1+z_{2e}} 
        \bracket{a_2+z_{134c}} 
        \bracket{a_3+z_{123d}} 
        \bracket{a_4+z_{4f}} 
        \bracket{a_5+z_{48}} 
        \bracket{a_6+z_{29}} 
        \\
        \times
        \frac{
            \Gamma(-z_1) 
            \cdots
            \Gamma(-z_f)
        }{\Gamma(a_1)\cdots\Gamma(a_6)\Gamma(z_{1234}+d/2)\Gamma(-z_{16})\Gamma(-z_{56})\Gamma(-z_{7b})\Gamma(-z_{37a})}
        \,,
    \end{multline}
which is called the \emph{presolution} of the Feynman integral.

\paragraph*{Rule D}
In the final step, the master formula (\Eref{eq:mbbrackets:master}) will be applied to the presolution (\Eref{eq:mbbrackets:presolution}).
Since \Eref{eq:mbbrackets:presolution} has 15 {\mbr} integrals but only 11 brackets, not all integrals can be solved with the master formula.
It is a matter of choice which {\mbr} integrals should remain.
Out of the ${15\choose11} = 1365$ possibilities, 868 lead to a singular matrix $A$ in \Eref{eq:mbbrackets:master} and must be rejected.
The remaining 497 choices all lead to possible {\mbr} representations.

Let us arrange the integration variables for which we want to apply the master formula into a vector $\vec z_1$ and the remaining ones into a vector $\vec z_2$.
With these definitions, the master formula reads
\begin{multline} \label{eq:mbbrackets:ruleD}
        \int \frac{\mathrm dz_1}{2\pi \mathrm i}
        \cdots
        \int \frac{\mathrm dz_J}{2\pi \mathrm i}
        \,
        \bracket{\beta_1+\vec\alpha_1\cdot\vec z_1+\vec\gamma_1\cdot\vec z_2 }
        \cdots
        \bracket{\beta_K+\vec\alpha_K\cdot\vec z_1+\vec\gamma_K\cdot\vec z_2}
        \,
        f(\vec z_1,\vec z_2)
        \\
        =
        \frac1{|\det A|}
        \int \frac{\mathrm dz_{K+1}}{2\pi \mathrm i}
        \cdots
        \int \frac{\mathrm dz_J}{2\pi \mathrm i}
        f(-A^{-1} \vec\beta - A^{-1} C\,\vec z_2,\vec z_2)\,,
   \end{multline}
where $A = (\vec\alpha_1, \cdots, \vec\alpha_K)^\mathrm{T}$ and $C = (\vec\gamma_1, \cdots, \vec\gamma_K)^\mathrm{T}$.

For the example, we choose the integrals over $z_2, z_4, z_6$, and $z_d$ to remain.
This choice leads to
\setcounter{MaxMatrixCols}{20}%
    \begin{gather*} 
        \vec z_1 = (z_1, z_3, z_5, z_7, z_8, z_9, z_a, z_b, z_c, z_e, z_f)^\mathrm{T}\,, \qquad
        \vec z_2 = (z_2, z_4, z_6, z_d)^\mathrm{T}\,, 
        \\
        f(\vec z_1, \vec z_2)
        =
        \frac{
            \left (-M^2 \right)^{z_{1234}}
           \left (m^2 \right )^{z_5}
            \Gamma(-z_1) 
            \cdots
            \Gamma(-z_f)
        }{\Gamma(a_1)\cdots\Gamma(a_6)\Gamma(z_{1234}+d/2)\Gamma(-z_{16})\Gamma(-z_{56})\Gamma(-z_{7b})\Gamma(-z_{37a})}\,,
        \\
        \vec\beta = \left(d/2, 0, 0, 0, 0, a_1, a_2, a_3, a_4, a_5, a_6\right)^\mathrm{T}\,, 
        \\
        A 
        =
        \begin{pmatrix}
            1 &  1 &  0 &  1 &  0 &  0 &  0 &  0 &  0 &  0 &  0 \\
            -1 &  0 &  0 &  0 &  1 &  1 &  0 &  0 &  0 &  0 &  0 \\
            0 &  0 &  -1 &  0 &  0 &  0 &  1 &  1 &  0 &  0 &  0 \\
            0 &  0 &  0 &  -1 &  0 &  0 &  0 &  -1 &  1 &  0 &  0 \\
            0 &  -1 &  0 &  -1 &  0 &  0 &  -1 &  0 &  0 &  1 &  1 \\
            0 &  0 &  0 &  0 &  0 &  0 &  0 &  0 &  0 &  1 &  0 \\
            1 &  1 &  0 &  0 &  0 &  0 &  0 &  0 &  1 &  0 &  0 \\
            1 &  1 &  0 &  0 &  0 &  0 &  0 &  0 &  0 &  0 &  0 \\
            0 &  0 &  0 &  0 &  0 &  0 &  0 &  0 &  0 &  0 &  1 \\
            0 &  0 &  0 &  0 &  1 &  0 &  0 &  0 &  0 &  0 &  0 \\
            0 &  0 &  0 &  0 &  0 &  1 &  0 &  0 &  0 &  0 &  0
        \end{pmatrix}
        \,, \qquad
        C 
        = 
        \begin{pmatrix}
            1 &  1 &  1 &  0 \\
            0 &  0 &  -1 &  0 \\
            0 &  0 &  -1 &  0 \\
            0 &  0 &  0 &  1 \\
            0 &  0 &  0 &  0 \\
            1 &  0 &  0 &  0 \\
            0 &  1 &  0 &  0 \\
            1 &  0 &  0 &  1 \\
            0 &  1 &  0 &  0 \\
            0 &  1 &  0 &  0 \\
            1 &  0 &  0 &  0
        \end{pmatrix} \,. 
    \end{gather*}
Substituting these into the right-hand side of \Eref{eq:mbbrackets:ruleD} yields the final four-dimensional {\mbr} represen\-tation
   \scriptsize
    \begin{multline*} 
        \left(m^2 \right)^{-a_{12456} + d}
       \left (-M^2 \right)^{-a_3}
        \int \frac{\mathrm dz_2}{2\pi \mathrm i}
        \int \frac{\mathrm dz_4}{2\pi \mathrm i}
        \int \frac{\mathrm dz_6}{2\pi \mathrm i}
        \int \frac{\mathrm dz_d}{2\pi \mathrm i}
        \left(-\frac{m^2}{M^2}\right)^{-z_4 + z_d}
        \frac{
            \Gamma(-z_2)
            \Gamma(-z_4)
            \Gamma(-z_6)
            \Gamma(-z_d)
        }{
            \Gamma(a_1)\cdots\Gamma(a_6)
        }
        \\ 
        \times
        \frac{
            \Gamma(a_1 + z_2)
            \Gamma(a_4 + z_4)
            \Gamma(a_5 + z_4)
            \Gamma(a_6 + z_2)
            \Gamma(a_{1456} - d/2 + z_{24})
            \Gamma(a_{12456} - d + z_4 - z_d)
        }{
            \Gamma(a_{14} + z_{24})
            \Gamma(a_{56} + z_{24})
        }
        \\
         \times
        \frac{
            \Gamma(a_3 - a_{56} - z_{46} + z_d)
            \Gamma(a_2 - a_3 - z_{2d} + z_4)
            \Gamma(a_2 - d/2 - z_{26d})
            \Gamma(-a_3 + d/2 + z_{46} - z_d)
            \Gamma(a_{56} + z_{246})
        }{
            \Gamma(a_2 - a_3 - z_2 + z_4 - 2 z_d)
            \Gamma(-a_3 + d/2 + z_4 - z_d)
            \Gamma(a_{12456} - d + z_4 - z_{6d})
        }
        \,,
    \end{multline*}        
\normalsize%
for the Feynman integral in \Fref{fig:mbbrackets:example}.

\subsection{Conclusion}
    In this contribution, we explained a new technique to construct {\mbr} representations with a non-trivial example.
    The approach is based on a reformulation of the method of brackets.
    Our modified method of brackets yields not only one but many possible {\mbr} representations, where every single one is a valid representation of the full Feynman integral. 

    A crucial part of the method is the optimization procedure.
    Here, one must analyse the graph polynomials for common subexpressions.
    With this optimization, the method is able to produce low-dimensional {\mbr} representations.


 \label{sec-mprausa}
\clearpage
\pagestyle{empty}
\cleardoublepage


\cleardoublepage

\section
[New approach to Mellin--Barnes integrals for massive one-loop 
Feynman inte\-grals \\ {\it J. Usovitsch, T. Riemann}]
{New approach to Mellin--Barnes integrals for massive one-loop 
Feynman inte\-grals  \label{sec-1loop}}

\pagestyle{fancy}
\fancyhead[LO]{}
\fancyhead[CO]{\thechapter.\thesection 
\hspace{1mm} 
New approach to Mellin--Barnes integrals for massive one-loop 
Feynman integrals}
\fancyhead[RO]{}
\fancyhead[LE]{}
\fancyhead[CE]{J. Usovitsch, T. Riemann}
\fancyhead[RE]{} 

\noindent 
{\bf Authors: Johann Usovitsch, Tord Riemann}
\\
Corresponding author: 
Johann Usovitsch {[jusovitsch@googlemail.com]}
\vspace*{.5cm}

\subsection*{Abstract}
Two methods are known for the infrared safe calculation 
of potentially arbitrary Feynman integrals: sector decomposition ({\sd}) and 
Mellin--Barnes ({\mbr}) representations. In recent years, in both approaches,  
mighty complementary tools have been developed for the numerical solution of 
Feynman integrals in arbitrary kinematics.
Unfortunately, the dimensionality of {\mbr} integrals  often increases 
quadratically with the number of scales, so that an evaluation becomes difficult 
or even impossible.
For the case of one-loop integrals, a dedicated choice of 
invariants leads to a linear growth of dimensionality. The most complicated 
term in a general one-loop Feynman-integral basis is the scalar box integral. Its 
dimensionality is reduced from nine to three with the new variables, in agreement with the dimensionality in the {\sd} 
method.
Further, Minkowskian kinematics shows no specific convergence problems compared with the Euclidean case.
A generalization of the approach to higher-loop cases would be extremely 
desirable.

\subsection{Introduction}
Equation (13) in Ref. \cite{Fleischer:2003rm}, the $d$-dependent recurrence relation for one-loop $n$-point Feynman integrals 
$J_n(d)$,  is used to derive solutions for the integrals $J_n(d)$ in terms of infinite sums of simpler 
functions $J_{n}(d-2), J_{n-1}(d-2)$, and in turn in terms of generalized hypergeometric functions:
\begin{eqnarray}\label{npointA}
 \left(d-1-n\right) ~ G_n ~ J_n(d) =  \left[2\lambda_n + \sum_{k=1}^n (\partial_k \lambda_n) {\bf 
k^-}\right] J_n(d-2)
 .
\end{eqnarray}
The relation was derived in Refs. \cite{Bern:1992em,Tarasov:1996br} for the case of canonical powers of propagators, 
$\nu_i=1$. The $G_n$ is the Gram determinant, the $\lambda_n$ the Cayley determinant, and further notation will be 
explained later.
In Refs. \cite{Bluemlein:2017rbi,Riemann:April2018}, another recurrence relation for 
$J_n(d)$, namely in terms of 
Mellin--Barnes integrals over dimensional shifts $s$ of simpler functions $J_{n-1}(d+2s)$ was derived and also solved.
Both methods give partially identical results, and the evaluation of the latter approach in terms of generalized 
hypergeometric functions has been described elsewhere. 

Here, we will use an intermediate result of Ref. \cite{Bluemlein:2017rbi}  to derive new compact Mellin--Barnes 
relations for one-loop $n$-point Feynman integrals $J_n(d)$, which extend the applicability of the {\mbr} method 
considerably. 

We start the derivations from the  well-known Feynman parameter representation 
for scalar one-loop $N$-point Feynman integrals:  
\begin{equation}\label{npoint}
J_n 
\equiv 
J_{n}(d; \{p_ip_j\}, \{m_i^2\})
 =
\int \dfrac{\mathrm{d}^d k}{\mathrm{i} \pi^{d/2}} \dfrac{1}{D_1^{\nu_1} D_2^{\nu_2}\cdots D_n^{\nu_n}}
,
\end{equation}
with inverse propagators 
\begin{eqnarray}\label{eq-Di}
D_i= (k+q_i)^2-m_i^2+\mathrm{i}\varepsilon.
\end{eqnarray}
In the following, we assume $d=4+2n-2\epsilon$ with integer $n\geq 0$, momentum 
conservation, and all momenta to be incoming, 
$\sum_{i}^np_i=0.$ 
The $q_i$ are loop momenta shifts and will be expressed for applications by the 
external momenta $p_i$.
Following Ref. \cite{Davydychev:1991va}, general tensor integrals 
with $n$ legs may be reduced to scalar integrals with up to four legs, but 
introducing higher space-time dimension $d \geq 4$ and higher propagator powers 
$D_i^{\nu_i}$.
Indices $\nu_i >1$ may be reduced by recursion relations given in Ref.
\cite{Fleischer:1999hq} or by a general-purpose implementation like that
in Ref. \cite{Maierhoefer:2017hyi}.
If we would reduce, in addition, the space-time dimension to $d=4-2\epsilon$, 
we would introduce inverse powers of potentially vanishing Gram determinants, 
see Refs. \cite{Fleischer:1999hq,Fleischer:2010sq} and references quoted therein.
For our derivations, this feature is unwanted, and we allow for a general 
space-time $d$.
To  summarize, by solving \Eref{npoint}, one covers arbitrary one-loop Feynman 
integrals. 

We use the Feynman parameter representation for the evaluation of 
\Eref{npoint}:
\begin{equation}\label{bh-1}
J_n
= 
(-1)^{n}
{\Gamma\left(n -d/2\right)}
   \int_0^1 \prod_{j=1}^n \mathrm{d}x_j
\delta \left( 1-\sum_{i=1}^n x_i \right)
\frac { 1 } {F_n(x)^{n-d/2}}
.
\end{equation}
Here, the $F$-function is the second Symanzik polynomial. It is  derived from 
the propagators 
$M^2 \equiv {x_1D_1+   \cdots + x_n D_n} ~=~ k^2 - 2Qk + J.$
Using  further the $\delta (1-\sum x_i)$ under the integral in order to 
transform in $F$ the linear terms in $x_i$ into bilinear ones, one obtains
\begin{equation}\label{eq-F-orig}
F_n(x) = Q^2 - J\times \left(\sum_{i=1}^n x_i\right) = \frac{1}{2} 
\sum_{i,j} x_i Y_{ij} x_j-\mathrm{i}\varepsilon
.
\end{equation}
This representation of $F$ relies on the elements of the Cayley 
matrix $Y$, which was introduced in Ref. \cite{Melrose:1965kb}:
\begin{equation}\label{eq-y}
\lambda
=
\begin{pmatrix}
Y_{11}  & Y_{12}  &\ldots & Y_{1n} \\
Y_{12}  & Y_{22}  &\ldots & Y_{2n} \\
\vdots  & \vdots  &\ddots & \vdots \\
Y_{1n}  & Y_{2n}  &\ldots & Y_{nn}
\end{pmatrix}
,         
\end{equation} 
with elements
\begin{equation}\label{eq-yij}
Y_{ij} = Y_{ji} = m_i^2+m_j^2-(q_i-q_j)^2 . 
\end{equation}
In a second preparational step after the elimination of linear terms in $x$, 
the $\delta$-function in \Eref{npoint} is eliminated by integrating over, 
\eg $x_n$; this re-introduces linear terms in $x$, as well as an inhomogeneity.
For later use, we split the expression into two pieces as follows:
\begin{eqnarray} \label{eq-Fyi}
F_n(x)=
 (x - y)^\mathrm{T} G (x-y)
 + r_{n}  - \mathrm{i} \varepsilon \equiv  \Lambda_{n}(x) +R_{n}
.
\end{eqnarray}
The Gram matrix $G$ is $(n-1)$-dimensional. It is independent of the 
propagator masses and depends in a symmetrical way on the internal momenta 
$q_1$ to $q_n$:
\begin{eqnarray}
 \label{Gram}
G 
= 
- 
\begin{pmatrix}
  (q_1-q_n)^2 
&\ldots & (q_1-q_n)(q_{n-1}-q_n) 
\\
  (q_1-q_n)(q_2-q_n) 
&\ldots 
& (q_2-q_n)(q_{n-1}-q_n) 
\\
  \vdots    
&\ddots   & \vdots 
\\
   (q_1-q_n)(q_{n-1}-q_n)    
&\ldots & (q_{n-1}-q_n)^2
\end{pmatrix}
.
\end{eqnarray}
Owing to relations like $q_i = p_{i+1}-p_{i}$, one may also write it in terms of 
the external momenta $p_{i}$.
At this stage, it is easy to prove that the isolated inhomogeneity part of 
$F(x)$ depends on two determinants: 
\begin{eqnarray}
R_n
\equiv r_n  - \mathrm{i} \varepsilon
= - \frac{\lambda_n}{G_n} - \mathrm{i} \varepsilon ,
\end{eqnarray}
where
\begin{eqnarray}
 \lambda_n = {\mathrm{det}} ( \lambda), \qquad G_n = {\mathrm{det}} ( G ).
\end{eqnarray}
The $(n-1)$ coefficients of the linear terms in \Eref{eq-Fyi} depend on the 
same determinants: 
\begin{equation} \label{eq-def-yi}
y_i = \dfrac{\partial r_{n}}{\partial m_i^2} 
=   - \frac{1}{G_{n}}~\dfrac{\partial \lambda_n}{\partial m_i^2} 
\equiv - \dfrac{1}{G_{n}}~\partial_i \lambda_n
 , \qquad 
i=1, \dots, n.
\end{equation} 
The auxiliary condition $\sum_i^n y_i =1$ is fulfilled.
The notations for the $F$-function are finally independent of the choice of 
variable, which was eliminated by use of the 
$\delta$-function in the integrand of the $x$-integral. 
The inhomogeneity $R_{n}$ 
is the only variable carrying the causal $\mathrm{i}\varepsilon$-prescription, while 
the matrices $\lambda$ and $G$, as well as $\Lambda(x)$ and the $y_i$, are, by 
definition, real quantities -- as long as we consider real masses and momenta. 

Representations like \Eref{eq-Fyi} for the $F$ polynomial may be found in 
the literature. Here,  \Eref{eq-Fyi} is the starting point for integrating out 
the $x$-variables and the introduction of a generic Mellin--Barnes integral.
This has been derived in Ref. \cite{Bluemlein:2017rbi}.

The use of Mellin--Barnes representations for solving Feynman integrals in
the form 
of Feynman representations was advocated in 
Ref. \cite{Smirnov:1999gc} and is being systematically worked out in the 
\textsc{Mathematica} package {\tt MBsuite} with the basic tools \ar{}/\mbr/MBnumerics 
\cite{Gluza:2007rt,Gluza:2010rn,Dubovyk:2015yba,Czakon:2005rk,Usovitsch:2018shx,Usovitsch:Jan2018}.
The idea is to replace sums by products \cite{Barnes:zbMATH02640947},
\begin{eqnarray}\label{eq-MB}
 \frac{1}{(A+B)^k} = \frac{B^{-k}}{2\pi \mathrm{i}} 
 \int\limits_{-\mathrm{i}\infty}^{+\mathrm{i}\infty}\mathrm{d}s  
  \dfrac{\Gamma(-s) \Gamma(k +s) } { \Gamma(k) }    
  \left[\dfrac{A}{B} \right]^s,
\end{eqnarray}
where the integration path separates the poles of $\Gamma(-s)$ from the poles 
of $\Gamma(k +s)$. Further, the condition $|\mathrm{Arg}(A/B)|<\pi$ must be 
fulfilled. 
Applying \Eref{eq-MB} to the $F$-function (\Eref{eq-F-orig}) in \Eref{bh-1}, 
$(x^\mathrm{T}Yx)^{d/2-n}$, leads, for our most general case 
of one-loop box functions with up to four masses and four 
virtualities, plus $s$ and $t$, to {\mbr} representations of dimensionality 
$D=9$, which are rather unsolvable. One has to compare this with the 
one other robust method to treat arbitrary infrared Feynman integrals: sector 
decomposition \cite{Hepp:1966eg,Heinrich:2008si,Freitas:2010nx}. Here, the 
dimensionality $D$ of (direct $x$-) integrations is just $(n-1)$; with $D=3$ 
for the box integral discussed. 
For higher-loop orders, the difference in dimensionality may become even more 
drastic.

Let us now follow Ref. \cite{Bluemlein:2017rbi} and apply an {\mbr} integration to the 
integrand of the $x$-integral, thus separating the $x$-dependent part 
from the isolated 
inhomogeneity:
\begin{eqnarray}
\label{c5}
\dfrac{1}{ \left [\Lambda_n(x) + R_n \right ]^{ n-\frac{d}{2} }}
= \dfrac{R_n^{\frac{d}{2}-n}}{2\pi \mathrm{i}} 
\int\limits_{-\mathrm{i}\infty}^{+\mathrm{i}\infty}\mathrm{d}s \; 
  \dfrac{\Gamma(-s)\;\Gamma \left ( n-\frac{d}{2} +s \right ) } { \Gamma
  \left (n-\frac{d}{2}
  \right ) }    
  \left[\dfrac{\Lambda_n(x)}{ R_n} \right]^s.
\end{eqnarray}
To solve the $x$-integral, the differential 
operator $\hat{P}_n$ \cite{Bernshtein1971-from-springer,Golubeva:1978},
\begin{equation}
\label{c7}
\frac{\hat{P}_n}{s} \left[\dfrac{\Lambda_n(x)}{R_n} \right]^s
\equiv
\sum_{i=1}^{n-1} 
   \frac{1}{2s} (x_i-y_i)\frac{\partial}{\partial x_i}
\left[\dfrac{\Lambda_n(x)}{R_n} \right]^s
   =    \left[\dfrac{\Lambda_n(x)}{R_n } \right]^s,
\end{equation}
is introduced in the integrand of the $x$-integral:
\begin{equation}
\label{c8}
\nonumber
K_n
= \frac{1}{s}
\int \mathrm{d}S_{n-1}\; \hat{P}_n 
    \left[ \dfrac{\Lambda_n(x)}{R_n} \right]^s
= \frac{1}{2s} \sum_{i=1}^{n-1} 
\prod_{k=1}^{n-1}
\int\limits_0^{u_k}  \mathrm{d}x'_k ~ 
   (x_i-y_i)\frac{\partial}{\partial x_i} 
   \left[\dfrac{\Lambda_n(x)}{R_n} \right]^s
.
\end{equation}

This trick allows us to rewrite, after application of a Barnes transformation, 
the $n$-point integral in $d$ space-time dimensions as an {\mbr} integral of an 
$(n-1)$-point integral in $d+2$ space-time dimensions:
\begin{multline} \label{JNJN1} 
J_n \left (d, \left \{q_i,m_i^2 \right \} \right )
= \frac{-1}{4\pi \mathrm{i}} 
\frac{1}{\Gamma(\frac{d-n+1}{2})}
\int\limits_{-\mathrm{i}\infty}^{+\mathrm{i}\infty}\mathrm{d}s  
 \Gamma(-s) \Gamma(s+1) \Gamma \left (s+\frac{d-n+1}{2} \right )      
     R_{n}^{-s}
\\
\times     \sum\limits_{k=1}^n 
     \left( \dfrac{1}{{r_{n}}} 
            \dfrac{\partial {r_n}}{\partial m_k^2} 
      \right) 
     {\bf k}^- J_n \left (d+2s \left \{q_i,m_i^2 \right \} \right ).
\end{multline}
The operator ${\bf k^{-}}$ reduces, notationally,  an $n$-point Feynman integral 
$J_n$ to an $(n-1)$-point integral $J_{n-1}$ 
by shrinking the $k$th propagator, $1/D_k$:
\begin{equation}\label{eq-def-operator-k-}
{\bf k^{-}} J_n 
= 
{\bf k^{-}} \int \dfrac{\mathrm{d}^d k}{\mathrm{i} \pi^{d/2}} \dfrac{1}{\prod_{j=1}^n D_j}
=
\int \dfrac{\mathrm{d}^d k}{\mathrm{i} \pi^{d/2}} \dfrac{1}{\prod_{j\neq k,j=1}^n D_j}
.
\end{equation}

Equation \eqref{JNJN1} is the starting point for our numerical approach.
It is an analogue to a similar relation derived 
in Ref. \cite{Bluemlein:2017rbi}, where a difference equation in $d$ was derived and 
solved iteratively for $J_n(d; \{p_ip_j\}, \{m_i^2\})$.
\subsection{\label{sec-MB}The \mbr iterations for the massive one-loop integrals}
In deriving \Eref{JNJN1}, it was assumed that $\Lambda_n(x)$ and $R_n$ are 
finite, and we will not discuss here the many special cases otherwise appearing.

The simplest case of a massive one-loop scalar Feynman integral is the 
one-point function or tadpole,
\begin{equation} \label{eq-tadpole}
J_1(d;m^2) 
= 
\int \dfrac{\mathrm{d}^d k}{\mathrm{i} \pi^{d/2}} 
\dfrac{1}{k^2-m^2+\mathrm{i} \varepsilon}
= \frac{-1}{4\pi \mathrm{i}}  
\frac{\Gamma( 1 -d/2)}
 {(m^2)^{1-d/2}}.
\end{equation}
For $m^2=0$, the tadpole vanishes, $J_{1}(d;0)=0$.
The two-point integral is: 
\begin{multline} \label{J2J1}
J_2(d;p^2,m_1^2,m_2^2)
 = 
   \left(\frac{-1}{4\pi \mathrm{i}}\right)^2 
\frac{1}{\Gamma \left (\frac{d-1}{2} \right )}
  \sum\limits_{k_1,k_2=1}^2 (1-\delta_{k_1k_2})
\left( \frac{1}{{r_{2}}} 
            \frac{\partial {r_2}}{\partial m_{k_2}^2} 
      \right)
(m_{k_1}^2)^{d/2-1}
\\ 
 \int\limits_{-\mathrm{i}\infty}^{+\mathrm{i}\infty}\mathrm{d}z_2  
      \left(\frac{m_{k_1}^2}{R_{2}}\right)^{z_2}       
     \Gamma(-z_2) \Gamma(z_2+1)  \Gamma \left (-z_2 -\frac{d-2}{2} \right ) 
\Gamma \left (z_2+\frac{d-1}{2} \right ).
   \end{multline}
The two-point function depends on the masses $m_1, m_2$ 
(owing to the 
tadpole basis) and on the ratio
\begin{eqnarray} \label{eq-r2}
 r_2 \equiv r_{12} \equiv r_2 \left (p^2,m_1^2,m_2^2 \right ) =  - \frac{\lambda_2}{g_2} 
=  
\frac{\lambda(p^2,m_1^2,m_2^2)}{4p^2},
\end{eqnarray}
where $\lambda(a,b,c)$ is the K\"{a}ll\'en function.
The vertex depends on the internal masses and the virtualities $p_i^2$.
A generic argument list, corresponding to 
$\{d,p_1^2,p_2^2,p_3^2,m_1^2,m_2^2,m_3^2\}$,  is 
$\{d,q_1,m_1^2,q_2,m_2^2,q_3,m_3^2\}$:
\begin{align} \label{J3J2}
J_3(d;\{p_i^2\},\{m_i^2\})
&= \left(\frac{-1}{4\pi \mathrm{i}}\right)^3  
\frac{1}{\Gamma \left (\frac{d-2}{2} \right )}
\sum\limits_{k_1,k_2,k_3=1}^3 D_{k_1k_2k_3}
     \left( \frac{1}{r_{3}} 
            \frac{\partial r_3}{\partial m_{k_3}^2} 
      \right)
                \left( \frac{1}{r_{k_2k_1}}
            \frac{\partial r_{k_2k_1}}{\partial m_{k_2}^2} 
      \right)
    \nonumber \\ \nonumber 
& \quad
 \left  (m_{k_1}^2 \right )^{d/2-1}
  \int\limits_{-\mathrm{i}\infty}^{+\mathrm{i}\infty}\mathrm{d}z_3
              \int\limits_{-\mathrm{i}\infty}^{+\mathrm{i}\infty}\mathrm{d}z_2      
       \left(\frac{m_{k_1}^2}{R_{3}}\right)^{z_3}
       \left(\frac{m_{k-1}^2}{R_{k_2k_1}}\right)^{z_2} 
     \Gamma(-z_3) \Gamma(z_3+1) 
  \\ \nonumber 
& \quad   
\times
\Gamma \left (z_2+z_3+\frac{d-1}{2} \right ) 
\Gamma \left (-z_2-z_3-\small{\frac{d+2}{2}} \right )
 \Gamma(-z_2) \Gamma(z_2+1) 
 \\  
& \quad
    \times~      
\frac{\Gamma\left (z_3+{{\frac{{d-2}}{2}}} \right )}{
\Gamma \left (z_3+\frac { d-1 } { 2 } \right  ) } 
,
    \end{align} 
with the condition $k_1 \neq k_2 \neq k_3$:
\begin{equation}\label{eq-del123}
 D_{k_1\cdots k_n} =  
 \prod_{\stackrel{i,j=1}{i\neq j}}^{n}  (1-\delta_{k_ik_j})
 .
 \end{equation}
Here, $r_3\equiv r_{123}$ and $r_{ij\cdots}=r_{\cdots j\cdots i\cdots}$.
One has to choose some conventions, \eg \Eref{eq-Di}, additionally 
setting $q_1=0, q_2=-p_2, q_3=p_1$. 
One often selects $p_3^2=s$.
We have to use, besides the three internal masses $m_k^2$ and 
the three parameters  $r_2(p_i^2,m_j^2,m_k^2)$, the additional  
variable 
\begin{eqnarray} \label{eq-r3}
 r_3 \left (p_1^2,p_2^2,p_3^2,m_1^2,m_2^2,m_3^2 \right ) \equiv  - \frac{\lambda_3}{g_3}.
 \end{eqnarray}
 For the chosen conventions, it is 
 \begin{align} \label{eq-lam3alt-g3}
\lambda_3 &= 
2\bigl( m_1^2 m_2^2 p_1^2 -  m_2^4 p_1^2 -  m_1^2 m_3^2 p_1^2  2 m_2^2 m_3^2 
p_1^2 - 
  m_2^2 p_1^4 -  m_1^2 m_2^2 p_2^2 +  m_1^2 m_3^2 p_2^2 +  m_2^2 m_3^2 
p_2^2  
\nonumber\\
& 
\qquad - 
  m_3^4 p_2^2
  +  m_2^2 p_1^2 p_2^2 +  m_3^2 p_1^2 p_2^2 -  m_3^2 p_2^4 - 
  m_1^4 p_3^2 + 2m_1^2 m_2^2 p_3^2 +  m_1^2 m_3^2 p_3^2 
 \nonumber\\
 & \qquad
 -  m_2^2 m_3^2 
p_3^2 + 
  m_1^2 p_1^2 p_3^2 +  m_2^2 p_1^2 p_3^2 +  m_1^2 p_2^2 p_3^2 +  m_3^2 
p_2^2 p_3^2 - 
  p_1^2 p_2^2 p_3^2 -  m_1^2 p_3^4\bigr),
\\
g_3 &= 2\lambda \left (p_1^2,p_2^2,p_3^2 \right ).
\end{align}
Specifying  a massive on-shell QED vertex with $p_3^2=s$, $p_1^2=p_2^2=M^2$ 
and $m_1=0, m_2=m_3=M$, one gets $\lambda_3=0$ and $g_3=-6M^2$.
This simple example is not covered by \Eref{JNJN1}.
It demonstrates that the various special 
cases of the approach must be worked out in order to cover all the 
frequently met physical situations. We do not go into details here because, \eg 
the massive on-shell QED vertex and box cases in the usual {\mbr}-approach of 
\ar{} will be represented by one- and two-dimensional {\mbr} integrals, so 
that there is no need to apply the present approach. 

Finally, we reproduce the box integral, dependent on $d$ and the internal variables 
$\{d,q_1,m_1^2,\allowbreak\cdots q_4,m_4^2\}$ or, equivalently, on a set of external 
variables, \eg $\{d,\{p_i^2\},\{m_i^2\},s,t\}$:
\begin{align} \label{J4J3}
J_4\left (d; \left \{p_i^2 \right \},s,t, \left \{m_i^2 \right \} \right )
&=  
 \left(\frac{-1}{4\pi \mathrm{i}}\right)^4  
\frac{1}{\Gamma \left (\frac{d-3}{2} \right )}
\sum_{k_1,k_2,k_3,k_4=1}^4 D_{k_1k_2k_3k_4}
\left( \frac{1}{r_{4}} \frac{\partial r_4}{\partial m_{k_4}^2} \right)
\nonumber \\ 
& \quad
\left( \frac{1}{r_{k_3k_2k_1}} \frac{\partial r_{k_3k_2k_1}}{\partial 
       m_{k_3}^2} \right)
\left( \frac{1}{r_{k_2k_1}} \frac{\partial r_{k_2k_1}}{\partial m_{k_2}^2} 
      \right)
      (m_{k_1}^2)^{d/2-1}
\nonumber \\
\nonumber 
& \quad       
\int\limits_{-\mathrm{i}\infty}^{+\mathrm{i}\infty}\mathrm{d}z_4
       \int\limits_{-\mathrm{i}\infty}^{+\mathrm{i}\infty}\mathrm{d}z_3
              \int\limits_{-\mathrm{i}\infty}^{+\mathrm{i}\infty}\mathrm{d}z_2  
          \left(\frac{m_{k_1}^2}{R_{4}}\right)^{z_4}          
       \left(\frac{m_{k_1}^2}{R_{k_3k_2k_1}}\right)^{z_3}
       \left(\frac{m_{k_1}^2}{R_{k_2k_1}}\right)^{z_2}
    \\ \nonumber 
& \quad    
\Gamma(-z_4) \Gamma(z_4+1) 
\frac  {\Gamma \left (z_4+\frac{d-3}{2} \right )}  {\Gamma \left (z_4+\frac{d-2}{2}
\right )}  
            \Gamma(-z_3) \Gamma(z_3+1) 
\frac{\Gamma \left (z_3+z_4+\frac{d-2}{2} \right )}{\Gamma \left (z_3+z_4+\frac{d-1}{2}
\right )} 
  \\  
  & \quad     
\Gamma \left (z_2+z_3+z_4+\frac{d-1}{2} \right ) 
\Gamma \left (-z_2-z_3-z_4-\frac{d+2}{2} \right )
     \Gamma(-z_2) \Gamma(z_2+1)    .   
\end{align}
Equations \eqref{J2J1}--\eqref{J4J3} can be treated by the 
 \textsc{Mathematica} packages {\mbr} and MBnumerics of  {\tt MBsuite}, replacing \ar{} by a derivative of MBnumerics 
\cite{Usovitsch:2018shx,Usovitsch:PhD2018}.
\subsection{The cases of the vanishing Cayley determinant
and the vanishing Gram determinant
}
We refer to two important special cases, where the general derivations cannot be applied.

In the case of a vanishing Cayley determinant, $\lambda_n=0$, we cannot introduce the inhomogeneity $R_n=-\lambda_n/G_n$ 
into the Symanzik polynomial $F_2$, see \Eref{eq-Fyi}.
Let us assume that it is  $G_n \neq 0$, so that $r_n=0$. 
A useful alternative representation to \Eref{JNJN1} is known from the literature, see, \eg Eq. (3) in Ref. 
\cite{Fleischer:2003rm}:
\begin{equation} \label{CayleyZero}
J_n(d) = \frac{1}{d-n-1} \sum_{k=1}^{n} \frac{\partial_k \lambda_n}{G_n} {\bf k^-} J_n(d-2).
\end{equation}

Another special case is a vanishing Gram determinant, $G_n=0$.
Here, again, one may use Eq. (3) of Ref. \cite{Fleischer:2003rm}; the result is (for $\lambda_n\neq 0$):
\begin{equation} \label{GramZero}
J_n(d) = - \sum_{k=1}^{n} \frac{\partial_k \lambda_n}{2 \lambda_n} {\bf k^-} J_n(d).
\end{equation}
The representation was, for the special case of the vertex function, also given in Eq. (46) of Ref. 
\cite{Devaraj:1997es}.

\subsection{Example: massive four-point function with vanishing Gram determinant}
As a very interesting non-trivial example, we study the numerics of  a  massive
four-point function with a small or 
vanishing Gram determinant.
The example has been taken from Appendix C of Ref. \cite{Fleischer:2010sq}.
The kinematics are:
\begin{align} \label{prd83-kinematics}
p_{1}^{2} & =p_{2}^{2} 
= \nonumber
m_{1}^{2}=m_{3}^{2}=m_{4}^{2}=0,
\\
\nonumber
m_{2}^{2} &= 
(911\,876/10\,000)^2,
\\ 
p_{3}^{2}&=s_3=s_{\nu u}=10\,000, \nonumber
\\
\nonumber
p_{4}^{2}&=t_\mathrm{ed}=-60\,000(1+x),
\\
\nonumber
s&=(p_{1}+p_{2})^2=s_{12}=t_{\mathrm{e}\mu}=-40\,000, 
\\
t&= (p_{2}+p_{3})^2=s_{23}=s_{\mu\nu \mathrm{u}}=20\,000.
\end{align}
The resulting Gram determinant is
\begin{eqnarray} \label{prd83-gram}
 G_4 = -2t_{\mathrm{e}\mu}  \left [s_{\mu\nu \mathrm{u}}^2 +s_{\nu \mathrm{u}}t_\mathrm{ed}-s_{\mu\nu \mathrm{u}} \left (s_{\nu \mathrm{u}}+t_\mathrm{ed}-t_{\mathrm{e}\mu} \right ) \right ].
\end{eqnarray}
The Gram determinant vanishes if 
\begin{eqnarray} \label{prd83-gramvanish}
 t_\mathrm{ed} \to t_\mathrm{ed,crit} = \frac{  s_{\mu\nu \mathrm{u}} (s_{\mu\nu \mathrm{u}}-s_{\nu \mathrm{u}} +t_{\mathrm{e}\mu})}
 {s_{\mu\nu \mathrm{u}} - s_{\nu \mathrm{u}}}
 .
\end{eqnarray}
Introducing a parameter $x$, we can describe the vanishing of $G_4$ by the limiting process $x\to 0$:
\begin{align} \label{prd83-x}
 t_\mathrm{ed} &= (1+x) t_\mathrm{ed,crit},
 \\
 G_4 &= -2 x s_{\mu\nu \mathrm{u}} t_{\mathrm{e}\mu} (s_{\mu\nu \mathrm{u}}-s_{\nu \mathrm{u}} +t_{\mathrm{e}\mu}).
\end{align}
For $x=0$, the Gram determinant vanishes.
Further, it is simple to calculate, 
\eg the value at $x=1$: 
$G_4(x=1) =-4.8 \times 10^{13}\UGeV^3$.
Moreover, if the Gram determinant vanishes exactly, a reduction of $J_4$ according to \Eref{npoint} is possible and 
allows a 
simple and very precise calculation. 
For small $x$, the calculations become, with the usual reductions \`a la Passarino and Veltman \cite{Passarino:1978jh}, unstable, owing to the 
occurrence of inverse Gram determinants.
Several solutions were obtained; we refer to Ref. \cite{AlcarazMaestre:2012vp}.
Here we will use the solution given in Ref. \cite{Fleischer:2010sq}, which is  implemented in the {\tt C++} 
program PJFry \cite{Fleischer:2011bi,Fleischer:2011zz,Fleischer:2012et,Yundin-phd:2012oai}. 

In the new approach presented here, a calculation of $J_4$ is possible, as follows.
First reduce the indices $\nu_i$ of the propagators, if any, to the canonical value $\nu_i=1$; then apply the 
{\mbr}-formula 
directly.
This has been done with the {\tt C++} package KIRA \cite{Maierhoefer:2017hyi},
without generating inverse powers of Gram determinants.
In fact, the procedure introduces, for a $J_4(d)$, a basis of functions $J_4(d+2n)$, where $n\geq 0$ is related to 
the indices $\nu_i \geq 1$.
We use here the short notation
\begin{eqnarray} \label{prd83-integral}
J_{4}(D,\nu_{1},\nu_{2},\nu_{3},\nu_{4})
=
I_{4}(D,p_{1}^{2},p_{2}^{2},p_{3}^{2},p_{4}^{2},s,t,m_{1}^{2},m_{2}^{2},m_{3}^{2},
m_{4}^{2})\left [\nu_{1},\nu_{2},\nu_{3},\nu_{4} \right ]
.
\end{eqnarray}
The numerical solution of \Eref{JNJN1} is also stable  in the Minkowskian case.
This contrasts with the usual {\mbr} representations derived using \ar{}.
A reason is that the instabilities in the \ar{} approach originate from $\Gamma$-functions from beta functions, which do 
not appear here.

Tables \ref{tab-d1111} and \ref{tab-d111} contain our numerical results, for two different cases, which were also discussed in Ref. \cite{Fleischer:2010sq}.
In LoopTools notation, they correspond to the tensor coefficients $D_{2222}$ and $D_{222}$.
The former  is numerically less stable then the second.
In both cases, we have a numerical agreement of more than 10 digits, although we performed no expansion here in the 
small parameter $x$. Such an expansion would considerably improve calculations.
Our results give an impression of the accuracy of the small Gram expansion in Ref. \cite{Fleischer:2010sq}, where an error 
propagation of the Pad\'e approach was not done. In all cases, Ref. \cite{Fleischer:2010sq} has at least 10 reliable digits.

\begin{table}
\caption{\label{tab-d1111}
  The Feynman integral $J_{4}(12-2\epsilon,1,5,1,1)$ 
  compared with 
numbers from Ref. \cite{Fleischer:2010sq}. 
The $I_{4,2222}^{[d+]^4}$ is the scalar integral where propagator 2 has index $\nu_2 = 1+(1+1+1+1) = 5$; the others 
have index 1.
The integral corresponds to $D_{1111}$ in the notation of LoopTools 
\cite{Hahn:1998yk}. For $x=0$, the Gram determinant vanishes.
We see an agreement of about 10 or 11 relevant digits. The deviations of the two calculations seem to stem from 
a limited accuracy of the Pad\'e approximations used in Ref. \cite{Fleischer:2010sq}.}
\centering
  \begin{tabular}[\linewidth]{lll}
\hline \hline
$x$ & Value for $4! \times ~ J_{4}(12-2\epsilon,1,5,1,1)$ \\\hline
\phantom{1}$0$       & $(2.05969289730 + 1.55594910118 \mathrm{i})10^{-10}$ & \cite{Fleischer:2010sq}
\\
\phantom{1}$0$      &   $(2.05969289730 + 1.55594910118 \mathrm{i})10^{-10}$ &
{MBOneLoop + Kira + MBnumerics}
\\
$10^{-8}$ &$(2.05969289342 + 1.55594909187 \mathrm{i})10^{-10}$ & \cite{Fleischer:2010sq}
\\
$10^{-8}$ &$(2.05969289363 + 1.55594909187 \mathrm{i})10^{-10}$ & {MBOneLoop + Kira + MBnumerics}
\\
$10^{-4}$ &$(2.05965609497 + 1.55585605343 \mathrm{i})10^{-10}$ & \cite{Fleischer:2010sq}
\\
$10^{-4}$ &$(2.05965609489 + 1.55585605343 \mathrm{i})10^{-10}$ & {MBOneLoop + Kira + MBnumerics}
\\\hline\hline
  \end{tabular}
\end{table}

\begin{table}
  \caption{\label{tab-d111} 
  The Feynman integral $J_{4}(10-2\epsilon,1,4,1,1)$ 
  compared with numbers from Ref. \cite{Fleischer:2010sq}. 
The $I_{4,222}^{[d+]^3}$ is the scalar integral where propagator 2 has index $\nu_2 = 1+(1+1+1)$ = 4; the others have 
index 1.
The integral corresponds to $D_{111}$ in the notation of LoopTools \cite{Hahn:1998yk}.
We see an agreement of about 11 or 12 relevant digits. The deviations of the two calculations seem to stem from 
a limited accuracy of the Pad\'e approximations used in Ref. \cite{Fleischer:2010sq}.}
\centering
\begin{tabular}[\linewidth]{lll}
  \hline \hline
  $x$ & Value for $3! \times ~ J_{4}(10-2\epsilon,1,4,1,1)$ \\\hline
\phantom{1}$0$       & $-(3.15407250453 + 3.31837792634\mathrm{i})10^{-10}$ & \cite{Fleischer:2010sq}
\\
\phantom{1}$0$      &   $-(3.15407250453 + 3.31837792634\mathrm{i})10^{-10}$ &
{MBOneLoop + Kira + MBnumerics}
\\
$10^{-8}$ & $-(3.15407250057 + 3.31837790700 \mathrm{i})10^{-10}$ & \cite{Fleischer:2010sq}\\
$10^{-8}$ & $-(3.15407250055 + 3.31837790700 \mathrm{i})10^{-10}$ & {MBOneLoop + Kira + MBnumerics}
\\
$10^{-4}$ & $-(3.15403282194 + 3.31818461838 \mathrm{i})10^{-10}$ & \cite{Fleischer:2010sq}\\
$10^{-4}$ & $-(3.15403282195 + 3.31818461838 \mathrm{i})10^{-10}$ & {MBOneLoop + Kira + MBnumerics}
\\\hline\hline
\end{tabular}
\end{table}

 \label{sec-mb1loop}
\cleardoublepage

\clearpage
\pagestyle{fancy}
\fancyhead[LO]{}
\fancyhead[RO]{}
\fancyhead[CO]{}
\fancyhead[CO]{\thechapter.\thesection
\hspace{1mm} 
In search of the optimal contour of integration in Mellin--Barnes integrals 
}
\fancyhead[LE]{}
\fancyhead[CE]{W. Flieger}
\fancyhead[RE]{}

\section[In search of the optimal contour of integration in Mellin--Barnes integrals \\ {\it W. Flieger}]
{In search of the optimal contour of integration in Mellin--Barnes integrals 
} \label{sec:mbth}
\noindent
{\bf Author: Wojciech Flieger} {~~[woj.flieger@gmail.com]}
\vspace*{.5cm}

\subsection{\label{s1}Introduction}
As is evident from a number of contributions to this report, Mellin--Barnes integrals (\mbr{}) are widely used in the computation of higher-order radiative corrections where more and more loops and scales are involved. 

Here, we will focus on the numerical computation of \mbr{} integrals. We aim at numerical calculations, as the complexity of the considered Feynman integrals increases with the number of loops and masses involved, so it is much harder to find analytical solutions. There has been substantial progress in recent years in the numerical treatment of \mbr{} integrals, mostly arising
from a better understanding of complex contours over which integrals are computed in the multidimensional complex plane. The main problem is the oscillatory character of the \mbr{} integrals in physical kinematic regions of calculations. This is not a problem in the Euclidean region; in fact, there is a public package \mbm{} \cite{Czakon:2005rk}, which works in this region by calculation of integrals over straight complex lines with fixed real values of integrands. A kind of real breakthrough brought an idea  \cite{Freitas:2010nx} to rotate contours of integration, which can damp oscillatory character of integrands in the Minkowskian region. Other ideas, such as shifted contours \cite{Gluza:2007rt,Gluza:2010rn,Dubovyk:2016ocz}, made it finally possible to build  a set of procedures and packages 
\cite{ambrewww,planarity,Bielas:2013rja,Czakon:2005rk,Smirnov:2009up,%
mbtools,Usovitsch:2018shx,Hahn:2004fe} where at least two-loop vertex Feynman diagrams represented by multidimensional \mbr{} integrals can be calculated in an efficient way with high precision, of several digits \cite{Dubovyk:2016aqv}.
 
However, there is always room for improvement. We would like to explore the geometrical aspect of \mbr{} integrals, \ie consider the optimal contour of integration. The biggest obstacle, from the numerical point of view, in the computation of Feynman integrals in the {\mbr} representation is their bad convergence due to their highly oscillatory behaviour. Thus, it would be a great improvement if we could find a contour on which that oscillatory behaviour would be alleviated. To do this, we will adopt a concept originated in asymptotic analysis \cite{wong2001asymptotic,temme2014asymptotic} called  the method of steepest descent.   

 This method extends the idea of Laplace's method to the domain of complex numbers. Therefore, we consider integrals of the form
 \begin{equation}\label{lap}
 J(\lambda)= \int_{\mathcal{C}}g(z)\mathrm{e}^{\lambda f(z)}\mathrm{d}z,
 \end{equation}
 where $\mathcal{C}$ is a contour in the complex plane, $f(z), \ g(z)$ are analytical functions of the complex variable $z$, and $\lambda$ is a real positive parameter. In this method, we take advantage of the Cauchy integral theorem  to deform  the contour of integration into a new contour $\mathcal{C}^{'}$ that makes integration as easy as possible. If we can do this in such a way that the imaginary part of the function $f(z)$ remains constant $\Im(f(z))=
\Im(f(z_{0}))= \mathrm{const}$ along the new contour, our integral will take the form
 \begin{equation}
 J(\lambda)= \mathrm{e}^{i\lambda \Im(f(z_{0}))}\int_{\mathcal{C}^{'}}g(z)\mathrm{e}^{\lambda \Re(f(z))}\mathrm{d}z.
 \end{equation}
 Thus, on this new contour, we are left with the Laplace's type of integral for which well-established methods exist.
 It is known that the oscillatory behaviour of the exponential function is mainly contained in the complex part of such an integrand.
 Therefore, if we are able to find a contour of integration in the {\mbr} representation along which an imaginary part is constant, the convergence problem can be reduced. This problem has been investigated recently for one-dimensional cases \cite{Gluza:2016fwh,Sidorov:2017aea} but going beyond is non-trivial and our present status is exploratory in character.
 
 \subsection{One dimension}
 To present the idea of the optimal contour of integration and its most important properties, we will start the discussion by considering a one-dimensional scenario. In this case, our integral has the form of the integral given by \Eref{lap}. We will say that a contour of integration is a contour of steepest descent $\mathcal{J}$ if it possesses the following three properties.
 \begin{enumerate}
 \item It goes through one or more critical points $z_{\alpha}$ of $f(z)$.
 \item The imaginary part of $f(z)$ is constant along $\mathcal{J}$.
 \item The real part of $f(z)$ decreases moving away from critical points.
 \end{enumerate}
 In the one-dimensional case, the contour of steepest descent is completely determined by these properties. Before we move on and try to generalize this idea to higher dimensions, let us look at the behaviour of the integrand $f(z)$ in the neighbourhood of its critical points. Since it is a holomorphic function, its imaginary and real parts are harmonic functions. Thus, by virtue of the maximum principle, none of them attains extremal values inside the domain of analyticity \cite{axler2001harmonic}. However, the imaginary part of $f(z)$ along the contour of steepest descent remains constant; therefore, we will look at a contour that goes from one valley into another on the landscape of the real part of $f(z)$ crossing the critical point, which, in this case, is a saddle point, see \Fref{fig:sad}.  Moreover, assuming that the function $f(z)$ has only non-degenerate critical points $z_{\alpha}$, \ie
 \begin{equation}
 \frac{\mathrm{d}^{2}f}{\mathrm{d}z^{2}}|_{z_{\alpha}} \neq 0,
 \end{equation}
 the behaviour of $f(z)$ near the critical point $z_{0}$ is mostly controlled by its second derivative
 \begin{equation}
 f(z)=f(z_{0})+\frac{1}{2}\frac{\mathrm{d}^{2}f}{\mathrm{d}z^{2}}|_{z_{0}}(z-z_{0})^{2}+
 \dots \Rightarrow f(z)-f(z_{0}) \approx \frac{1}{2}\frac{\mathrm{d}^{2}f}{\mathrm{d}z^{2}}|_{z_{0}}(z-z_{0})^{2}
 \, .
 \end{equation}
 
  \begin{figure}
 \centering
 \includegraphics[scale=0.3]{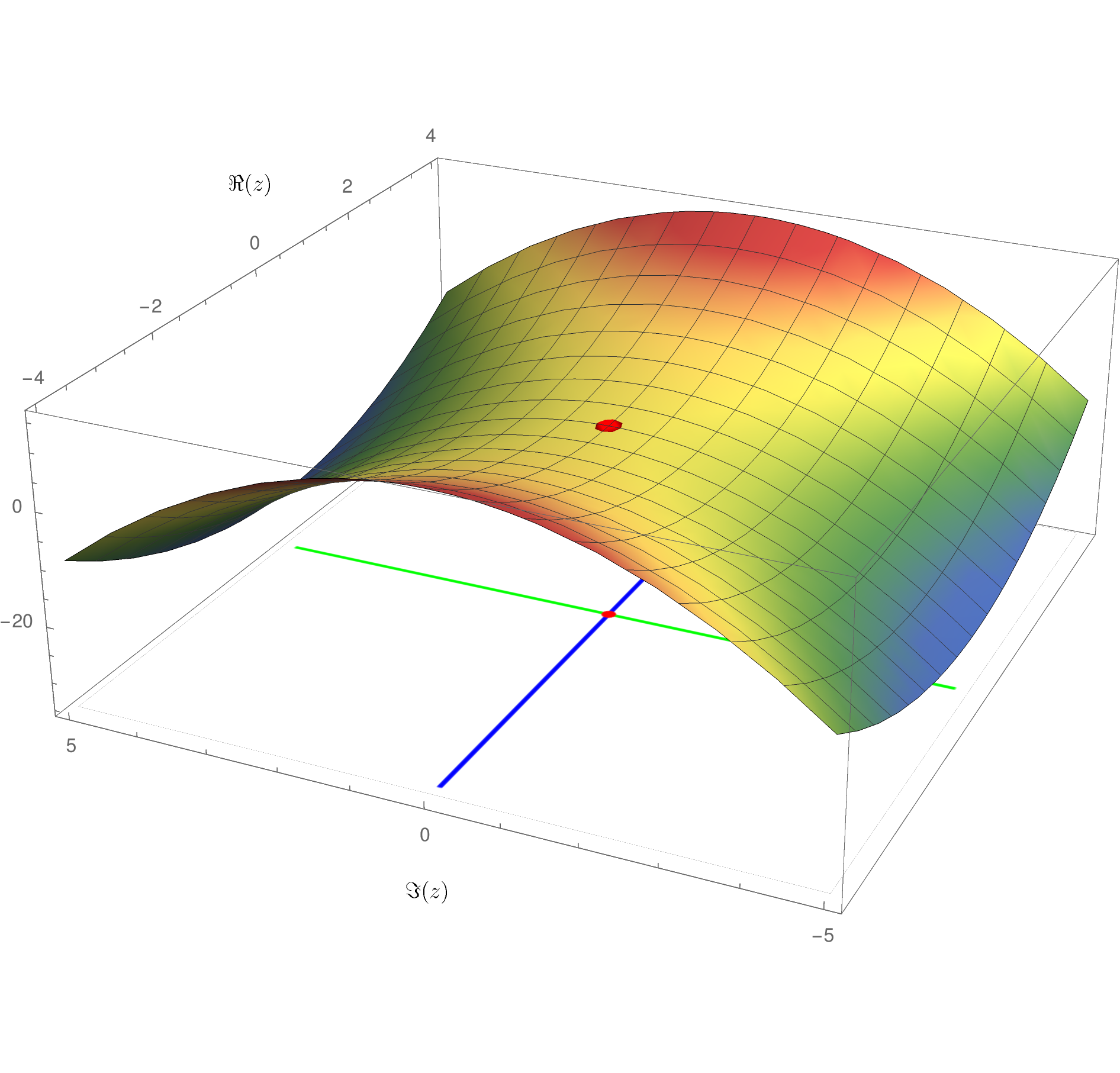} 
 \caption{ The real part of the function $f(z)=z^{2}$ with a marked critical value $f(z_{0})=0$ at the critical point $z_{0}=(0,0)$. 
 Below the surface of $\Re(z^{2})$, note the level set curves of $\Im(f(z))=\Im(f(z_{0}))$, which correspond to the contours of steepest descent (green) and steepest ascent (blue).}
 \label{fig:sad}
 \end{figure}
 
 Let us write the right-hand side of the second equation as follows:
 \begin{align}
 \frac{d^{2}f}{\mathrm{d}z^{2}}  \bigg \rvert_{z_{0}} & =r\mathrm{e}^{\mathrm{i}\theta}, \\
 (z-z_{0})^{2} & =R\mathrm{e}^{\mathrm{i}\varphi}.
 \end{align}
 This gives 
 \begin{equation}
 \frac{\mathrm{d}^{2}f}{\mathrm{d}z^{2}} \bigg \rvert_{z_{0}}(z-z_{0})^{2}=rR^{2}\mathrm{e}^{\mathrm{i}(\theta+2\varphi)}.
 \end{equation}
 The direction of steepest descent is given when the value of $\Re[f(z)-f(z_{0})]$ is negative and the imaginary part of this difference is equal to zero. This occurs when   
 \begin{equation}
 \varphi= -\frac{\theta}{2}+\frac{\pi}{2}.
 \end{equation}  
 However, since the critical point is a saddle point, there is also a direction of steepest ascent that is perpendicular to the direction of steepest descent. In this case, $\Re[f(z)-f(z_{0})]$ is positive and we have
 \begin{equation}
 \varphi= -\frac{\theta}{2}.
 \end{equation} 
 We can see that the direction of steepest descend is determined by the argument of $\frac{\mathrm{d}^{2}f}{\mathrm{d}z^{2}}|_{z_{0}}$. If it happens  that the second derivative vanishes, we must look at higher derivatives.

 \subsection{Extension to higher dimensions}
 Let us extend the idea of the contour of steepest descent to an arbitrary $n$-complex dimensional space, where the integrand takes the form $f(z)=f(z_{1},\dots,z_{n})$. We are looking for an universal parametrization of integration surfaces. 
Learning from the one-dimensional case, we know, that along the contour of steepest descent, the imaginary part of the integrand is constant and that this contour goes through saddle points of $f(z)$.  This is sufficient information in the one-dimensional scenario. However, the first condition,
 \begin{equation}
 \Im(f(z))-\Im(f(z_{0}))=0,
 \end{equation}
 is not sufficient in higher dimensions since, in the general $n$-complex space $\mathbb{C}^{n}$, this condition constrains only one real dimension. Thus, we are left with a $2n-1$ real dimensional variety. Therefore, either we find additional constraints or we  find a way to parametrize the surface of steepest descent. We will discuss the second option and try to use the one-dimensional definition and adapt it to higher dimensions. To begin, let us define the set of curves of steepest descent as the solution of the flow equation

 \begin{equation}\label{grad}
 \left(\frac{\mathrm{d} x_{1}}{\mathrm{d}t},\frac{\mathrm{d} y_{1}}{\mathrm{d}t},
 \dots ,\frac{\mathrm{d} x_{n}}{\mathrm{d}t},\frac{\mathrm{d} y_{n}}{\mathrm{d}t}\right)=-\nabla \Re[f(x_{1},y_{1},\dots ,x_{n},y_{n})] \, ,
 \end{equation}
 where $z_{k}=x_{k}+\mathrm{i}y_{k}$ for $k=1,2,\dots,n$. Taking $t$ as a nonnegative real parameter, this equation ensures that the real part of $f(z)$ decreases along these curves and that the imaginary part remains constant. The approach involving this flow equation as the parametrization of the surface of steepest descent as a whole is considered in Ref. \cite{Kaminski}. However, instead of looking for the whole surface of integration, we can use the idea of geometrical objects called Lefschetz thimbles $\mathcal{J}$, each associated with one critical point of $f(z)$. Equation \eqref{grad} can be written in the  form 
 \begin{equation}\label{flow}
 \frac{\mathrm{d} z_{i}}{\mathrm{d} t}=-\left(\frac{\partial f}{\partial z_{i}}\right)^{*}, \qquad i=1,\dots,n.
 \end{equation}
 This allows us to define Lefschetz thimbles associated with a given critical point as
 \begin{equation}
 \mathcal{J}_{\alpha}=\lbrace z\in \mathbb{C}^{n}: \lim_{t\rightarrow \infty} z(t)=z_{\alpha} \rbrace,
 \end{equation}
 where $z(t)$ is the solution of the flow equation (\Eref{flow}).
 Thus, the Lefschetz thimble is a collection of all curves flowing into the direction of steepest descent that asymptotically reach the corresponding critical point. Note that the real parts of the integrand for critical points are at the same time saddle points. There exists an alternative direction, in which the real part changes the most, \ie the direction of steepest ascent. With this direction are associated geometrical structures similar to the Lefschetz thimbles, called antithimbles, which satisfy the system of upward flow differential equations
 \begin{equation}
 \mathcal{K}_{\alpha}= \lbrace z \in \mathbb{C}^{n} : \lim_{t\rightarrow \infty} \bar{z}(t)=z_{\alpha} \rbrace,
 \end{equation}
 where $\bar{z}(t)$ are curves that satisfy the following system of upward flow equations:
 \begin{equation}
 \frac{\mathrm{d} z_{i}}{\mathrm{d} t}=\left(\frac{\partial f}{\partial z_{i}}\right)^{*}, \qquad i=1, \dots,n.
 \end{equation}
 The idea of thimbles as a surface of steepest descent appeared in Refs. \cite{Pham1,Pham2}. It has been shown there that, within homology theory, an initial contour of integration $\mathcal{C}$ can be changed into a finite sum of thimbles, each associated with one critical point,
 \begin{equation}
 \mathcal{C}=\sum_{\alpha}N_{\alpha}\mathcal{J}_{\alpha}.
 \label{split}
 \end{equation}
 The sum in this equation goes over critical points and $N_{\alpha} \in \mathbb{Z}$. In this way, we can change the integration over one surface into the sum of integrals over thimbles.
 In practical implementations of the thimble method, it is convenient to take, as the initial point, a point near the critical point, since none of the points on a thimble is known in the beginning, except for the critical point. Then instead of solving downward flow equations, which, in this situation, will result in a short line segment connecting the initial point to the critical one, we move in the reverse direction, \ie we solve upward flow equations. For the initial condition, it is also important to set a proper initial direction towards the steepest descent path. To do this, we mimic the method for the
one-dimensional case. Therefore, we look at the behaviour of  $f(z)$ in the neighbourhood of the critical point. In the general $n$-dimensional space, this behaviour is governed by the Hessian matrix $\mathcal{H}$ evaluated at the critical point, \ie the matrix of partial second derivatives of $f(z)$
 \begin{equation}
 f(z)=f(z_{0})+\frac{1}{2}(z-z_{0})^\mathrm{T}\mathcal{H}|_{z_{0}}(z-z_{0})+\dots
 \, ,
 \end{equation} 
 with
 \begin{equation}
 \mathcal{H}=
  \begin{pmatrix}
 \frac{\partial^{2}f}{\partial z_{1}^{2}} & \frac{\partial^{2}f}{\partial z_{1}z_{2}} & \cdots & \frac{\partial^{2}f}{\partial z_{1}z_{n}} \\
 \frac{\partial^{2}f}{\partial z_{2}z_{1}} & \frac{\partial^{2}f}{\partial z_{2}^{2}} & \cdots & \frac{\partial^{2}f}{\partial z_{2}z_{n}} \\
 \vdots & \vdots & \ddots & \vdots \\
 \frac{\partial^{2}f}{\partial z_{n}z_{1}} & \frac{\partial^{2}f}{\partial z_{n}z_{2}} & \cdots & \frac{\partial^{2}f}{\partial z_{n}^{2}}
 \end{pmatrix}
.
 \end{equation} 
 The direction of steepest ascent or descent is then determined by eigenvectors of this matrix. 

There  exists  another approach of the Lefschetz thimbles parametrization. This approach appears in Ref. \cite{Pham1} and is based on cuts $L_{\alpha}$, \ie half-lines attached to the value of $f(z)$ at the critical points and going towards infinity in the direction of the positive real axis
 \begin{equation}
 L_{\alpha}=(f(\alpha),t_{0}),
 \end{equation}
 where $t_{0}=f(\alpha)+r_{0}\mathrm{e}^{\mathrm{i}\phi}$, where $\phi$ is the phase of the $\lambda$ parameter and $r_{0}$ denotes a small radius of the neighbourhood of a given critical point.
 The relation between thimbles and cuts follows from the behaviour of the function $f(z)$ near the critical point, which tells us that thimbles are mapped by $f(z)$ into cuts
 \begin{equation}
 \mathcal{J}_{\alpha} \xrightarrow{\quad f \quad} L_{\alpha}.
 \end{equation}
 Thus, to find thimbles starting from cuts, we must consider the inverse function $f^{-1}$
\begin{equation}
 L_{\alpha} \xrightarrow{\quad f^{-1} \quad} \mathcal{J}_{\alpha}.
 \end{equation}
 A detailed discussion of this method, with an example, can be found in Ref. \cite{Delabaere_globalasymptotics}.
 
 \subsection{Application to physics and \mbr{} trials}
 The idea of choosing the initial contour of integration  as the contour of steepest descent was considered in a study of the evolution of parton distribution in QCD \cite{Kosower:1997hg}. Recently, Witten used geometric and topological properties of the Lefschetz thimbles to prove analytical continuation of the Chern--Simons theory \cite{Witten:2010zr,Witten:2010cx}. These works also contain a nice introduction to  Morse theory and Lefschetz thimbles in the context of physics. The methods of Lefschetz thimbles and contours of steepest descent are also used  in calculations in the context of the sign problem in QCD, which is connected to cancellations occurring in oscillatory lattice integrals. Here, both  theoretical and practical aspects of thimbles are studied. Much effort has been put into the development of Monte
Carlo algorithms to compute system dynamics on thimbles \cite{Cristoforetti:2012su, Fujii:2013sra, Alexandru:2015xva, Alexandru:2015sua, Tanizaki:2017yow, Alexandru:2017czx, Bluecher:2018sgj}.
  In the case of Mellin--Barnes integrals, both exact contours of steepest descent as a solution of flow equations and some useful approximations have
been studied \cite{Gluza:2016fwh, Sidorov:2017aea}.
  
Despite the fact that thimbles are described by well-established mathematical theory \cite{nicolaescu2007invitation}, a practical parametrization of them is a very challenging task.  
In the case of \mbr{} integrals, we must face the following issues:  
\begin{itemize}
  
\item since the gamma function has infinitely many critical points,  how many of them are necessary to match the assumed numerical precision;
  
\item the parametrization of thimbles by solving flow equations for each relevant critical point: this stage also requires  calculation of the Hessian matrix and the associated eigensystem;
  
\item  knowledge of singularities of the {\mbr} integrand, which is necessary to avoid them by thimbles; 
  
\item  determination of the inverse function for the cuts approach;
  
\item summation of thimbles with suitable coefficients in \Eref{split}, to reconstruct the whole integral. 
  
\end{itemize}

Therefore, the  work done so far concerns  one-dimensional problems, with very few exceptions, such as the two-dimensional case discussed recently in Ref. \cite{Alexandru:2016ejd}.  
 
 \subsection{Summary}
 The Mellin--Barnes representation is one of the methods for the computation
of  Feynman integrals. The main problem with numerical integration of these integrals is bad convergence, owing to highly oscillatory behaviour of integrands. Fortunately, since they are defined in the complex domain, we can change the contour of integration and find the optimal contour of integration along which the imaginary part of the function is constant, defining the contour of steepest descent. In the one-dimensional case, the condition that the imaginary part of the integrand must be constant, together with a requirement that the contour must go through the critical point is sufficient to parametrize it. This idea can be extended to higher dimensions  with the help of geometric structures called Lefschetz thimbles, where each thimble is associated with one critical point of the integrand. However, the practical implementation of this idea to a particular problem is  very difficult. We discussed two known approaches to this issue.  Namely, thimbles can be parametrized as the union of curves given by the solution of flow differential equations. Alternatively, a correspondence between thimbles and half-lines connected to the value of a function at the critical point can be used. Both approaches require analytical knowledge of the integrand at all its critical points. Moreover, the singularity structure of the function is very important, since the new contour cannot cross any singularity. Knowledge of singularities is also needed  in the second approach, where the inverse function is required. A numerical solution of the flow equations requires  calculation of all critical points, the Hessian matrix, and its eigensystem to make the construction of thimbles efficient. Currently, the methods of Lefschetz thimbles is developed in the lattice QCD and in the context of Mellin--Barnes integrals. However, only simple one-dimensional problems are solved. Multidimensional cases are still in the exploratory stage and more work is needed. Though we are aiming at numerical applications, the method of Lefschetz thimbles and the homology theory, singularity theory, or multidimensional complex analysis required to study  complex contours can be helpful in obtaining a deeper understanding of the analytical properties of  Feynman integrals.

 \label{sec-mbth}
\cleardoublepage
\cleardoublepage

\section
[Differential equations for multiloop integrals \\ {\it R. Lee}]
{Differential equations for multiloop integrals} \label{sec:rlee}

\pagestyle{fancy}
\fancyhead[CO]{\thechapter.\thesection 
\hspace{1mm}
Differential equations for multiloop integrals
}
\fancyhead[RO]{}
\fancyhead[LE]{}
\fancyhead[CE]{R. Lee}
\fancyhead[RE]{} 

\noindent
{\bf Author: Roman Lee} {~~[r.n.lee@inp.nsk.su]}
\vspace*{.5cm}

\noindent The method of differential equations is one of the most powerful methods available for  multiloop calculations.
First introduced by Kotikov in Refs. \cite{Kotikov:1990kg,Kotikov1991a,Kotikov1991b},
and later generalized by Remiddi \cite{Remiddi:1997ny} for  differentiation with respect to kinematic invariants,  this technique relies on the IBP reduction procedure, another powerful technique suggested by Chetyrkin and Tkachov  \cite{Chetyrkin:1981qh,Tkachov:1981wb}. In this contribution, we will review the recent advances and the present status of the differential equation method.

\subsection{Obtaining differential equations}

Integration-by-part identities originally relied on the momentum-space representation of the multiloop integral in $d$ dimensions
\begin{equation}\label{eq:loop_integral}
 j(\boldsymbol {n}) =  j(n_1,\ldots, n_N) = \int \frac{\mathrm{d}^dl_1\cdots \mathrm{d}^dl_L}{(\mathrm{i}\pi^{d/2})^LD_1^{n_1}\cdots D_N^{n_N}}\,.
\end{equation}
Here, the `denominators' $D_\alpha$ depend linearly on the scalar products $s_{ij}=l_i\cdot q_j$ involving loop momenta $l_i$ ($q_j$ is either loop or external momentum). We assume that these denominators constitute a basis, in the sense that any scalar product $s_{ij}$ can be uniquely  expressed as a linear combination of $D_1,\ldots,D_N$ and a free term. In particular, $N=L(L+1)/2+LE$, where $L$ is a number of loops and $E$ is a number of linearly independent external momenta.

The IBP identities have the form
\begin{equation}
   0 = \int \frac{\mathrm{d}^dl_1\cdots \mathrm{d}^dl_L}{(\mathrm{i}\pi^{d/2})^L}\partial_i\cdot q_j D_1^{-n_1}\cdots D_N^{-n_N}\,.
\end{equation}
Performing explicit differentiation, we express the right-hand side in terms of the integrals of Eq. (\ref{eq:loop_integral}) with shifted indices. It is convenient to represent the resulting identities with the help of operators $A_\alpha$ and $B_\alpha$,
\begin{align}
  (A_\alpha f)(\ldots,n_\alpha,\ldots) &= n_i f(\ldots,n_\alpha+1,\ldots)\nonumber\\
  (B_\alpha f)(\ldots,n_i,\ldots) &= f(\ldots,n_\alpha-1,\ldots) \, .
\end{align}
Then the IBP identities have the structure
\begin{eqnarray}
  0 = \mathcal{O}_{ij}j(\boldsymbol {n}) = (a_{ij}^{\alpha\beta}A_\alpha B_\beta+b_{ij}^{\alpha}A_\alpha + d\delta_{ij})j(\boldsymbol {n})\,,
\end{eqnarray}
where the explicit form of the coefficients $a_{ij}^{\alpha\beta}$ and $b_{ij}^{\alpha}$ is determined by that of the functions $D_\alpha$. Solving these identities for different $\boldsymbol {n}, i,j$  with respect to the most complex integrals, we can collect a database of rules, which reduce all integrals (\Eref{eq:loop_integral}) to a finite set of what are called \emph{master integrals}
\begin{equation}
  \boldsymbol {j} =(j_1,\ldots,j_r)^\mathrm{T} = \left(j(\boldsymbol {n}^{(1)}),\ldots,j(\boldsymbol {n}^{(r)})\right)^\mathrm{T}
\end{equation}
Note that the set of master integrals depends on the chosen criterion of complexity, which can vary within certain limits. In this review, we will not discuss several important properties of the IBP reduction, such as the Lie algebra of the IBP operators $\mathcal{O}_{ij}$, the proof of finiteness of the set of master integrals, the connection between the number of master integrals and topological invariants of Feynman and Baikov polynomials, or
 the derivation of IBP identities and differential equations directly in Feynman and Baikov representation using syzygies and other algebraic geometry tools, and refer the interested reader to Refs. 
\cite{Lee2008,SmirPet2010,LeePomeransky2013,Lee2014a,Larsen:2015ped,Bosma:2017hrk,Boehm2017,Bitoun2017,Boehm:2018fpv}.

Although the original idea behind the IBP reduction was to reduce the number of integrals to be calculated, in no way less important is the possibility of using IBP reduction to obtain differential and difference systems for a finite number of master integrals.

Let the integrals depend on some external parameter $x$. In particular, $x$ may be some mass or kinematic invariant constructed of the external momenta. Then we can express the derivative of the master integrals via the integrals of the same form (\Eref{eq:loop_integral}). If $x$ is mass, this is obvious, as one may directly differentiate the integrand. If $x$ is a kinematic invariant, one should take into account the fact that the integrand depends on external momenta but, in general, not on their invariants. One may use the formulae
\begin{align}
\frac{\partial}{\partial\left(p_{1}\cdot p_{2}\right)}j\left(\boldsymbol {n}\right) & =\sum\left[\mathbb{G}^{-1}\right]_{i2}p_{i}\cdot\partial_{p_{1}}j\left(\boldsymbol {n}\right), \nonumber \\
\frac{\partial}{\partial\left(p_{1}^{2}\right)}j\left(\boldsymbol {n}\right) & =\frac{1}{2}\sum\left[\mathbb{G}^{-1}\right]_{i1}p_{i}\cdot\partial_{p_{1}}j\left(\boldsymbol {n}\right),\label{eq:Dinv}
\end{align}
where $\mathbb{G}=\{p_{i}\cdot p_j\}_{i,j=1\ldots E}$ is a Gram matrix of the external momenta $p_i$. The operators on the right-hand side can act directly on the integrand.

Using IBP reduction, we may reduce the result of differentiation to the same set of master integrals. Thus, we obtain the system of the first-order differential equations, which is natural to represent in matrix form
\begin{equation}\label{eq:DE}
  \partial_x {\boldsymbol j} = M  {\boldsymbol j}\,,
\end{equation}
where $M$ is an $r\times r$ matrix with entries being the rational functions of $x$ and $d$.

The idea of the differential equation method is that, instead of performing the explicit loop integration in \Eref{eq:loop_integral}, one may search the master integrals as solutions of  the differential equations (\Eref{eq:DE}). This method appears to be superior in the majority of applications. Also, many properties of the loop integrals can  easily be read from the differential system even before solving it. In particular, the branching points of the integrals necessarily coincide with the singular points of the differential system (the poles of $M$).

Let us note that, for the cases when the loop integrals depend only on one dimensionful parameter, the differential equations cannot help, as the system trivializes to a set of decoupled equations of the form $\partial_x  j_k =\nu_k x^{-1}  j_k$, where $\nu_k$ is the physical dimension of $j_k$ defined as $[j_k]=[x^{\nu_k}]$. In this case, the differential equations may only serve as a check of completeness of the IBP reduction.

It is very instructive to use the differential equations for the cases when there are several dimensionful parameters $x_1,\ldots x_k$. Then we have one differential system for each  parameter $x_a$:
\begin{equation}
  \partial_a {\boldsymbol j} = M_a  {\boldsymbol j}\,,
\end{equation}
where now each matrix $M_a$, in addition to $d$, depends, in general, on all variables $\boldsymbol x$. There is a non-trivial compatibility condition:
\begin{equation}\label{eq:compatibility}
  \partial_b M_a - \partial_a M_b + [M_a,M_b]=0\,,
\end{equation}
which can be thought of as the flatness of the connection $\partial_a-M_a$. In this review, we will mostly concentrate on the differential system with respect to one variable and discuss the multivariate case at the end.

\subsection{$\epsilon$-expansion of differential equations}

Let us assume that $M$, as a function of $x$, does not have singular points that depend on $\epsilon$. Strictly speaking, this might not be the case. Even worse: in the denominators of entries of $M$, there might appear powers of ugly irreducible polynomials depending simultaneously on $x$ and $\epsilon$. Fortunately, such singularities must be apparent; this can be understood on physical grounds and also strictly proved from the properties of Feynman parametrization (or Baikov representation). Such apparent singularities can be removed in a systematic way that does not require explicit manipulations with roots of these polynomials. In particular, to reduce the powers to, at most, one, it is instructive to use the algorithm of Ref. \cite{Barkatou1995}. The algorithm of eliminating first powers of irreducible polynomials in the denominators will be presented elsewhere. Thus, from now on, we assume that all singularities of $M$, as function of $x$, are independent of $\epsilon$.

In general, the solution of the differential equations exact in $d$ cannot be found in a close form. Fortunately, we typically need the expansion of the loop integral in $\epsilon=(4-d)/2$. Therefore,  knowledge of the exact-in-$\epsilon$ solution  is not necessary for physical applications (and may not even be sufficient given the complexity of series expansion of some emerging functions in $\epsilon$). Assuming that the solution of Eq. (\ref{eq:DE}) can be expanded in Laurent series\footnote{Note that the existence of the Laurent series does not follow solely from the rationality of $M$ (consider, e.g, $\partial_x f= f / {\epsilon x}$), but  rather can be safely assumed and checked \emph{a posteriori} for all differential systems that appear in multiloop calculations.} in $\epsilon$,
\begin{equation}\label{eq:solution_epsilon_expansion}
  \boldsymbol j = \sum \boldsymbol j^{(n)}\epsilon^{n} \, ,
\end{equation}
we can search for the coefficients $\boldsymbol j^{(n)}$ by expanding both sides of Eq. (\ref{eq:DE}) in $\epsilon$.

This is, of course, valid when we hold $x$ to be general (\ie as a symbol), but we should be careful when applying boundary conditions in a singular point of the differential equations. If $x_0$ is a singular point of the matrix $M$, one should bear in mind that $\epsilon$-expansion, in general, does not commute with the expansion in $x-x_0$. In particular, it is not safe to put $x=x_0$ in the $d$-dimensional integral and then expand in $\epsilon$ to obtain the boundary conditions. The irony is that  we usually do need to fix boundary conditions at a singular point of $M$. One should, therefore, consider the asymptotics of the integrals when $x\to x_0$ and use the expansion-by-regions method \cite{BeneSmi1998} to obtain contributions of different regions.

\subsection{$\epsilon$-form of the differential systems}

In Ref. 
\cite{Henn:2013pwa}, a remarkable observation is  made. It appears that, in many cases, a suitable choice of the master integrals exists, such that the matrix $M$ in  \Eref{eq:DE} is plainly proportional to $\epsilon$, \ie when $M(x,\epsilon)=\epsilon S(x)$, so that
\begin{equation}\label{eq:DEeps}
\partial_x {\boldsymbol J} = \epsilon S (x) {\boldsymbol J}\,,
\end{equation}
Here, we intentionally changed notation ${\boldsymbol j}\to {\boldsymbol J}$ in order to have a reference both to generic masters and to the `canonical' masters.
This form (we will refer to it as the \emph{$\epsilon$-form}) makes the $\epsilon$-expansion of Eq. (\ref{eq:DE}) trivial, and one can obtain as many terms of \Eref{eq:solution_epsilon_expansion} as is needed by successive integration, using
\begin{equation}
  \boldsymbol J^{(n)}(x) = \int \mathrm{d}x S(x)\boldsymbol J^{(n-1)} (x)
  \, .
\end{equation}
Immediately after  Ref. 
\cite{Henn:2013pwa} appeared, the $\epsilon$-form of the differential equations was used in many practical applications, Refs.
\cite{HennSmirnov2013,HennSmirnovSmirnov2013,Henn:2014lfa,Henn:2013nsa},
to mention a few.

It is instructive to write the general solution of the differential system (\Eref{eq:DEeps}) in the form
$\boldsymbol J(x)=F(x,x_0)\boldsymbol J(x_0)$, where the fundamental matrix $F(x,x_0)$ is expressed via the path-ordered exponent:
\begin{equation}
        F(x,x_0)=\Pexp\left[\epsilon \int_{x_0}^{x} \mathrm{d}x'M(x')\right]\,.
\end{equation}
We will refer to $F(x,x_0)$ as an evolution operator from $x_0$ to $x$. Then, using the well-known form of the perturbative expansion of the path-ordered exponent, we obtain
\begin{equation}\label{eq:PexpExpansion}
F(x,x_0)=\sum_{n=0}^{\infty}\epsilon^n F^{(n)}(x,x_0)=\sum_{n=0}^{\infty}\epsilon^n \mathop{\int\cdots \int}\limits_{x_0<x_1\dots<x_n<x} \mathrm{d}x_n\cdots \mathrm{d}x_1 S(x_n)\cdots S(x_1) \, .
\end{equation}
If the entries of the matrix $S(x)$ are rational functions, the integrals in this equation are expressed in terms of multiple polylogarithms, Ref. \cite{Goncharov:1998kja},
or equivalent functions $G$, see the following.

Let us explain how the boundary conditions should be fixed when $x_0$ tends to a singular point of the differential system. Suppose that $x=0$ is a singular point of the matrix $S$ and $x_0\to 0$. Then, obviously, $\lim_{x_0\to 0}F(x,x_0)$ does not exist, as in Eq. \eqref{eq:PexpExpansion} the integrals over $x_1$ diverge at the lower limit. 
We will assume that $x=0$ is a simple pole of $S(x)$, \ie that 
\begin{equation}
        S(x)=\frac{A}{x}+B(x)\,,
\end{equation}
where $B(x)$ is regular at $x=0$. Then we define a reduced evolution operator
\begin{equation}
        \tilde{F}(x,0)=\lim_{x_0\to 0}F(x,x_0)x_0^{\epsilon A}\,.
\end{equation}
Here and in the following we assume that we have fixed a path in the complex plane starting from $0$ and containing both $x$ and $x_0$, and that the limits $x_0\to 0$ or $x\to 0$ are understood in the sense that the corresponding variable travels along this path towards zero. Note that if $x=0$ is a regular point of the differential system, we have $A=0$ and, therefore, $\tilde{F}(x,0)={F}(x,0)$. It can be understood that the thus-defined function $\tilde{F}$ has the asymptotics
\begin{equation}
        \tilde{F}(x,0)\stackrel{{x\to 0}}{\longrightarrow}x^{\epsilon A}\,.
\end{equation}
The convenience of our definition for $\tilde F$ becomes especially obvious if the matrix $S(x)$ is rational, decays at infinity, and has only simple poles. Then, we have a symbolic identity
\begin{equation}
        \tilde{F}^{(n)}(x,0)= S(x_n)\cdots S(x_1)|_{\frac1{x_n-a_n}\cdots\frac1{x_1-a_1}\to G(a_n,\ldots, a_1|x)}\,,
\end{equation}
where $G$ is defined recursively as 
\cite{Vollinga:2004sn}
\begin{align}
        G(\underbrace{0,\ldots,0}_n|x)&= \frac{\ln^nx}{n!}\,,\nonumber\\
        G(a_n,\ldots,a_1|x)&= \int\limits_{0}^{x} \frac{\mathrm{d}x_n}{x_n-a_n}G(a_{n-1},\ldots,a_1|x_n)\,.
\end{align}

To fix the boundary conditions, we write the general solution as 
\begin{equation}
        \boldsymbol J(x)=\tilde{F}(x,0)\boldsymbol{J}_0\,,
        \label{eq:DEsolution}
\end{equation}
where $\boldsymbol{J}_0$ is a column of constants. These constants can be related to the specific coefficients in the asymptotic expansion of $J(x)$ defined by that of $\tilde{F}(x,0)$
\begin{equation}
        \tilde{F}(x,0)=\sum_{\lambda\in \mathcal{S}}\sum_{n=0}^{\infty}\sum_{k=0}^{K_{\lambda}}\frac1{k!}C\left(n+\lambda,k\right)x^{n+\lambda}\ln^{k}x\,.
        \label{evolution_expansion}
\end{equation}
We refer the reader to Ref. \cite{Lee2018a} for details on how to find this expansion in an algorithmic way. Here, we only note that the described method is very economical, in the sense that it enables determination of  the minimal set of asymptotic contributions needed to fix the constants $\boldsymbol J_0$.

\subsection{Reducing differential systems to $\epsilon$-form}

As we have seen in the previous section, casting differential equations in $\epsilon$-form is very desirable. This raises the question of how to find the appropriate transformation. One possible approach is to use some additional information that is not contained in the differential systems. Namely, one might try to find the set of master integrals, having the homogeneous transcendentality property, by means of the constant leading singularity method. This approach is supported in Ref. 
\cite{Henn:2013pwa}  and has been used in many applications. However, it is very desirable to be able to reduce a given differential system to $\epsilon$-form relying only on its form. This is especially required for the complicated cases, in particular, for the massive diagrams where the required transformation is no longer rational  in terms of the original kinematic parameters. 

Thus, we would like to find the transformation matrix $T$ that connects the initial functions $\boldsymbol j$ and new functions $\boldsymbol J$ by
\begin{equation}
        \boldsymbol j = T\boldsymbol J\,,
\end{equation}
such that $\boldsymbol J$ satisfies Eq. \eqref{eq:DEeps}.

The algorithm of finding the rational transformation $T$ is  presented in Ref.  \cite{Lee:2014ioa}.
There are now two public implementations of the algorithm of Ref. \cite{Lee:2014ioa}, 
\texttt{Fuchsia} 
\cite{Gituliar:2017vzm} and \texttt{epsilon} 
\cite{Prausa:2017ltv}. The algorithm consists of a series of transformations, $T=T_1T_2\ldots T_K$, each `improving' the properties of the system. The process of reduction can be approximately split into three stages.
\setlist[description]{font=\normalfont\itshape}
\begin{description}
        \item[Fuchsification.] The multiple poles and polynomial part of $M$ are removed.
        \item[Normalization.]  The eigenvalues of the matrix residues are normalized to be proportional to $\epsilon$.
        \item [Factoring out $\epsilon$.] A constant transformation depending on $\epsilon$ is sought to factor out $\epsilon$.
\end{description}

\subsubsection{Fuchsification}
Reducing the differential system to Fuchsian form is a classical problem of ordinary differential equation theory; the algorithms to accomplish this task have been known long before Ref. 
\cite{Lee:2014ioa}. In particular, the algorithm of Barkatou and Pfl\"ugel, Refs. \cite{Barkatou2007,barkatou2009moser}, allows one to get rid of all multiple poles. However, the polynomial part of the matrix, in general, explodes after applying this algorithm. The reason is that each separate step of the Barkatou--Pfl\"ugel algorithm can increase the order of the polynomial part by one. A possible cause of such an inaccurate treatment of the singularity at infinity is that, in general, it is not possible to get rid of both the multiple poles and the polynomial part. This impossibility is closely related to the negative solution of the $21$st Hilbert problem,\footnote{Formulation of the 21st Hilbert problem: to show that there always exists a linear differential system of the Fuchsian class, with given singular points and monodromic group.} given by Bolibrukh \cite{Bolibrukh1989}. 

In Ref. 
\cite{Lee:2014ioa}, transformations similar to those of Refs. \cite{Barkatou2007,barkatou2009moser} are used. However, they are additionally adjusted to conserve, if possible, the properties of the system in the second point (always chosen to be $\infty$ in the  Barkatou--Pfl\"ugel algorithm). To explain one step of the `Fuchsification' stage, let us suppose that $x_1\neq\infty$ and $x_2\neq\infty$ are singular points of the differential system (\Eref{eq:DE}). Near these points, the matrix $M$ has the expansions:
\begin{align}
        M(x)&=\frac{A_0}{(x-x_1)^{p+1}} + \frac{A_1}{(x-x_1)^{p}}+O\left((x-x_1)^{1-p}\right)\,,\\
        M(x)&=\frac{B_0}{(x-x_2)^{q+1}} +O\left((x-x_2)^{-q}\right)\,,
\end{align}
where $p>0$ and $q\geqslant0$ are the \emph{Poincar\'e ranks} of the system at $x=x_1$ and $x=x_2$. The algorithm of Ref. 
\cite{Lee:2014ioa} consists in finding the \emph{balance} transformation
\begin{equation}\label{eq:balance}
        T=\mathcal{B}(P,x_1,x_2|x) = \bar{P}+\frac{x-x_2}{x-x_1}P\,,
\end{equation}
where $P=P^2$ is some projector, $\bar{P}=1-P$, and the following properties of $P$ are required. Let 
\begin{align*}
U&=\Ima P=\{\boldsymbol u|\ \exists \tilde{\boldsymbol u}: \boldsymbol u = P\tilde{\boldsymbol u}\}
\intertext{and}
V^{\intercal}&=\coIma P=\{\boldsymbol v^{\intercal}|\ \exists \tilde{\boldsymbol v}^{\intercal}:\boldsymbol v^{\intercal} = \tilde{\boldsymbol v}^{\intercal}P\} 
\end{align*}
be the image and coimage of $P$. Then we require
\begin{align}
\mathrm{1.}\ & A_{0}U = 0,\nonumber\\
\mathrm{2.}\ & A_{1}U \subseteq  \Ima A_{0}+U,\nonumber\\
\mathrm{3.}\ & U \cap \Ima A_{0}>\{0\},\label{eq:conds}
\intertext{and}
\mathrm{4.}\ & V^{\intercal}B_{0}\subseteq V^{\intercal}\,.
\label{eq:cond4}
\end{align}
The first three conditions concern the image of $P$ and are basically accounted for in the Barkatou--Pfl\"ugel algorithm. The last condition secures that  \Eref{eq:balance} does not increase the Poincar\'e rank at $x=x_2$. More specifically, it is straightforward to check that the transformed matrix $\tilde{M}=T^{-1}\left(MT-\partial_xT\right)$ has the expansions 
\begin{align}
\tilde{M}(x)&=\frac{\tilde A_0}{(x-x_1)^{p+1}} + O\left((x-x_1)^{-p}\right)\,,\\
\tilde{M}(x)&=\frac{\tilde B_0}{(x-x_2)^{q+1}} +O\left((x-x_2)^{-q}\right)\,,
\end{align}
where 
\begin{equation}
        \tilde{A}_{0}=\bar{P} \left(A_{0}+(x_1-x_2)A_{1}P\right)\,.
\end{equation}
The first expansion is valid thanks to condition 1 ($A_{0}U=0$), while the second is due to condition 4. It is easy to show that $\rank \tilde{A}_0<\rank A_0$. Indeed, from condition 2, we have $\Ima(A_{0}+(x_1-x_2)A_{1}P)\subseteq \Ima A_{0}+U$, and from condition 3 it follows that $\bar{P}(\Ima A_{0}+U)\subsetneq \Ima A_{0}$ (strict inclusion). Therefore, $\Ima \tilde A_{0} \subsetneq \Ima A_{0}$ which implies $\rank \tilde{A}_0<\rank A_0$.

In Refs. \cite{Barkatou2007,barkatou2009moser,Lee:2014ioa},
finding the subspace $U$ with  properties 1 to 3 relies on quite technical issues (in particular, Claim 1 and Algorithm 1  of Ref. \cite{Lee:2014ioa}).
Here, we will present a much  simpler algorithm to find $U$. We start from  Moser's necessary criterion of reducibility \cite{Moser1959}:
\begin{equation}\label{eq:crit1}
\left.x^r \det(A_0/x+A_1-\lambda)\right|_{x=0}=0\,,
\end{equation}
where $r=\rank A_0$. Let us note that this criterion can be reformulated as
\begin{equation}\label{eq:crit2}
        \dim \ker \begin{pmatrix}
        A_{0} & A_{1}-\lambda\\
        0 & A_{0}
        \end{pmatrix} > \dim \ker A_0\,.
\end{equation}
To understand the equivalence of Eqs. \eqref{eq:crit1} and \eqref{eq:crit2}, it is instructive to rely on the basis where $A_0$ is in Jordan form. Then, striking out the rows and columns that contain ones in the upper-left and lower-right blocks (blocks corresponding to $A_0$), we can easily identify the matrix $L(A,\lambda)$ from Ref. \cite{Barkatou2007} standing in the upper-right block. Equation \eqref{eq:crit2} simply states that the number of null eigenvectors of the matrix
\[
 \mathcal{A}= \begin{pmatrix}
A_{0} & A_{1}-\lambda\\ 0 & A_{0}\end{pmatrix}
\]
 should be larger than that of $A_0$. Let us note that if $u$ is a null eigenvector of $A_0$ then $u \choose 0 $ is that of $\mathcal{A}$, and {vice versa}. Therefore,  \Eref{eq:crit2} states that there must be at least one null eigenvector of $\mathcal{A}$ of the form $w(\lambda)\choose u(\lambda)$ with $u(\lambda)\neq0$. The null eigenvectors can be routinely found  and their components are, in general,  rational functions of $\lambda$. As one can always get rid of a
common denominator, we can assume that these components are polynomials in $\lambda$ and
\begin{equation}
        u(\lambda)=u_0+u_1\lambda +\cdots + u_k \lambda^k
\end{equation}

Now, we can easily check that $U=\{c_0u_0+\dots + c_ku_k|\ c_i\in \mathbb{C}\}$ satisfies conditions 1 to 3 in \Eref{eq:conds}. We leave this as a simple exercise for the reader. 

Once we have found subspaces $U$ and $V^{\intercal}$ satisfying Eqs. \eqref{eq:conds} and \eqref{eq:cond4} and having equal dimensionality, it is easy to reconstruct the projector $P$ with $\Ima P=U$ and $\coIma P=V^{\intercal}$. To do this, it is convenient to slightly abuse the notation by denoting as $U$ not the image of $P$ itself, but the matrix whose columns constitute the basis of the image. (Note: in general, $U$ is a rectangular matrix.) Equivalently, $V^{\intercal}$ will denote the matrix whose columns constitute the basis of the coimage. Then 
\begin{equation}
        P=U(V^{\intercal}U)^{-1}V^{\intercal} \, .
\end{equation}
This form make obvious one possible obstruction to the construction of the projector: the matrix $V^{\intercal}U$ may be singular. If, among many possible choices of $U$ and $V$ pairs, we do not manage to find one suitable for the construction of the projector, this is a strong indication of the irreducibility to Fuchsian form (so to speak, the `Bolibrukh counterexample').

Repetitive application of  \Eref{eq:balance} with the projector $P$ satisfying \Eref{eq:conds} reduces the matrix rank of the leading coefficient $A_0$ to zero (which means that the coefficient itself becomes zero). If, in addition, on each step we manage to find a suitable singular point $x_2$, such that  \Eref{eq:cond4} can be fulfilled, we can reduce the Poincar\'e rank at $x=x_1$ to zero. Acting in the same way for all singular points, we can eliminate higher-order poles and polynomial terms in $M$. The point $x=\infty$ does not make too much difference as, \eg it can be mapped to $y=0$ by the variable change $x=y^{-1}$.

\subsubsection{Normalization of matrix residues and choice of variable}
Suppose now that we have successfully achieved the global Fuchsian form of the differential system, \ie the matrix $M$ has the form
\begin{equation}
        M=\sum_{i=1}^{k} \frac{R_i(\epsilon)}{x-x_i}.
\end{equation}
The next step of the algorithm of Ref. \cite{Lee:2014ioa} 
is the normalization of the matrix residues $R_i(\epsilon)$. We want to find the rational transformation that makes all eigenvalues of all matrix residues to be proportional to $\epsilon$. Since the rational transformations do not change the monodromy group, they can only shift the eigenvalues by integer values. Therefore, if not all eigenvalues have the form 
\begin{equation}\label{eq:evs}
        n+ k\epsilon \, ,
\end{equation}
where $n$ is an integer, there is no rational transformation that normalizes the matrix residues. One might try to extend the class of `allowed' transformations by considering those that are rational in a new variable $y$ connected with $x$ by 
\begin{equation}\label{eq:xviay}
        x=f(y)\,,
\end{equation}
 where $f(y)$ is some rational function.  Indeed, such transformations might help. We stress that this kind of variable change retains Fuchsian form, as can be checked explicitly.
 
Suppose that the eigenvalues of the matrix residue at point $x=x_1$ have the form of \Eref{eq:evs} but with rational $n$. Let $b$ be the common denominator of the eigenvalues at $\epsilon\to 0$. Then the necessary condition is that the function $f^{-1}_{y^*}(x)$ has the $B$-root branching point at $x=x_1$, where $B$ is a multiple of $b$. By $y^*$, we denote a separate pre-image of $x_1$, so that $f(y^*)=x_1$, and $f^{-1}_{y^*}(x)$ denotes a local inversion of the function $f$ in the vicinity of $y^*$. In Ref. \cite{Lee:2017oca}, 
the frequently encountered case $b=2$ (in the following, we refer to singular points with $b=2$ as \emph{square root branching points}) has been considered in detail. The results of these considerations are as follows.
\begin{itemize}
        \item If there are two square root branching points, one should map them to $0$ and $\infty$ by M\"obius transformation and make the canonical\footnote{Here, by `canonical' variable change, we mean that once this variable change is tried, any other may be spared; this variable change is definitely not unique, for example, different numeration of the points result in different variable changes.} variable change $x=y^2$ (for two finite points $x_1$ and $x_2$ this amounts to $x={(x_2 y^2-x_1)} / {(y^2-1)}$). Normalization (with rational-in-$y$ transformation) is either possible after this, or impossible for any variable change (\Eref{eq:xviay}) at all.
        \item If there are three square root branching points, one should map them to $0$, $1$, and $\infty$ by M\"obius transformation and make the canonical variable change
\[
 x=\left(\frac{y^2+1}{2y}\right)^2 \, .
\]
 For three finite points $x_1$, $x_2$, and $x_3$, this amounts to
\[
  x=\frac{x_1 \left(4 x_2 y^2+x_3 \left(y^2-1\right)^2\right)-x_2 x_3 \left(y^2+1\right)^2}{-4 x_3 y^2+x_1 \left(y^2+1\right)^2-x_2 \left(y^2-1\right)^2} \, .
\]
Normalization (with rational-in-$y$ transformation) is either possible after this, or impossible for any variable change (\Eref{eq:xviay}) at all.
        \item  If there are four or more square root branching points with $b=2$, there is no suitable variable change. Normalization (and, therefore, $\epsilon$-form) is impossible by any transformation rational in $y$ (connected with $x$ by rational function $f$, Eq. \eqref{eq:xviay}).
        \item If there is one square root branching point, there is no canonical variable change, but rather a wide class of suitable transformations $x={p(y)^2}
/ {q(y)}$, where $p$ and $q$ are polynomials.
\end{itemize}
Apart from the rarely encountered latter case, this allows one to define,
unambiguously, the suitable variable, or understand that $\epsilon$-form is impossible.

Suppose now, that we have managed to find a proper variable, in terms of which all eigenvalues of all matrix residues have the form of \Eref{eq:evs} with integer $n$. Then the prescription of Ref. \cite{Lee:2014ioa} 
is to apply transformations of the form of \Eref{eq:balance} with the projector 
\begin{equation}
        P=\frac{u v^{\intercal}}{(v^{\intercal}u)}\,,
\end{equation}
where $u$ and $v^\intercal$ are the eigenvectors of $R_1$ and $R_2^\intercal$, \ie 
\begin{align}
        R_1u=\lambda u\,,\qquad v^\intercal R_2=\mu v^\intercal\,.
\end{align} This transformation shifts $\lambda\to \lambda+1$, $\mu\to \mu-1$. Again, there might be an obstruction $(v^{\intercal}u)=0$ to the construction of such a transformation. Therefore, successive application of such transformations can eliminate integer constants $n$ in the eigenvalues of the form of \Eref{eq:evs}.

\subsubsection{Factorizing $\epsilon$} Suppose now that we have successfully normalized all matrix residues. Then the final stage involves finding an $x$-independent transformation $T(\epsilon)$, such that the transformed matrix has the form $\tilde{M}(x,\epsilon)=\epsilon A(x)$. This step can be reduced to solving the linear system of equations
\begin{align*}
        \epsilon^{-1}R_1(\epsilon)T &= \mu^{-1}R_1(\mu)T\\
       & ~~  \vdots\\
        \epsilon^{-1}R_k(\epsilon)T &= \mu^{-1}R_k(\mu)T
\end{align*}
for sufficiently generic $\mu$, treating elements of $T$ as unknowns (see Ref. \cite{Lee:2014ioa} 
for details).

\subsubsection{Criterion of (ir)reducibility}
It is important to have both the effective reduction algorithm and the strict criterion of irreducibility. The latter is presented in Ref. \cite{Lee:2017oca} 

A simple but important proposition from  Ref. \cite{Lee:2017oca} 
states that the normalization at any given point cannot survive the transformation, which is singular in this point. This, in particular, means that once we have achieved a global normalized Fuchsian form, we are allowed to make only transformations independent on $x$. Now, by examining the steps of the reduction algorithm, one may see that it is always possible to reduce the system to normalized Fuchsian form everywhere, except, maybe, one singular point (called next the \emph{exceptional}), chosen arbitrarily. The recipe of Ref. \cite{Lee:2017oca}  
is to reduce the system to normalized Fuchsian form, first choosing the exceptional point to be $x=\infty$, and then to be $x=0$.\footnote{By M\"obius transformations, we can always map any two points to $\infty$ and $0$.} Let $M$ correspond to the first choice and 
\begin{equation}\label{eq:M_U}
\tilde{M}=M_U=U^{-1}(MU -\partial_x U)
\end{equation} 
correspond to the second. Now we note that, if $T$ exists, such that $M_T=T^{-1}(MT -\partial_x T)$ is normalized everywhere, then both $T$ and $T^{-1}$ should be polynomial in $x$, and $S=U^{-1}T$ is necessarily polynomial in $x^{-1}$ with constant determinant. These are direct consequences of the proposition formulated previously. Therefore, there should exist a decomposition
\begin{equation}\label{eq:decomposition}
U =T(x)S^{-1}\left(x^{-1}\right)
\end{equation}
where both $T$ and $S$ are polynomial matrices of their arguments ($x$ and $x^{-1}$, respectively), with the determinants independent of $x$. This decomposition explicitly demonstrates that $\det U$ should be independent of $x$ and that $U$ and $U^{-1}$ are necessarily regular in all points but two: $x=0$ and $x=\infty$, so their entries are  Laurent polynomials in $x$. Let us note that if the decomposition exists, it is unique, up to the transformations $T(x)\to T(x) L\,,\ S(x^{-1})\to S(x^{-1}) L$, where $L$ is a constant matrix. We refer the reader to Ref. \cite{Lee:2017oca} 
for an algorithm that finds the decomposition (\Eref{eq:decomposition}) or proves that it does not exist. 

Table \ref{table:Fred} is a `troubleshooting' table for the reduction to $\epsilon$-form.

\begin{table} 
\caption{`Troubleshooting' table for the reduction to $\epsilon$-form}
\label{table:Fred}
\centering
\begin{tabular}{p{0.48\textwidth} p{0.48\textwidth}}
\hline \hline
{Case} & {Explanation}\\
\hline
It appears that there are some irreducible denominators depending both on $x$ and $\epsilon$, which cannot be eliminated by means of the algorithm of Ref. \cite{Barkatou1995}. & 
The system is obviously irreducible to $\epsilon$-form by any transformations. Almost certainly, the IBP reduction is not complete.\\  
After the reduction to Fuchsian form, there are some square root branching points (say, $m$ points), \ie some matrix residues have eigenvalues of the form $n+\tfrac12+k \epsilon$. &
If $m\leqslant 3$, make the variable change \cite{Lee:2017oca} 
described  (in particular, for 2 and 3 points, try only a canonical change). If $m\geqslant 4$, the system cannot be reduced to $\epsilon$-form.\\
It seems to be impossible to find a pair $(U,V^\intercal)$ ($(u,v^\intercal)$) on the first (second) stage, such that $V^\intercal U$ is invertible ($v^\intercal u \neq 0$). &
Most probably, this is an irreducible case. To strictly prove it 
(or disprove and find the required transformation), follow the algorithm of Ref.\cite{Lee:2017oca}. 
\\

The $\epsilon$-factorization stage does not give an  invertible matrix $T$. &
Factorization is not possible \cite{Lee:2017oca}.
\\
\hline  \hline
\end{tabular}
\end{table}

Let us comment about irreducible cases (also known as `elliptic'). A natural generalization of $\epsilon$-form is the linear-in-$\epsilon$ form
\begin{equation}\label{eq:eplin}
        \partial_x \boldsymbol J = \left[A(x)+ \epsilon B(x)\right]\boldsymbol J\,.
\end{equation}
If this form is obtained, one can achieve $\epsilon$-form by `integrating
out' the $\epsilon^0$ term \cite{Argeri:2014qva,Adams:2018yfj},
\ie by passing to $\widetilde{\boldsymbol J}$ via $\boldsymbol J= F_0 \widetilde{\boldsymbol J}$, where $F_0$ is the solution of the equation 
\begin{equation}\label{eq:DE0}
        \partial_x F_0 = A(x)F_0 \, .
\end{equation}
One has
\begin{equation}
\partial_x \widetilde{\boldsymbol J} = \epsilon \widetilde{B}(x)\widetilde{\boldsymbol J}\,,
\end{equation}
where $\widetilde{B}(x) = F_0^{-1}(x)B(x)F_0(x)$. While being natural and, in principle, quite promising, this approach has weaknesses connected with the necessity to solve Eq. \eqref{eq:DE0} on a case-by-case basis and with our poor understanding of the class of functions that may appear in the general solution of Eq. \eqref{eq:DE0}. We refer the reader to Refs.
\cite{Adams2015a,Adams:2016xah,Remiddi:2016gno,Adams:2017xsu,Primo:2017ipr,Primo:2016ebd,Remiddi:2017har}
for further information about known elliptic cases. From a practical point of view, there is also a problem of efficient numerical calculation of the iterative integrals
\[
 \mathop{\int\cdots \int}\limits_{x_0<x_1\cdots<x_n<x} \mathrm{d}x_n\cdots \mathrm{d}x_1 \widetilde{B}(x_n)\cdots \widetilde{B}(x_1) \, ,
\]
 which appear in the $\epsilon$-expansion of $\widetilde{\boldsymbol J} $.

By contrast, the problem of finding the transformation to the form of \Eref{eq:eplin} seems to be quite approachable, thanks to the algorithms reviewed here. The reason is that, in many cases (in fact, in all known to the author), it is possible to reduce the `elliptic' systems to $\epsilon$-form near an odd dimensionality. (We remark here that once $\epsilon$-form exists for $d=n-2\epsilon$, it automatically exists for  $d=n+2k-2\epsilon$, where $k$ is an integer. This is a trivial consequence of the dimensional recurrences. However, shifting $d$ by one is not trivial.) This means that we can achieve the even more specific form (naturally referred to as the $(\epsilon+1/2)$-form)
\begin{equation}\label{eq:ephalf}
\partial_x \boldsymbol J = (\epsilon+1/2)B(x)\boldsymbol J\,.
\end{equation}

Let us briefly discuss the multivariate case. First, we note that, for the multivariate case, the $\epsilon$-form of the differential equations is even more profitable than for the case of the single variable. Indeed, \Eref{eq:compatibility} becomes
\begin{equation}\label{eq:compatibility1}
\epsilon\left(\partial_b S_a - \partial_a S_b\right) + \epsilon^2[S_a,S_b]=0\,,
\end{equation}
and, therefore, coefficients in front of $\epsilon$ and $\epsilon^2$ should vanish separately 
\cite{Henn:2013pwa}. The first condition $\partial_b S_a - \partial_a S_b=0$ indicates that there should be a matrix $S$, such that $S_a=\partial_a S$ and the systems with respect to different variables can be unified in what is called d\,log form:
\begin{equation}
        \mathrm{d}\boldsymbol J=\epsilon (\mathrm{d}S) \boldsymbol J\,.
\end{equation}

It is important to realize that the algorithms presented here are sufficient for the multivariate reduction once the variables are fixed. To use them, one should simply make the reduction with respect to each variable in turn. The important consequence of the proposition proved in Ref. \cite{Lee:2017oca} 
(and presented here) is that, after passing to the next variable, one should consider only the transformations, independently of the variables already processed. We refer the reader to Ref. \cite{Meyer:2017joq}
for an alternative approach to multivariate reduction to $\epsilon$-form. Unfortunately, the choice of `correct' variables for multivariate set-up lacks systematic treatment so far and mostly remains a matter of art and luck.

\section*{Acknowledgements}
Roman Lee is grateful to Andrei Pomeransky for constant interest and
numerous fruitful discussions.

 \label{sec-rlee}
\clearpage
\pagestyle{empty}
\cleardoublepage


\section
[About cuts of Feynman integrals and differential equations \\ {\it C.G. Papadopoulos}]
{About cuts of Feynman integrals and differential equations} \label{sec:costas}

\pagestyle{fancy}
\fancyhead[LO]{}
\fancyhead[RO]{}
\fancyhead[CO]{}
\fancyhead[CO]{\thechapter.\thesection 
\hspace{1mm}
About cuts of Feynman integrals and differential equations
}
\fancyhead[LE]{}
\fancyhead[CE]{C.G. Papadopoulos}
\fancyhead[RE]{} 

\noindent
{\bf Author: Costas~G.~Papadopoulos} {~~[Costas.Papadopoulos@cern.ch]} 
\vspace*{.5cm}

\subsection{Introduction}
\label{sIntro}

It is almost 70\,years from the time Feynman integrals  were first introduced~\cite{Feynman:1949zx,Dyson:1949bp,Dyson:1949ha} and 45\,years 
since  dimensional regularization~\cite{tHooft:1972tcz} set up the framework for an efficient use of loop integrals 
in computing scattering matrix elements, and still the frontier of multiscale multiloop integral calculations (maximal in both number of scales and number of loops) is determined by the planar five-point two-loop on-shell massless integrals~\cite{Gehrmann:2015bfy,Papadopoulos:2015jft}, 
recently computed.\footnote{Complete results, including physical region kinematics, are presented in~\cite{Papadopoulos:2015jft}: notice that numerical codes, 
 for instance {\tt SecDec}~\cite{Borowka:2015mxa}, can reproduce analytical results only at Euclidean region kinematics; results for physical region kinematics are not supported, owing to poor numerical convergence.}  
Conversely,  to keep up with the increasing experimental accuracy as more data are collected at the LHC, more precise theoretical predictions and higher-loop calculations are required~\cite{Andersen:2014efa}.

In the last  years, our understanding of the reduction of one-loop amplitudes to a set of master integrals, a minimal set of Feynman integrals that form a basis, 
either based on unitarity methods~\cite{Bern:1994cg,Bern:1994zx,Berger:2008sj} 
or at the integrand level via the OPP method~\cite{Ossola:2006us,Ossola:2008xq}, has drastically changed the way one-loop calculations are performed, resulting in many fully automated numerical tools (some reviews on the topic are Refs.~\cite{AlcarazMaestre:2012vp,Ellis:2011cr,vanDeurzen:2015jmn}), making the next-to-leading-order (NLO) approximation the default precision for theoretical predictions at the LHC.
In  recent years, progress has also been made  towards the extension of these reduction methods for two-loop amplitudes at the integral~\cite{Gluza:2010ws,Kosower:2011ty,CaronHuot:2012ab,Johansson:2012zv,Johansson:2012sf,Johansson:2013sda,Sogaard:2013fpa,Ita:2015tya,Johansson:2015ava,Mastrolia:2016dhn} 
as well as the integrand~\cite{Mastrolia:2011pr,Badger:2012dp,Mastrolia:2012wf,Badger:2013gxa,Papadopoulos:2013hra,Badger:2015lda} level. 
The master equation at the integrand level can be given schematically as follows~\cite{Papadopoulos:2013hra}:
\begin{equation}
\frac{N\left(l_1,l_2;\left\{p_i\right\}\right)}{D_1 D_2 \cdots D_n}=\sum_{m=1}^{\min(n,8)} \sum_{S_{m;n}} \frac{\Delta_{{i_1}{i_2} \cdots {i_m}}\left(l_1,l_2;\left\{p_i\right\}\right)}{D_{i_1} D_{i_2} \cdots D_{i_m}}.
\label{oppmaster2}
\end{equation}
where an arbitrary contribution to the two-loop amplitude (left), can be reduced to a sum of terms (right) of all partitions $S_{m;n}$, with up to eight denominators; $l_1,l_2$ are the loop momenta, $D_i$ are the inverse scalar Feynman propagators, $N\left(l_1,l_2;\left\{p_i\right\}\right)$ is a general numerator polynomial, and \hfill \\ $\Delta_{{i_1}{i_2} \cdots {i_m}}\left(l_1,l_2;\left\{p_i\right\}\right)$ are the residues of multivariate polynomial division. In addition, $R_2$ terms~\cite{Ossola:2008xq} must be studied at two loops in order to achieve a complete framework. 

Two-loop master integrals are defined 
using the {\it integration-by-parts} (IBP) identities~\cite{Chetyrkin:1981qh,Tkachov:1981wb,Laporta:2001dd}, an indispensable tool beyond one loop.
Contrary to the one-loop case, where master integrals have  been known for a long time~\cite{'tHooft:1978xw}, a complete library of master integrals at two loops is still missing.
At the moment, this is the main obstacle to obtaining a fully automated NNLO calculation framework, similar to the one-loop one, that will satisfy the precision requirements at the LHC~\cite{Andersen:2014efa}.

Many methods have been introduced  to compute master integrals~\cite{Smirnov:2012gma}. 
The  most successful one, overall, is based on expressing the Feynman integrals in terms of an integral representation over Feynman parameters, involving the two well-known Symanzik polynomials $U$ and $F$~\cite{Bogner:2010kv}.
The introduction of the sector decomposition~\cite{Hepp:1966eg,Roth:1996pd,Binoth:2000ps,Binoth:2003ak,Bogner:2007cr} method resulted in a powerful computational framework for the numerical evaluation of Feynman integrals, 
see, for instance, {\tt SecDec}~\cite{Borowka:2015mxa}. An alternative is based on Mellin--Barnes representation~\cite{Smirnov:1999gc,Tausk:1999vh}, implemented in Ref. \cite{Czakon:2005rk}.\footnote{See also Ref. \cite{MBasymptotics}.} Nevertheless, the most successful method for calculating multiscale multiloop Feynman integrals is, for the time being,  the differential equations approach~\cite{Kotikov:1990kg,Kotikov1991b,Bern:1992em,Remiddi:1997ny,Gehrmann:1999as}, which has been used in the past two decades to calculate various master integrals at two loops and beyond. 
Following the work of Refs.~\cite{Goncharov:1998kja,Remiddi:1999ew,Goncharov:2001iea}, there has been increasing consensus that the so-called {\it Goncharov polylogarithms}  form a functional basis for many master integrals. 
The so-called canonical form of differential equations, introduced by Henn~\cite{Henn:2013pwa}, manifestly results in master integrals expressed in terms of Goncharov polylogarithms.\footnote{For an alternative method in the single-scale case, see also Ref.~\cite{Ablinger:2015tua}.} 
Nevertheless, the reduction of a given differential equation to a canonical form is by no means
fully understood. First of all, despite recent efforts~\cite{Lee:2014ioa,Meyer:2017joq,Gituliar:2017vzm}, 
and the existence of {\it sufficient} conditions that a given master integral can be expressed in terms of Goncharov polylogarithms, no criterion, with practical applicability, that is at the same time {\it necessary and sufficient} 
has been introduced so far. Moreover, it is well-known that when, for instance, enough internal masses
are introduced, master integrals are no longer expressible in terms of Goncharov polylogarithms; in fact, a new class of functions involving elliptic integrals is needed~\cite{Adams:2015gva,Bonciani:2016qxi,Ablinger:2017bjx,Bourjaily:2017bsb,Broedel:2017kkb}. 

In this contribution, we start, in Section E.\ref{SDE}, by briefly reviewing the simplified differential equations approach. In Section E.\ref{5p}, we present the results 
of the most advanced calculation achieved nowadays, namely the calculation of planar pentabox master integrals. The Baikov representation is introduced 
in Section E.\ref{sBaikov}. In Section E.\ref{sCUT}, we give the definition of the cut~\cite{Frellesvig:2017aai} of a  Feynman integral based on Baikov representation~\cite{Baikov:1996iu,Baikov:1996rk,Smirnov:2003kc,Lee:2010wea,Grozin:2011mt,LeePomeransky2013}. Finally, in Section E.\ref{Sdisc}, we discuss possible future directions.

\subsection{The simplified differential equations approach}
\label{SDE}

Assume that one is interested in calculating an $l$-loop Feynman integral with external momenta $\{p_j\}$ and internal propagators that are massless.
Any $l$-loop  Feynman integral can  then be written as
\begin{equation}
G_{a_1\cdots a_n}(\{p_j\},\epsilon)=\int\left(\prod_{r=1}^l \frac{\mathrm{d}^dk_r}{\mathrm{i}\pi^{d/2}}\right)\frac{1}{D_1^{a_1}\cdots D_n^{a_n}}, \qquad
D_i=\left(c_{ij}k_j+d_{ij}p_j\right)^2,\qquad d=4-2\epsilon
\label{eq:loopgen}
\end{equation}
with matrices $\{c_{ij}\}$ and $\{d_{ij}\}$ determined by the topology and the momentum flow of the graph, and
the denominators  defined in such a way that all scalar product invariants can be written as a linear combination of them. The exponents $a_i$ are integers and may be negative in order to accommodate irreducible numerators.
Any integral $G_{a_1\cdots a_n}$ can be written as a linear combination of a finite subset of such integrals, called master integrals,  
with coefficients depending on the independent scalar products, $s_{ij}=p_i\cdot p_j$, and space-time dimension $d$, by the use of {\it integration-by-parts}  identities~\cite{Chetyrkin:1981qh,Tkachov:1981wb}.  

The simplified differential equations approach~\cite{Papadopoulos:2014lla} is an attempt not only to simplify, but also to systemize, as much as possible, the derivation of the appropriate system of differential equations satisfied by the master integrals. 
To this end, 
the external incoming momenta are {\it parametrized} linearly in terms of $x$, as $p_i(x)=p_i+(1-x)q_i$, where the $q_i$ are a linear combination of the  momenta $\{p_i\}$, such that $\sum_iq_i=0$. If $p_i^2=0$, the parameter $x$ captures the off-shell-ness of the external legs. 
The class of Feynman integrals in \Eref{eq:loopgen} are now dependent on $x$ through the external momenta:
\begin{equation}
G_{a_1\cdots a_n}(\{s_{ij}\},\epsilon;x)=\int\left(\prod_{r=1}^l \frac{\mathrm{d}^dk_r}{\mathrm{i}\pi^{d/2}}\right)\frac{1}{D_1^{a_1}\cdots D_n^{a_n}}, 
\qquad
D_i=\left( c_{ij}k_j+d_{ij}p_j(x) \right)^2.
\label{eq:loopgenx}
\end{equation}
By introducing the dimensionless parameter $x$, the vector of master integrals $\vec{G}^\mathrm{MI}(\{s_{ij}\},\epsilon;x)$, which now depends on $x$, satisfies
\begin{equation}
\frac{\partial}{\partial x} \vec{G}^\mathrm{MI}(\{s_{ij}\},\epsilon;x)=\mathbf{H}(\{ s_{ij}\},\epsilon;x)\vec{G}^\mathrm{MI}(\{s_{ij}\},\epsilon;x) \, ,\label{eq:DEx}
\end{equation}
a system of differential equations in one independent variable. 
Experience up to now shows that this simple parametrization can be used universally to deal with up to six kinematic scales~\cite{Papadopoulos:2014lla,Papadopoulos:2014hla,Papadopoulos:2015jft}.
 The expected benefit is that the integration of the differential equations naturally captures the expressibility of master integrals in terms of Goncharov polylogarithms and, more importantly, makes the problem 
 {\it independent of the number of kinematic scales} (independent invariants) involved. 
Note that as $x\rightarrow 1$, the original configuration of the loop integrals (\Eref{eq:loopgen}) is reproduced, which eventually corresponds to a simpler one with one scale less.

\subsection{Massless pentabox master integral with up to one off-shell leg} 
\label{5p}

For the massless pentabox master integrals with one off-shell leg, there are, in total, three families of planar master integrals, whose members, with the maximum amount of denominators, namely eight, are  
shown in \Fref{fig:param-P}. 
We have checked that the other five-point integrals with one massive external leg are reducible to master integrals in one of these eight master integral families. 
In Ref.~\cite{Papadopoulos:2015jft}, we have recently completed  the calculation of the $P_1$ family (\Fref{fig:param-P}). In fact, by taking the limit $x\to 1$, all planar graphs for massless on-shell external momenta have been derived as well. 
We have used the {\tt c++} implementation of the program {\bf FIRE}~\cite{Smirnov:2014hma} to perform the IBP reduction to the set of master integrals in $P_1$.

\begin{figure}
\centering
\includegraphics[width=0.25 \linewidth]{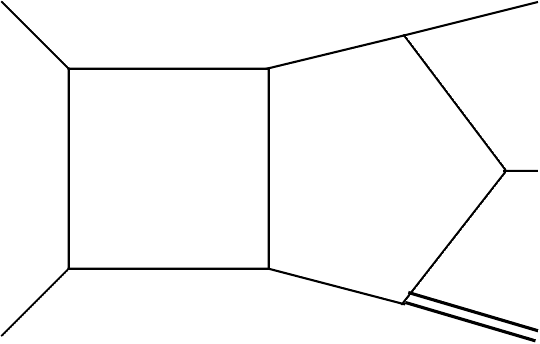} \hspace{0.4 cm}
\includegraphics[width=0.25 \linewidth]{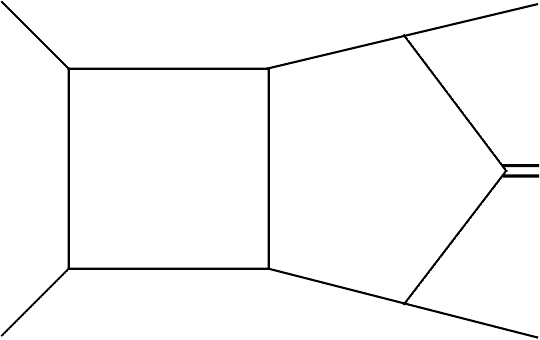} \hspace{0.4 cm}
\includegraphics[width=0.25 \linewidth]{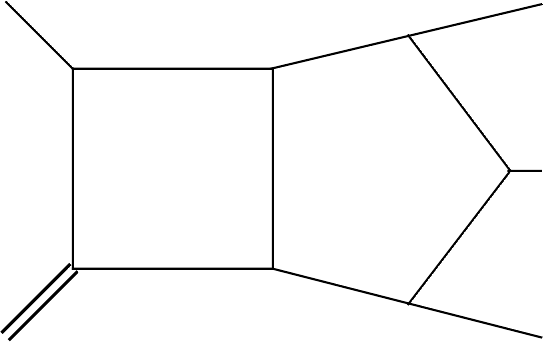}
  \caption{The three planar pentaboxes of the families $P_1$ (left), $P_2$ (middle), and $P_3$ (right), with one external massive leg.}
  \label{fig:param-P}
 \end{figure}

For the family of integrals $P_1$, the external momenta are parametrized in $x$, as shown in \Fref{fig:xparam-P1}. The parametrization is chosen such that the double-box master integral with two massive external legs that is contained in the family $P_1$ has exactly the same parametrization as that  chosen in Ref.~\cite{Papadopoulos:2014hla}, \ie two massless external momenta, $xp_1$ and $xp_2$, and two massive external momenta, $p_{123}-xp_{12}$ and $-p_{123}$. The master integrals in the family $P_1$ are, therefore, a function of a parameter $x$ and the following five invariants:
$s_{12}:=p_{12}^2$,  $s_{23}:=p_{23}^2$,  $s_{34}:=p_{34}^2$,  $s_{45}:=p_{45}^2=p_{123}^2$,  $s_{51}:=p_{15}^2=p_{234}^2$,  $p_i^2=0$,
where the notation $p_{i\cdots j}=p_i+\cdots +p_j$ is used and $p_5:=-p_{1234}$. As the parameter $x\rightarrow 1$, the external momentum $q_3:= p_{123}-x p_{12}$ becomes massless, such that our parametrization also captures the on-shell case. 


\begin{figure}
\centering
\includegraphics[width=0.30 \linewidth]{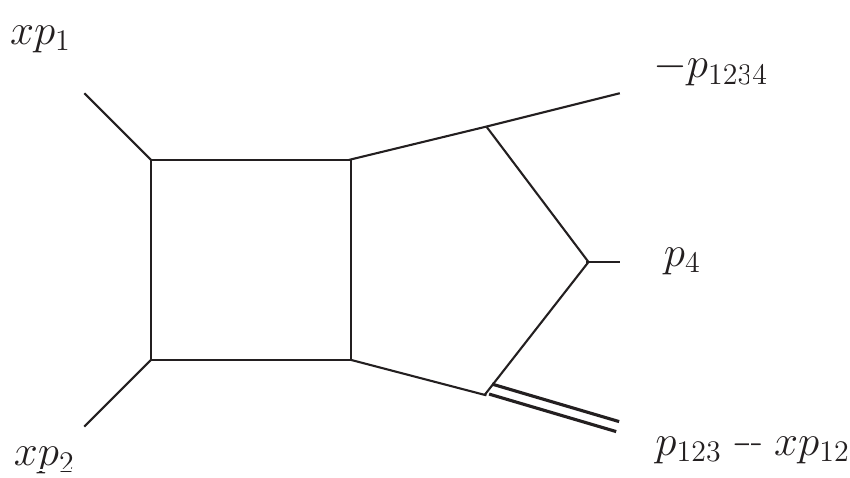}
  \caption{Parametrization of external momenta in terms of $x$ for the planar pentabox of the family $P_1$. All external momenta are incoming.}
  \label{fig:xparam-P1}
  \vspace{-12pt}
\end{figure}

The resulting differential equation in matrix form can be written as 
\begin{equation}
{\partial _x}{\bf{G}} = {\bf{M}}\left( {\left\{ {{s_{ij}}} \right\},\varepsilon ,x} \right){\bf{G}} \, ,
\label{eq:ini}
\end{equation}
where ${\bf{G}}$ stands for the array of the 74 master integrals involved in the family $P_1$.  
The 20 letters $l_i$  involved are given in Ref. \cite{Papadopoulos:2015jft}.
Although the differential equations can be solved starting from, \eg \Eref{eq:ini},  and the result can be expressed as a sum of Goncharov polylogarithms with argument $x$ and weights given by the letters $l_i$, it is more elegant and easy-to-solve to derive 
a Fuchsian system of equations~\cite{Henn:2014qga}, where only single poles in the variable $x$ will appear. 
A series of transformations, described in Ref. \cite{Papadopoulos:2015jft}, brings the system into the form
\begin{equation}
{\partial _x}{\bf{G}} = \left( {\varepsilon \sum\limits_{i = 1}^{19} {\frac{{{{\bf{M}}_i}}}{{\left( {x - {l_i}} \right)}}} } \right){\bf{G}}
\end{equation}
where the residue matrices ${{{\bf{M}}_i}}$ are independent of $x$ and $\varepsilon$. 
The result can be given straightforwardly,  as 
\small{
\begin{align}
\label{eq:solx}
{\bf{G}} &= {\varepsilon ^{ - 2}}{\bf{b}}_0^{\left( { - 2} \right)} 
+ {\varepsilon ^{ - 1}}\left( {\sum {{{\cal G}_a}{{\bf{M}}_a}{\bf{b}}_0^{\left( 
{ - 2} \right)}}  + {\bf{b}}_0^{\left( { - 1} \right)}} \right) 
+ \left( {\sum {{{\cal G}_{ab}}{{\bf{M}}_a}{{\bf{M}}_b}} {\bf{b}}_0^{\left( { - 
2} \right)} + \sum {{{\cal G}_a}{{\bf{M}}_a}{\bf{b}}_0^{\left( { - 1} \right)}} 
 + {\bf{b}}_0^{\left( 0 \right)}} \right)
\nonumber \\
 & \quad + \varepsilon \left( {\sum {{{\cal G}_{abc}}{{\bf{M}}_a}{{\bf{M}}_b}{{\bf{M}}
_c}} b_0^{\left( { - 2} \right)} + \sum {{{\cal G}_{ab}}{{\bf{M}}_a}{{\bf{M}}_b}
{\bf{b}}_0^{\left( { - 1} \right)}}  + \sum {{{\cal G}_a}{{\bf{M}}_a}{\bf{b}}_0^{\left( 0 \right)} + } {\bf{b}}_0^{\left( 1 \right)}} \right)\nonumber\\
 & \quad + {\varepsilon ^2}\left( \sum {{{\cal G}_{abcd}}{{\bf{M}}_a}{{\bf{M}}_b}{{\bf{M}}_c}{{\bf{M}}_d}} {\bf{b}}_0^{\left( { - 2} \right)} + \sum {{{\cal G}_{abc}}{{\bf{M}}_a}{{\bf{M}}_b}{{\bf{M}}_c}{\bf{b}}_0^{\left( { - 1} \right)}}  \right.
\nonumber  \\
 & \qquad \qquad + \left. \sum{{{\cal G}_{ab}}{{\bf{M}}_a}{{\bf{M}}_b}{\bf{b}}_0^{\left( 0 \right)}} + \sum {{{\cal G}_a}{{\bf{M}}_a}{\bf{b}}_0^{\left( 1 \right)}} + {\bf{b}}_0^{\left( 2 \right)} \right)
\end{align}
}%
\noindent with the arrays ${\bf{b}}_0^{\left( {k} \right)}$, $k=-2,\dots,2$ representing the $x$-independent boundary terms in the limit $x=0$ at order $\varepsilon^k$. The expression is in terms of Goncharov polylogarithms, 
${{\cal G}_{ab \ldots }} = {\cal G}\left( {{l_a},{l_b}, \ldots ;x} \right)$. Details of the calculation of boundary terms and of the $x\to 1$ limit can be found in~Ref. \cite{Papadopoulos:2015jft}.

The solution for all 74 master integrals contains $O(3,000)$ Goncharov polylogarithms, which is roughly six times more than the corresponding double-box with two off-shell legs planar master integrals.
 We have made several numerical checks of all our calculations. 
The numerical results, also included in the ancillary files\cite{Dick} 
have been obtained using the {\bf GiNaC} library~\cite{Vollinga:2004sn} and compared with those provided by the numerical code {\bf SecDec}~\cite{Binoth:2000ps,Heinrich:2008si,Borowka:2012yc,Borowka:2015mxa} in the Euclidean region for all master integrals and in the physical region whenever possible
(owing to CPU time limitations in using {\bf SecDec}) and found perfect agreement. For the physical region, we are using the analytical continuation, as described in the previous section, as well as in Ref.~\cite{Papadopoulos:2014hla}.
At this stage, we are not setting a fully fledged numerical implementation, which will be done when all families have been computed. Our experience with double-box computations shows that using, for instance, {\tt HyperInt}~\cite{Panzer:2014caa}
 to bring all Goncharov polylogarithms in their range of convergence, {\it before} evaluating them numerically by {\bf GiNaC}, 
increases efficiency by two orders of magnitude.  Moreover, expressing Goncharov polylogarithms in terms of classical polylogarithms and $\mathrm{Li}_{2,2}$, could also  substantially reduce the CPU time~\cite{Gehrmann:2015ora}. 
Based on these findings, we estimate that a target of  $O\left( {{{10}^2-{10}^3}} \right)$\,ms can be achieved.

\subsection{The Baikov representation}
\label{sBaikov}

An $L$-loop Feynman integral with $E+1$ external lines can be written in the form
\begin{equation}
{F_{{\alpha_1}\dots{\alpha_N}}} = \int {\left( {\prod\limits_{i = 1}^L {\frac{{{\mathrm{d}^d}{k_i}}}{{\mathrm{i}{\pi ^{d/2}}}}} } \right)} \frac{1}{{D_1^{{\alpha_1}}\dots D_N^{{\alpha_N}}}}
\label{FI}
\end{equation}
with $N = {{L\left( {L + 1} \right)}}/{2} + LE$, $\alpha_i$ arbitrary integers, and $D_a$, $a=1,\dots,N$, inverse Feynman propagators,
\begin{equation}
{D_a} = \sum\limits_{i = 1}^L {\sum\limits_{j = i}^M {A_{_a}^{ij}{s_{ij}} + {f_a}} }  = \sum\limits_{i = 1}^L {\sum\limits_{j = i}^L {A_{_a}^{ij}{k_i} \cdot {k_j}} }  + \sum\limits_{i = 1}^L {\sum\limits_{j = L+1}^M {A_{_a}^{ij}{k_i} \cdot {p_{j-L}}} }  + {f_a}, \qquad a = 1, \ldots ,N \, ,
\label{eq:x-def}
\end{equation}
where $q_i=k_i, (i=1,\dots,L)$, the loop momenta, and $q_{L+i}=p_i, (i=1,\dots,E)$, the independent external momenta, $M=L+E$, $s_{ij}=q_i\cdot q_j$, and $f_a$ depend on external kinematics and internal masses.
$A_{_a}^{ij}$ can be understood as an $N\times N$ matrix, with $a$ running obviously from $1$ to $N$ and with $(ij)$ taking also $N$ values as $i=1,\ldots,L$ and $j=i,\ldots,M$.
The elements of the matrix $A_{_a}^{ij}$ are integer numbers taken  from the set $\left\{-2,-1,0,+1,+2 \right\}$. This matrix is characteristic of the corresponding Feynman graph and can, in a loose sense, 
be associated with the `topology' of the graph. 
Then, by projecting each of the loop momenta $q_i=k_i, (i=1,\dots,L)$ with respect to the space spanned by the external momenta involved plus a transverse component (for details, see Ref.~\cite{Grozin:2011mt}), we may write
\begin{equation}
{F_{{\alpha_1}\cdots{\alpha_N}}} = C_{N}^L{\left( {G\left( {{p_1},\dots,{p_E}} \right)} \right)^{\left( { - d + E + 1} \right)/2}}\int {\frac{{\mathrm{d}{x_1}\cdots
\mathrm{d}{x_N}}}{{x_1^{{\alpha_1}}\cdots
x_N^{{\alpha_N}}}}} P_N^L{\left( {{x_1} - {f_1},\dots,{x_N} - {f_N}} \right)^{\left( {d - M - 1} \right)/2}}
\label{BR}
\end{equation}
with 
\begin{equation}
C_{N}^L = \frac{{{\pi ^{ - L\left( {L - 1} \right)/4 - LE/2}}}}{{\prod\nolimits_{i = 1}^L {\Gamma \left( {\frac{{d - M + i}}{2}} \right)} }}\det \left( {A_{ij}^a} \right)
\end{equation}
and 
\[P_N^L\left( {{x_1},{x_2},\dots,{x_N}} \right) = { { {{{ {G\left( {{k_1},\dots,{k_L},{p_1},\dots,{p_E}} \right)} }}} } \Big|_{{s_{ij}} = \sum\limits_{a = 1}^N {A_{ij}^a{x_a}\,\,\& \,\,\,{s_{ji}} = {s_{ij}}} }} \, ,
\]
where $G$ represents the Gram determinant, $G\left( {{q_1}, \ldots ,{q_n}} \right) = \det \left( {{q_i} \cdot {q_j}} \right)$ and $A_{ij}^a$ is the inverse of the topology matrix $A_{_a}^{ij}$. 
An alternative derivation of the Baikov representation for one- and two-loop Feynman integrals, as well as the {\it loop-by-loop representation}, can be found in Appendix A of Ref.~\cite{Frellesvig:2017aai}.
The derivation of the Baikov representation can easily be implemented in a computer algebra code.\footnote{A \textsc{Mathematica} script, {\tt Baikov.m}, is provided as an attachment in Ref.~\cite{Frellesvig:2017aai}.} 

We conclude this section by elaborating on the limits of the $x_a$-integrations in Eq. (\ref{BR}). 
To simplify the discussion, let us start with a generic one-loop configuration defined by 
\[{x_1} =   {k^2} - m_1^2\,, \qquad {x_2} = \left( {k + {p_1}} \right)^2 - m_2^2\,, \qquad \ldots\,, \qquad {x_N} = \left( {k + {p_1} + \dots + {p_{N - 1}}} \right)^2 - m_N^2 \, .\]
Then consider the generic integral ($\alpha_i \ge 0$),
\begin{align}
F_{\alpha_1 \cdots \alpha_N} & = C^1_N G \! \left( p_1, \ldots, p_{N-1} \right)^{(N-d)/2} \int {\frac{{\mathrm{d}{x_1}\cdots \mathrm{d}{x_N}}}{{x_1^{{\alpha_1}}\cdots
x_N^{{\alpha_N}}}}} {P^1_N}^{\left( {d - N - 1} \right)/2}
\\
C^1_N & = \frac{{{\pi ^{ - (N-1)/2}}}}{{{\Gamma \left( {\frac{{d - N + 1}}{2}} \right)} }}\left(\frac{1}{2}\right)^{N-1}
\end{align}
It is easy to verify that ${P^1_N}$ is a polynomial that is quadratic in the variables $x_a$~\cite{Smirnov:2003kc}, and that obviously when $\alpha_N=0$, the external momentum $p_{N-1}$ decouples, so that
\begin{align}
{F_{{\alpha_1}\cdots{\alpha_{N-1}}{0}}}& = {C^1_N} G  \left( p_1, \ldots, p_{N-1} \right)^{(N-d)/2} \int {\frac{{\mathrm{d}{x_1}\cdots \mathrm{d}{x_{N - 1}}}}{{x_1^{{\alpha _1}}\cdots x_{N - 1}^{{\alpha _{N - 1}}}}}\int\limits_{x_N^ - }^{x_N^ + } {\mathrm{d}{x_N}} } {P^1_N}^{\left( {d - N - 1} \right)/2}
\nn \\
&= C^1_{N-1} G  \left( p_1, \ldots, p_{N-2} \right)^{(N-1-d)/2}  \int {\frac{{\mathrm{d}{x_1}
\cdots \mathrm{d}{x_{N-1}}}}{{x_1^{{\alpha_1}}\dots \,\,x_{N-1}^{{\alpha_{N-1}}}}}} {P^1_{N-1}}^{\left( {d - \left(N-1\right) - 1} \right)/2}
\end{align}
where ${P^1_N}\left( {x_N^ + } \right) = {P^1_N}\left( {x_N^ - } \right) = 0$ and 
\begin{multline}
\int\limits_{x_N^ - }^{x_N^ + } {\mathrm{d}{x_N}{P^1_N}^{\left( {d - N - 1} \right)/2}}  = \frac{{2{\pi ^{1/2}}\Gamma \left( {\frac{{d - N + 1}}{2}} \right)}}{{\Gamma \left( {\frac{{d - N + 2}}{2}} \right)}}
G \left( p_1, \ldots, p_{N-1} \right)^{(d-N)/2}
G \left( p_1, \ldots, p_{N-2} \right)^{(N-1-d)/2}
\nn\\
\times 
{P^1_{N - 1}}^{\left( {d - \left( {N - 1} \right) - 1} \right)/2}
\nn
\end{multline}
using $P_N^1 = \frac{1}{4}G\left( {{p_1}, \ldots ,{p_{N - 2}}} \right)\left( {x_N^ +  - {x_N}} \right)\left( {{x_N} - x_N^ - } \right)$ and
\[
 {\left( {x_N^ +  - x_N^ - } \right)^2} = 16\frac{{G\left( {{p_1}, \ldots ,{p_{N - 1}}} \right)}}{{G{{\left( {{p_1}, \ldots ,{p_{N - 2}}} \right)}^2}}}P_{N - 1}^1\, .
\]
This can be repeated straightforwardly for all variables except $x_1=k^2-m_1^2$, whose integration limits are simply derived from the  $k$-modulus integration limits. The generalization to the two-loop case is straightforward, with the integration at each step
performed over the $x$-variables involving a given external momentum, and the last ones derived by the corresponding $k_1$- and $k_2$-modulus integration limits. We have checked, both analytically and numerically, that
the limits, as defined here, reproduce  the known results for several examples at one and two loops.

\subsection{Cutting Feynman integrals}
\label{sCUT}

Cutting Feynman integrals in the Baikov representation has a very natural definition. Indeed we define an $n$ cut as 
\begin{equation}
{F}_{{\alpha _1}\cdots {\alpha _N}}|_{n \times \text{cut}} \equiv C_N^L{\left( G \right)^{\left( { - d + E + 1} \right)/2}}\left( {\prod\limits_{a = n+1}^{N} {\int {\mathrm{d}{x_a}} } } \right)\left( {\prod\limits_{c = 1}^n {\oint\limits_{{x_{c = 0}}} {\mathrm{d}{x_c}} } } \right)\frac{1}{{x_1^{{\alpha _1}} \cdots x_N^{{\alpha _N}}}} {P_N^L}^{\left( {d - M - 1} \right)/2} \, ,
\label{cut}
\end{equation}
where the Baikov variables $\{x_a:a=1,\dots ,N\}$ have been divided in two subsets, containing $n$ cut propagators and $(N-n)$ uncut ones. The cut operation defined here is operational in any space-time dimension $d$
and for any Feynman integrals given by Eq.~(\ref{FI}). Notice that the definition of the cut, Eq.~(\ref{cut}), is not identical to the traditional unitarity cut, see, for instance, Section 8.4 of Ref.~\cite{Veltman:1994wz}, owing to the lack of the
$\theta$-function constraint on the energy, and, therefore, it is not directly related to the discontinuity of the Feynman integrals~\cite{Abreu:2014cla,Abreu:2015zaa}.

Let us now consider a set of master integrals, ${F_i} \equiv {F_{\alpha _1^{\left( i \right)}\cdots \alpha _N^{\left( i \right)}}}$, $i=1,\dots ,I$, satisfying a system of differential equations, with respect to variables $X_j$,
\begin{equation}
\frac{\partial }{{\partial {X_j}}}{F_i} = \sum\limits_{l = 1}^I {M_{il}^{\left( j \right)}{F_l}} 
\label{fullDE}
\end{equation}
with matrices $M^{(j)}$ depending on kinematic variables, internal masses, and the space-time dimension, $d$. 
Since the derivation of differential equations~\cite{Frellesvig:2017aai} is insensitive to the cut operation, as defined in Eq.~(\ref{cut}), we may immediately write 
\begin{align}
\frac{\partial }{\partial X_j} F_i |_{n \times \text{cut}} = \sum_{l = 1}^I M_{il}^{(j)} F_l |_{n \times \text{cut}}
\label{cutDE}
\end{align}
with $F |_{n \times \text{cut}}$ representing an arbitrary $n$ cut: in other words, the cut integrals satisfy the same differential equations as the uncut ones.\footnote{See also Refs.~\cite{Anastasiou:2002yz,Lee:2012te,Larsen:2015ped} for related considerations.} Of course, for a given $n$ cut, many of the master integrals
that are not supported on the corresponding cut vanish identically. Nevertheless, Eq.~(\ref{cutDE}) remains valid. In particular, for the maximally cut integrals defined so that $n$ is equal to the number of
propagators (with $\alpha_i > 0$) of the integral, all integrals not supported on the cut vanish and the resulting differential equation is restricted to its homogeneous 
part. Evaluating the maximally cut master integrals provides, therefore, a solution to the homogeneous equation~\cite{Henn:2014qga,Primo:2016ebd}. Non-maximally cut integrals, however, can resolve non-homogeneous parts
of the differential equation as well~\cite{Henn:2014qga}. 

One important implication is that cut and uncut integrals, although very different in many respects, such as, for instance, their structure in $\epsilon$-expansion ($\epsilon\equiv (4-d)/2$), they are expressed in terms of the same class of functions.\footnote{See also, the related discussion in Section 3.4.1.
of Ref.~\cite{CaronHuot:2012ab}.} This 
is particularly important if we want to know, {\it a priori}, whether a system of differential equations can be solved, for instance, in terms of Goncharov polylogarithms, or whether the solution contains a larger
class of functions, including, for instance elliptic integrals. 

Several results of maximally cut master integrals, expressed either in terms of polylogarithmic functions or in terms of elliptic integrals, can be found in  Appendix B of Ref.~\cite{Frellesvig:2017aai} as well as in Refs.~\cite{Bosma:2017ens,Harley:2017qut}. 


\subsection{Discussion and outlook}
\label{Sdisc}

In this contribution, we have presented recent results for the calculation of two-loop five-point Feynman integrals, based on the simplified differential equations approach. 
We have also presented the  Baikov representation of Feynman integrals. We have shown how to determine the limits of integration and how to obtain differential equations with respect 
to external kinematics and internal masses. 
Then we provided a definition of a cut integral, operational in $d$ dimensions, and showed that a cut integral satisfies the same system of differential equations as the 
uncut, original integral. 

Based on the fact  that cut integrals satisfy the same system of differential equations as the full, uncut integrals, we have verified that
their analytical expressions are given in terms of the same class of functions, such as Goncharov polylogarithms or elliptic integrals. 
We have, therefore, arrived at the conclusion that, in a family of master integrals
satisfying a given system of differential equations, the study of the maximally cut integrals for all its members 
can provide a {\it necessary and sufficient} criterion for the existence of a canonical form of the differential equations, and that, in the case when such a canonical form does not exist,
it provides solutions of the homogeneous
parts of the system of differential equations (see also Refs.~\cite{Henn:2014qga,Primo:2016ebd}). 

By completing the calculation of non-planar pentabox integral families, we expect that, in the near future, all five-point massless two-loop Feynman integrals
will become available. Completion of the full list of two-loop master integrals with up to eight denominators and vanishing internal masses seems 
feasible for the years to come. Nevertheless,  calculation of the full list of  two-loop master integrals with up to eight denominators and non-vanishing internal masses
requires further understanding of the class of functions involved. Combination of analytical and numerical approaches may be the final solution.  
Once this problem is solved, the efficient computation of arbitrary two-loop scattering amplitudes will be realized, providing the basis of 
the most precise theoretical predictions for current and future high-energy collider experiments.


 \label{sec-costas}
\clearpage
\pagestyle{empty}
\cleardoublepage

\cleardoublepage
\pagestyle{empty}
\cleardoublepage
\pagestyle{empty}
\section
[Exploring the function space of Feynman integrals \\ {\it S. Weinzierl}]
{Exploring the function space of Feynman integrals 
} \label{sec:sweinz}

\pagestyle{fancy}
\fancyhead[CO]{\thechapter.\thesection 
\hspace{1mm}
Exploring the function space of Feynman integrals 
}
\fancyhead[RO]{}
\fancyhead[LO]{}
\fancyhead[LE]{}
\fancyhead[CE]{S. Weinzierl}
\fancyhead[RE]{} 

\noindent
{\bf Author: Stefan~Weinzierl} {~~[weinzierl@uni-mainz.de]}
\vspace*{.5cm}

\subsection{\label{weinzierl:sect:intro}Introduction: precision calculations for the FCC-ee}

The physics programme of a future circular collider operating in electron--positron annihilation mode
will be centred around a detailed study of the heavy particles of the Standard Model:
the Z and W bosons, the Higgs boson, and the top quark.
Precision studies of these particles are sensitive to contributions from new physics
at higher mass scales reaching the order of $100\UTeV$.
However,  to extract these effects, the experimental precision must be matched with
the same precision in the theoretical calculations.
This requires the computation of quantum corrections at the two- or three-loop order.
There is a class of Feynman integrals, which evaluate to multiple polylogarithms.
This class of Feynman integrals is  now quite well understood and 
captures a significant part of the Feynman integrals occurring in massless quantum field theories.
However, for the processes relevant to the FCC-ee we do not want to neglect the masses 
of the heavy particles of the Standard Model (\ie the Z, W, Higgs, and top masses).
With non-zero internal masses, we already leave  the function space of multiple polylogarithms at two loops.
The simplest Feynman integral, which cannot be expressed in terms of multiple polylogarithms, is given
by the two-loop sunrise integrals with equal non-zero masses.
The systematic study of these functions, to which these more complicated Feynman integrals evaluate, is now an
active field of study.
In this section, I will summarize the current state of the art, based on
Ref. \cite{Adams:2017xsu}.
As it is a field under active development, it is likely that this survey will be outdated in a few years.
To put it differently, the need for precision (as imposed by the FCC-ee programme)
triggers research in this direction; it is not unlikely that we will see substantial progress in the years to come.


\subsection{Differential equations and multiple polylogarithms}

Let us start with a review of differential equations and multiple polylogarithms.
The method of differential 
equations \cite{Kotikov:1990kg,Kotikov1991b,Remiddi:1997ny,Gehrmann:1999as,Argeri:2007up,MullerStach:2012mp,Henn:2013pwa,Henn:2014qga,Ablinger:2015tua,Bosma:2017hrk}
is a powerful tool for tackling Feynman integrals.
Let $t$ be an external invariant (\eg $t=(p_i+p_j)^2$) or an internal mass and let $I_i \in \{I_1,\dots ,I_N\}$ be a master integral.
Carrying out the derivative $\partial I_i/\partial t$ under the integral sign and using 
integration-by-parts identities allows us to express the derivative as a linear combination of the master integrals:
\begin{equation}
 \frac{\partial}{\partial t} I_i
 = 
 \sum\limits_{j=1}^N a_{ij} I_j
\end{equation}
More generally, let us denote by $\vec{I} = \left(I_1,\dots,I_N\right)$ the vector of the master integrals,
and by $\vec{x} = \left(x_1,\dots,x_n\right)$ the vector of kinematic variables
on which the master integrals depend.
Repeating this procedure for every master integral and every kinematic variable, we obtain a system of differential equations of Fuchsian type,
\begin{equation}
\label{weinzierl:diff_eq}
 \mathrm{d}\vec{I}  =  A \vec{I},
\end{equation}
where $A$ is a matrix-valued one-form
\begin{equation}
 A  =  
 \sum\limits_{i=1}^n A_i \mathrm{d}x_i.
\end{equation}
The matrix-valued one-form $A$ satisfies the integrability condition $\mathrm{d}A - A \wedge A = 0$.

There is a class of Feynman integrals that may be expressed in terms of multiple polylogarithms.
Multiple polylogarithms are defined by the nested sum 
\cite{Goncharov:1998kja,Goncharov:2001iea,Borwein,Moch:2001zr}
\begin{equation}
 \mbox{Li}_{m_1,m_2,\dots,m_k}(x_1,x_2,\dots,x_k)
  =  
 \sum\limits_{n_1 > n_2 > \dots > n_k > 0}^\infty 
 \;\;\;
 \frac{x_1^{n_1}}{n_1^{m_1}} \cdot \frac{x_2^{n_2}}{n_2^{m_2}} \cdot \ldots \cdot \frac{x_k^{n_k}}{n_k^{m_k}}.
\end{equation}
There is an alternative definition, based on iterated integrals
\begin{equation}
 G(z_1,\dots,z_k;y) 
  =  
 \int\limits_0^y \frac{\mathrm{d}t_1}{(t_1-z_1)}
 \int\limits_0^{t_1} \frac{\mathrm{d}t_2}{(t_2-z_2)} \dots
 \int\limits_0^{t_{k-1}} \frac{\mathrm{d}t_k}{(t_k-z_k)}.
\end{equation}
The two notations are related by
\begin{equation}
\mbox{Li}_{m_1,\dots,m_k}(x_1,\dots,x_k) 
 = 
 (-1)^k 
 G_{m_1,\dots,m_k}\left( \frac{1}{x_1}, \frac{1}{x_1 x_2},\dots, \frac{1}{x_1\cdots
 x_k};1 \right),
\end{equation}
where
\begin{equation}
G_{m_1,\dots,m_k}(z_1,\dots,z_k;y)  = 
 G(\underbrace{0,\dots,0}_{m_1-1},z_1,\dots,z_{k-1},\underbrace{0, \dots,0}_{m_k-1},z_k;y).
\end{equation}
Let us return to the differential equation (\Eref{weinzierl:diff_eq}). If we change the basis of the master
integrals $\vec{J} = U \vec{I}$, where $U$ is a $N \times N$ matrix, whose entries may depend on $\vec{x}$ and 
the dimensional regularization parameter $\varepsilon$, the differential equation becomes
\begin{equation}
 \mathrm{d} \vec{J} = A' \vec{J}, \qquad
   A' = U A U^{-1} - U \mathrm{d} U^{-1}.
\end{equation}
Suppose further that one finds a transformation matrix $U$, such that
\begin{equation}
 A'  =  \varepsilon \sum\limits_j  C_j  \mathrm{d}\ln p_j(\vec{x}),
\end{equation}
where the dimensional regularization parameter $\varepsilon$ appears only as a prefactor,
the $C_j$ are matrices with constant entries, and  the 
$p_j(\vec{x})$ are polynomials in the external variables; then the system of differential equations 
is easily solved in terms of multiple polylogarithms \cite{Henn:2013pwa,Henn:2014qga}.
To obtain the $\varepsilon$-form, we may perform a transformation of the kinematic variables
\begin{equation}
 (x_1,\dots,x_n)  \rightarrow  (x_1',\dots,x_n').
\end{equation}
This corresponds to a change of variables in the base manifold.
Quite often, the transformation is rational or algebraic.
A change of kinematic variables can be made to absorb square roots for massive integrals.
For example, the transformation ($x=s/m^2$)
\begin{equation}
 \frac{s}{m^2}  =  - \frac{\left(1-x'\right)^2}{x'},
\end{equation}
rationalizes the typical square root occurring in two-particle cuts:
\begin{equation}
 \frac{\mathrm{d}s}{\sqrt{-s\left(4m^2-s\right)}}
  = 
 \frac{\mathrm{d}x'}{x'}.
\end{equation}
In addition, we may change the basis of the master integrals
\begin{equation}
\label{weinzierl:fibre_trafo}
 \vec{I}  \rightarrow  U \vec{I}.
\end{equation}
Quite often, $U$ is rational in the kinematic variables.
A transformation of the form of Eq.~(\ref{weinzierl:fibre_trafo}) corresponds to a change of basis in the fibre.
Methods to find the right transformation have been
discussed in Refs. \cite{Gehrmann:2014bfa,Argeri:2014qva,Lee:2014ioa,Prausa:2017ltv,Gituliar:2017vzm,Meyer:2016slj,Adams:2017tga,Lee:2017oca,Meyer:2017joq,Becchetti:2017abb}.

At the end of the day, we would like to evaluate the multiple polylogarithms numerically, taking into account that 
the multiple polylogarithms $\mbox{Li}_{m_1,m_2,\dots,m_k}(x_1,x_2,\dots,x_k)$ have branch cuts as a function of the $k$ complex variables $x_1, x_2, \dots, x_k$.
The numerical evaluation can be achieved as follows. One uses a truncation of the sum representation within the region of convergence.
The integral representation is used to map the arguments into the region of convergence. In addition,
acceleration techniques are used to speed up the computation \cite{Vollinga:2004sn}.


\subsection{Beyond multiple polylogarithms: single-scale integrals}

Starting from two loops, there are integrals that cannot be expressed in terms of multiple polylogarithms.
The simplest example is given by the 
two-loop sunrise integral \cite{Broadhurst:1993mw,Berends:1993ee,Bauberger:1994nk,Bauberger:1994by,Bauberger:1994hx,Caffo:1998du,Laporta:2004rb,Kniehl:2005bc,Groote:2005ay,Groote:2012pa,Bailey:2008ib,MullerStach:2011ru,Adams:2013nia,Bloch:2013tra,Adams:2014vja,Adams:2015gva,Adams:2015ydq,Remiddi:2013joa,Bloch:2016izu}
with equal masses.
A slightly more complicated integral is the 
two-loop kite integral \cite{SABRY1962401,Remiddi:2016gno,Adams:2016xah,Adams:2017ejb,Bogner:2017vim,Adams:2018yfj},
which contains the sunrise integral as a subtopology.
Both integrals depend on a single dimensionless variable $t/m^2$. In the following, we will change the variable from $t/m^2$
to the nome $q$ of an elliptic curve or the parameter $\tau$, related
to the nome by $q=\mathrm{e}^{\mathrm{i} \pi \tau}$.
Before giving a definition of these new variables, let us first see how an elliptic curve emerges. For the sunrise integral, there are two possibilities.
The first option reads off an elliptic curve from the Feynman graph polynomial
\begin{equation}
\label{weinzierl:elliptic_curve}
 E_{\mathrm{graph}}
 : 
 - x_1 x_2 x_3 t + m^2 \left( x_1 + x_2 + x_3 \right) \left( x_1 x_2 + x_2 x_3 + x_3 x_1 \right) 
  =  0 \, ;
\end{equation}
 the second option obtains an elliptic curve from the 
maximal cut \cite{Baikov:1996iu,Lee:2009dh,Kosower:2011ty,CaronHuot:2012ab,Frellesvig:2017aai,Bosma:2017ens,Harley:2017qut} of the sunrise integral
\begin{equation}
 E_{\mathrm{cut}}
  : 
 y^2 
 -
 \left(x - \frac{t}{m^2} \right) 
 \left(x + 4 - \frac{t}{m^2} \right) 
 \left(x^2 + 2 x + 1 - 4 \frac{t}{m^2} \right)
  =  0.
\end{equation}
In the following, we will consider the elliptic curve of Eq.~(\ref{weinzierl:elliptic_curve}).
The periods $\psi_1$, $\psi_2$ of the elliptic curve are solutions of the homogeneous differential equation \cite{Adams:2013nia}.
In general, the maximal cut of a Feynman integral is a solution of the homogeneous differential equation for this Feynman integral \cite{Primo:2016ebd}.
We define the new variables $\tau$ and $q$ by
\begin{equation}
 \tau  =  \frac{\psi_2}{\psi_1},
 \qquad
 q  =  \mathrm{e}^{\mathrm{i} \pi \tau}.
\end{equation}
Let us now turn to the transcendental functions, in which we may express the sunrise and  kite integrals.
We remind the reader of the definition of the classical polylogarithms
\begin{equation}
 \mathrm{Li}_n\left(x\right)  =  \sum\limits_{j=1}^\infty \; \frac{x^j}{j^n}.
\end{equation}
Starting from this expression, we consider a generalization with two sums, which are coupled through the variable $q$:
\begin{equation}
 \mathrm{ELi}_{n;m}\left(x;y;q\right)  =  
 \sum\limits_{j=1}^\infty \sum\limits_{k=1}^\infty \; \frac{x^j}{j^n} \frac{y^k}{k^m} q^{j k}.
\end{equation}
The elliptic dilogarithm is a linear combination of these functions and the classical dilogarithm:
\begin{equation}
 \mathrm{E}_{2;0}\left(x;y;q\right)
  = 
 \frac{1}{\mathrm{i}}
 \left[
 \frac{1}{2} \mathrm{Li}_2\left( x \right) 
 - \frac{1}{2} \mathrm{Li}_2\left( x^{-1} \right)
 + \mathrm{ELi}_{2;0}\left(x;y;q\right)
 - \mathrm{ELi}_{2;0}\left(x^{-1};y^{-1};q\right)
 \right].
\end{equation}
In the mathematical literature, there exist various slightly different definitions 
of elliptic polylogarithms \cite{Beilinson:1994,Levin:1997,Levin:2007,Enriquez:2010,Brown:2011,Wildeshaus,Bloch:2013tra,Bloch:2014qca,Remiddi:2017har,Broedel:2017kkb,Broedel:2017siw,Broedel:2018iwv}. To express the sunrise and the kite integral
to all orders in $\varepsilon$, we introduce the functions
\begin{multline}
\label{weinzierl:def_ELi}
 \mathrm{ELi}_{n_1,\dots,n_l;m_1,\dots,m_l;2o_1,\dots,2o_{l-1}}\left(x_1,\dots,x_l;y_1,\dots,y_l;q\right)
\\
  = 
 \sum\limits_{j_1=1}^\infty \dots \sum\limits_{j_l=1}^\infty
 \sum\limits_{k_1=1}^\infty \dots \sum\limits_{k_l=1}^\infty
  \frac{x_1^{j_1}}{j_1^{n_1}} \cdots \frac{x_l^{j_l}}{j_l^{n_l}}
  \frac{y_1^{k_1}}{k_1^{m_1}} \cdots \frac{y_l^{k_l}}{k_l^{m_l}}
  \frac{q^{j_1 k_1 + \dots + j_l k_l}}{\prod\limits_{i=1}^{l-1} \left(j_i k_i + \dots + j_l k_l \right)^{o_i}}.
\end{multline}
Let us write the Taylor expansion of the sunrise integral around $D=2-2\varepsilon$ as
\begin{equation}
 S  = 
 \frac{\psi_1}{\pi}
 \sum\limits_{j=0}^\infty \varepsilon^j E^{(j)}.
\end{equation}
Each term in this $\varepsilon$-series is of the form
\begin{equation}
 E^{(j)}  \sim 
 \mbox{linear combination of} \;
 \mathrm{ELi}_{n_1,\dots,n_l;m_1,\dots,m_l;2o_1,\dots,2o_{l-1}}
 \quad \mbox{and} \quad \mathrm{Li}_{n_1,\dots,n_l}.
\end{equation}
Using dimensional-shift relations, this translates to the expansion around $D=4-2\varepsilon$.
Thus, we find that the functions of Eq.~(\ref{weinzierl:def_ELi}), together with the multiple polylogarithms, are the class of functions
that express the equal-mass sunrise graph and the kite integral to all orders in $\varepsilon$ \cite{Adams:2015ydq,Adams:2016xah}.

The functions in Eq.~(\ref{weinzierl:def_ELi}) are defined as multiple sums. We may ask whether every term in the $\varepsilon$-expansion
can be expressed in terms of iterated integrals.
For the equal-mass sunrise integral and the kite integral, this is indeed the case and  these Feynman integrals are related to modular forms \cite{Adams:2017ejb}.
A function $f(\tau)$ on the complex upper half plane is a modular form of weight $k$ for $\mathrm{SL}_2(\mathbb{Z}$) if
$f$ transforms under M\"obius transformation as
\begin{equation}
 f\left( \frac{a\tau+b}{c\tau+d} \right) = (c\tau+d)^k \cdot f(\tau) 
 \qquad \mbox{for} \qquad  \begin{pmatrix}
a & b \\ 
c & d
\end{pmatrix}  \in \mathrm{SL}_2(\mathbb{Z}).
\end{equation}
In addition, $f$ is required to be holomorphic on the complex upper half plane and at $\tau=\mathrm{i} \infty$.
Furthermore, there are modular forms for congruence subgroups of $\mathrm{SL}_2(\mathbb{Z})$. We introduce iterated integrals of modular forms
\begin{equation}
 I\left(f_1,f_2,\dots,f_n;\bar{q}\right)
  = 
 \left(2 \pi \mathrm{i} \right)^n
 \int\limits_{\tau_0}^{\tau} \mathrm{d}\tau_1
 f_1\left(\tau_1\right)
 \int\limits_{\tau_0}^{\tau_1} \mathrm{d}\tau_2
 f_2\left(\tau_2\right)
\dots
 \int\limits_{\tau_0}^{\tau_{n-1}} \mathrm{d}\tau_n
 f_n\left(\tau_n\right),
 \qquad
 \bar{q} = \mathrm{e}^{2\pi \mathrm{i} \tau}.
\end{equation}
As base point, it is convenient to take $\tau_0=\mathrm{i} \infty$.
Repeated sequences of letters are abbreviated as in $\{f_{1},f_{2}\}^3 = f_{1},f_{2},f_{1},f_{2},f_{1},f_{2}$.
With the help of the iterated integrals of modular forms, one finds a compact all-order expression for the equal-mass sunrise integral
around $D=2-2\varepsilon$ dimensions:
\begin{multline}
 S 
 =  
 \frac{\psi_1}{\pi}
 \mathrm{e}^{-\varepsilon I(f_2;q) + 2\sum\limits_{n=2}^\infty \frac{\left(-1\right)^n}{n} \zeta_n \varepsilon^n} 
 \left\{
  \sum\limits_{j=0}^\infty 
   \varepsilon^j 
   \sum\limits_{k=0}^{\lfloor \frac{j}{2} \rfloor} I\left( \left\{1,f_4\right\}^k, 1, f_3, \left\{f_2\right\}^{j-2k}; q\right)
 \right.
 \\
 \left.
 +
 \left[
 \sum\limits_{j=0}^\infty 
 \left(
 \varepsilon^{2j} I\left(\left\{1,f_4\right\}^j;q \right)
 -
 \frac{1}{2} \varepsilon^{2j+1} I\left(\left\{1,f_4\right\}^j,1;q \right)
 \right)
 \right]
 \sum\limits_{k=0}^\infty \varepsilon^k B^{(k)}
 \right\},
\end{multline}
where the $B^{(k)}$ are boundary constants.
This expression has uniform depth, \ie at order $\varepsilon^j$ one has exactly $(j+2)$ iterated integrations.
The alphabet is given by four modular forms $1$, $f_2$, $f_3$, $f_4$.
To give an example, the modular form $f_3$ is given by
\begin{equation}
 f_3 
 = 
 -
 \frac{1}{24}
 \left( \frac{\psi_1}{\pi} \right)^3
 \frac{t\left( t - m^2 \right)\left( t - 9 m^2 \right)}{m^6}.
\end{equation}
The letters $1$, $f_2$, $f_3$, and $f_4$ may be expressed as a linear combination of generalized Eisenstein series, which makes the property of being a modular form manifest.

Let us now return to the question of whether there is an $\varepsilon$-form for the differential equations for the sunrise and kite integrals.
It is not possible to obtain an $\varepsilon$-form by an algebraic change of variables or an algebraic transformation of the basis of master integrals.
However, by the (non-algebraic) change of variables from $t$ to $\tau$ and by factoring off the (non-algebraic) expression $\psi_1/\pi$ 
from the master integrals in the sunrise sector, one obtains an $\varepsilon$-form for the kite--sunrise family \cite{Adams:2018yfj}:
\begin{equation}
 \frac{\mathrm{d}}{\mathrm{d}\tau} \vec{I}
  = 
 \varepsilon A(\tau)  \vec{I},
\end{equation}
where $A(\tau)$ is an $\varepsilon$-independent $8 \times 8$-matrix whose entries are modular forms.

Let us turn to the numerical evaluation.
The complete elliptic integrals entering $\psi_1$ can be computed efficiently from the arithmetic--geometric mean.
The numerical evaluation of the ${\mathrm{ELi}}$-functions is straightforward in the region where the sum converges:
one simply truncates the $q$-series at a certain order, such that the desired numerical precision is reached.
Methods of mapping the arguments outside the region of convergence into this region have been discussed in Ref, \cite{Passarino:2017EPJC}.
It turns out that for the sunrise integral and the kite integral the $q$-series converges for 
all $t \in {\mathbb R}\backslash \{m^2, 9m^2, \infty\}$; in particular, there is no need to distinguish 
the cases $t<0$, $0<t<m^2$, $m^2<t<9m^2$, or $9m^2<t$ \cite{Bogner:2017vim}.
Our default choice of the periods corresponds to $q=0$ for $t=0$.
We may use a $\mathrm{SL}_2(\mathbb{Z})$-transformation so  that $q$ vanishes at a chosen singular point $t \in \{0,m^2,9m^2,\infty\}$ of the differential equation.
This gives a convergent $q$-series around the chosen singular point, and allows a numerical evaluation for all $t \in {\mathbb R}$.


\subsection{Towards multiscale integrals beyond multiple polylogarithms}

Let us now turn from single-scale integrals to multiscale integrals.
We are interested in those that are not expressible in terms of 
multiple polylogarithms \cite{Adams:2014vja,Sogaard:2014jla,Adams:2015gva,Bonciani:2016qxi,vonManteuffel:2017hms,Primo:2017ipr,Ablinger:2017bjx,Bourjaily:2017bsb,Hidding:2017jkk}
but are expressible in terms of elliptic generalizations of these functions.
Therefore, we expect  irreducible second-order factors in the differential equation for a given master integral.
A system of first-order differential equations is easily converted to a higher-order differential equation for a single master integral.
We may work modulo subtopologies; therefore, the order of the differential equation is given by the number of master integrals in this sector.
The number of master integrals in a given sector may be greater than two and we face the question of how to transform to a suitable basis of master integrals,
which decouples the original system of differential equations at order $\varepsilon^0$ to a system of maximal block size of two.
This can be done by exploiting the factorization properties of the Picard--Fuchs operator\cite{Adams:2017tga}.
With this aim, one first projects the problem to a single-scale problem by setting
$x_i\left(\lambda\right) = \alpha_i \lambda$ with $\alpha=[\alpha_1:\dots:\alpha_n] \in {\mathbb C} {\mathbb P}^{n-1}$ and
by viewing the master integrals as functions of $\lambda$. For the derivative with respect to $\lambda$, we have
\begin{equation}
 \frac{\mathrm{d}}{\mathrm{d}\lambda} \vec{I}
  = 
 B \vec{I},
 \qquad
 B  = 
 \sum\limits_{i=1}^n \alpha_i A_i,
 \qquad 
 B  =  B^{(0)} + \sum\limits_{j>0} \varepsilon^j B^{(j)}.
\end{equation}
To find the required transformation, we may work modulo $\varepsilon$-corrections, \ie
we focus on $B^{(0)}$. Let $I$ be one of the master integrals $\{I_1,\dots,I_N\}$.
We determine the largest number $r$, such that the matrix that expresses 
$I$, $(\mathrm{d}/\mathrm{d}\lambda)I$, \dots, $(\mathrm{d}/\mathrm{d}\lambda)^{r-1}I$ in terms of the original set $\{I_1,\dots ,I_N\}$ has full rank.
It follows that $(\mathrm{d}/\mathrm{d}\lambda)^rI$ can be written as a linear combination of $I, \dots, (\mathrm{d}/\mathrm{d}\lambda)^{r-1}I$.
This defines the Picard--Fuchs operator $L_r$ for the master integral $I$ with respect to $\lambda$:
\begin{equation}
 L_{r} I  =  0,
 \qquad
 L_r  =  \sum\limits_{k=0}^r R_k \frac{\mathrm{d}^k}{\mathrm{d}\lambda^k}.
\end{equation}
$L_r$ is easily found by transforming to a basis that contains $I, \dots, (\mathrm{d}/\mathrm{d}\lambda)^{r-1}I$.
We may factor the differential operator into irreducible factors \cite{vanHoeij:1997},
\begin{equation}
 L_r
  = 
 L_{1,r_1} L_{2,r_2} \cdots L_{s,r_s}, 
\end{equation}
where $L_{i,r_i}$ denotes a differential operator of order $r_i$.
We may then convert the system of differential equations at order $\varepsilon^0$ into a block triangular form with blocks of size $r_1$, $r_2$, \dots, $r_s$.
A basis for block $i$ is given by
\begin{equation}
 J_{i,j}  = 
 \frac{\mathrm{d}^{j-1}}{\mathrm{d}\lambda^{j-1}} L_{i+1,r_{i+1}} \cdots
  L_{s,r_s} I,
 \qquad 1 \le j \le r_i.
\end{equation}
This decouples the original system into subsystems of size $r_1, r_2, \dots
, r_s$.
Let us write the transformation to the new basis as
$\vec{J} = V\left(\alpha_1,\dots ,\alpha_{n-1},\lambda\right) \vec{I}$.
Setting 
\begin{equation}
 U  =  V\left(\frac{x_1}{x_n},\dots ,\frac{x_{n-1}}{x_n},x_n\right)
\end{equation}
gives a transformation in terms of the original variables $x_1, \dots , x_n$.
Terms in the original matrix $A$ of the form $\mathrm{d} \ln Z(x_1,\dots
,x_n)$,
where $Z(x_1,\dots ,x_n)$ is a rational function in $(x_1,\dots ,x_n)$ and homogeneous of degree zero in $(x_1,\dots ,x_n)$, map to zero in the matrix $B$.
These terms are, in many cases, easily removed by a subsequent transformation.
Let us look at an example. For the planar double-box integral for $\mathrm{t}\bar{{\mathrm{t}}}$-production with a closed top loop, one finds, in the top sector,
five master integrals.
These may be decoupled as
\begin{equation}
 5  =  1 + 2 + 1 + 1.
\end{equation}
Thus, we need to solve only two coupled equations, not five.


\subsection{Conclusions}

There is rapid progress on Feynman integrals, which cannot be expressed in terms of multiple polylogarithms.
The function space of these Feynman integrals leads to elliptic generalizations of multiple polylogarithms.
The simplest examples of these are the Feynman integrals belonging to the families of the
equal-mass sunrise integral and the kite integral.
At present, we have quite a good understanding of the Feynman integrals from this family:
with a non-algebraic change of variables and a non-algebraic basis transformation, it is possible to transform
the differential equation to an $\varepsilon$-form.
This allows us to obtain the analytical expression to any order in the dimensional regularization parameter.
At each order $\varepsilon$, the solution is  given either as a sum representation ($\mathrm{ELi}$-representation)
or as an iterated integral representation (iterated integrals of modular forms).
There exist fast and efficient methods for the numerical evaluation.

In  future, the focus will be on multiscale integrals beyond multiple polylogarithms.
These will be needed for precision calculations within the FCC-ee physics programme.
We expect that sufficient progress will be made within the coming years.

 \label{sec-sweinz}
\clearpage
\pagestyle{empty}
\cleardoublepage

\cleardoublepage
\section
[Direct calculation of multiloop integrals in $d= 4$ with the 
four-dimensional regularization/re\-normalization approach (FDR) 
\\ {\it R. Pittau}]
{Direct calculation of multiloop integrals in $d= 4$ with the 
four-dimensional regularization/renormalization approach (FDR)
\label{sec:rp}
}

\pagestyle{fancy}
\fancyhead[LO]{}
\fancyhead[CO]{\thechapter.\thesection 
\hspace{1mm} 
Direct calculation of multiloop integrals in $d= 4$ with FDR}
\fancyhead[RO]{}
\fancyhead[LE]{}
\fancyhead[CE]{R. Pittau}
\fancyhead[RE]{} 

\noindent
{\bf Author: Roberto Pittau} {~~[pittau@ugr.es]}
\vspace*{.5cm}

\noindent In view of the increasing complexity of the perturbative calculations needed to cope with the precision requirements of  future FCC-ee experiments, it is appropriate to investigate new methods for computing radiative corrections.  Here, we report on the main features of the FDR \cite{Pittau:2012zd} approach to multiloop calculus in the presence of UV and IR divergences.

In FDR, the UV subtraction is embedded in a new definition of the loop integration, allowing one to compute renormalized Green's functions directly in four dimensions. Note that this differs from other four-dimensional techniques \cite{Sborlini:2016gbr,Seth:2016hmv}, where dimensional regularization (DReg \cite{tHooft:1972tcz}) is implicitly assumed.

The advantage of working in four dimensions is that it is expected to lead to considerable simplification, especially in connection with numerical techniques. In addition, this approach is also attractive  in supersymmetric calculations, where the fermionic and bosonic sectors must share the same number of degrees of freedom.

\subsection{Ultraviolet divergent loop integrals}
The FDR loop integration over UV divergent integrands is defined in such a way as to preserve the fundamental properties of the gauge theories at the quantum level.
\bea
\label{eq:rpa}
&&\mbox{Shift invariance of the loop integrals.} \\
\label{eq:rpb}
&&\mbox{The possibility of simplifying, at the integrand level, reconstructed propagators and denominators.}\nonumber \\  \\
&&\mbox{The possibility of inserting subloop expressions in higher-loop calculations (subintegration consistency).} \nonumber\\
&&\label{eq:rpc}
\eea
The first condition ensures routing invariance, the second requirement maintains the necessary gauge cancellations, and the third property is essential to keep the theory unitary.\footnote{The unitarity equation $T-T^\dag= \mathrm{i} T^\dag T$ 
mixes different loop orders, so that it is essential that the result of a subloop integration also stays the same when embedded in higher-loop computations.} 

As an illustrative example, we consider the FDR loop integration over the  one-loop integrand
\begin{equation}
J^{\mu \nu} = \frac{l^2 g^{\mu \nu}-4 l^{\mu} l^{\nu}}{(l^2-M^2)^3},
\end{equation}
which is logarithmic divergent by power counting. This integral is responsible for the no-decoupling properties of the $\mathrm{H} \to \gamma \gamma$ amplitude when $m_\mathrm{H} \to \infty$, giving rise to a certain debate in the recent literature
\cite{Gastmans:2011wh,Melnikov:2016nvo}.

The UV behaviour is extracted by making the replacement
\begin{equation}
\label{eq:rp1}
l^2 \to \bar l^2 \equiv l^2-\mu^2
\end{equation}
%
and expanding
\begin{equation}
\label{eq:rp2}
\frac{1}{(\bar l^2-M^2)^3}= \frac{1}{\bar l^6}+M^2
\left(
 \frac{1}{(\bar l^2-M^2)^3 \bar l^2}
+\frac{1}{(\bar l^2-M^2)^2 \bar l^4}
+\frac{1}{(\bar l^2-M^2)^3 \bar l^6}
\right).
\end{equation}
This results in
\begin{equation}
J^{\mu \nu} \to \bar J^{\mu \nu} \equiv \frac{\bar l^2 g^{\mu \nu}-4 l^{\mu} l^{\nu}}{\bar l^6}
+ M^2(\bar l^2 g^{\mu \nu}-4 l^{\mu} l^{\nu})
\left(
 \frac{1}{(\bar l^2-M^2)^3 \bar l^2}
+\frac{1}{(\bar l^2-M^2)^2 \bar l^4}
+\frac{1}{(\bar l^2-M^2)^3 \bar l^6}
\right).
\end{equation}
Only the first term contains UV divergent integrands parametrized in terms of the unphysical scale $\mu$. It acts as a natural counterterm, cancelling the UV behaviour directly at the integrand level. Thus,
the FDR integration over $J^{\mu \nu}$ is {\em defined} as
\begin{equation}
\label{eq:rp3}
\int \left [\mathrm{d}^4 l \right ] \bar J^{\mu \nu} \equiv \lim_{\mu \to 0}
\int \mathrm{d}^4l \left(\bar J^{\mu \nu} -\frac{\bar l^2 g^{\mu \nu}-4 l^{\mu} l^{\nu}}{\bar l^6}\right).
\end{equation}
Finally, by tensor decomposition, one easily computes
\begin{equation}
\label{eq:rp4}
\int \left [\mathrm{d}^4 l \right ] \bar J^{\mu \nu}= g^{\mu \nu} \lim_{\mu \to 0}
\mu^2 \int \mathrm{d}^4l \frac{1}{\bar l^6}= -g^{\mu \nu} \frac{\mathrm{i} \pi^2}{2},
\end{equation}
which implies that there is a no-decoupling limit of $\mathrm{H} \to \gamma \gamma$.

The main properties of the FDR integration are as follows.
\begin{itemize}
\item Expansions as in~\Eqn{eq:rp2} can be performed at any loop order.
A two-loop example with  $
\bar D_1   = \bar{l}_1^2-m_1^2$, $\bar D_2   = \bar{l}_2^2-m_2^2$, $\bar D_{12} = \bar{l}_{12}^2-m_{12}^2$, and $l_{12}= l_1+l_2$ reads
\begin{multline}
\label{eq:rp4a}
\frac{1}{\bar D_1\bar D_2\bar D_{12}} =
{ \left[\frac{1}{\bar{l}_1^2\bar{l}_2^2 \bar{l}_{12}^2}\right]} 
+ m_1^2
{ \left[
\frac{1}{\bar{l}_1^4\bar{l}_2^2 \bar{l}_{12}^2}
 \right]}
+ m_2^2
{ \left[
\frac{1}{\bar{l}_1^2 \bar{l}_2^4\bar{l}_{12}^2}
 \right]}
+
 m_{12}^2
{
\left[
\frac{1}{\bar{l}_1^2 \bar{l}_2^2\bar{l}_{12}^4}
 \right]}
\\
+
\frac{m_1^4}{ \left (\bar D_1\bar{l}_1^4 \right )}
{ \left[\frac{1}{\bar{l}_2^4} \right]}
+ 
\frac{m_2^4}{ \left (\bar D_2\bar{l}_2^4 \right )}
{ \left[\frac{1}{\bar{l}_1^4} \right]}
+
 \frac{m_{12}^4}{ \left (\bar D_{12}\bar{l}_{12}^4 \right )}
{ \left[\frac{1}{\bar{l}_1^4} \right]} +J_{{\rm F}}(l_1,l_2) \,,
\end{multline}
where $J_{{\rm F}}(l_1,l_2)$ is UV convergent and UV divergent integrands are written between square brackets. Notice the appearance of factorized subdivergences beyond one loop. Thus,
\begin{equation}
\int \left [\mathrm{d}^4l_1 \right ]  \left [\mathrm{d}^4l_2 \right ] \frac{1}{\bar D_1\bar D_2\bar D_{12}} \equiv
\lim_{\mu \to 0} \int \mathrm{d}^4l_1 \mathrm{d}^4l_2  J_{{\rm F}}(l_1,l_2).
\end{equation}
A few more two-loop examples are reported in  Appendix C of Ref. \cite{Donati:2013voa}. 
\item In general, a logarithmic dependence of $\mu$ is left after the $\mu \to 0$ limit. In this case, $\mu$ is interpreted as the renormalization scale $\mu_{\mbox{\tiny R}}$. For instance \cite{Pittau:2012zd},
\begin{equation}
\label{eq:rp4b}
\int \left [\mathrm{d}^4 l \right ] \frac{1}{(\bar l^2-M^2)^2}= -\mathrm{i} \pi^2 \ln\frac{M^2}{\mu^2_{\mbox{\tiny R}}}.  
\end{equation}
\item In the absence of IR singularities,
$\lim_{\mu \to 0} \int \mathrm{d}^4l \bar J^{\mu \nu}= \int \mathrm{d}^4l J^{\mu \nu}$ in~\Eqn{eq:rp3}. 
\item UV counterterms, extracted as in Eqs.~\eqref{eq:rp2} and~\eqref{eq:rp4a},
should always be subtracted from the original integrands. As a consequence, UV divergent integrals coinciding with their own counterterms vanish. These are sometimes refereed as {\em vacuum integrals}. For instance,
  \begin{equation}
\int \left [\mathrm{d}^4l \right  ] \frac{1}{\bar l^4} = 
\int \left [\mathrm{d}^4l_1 \right ] \left [\mathrm{d}^4l_2 \right ]  \frac{1}{\bar{l}_1^2\bar{l}_2^2 \bar{l}_{12}^2}= 
\int \left [\mathrm{d}^4l_1 \right ] \left [\mathrm{d}^4l_2 \right ]  \frac{1}{\bar{l}_1^4\bar{l}_2^2 \bar{l}_{12}^2}= 0. 
\end{equation}
\item  In UV convergent integrals, the subtraction term  is zero, so that FDR and ordinary integration coincide, as should be the case.\item This definition fulfils \Eref{eq:rpa} because the same UV counterterm is shared by the shifted and  unshifted versions of any loop integral.   
\item To keep \Eref{eq:rpb}, the replacement of~\Eqn{eq:rp1} should also
be performed in the numerator when $l^2$ comes from  Feynman rules. We dub this the {\em global prescription}.
\item The replacement of~\Eqn{eq:rp1} {\em should not} be performed in the numerator when $l^2$ is generated by tensor decomposition. (UV divergent tensors
are defined by subtracting their corresponding UV counterterms, such as
${l^{\mu} l^{\nu}} / {\bar l^6}$ in~\Eqn {eq:rp3}. Replacing
$l^2$ with $\bar l^2$ when reducing them would change this definition. In practice, this means that the replacement of~\Eqn{eq:rp1} should be performed before tensor reduction.)
This difference is parametrized by the introduction of extra integrals of the kind
\begin{equation}
\int \left [\mathrm{d}^4 l \right ] \frac{l^2-\bar l^2}{(\bar l^2-M^2)^3}= 
\int \left [\mathrm{d}^4 l \right ] \frac{\mu^2}{(\bar l^2-M^2)^3}, 
\end{equation}
where the same UV subtraction should be performed as if $\mu^2= l^2$.
Using~\Eqn{eq:rp2} gives the constant
\begin{equation}
\label{eq:rpexi}
\int \left [\mathrm{d}^4 l \right ] \frac{\mu^2}{(\bar l^2-M^2)^3}= -\lim_{\mu \to 0}
\mu^2 \int \mathrm{d}^4l \frac{1}{\bar l^6}= \frac{\mathrm{i} \pi^2}{2},
\end{equation}
 which causes a non-zero value in the right-hand side of~\Eqn{eq:rp4}.
\item It is possible to use integrand manipulations, tensor reduction, and integration by parts \cite{Pittau:2014tva} directly on FDR integrals {\em before} using their explicit definition.
\end{itemize}
\subsection{Keeping unitarity}
With more than one loop, the requirement in \Eref{eq:rpb} may clash with the subintegration consistency of \Eref{eq:rpc}. In fact, the
global prescription at the level of a multiloop diagram may be incompatible with the global prescription needed in each of its UV divergent subloops. This is fixed by adding additional FDR integrals (called EEIs), which restore the correct behaviour. Such integrals can be inferred by directly considering the multiloop diagram and do not require the introduction of counterterms in the Lagrangian. This strategy has been proved to work at two loops in QCD \cite{Page:2015zca}.

It is intriguing  that the EEIs also provide a fix to two-loop `{\em naive}' FDH in DReg \cite{Page:2015zca}. By dubbing $G^{(\text{2-loop})}$ a generic two-loop QCD correlator, one finds that the replacement
\begin{equation}
\label{eq:rp5}
G^{(\text{2-loop})}_{\rm{bare,\,DReg}} \big \rvert_{n_s=4} 
\rightarrow 
G^{(\text{2-loop})}_{\rm{bare,\,DReg}} \big \rvert_{n_s=4} +   {\textstyle \sum_{\rm{Diag}}} 
{\rm EEI}_b|_{n_s=4}
\end{equation}
produces correlators with the right behaviour under renormalization.
In~\Eqn{eq:rp5}, $n_s= \gamma_\mu \gamma^\mu= g_{\mu \nu}g^{\mu \nu}$ is the number of spin degrees of freedom and the EEI$_b$s are DReg integrals obtained from the EEIs by dropping the subtraction term, such as
\begin{equation}
\label{eq:rp6}
\mbox{EEI} =  \int \left [\mathrm{d}^4l \right ] 
\frac{1}{(\bar l^2 -M^2)^2} \to  \int \mathrm{d}^dl
\frac{1}{(l^2 -M^2)^2} = \mbox{EEI$_b$}.
\end{equation}
To put it another way, the EEI$_b$s reproduce the effect of the evanescent operators needed in FDH and dimensional reduction to restore renormalizability, at least off-shell.
A two-loop study of the 
$\gamma^\ast \to \mathrm{q} \bar {\mathrm{q}}$ and $\mathrm{H} \to \mathrm{q} \bar {\mathrm{q}}$ QCD vertices
indicates that the same pattern is observed on-shell.\footnote{C.~Gnendiger and A.~Signer, private communication.}

\subsection{Infrared singularities}
In massless calculations, IR divergences are present  in both loop and phase
space integrals.
The replacement of~\Eqn{eq:rp1} in the loop propagators regulates soft and collinear virtual singularities. For instance, the massless one-loop triangle is defined as \cite{Pittau:2013qla}
\begin{equation}
\label{eq:rp7}
\int \left [\mathrm{d}^4l \right]\frac{1}{\bar l^2 \bar D_{p_1} \bar D_{p_2}} 
\equiv \lim_{\mu \to 0} \int \mathrm{d}^4l \frac{1}{\bar l ^2 \bar D_{p_1} \bar D_{p_2}}=
\frac{\mathrm{i} \pi^2}{2s} \ln^2\left(\frac{\mu^2}{-s-\mathrm{i}0} \right),
\end{equation}
where $\bar D_{p_i}= (l+p_i)^2-\mu^2$, $p_i^2= 0$, and  $s= (p_1-p_2)^2= -2 (p_1\cdot p_2)$.
Thus, the IR behaviour is parametrized in terms of logarithms of $\mu$.
These have to be compensated by  phase space integration over the real radiation.\footnote{Or are absorbed in the initial-state parton densities.} The
underlying mechanism for this cancellation is provided by the cutting rule
\begin{equation}
\label{eq:rpcutrule}
\frac{\mathrm{i}}{\bar l^2 + \mathrm{i}0^+} \to (2\pi) \delta_+({\bar l^2})
\end{equation}
depicted in \Fref{fig:rp1}. Uncut thick lines represent virtual propagators, replaced according to~\Eqn{eq:rp1}, while thick cut lines refer to external
particles with $\bar k_{i,j}^2= \mu^2$, dubbed `$\mu$-massive' particles.  For instance,~\Eqn{eq:rp7} can be rewritten as an eikonal integrated over a $\mu$-massive three-body phase space $\bar \Phi_3$
\begin{equation}
\label{eq:rpcorr}
\int_{\Phi_2} \Re \left(\int \left [\mathrm{d}^4l \right ]\frac{1}{\bar l^2 \bar D_{p_1} \bar D_{p_2}} \right)
= \lim_{\mu \to 0}\int_{\bar \Phi_3} \frac{1}{\bar s_{13} \bar s_{23}}~~~~\begin{cases}
\bar s_{ij}= (\bar k_i+\bar k_j)^2 \\
\bar k^2_{i,j}=\mu^2 
\end{cases}, 
\end{equation}
which acts as a counterterm for the real radiation.

\begin{figure}
\vspace*{-1.9cm}
\centering
\includegraphics[width=0.5\textwidth]{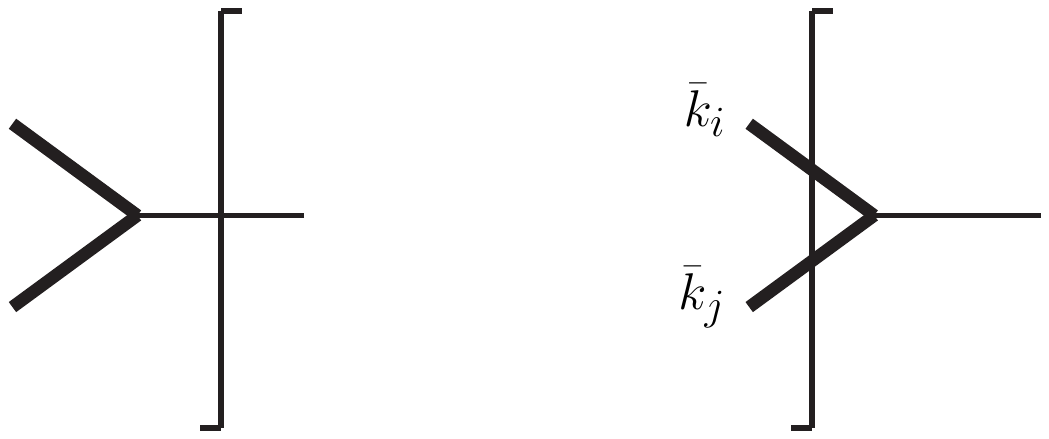}
\vspace*{-1.3cm}
\caption{IR-divergent splitting regulated by $\mu$-massive (thick) unobserved particles. The cut on the left represents the virtual part. The two-particle cut on the right contributes to the real radiation. The sum of the two is free of IR divergences.}
\label{fig:rp1}
\end{figure}

To keep gauge-invariance, numerator or denominator cancellations should be maintained when replacing
\begin{equation}
\label{eq:rp8}
k_i \to \bar k_i
\end{equation}
in the real matrix element before integrating it over the $\mu$-massive phase
space. This can be achieved by performing the massless calculation and then replacing everywhere:
\begin{equation}
s_{i \cdots j}= (k_i + \dots + k_j)^2 \to \bar s_{i \cdots j}= (\bar k_i + \dots + \bar k_j)^2. 
\end{equation}
As in the loop case, tensor reduction should be performed {\em after} the replacement of~\Eqn{eq:rp8}. This is usually immaterial in NLO computations but relevant at NNLO. Take, for instance, 
\begin{equation}
\gamma^\ast(P) \to q(k_1) \bar q(k_2) q^\prime(k_3) \bar q^\prime (k_4) \,
,
\end{equation}
considered  a NNLO QCD correction to $\gamma^\ast \to$ 2 jets.
A massless calculation generates the irreducible tensor
\begin{equation}
\frac{k_3^\mu k_3^\nu}{s_{134} s_{234} s^2_{34}}.
\end{equation}
The phase space integral
\begin{equation}
\label{eq:rp8a}
R^{\mu \nu} \equiv \int \mathrm{d} \Phi_4(P \to k_1 + k_2 + k_3 + k_4)
\frac{k_3^\mu k_3^\nu}{s_{134} s_{234} s^2_{34}}
\end{equation}
is IR-divergent, so that one must consider
\begin{equation}
\label{eq:rp9}
\bar R^{\mu \nu} \equiv \lim_{\mu \to 0}\int \mathrm{d} \bar \Phi_4 \left(P \to \bar k_1\!+\!\bar k_2\!+\!\bar k_3\!+\!\bar k_4 \right)
\frac{\bar k_3^\mu \bar k_3^\nu}{\bar s_{134} \bar s_{234} \bar s^2_{34}}.
\end{equation}
Only after the integral is defined, as in~\Eqn{eq:rp9}, is tensor reduction allowed. Splitting the four-body phase space gives
\begin{equation}
\label{eq:rp10}
\bar R^{\mu \nu} = \lim_{\mu \to 0}
\int_{4 \mu^2}^{(\sqrt{s}-2 \mu)^2} \mathrm{d} \bar s_{34}
\int \mathrm{d} \bar \Phi_3 \left (P \to \bar k_1 + \bar k_2 + \bar k_{34}
\right )
\frac{1}{\bar s_{134} \bar s_{234} \bar s^2_{34}}
\int \mathrm{d} \bar \Phi_2 \left (\bar k_{34} \to \bar k_3 + \bar k_4 \right
) \bar k_3^\mu \bar k_3^\nu,
\end{equation}
and Lorentz invariance dictates
\begin{equation}
\label{eq:rp11}
\int \mathrm{d} \bar \Phi_2 \left (\bar k_{34} \to \bar k_3 + \bar k_4
\right ) \bar k_3^\mu \bar k_3^\nu=
\left[
- g^{\mu \nu}\frac{\bar s_{34}}{12}
\left(1-4 \frac{\mu^2}{\bar s_{34}} \right)
+ \frac{\bar k^\mu_{34} \bar k^\nu_{34}}{3}
\left(1-  \frac{\mu^2}{\bar s_{34}} \right)
\right]
\int \mathrm{d} \bar \Phi_2 \left (\bar k_{34} \to \bar k_3 + \bar k_4 \right). \end{equation}
Despite the $\mu \to 0$ limit, the ${\mu^2}/{\bar s_{34}}$ terms in~\Eqn{eq:rp11} produce a non-vanishing $\mu^2/\mu^2$ contribution when inserted in~\Eqn{eq:rp10}. Such terms are missed if one first reduces~\Eqn{eq:rp8a} and then
uses~\Eqn{eq:rp8}.

\subsection{Scaleless integrals}
As in DReg, scaleless FDR integrals which are both UV and logarithmically IR-divergent vanish.  This is a direct consequence of their definition.
For example, if $p^2= 0$, 
\begin{equation}
\int \left [\mathrm{d}^4l \right ] \frac{1}{\bar l^2 (\bar l^2+2(l \cdot p))}=
\int \left [\mathrm{d}^4l \right ] \left(
  \frac{1}{\bar l^4}
 -\frac{2(l \cdot p)}{\bar l^4 (\bar l^2+2(l \cdot p))}
\right)= 0,
\end{equation}
because the first term in the right-hand side is a vacuum integral and the last contribution gives zero by tensor decomposition.
Analogously, when $p_1^2 =p_2^2= 0$, 
\begin{equation}
\int \left [\mathrm{d}^4l_1 \right] \left [\mathrm{d}^4l_2 \right] \frac{(l_2 \cdot p_1)(l_2 \cdot p_2)}{\bar l_1^4 \left (\bar l_1^2+2(l_1 \cdot p_1)
\right ) \bar l_2^2  \left ((l_1+l_2)^2 -\mu^2 \right )}= 0.
\end{equation}
This facilitates the computation, since the wave function renormalization of massless external states never contributes. 

\subsection{Renormalization}
Unlike in the customary approach \cite{Bogoliubov:1957gp,Hepp:1966eg,Zimmermann:1969jj}, no order-by-order renormalization is necessary in FDR. The reason is that UV subdivergences are already subtracted by the definition of FDR integration, as in~\Eqn{eq:rp4a}.
This means that no UV counterterms need to be introduced in the Lagrangian ${\cal L}(p_i,\ldots,p_m)$.  The bare parameters $p_i$ in ${\cal L}(p_i,\ldots,p_m)$ are directly linked to physical observables by means of $m$ measurements
\begin{equation}
\label{eq:rp12}
{\cal O}_i^{\rm EXP}= {\cal O}_i^{\rm TH, \ell\text{- loop}}(p_i,\ldots,p_m), 
\end{equation}
which determine them globally in terms of measured observables ${\cal O}_i^{\rm EXP}$ and corrections computed at the perturbative level $\ell$ at which
one is working.
Inverting~\Eqn{eq:rp12} gives
\begin{equation}
\label{eq:rp13}
p_i^{\ell\text{- loop}} ({\cal O}_i^{\rm EXP},\ldots,{\cal O}_m^{\rm EXP}) \equiv \hat p_i,
\end{equation}
where the $\hat p_i$ do not have to be calculated iteratively in the perturbative expansion. Note that $p_i$ and $\hat p_i$ are both finite, so that we refer to~\Eqn{eq:rp13} as a global finite renormalization. 

In the presence of IR divergences, one has to disentangle logarithms of $\mu$ of IR origin from logs of the renormalization scale $\mu_{\mbox{\tiny R}}$
(see, \eg \Eqn{eq:rp4b}). The former compensate the divergent real radiation contribution, or are absorbed in the parton densities. The latter contribute to the running of the $\hat p_i$.

As an example, in massless QCD, the only parameter to be fixed is the strong coupling constant, which can be determined by means of $\mathrm{e}^+ \mathrm{e}^- \to \mathrm{hadrons}$, computed at the scale $Q^2$. In this case,~\Eqn{eq:rp13} reads, at one loop  
\begin{equation}
\hat p_1 \equiv \alpha_S^{{1\text{-loop}}} \left (\alpha_S^{\rm FDR} \left
(Q^2 \right )
\right )=
\frac{\alpha_S^{\rm FDR}(Q^2)}{1+\frac{\alpha_S^{\rm FDR}(Q^2)}{4 \pi}\left(11-2N_f/3\right)
 \ln\frac{\mu^2_{\mbox{\tiny R}}}{Q^2}}.
\end{equation}
As for the numerical value of the strong coupling constant, the shift between $\alpha_S^{\rm FDR}$ and $\alpha_S^{\rm \overline{MS}}$ is known up to two loops \cite{Page:2015zca}:
\begin{equation}
    \frac{\alpha_S^{\text{FDR}}}{\alpha_S^{\overline{\text{MS}}}}
    = 
        1 +  \left(\frac{\alpha_S^{\overline{\text{MS}}}}{4\pi}\right)\frac{N_c}{3}
      +  \left(\frac{\alpha_S^{\overline{\text{MS}}}}{4\pi}\right)^2
        \left \{
            \frac{89}{18} N_c^2 + 8 N_c^2 f 
            +N_f
            \left[
                N_c - \frac{3}{2}C_F - 
                f 
                \left(
                    \frac{2}{3}N_c + \frac{4}{3}C_F
                \right)
            \right]
        \right \},     \label{eq:rp14}
\end{equation}
where
\begin{equation}
\label{eq:rpf}
f = \int_0^1 \frac{\ln(x)}{1-x(1-x)}= -\frac{2}{\sqrt{3}}{\rm Cl}_2\left(\frac{\pi}{3}\right).
\end{equation}
Note that $\alpha_S^{\text{FDR}}=\alpha_S^{\text{FDH}}$ at one loop.
In massive QCD, the quark mass shift reads \cite{Page:2015zca}
\begin{equation}
    \frac{m_\mathrm{q}^{\text{FDR}}}{m_\mathrm{q}^{\overline{\text{MS}}}} = 
        1 - C_F
        \left(\frac{\alpha_S^{\overline{\text{MS}}}}{4\pi}\right) 
        + C_F
        \left(\frac{\alpha_S^{\overline{\text{MS}}}}{4\pi}\right)^2
        \left \{
            \frac{77}{24} N_c - \frac{5}{8}C_F + 
            f\left(9 N_c + \frac{11}{3} C_F \right)
                       + N_f \left(\frac{1}{4} - \frac{2}{3} f\right)
        \right \}.
       \label{eq:rp15}
\end{equation}
Equations (\ref{eq:rp14}) and (\ref{eq:rp15}) provide the transition rules from IR-finite QCD quantities computed in FDR and their analogue in $\overline{\rm MS}$.
\subsection{Making contact with other methods}
FDR is a four-dimensional renormalization approach independent of DReg. This is in contrast with other four-dimensional methods \cite{Sborlini:2016gbr,Seth:2016hmv}, where DReg is implicitly assumed but convenient subtractions are performed in order to directly set  $d \to 4$ in particular combinations of integrals.
The mechanism of avoiding an order-by-order renormalization by means of the EEIs described in Ref. \cite{Page:2015zca}  is also peculiar to FDR. To our knowledge, this represents a new approach to renormalization \cite{Pittau:2013ica}.
Nevertheless,  a one-to-one correspondence exists at NLO between DReg and FDR integrals \cite{Gnendiger:2017pys}.
As for one-loop integrals, one finds
\begin{equation}
{\Gamma(1-\epsilon) {\pi^\epsilon}} \int \frac{\mathrm{d}^nl}{\mu_{\mbox{\tiny R}}^{-2 \epsilon}}
F(l^2,l) \Bigg|_{\mu_{\mbox{\tiny R}}= \mu{\mbox{\,\,and $\frac{1}{\epsilon^i}$}=
0}}= \int \left [\mathrm{d}^4l \right ]~F(\bar l^2,l),
\end{equation}
while the connection between real radiation integrals reads
\begin{multline}
\label{eq:rpreal}
\left(\frac{\mu_{\mbox{\tiny R}}^2}{s}\right)^\epsilon \int_{\phi_3} \mathrm{d}x \mathrm{d}y \mathrm{d}z
( \cdots )
F (x,y,z)
\delta(1-x-y-z) (xyz)^{-\epsilon}\Bigg|_{\mu_{\mbox{\tiny R}}= \mu{\mbox{\,\,and $\frac{1}{\epsilon^i}$}=\, 0}}
 \\
= 
\int_{\bar \phi_3} \mathrm{d}\bar x \mathrm{d}\bar y \mathrm{d}\bar z
F (\bar x,\bar y,\bar z)
\delta(1-\bar x-\bar y-\bar z+3 \mu^2/s), 
\end{multline}
where $\phi_3$ and $\bar \phi_3$ are massless and $\mu$-massive three-body phase spaces, and $\bar x= \bar s_{12}/s$, $\bar y= \bar s_{13}/s$, and $\bar z= \bar s_{23}/s$.

To our knowledge, a direct correspondence between FDR and DReg integrals
cannot be found beyond one loop. This is probably because of the absence of counterterms in the FDR approach. The equivalence between DReg and FDR is restored only once all parts of the calculation are collected together to produce physical predictions.

\subsection{Results}
\label{sec:rpres}
In this section, we present a few selected results obtained in the framework of FDR.

\subsubsection{Internal consistency}

The consistency of the FDR approach up to NNLO has been proven by using off-shell QCD as a test case~\cite{Page:2015zca}.
An explicit calculation of the correlators in~\Fign{fig:rp2} shows that the FDR results can be mimicked by a particular choice of renormalization scheme in DReg, so that FDR produces the same physical predictions of DReg. The necessary shifts between the two schemes have been presented in Eqs.~(\ref{eq:rp14}) and (\ref{eq:rp15}).
 
\begin{figure}
\vspace*{-1.9cm}
\centerline{
\includegraphics[width=0.5\textwidth,angle=-90]{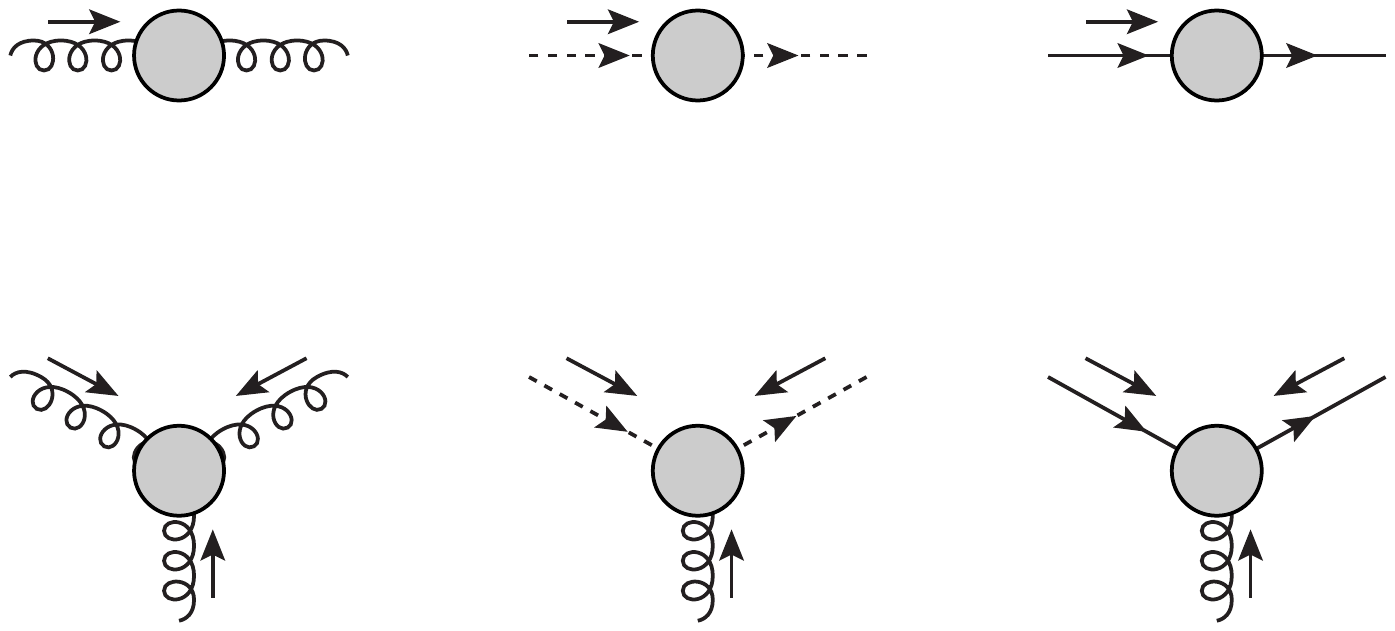}
}
\vspace*{-1.6cm}
\caption{
Irreducible QCD Green's functions. The grey blobs denote the sum of all possible two-loop Feynman diagrams.
}
\label{fig:rp2}
\end{figure}

It is also interesting to study how the correct value of the Adler--Bell--Jackiw anomaly is produced in FDR \cite{Pittau:2012zd}.
The contributing diagrams are given in~\Fign{fig:rpanomaly}. In the massless case, a convenient way of implementing the global prescription is to replace the 
fermion propagators, as follows:
\bea
\label{eq:rpshi}
\frac{1}{             \rlap/l} \to \frac{1}{            \rlap/l -\mu},~~~~~
\frac{1}{\rlap/ p_1 + \rlap/l} \to \frac{1}{\rlap/ p_1 + \rlap/l -\mu},~~~~~
\frac{1}{\rlap/ p_2 + \rlap/l} \to \frac{1}{\rlap/ p_2 + \rlap/l -\mu}.
\eea
Contracting with $p = p_1-p_2$ gives a result completely proportional to
a FDR extra integral:
\begin{equation}
p^\alpha \left(T^{(1)}_{\alpha \nu \lambda}+T^{(2)}_{\alpha \nu \lambda}\right)
= -\mathrm{i} \frac{e^2}{4\pi^4} 
{\rm Tr} \left [\gamma_5 \rlap/p_2 \gamma_\lambda \gamma_\nu \rlap/p_1 \right]
\int \left [\mathrm{d}^4l \right ] 
\mu^2
 \frac{1}{\bar l^2 ((l+p_1)^2-\mu^2) ((l+p_2)^2-\mu^2)}\,.
\end{equation}
Computing this as in~\Eqn{eq:rpexi} leads to the right answer.
Note that, before performing the shift of~\Eqn{eq:rpshi}, $\gamma_5$ should be put in the vertex in which the current is not conserved.

\begin{figure}
  \vspace*{-8.9cm}
\hspace*{-1.1cm} 
 \centerline{
\includegraphics[width=1.7\textwidth]{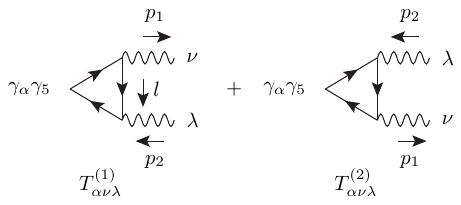}
}
\vspace*{-8.6cm}
\caption{The two diagrams generating the Adler--Bell--Jackiw anomaly
}
\label{fig:rpanomaly}
\end{figure}

\subsubsection{Gluonic NLO corrections to $\mathrm{H} \to \mathrm{gg}$ in the large top mass limit}

The well-known fully inclusive result
\begin{equation}
\Gamma(\mathrm{H} \to \mathrm{gg}) = \Gamma^{(0)}(\alpha_S(M_\mathrm{H}^2))
\left[ 
1+\frac{95}{4}\,\frac{\alpha_S}{\pi}
\right]
\end{equation}
has been re-derived in Ref. \cite{Pittau:2013qla}. 
The calculation of the diagrams in~\Fign{fig:rp2a} involves all key ingredients of QCD, namely ultraviolet, infrared and collinear divergences, besides $\alpha_S$ renormalization, showing that FDR can be used successfully in massless NLO computations.

\begin{figure}
\vspace*{-2.5cm}
\centerline{
\includegraphics[width=0.7\textwidth,angle= -90]{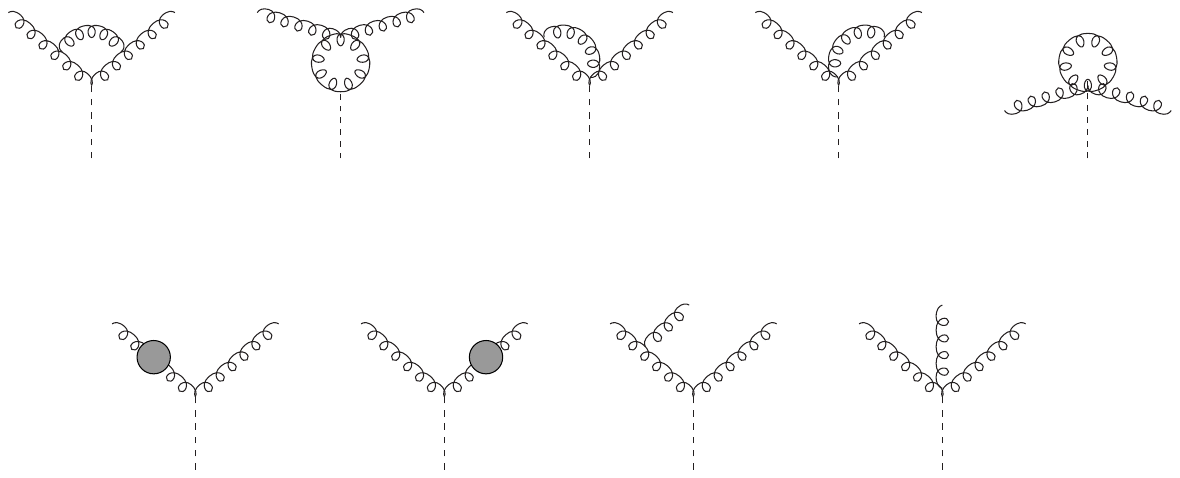}
}
\vspace*{-4.3cm}
\caption{Virtual and real diagrams contributing to $\mathrm{H} \to \mathrm{g} \mathrm{g} (\mathrm{g})$
 at ${\cal O}(\alpha_S^3)$}
\label{fig:rp2a}
\end{figure}

\subsubsection{$\mathrm{H} \to \gamma \gamma$ at two loops}

The leading QCD corrections to the amplitude $\cal M^{\text{2-loop}}(\mathrm{H} \to \gamma \gamma)$ have been computed  \cite{Donati:2013voa} by considering the $m_{\rm top} \to \infty$ limit of the two-loop diagrams depicted in~\Fign{fig:rp3}.
The known result
\begin{equation}
\label{eq:rp16}
\cal M^{\text{2-loop}}= \cal M^{\text{1-loop}}\left(1-\frac{\alpha_S}{\pi} \right) 
+ {\cal O}\left(\frac{m^2_{\rm H}}{m^2_{\rm top}} \right)
\end{equation}
is reproduced. It is worth mentioning that there is no need to renormalize, at one loop, the top mass and the Yukawa coupling. In fact, when $m_{\rm top} \to \infty$, no parameter is left to be fixed by means of~\Eqn{eq:rp13}, so that
the UV finite result in~\Eqn{eq:rp16} follows directly  from the two-loop computation. By contrast, one-loop renormalization is necessary in DReg to subtract the subdivergences in the diagrams of~\Fign{fig:rp3}.

\begin{figure}
\vspace*{-6.cm}
\centerline{
 \hspace*{-1.5cm} 
\includegraphics[width=1.3\textwidth]{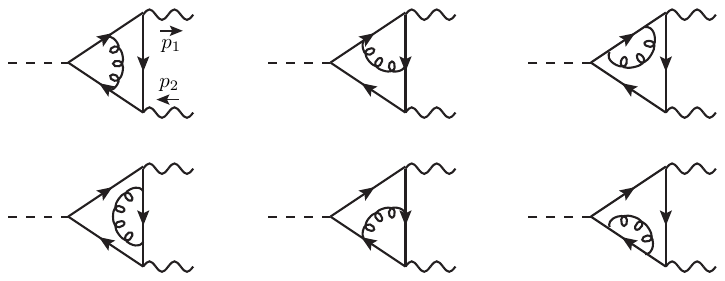}
}
\vspace*{-6cm}
\caption{
Feynman diagrams contributing to the QCD corrections of the top-loop-mediated Higgs decay into two photons. The same diagrams with the electric charge flowing counterclockwise also contribute.}
\label{fig:rp3}
\end{figure}

\subsubsection{The three-loop beta function}

By considering the difference between $\alpha_S^{\overline{\text{MS}}}$ and
$\alpha_S^{\rm FDR}$, it is possible to extract the three-loop beta function
in FDR:
\begin{equation}
    \beta^{\text{FDR}} = \mu \frac{\mathrm{d}}{\mathrm{d}\mu} \frac{\alpha_S^{\text{FDR}}}{4 \pi} = 
    \left(\frac{\alpha_S^{\text{FDR}}}{4 \pi}\right)^2
      \left[ 
          b_0^{\text{FDR}} + b_1^{\text{FDR}} \left( \frac{\alpha_S^{\text{FDR}}}{4 \pi} \right)
        + b_2^{\text{FDR}} \left(\frac{\alpha_S^{\text{FDR}}}{4 \pi}\right)^2
        +{\cal O} \left(\frac{\alpha_S^{\text{FDR}}}{4 \pi}\right)^3
      \right].
\end{equation}
One finds \cite{Page:2015zca}
\begin{multline}
b_2^{\text{FDR}} = N_c ^ 3 \left( -\frac{3610}{27}  -\frac{176}{3} f \right) 
+ N_f ^ 2 \left(
   -\frac{40}{9} C_F - \frac{43}{27} N_c + f \left( -\frac{16}{9} C_F -
   \frac{8}{9} N_c \right) \right)
\\
+ N_f \left( \frac{1331}{27} N_c ^ 2 +
   \frac{292}{9} N_c C_F - 2 C_F ^ 2 + f \left( \frac{140}{9} N_c ^ 2 +
   \frac{88}{9} N_c C_F \right) \right),
\end{multline}
with $f$ given in~\Eqn{eq:rpf}.

\subsubsection{Numerical evaluation of two-loop FDR integrals}

The ${\cal O}(G_F \alpha_s)$ corrections to the $\rho$ parameter in the $m_{\rm top} \to \infty$ limit have been determined numerically by Tom Zirke \cite{Zirke:2015spg} by analytically extracting the logarithmic $\mu$ dependence from the two-loop integrals presented in~\Fign{fig:rp4}. The comparison with the $\overline{\rm MS}$ result requires the use of~\Eqn{eq:rp15} at one loop.
  
\begin{figure}
\vspace*{-8.cm}
\centerline{
\includegraphics[width=1.5\textwidth]{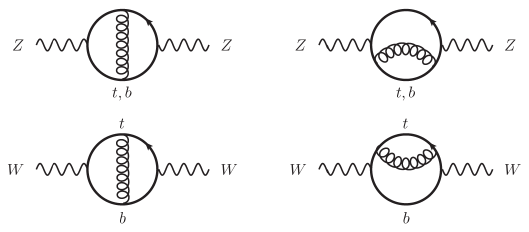}
}
\vspace*{-8.3cm}
\caption{Heavy quark corrections to the W and Z propagator contributing to
the $\rho$ parameter at ${\cal O}(G_F\alpha_S)$ (from Ref. \cite{Zirke:2015spg}).}
\label{fig:rp4}
\end{figure}

Finally, the result in~\Eqn{eq:rp16} is reproduced by means of numerical techniques.

\subsubsection{Local IR subtraction at one loop}

It is possible to set up a local FDR subtraction of the final-state IR infinities by rewriting the virtual logarithms as counterterms to be added to the real radiation \cite{Gnendiger:2017pys}, in the same spirit as~\Eqn{eq:rpcorr}. Schematically,  
\begin{align}
\label{eqn:snlo}
\sigma_{\rm{NLO} } &= 
\int_{\Phi_2}
\left(
|M|^2_{\rm{Born} } + 
\underbrace{|M|^2_{\rm{Virt}}}_{\rm{devoid~of~logs~of~\mu}}
\right)  F_\mathrm{J}^{\rm{\tiny(2)}}(k_1,k_2) \nonumber \\
& \quad +
\underbrace{\int_{\Phi_3}}_{\rm{\mu \to 0~here}} 
\left(
 |M|^2_{\rm{Real}}\, F_\mathrm{J}^{\rm{(3)}}(k_1,k_2,k_3) 
-|M|^2_{\rm{CT}}  
F_\mathrm{J}^{\rm{(2)}}\left(\underbrace{\hat k_1,\hat k_2}_{\rm{mapped~kinematics}}
\right)
\right ),
\end{align}
where $F_\mathrm{J}$ are jet functions.
For instance, in the case of $\mathrm{e}^+ \mathrm{e}^- \to \gamma^\ast \to \mathrm{q} \bar {\mathrm{q}}(\mathrm{g})$, the explicit form of the local counterterm is
\begin{equation}
\label{eq:ct}
|M|^2_{\rm{CT}} = \frac{16 \pi \alpha_s}{s} C_F
|M|^2_{\rm{Born}} \left ({\hat{k}_1,\hat{k}_2} \right ) 
\Biggl(
\frac{s^2}{ s_{13}  s_{23}}
-\frac{s}{ s_{13}}
-\frac{s}{ s_{23}} 
+ \frac{ s_{13}}{2 s_{23}}
+ \frac{ s_{23}}{2 s_{13}}
-\frac{17}{2}
\Biggr),
\end{equation}
and the mapping reads
\begin{equation}
\hat{k}^{\alpha}_1= \kappa \Lambda^{\alpha}_{\beta} k^\beta_1
\left(1+\frac{s_{23}}{s_{12}}\right), \qquad
\hat{k}^{\alpha}_2= \kappa \Lambda^{\alpha}_{\beta} k^\beta_2
\left(1+\frac{s_{13}}{s_{12}}\right),
\end{equation}
where
\[
 \kappa= \sqrt{\frac{s s_{12}}{(s_{12}+s_{13})(s_{12}+s_{23})}}
\]
  and 
$\Lambda^{\!\alpha}_{\,\beta}$ is the boost that brings  the sum of 
$\hat{k}_1$ and $\hat{k}_2$ back to the centre-of-mass frame:
$\hat{k}_1+\hat{k}_2= (\sqrt{s},0,0,0)$.

The inclusive cross-section
\[
 \sigma_{\rm  \mbox{\tiny NLO}}= \sigma_0 
\left(1+C_{\mbox{\tiny \rm F}} \frac{3}{4}
\frac{\alpha_s}{\pi}\right)
\]
 is reproduced by a numerical implementation of \Eref{eqn:snlo}. In addition, successful comparisons\footnote{M.~Moretti and R.~Pittau, in preparation.} with  
{\tt MadGraph5\_aMC@NLO} \cite{Alwall:2014hca} interfaced with {\tt FastJet} \cite{Cacciari:2011ma} have been attained for realistic jet observables.

\subsection{Outlook}
The results collected so far show that FDR is becoming a competitive tool for computing radiative corrections.
The UV subtraction is incorporated in the definition of the loop integration.
As a consequence, one deals directly  with four-dimensional integrals and avoids the customary order-by-order insertion of UV counterterms in the Lagrangian. 
This has been proven to be a workable alternative to DReg up to two loops for off-shell quantities.

The FDR regularization of the IR divergences is well understood at
NLO, and a completely local subtraction of final-state IR infinities has been
determined for two-jet cross-sections. Initial-state IR singularities have not yet been studied. However, the one-to-one correspondence in~\Eqn{eq:rpreal} makes it plausible that it should not be difficult to accommodate them in FDR, at least at NLO.

Going on-shell at two loops looks feasible. A complete NNLO FDR calculation of
the processes $\gamma^\ast,\mathrm{H} \to \mathrm{q} \bar {\mathrm{q}} ( \mathrm{q}^\prime \bar {\mathrm{q}}^\prime )$ is ongoing. In particular, it is interesting to study the correspondence in~\Eqn{eq:rpcutrule} beyond one loop. 
Ideally, one would like to exploit it to establish a general method to extract the NNLO IR behaviour directly from the virtual integrals, to be rewritten in the form of local counterterms for the real radiation without the need of computing them, as in~\Eqn{eq:rpcorr}.
If possible, this would pave the way for a completely numerical approach, considerably simplifying the computation of radiative corrections beyond NLO.

 \label{sec-rp}
\clearpage
\pagestyle{empty}
\cleardoublepage

\cleardoublepage

\pagestyle{fancy}
\fancyhead[LO]{}
\fancyhead[CO]{\thechapter.\thesection 
\hspace{1mm}
Subtractions versus unsubtractions of IR singularities at higher orders
}
\fancyhead[RO]{}
\fancyhead[LE]{}
\fancyhead[CE]{G. Rodrigo}
\fancyhead[RE]{} 

\section
[Subtractions versus unsubtractions of IR singularities at higher orders \\ {\it G. Rodrigo}]
{Subtractions versus unsubtractions of IR singularities at higher orders \label{sec:gr}
}

\noindent
{\bf Author: German Rodrigo} {~~[german.rodrigo@csic.es]}
\vspace*{.5cm}

\noindent One of the main difficulties in perturbative calculations at higher orders in quantum field theory (QFT) is the requirement to cancel
the unphysical soft and collinear singularities. Those singularities are the consequence of treating the quantum state with $N$ external 
partons as different from the quantum state with emission of extra massless particles at zero energy, such as photons and gluons,
and the possibility of emitting particles in exactly the same direction. Moreover, loops in QFT implicitly extrapolate the validity of the Standard Model  
to infinite energies, much above the Planck scale. 

The traditional approach to solving this problem and extracting physical predictions from the theory involves altering 
the dimensions of the space-time to, \eg $d=4-2\varepsilon$. In dimensional regularization (DReg)~\cite{Bollini:1972ui,tHooft:1972tcz,Cicuta:1972jf,Ashmore:1972uj,Wilson:1972cf},
the singularities appear as explicit poles in $1/\varepsilon$ through integration of the loop momenta and the phase space of real radiation.
After renormalization of the ultraviolet (UV) singularities, virtual and real quantum corrections contribute with poles in $1/\varepsilon$ of opposite 
sign, such that the total result is finite. Although this procedure efficiently transforms the theory into a calculable and well-defined mathematical framework, 
 much effort needs to be invested in evaluating loop integrals in higher space-time dimensions and in adequately subtracting the singularities of the 
real radiation contributions, particularly at higher perturbative orders. 

The general idea of subtraction involves introducing counterterms that mimic the local IR behaviour
of the real components and that can easily be integrated analytically in $d$ dimensions. In this way, the integrated
form is combined with the virtual component, while the unintegrated counterterm cancels the IR
poles originating from the phase space integration of the real radiation contribution.
Part of the success of the so-called NLO revolution, which led to the automation of radiative corrections 
at second-order in Monte Carlo event generators, was due, as well as to a better understanding of the 
mathematical beauty of scattering amplitudes, to the existence of general-purpose subtraction algorithms
at NLO~\cite{Kunszt:1992tn,Frixione:1995ms,Catani:1996jh,Catani:1996vz}.

In the last two years, there has also been  quite stunning progress at NNLO, with lots of new calculations of $2\to 2$ LHC processes. 
This progress has been possible thanks to the effort of many different groups, each one  developing different working subtraction 
methods at NNLO~\cite{GehrmannDeRidder:2005cm,Catani:2007vq,Czakon:2010td,Bolzoni:2010bt,Boughezal:2015dva,Gaunt:2015pea,DelDuca:2016ily,Caola:2017dug,Magnea:2018hab}. Unfortunately, there is not yet a general working method, as at NLO, able to overpass
the current frontier for $2\to 3$ processes at NNLO. 

An alternative approach is introduced by the four-dimensional unsubtraction ~\cite{Sborlini:2016gbr,Hernandez-Pinto:2015ysa,Sborlini:2016hat,Driencourt-Mangin:2017gop,Ramirez-Uribe:2017gbf}, which is based on  loop--tree duality (LTD)~\cite{Catani:2008xa,Bierenbaum:2010cy,Bierenbaum:2012th,Buchta:2014dfa,Buchta:2015wna}. 
The idea behind four-dimensional unsubtraction is to exploit a suitable mapping of momenta between the virtual and real kinematics in such a way that the summation 
over the degenerate soft and collinear quantum states is performed locally at integrand level without the necessity of introducing infrared (IR) subtractions. 
Suitable counterterms are used to cancel, also locally,  UV singularities, in such a way that calculations can be made without 
altering the dimensions of the space-time. The method should improve the efficiency of Monte Carlo event generators because it is meant for 
integrating simultaneously real and virtual contributions. 
The LTD formalism, or a similar framework, has also been used to derive causality and unitarity constraints~\cite{Tomboulis:2017rvd}, 
and to integrate numerically subtraction 
terms~\cite{Seth:2016hmv}, and can be related to the forward limit of scattering amplitudes~\cite{Catani:2008xa,CaronHuot:2010zt}. 
It has also been used in the framework of  colour-kinematics duality~\cite{Jurado:2017xut}.

There has also been lot of effort in the community to define perturbative methods directly in $d=4$ space-time dimensions, in order 
to avoid the complexity of working in a non-physical multidimensional space. Examples of these methods are the
four-dimensional formulation~\cite{Fazio:2014xea} of the four-dimensional helicity scheme,
the six-dimensional formalism ~\cite{Bern:2002zk}, implicit regularization ~\cite{Battistel:1998sz}, 
and four-dimensional regularization/renormalization (FDR)~\cite{Donati:2013voa}, see Section \ref{chmt}.\ref{sec:rp}.
For a review of all these methods see, \eg Ref.~\cite{Gnendiger:2017pys}.


Loop--tree duality~\cite{Catani:2008xa,Bierenbaum:2010cy,Bierenbaum:2012th} transforms any loop integral or loop scattering 
amplitude into a sum of tree-level-like objects that are constructed by setting on-shell a number of internal loop propagators
equal to the number of loops. Explicitly, LTD is realized by modifying the ${\mathrm{i}0}$ prescription of the Feynman propagators
that remain off-shell 
\begin{equation}
G_F(q_j) = \frac{1}{q_j^2-m_j^2+\mathrm{i} 0} \qquad \to \qquad G_D(q_i;q_j) = \left. \frac{1}{q_j^2-m_j^2-\mathrm{i} 0 \, \eta k_{ji}} \right|_{G_F(q_i) ~ \text{on-shell}}~,
\end{equation}
with $k_{ji} = q_j-q_i$, and $\eta^\mu$ an arbitrary future-like vector. The most convenient choice, however, is $\eta^\mu=(1,\bf{0})$, which is equivalent
to integrating out the loop energy components of the loop momenta through the Cauchy residue theorem. The left-over integration is
then restricted to the Euclidean space of the  loop three momenta. The dual prescription can hence be $+\mathrm{i} 0$ for some dual propagators, and
$-\mathrm{i} 0$ for others, and encodes, in a compact and elegant way, the contribution of the multiple cuts that are 
introduced by the Feynman tree theorem~\cite{Feynman:1963ax}. The on-shell condition is given by 
$\tilde \delta(q_i) = \mathrm{i} \, 2 \pi \, \theta(q_{i,0}) \, \delta(q_i^2-m_i^2)$, and determines that the
loop integration is restricted to the positive energy modes of the on-shell hyperboloids (light-cones for massless particles).
 
It is interesting to note that, although the on-shell loop three momenta are unrestricted, after analysing 
the singular behaviour of the loop integrand, one realizes that, thanks to a partial cancellation of 
singularities among different dual components, all the physical threshold and 
IR singularities are restricted to a compact region of the loop three momentum~\cite{Buchta:2014dfa}.
This relevant fact allows  the construction of mappings   between the virtual and real kinematics 
based on the factorization properties of QCD 
and then the summation over degenerate soft and collinear states.

As usual, the NLO cross-section is constructed in four-dimensional unsubtraction from the one-loop virtual correction with $N$ external partons 
and the exclusive real cross-section with $N+1$ partons 
\begin{equation}
\sigma^{\rm NLO} = \int_{N} \mathrm{d}\sigma_{{\rm V}}^{(1,{\rm R})}+ \int_{N+1} \mathrm{d}\sigma_{{\rm R}}^{(1)}~,
\end{equation}
integrated over the corresponding phase space, $\int_N$ and $\int_{N+1}$. The virtual contribution is obtained from its LTD representation 
\begin{equation}
 \int_N \mathrm{d}\sigma_{\rm V}^{(1,{\rm R})} = \int_N \int_{\vec \ell_1}   2 \Re  \langle {\cal M}^{(0)}_N|\bigg(\sum_i{\cal M}^{(1)}_N(\tilde \delta(q_i)) \bigg)- {\cal M}^{(1)}_{\rm UV} (\tilde \delta(q_{\rm UV})) \rangle   \hat{\cal O}(\{p_k\}_N)~.
 \label{eq:nlov}
\end{equation}
In Eq.~(\ref{eq:nlov}), ${\cal M}^{(0)}_N$ is the $N$-leg scattering amplitude at LO and ${\cal M}^{(1)}_N (\tilde \delta(q_i))$ 
is the dual representation of the unrenormalized one-loop scattering amplitude with the internal momentum 
$q_i$ set on-shell. The integral is weighted with the function $\hat{\cal O}(\{p_k\}_N)$ that defines a given observable, 
for example the jet cross-section in the $k_T$-algorithm. 
The expression includes appropriate counterterms that implement renormalization by subtracting the UV singularities 
locally, as discussed in Ref.~\cite{Sborlini:2016gbr,Ramirez-Uribe:2017gbf}, 
including UV singularities of degree higher than logarithmic that integrate to zero.

By means of an appropriate mapping between the real and virtual kinematics\cite{Sborlini:2016hat,Sborlini:2016gbr},
\begin{equation}
\{p_j'\}_{N+1} \to (q_i, \{p_k\}_N)~,
\label{mapping}
\end{equation}
the real phase space is rewritten in terms of the virtual phase space and the loop three momentum
\begin{equation}
\int_{N+1}  =  \int_{N} \, \int_{\vec{\ell_1}} \, \sum_i  {\cal J}_i(q_i)  \, {\cal R}_i(\{p_j'\}_{N+1})~,
\end{equation}
where ${\cal J}_i(q_i)$ is the Jacobian of the transformation with $q_i$ on-shell, and ${\cal R}_i(\{p_j'\}_{N+1})$ defines 
a complete partition of the real phase space
\begin{equation}
\sum_i {\cal R}_i (\{p_j'\}_{N+1}) = 1~.
\end{equation} 
In this way, the NLO cross-section can be cast into a single integral in the Born or virtual phase space and the loop three momentum
\begin{multline}
\sigma^{\rm NLO} =  \int_N \int_{\vec \ell_1} \bigg[  2  \Re  \langle {\cal M}^{(0)}_N|\bigg(\sum_i{\cal M}^{(1)}_N(\tilde \delta(q_i)) \bigg)- {\cal M}^{(1)}_{\rm UV} (\tilde \delta(q_{\rm UV})) \rangle    \hat{\cal O}(\{p_k\}_N)  \\
+ \sum_i {\cal J}_i(q_i)  {\cal R}_i(\{p_j'\}_{N+1})  |{\cal M}^{(0)}_{N+1}(\{p_j'\}_{N+1})|^2  \hat{\cal O}(\{p_j'\}_{N+1}) \bigg] \,.
\label{eq:nlovr}
\end{multline} 
The NLO cross-section defined in Eq.~(\ref{eq:nlovr}) has a smooth four-dimensional 
limit and can be evaluated directly in four space-time dimensions. DReg is only necessary to fix the UV renormalization 
counterterms in order to define the cross-section in, \eg the $\overline{\mathrm{MS}}$ scheme; the rest of the calculation 
is feasible directly at $d=4$.  Equation (\ref{eq:nlovr}) also exhibits  a smooth massless limit for massive partons if the mapping 
in Eq.~(\ref{mapping})  conveniently maps the quasicollinear configurations~\cite{Sborlini:2016gbr}. This is another advantage of the formalism because
it allows  description with a single implementation of the same process with either massless or massive partons. 


To extend LTD to higher orders~\cite{Bierenbaum:2010cy}, we need to introduce the following functions: 
\begin{equation}
G_F(\alpha_k) = \prod_{i\in \alpha_k} G_F(q_i)~, \qquad
G_D(\alpha_k) = \sum_{i\in \alpha_k} \tilde \delta(q_i) \,
\prod_{j\in \alpha_k, j\ne i} G_D(q_i;q_j)~, 
\label{eq:fdual}
\end{equation}
where $\alpha_k$ labels all the internal propagators, Feynman or dual, of a given subset. 
An interesting identity fulfilled by these functions is the following: 
\begin{equation}
\label{eq:split}
G_D(\alpha_i \cup \alpha_j) = G_D(\alpha_i) \, G_D(\alpha_j) + G_D(\alpha_i) \, G_F(\alpha_j) + G_F(\alpha_i) \, G_D(\alpha_j)~,
\end{equation}
involving the union of two subsets $\alpha_i$ and $\alpha_j$. 
These are all the ingredients necessary to extend LTD iteratively
to two loops and beyond. 
For example, a one-loop scattering amplitude with $N$ external partons gets the form
\begin{equation}
\label{eq:dualoneloop}
{\cal A}^{(1)}_N = \int_{\ell_1} {\cal N}(\ell_1,\{p_k\}_N) \, G_F(\alpha_1) = - \int_{\ell_1} {\cal N}(\ell_1,\{p_k\}_N) \, G_D(\alpha_1)~,
\end{equation}
where ${\cal N}(\ell_1,\{p_k\}_N)$ is the numerator. 

At two loops, all the internal propagators can be classified into three different subsets. 
Starting from the Feynman representation of a two-loop scattering amplitude,
\begin{equation}
\label{Ln}
{\cal A}^{(2)}_N =
\int_{\ell_1} \, \int_{\ell_2} {\cal N} (\ell_1, \ell_2,\{p_k\}_N) \, G_F(\alpha_1\cup \alpha_2 \cup \alpha_3)~, 
\end{equation}
we obtain, in a first step, by applying LTD to one of the loops (Eq.~(\ref{eq:dualoneloop})),
\begin{equation}
{\cal A}^{(2)}_N =
- \int_{\ell_1} \, \int_{\ell_2} {\cal N} (\ell_1, \ell_2,\{p_k\}_N) \, G_F(\alpha_1) \, G_D(\alpha_2 \cup \alpha_3)~.
\end{equation}
Before applying LTD to the second loop, it is necessary to use Eq.~(\ref{eq:split}) to express the dual function $G_D(\alpha_2 \cup \alpha_3)$
in a suitable form, where 
the dual integrand is split into a first term that contains two dual functions, and therefore two internal lines on-shell, and two 
more terms with the dual function of one of the subsets and Feynman propagators of the other, to which we can again 
apply LTD. The final dual representation of the two-loop amplitude in Eq.~(\ref{Ln}) is
\begin{multline}
{\cal A}^{(2)}_N  = 
\int_{\ell_1}  \int_{\ell_2} {\cal N} (\ell_1, \ell_2,\{p_k\}_N)  \otimes  \bigg\{  G_D(\alpha_2)  G_D(\alpha_1 \cup \alpha_3)  \\
 + 
G_D(-\alpha_2 \cup \alpha_1)   G_D(\alpha_3)  
-  G_F(\alpha_1)   G_D(\alpha_2)   G_D(\alpha_3) \bigg\}~.
\label{twoloopduality}
\end{multline} 
In Eq.~(\ref{twoloopduality}), it is necessary to take into account the fact
that the momentum flow in the loop formed by the union 
of $\alpha_1$ and $\alpha_2$ occurs in opposite directions. Therefore, it is necessary to change the direction 
of the momentum flow in one of the two sets. This is represented by adding a sign in front of, \eg $\alpha_2$;
namely, we have written
\begin{equation}
\int_{\ell_1} \int_{\ell_2} \, G_F(\alpha_1 \cup \alpha_2)  = - \int_{\ell_1} \int_{\ell_2} \, G_D(-\alpha_2 \cup \alpha_1)~.
\end{equation}
Changing the momentum flow is equivalent to selecting the negative energy modes. For the internal momenta 
in the set $\alpha_2$, this means 
\begin{equation}
\tilde \delta(-q_j) = \frac{\mathrm{i}  \pi}{q_{j,0}^{(+)}}  \delta \left
(q_{j,0}+q_{j,0}^{(+)} \right)~, \qquad j \in \alpha_2~.
\end{equation}

The dual representation of Eq.~(\ref{twoloopduality}) assumes that there are only single powers of the Feynman propagators. 
This restriction can no longer be avoided  at two loops where, for example, self-energy insertions in internal lines 
 automatically lead  to double powers of one propagator. However, all the double poles can be included with  clever labelling 
of the internal momenta in the set $\alpha_1$, exclusively, which is not integrated in the first instance. 
Therefore, the numerator will not be affected by the first application of LTD, and Eq.~(\ref{eq:split}) will still be valid, regardless of the numerator. 
The calculation of the residues of double poles to obtain the LTD representation requires the participation of the numerator. 
This is represented in Eq.~(\ref{twoloopduality}) by ${\cal N} (\ell_1, \ell_2,\{p_i\}_N) \, \otimes \,G_D(\pm \alpha_i\cup \alpha_1)$. 
A similar iterative procedure would extend LTD to higher orders.


Once we have obtained the dual representation of the two-loop scattering amplitude, we can outline how to extend 
four-dimensional unsubtraction at NNLO or higher orders. 
Analogously to the NLO case, at NNLO, the total cross-section consists of three contributions, 
\begin{equation}
\sigma^{\rm NNLO} = \int_{N} \mathrm{d}\sigma_{{\rm V}{\rm V}}^{(2)} + \int_{N+1} \mathrm{d}\sigma_{{\rm V}{\rm R}}^{(2)} + \int_{N+2} \mathrm{d}\sigma_{{\rm R}{\rm R}}^{(2)}~,
\end{equation}
where the double virtual cross-section $\mathrm{d}\sigma_{{\rm V}{\rm V}}^{(2)}$ receives contributions from 
the interference of the two-loop with the Born scattering amplitudes and the square 
of the one-loop scattering amplitude with $N$ external partons, the virtual-real cross-section 
$\mathrm{d}\sigma_{{\rm V}{\rm R}}^{(2)}$ includes the contributions from the interference of one-loop and tree-level scattering 
amplitudes with one extra external particle, and the double real cross-section $\mathrm{d}\sigma_{{\rm R}{\rm R}}^{(2)}$ comprises 
tree-level contributions with the emission of two extra particles. 
The LTD representation of the two-loop scattering amplitude is obtained by setting two 
internal lines on-shell~\cite{Bierenbaum:2010cy}. This leads to the two-loop dual components
$\langle {\cal M}^{(0)}_N|{\cal M}^{(2)}_N(\tilde \delta(q_i,q_j))\rangle$, while the two-loop momenta of the squared one-loop 
amplitude are independent and generate dual contributions of the type 
$\langle {\cal M}^{(1)}_N(\tilde \delta(q_i))|{\cal M}^{(1)}_N(\tilde \delta(q_j))\rangle$. In both cases, there are two independent
loop three momenta and $N$ external momenta, from which we can reconstruct the 
kinematics of the tree-level corrections entering $\mathrm{d}\sigma_{{\rm R}{\rm R}}^{(2)}$
and the one-loop corrections in $\mathrm{d}\sigma_{{\rm V}{\rm R}}^{(2)}$:
\begin{equation}
\{p_r''\}_{N+2} \to (q_i, q_j, \{p_k\}_N)~, \qquad 
(q_l' , \{p_s'\}_{N+1} )  \to (q_i, q_j, \{p_k\}_N)~,
\end{equation}

In summary, the bottleneck in higher-order perturbative calculations is not only the evaluation of multiloop Feynman 
diagrams, but also the gathering of all the quantum corrections from different loop orders (and thus different numbers of final-state partons). 
To match the expected experimental accuracy at the LHC, particularly in the high-luminosity phase, and at future colliders, 
new theoretical efforts are still needed to overcome the current precision frontier.

 \label{sec-gr}
\clearpage
\pagestyle{empty}
\cleardoublepage

\cleardoublepage

\pagestyle{fancy}
\fancyhead[CO]{\thechapter.\thesection 
\hspace{1mm}
Numerical integration with the \textsc{Cuba} Library}
\fancyhead[RO]{}
\fancyhead[LO]{}
\fancyhead[LE]{}
\fancyhead[CE]{T. Hahn}
\fancyhead[RE]{} 

\section
[Numerical integration with the \textsc{Cuba} Library \\ {\it T. Hahn}]
{Numerical integration with the \textsc{Cuba} Library \label{sec:hcuba}
}

\noindent
{\bf Author: Thomas Hahn} {~~[hahn@mpp.mpg.de]}
\vspace*{.5cm}








\subsection{Overview}

Concepts and implementation of the \textsc{Cuba} library for multidimensional
numerical integration are elucidated, with special emphasis on the
parallelization features.

The \textsc{Cuba} library 
\cite{Hahn:2004fe,Agrawal:2011tm,Nejad:2013ina,Hahn:2014fua} offers four 
algorithms (\Tref{table:Tom}) for multidimensional numerical integration.  All four can 
integrate vector integrands and  have interfaces for Fortran, C/{\tt C++}, 
and \textsc{Mathematica}.  Third-party interfaces exist for Maple \cite{MapleCuba},
R \cite{R2Cuba}, REDUCE \cite{REDUCECuba}, Python \cite{PythonCuba}, and
Julia \cite{JuliaCuba}, though none of these use \textsc{Cuba}'s parallel features.

The terms in \Tref{table:Tom} are explained in the following sections on concepts and algorithms.  Theoretical understanding is of limited use in 
picking the optimal integration routine in most practical applications, 
however.  Performance depends highly on the integrand and there are always 
cases, and not just academic ones, in which one routine outperforms the 
others or, conversely, in which one routine simply gives wrong results.  
This, of course, is the main reason why there are four independent and 
easily interchangeable algorithms in the \textsc{Cuba} library.

Numerical integration is a fairly obvious candidate for distributed 
computing.  Indeed, in Fortran and C/{\tt C++}, the \textsc{Cuba} routines by default 
automatically parallelize the sampling.  The details (and caveats) are 
discussed in the section on parallelization.


\subsection{Concepts}

\subsubsection{Deterministic versus Monte Carlo}

\textsc{Cuba} contains both deterministic and Monte Carlo integration methods.  
The deterministic approach is based on \emph{cubature rules},
\begin{equation}
\opI f\approx
\opQ_n f := \sum_{i = 1}^n w_i f(\vec x_i) \, ,
\end{equation}
with specially chosen \emph{nodes} $\vec x_i$ and \emph{weights} $w_i$.
Error estimation is achieved, \eg by null rules $\opN_m$ ($m < n$), which are 
constructed to give zero for functions integrated exactly by $\opQ_n$ 
and thus measure errors due to `higher terms'.

\begin{table}
\caption{Algorithms for multidimensional numerical integration in the \textsc{Cuba} library}
\label{table:Tom}
\centering
\begin{tabular}{llll}
\hline \hline
Routine  &
        Basic integration method &
        Type &
        Variance reduction \\ \hline 
Vegas &
        Sobol sample &
        Quasi-Monte Carlo &
        Importance sampling \\
&
        \textit{or} Mersenne Twister sample &
        Pseudo-Monte Carlo & \\
&
        \textit{or} Ranlux sample &
        Pseudo-Monte Carlo & \\
Suave &
        Sobol sample &
        Quasi-Monte Carlo &
        Globally adaptive subdivision \\
&
        \textit{or} Mersenne Twister sample &
        Pseudo-Monte Carlo &
        \quad + importance sampling \\
&
        \textit{or} Ranlux sample &
        Pseudo-Monte Carlo \\
Divonne &
        Korobov sample &
        Lattice method &
        Stratified sampling, \\
&
        \textit{or} Sobol sample &
        Quasi-Monte Carlo &
        \quad aided by methods from \\
&
        \textit{or} Mersenne Twister sample &
        Pseudo-Monte Carlo &
        \quad numerical optimization \\
&
        \textit{or} Ranlux sample &
        Pseudo-Monte Carlo \\
&
        \textit{or} cubature rules &
        Deterministic \\
Cuhre &
        Cubature rules &
        Deterministic &
        Globally adaptive subdivision 
\\ \hline \hline
\end{tabular}
\end{table}

The Monte Carlo estimate, although quite similar in form,
\begin{equation}
\opI f\approx \opM_n f := \frac 1n\sum_{i = 1}^n f(\vec x_i)\,,
\end{equation}
is conceptually very different, as this formula denotes the
\emph{statistical average} over independent and identically distributed
random samples $\vec x_i$.  In this case, the standard deviation 
furnishes a probabilistic estimate of the integration error:
\begin{equation}
\sigma(\opM_n f) = \sqrt{\opM_n f^2 - \opM_n^2 f}\,.
\end{equation}


\subsubsection{Construction of cubature rules}

Starting from an orthogonal basis of functions $\{b_1, \dots, b_m\}$ -- 
usually monomials -- with which most $f$ can (hopefully) be approximated
sufficiently well, one imposes the condition that each $b_i$ be integrated exactly by
$\opQ_n$: $\opI\,b_i \,\smash{\stackrel{\lower 3pt\hbox{\tiny !}}
{=}}\, \opQ_n b_i$.  There are $m$ moment equations
\begin{equation}
\sum_{k = 1}^n w_k b_i(\vec x_k) = \int_0^1\rd^d x\,b_i(\vec x)
\end{equation}
for $n d + n$ unknowns $\vec x_i$ and $w_i$.  They pose a formidable, 
in general non-linear, system of equations.  Additional assumptions, \eg 
symmetries, are usually necessary to solve this system.  \textsc{Cuba} employs 
the Genz--Malik rules \cite{Genz:1983}, constructed from a symmetric 
monomial basis.


\subsubsection{Globally adaptive subdivision}

Once an error estimate for the integral is available, global 
adaptivity is easy to implement.

\begin{enumerate}
\item Integrate the entire region: $\Itot\pm\Etot$.     \\[-4ex]
\item while $\Etot > \max(\epsrel\Itot, \epsabs)$       \\[-4ex]
\item\quad Find the region $r$ with the largest error.  \\[-4ex]
\item\quad Bisect (or otherwise cut up) $r$.            \\[-4ex]
\item\quad Integrate each subregion of $r$ separately.  \\[-4ex]
\item\quad $\Itot = \sum I_i$,~~
           $\Etot = \sqrt{\sum E_i^2}$.                 \\[-4ex]
\item end while
\end{enumerate}

A remark is in order here about the two precisions, $\epsrel$ and
$\epsabs$.  Naively, what one imposes is the relative precision: the
result is supposed to be accurate to, say, one part in a thousand, \ie
$\epsrel = 10^{-3}$.  For integral values approaching zero, however,
this goal becomes harder and harder to reach; so as not to spend
inordinate amounts of time in such cases, an absolute precision
$\epsabs$ can be prescribed, where typically $\epsabs\ll\epsrel$.


\subsubsection{Importance sampling}

Importance sampling introduces a weight function into the integral:
\begin{equation}
\opI f = \int_0^1\rd^d x 
  w(\vec x) \frac{f(\vec x)}{w(\vec x)}\,,
\qquad w(\vec x) > 0\,,
\qquad \opI w = 1\,,
\end{equation}
with two requirements:
(a) one must be able to sample from the distribution $w(\vec x)$ and
(b) $f/w$ should be `smooth' in the sense that
   $\sigma_w(f/w) < \sigma(f)$, \eg $w$ and $f$ should have
   the same peak structure.
The ideal choice is known to be $w(\vec x) = |f(\vec x)|/\opI f$, which
has $\sigma_w(f/w) = 0$, but is of little use, as it requires \textit{a-priori}
knowledge of the integral value.


\subsubsection{Stratified sampling}

Stratified sampling works by sampling subregions.  Consider a total of
$n$ points sampled in a region $r = r_a + r_b$, as opposed to\ $n/2$ points sampled
in $r_a$ and $n/2$ in $r_b$.  In the latter case, the variance is
\begin{equation}
\frac 14\left(\frac{\sigma_a^2 f}{n/2} +
              \frac{\sigma_b^2 f}{n/2}\right)
= \frac{\sigma_a^2 f + \sigma_b^2 f}{2n} \, ,
\end{equation}
whereas in the former case, it can be written as
\begin{equation}
\frac{\sigma^2 f}{n} =
  \frac{(\sigma_a^2 f + \sigma_b^2 f)}{2n} +
  \frac{(\opI_a f - \opI_b f)^2}{4n}\,.
\end{equation}
Even in this simple example, the latter variance is, at best, equal to the
former one, and only if the integral values are identical.  The optimal
reduction of variance can be shown to occur for $n_a/n_b = \sigma_a
f/\sigma_b f$ \cite{Press:1992zz}.  The recipe is thus to split up the
integration region into parts with equal variance, and then sample all
parts with the same number of points.


\subsubsection{Quasi-Monte Carlo methods}

Quasi-Monte Carlo methods are based on the Koksma--Hlawka inequality,
which gives an upper bound on the error of a cubature formula 
$\opQ_n f = \frac 1n\sum_{i = 1}^n f(\vec x_i)$,
\begin{equation}
|\opQ_n f - \opI f|\leqslant V(f)\,D^*(\vec x_1, \dots, \vec x_n)\,.
\end{equation}
Apart from choosing a different integrand, there is little one can do 
about $V(f)$, the `variation in the sense of Hardy and Krause'.
The \emph{discrepancy} $D^*$ of a sequence $\vec x_1, \dots, \vec x_n$
is defined as
\begin{equation}
D^* = \sup_{r\,\in\,[0, 1]^d}
  \left|\frac{\nu(r)}{n} - \Vol r\right|,
\end{equation}
where $\nu(r)$ counts the $\vec x_i$ that fall into $r$.  For an
equidistributed sequence, $\nu(r)$ should be proportional to $\Vol r$.
Quasi-random sequences can be constructed with a substantially lower
discrepancy than (pseudo-) \mbox{random} numbers.  A Monte Carlo algorithm based
on these sequences typically achieves convergence rates of $\ord(\log^{d
- 1} n/n)$, rather than the usual $\ord(1/\sqrt n)$.

\textsc{Cuba} offers a choice of quasi-random Sobol sequences 
\cite{Bratley:1988:AIS:42288.214372}, or  pseudo-random Mersenne Twister
\cite{Matsumoto:1998:MTE:272991.272995} or RANLUX 
\cite{Luscher:1993dy,James:1993np} sequences for all Monte Carlo 
algorithms.  Figure \ref{fig:quasipseudo} shows that quasi-random 
numbers cover the plane much more homogeneously than pseudo-random 
numbers.

\begin{figure}
\begin{minipage}{.5\hsize}
\centering
\includegraphics[width=.75\hsize]{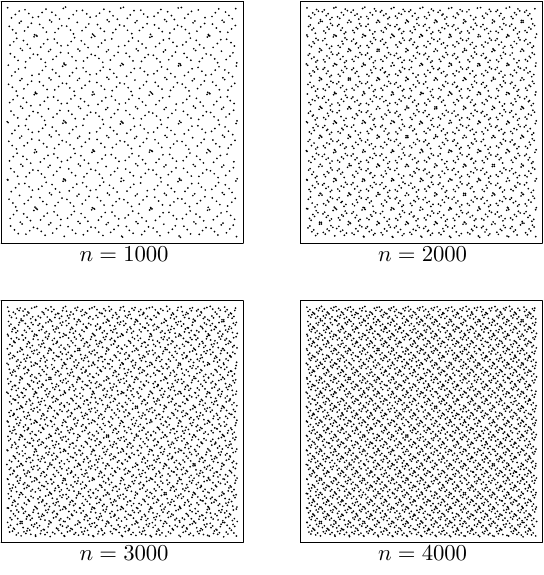} \\
Sobol quasi-random numbers
\end{minipage}\begin{minipage}{.5\hsize}
\centering
\includegraphics[width=.75\hsize]{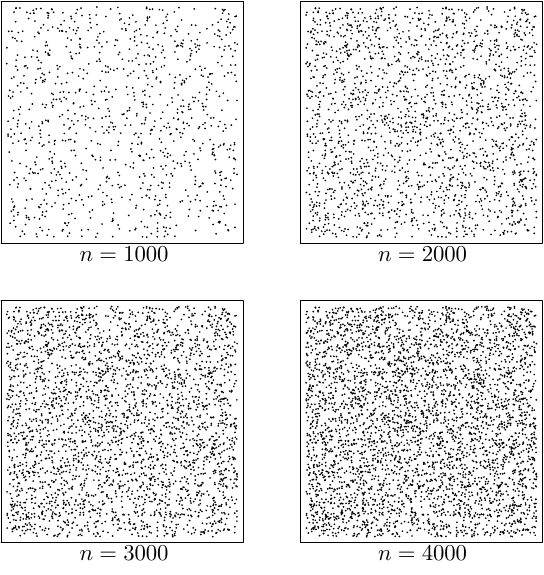} \\
Mersenne Twister pseudo-random numbers \\
\end{minipage}
\caption{\label{fig:quasipseudo}Comparison of sequences}
\end{figure}


\subsubsection{Lattice methods}

Lattice methods require a periodic integrand, usually obtained by
applying a \emph{periodizing transformation} (\textsc{Cuba}'s Divonne uses
$x\to |2x - 1|$).  Sampling is done on an \emph{integration lattice $L$} 
spanned by a carefully selected integer vector $\vec z$:
\begin{equation}
\opQ_n f = \frac 1n\sum_{i = 0}^{n - 1}
  f\bigl(\{\tfrac in\vec z\,\}\bigr)\,,
\qquad \{x\} = \text{fractional part of }x\,.
\end{equation}
where $\vec z$ is chosen (by extensive computer searches) to knock out as 
many low-order `Bragg reflections', as possible in the error term 
(see, \eg \cite{Keng:1981}):
\begin{equation}
\opQ_n f - \opI f
= \sum_{\vec k\in\mathbb{Z}^d}
  \tilde f(\vec k)\,\opQ_n\re^{2\pi\ri\,\vec k\cdot\vec x} -
  \tilde f(\vec 0)
= \sum_{\vec k\in L^\perp,\,\vec k\neq\vec 0} \tilde f(\vec k)\,,
\end{equation}
where $L^\perp = \{\vec k\in\mathbb{Z}^d: \vec k\cdot\vec z =
0\pmod{n}\}$ is the reciprocal lattice.


\subsection{Algorithms}

\subsubsection{Vegas}

Vegas is Lepage's classic Monte Carlo algorithm \cite{Lepage:1980dq,Lepage:1977sw}.  
It uses importance sampling for variance reduction, for which it iteratively
builds up a piecewise constant weight function, represented on a
rectangular grid.  Each iteration consists of a sampling step followed
by a refinement of the grid.

In \textsc{Cuba}'s implementation, Vegas can memorize its grid for subsequent
invocations and can save its internal state to a file from which it can 
be used in a different integration.


\subsubsection{Suave}

Suave is a cross-breed of Vegas and Miser \cite{Press:1989vk}, a Monte Carlo
algorithm that combines Vegas-style importance sampling with globally
adaptive subdivision.

The algorithm works as follows. Until the requested accuracy is reached,
bisect the region with the largest error along the axis in which the
fluctuations of the integrand are reduced most.  Prorate the number of
new samples in each half for its fluctuation.

The Vegas grid is kept across divisions, \ie a region that is the
result of $n - 1$ subdivisions has had $n$ Vegas iterations performed on
it.  On the downside, Suave is somewhat memory-hungry, as it needs to 
retain samples for later use.


\subsubsection{Divonne}
\label{sect:divonne}

Divonne is a significantly extended version of CERNlib's Algorithm D151
\cite{Friedman:1981ak,Friedman:1981:NPP:355934.355939}.  It is, essentially, 
a Monte Carlo algorithm but it also has cubature rules built in for comparison.  
Variance reduction is by stratified sampling, which is aided by methods 
from numerical optimization.  Divonne has a three-phase algorithm.
\begin{itemize}
\item[]\textnormal{\textit{Phase 1: Partitioning.}}

The integration region is split into subregions of (approximately) 
equal spread $s$, defined as
\begin{equation}
s(r) = \frac{\Vol r}{2}
  \left (\max_{\vec x\in r} f(\vec x) -
        \min_{\vec x\in r} f(\vec x)\right ).
\end{equation}
The minimum and maximum of each subregion are sought using methods from 
numerical optimization (basically, a quasi-Newton search).  Then
`dividers' are moved around   to find the optimal 
splitting, see Fig.~\ref{figdivid}.  This latter procedure can cleverly be translated into the 
solution of a linear system and is hence quite fast (for details see Ref. 
\cite{Friedman:1981:NPP:355934.355939}).

\begin{figure}
\centering
\includegraphics[width=.25\hsize]{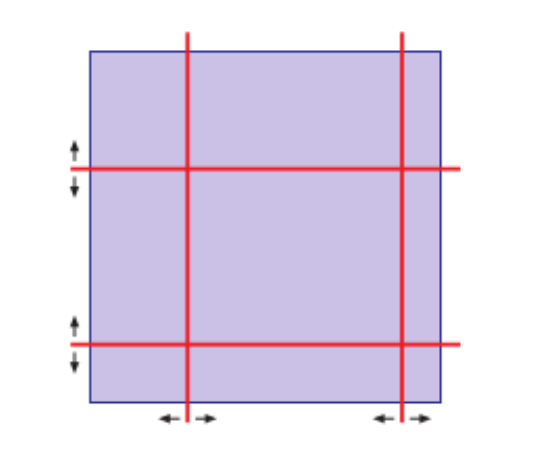
}
\caption{`Dividers' are moved around to find the optimal splitting \label{figdivid}}
\end{figure}

\item[]\textnormal{\textit{Phase 2: Sampling.}}

The subregions determined in Phase 1 are independently sampled with the 
same number of points each.  The latter is extrapolated from the results 
of Phase 1.

\item[]\textnormal{\textit{Phase 3: Refinement.}}

Regions whose results from Phase 1 and 2 do not agree within their
errors are subdivided or sampled again.  This phase is an addition to
the original algorithm, since it was often found that  the error
estimate, or even the integral value, was  off because characteristics of
the integrand had not been found in Phase 1.
\end{itemize}

Two important features have been added in the \textsc{Cuba} implementation.
\begin{itemize}
\item The user can point out extrema for tricky integrands.

\item For integrands that cannot be sampled too close to the border,
      a `safety distance' can be prescribed, in which values will be
      extrapolated from two points in the interior.
\end{itemize}


\subsubsection{Cuhre}

Cuhre is a deterministic algorithm.  It uses the Genz--Malik cubature
rules \cite{Genz:1983} in a globally adaptive subdivision scheme.  The
algorithm is thus: until the requested accuracy is reached, bisect the
region with the largest error along the axis with the largest fourth
difference.

Cuhre has been re-implemented in \textsc{Cuba}, mostly for a consistent
interface; it is the same as the original DCUHRE subroutine
\cite{Berntsen:1991:ADA:210232.210234}.


\subsection{\textsc{Mathematica} interface}

The \textsc{Mathematica} interface deserves a special mention, as it is not a
library in the proper sense.  It is, rather, four executables, which
communicate with \textsc{Mathematica} via the MathLink API (\Fref{fig:Fred}).

\begin{figure}
\centering
\includegraphics[scale=.5]{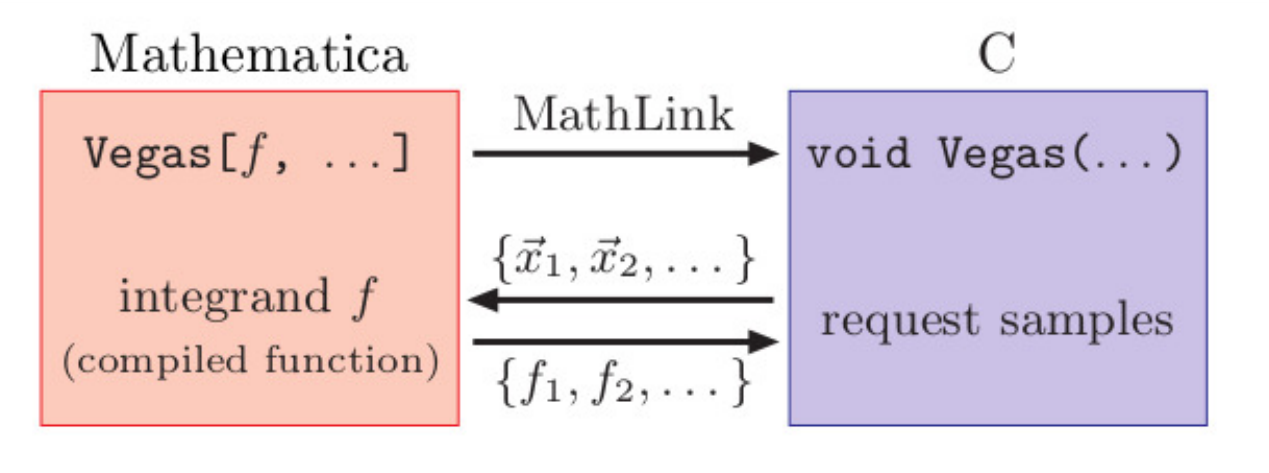}
\caption{The \textsc{Mathematica} interface}
\label{fig:Fred}
\end{figure}



After loading the appropriate MathLink executable, \eg with
\texttt{Install["Vegas"]}, the corresponding routine can be used almost
like \textsc{Mathematica}'s native \texttt{NIntegrate}.  The integrand is
evaluated completely in \textsc{Mathematica}, which means that one can do things
like
\begin{verbatim}
   Cuhre[Zeta[x y], {x,2,3}, {y,4,5}]
\end{verbatim}


\subsection{Parallelization}

Numerical integration is particularly suited for parallelization, which 
can significantly speed up the computation, as it generally incurs only a 
small overhead.  The following design considerations were relevant for the
parallelization of the \textsc{Cuba} routines.
\begin{enumerate}
\item \textit{No additional software shall be needed.}
Parallelization thus proceeds through operating-system functions only; 
no external message-passing interface is used.  This, of course, only fills the 
internal cores (typically four to eight), \ie parallelization across the 
network is not possible.  However, the speed-ups cannot be 
expected to increase linearly, so utilizing many more cores is more of a 
theoretical option anyway.

\item \textit{Parallelization shall work for any integrand.}
The programmer shall not need to rewrite the integrand function for 
parallel execution, which means, in particular, that reentrancy cannot be 
taken for granted.  \textsc{Cuba} uses \Code{fork}/\Code{wait} rather than the 
\Code{pthread*} functions, since the former create a completely
independent copy of the running process and thus work for any integrand.

\item \textit{Parallelization shall work `automatically'.}
The number of cores to use is determined through an environment variable 
and hence requires no re-compile.  More importantly, its default value 
is automatically set to the number of idle cores on the present machine, 
so a program using \textsc{Cuba} will automatically take advantage of the 
system's capabilities.

\item \textit{Parallelization shall be available on all platforms.}
The POSIX functions \Code{fork} and \Code{wait} are unavailable on 
native Windows, but Cygwin provides a suitably licensed (GPL) emulation.  
The emulated \Code{fork} is rather slow, however, so \textsc{Cuba} keeps the 
\Code{fork} calls to a minimum.
\end{enumerate}

\subsubsection{Parallelization model}

\textsc{Cuba} operates in a master--worker model.  The master process 
orchestrates the parallelization but does not count towards the number 
of cores, \eg \Code{CUBACORES = 4} means four workers and one master.  
Most importantly, the samples are generated by the master process only 
and distributed to the workers, such that random numbers are never used 
more than once.

The parallelization of \textsc{Cuba} centres on the main sampling routine 
\Code{DoSample}, which, in principle, parallelizes straightforwardly on 
$N$ cores:
\begin{quote}
\begin{tabular}{ll|ll}
\textit{Serial version:} &&&
\textit{Parallel version:} \\
        sample $n$ points. &&&
        sample $n/N$ points on core 1, \\[-.5ex]
&&&     \quad $\vdots$ \\[-.3ex]
&&&     sample $n/N$ points on core $N$.
\end{tabular}
\end{quote}
The actual distribution strategy is somewhat more involved and is
described in Section E.\ref{sect:cores}.

Speed-ups measured with Divonne by parallelizing \Code{DoSample} alone 
were generally unsatisfactory and significantly below those of the other 
integrators, \eg $\lesssim$ 1.5 on four cores.  It turned out that the 
partitioning phase (see Section E.\ref{sect:divonne}) was crucial for 
attaining reasonable speed-ups and needed special treatment.
Firstly, the original partitioning algorithm divided the regions 
recursively (and with a minimum recursion depth, too) and had to be 
`un-recursed', mainly by better bookkeeping of the subregions.
Secondly, the partitioning phase was modified, such that each core 
receives an entire region to subdivide, not just a list of points (as 
\Code{DoSample} does).  In particular, the minimum--maximum search, during 
which only one point at a time is sampled, is distributed much more 
efficiently this way.  The other two phases were not so critical, 
precisely because they sample more points per region.


\subsubsection{Parallelization in \textsc{Mathematica}}

The parallelization procedure is rather different in Fortran/C/{\tt C++} and 
in \textsc{Mathematica}; much of this has to do with the way in which the MathLink 
protocol works. Like all MathLink programs, \textsc{Cuba}'s run independently of 
\textsc{Mathematica}, so all parallelization must be done by 
\textsc{Mathematica} means.  (This is a clever way, also, of preventing the user from 
acquiring more licences than paid for.)

Sampling in \textsc{Mathematica} is handled through a function \Code{MapSample}, 
which, by default, is identical to \Code{Map}, that is,  the serial version.  
To parallelize sampling, one merely needs to redefine \Code{MapSample} \Code{=} \Code{ParallelMap} (after loading \textsc{Cuba}).
If the integrand depends on user-defined symbols or functions, their 
definitions must be distributed to the workers beforehand using 
\Code{DistributeDefinitions}; likewise, required packages must be 
loaded with \Code{ParallelNeeds} instead of \Code{Needs}.  This is 
standard procedure, however, and is explained in detail in the 
\textsc{Mathematica} manual.


\subsubsection{Parallelization in Fortran and C/{\tt C++}}

In Fortran and C/{\tt C++} the \textsc{Cuba} library can (and usually does) 
automatically parallelize the sampling of the integrand.  It 
parallelizes through \Code{fork} and \Code{wait}, which, though slightly 
less performant than pthreads, do not require reentrant code.  
(Reentrancy may not even be under the full control of the programmer, for 
example, Fortran's I/O is usually non-reentrant.)  Worker processes are 
started and shut down  as few times as possible, however, so the 
performance penalty is really quite minor, even for non-native fork 
implementations, such as Cygwin's.  Parallelization is not available on 
native Windows, for lack of the \Code{fork} function.

The communication of samples to and from the workers is done through IPC 
shared memory (\Code{shmget}, etc.) or, if that is not available, through 
a \Code{socketpair} (two-way pipe).  The control information is always 
transferred via \Code{socketpair}, since this elegantly and portably 
solves the synchronization issues.  Remarkably, shared memory's 
anticipated performance advantage turned out to be hardly perceptible.  
Possibly there are cache-coherence issues introduced by several workers 
writing simultaneously to the same shared-memory area.


\subsubsection{Starting and stopping the workers}
\label{sect:spinning}

The workers are usually started and stopped automatically by \textsc{Cuba}'s 
integration routines, but the user may choose to start them manually or 
keep them running after one integration and shut them down later, \eg at 
the end of the program, which can be slightly more efficient.  The 
latter mode is referred to as `spinning cores' and must be employed with 
certain care, for running workers will not `see' subsequent changes in 
the main program's data (\eg global variables, common blocks) or code 
(\eg via \Code{dlsym}) unless special arrangements are made (\eg shared 
memory).


\subsubsection{Accelerators and cores}
\label{sect:cores}

Based on the strategy used to distribute samples, \textsc{Cuba} distinguishes two 
kinds of worker.

Workers of the first kind are referred to as `accelerators', even though 
\textsc{Cuba} does not actually send anything to a GPU or accelerator in the 
system by itself -- this can only be done by the integrand routine.  The 
assumption behind this strategy is that the integrand evaluation is 
running on a device so highly parallel that the sampling time is more or 
less independent of the number of points, up to the number of threads 
$p\accel$ available in the hardware.  \textsc{Cuba} tries to send exactly $p\accel$ 
points to each core -- never more, less only for the last batch.  To 
sample, \eg 2400 points on three accelerators with $p\accel = 1000$, \textsc{Cuba} 
sends batches of 1000/1000/400 and not, for example, 800/800/800 or 
1200/1200.

CPU-bound workers are just called `cores'.  Their distribution strategy 
is different, in that all available cores are used and points are 
distributed evenly.  In the previous example, the batches would thus be 
800/800/800.  Each core receives at least 10 points, otherwise fewer 
cores are used.  If no more than 10 points are requested in total, \textsc{Cuba} 
uses no workers at all but lets the master sample those few points.  
This happens during the partitioning phase of Divonne, for instance, 
where only single points are evaluated in the minimum--maximum search. 
Conversely, if the division of points by cores does not come out even, 
the remaining few points ($< n\cores$) are simply added to the existing 
batches, to avoid an extra batch because of rounding.  Sampling 2001 
points on two cores with $p\cores = 1000$ will hence give two 
batches 1001/1000, not three batches 1000/1000/1.


\subsubsection{Concurrency issues}

By creating a new process image, \Code{fork} circumvents all memory 
concurrency, to wit: each worker modifies only its own copy of the 
parent's memory and never overwrites any other's data.  The programmer 
should be aware of a few potential problems, nevertheless:
\begin{itemize}
\item Communicating back results other than the intended output from the 
integrand to the main program or exchanging data between workers is not 
straightforward because, by the same token, each worker modifies only 
its own memory space.

\item \Code{fork} does not guard against competing use of other common 
resources.  For example, if the integrand function writes to a file 
(debug output, say), there is no telling in which order the lines will 
end up in the file, or even if they will end up as complete lines at 
all.

\item Fortran users should flush (or close) open files before calling 
\textsc{Cuba}, \ie \Code{call flush(\Var{unit})}, because the workers inherit all 
file buffers, and \emph{each} will write out the buffer content at 
exit.
\end{itemize}


\subsubsection{Vectorization}

Vectorization means evaluating the integrand function for several points 
at once.  This is different from ordinary parallelization, where 
independent threads are executed concurrently.  Vector instructions are 
commonly available in hardware; it is usually possible to employ 
vectorization on top of parallelization.

\textsc{Cuba} cannot automatically vectorize the integrand function, of course, 
but it does pass (up to) \Code{nvec} points per integrand call.  This 
value need not correspond to the hardware vector length -- computing 
several points in one call can also make sense, \eg if the computations 
have significant intermediate results in common.  The actual number of 
points passed is indicated through the corresponding \Code{nvec} 
argument of the integrand.


\subsection{Summary}

\textsc{Cuba} is a library for multidimensional numerical integration written
in C.  It contains four independent algorithms: Vegas, Suave, Divonne,
and Cuhre, which have similar invocations and can be exchanged easily for
comparison.  All routines can integrate vector integrands and have a
Fortran, C/{\tt C++}, and \textsc{Mathematica} interface.  

Concurrent sampling achieves significant speed-ups and is turned on by 
default for Fortran and C/{\tt C++} (but can be controlled through API calls or 
the environment).  No extra software needs to be installed, since only 
operating-system functions are used.  No reentrancy is required for the 
integrand function since \Code{fork}/\Code{wait} is applied.

\textsc{Cuba} is available from \Code{http://feynarts.de/cuba}, is licensed under 
the LGPL, and is easy to build (autoconf).  The download includes a manual ,
which gives full details on installation and usage.

 \label{sec-hcuba}
\clearpage \pagestyle{empty} 
\chapter*{Acknowledgements}
\phantomsection       
\addcontentsline{toc}{chapter}{Acknowledgements}


\vspace{2mm}\noindent
{The work of \textit{T.R.}\ was supported in part by an Alexander von Humboldt Polish Honorary Research Fellowship.

  The work of \textit{I.D.}\ was supported by a research grant from the Deutscher Akademischer Austauschdienst (DAAD) and by  Deutsches Elektronen-Sychrotron (DESY).

The work of \textit{W.F.} and \textit{J.G.} was supported by the Polish National Science Centre (NCN) (grant agreement 2017/25/B/ST2/01987).

  The work of \textit{A.F.}\  was supported in part by the National Science Foundation (grant no.\ PHY-1519175).
 
The work of \textit{R.P.} was supported by  MECD project FPA2016-78220-C3-3-P.

The work of \textit{M.P.} was supported by BMBF contract 05H15PACC1. 

The work of \textit{G.R.} was supported by the Spanish government and ERDF funds from the European Commission 
(grant nos FPA2014-53631-C2-1-P and SEV-2014-0398), Generalitat Valenciana (grant no.\ PROMETEO/2017/053), and
Consejo Superior de Investigaciones Cient\'{\i}ficas (grant no.\ PIE-201750E021).

The work of \textit{R.L.} was supported by the `Basis' foundation for theoretical physics and mathematics.

The work of \textit{R.B.} was supported by the German Science Foundation (DFG) within the Collaborative Research Center 676 \textit{Particles, Strings and the Early Universe}. 

The work of \textit{S.B.} was supported by the ERC Starting Grant `MathAm' (grant no.\ 39568).

\textit{G.Y.} was supported in part by the Chinese Academy of Sciences (CAS) Hundred-Talent Program, by the Key Research Program of Frontier Sciences of CAS, and by Project 11747601, supported by the National Natural Science Foundation of China.

The work of \textit{J.U.}  received funding from the European Research Council (ERC) under the European Union{'}s Horizon 2020 research and innovation programme (grant no.\ 647356 (CutLoops)). It was also supported by 
Graduiertenkolleg 1504 \textit{Masse, Spektrum, Symmetrie} of the Deutsche Forschungsgemeinschaft (DFG).

The work of \textit{S.J.} was partly supported by
 the Polish National Science Center (grant no.\ 2016/23/B/ST2/ 03927).

The report was partly supported by COST (European Cooperation in Science and Technology) Action CA16201 PARTICLEFACE.

\begin{flushleft}
\bibliographystyle{elsarticle-num}

\begin{thebibliography}{100}
\expandafter\ifx\csname url\endcsname\relax
  \def\url#1{\texttt{#1}}\fi
\expandafter\ifx\csname urlprefix\endcsname\relax\def\urlprefix{URL }\fi
\expandafter\ifx\csname href\endcsname\relax
  \def\href#1#2{#2} \def\path#1{#1}\fi

\makeatletter
    \clubpenalty10000
    \@clubpenalty \clubpenalty
    \widowpenalty10000
\interlinepenalty=10000
\makeatother




\bibitem{strat}
European Strategy for Particle Physics by the European Strategy Group for
  Particle Physics, CERN's white paper for the 15th European Strategy Session
  of the Council,
  \url{https://web.archive.org/web/20180704110504/https://council.web.cern.ch/en/content/european-strategy-particle-physics}.

\bibitem{fcc}
\url{https://fcc.web.cern.ch}, last accessed May 17th 2019.

\bibitem{tlep}
M.~Bicer \textit{et~al.}, {First look at the physics case of TLEP}, JHEP 01 (2014) 164,
  {Proc. {\it } 2013 Community Summer Study on the Future of
  U.S. Particle Physics: Snowmass on the Mississippi (CSS2013), Minneapolis,
  MN, USA,  2013},
  \href {http://arxiv.org/abs/1308.6176} {\path{arXiv:1308.6176}},
  \href {http://dx.doi.org/10.1007/JHEP01(2014)164}
  {\path{doi:10.1007/JHEP01(2014)164}}.

\bibitem{Moortgat-Picka:2015yla}
A.~Arbey \textit{et~al.}, \textit{Eur. Phys. J.}  \textbf{C75} (2015) 371.
  \href {http://arxiv.org/abs/1504.01726} {\path{arXiv:1504.01726}},
  \href {http://dx.doi.org/10.1140/epjc/s10052-015-3511-9}
  {\path{doi:10.1140/epjc/s10052-015-3511-9}}.

\bibitem{clic}
  \url{http://clic-study.web.cern.ch/}, last accessed May 17th 2019.

\bibitem{cepc}
  \url{http://cepc.ihep.ac.cn/}, last accessed May 17th 2019.

\bibitem{Abe:2005nqa}
K.~Abe \textit{et~al.},  \textit{Phys. Rev.} \textbf{D71} (2005) 112004.
  \href {http://arxiv.org/abs/hep-ex/0503005}
  {\path{arXiv:hep-ex/0503005}}, \href
  {http://dx.doi.org/10.1103/PhysRevD.71.112004}
  {\path{doi:10.1103/PhysRevD.71.112004}}.


\bibitem{Altarelli:1989YRFred}
{G. Altarelli \textit{et al.} (eds.)}, {Z physics at LEP 1, Yellow
  Report CERN \mbox{89-08} (1989)}: vol.~1: Standard physics,
  \url{http://dx.doi.org/10.5170/CERN-1989-008-V-1}.

\bibitem{Altarelli:1989YRTom}
{G. Altarelli \textit{et al.} (eds.)}, {Z physics at LEP 1, Yellow
  Report CERN \mbox{89-08} (1989)}:  vol.~2: Higgs search and
  new physics, \url{http://dx.doi.org/10.5170/CERN-1989-008-V-2}.
  
  \bibitem{Altarelli:1989YRDick}
{G. Altarelli \textit{et al.} (eds.)}, {Z physics at LEP 1, Yellow
  Report CERN \mbox{89-08} (1989)}:  vol.~3: Event
  generators and software, \url{http://dx.doi.org/10.5170/CERN-1989-008-V-3}.

\bibitem{Bardin:1995XX}
{D. Bardin \textit{et al.} (eds.)}, {Reports of the working group on
  precision calculations for the Z resonance, Yellow Report CERN 95-03
  (1995): parts I to III, 
  \url{http://cds.cern.ch/record/280836/files/CERN-95-03.pdf} }.
  \href {http://dx.doi.org/10.5170/CERN-1995-003}
  {\path{doi:10.5170/CERN-1995-003}}.

\bibitem{Bardin:1997xq}
D.{\relax Y}. Bardin \textit{et~al.}, {Electroweak working group report},   {Workshop Group on Precision Calculations for the Z Resonance (2nd meeting
  held Mar 31, 3rd meeting held Jun 13) Geneva, Switzerland, January 14, 1994},
  1997, p. 7,
  \href {http://arxiv.org/abs/hep-ph/9709229}
  {\path{arXiv:hep-ph/9709229}}.

\bibitem{mini}
A.~Blondel \textit{et~al.}, Mini~Workshop: {Precision EW and QCD
  Calculations for the FCC Studies: Methods and Tools},  2018
  (CERN, Geneva, Switzerland), 
  \url{https://indico.cern.ch/event/669224/}.

\bibitem{Awramik:2006ar}
M.~Awramik \textit{et~al.},  \textit{Phys. Lett.} \textbf{B642} (2006) 563.
  \href {http://arxiv.org/abs/hep-ph/0605339}
  {\path{arXiv:hep-ph/0605339}}, \href
  {http://dx.doi.org/10.1016/j.physletb.2006.07.035}
  {\path{doi:10.1016/j.physletb.2006.07.035}}.

\bibitem{Awramik:2006uz}
M.~Awramik \textit{et~al.},  \textit{J. High Energy Phys.} \textbf{0611} (2006) 048.
  \href {http://arxiv.org/abs/hep-ph/0608099}
  {\path{arXiv:hep-ph/0608099}}, \href
  {http://dx.doi.org/10.1088/1126-6708/2006/11/048}
  {\path{doi:10.1088/1126-6708/2006/11/048}}.

\bibitem{ALEPH:2005ab}
{ALEPH Collaboration \textit{et~al.}}, 
  \textit{Phys. Rep.} \textbf{427} (2006) 257.
  \href {http://arxiv.org/abs/hep-ex/0509008}
  {\path{arXiv:hep-ex/0509008}}, \href
  {http://dx.doi.org/10.1016/j.physrep.2005.12.006}
  {\path{doi:10.1016/j.physrep.2005.12.006}}.

\bibitem{ATLAS-CONF-2018-037}
ATLAS Collaboration, Measurement of the effective leptonic weak mixing
  angle using electron and muon pairs from Z-boson decay in the ATLAS
  experiment at $\sqrt{s} = 8\UTeV$, Atlas Conf Note, ATL-CONF-2018-037, (11 July 2018),
  \url{https://cds.cern.ch/record/2630340/files/ATLAS-CONF-2018-037.pdf}.

\bibitem{Baer:2013cma}
H.~Baer \textit{et al.}, The International Linear
  Collider technical design report -- volume 2: physics, \href
  {http://arxiv.org/abs/1306.6352} {\path{arXiv:1306.6352}}.

\bibitem{Erler:2000jg}
J.~Erler \textit{et al.},  \textit{Phys. Lett.} \textbf{B486} (2000) 125.
  \href {http://arxiv.org/abs/hep-ph/0005024}
  {\path{arXiv:hep-ph/0005024}}, \href
  {http://dx.doi.org/10.1016/S0370-2693(00)00749-8}
  {\path{doi:10.1016/S0370-2693(00)00749-8}}.

\bibitem{Dubovyk:2016aqv}
I.~Dubovyk \textit{et al.}, \textit{Phys.
  Lett.} \textbf{B762} (2016) 184.
  \href {http://arxiv.org/abs/1607.08375} {\path{arXiv:1607.08375}},
  \href {http://dx.doi.org/10.1016/j.physletb.2016.09.012}
  {\path{doi:10.1016/j.physletb.2016.09.012}}.

\bibitem{Dubovyk:2018rlg}
I.~Dubovyk \textit{et al.},  \textit{Phys.
  Lett.} \textbf{B783} (2018) 86.
  \href {http://arxiv.org/abs/1804.10236} {\path{arXiv:1804.10236}},
  \href {http://dx.doi.org/10.1016/j.physletb.2018.06.037}
  {\path{doi:10.1016/j.physletb.2018.06.037}}.

\bibitem{Gluza:Jan2018}
J. Gluza \textit{et al.},  Z-boson physics in the context of FCC-ee: theory,  Mini~Workshop: {Precision EW and QCD
  Calculations for the FCC Studies: Methods and Tools},  2018,
  (CERN, Geneva, Switzerland),
  \url{https://indico.cern.ch/event/669224/contributions/2805413/attachments/1581532/2499590/FCC_gluza_TheoryStatus.pdf}.

\bibitem{Jadach:April2018}
I.~Dubovyk \textit{et al.}, {Precision} calculations for the Z line shape
at the FCC-ee,   FCC Week,  Amsterdam, 2018,
  \url{https://indico.cern.ch/event/656491/contributions/2947663/attachments/1622685/2582801/Poster-FCC-Amsterdam_SJadach_et_al.pdf}.

\bibitem{Gluza:April2018}
J. Gluza \textit{et al.},  {The Z} boson
  resonance at two loops,  XIV `Loops and Legs',    Sankt Goar, Germany,
2018,   \url{https://indico.desy.de/indico/event/16613/session/4/contribution/22/material/slides/0.pdf}.

\bibitem{Czakon:1999ha}
M.~Czakon \textit{et al.},  \textit{Eur. Phys. J.} \textbf{C13} (2000)
  275.
  \href {http://arxiv.org/abs/hep-ph/9909242}
  {\path{arXiv:hep-ph/9909242}}, \href
  {http://dx.doi.org/10.1007/s100520000278} {\path{doi:10.1007/s100520000278}}.

\bibitem{Grzadkowski:2010es}
B. Grzadkowski \textit{et al.},   \textit{J. High Energy Phys.}  \textbf{10} (2010) 085.
  \href {http://arxiv.org/abs/1008.4884} {\path{arXiv:1008.4884}},
  \href {http://dx.doi.org/10.1007/JHEP10(2010)085}
  {\path{doi:10.1007/JHEP10(2010)085}}.

\bibitem{Giudice:2007fh}
G.F. Giudice \textit{et al.},   \textit{J. High Energy Phys.} \textbf{06} (2007) 045.
  \href {http://arxiv.org/abs/hep-ph/0703164}
  {\path{arXiv:hep-ph/0703164}}, \href
  {http://dx.doi.org/10.1088/1126-6708/2007/06/045}
  {\path{doi:10.1088/1126-6708/2007/06/045}}.

\bibitem{Contino:2013kra}
R.~Contino \textit{et al.},  \textit{J. High Energy Phys.} \textbf{07} (2013) 035.
  \href {http://arxiv.org/abs/1303.3876} {\path{arXiv:1303.3876}},
  \href {http://dx.doi.org/10.1007/JHEP07(2013)035}
  {\path{doi:10.1007/JHEP07(2013)035}}.

\bibitem{Bakshi:2018ics}
S.~Das~Bakshi \textit{et al.},   \textit{Eur. Phys. J.} \textbf{C79}  (2019) 21.
  \href {http://arxiv.org/abs/1808.04403} {\path{arXiv:1808.04403}},
  \href {http://dx.doi.org/10.1140/epjc/s10052-018-6444-2}
  {\path{doi:10.1140/epjc/s10052-018-6444-2}}.

\bibitem{Dawson:2018jlg}
S.~Dawson and A.~Ismail,  \textit{Phys.
  Rev.} \textbf{D98} (2018) 093003.
  \href {http://arxiv.org/abs/1808.05948} {\path{arXiv:1808.05948}},
  \href {http://dx.doi.org/10.1103/PhysRevD.98.093003}
  {\path{doi:10.1103/PhysRevD.98.093003}}.

\bibitem{deBlas:2017wmn}
J.~de~Blas \textit{et al.},  \textit{Proc. Sci.}  \textbf{314} (2017) 467.
  \href {http://arxiv.org/abs/1710.05402} {\path{arXiv:1710.05402}},
  \href {http://dx.doi.org/10.22323/1.314.0467}
  {\path{doi:10.22323/1.314.0467}}.

\bibitem{Bardin:1999yd}
D.{\relax Y}. Bardin \textit{et al.},   \textit{Comput. Phys. Commun.} \textbf{133} (2001)
  229.
  \href {http://arxiv.org/abs/hep-ph/9908433}
  {\path{arXiv:hep-ph/9908433}}, \href
  {http://dx.doi.org/10.1016/S0010-4655(00)00152-1}
  {\path{doi:10.1016/S0010-4655(00)00152-1}}.

\bibitem{Burgers:1985qg}
G.J.H. Burgers,  \textit{Phys. Lett.} \textbf{164B} (1985) 167.
  \href {http://dx.doi.org/10.1016/0370-2693(85)90053-X}
  {\path{doi:10.1016/0370-2693(85)90053-X}}.

\bibitem{Jadach:1999vf}
S.~Jadach \textit{et al.},   \textit{Comput. Phys. Commun.}
  \textbf{130} (2000) 260. 
   \href {http://arxiv.org/abs/hep-ph/9912214}
  {\path{arXiv:hep-ph/9912214}}, \href
  {http://dx.doi.org/10.1016/S0010-4655(00)00048-5}
  {\path{doi:10.1016/S0010-4655(00)00048-5}}.

\bibitem{Bardin:1989tq}
D.Y. Bardin \textit{et al.}, 
   \textit{Comput. Phys. Commun.} \textbf{59} (1990) 303.
  \href {http://dx.doi.org/10.1016/0010-4655(90)90179-5}
  {\path{doi:10.1016/0010-4655(90)90179-5}}.

\bibitem{Leike:1991pq}
A.~Leike \textit{et al.}, \textit{Phys.
  Lett.} \textbf{B273} (1991) 513.
  \href {http://arxiv.org/abs/hep-ph/9508390}
  {\path{arXiv:hep-ph/9508390}}, \href
  {http://dx.doi.org/10.1016/0370-2693(91)90307-C}
  {\path{doi:10.1016/0370-2693(91)90307-C}}.

\bibitem{Riemann:1992gv}
T.~Riemann,  \textit{Phys. Lett.} \textbf{B293}
  (1992) 451.
  \href {http://arxiv.org/abs/hep-ph/9506382}
  {\path{arXiv:hep-ph/9506382}}, \href
  {http://dx.doi.org/10.1016/0370-2693(92)90911-M}
  {\path{doi:10.1016/0370-2693(92)90911-M}}.

\bibitem{Misiak:2010sk}
M.~Misiak and M.~Steinhauser,  \textit{Nucl. Phys.} \textbf{B840} (2010) 271.
  \href {http://arxiv.org/abs/1005.1173} {\path{arXiv:1005.1173}},
  \href {http://dx.doi.org/10.1016/j.nuclphysb.2010.07.009}
  {\path{doi:10.1016/j.nuclphysb.2010.07.009}}.

\bibitem{Misiak:2017woa}
M.~Misiak \textit{et al.}, \textit{Phys. Lett.} \textbf{B770}
  (2017) 431.
  \href {http://arxiv.org/abs/1702.07674} {\path{arXiv:1702.07674}},
  \href {http://dx.doi.org/10.1016/j.physletb.2017.05.008}
  {\path{doi:10.1016/j.physletb.2017.05.008}}.

\bibitem{Riemann:Jan2018}
T. Riemann, The Z boson line shape at the FCCee: from 1.5
  loops at LEP to 2.5 loops in future, Mini~Workshop: {Precision EW and QCD
  Calculations for the FCC Studies: Methods and Tools},  2018,
  (CERN, Geneva, Switzerland),
  \url{
  https://indico.cern.ch/event/669224/contributions/2805418/attachments/1581648/2504620/riemann-FCCeeminiCERN-short.pdf}.

\bibitem{Arbuzov:2005ma}
A.~Arbuzov \textit{et al.},  \textit{Comput. Phys. Commun.} \textbf{174} (2006) 728.
  \href {http://arxiv.org/abs/hep-ph/0507146}
  {\path{arXiv:hep-ph/0507146}}, \href
  {http://dx.doi.org/10.1016/j.cpc.2005.12.009}
  {\path{doi:10.1016/j.cpc.2005.12.009}}.

\bibitem{Chetyrkin:1994jsFred}
K.G. Chetyrkin \textit{et al.},  {QCD corrections to the $\mathrm{e}^{+}
  \mathrm{e}^{-}$ cross-section and the Z boson decay rate}, CERN 95-03, (CERN,
Geneva, 1995).
 
\bibitem{Chetyrkin:1994jsTom}
K.G. Chetyrkin \textit{et al.},   \textit{Phys. Rep.} \textbf{277} (1996) 189.
  \href {http://arxiv.org/abs/hep-ph/9503396}
  {\path{arXiv:hep-ph/9503396}}, \href
  {http://dx.doi.org/10.1016/S0370-1573(96)00012-9}
  {\path{doi:10.1016/S0370-1573(96)00012-9}}.

\bibitem{Awramik:2008gi}
M.~Awramik \textit{et al.},   \textit{Nucl. Phys.} \textbf{B813}
  (2009) 174.
  \href {http://arxiv.org/abs/0811.1364} {\path{arXiv:0811.1364}},
  \href {http://dx.doi.org/10.1016/j.nuclphysb.2008.12.031}
  {\path{doi:10.1016/j.nuclphysb.2008.12.031}}.

\bibitem{Freitas:2014hra}
A.~Freitas,  \textit{J. High Energy Phys.} \textbf{2014} (2014) 070.
  \href {http://arxiv.org/abs/1401.2447} {\path{arXiv:1401.2447}},
  \href {http://dx.doi.org/10.1007/JHEP04(2014)070}
  {\path{doi:10.1007/JHEP04(2014)070}}.

\bibitem{Alain:Jan2018}
 A.~Blondel, {Motivation: experimental capabilities and requirements}, Mini~Workshop: {Precision EW and QCD
  Calculations for the FCC Studies: Methods and Tools},  2018,
  (CERN, Geneva, Switzerland),
  \url{https://indico.cern.ch/event/669224/contributions/2805398/attachments/1581811/2499870/Blondel-FCC-ee-physics_case_and_theory_errors-v2.pdf}.

\bibitem{Akhundov:1985fc}
A.~Akhundov \textit{et al.},  \textit{Nucl. Phys.} \textbf{B276} (1986) 1.   \href {http://dx.doi.org/10.1016/0550-3213(86)90014-3}
  {\path{doi:10.1016/0550-3213(86)90014-3}}.

\bibitem{Patrignani:2016xqp}
C.~Patrignani \textit{et al.},  \textit{Chin. Phys.} \textbf{C40} 
  (2016) 100001.
  \href {http://dx.doi.org/10.1088/1674-1137/40/10/100001}
  {\path{doi:10.1088/1674-1137/40/10/100001}}.

\bibitem{Steinhauser:1998rq-new}
M.~Steinhauser,  \textit{Phys. Lett.} \textbf{B429} (1998) 158.
  \href {http://arxiv.org/abs/hep-ph/9803313}
  {\path{arXiv:hep-ph/9803313}}, \href
  {http://dx.doi.org/10.1016/S0370-2693(98)00503-6}
  {\path{doi:10.1016/S0370-2693(98)00503-6}}.

\bibitem{Davier:2010nc}
M.~Davier \textit{et al.},   \textit{Eur. Phys. J.} \textbf{C71} (2011)
  1515 [Erratum: \textit{Eur. Phys. J.} \textbf{C72} (2012) 1874].
  \href {http://arxiv.org/abs/1010.4180} {\path{arXiv:1010.4180}},
  \href {http://dx.doi.org/10.1140/epjc/s10052-010-1515-z}
  {\path{doi:10.1140/epjc/s10052-010-1515-z}}.


\bibitem{Barbieri:1992nz}
R.~Barbieri \textit{et al.},    \textit{Phys. Lett.} \textbf{B288} (1992) 95
  [Erratum: \textit{Phys. Lett.} \textbf{B312} (1993) 511].
  \href {http://arxiv.org/abs/hep-ph/9205238}
  {\path{arXiv:hep-ph/9205238}}, \href
  {http://dx.doi.org/10.1016/0370-2693(92)91960-H}
  {\path{doi:10.1016/0370-2693(92)91960-H}}.

\bibitem{Barbieri:1992dq}
R.~Barbieri \textit{et al.},   \textit{Nucl. Phys.} \textbf{B409} (1993) 105.
  \href {http://dx.doi.org/10.1016/0550-3213(93)90448-X}
  {\path{doi:10.1016/0550-3213(93)90448-X}}.

\bibitem{Fleischer:1993ub}
J.~Fleischer \textit{et al.},    \textit{Phys.
  Lett.} \textbf{B319} (1993) 249.
  \href {http://dx.doi.org/10.1016/0370-2693(93)90810-5}
  {\path{doi:10.1016/0370-2693(93)90810-5}}.

\bibitem{Fleischer:1994cb}
J.~Fleischer \textit{et al.},    \textit{Phys. Rev.} \textbf{D51} (1995) 3820.
  \href {http://dx.doi.org/10.1103/PhysRevD.51.3820}
  {\path{doi:10.1103/PhysRevD.51.3820}}.

\bibitem{Djouadi:1987gn}
A.~Djouadi and C.~Verzegnassi,  \textit{Phys. Lett.} \textbf{B195} (1987) 265.
  \href {http://dx.doi.org/10.1016/0370-2693(87)91206-8}
  {\path{doi:10.1016/0370-2693(87)91206-8}}.

\bibitem{Djouadi:1987di}
A.~Djouadi, \textit{Nuovo Cim.} \textbf{A100} (1988) 357.
  \href {http://dx.doi.org/10.1007/BF02812964}
  {\path{doi:10.1007/BF02812964}}.

\bibitem{Kniehl:1989yc}
B.A. Kniehl,  \textit{Nucl. Phys.} \textbf{B347} (1990) 86.
  \href {http://dx.doi.org/10.1016/0550-3213(90)90552-O}
  {\path{doi:10.1016/0550-3213(90)90552-O}}.

\bibitem{Kniehl:1991gu}
B.A. Kniehl and A.~Sirlin,  \textit{Nucl. Phys.} \textbf{B371} (1992) 141.
  \href {http://dx.doi.org/10.1016/0550-3213(92)90232-Z}
  {\path{doi:10.1016/0550-3213(92)90232-Z}}.

\bibitem{Djouadi:1993ss}
A.~Djouadi and P.~Gambino,  \textit{Phys. Rev.} \textbf{D49} (1994) 3499 [Erratum: \textit{Phys.
  Rev.} \textbf{D53} (1996) 4111].
  \href {http://arxiv.org/abs/hep-ph/9309298}
  {\path{arXiv:hep-ph/9309298}}, \href
  {http://dx.doi.org/10.1103/PhysRevD.53.4111}
  {\path{doi:10.1103/PhysRevD.53.4111}}.

\bibitem{Avdeev:1994db}
L.~Avdeev \textit{et al.},   \textit{Phys. Lett.} \textbf{B336} (1994)
  560 [Erratum: \textit{Phys. Lett.} \textbf{B349} (1995) 597].
  \href {http://arxiv.org/abs/hep-ph/9406363}
  {\path{arXiv:hep-ph/9406363}}, \href
  {http://dx.doi.org/10.1016/0370-2693(95)00269-Q}
  {\path{doi:10.1016/0370-2693(95)00269-Q}}.

\bibitem{Chetyrkin:1995ix}
K.G. Chetyrkin  \textit{et al.},  \textit{Phys. Lett.} \textbf{B351} (1995)
  331.
  \href {http://arxiv.org/abs/hep-ph/9502291}
  {\path{arXiv:hep-ph/9502291}}, \href
  {http://dx.doi.org/10.1016/0370-2693(95)00380-4}
  {\path{doi:10.1016/0370-2693(95)00380-4}}.

\bibitem{vanderBij:2000cg}
J.J. van~der Bij \textit{et al.},  
   \textit{Phys.
  Lett.} \textbf{B498} (2001) 156.
  \href {http://arxiv.org/abs/hep-ph/0011373}
  {\path{arXiv:hep-ph/0011373}}, \href
  {http://dx.doi.org/10.1016/S0370-2693(01)00002-8}
  {\path{doi:10.1016/S0370-2693(01)00002-8}}.

\bibitem{Faisst:2003px}
M.~Faisst \textit{et al.},  \textit{Nucl. Phys.} \textbf{B665} (2003) 649.
  \href {http://arxiv.org/abs/hep-ph/0302275}
  {\path{arXiv:hep-ph/0302275}}, \href
  {http://dx.doi.org/10.1016/S0550-3213(03)00450-4}
  {\path{doi:10.1016/S0550-3213(03)00450-4}}.

\bibitem{Schroder:2005db}
Y.~Schr{\"o}der and M.~Steinhauser, \textit{Phys. Lett.} \textbf{B622} (2005) 124.
  \href {http://arxiv.org/abs/hep-ph/0504055}
  {\path{arXiv:hep-ph/0504055}}, \href
  {http://dx.doi.org/10.1016/j.physletb.2005.06.085}
  {\path{doi:10.1016/j.physletb.2005.06.085}}.

\bibitem{Chetyrkin:2006bj}
K.G. Chetyrkin \textit{et al.},    \textit{Phys. Rev. Lett.} \textbf{97} (2006) 102003.
  \href {http://arxiv.org/abs/hep-ph/0605201}
  {\path{arXiv:hep-ph/0605201}}, \href
  {http://dx.doi.org/10.1103/PhysRevLett.97.102003}
  {\path{doi:10.1103/PhysRevLett.97.102003}}.

\bibitem{Boughezal:2006xk}
R.~Boughezal and M.~Czakon,  \textit{Nucl. Phys.} \textbf{B755} (2006) 221.
  \href {http://arxiv.org/abs/hep-ph/0606232}
  {\path{arXiv:hep-ph/0606232}}, \href
  {http://dx.doi.org/10.1016/j.nuclphysb.2006.08.007}
  {\path{doi:10.1016/j.nuclphysb.2006.08.007}}.

\bibitem{Czarnecki:1996ei}
A.~Czarnecki and J.H. K\"uhn,  \textit{Phys. Rev. Lett.} \textbf{77} (1996) 3955.
  \href {http://arxiv.org/abs/hep-ph/9608366}
  {\path{arXiv:hep-ph/9608366}}, \href
  {http://dx.doi.org/10.1103/PhysRevLett.77.3955}
  {\path{doi:10.1103/PhysRevLett.77.3955}}.

\bibitem{Harlander:1997zb}
R.~Harlander \textit{et al.},   \textit{Phys.
  Lett.} \textbf{B426} (1998) 125.
  \href {http://arxiv.org/abs/hep-ph/9712228}
  {\path{arXiv:hep-ph/9712228}}, \href
  {http://dx.doi.org/10.1016/S0370-2693(98)00220-2}
  {\path{doi:10.1016/S0370-2693(98)00220-2}}.

\bibitem{Fleischer:1992fq}
J.~Fleischer  \textit{et al.},    \textit{Phys. Lett.} \textbf{B293} (1992)
  437.
  \href {http://dx.doi.org/10.1016/0370-2693(92)90909-N}
  {\path{doi:10.1016/0370-2693(92)90909-N}}.

\bibitem{Buchalla:1992zm}
G.~Buchalla and A.J. Buras,  \textit{Nucl. Phys.} \textbf{B398} (1993) 285.
  \href {http://dx.doi.org/10.1016/0550-3213(93)90110-B}
  {\path{doi:10.1016/0550-3213(93)90110-B}}.

\bibitem{Degrassi:1993ij}
G.~Degrassi,  \textit{Nucl. Phys.} \textbf{B407} (1993) 271.
  \href {http://arxiv.org/abs/hep-ph/9302288}
  {\path{arXiv:hep-ph/9302288}}, \href
  {http://dx.doi.org/10.1016/0550-3213(93)90058-W}
  {\path{doi:10.1016/0550-3213(93)90058-W}}.

\bibitem{Chetyrkin:1993jp}
K.~Chetyrkin \textit{et al.},    \textit{Mod. Phys. Lett.} \textbf{A8} (1993) 2785.
   \href {http://dx.doi.org/10.1142/S0217732393003172}
  {\path{doi:10.1142/S0217732393003172}}.

\bibitem{Dubovyk:Jan2018}
I.~Dubovyk \textit{et al.},  MB-suite 1: AMBRE news:
  non-planar 3-loop vertices, Mini~Workshop: {Precision EW and QCD
  Calculations for the FCC Studies: Methods and Tools},  2018,
  (CERN, Geneva, Switzerland),
  \url{https://indico.cern.ch/event/669224/overview}.

\bibitem{Borowka:Jan2018}
S.~Borowka, {pySecDec for phenomenological predictions}, Mini~Workshop: {Precision EW and QCD
  Calculations for the FCC Studies: Methods and Tools},  2018,
  (CERN, Geneva, Switzerland),
  \url{https://indico.cern.ch/event/669224/contributions/2805456/attachments/1582115/2500426/miniWS2018_sophia.pdf}.

\bibitem{Freitas:2016sty}
A.~Freitas,  \textit{Prog. Part.
  Nucl. Phys.} \textbf{90} (2016) 201.
  \href {http://arxiv.org/abs/1604.00406} {\path{arXiv:1604.00406}},
  \href {http://dx.doi.org/10.1016/j.ppnp.2016.06.004}
  {\path{doi:10.1016/j.ppnp.2016.06.004}}.

\bibitem{Smirnov:2013eza}
A.V. Smirnov,  \textit{Comput. Phys. Commun.} \textbf{185} (2014)
  2090.
  \href {http://arxiv.org/abs/1312.3186} {\path{arXiv:1312.3186}},
  \href {http://dx.doi.org/10.1016/j.cpc.2014.03.015}
  {\path{doi:10.1016/j.cpc.2014.03.015}}.

\bibitem{Borowka:2015mxa}
S.~Borowka \textit{et al.}, 
   \textit{Comput. Phys. Commun.} \textbf{196} (2015) 470.
  \href {http://arxiv.org/abs/1502.06595} {\path{arXiv:1502.06595}},
  \href {http://dx.doi.org/10.1016/j.cpc.2015.05.022}
  {\path{doi:10.1016/j.cpc.2015.05.022}}.

\bibitem{Gluza:2007rt}
J.~Gluza \textit{et al.},   \textit{Comput.
  Phys. Commun.} \textbf{177} (2007) 879.
  \href {http://arxiv.org/abs/0704.2423} {\path{arXiv:0704.2423}},
  \href {http://dx.doi.org/10.1016/j.cpc.2007.07.001}
  {\path{doi:10.1016/j.cpc.2007.07.001}}.

\bibitem{Gluza:2010rn}
J.~Gluza \textit{et al.},   \textit{Eur. Phys. J.} \textbf{C71} (2011) 1516.
  \href {http://arxiv.org/abs/1010.1667} {\path{arXiv:1010.1667}},
  \href {http://dx.doi.org/10.1140/epjc/s10052-010-1516-y}
  {\path{doi:10.1140/epjc/s10052-010-1516-y}}.

\bibitem{Bielas:2013rja}
K.~Bielas \textit{et al.},   \textit{Acta Phys. Pol.} \textbf{B44} (2013) 2249.
  \href {http://arxiv.org/abs/1312.5603} {\path{arXiv:1312.5603}},
  \href {http://dx.doi.org/10.5506/APhysPolB.44.2249}
  {\path{doi:10.5506/APhysPolB.44.2249}}.

\bibitem{Dubovyk:2015yba}
I.~Dubovyk \textit{et al.},   \textit{J. Phys.  Conf. Ser.} \textbf{608} (2015) 012070.
  \href {http://dx.doi.org/10.1088/1742-6596/608/1/012070}
  {\path{doi:10.1088/1742-6596/608/1/012070}}.

\bibitem{Dubovyk:2016ocz}
I.~Dubovyk \textit{et al.},   \textit{Proc. Sci.} \textbf{260}
  (2016) 034.
  \href {http://arxiv.org/abs/1607.07538} {\path{arXiv:1607.07538}},
  \href {http://dx.doi.org/10.22323/1.260.0034}
  {\path{doi:10.22323/1.260.0034}}.

\bibitem{Dubovyk:2017cqw}
I.~Dubovyk \textit{et al.},  \textit{Acta Phys. Pol.} \textbf{B48} (2017) 995.
  \href {http://arxiv.org/abs/1704.02288} {\path{arXiv:1704.02288}},
  \href {http://dx.doi.org/10.5506/APhysPolB.48.995}
  {\path{doi:10.5506/APhysPolB.48.995}}.

\bibitem{Jadach:Jan2018}
S.~Jadach, How to calculate QED higher orders in a form useful for MC event
  generators with soft photon and/or collinear resummation?, Mini~Workshop: {Precision EW and QCD
  Calculations for the FCC Studies: Methods and Tools},  2018,
  (CERN, Geneva, Switzerland), \url{https://indico.cern.ch/event/669224/overview}.

\bibitem{Bardin:1988xt}
D.Y. Bardin \textit{et al.}, \textit{Phys. Lett.} \textbf{B206}
  (1988) 539.
  \href {http://dx.doi.org/10.1016/0370-2693(88)91625-5}
  {\path{doi:10.1016/0370-2693(88)91625-5}}.

\bibitem{Bardin:1989cw}
D.Y. Bardin \textit{et al.},  \textit{Phys. Lett.} \textbf{B229} (1989) 405.
  \url{http://cds.cern.ch/record/198110/files/198906505.pdf}.
  \href {http://dx.doi.org/10.1016/0370-2693(89)90428-0}
  {\path{doi:10.1016/0370-2693(89)90428-0}}.

\bibitem{Bardin:1989di}
D.~Bardin \textit{et al.}, \textit{Z. Phys.} \textbf{C44} (1989) 493.
  \url{https://lib-extopc.kek.jp/preprints/PDF/1989/8906/8906215.pdf}.
  \href {http://dx.doi.org/10.1007/BF01415565}
  {\path{doi:10.1007/BF01415565}}.

\bibitem{Bardin:1990fu}
D.Y. Bardin \textit{et al.},   \textit{Nucl. Phys. B} \textbf{351} (1991)
  1.
  \href {http://arxiv.org/abs/hep-ph/9801208}
  {\path{arXiv:hep-ph/9801208}}, \href
  {http://dx.doi.org/10.1016/0550-3213(91)90080-H}
  {\path{doi:10.1016/0550-3213(91)90080-H}}.

\bibitem{Bardin:1990de}
D.~Bardin \textit{et al.},   \textit{Phys. Lett.} \textbf{B255} (1991) 290.
  \href {http://arxiv.org/abs/hep-ph/9801209}
  {\path{arXiv:hep-ph/9801209}}, \href
  {http://dx.doi.org/10.1016/0370-2693(91)90250-T}
  {\path{doi:10.1016/0370-2693(91)90250-T}}.

\bibitem{Bardin:1992jc}
D.~Bardin \textit{et al.},   {{ZFITTER} -- an analytical program for fermion pair
  production in $\mathrm{e}^+ \mathrm{e}^-$ annihilation},   CERN/TH. 6443, (CERN, Geneva, 1992),  
  \url{http://cds.cern.ch/record/265101/files/th-6443-92.ps.gz?version=1}.
  \href {http://arxiv.org/abs/hep-ph/9412201}
  {\path{arXiv:hep-ph/9412201}}.

\bibitem{Christova:1999cc}
P.C. Christova \textit{et al.},   \textit{Phys. Lett.} \textbf{B456} (1999) 264.
  \href {http://arxiv.org/abs/hep-ph/9902408}
  {\path{arXiv:hep-ph/9902408}}, \href {http://dx.doi.org/10.1016/S0370-2693(99)00528-6}
  {\path{doi:10.1016/S0370-2693(99)00528-6}}.

\bibitem{Akhundov:2013ons}
A.~Akhundov \textit{et al.},   \textit{Phys.
  Part. Nucl.} \textbf{45} (2014) 529.
  \href {http://arxiv.org/abs/1302.1395} {\path{arXiv:1302.1395}},
  \href {http://dx.doi.org/10.1134/S1063779614030022}
  {\path{doi:10.1134/S1063779614030022}}.

\bibitem{web-sanc.zfitter:2016}
  \href{http://sanc.jinr.ru/users/zfitter}{http://sanc.jinr.ru/users/zfitter},
  last accessed May 20th. 2019.

\bibitem{Schael:2013ita}
ALEPH Collaboration \textit{et al.},    \textit{Phys. Rep.}
  \textbf{532} (2013) 119.
  \href {http://arxiv.org/abs/1302.3415} {\path{arXiv:1302.3415}},
  \href {http://dx.doi.org/10.1016/j.physrep.2013.07.004}
  {\path{doi:10.1016/j.physrep.2013.07.004}}.

\bibitem{Tenchini:April2018}
R.~Tenchini, {EW measurements at FCC},  FCC Week,  Amsterdam, 2018,
  \url{https://indico.cern.ch/event/656491/contributions/2945302/attachments/1630715/2599367/FCC-EWK_Amsterdam2018.pdf}.

\bibitem{Greco:1975rm}
M.~Greco \textit{et al.},   \textit{Nucl. Phys.} \textbf{B101} (1975) 234.
  \href {http://dx.doi.org/10.1016/0550-3213(75)90304-1}
  {\path{doi:10.1016/0550-3213(75)90304-1}}.

\bibitem{Bonneau:1971mk}
G.~Bonneau and F.~Martin,  \textit{Nucl.
  Phys.} \textbf{B27} (1971) 381.
  \href {http://dx.doi.org/10.1016/0550-3213(71)90102-7}
  {\path{doi:10.1016/0550-3213(71)90102-7}}.

\bibitem{Montagna:1993ai}
G.~Montagna \textit{et al.},   \textit{Comput. Phys. Commun.} \textbf{76}
  (1993) 328.
  \href {http://dx.doi.org/10.1016/0010-4655(93)90060-P}
  {\path{doi:10.1016/0010-4655(93)90060-P}}.

\bibitem{Montagna:1995ja}
G.~Montagna \textit{et al.},  
  \textit{Comput. Phys. Commun.} \textbf{93} (1996) 120.
  \href {http://arxiv.org/abs/hep-ph/9506329}
  {\path{arXiv:hep-ph/9506329}}, \href
  {http://dx.doi.org/10.1016/0010-4655(95)00127-1}
  {\path{doi:10.1016/0010-4655(95)00127-1}}.

\bibitem{Montagna:1998kp}
G.~Montagna \textit{et al.},   \textit{Comput. Phys. Commun.} \textbf{117} (1999) 278.
  \href {http://arxiv.org/abs/hep-ph/9804211}
  {\path{arXiv:hep-ph/9804211}}, \href
  {http://dx.doi.org/10.1016/S0010-4655(98)00080-0}
  {\path{doi:10.1016/S0010-4655(98)00080-0}}.


\bibitem{Bilenky:1989zg}
M.~Bilenky and A.~Sazonov, {QED} corrections at $\mathrm{Z}^0$ pole with realistic
  kinematical cuts,  JINR  E2-89-792, (1989).
  \url{https://lib-extopc.kek.jp/preprints/PDF/1990/9003/9003360.pdf}.

\bibitem{Jack:2000as}
M.A. Jack, Ph.D. thesis, Humboldt-Universit{\"a}t zu Berlin,  2000.
  \href {http://arxiv.org/abs/hep-ph/0009068}
  {\path{arXiv:hep-ph/0009068}}.

\bibitem{Fleischer:2003kk}
J.~Fleischer \textit{et al.},  \textit{Eur. Phys. J.} \textbf{C31} (2003) 37.
  \href {http://arxiv.org/abs/hep-ph/0302259}
  {\path{arXiv:hep-ph/0302259}}, \href
  {http://dx.doi.org/10.1140/epjc/s2003-01263-8}
  {\path{doi:10.1140/epjc/s2003-01263-8}}.

\bibitem{Bardin:2000kn}
D.Y. Bardin \textit{et al.},  {An electroweak library for the
  calculation of {EWRC} to $\mathrm{e}^+ \mathrm{e}^- \to \mathrm{f} {\bar {\mathrm{f}}}$ within the topfit
  project}, \href {http://arxiv.org/abs/hep-ph/0012080}
  {\path{arXiv:hep-ph/0012080}}.

\bibitem{Fleischer:2002nn}
J.~Fleischer \textit{et al.},  {One-loop corrections to the process $\mathrm{e}^+ \mathrm{e}^- \to \mathrm{t} {\bar{\mathrm{t}}}$
  including hard bremsstrahlung,   Second Symposium on
  Computational Particle Physics, Tokyo,  2001,  Ed. Y. Kurihara (KEK Proc.
  2002-11 (2002) p. 153),} \href {http://arxiv.org/abs/hep-ph/0203220}
  {\path{arXiv:hep-ph/0203220}}.

\bibitem{FRW:2002sw}
J.~Fleischer \textit{et al.},  Fortran program {\tt
  topfit.F} 0.91 (06 March 2002),
  \url{https://www-zeuthen.desy.de/~riemann/doc/topfit/topfit.html}.

\bibitem{Fleischer:2003aa}
J.~Fleischer \textit{et al.},  Fortran program
  {\tt topfit.F} 0.92 (01 July 2003),
  \url{https://www-zeuthen.desy.de/~riemann/doc/topfit/topfit.html}.

\bibitem{Hahn:2003ab}
T.~Hahn \textit{et al.},  ${O}(\alpha)$
  electroweak corrections to the processes $\mathrm{e}^+ \mathrm{e}^-$ $\to$ $\tau^- \tau^+$, $\mathrm{c}
  {\bar {\mathrm{c}}}$, $\mathrm{b} {\bar {\mathrm{b}}}$, $\mathrm{t} {\bar {\mathrm{t}}}$: a
  comparison, \href{http://www-flc.desy.de/lcnotes/notes/LC-TH-2003-083.pdf}{http://www-flc.desy.de/lcnotes/notes/LC-TH-2003-083.pdf},
  \href {http://arxiv.org/abs/hep-ph/0307132}
  {\path{arXiv:hep-ph/0307132}}.

\bibitem{Gluza:2004tq}
J.~Gluza \textit{et al.},   \textit{Nucl. Instrum. Methods} \textbf{A534} (2004) 289.
  \href {http://arxiv.org/abs/hep-ph/0409011}
  {\path{arXiv:hep-ph/0409011}}, \href
  {http://dx.doi.org/10.1016/j.nima.2004.07.103}
  {\path{doi:10.1016/j.nima.2004.07.103}}.

\bibitem{Lorca:2004dk}
A.~Lorca and T.~Riemann,  \textit{Nucl. Phys. Proc. Suppl.} \textbf{135} (2004) 328.
  \href {http://arxiv.org/abs/hep-ph/0407149}
  {\path{arXiv:hep-ph/0407149}}.

\bibitem{Lorca:2004fg}
A.~Lorca and T.~Riemann,   \textit{Comput. Phys. Commun.} \textbf{174} (2006) 71.
  \href {http://arxiv.org/abs/hep-ph/0412047}
  {\path{arXiv:hep-ph/0412047}}, \href
  {http://dx.doi.org/10.1016/j.cpc.2005.09.003}
  {\path{doi:10.1016/j.cpc.2005.09.003}}.

\bibitem{Fleischer:2006ht}
J.~Fleischer \textit{et al.},  \textit{Eur. J. Phys.} \textbf{48} (2006)
  35.
  \href {http://arxiv.org/abs/hep-ph/0606210}
  {\path{arXiv:hep-ph/0606210}}, \href
  {http://dx.doi.org/10.1140/epjc/s10052-006-0008-6}
  {\path{doi:10.1140/epjc/s10052-006-0008-6}}.

\bibitem{Sirlin:1980nh}
A.~Sirlin,  \textit{Phys. Rev.} \textbf{D22} (1980) 971.
  \href {http://dx.doi.org/10.1103/PhysRevD.22.971}
  {\path{doi:10.1103/PhysRevD.22.971}}.

\bibitem{Bardin:1999gt}
D.Y. Bardin \textit{et al.},  {Precision calculation
  project report}, \href {http://arxiv.org/abs/hep-ph/9902452}
  {\path{arXiv:hep-ph/9902452}}.

\bibitem{Marciano:1980pb}
W.J. Marciano and A.~Sirlin,  \textit{Phys. Rev.} \textbf{D22} (1980)
  2695.
  \href {http://dx.doi.org/10.1103/PhysRevD.22.2695}
  {\path{doi:10.1103/PhysRevD.22.2695}}.

\bibitem{Teubner:2012qb}
T.~Teubner \textit{et al.},  
  \textit{Nucl. Phys.
  Proc. Suppl.} \textbf{225--227} (2012) 282.
  \href {http://dx.doi.org/10.1016/j.nuclphysbps.2012.02.059}
  {\path{doi:10.1016/j.nuclphysbps.2012.02.059}}.

\bibitem{Davier:2017zfy}
M.~Davier \textit{et al.},  
  \textit{Eur. Phys. J.} \textbf{C77} (2017) 827.
  \href {http://arxiv.org/abs/1706.09436} {\path{arXiv:1706.09436}},
  \href {http://dx.doi.org/10.1140/epjc/s10052-017-5161-6}
  {\path{doi:10.1140/epjc/s10052-017-5161-6}}.

\bibitem{Jegerlehner:2017zsb}
F.~Jegerlehner, {Variations on photon vacuum polarization}, \href
  {http://arxiv.org/abs/1711.06089} {\path{arXiv:1711.06089}}.

\bibitem{Keshavarzi:2018mgv}
A.~Keshavarzi \textit{et al.},  \textit{Phys. Rev.} \textbf{D97} (2018) 114025.
  \href {http://arxiv.org/abs/1802.02995} {\path{arXiv:1802.02995}},
  \href {http://dx.doi.org/10.1103/PhysRevD.97.114025}
  {\path{doi:10.1103/PhysRevD.97.114025}}.

\bibitem{Bardin:1988by}
D.~Bardin \textit{et al.},   \textit{Z. Phys.} \textbf{C42} (1989) 679.
  \url{https://lib-extopc.kek.jp/preprints/PDF/1989/8904/8904210.pdf}.
  \href {http://dx.doi.org/10.1007/BF01557676}
  {\path{doi:10.1007/BF01557676}}.

\bibitem{Bardin:1982sv}
D.Y. Bardin \textit{et al.},   \textit{Nucl. Phys.} \textbf{B197} (1982) 1.
  \href {http://dx.doi.org/10.1016/0550-3213(82)90152-3}
  {\path{doi:10.1016/0550-3213(82)90152-3}}.

\bibitem{Bardin:1980fe}
D.~Bardin \textit{et al.},  
  \textit{Nucl. Phys.} \textbf{B175} (1980) 435.
  \href {http://dx.doi.org/10.1016/0550-3213(80)90021-8}
  {\path{doi:10.1016/0550-3213(80)90021-8}}.

\bibitem{Borrelli:1989bd}
A.~Borrelli \textit{et al.},   \textit{Nucl. Phys.} \textbf{B333} (1990) 357.
  \href {http://dx.doi.org/10.1016/0550-3213(90)90042-C}
  {\path{doi:10.1016/0550-3213(90)90042-C}}.

\bibitem{Stuart:1991xk}
R.G. Stuart,  \textit{Phys. Lett.} \textbf{B262} (1991) 113.
  \href {http://dx.doi.org/10.1016/0370-2693(91)90653-8}
  {\path{doi:10.1016/0370-2693(91)90653-8}}.

\bibitem{Stuart:1991cc}
R.G. Stuart,  \textit{Phys. Lett.} \textbf{B272} (1991) 353.
  \href {http://dx.doi.org/10.1016/0370-2693(91)91842-J}
  {\path{doi:10.1016/0370-2693(91)91842-J}}.

\bibitem{Kirsch:1994cf}
S.~Kirsch and T.~Riemann,  \textit{Comput. Phys. Commun.} \textbf{88} (1995) 89.
  \href {http://arxiv.org/abs/hep-ph/9408365}
  {\path{arXiv:hep-ph/9408365}}, \href
  {http://dx.doi.org/10.1016/0010-4655(95)00016-9}
  {\path{doi:10.1016/0010-4655(95)00016-9}}.

\bibitem{Riemann:2015wpn}
T.~Riemann,  \textit{Acta Phys. Pol.}
  \textbf{B46} (2015) 2235.  \href {http://arxiv.org/abs/1610.04501} {\path{arXiv:1610.04501}},
  \href {http://dx.doi.org/10.5506/APhysPolB.46.2235}
  {\path{doi:10.5506/APhysPolB.46.2235}}.

\bibitem{Willenbrock:1991hu}
S.~Willenbrock and G.~Valencia,  \textit{Phys.
  Lett.} \textbf{B259} (1991) 373.
  \href {http://dx.doi.org/10.1016/0370-2693(91)90843-F}
  {\path{doi:10.1016/0370-2693(91)90843-F}}.

\bibitem{Sirlin:1991fd}
A.~Sirlin,  \textit{Phys. Rev.
  Lett.} \textbf{67} (1991) 2127.
  \href {http://dx.doi.org/10.1103/PhysRevLett.67.2127}
  {\path{doi:10.1103/PhysRevLett.67.2127}}.

\bibitem{Veltman:1992tm}
H.G.J. Veltman, \textit{Z. Phys.} \textbf{C62}
  (1994) 35.
  \href {http://dx.doi.org/10.1007/BF01559523}
  {\path{doi:10.1007/BF01559523}}.

\bibitem{Passera:1998uj}
M.~Passera and A.~Sirlin,  \textit{Phys. Rev.} \textbf{D58} (1998) 113010.
  \href {http://arxiv.org/abs/hep-ph/9804309}
  {\path{arXiv:hep-ph/9804309}}, \href
  {http://dx.doi.org/10.1103/PhysRevD.58.113010}
  {\path{doi:10.1103/PhysRevD.58.113010}}.

\bibitem{Gambino:1999ai}
P.~Gambino and P.A. Grassi,  \textit{Phys. Rev.} \textbf{D62} (2000) 076002.
  \href {http://arxiv.org/abs/hep-ph/9907254}
  {\path{arXiv:hep-ph/9907254}}, \href
  {http://dx.doi.org/10.1103/PhysRevD.62.076002}
  {\path{doi:10.1103/PhysRevD.62.076002}}.

\bibitem{Bohm:2000jw}
A.R. Bohm and N.L. Harshman,  \textit{Nucl. Phys.} \textbf{B581} (2000) 91.
  \href {http://arxiv.org/abs/hep-ph/0001206}
  {\path{arXiv:hep-ph/0001206}}, \href
  {http://dx.doi.org/10.1016/S0550-3213(00)00249-2}
  {\path{doi:10.1016/S0550-3213(00)00249-2}}.

\bibitem{gruenewald-smatasy:2005}
M. Gr{\"u}newald \textit{et al.}, Fortran package SMATASY 6.42 (2 June
  2005),
  \url{http://www.cern.ch/Martin.Grunewald/afs/public/smatasy/smata6_42.fortran}.

\bibitem{Cahn:1986qf}
R.N. Cahn,  \textit{Phys. Rev.} \textbf{D36} (1987) 2666.
  \href {http://dx.doi.org/10.1103/PhysRevD.36.2666}
  {\path{doi:10.1103/PhysRevD.36.2666}}.

\bibitem{Alexander:1996ha}
G.~Alexander \textit{et al.},   \textit{Z. Phys.} \textbf{C72} (1996) 365.
 \href {http://dx.doi.org/10.1007/s002880050257}
  {\path{doi:10.1007/s002880050257}}.

\bibitem{Riemann:2010zz}
S.~Riemann, \textit{Rep. Prog. Phys.}
  \textbf{73} (2010) 126201.
 \href {http://dx.doi.org/10.1088/0034-4885/73/12/126201}
  {\path{doi:10.1088/0034-4885/73/12/126201}}.

\bibitem{Freitas:2013dpa}
A.~Freitas,  \textit{Phys. Lett.} \textbf{B730} (2014) 5052.
\href {http://arxiv.org/abs/1310.2256} {\path{arXiv:1310.2256}},
  \href {http://dx.doi.org/10.1016/j.physletb.2014.01.017}
  {\path{doi:10.1016/j.physletb.2014.01.017}}.

\bibitem{Jegerlehner:1988ak}
F.~Jegerlehner, Precision tests of electroweak interaction parameters,  Proc. 11th Int. School of Theoretical
  Physics, Testing the Standard Model, Szczyrk, Poland,  1987, Eds.
  R.
  Manka and M. Zralek, (World Scientific, Singapore, 1988), p. 33, Bielefeld preprint
  BI-TP-87/16.
  \url{https://lib-extopc.kek.jp/preprints/PDF/1988/8801/8801263.pdf}.

\bibitem{Bernabeu:1987me}
J.~Bernabeu \textit{et al.},  \textit{Phys. Lett.} \textbf{B200} (1988) 569.
  \href {http://dx.doi.org/10.1016/0370-2693(88)90173-6}
  {\path{doi:10.1016/0370-2693(88)90173-6}}.

\bibitem{Beenakker:1988pv}
W.~Beenakker and W.~Hollik,  \textit{Z. Phys.} \textbf{C40} (1988) 141.
 \href {http://dx.doi.org/10.1007/BF01559728}
  {\path{doi:10.1007/BF01559728}}.

\bibitem{Awramik:2004ge}
M.~Awramik \textit{et al.},   \textit{Phys. Rev. Lett.} \textbf{93} (2004) 201805.
 \href {http://arxiv.org/abs/hep-ph/0407317}
  {\path{arXiv:hep-ph/0407317}}, \href
  {http://dx.doi.org/10.1103/PhysRevLett.93.201805}
  {\path{doi:10.1103/PhysRevLett.93.201805}}.

\bibitem{Freitas:2012sy}
A.~Freitas and Y.-C. Huang,  \textit{J. High Energy Phys.} \textbf{08} (2012) 050 [Erratum:  \textit{J. High Energy Phys.} \textbf{10} (2013) 044].
  \href {http://arxiv.org/abs/1205.0299} {\path{arXiv:1205.0299}},
  \href {http://dx.doi.org/10.1007/JHEP05(2013)074}
  {\path{doi:10.1007/JHEP05(2013)074}}.

\bibitem{Bohm:2004zi}
A.R. B{\"o}hm and Y.~Sato,  \textit{Phys. Rev.} \textbf{D71} (2005)
  085018.
  \href {http://arxiv.org/abs/hep-ph/0412106}
  {\path{arXiv:hep-ph/0412106}}, \href
  {http://dx.doi.org/10.1103/PhysRevD.71.085018}
  {\path{doi:10.1103/PhysRevD.71.085018}}.

\bibitem{Penin:2005kf}
A.A. Penin,  \textit{Phys. Rev. Lett.} \textbf{95}
  (2005) 010408.
  \href {http://arxiv.org/abs/hep-ph/0501120}
  {\path{arXiv:hep-ph/0501120}}, \href
  {http://dx.doi.org/10.1103/PhysRevLett.95.010408}
  {\path{doi:10.1103/PhysRevLett.95.010408}}.

\bibitem{Penin:2005eh}
A.~Penin,  \textit{Nucl.
  Phys.} \textbf{B734} (2006) 185.
  \href {http://arxiv.org/abs/hep-ph/0508127}
  {\path{arXiv:hep-ph/0508127}}, \href
  {http://dx.doi.org/10.1016/j.nuclphysb.2005.11.016}
  {\path{doi:10.1016/j.nuclphysb.2005.11.016}}.

\bibitem{Kuhn:2001hz}
J.H. K{\"u}hn \textit{et al.},   \textit{Nucl. Phys.}
  \textbf{B616} (2001) 286.
  \href {http://arxiv.org/abs/hep-ph/0106298}
  {\path{arXiv:hep-ph/0106298}}, \href
  {http://dx.doi.org/10.1016/S0550-3213(01)00454-0}
  {\path{doi:10.1016/S0550-3213(01)00454-0}}.

\bibitem{Jadach:1998jb}
S.~Jadach \textit{et al.},   \textit{Phys. Lett.} \textbf{B449} (1999) 97.
  \href {http://arxiv.org/abs/hep-ph/9905453}
  {\path{arXiv:hep-ph/9905453}}, \href
  {http://dx.doi.org/10.1016/S0370-2693(99)00038-6}
  {\path{doi:10.1016/S0370-2693(99)00038-6}}.

\bibitem{Jadach:2000ir}
S.~Jadach \textit{et al.},  \textit{Phys. Rev.} \textbf{D63} (2001) 113009.
  \href {http://arxiv.org/abs/hep-ph/0006359}
  {\path{arXiv:hep-ph/0006359}}, \href
  {http://dx.doi.org/10.1103/PhysRevD.63.113009}
  {\path{doi:10.1103/PhysRevD.63.113009}}.

\bibitem{Greco:1975ke}
M.~Greco  \textit{et al.},   \textit{Phys. Lett.} \textbf{B56}
  (1975) 367.
  \href {http://dx.doi.org/10.1016/0370-2693(75)90321-4}
  {\path{doi:10.1016/0370-2693(75)90321-4}}.

\bibitem{Greco:1975wq}
M.~Greco and A.F. Grillo,  \textit{Lett. Nuovo Cim.} \textbf{15} (1976) 174.
  \href {http://dx.doi.org/10.1007/BF02727477}
  {\path{doi:10.1007/BF02727477}}.

\bibitem{Greco:1980mh}
M.~Greco \textit{et al.},   \textit{Nucl. Phys.} \textbf{B171} (1980) 118 [Erratum: \textit{Nucl.
  Phys.} \textbf{B197} (1982) 543].
  \href {http://dx.doi.org/10.1016/0550-3213(80)90363-6}
  {\path{doi:10.1016/0550-3213(80)90363-6}}.

\bibitem{Consoli:1982ib}
M.~Consoli \textit{et al.},   
  \textit{Phys. Lett.} \textbf{113B} (1982) 415.
  \href {http://dx.doi.org/10.1016/0370-2693(82)90776-6}
  {\path{doi:10.1016/0370-2693(82)90776-6}}.

\bibitem{Greco:1986dc}
M.~Greco,  \textit{Phys. Lett.} \textbf{B177} (1986)
  97.
  \href {http://dx.doi.org/10.1016/0370-2693(86)90023-7}
  {\path{doi:10.1016/0370-2693(86)90023-7}}.

\bibitem{Nicrosini:1986sm}
O.~Nicrosini and L.~Trentadue, \textit{Phys. Lett.} \textbf{B196} (1987) 551.
  \href {http://dx.doi.org/10.1016/0370-2693(87)90819-7}
  {\path{doi:10.1016/0370-2693(87)90819-7}}.

\bibitem{Nicrosini:1987sw}
O.~Nicrosini and L.~Trentadue,  \textit{Z. Phys.} \textbf{C39} (1988) 479.
  \href {http://dx.doi.org/10.1007/BF01555976}
  {\path{doi:10.1007/BF01555976}}.

\bibitem{Aversa:1991rw}
F.~Aversa and M.~Greco,  \textit{Phys. Lett.} \textbf{B271} (1991) 435.
  \href {http://dx.doi.org/10.1016/0370-2693(91)90114-6}
  {\path{doi:10.1016/0370-2693(91)90114-6}}.

\bibitem{Fadin:1992ue}
V.S. Fadin \textit{et al.},   {Small
  angle {Bhabha} scattering: two loop approximation}, jINR-E2-92-577.

\bibitem{Fadin:1994ny}
V.S. Fadin \textit{et al.},  {Forward
  per mille {Bhabha} scattering},  Tennessee International
  Symposium on Radiative Corrections: Status and Outlook, Gatlinburg, TN,
USA,  1994.

\bibitem{Hahn:2000kx}
T.~Hahn, \textit{Comput.
  Phys. Commun.} \textbf{140} (2001) 418.
  \href {http://arxiv.org/abs/hep-ph/0012260}
  {\path{arXiv:hep-ph/0012260}}, \href
  {http://dx.doi.org/10.1016/S0010-4655(01)00290-9}
  {\path{doi:10.1016/S0010-4655(01)00290-9}}.

\bibitem{Nejad:2013ina}
B.~Chokoufe~Nejad  \textit{et al.},  \textit{J. Phys. Conf. Ser.} \textbf{523} (2014) 012050.
  \href {http://arxiv.org/abs/1310.0274} {\path{arXiv:1310.0274}},
  \href {http://dx.doi.org/10.1088/1742-6596/523/1/012050}
  {\path{doi:10.1088/1742-6596/523/1/012050}}.

\bibitem{Ossola:2007ax}
G.~Ossola \textit{et al.},  \textit{J. High Energy Phys.} \textbf{03} (2008) 042.
  \href {http://arxiv.org/abs/0711.3596} {\path{arXiv:0711.3596}},
  \href {http://dx.doi.org/10.1088/1126-6708/2008/03/042}
  {\path{doi:10.1088/1126-6708/2008/03/042}}.

\bibitem{Berger:2008sj}
C.F. Berger \textit{et al.},  \textit{Phys. Rev.} \textbf{D78} (2008) 036003.
  \href {http://arxiv.org/abs/0803.4180} {\path{arXiv:0803.4180}},
  \href {http://dx.doi.org/10.1103/PhysRevD.78.036003}
  {\path{doi:10.1103/PhysRevD.78.036003}}.

\bibitem{Kanaki:2000ey}
A.~Kanaki and C.G. Papadopoulos,  \textit{Comput. Phys. Commun.} \textbf{132} (2000) 306.
  \href {http://arxiv.org/abs/hep-ph/0002082}
  {\path{arXiv:hep-ph/0002082}}, \href
  {http://dx.doi.org/10.1016/S0010-4655(00)00151-X}
  {\path{doi:10.1016/S0010-4655(00)00151-X}}.

\bibitem{vanHameren:2009dr}
A.~van Hameren \textit{et al.},   \textit{J. High Energy Phys.} \textbf{09} (2009) 106.
  \href {http://arxiv.org/abs/0903.4665} {\path{arXiv:0903.4665}},
  \href {http://dx.doi.org/10.1088/1126-6708/2009/09/106}
  {\path{doi:10.1088/1126-6708/2009/09/106}}.

\bibitem{Mastrolia:2010nb}
P.~Mastrolia \textit{et al.},    \textit{J. High Energy Phys.}  \textbf{08} (2010)
  080.
  \href {http://arxiv.org/abs/1006.0710} {\path{arXiv:1006.0710}},
  \href {http://dx.doi.org/10.1007/JHEP08(2010)080}
  {\path{doi:10.1007/JHEP08(2010)080}}.

\bibitem{Hirschi:2011pa}
V.~Hirschi \textit{et al.}, 
  \textit{J. High Energy Phys.} \textbf{05} (2011) 044.
  \href {http://arxiv.org/abs/1103.0621} {\path{arXiv:1103.0621}},
  \href {http://dx.doi.org/10.1007/JHEP05(2011)044}
  {\path{doi:10.1007/JHEP05(2011)044}}.

\bibitem{Cullen:2011ac}
G.~Cullen \textit{et al.},  \textit{Eur.
  Phys. J.} \textbf{C72} (2012) 1889.
  \href {http://arxiv.org/abs/1111.2034} {\path{arXiv:1111.2034}},
  \href {http://dx.doi.org/10.1140/epjc/s10052-012-1889-1}
  {\path{doi:10.1140/epjc/s10052-012-1889-1}}.

\bibitem{Cascioli:2011va}
F.~Cascioli \textit{et al.},  \textit{Phys. Rev. Lett.} \textbf{108} (2012) 111601.
  \href {http://arxiv.org/abs/1111.5206} {\path{arXiv:1111.5206}},
  \href {http://dx.doi.org/10.1103/PhysRevLett.108.111601}
  {\path{doi:10.1103/PhysRevLett.108.111601}}.

\bibitem{Actis:2016mpe}
S.~Actis \textit{et al.},   \textit{Comput. Phys. Commun.} \textbf{214}
  (2017) 140.
  \href {http://arxiv.org/abs/1605.01090} {\path{arXiv:1605.01090}},
  \href {http://dx.doi.org/10.1016/j.cpc.2017.01.004}
  {\path{doi:10.1016/j.cpc.2017.01.004}}.

\bibitem{Fleischer:2007ph}
J.~Fleischer \textit{et al.},  \textit{Acta Phys. Pol.} \textbf{B38} (2007) 3529.
  \url{http://www.actaphys.uj.edu.pl/_cur/store/vol38/pdf/v38p3529.pdf}.
  \href {http://arxiv.org/abs/0710.5100} {\path{arXiv:0710.5100}}.

\bibitem{Diakonidis:2008ij}
T.~Diakonidis \textit{et al.},  \textit{Phys. Rev.}
 \textbf{D80} (2009) 036003.
  \href {http://arxiv.org/abs/0812.2134} {\path{arXiv:0812.2134}},
  \href {http://dx.doi.org/10.1103/PhysRevD.80.036003}
  {\path{doi:10.1103/PhysRevD.80.036003}}.

\bibitem{Fleischer:2010sq}
J.~Fleischer and T.~Riemann,  \textit{Phys. Rev.} \textbf{D83} (2011) 073004.
  \href {http://arxiv.org/abs/1009.4436} {\path{arXiv:1009.4436}},
  \href {http://dx.doi.org/10.1103/PhysRevD.83.073004}
  {\path{doi:10.1103/PhysRevD.83.073004}}.

\bibitem{Fleischer:2012et}
J.~Fleischer \textit{et al.},   \textit{Proc. Sci.} \textbf{LL2012} (2012) 020.
  \href {http://arxiv.org/abs/1210.4095} {\path{arXiv:1210.4095}},
  \href {http://dx.doi.org/10.22323/1.151.0020}
  {\path{doi:10.22323/1.151.0020}}.

\bibitem{Fleischer:2011zz}
J.~Fleischer \textit{et al.},  {PJFry: a C++ package for tensor reduction
  of one-loop Feynman integrals.}  The SM and NLO Multileg and SM MC Working
Groups: Summary Report,   DESY 11-252, (2011). 
  \href{http://www-library.desy.de/cgi-bin/showprep.pl?desy11-252} {
  http://www-library.desy.de/cgi-bin/showprep.pl?desy11-252}.

\bibitem{pjfry-project}
V. Yundin, C++ package PJFry, 
  \url{https://github.com/Vayu/PJFry/}.

\bibitem{Denner:2016kdg}
A.~Denner \textit{et al.},   \textit{Comput. Phys. Commun.} \textbf{212} (2017)
  220.
  \href {http://arxiv.org/abs/1604.06792} {\path{arXiv:1604.06792}},
  \href {http://dx.doi.org/10.1016/j.cpc.2016.10.013}
  {\path{doi:10.1016/j.cpc.2016.10.013}}.

\bibitem{AlcarazMaestre:2012vp}
J. Alcaraz Maestre \textit{et al.},  {The SM and
  NLO Multileg and SM MC Working Groups: Summary Report.} \href
  {http://arxiv.org/abs/1203.6803} {\path{arXiv:1203.6803}}.

\bibitem{Actis:2010gg}
S.~Actis \textit{et al.},   \textit{Eur. Phys. J.} \textbf{C66} (2010)
  585.
  \href {http://arxiv.org/abs/0912.0749} {\path{arXiv:0912.0749}},
  \href {http://dx.doi.org/10.1140/epjc/s10052-010-1251-4}
  {\path{doi:10.1140/epjc/s10052-010-1251-4}}.

\bibitem{CarloniCalame:2011zq}
C.~Carloni~Calame \textit{et al.},   \textit{J. High Energy Phys.} \textbf{07}
  (2011) 126.
  \href {http://arxiv.org/abs/1106.3178} {\path{arXiv:1106.3178}},
  \href {http://dx.doi.org/10.1007/JHEP07(2011)126}
  {\path{doi:10.1007/JHEP07(2011)126}}.

\bibitem{CarloniCalame:2011aa}
C.M. Carloni~Calame \textit{et al.},   \textit{Nucl. Phys. Proc.
  Suppl.} \textbf{225--227} (2012) 293.
  \href {http://arxiv.org/abs/1112.2851} {\path{arXiv:1112.2851}},
  \href {http://dx.doi.org/10.1016/j.nuclphysbps.2012.02.061}
  {\path{doi:10.1016/j.nuclphysbps.2012.02.061}}.

\bibitem{Campanario:2013uea}
F.~Campanario \textit{et al.},  
  \textit{J. High Energy Phys.} \textbf{1402} (2014) 114.
  \href {http://arxiv.org/abs/1312.3610} {\path{arXiv:1312.3610}},
  \href {http://dx.doi.org/10.1007/JHEP02(2014)114}
  {\path{doi:10.1007/JHEP02(2014)114}}.

\bibitem{Gluza:2012yz}
J.~Gluza \textit{et al.},  \textit{Proc. Sci.} \textbf{RADCOR2011} (2011)
034.
  \href {http://arxiv.org/abs/1201.0968} {\path{arXiv:1201.0968}},
  \href {http://dx.doi.org/10.22323/1.145.0034}
  {\path{doi:10.22323/1.145.0034}}.

\bibitem{Becher:2007cu}
T.~Becher and K.~Melnikov,  \textit{J. High Energy Phys.}
  \textbf{06} (2007) 084.
  \href {http://arxiv.org/abs/0704.3582} {\path{arXiv:0704.3582}},
  \href {http://dx.doi.org/10.1088/1126-6708/2007/06/084}
  {\path{doi:10.1088/1126-6708/2007/06/084}}.

\bibitem{Bonciani:2004gi}
R.~Bonciani \textit{et al.}, 
   \textit{Nucl.
  Phys.} \textbf{B701} (2004) 121.
  \href {http://arxiv.org/abs/hep-ph/0405275}
  {\path{arXiv:hep-ph/0405275}}, \href
  {http://dx.doi.org/10.1016/j.nuclphysb.2004.09.015}
  {\path{doi:10.1016/j.nuclphysb.2004.09.015}}.

\bibitem{Actis:2007gi}
S.~Actis \textit{et al.},   \textit{Nucl. Phys.} \textbf{B786} (2007) 26.
  \href {http://arxiv.org/abs/0704.2400v2} {\path{arXiv:0704.2400v2}},
  \href {http://dx.doi.org/10.1016/j.nuclphysb.2007.06.023}
  {\path{doi:10.1016/j.nuclphysb.2007.06.023}}.

\bibitem{Actis:2007pn}
S.~Actis \textit{et al.},   \textit{Acta Phys. Pol.} \textbf{B38} (2007) 3517.
  \url{http://www.actaphys.uj.edu.pl/fulltext?series=Reg&vol=38&page=3517}.
  \href {http://arxiv.org/abs/0710.5111 [hep-ph]}
  {\path{arXiv:0710.5111 [hep-ph]}}.

\bibitem{Actis:2007fs}
S.~Actis \textit{et al.},   \textit{Phys. Rev. Lett.} \textbf{100} (2008) 131602.
  \href {http://arxiv.org/abs/0711.3847} {\path{arXiv:0711.3847}},
  \href {http://dx.doi.org/10.1103/PhysRevLett.100.131602}
  {\path{doi:10.1103/PhysRevLett.100.131602}}.

\bibitem{Bonciani:2008zz}
R.~Bonciani,  \textit{J. Phys.
  Conf. Ser.} \textbf{110} (2008) 042004.
  \href {http://dx.doi.org/10.1088/1742-6596/110/4/042004}
  {\path{doi:10.1088/1742-6596/110/4/042004}}.

\bibitem{Bonciani:2008ep}
R.~Bonciani \textit{et al.},  \textit{J. High Energy Phys.}  \textbf{0802} (2008) 080.
  \href {http://arxiv.org/abs/0802.2215} {\path{arXiv:0802.2215}},
  \href {http://dx.doi.org/10.1088/1126-6708/2008/02/080}
  {\path{doi:10.1088/1126-6708/2008/02/080}}.

\bibitem{Actis:2008br}
S.~Actis \textit{et al.},   \textit{Phys. Rev.} \textbf{D78} (2008)
  085019.
  \href {http://arxiv.org/abs/0807.4691} {\path{arXiv:0807.4691}},
  \href {http://dx.doi.org/10.1103/PhysRevD.78.085019}
  {\path{doi:10.1103/PhysRevD.78.085019}}.

\bibitem{Kuhn:2008zs}
J.H. K{\"u}hn and S.~Uccirati,  \textit{Nucl. Phys.} \textbf{B806} (2009) 300.
  \href {http://arxiv.org/abs/0807.1284} {\path{arXiv:0807.1284}},
  \href {http://dx.doi.org/10.1016/j.nuclphysb.2008.08.002}
  {\path{doi:10.1016/j.nuclphysb.2008.08.002}}.

\bibitem{Jadach:1992aa}
S.~Jadach \textit{et al.},  \textit{Phys. Lett.} \textbf{B280} (1992) 129.
  \href {http://dx.doi.org/10.1016/0370-2693(92)90786-4}
  {\path{doi:10.1016/0370-2693(92)90786-4}}.

\bibitem{Jadach:2018lwm}
S.~Jadach and S.~Yost, {QED interference in charge asymmetry near the Z resonance
  at future electron--positron colliders,} \textit{Phys. Lett. B.} (in press).
  \href {http://arxiv.org/abs/1801.08611} {\path{arXiv:1801.08611}}.

\bibitem{Jadach:1999gz}
S.~Jadach \textit{et al.},  \textit{Phys. Lett.} \textbf{B465} (1999) 254.
  \href {http://arxiv.org/abs/hep-ph/9907547}
  {\path{arXiv:hep-ph/9907547}}, \href
  {http://dx.doi.org/10.1016/S0370-2693(99)01047-3}
  {\path{doi:10.1016/S0370-2693(99)01047-3}}.

\bibitem{Yennie:1961ad}
D.R. Yennie \textit{et al.},   \textit{Ann. Phys.} \textbf{13} (1961) 379.
  \href {http://dx.doi.org/10.1016/0003-4916(61)90151-8}
  {\path{doi:10.1016/0003-4916(61)90151-8}}.



\bibitem{web-zfitter.com:2015}
  \href{http://zfitter.com}{http://zfitter.com}, last accessed May 20th 2019.

\bibitem{web-zfitter.education:2015}
  \href{http://zfitter.education}{http://zfitter.education}, last accessed May 20th 2019.

\bibitem{CALC2018:2018}
Conf. CALC2018, JINR, Dubna, Russia,  2018,
  \url{https://indico.jinr.ru/conferenceDisplay.py?ovw=True&confId=418}.

\bibitem{RiemannTord:CALC2018}
T. Riemann,  A legacy of Dima Bardin: ZFITTER and the future, Conf. CALC2018, JINR, Dubna, Russia,  2018,
  \url{https://indico.jinr.ru/getFile.py/access?contribId=19&sessionId=0&resId=0&materialId=slides&confId=418}.

\bibitem{Bardin:1986fi}
D.~Bardin \textit{et al.},  \textit{Z. Phys.} \textbf{C32} (1986) 121.
  \url{https://lib-extopc.kek.jp/preprints/PDF/1986/8607/8607019.pdf}.
  \href {http://dx.doi.org/10.1007/BF01441360}
  {\path{doi:10.1007/BF01441360}}.

\bibitem{Fred}
\url{http://project-gfitter.web.cern.ch/project-gfitter/}, last accessed
May 9th 2019.

\bibitem{Tom}
H. Fl\"acher \textit{et al.} \textit{Eur. Phys. J. C} \textbf{60} (2009)
543.

\bibitem{web-desy-zfitter-gfitter:2014-c}
  \href{http://zfitter-gfitter.desy.de}{http://zfitter-gfitter.desy.de},
  last accessed March 22nd 2014.

\bibitem{Aushev:2010bq}
T.~Aushev \textit{et al.}, {Physics at Super B Factory,} \href
  {http://arxiv.org/abs/1002.5012} {\path{arXiv:1002.5012}}.

\bibitem{Ferber:dfg2015}
T.~Ferber, {Towards first physics at Belle II.  Spring Conf.
  DPG, Wuppertal, Germany},
2015,  \url{http://www.staff.uni-giessen.de/~gd1472/belle/dpg2015_torbenferber.pdf}.

\bibitem{vanderBij:2014mxa}
J.~van~der Bij \textit{et al.}, 
  {Mini-Proc., 15th Meeting of the Working Group on Radiative Corrections and
  MC Generators for Low Energies,} \href {http://arxiv.org/abs/1406.4639}
  {\path{arXiv:1406.4639}}.

\bibitem{Fedorenko:1986hw}
O.~Fedorenko and T.~Riemann,  \textit{Acta Phys. Pol.} \textbf{B18} (1987) 761.   \url{http://www.actaphys.uj.edu.pl/fulltext?series=Reg&vol=18&page=761}.

\bibitem{Stuart:1991rv}
R.G. Stuart, {Gauge-invariant perturbation theory near a gauge resonance},   {Workshop on High Energy Phenomenology (CINVESTAV), Mexico City, Mexico}, 1991, p 331, CERN-TH.6261/91,
  \url{https://lib-extopc.kek.jp/preprints/PDF/1992/9201/9201503.pdf}.

\bibitem{Stuart:1992jf}
R.G. Stuart,  \textit{Phys. Rev. Lett.} \textbf{70} (1993) 3193.
  \href {http://dx.doi.org/10.1103/PhysRevLett.70.3193}
  {\path{doi:10.1103/PhysRevLett.70.3193}}.

\bibitem{Stuart:1995zr}
R.G. Stuart,  \textit{Nucl. Phys.}
  \textbf{B498} (1997) 28.
  \href {http://arxiv.org/abs/hep-ph/9504215}
  {\path{arXiv:hep-ph/9504215}}, \href{http://dx.doi.org/10.1016/S0550-3213(97)00276-9}
  {\path{doi:10.1016/S0550-3213(97)00276-9}}.

\bibitem{Adriani:1993sx}
O.~Adriani \textit{et al.},  \textit{Phys. Lett.} \textbf{B315}
  (1993) 494.
  \href {http://dx.doi.org/10.1016/0370-2693(93)91646-5}
  {\path{doi:10.1016/0370-2693(93)91646-5}}.

\bibitem{Buskulic:1996ua}
ALEPH Collaboration \textit{et al.},  \textit{Phys. Lett.} \textbf{B378} (1996) 373.
  \href {http://dx.doi.org/10.1016/0370-2693(96)00504-7}
  {\path{doi:10.1016/0370-2693(96)00504-7}}.

\bibitem{Abbiendi:2000hu}
OPAL Collaboration  \textit{et al.},  \textit{Eur. Phys. J.} \textbf{C19} (2001) 587.
  \href {http://arxiv.org/abs/hep-ex/0012018}
  {\path{arXiv:hep-ex/0012018}}, \href
  {http://dx.doi.org/10.1007/s100520100627} {\path{doi:10.1007/s100520100627}}.

\bibitem{Sachs:2003ja}
K.~Sachs, {Standard model at LEP II,} Proc.  XXXVIII Rencontres de Moriond:
  Electroweak Interactions and Unified Theories, Les Arcs, 2003,
  \href {http://arxiv.org/abs/hep-ex/0307009}
  {\path{arXiv:hep-ex/0307009}}.

\bibitem{Holt:2014moa}
P.J. Holt,  \textit{Proc. Sci.} \textbf{hep2001}
  (2001) 115.
  \href {http://dx.doi.org/10.22323/1.007.0115}
  {\path{doi:10.22323/1.007.0115}}.

\bibitem{Yusa:1999dx}
{VENUS Collaboration} \textit{et al.},  \textit{Phys.
  Lett.} \textbf{B447} (1999) 167.
  \href {http://dx.doi.org/10.1016/S0370-2693(98)01560-3}
  {\path{doi:10.1016/S0370-2693(98)01560-3}}.

\bibitem{Miyabayashi:1994ej}
TOPAZ Collaboration \textit{et al.},  \textit{Phys. Lett.} \textbf{B347} (1995) 171.
  \href {http://dx.doi.org/10.1016/0370-2693(95)00038-M}
  {\path{doi:10.1016/0370-2693(95)00038-M}}.

\bibitem{Leike:1992uf}
A.~Leike \textit{et al.}, \textit{Phys. Lett.} \textbf{B291} (1992) 187.
  \href {http://arxiv.org/abs/hep-ph/9507436}
  {\path{arXiv:hep-ph/9507436}}, \href
  {http://dx.doi.org/10.1016/0370-2693(92)90142-Q}
  {\path{doi:10.1016/0370-2693(92)90142-Q}}.

\bibitem{Bardin:1987hv}
D.~Bardin \textit{et al.}, {The} electromagnetic $\alpha^3$
  contributions to $\mathrm{e}^+ \mathrm{e}^-$ annihilation into fermions in the electroweak
  theory: total cross-section {$\sigma_T$} and integrated asymmetry
  {${A}_\mathrm{FB}$}.  JINR preprint E2-87-663, (1987),
  \url{https://lib-extopc.kek.jp/preprints/PDF/1988/8801/8801179.pdf}.



\bibitem{Bardin:1988ze}
D.~Bardin \textit{et al.}, {The} electromagnetic
  $\alpha^3$ contributions to $\mathrm{e}^+ \mathrm{e}^-$ annihilation into fermions in the
  electroweak theory: total cross-section $\sigma_T$ and integrated asymmetry
  ${A}_\mathrm{FB}$.  JINR preprint E2-88-324, (1988).
  \url{https://lib-extopc.kek.jp/preprints/PDF/1988/8808/8808103.pdf}.

\bibitem{Berends:1982ie}
F.A. Berends \textit{et al.}, 
  \textit{Nucl. Phys.} \textbf{B202} (1982) 63.
  \href {http://dx.doi.org/10.1016/0550-3213(82)90221-8}
  {\path{doi:10.1016/0550-3213(82)90221-8}}.

\bibitem{Berends:1983mi}
F.A. Berends \textit{et al.},   \textit{Comput. Phys. Commun.} \textbf{29} (1983) 185.
   \href {http://dx.doi.org/10.1016/0010-4655(83)90073-5}
  {\path{doi:10.1016/0010-4655(83)90073-5}}.

\bibitem{Riemann:1988gy}
T.~Riemann \textit{et al.},  The {Z} boson line shape at {LEP},
   XI Warsaw Symposium on Elementary Particle Physics: New
  Theories in Physics,   Kazimierz, Poland, 1998, Eds.
  Z. Ajduk \textit{et al.},  (World Scientific, Teaneck, NJ, USA, 1988),
  p. 238.
  \url{https://lib-extopc.kek.jp/preprints/PDF/1989/8901/8901077.pdf}.

\bibitem{Leike:1989ah}
A.~Leike \textit{et al.},  \textit{Phys. Lett.} \textbf{B241} (1990) 267.
  \href {http://dx.doi.org/10.1016/0370-2693(90)91291-I}
  {\path{doi:10.1016/0370-2693(90)91291-I}}.

\bibitem{Passarino:1982zp}
G.~Passarino,  \textit{Nucl. Phys.} \textbf{B204} (1982) 237.
  \href {http://dx.doi.org/10.1016/0550-3213(82)90147-X}
  {\path{doi:10.1016/0550-3213(82)90147-X}}.

\bibitem{Bardin:1976qa}
D.~Bardin and N.~Shumeiko,  \textit{Nucl. Phys.} \textbf{B127}
  (1977) 242.   \href {http://dx.doi.org/10.1016/0550-3213(77)90213-9}
  {\path{doi:10.1016/0550-3213(77)90213-9}}.

\bibitem{Akhundov:1984mm}
A.~Akhundov \textit{et al.},  Some integrals for exact
  calculation of {QED} bremsstrahlung, {Dubna} preprint JINR-E2-84-777,
   \url{https://lib-extopc.kek.jp/preprints/PDF/1985/8504/8504203.pdf}.

\bibitem{Bardin:1987ht}
D.Y. Bardin \textit{et al.},  Some integrals for
  analytic bremsstrahlung calculation: photon and $\mathrm{Z}^0$ boson exchange in the
  ultrarelativistic
  limit, JINR-E2-87-664, \url{https://lib-extopc.kek.jp/preprints/PDF/1988/8801/8801180.pdf}.

\bibitem{CarloniCalame:2000pz}
C.M. {Carloni Calame} \textit{et al.}, 
  \textit{Nucl.
  Phys.} \textbf{B584} (2000) 459.
  \href {http://arxiv.org/abs/hep-ph/0003268}
  {\path{arXiv:hep-ph/0003268}}, \href
  {http://dx.doi.org/10.1016/S0550-3213(00)00356-4}
  {\path{doi:10.1016/S0550-3213(00)00356-4}}.

\bibitem{CarloniCalame:2001ny}
C.M. {Carloni Calame},  \textit{Phys.
  Lett.} \textbf{B520} (2001) 16.
  \href {http://arxiv.org/abs/hep-ph/0103117}
  {\path{arXiv:hep-ph/0103117}}, \href
  {http://dx.doi.org/10.1016/S0370-2693(01)01108-X}
  {\path{doi:10.1016/S0370-2693(01)01108-X}}.

\bibitem{CarloniCalame:2003yt}
C.M. {Carloni Calame} \textit{et al.},  \textit{Nucl. Phys. Proc. Suppl.} \textbf{131} (2004) 48.
  \href {http://arxiv.org/abs/hep-ph/0312014}
  {\path{arXiv:hep-ph/0312014}}, \href
  {http://dx.doi.org/10.1016/j.nuclphysbps.2004.02.008}
  {\path{doi:10.1016/j.nuclphysbps.2004.02.008}}.

\bibitem{Balossini:2006wc}
G.~Balossini \textit{et al.}, 
   \textit{Nucl. Phys.} \textbf{B758} (2006) 227.
  \href {http://arxiv.org/abs/hep-ph/0607181}
  {\path{arXiv:hep-ph/0607181}}, \href
  {http://dx.doi.org/10.1016/j.nuclphysb.2006.09.022}
  {\path{doi:10.1016/j.nuclphysb.2006.09.022}}.

\bibitem{Balossini:2008xr}
G.~Balossini \textit{et al.},   \textit{Phys. Lett.} \textbf{B663} (2008) 209.
  \href {http://arxiv.org/abs/0801.3360} {\path{arXiv:0801.3360}},
  \href {http://dx.doi.org/10.1016/j.physletb.2008.04.007}
  {\path{doi:10.1016/j.physletb.2008.04.007}}.

\bibitem{Jadach:1995nk}
S.~Jadach \textit{et al.},  
  \textit{Phys. Lett.} \textbf{B390} (1997) 298.
  \href {http://arxiv.org/abs/hep-ph/9608412}
  {\path{arXiv:hep-ph/9608412}}, \href
  {http://dx.doi.org/10.1016/S0370-2693(96)01382-2}
  {\path{doi:10.1016/S0370-2693(96)01382-2}}.

\bibitem{Arbuzov:2005pt}
A.B. Arbuzov \textit{et al.}, 
   \textit{Eur. Phys. J.} \textbf{C46} (2006) 689.
  \href {http://arxiv.org/abs/hep-ph/0504233}
  {\path{arXiv:hep-ph/0504233}}, \href
  {http://dx.doi.org/10.1140/epjc/s2006-02532-8}
  {\path{doi:10.1140/epjc/s2006-02532-8}}.

\bibitem{Berends:1983fs}
F.~Berends and R.~Kleiss, 
  \textit{Nucl. Phys.} \textbf{B228} (1983) 537.
  \href {http://dx.doi.org/10.1016/0550-3213(83)90558-8}
  {\path{doi:10.1016/0550-3213(83)90558-8}}.

\bibitem{Berends:1980px}
F.A. Berends and R.~Kleiss,  \textit{Nucl. Phys.} \textbf{B186} (1981) 22.
  \href {http://dx.doi.org/10.1016/0550-3213(81)90090-0}
  {\path{doi:10.1016/0550-3213(81)90090-0}}.

\bibitem{Bonciani:2004qt}
R.~Bonciani \textit{et al.}, 
  \textit{Nucl. Phys.} \textbf{B716} (2005)
  280.
  \href {http://arxiv.org/abs/hep-ph/0411321}
  {\path{arXiv:hep-ph/0411321}}, \href
  {http://dx.doi.org/10.1016/j.nuclphysb.2005.03.010}
  {\path{doi:10.1016/j.nuclphysb.2005.03.010}}.

\bibitem{Bonciani:2003cj}
R.~Bonciani \textit{et al.},  \textit{Nucl. Phys.} \textbf{B681} (2004) 261.
  \href {http://arxiv.org/abs/hep-ph/0310333}
  {\path{arXiv:hep-ph/0310333}}, \href
  {http://dx.doi.org/10.1016/j.nuclphysb.2004.08.003}
  {\path{doi:10.1016/j.nuclphysb.2004.08.003}}.

\bibitem{Bonciani:2003ai}
R.~Bonciani \textit{et al.},  
  \textit{Nucl. Phys.} \textbf{B676} (2004) 399.
  \href {http://arxiv.org/abs/hep-ph/0307295}
  {\path{arXiv:hep-ph/0307295}}, \href
  {http://dx.doi.org/10.1016/j.nuclphysb.2003.10.031}
  {\path{doi:10.1016/j.nuclphysb.2003.10.031}}.

\bibitem{Bonciani:2003te}
R.~Bonciani \textit{et al.},   \textit{Nucl. Phys.} \textbf{B661} (2003) 289   [Erratum: \textit{Nucl. Phys.} \textbf{B702} (2004) 359].
  \href {http://arxiv.org/abs/hep-ph/0301170}
  {\path{arXiv:hep-ph/0301170}}, \href
  {http://dx.doi.org/10.1016/j.nuclphysb.2004.08.009}
  {\path{doi:10.1016/j.nuclphysb.2004.08.009}}.

\bibitem{Bonciani:2005im}
R.~Bonciani and A.~Ferroglia,  \textit{Phys. Rev.} \textbf{D72}
  (2005) 056004.
  \href {http://arxiv.org/abs/hep-ph/0507047}
  {\path{arXiv:hep-ph/0507047}}, \href
  {http://dx.doi.org/10.1103/PhysRevD.72.056004}
  {\path{doi:10.1103/PhysRevD.72.056004}}.

\bibitem{Czakon:2006pa}
M.~Czakon \textit{et al.},   \textit{Nucl. Phys.} \textbf{B751} (2006) 1.
  \href {http://arxiv.org/abs/hep-ph/0604101}
  {\path{arXiv:hep-ph/0604101}}, \href
  {http://dx.doi.org/10.1016/j.nuclphysb.2006.05.033}
  {\path{doi:10.1016/j.nuclphysb.2006.05.033}}.

\bibitem{Czakon:2004wm}
M.~Czakon \textit{et al.}, \textit{Phys. Rev.} \textbf{D71} (2005) 073009.
  \href {http://arxiv.org/abs/hep-ph/0412164}
  {\path{arXiv:hep-ph/0412164}}, \href
  {http://dx.doi.org/10.1103/PhysRevD.71.073009}
  {\path{doi:10.1103/PhysRevD.71.073009}}.

\bibitem{Barze:2010pf}
L.~Barze \textit{et al.},   \textit{Eur. Phys. J.} \textbf{C71} (2011) 1680.
  \href {http://arxiv.org/abs/1007.4984} {\path{arXiv:1007.4984}},
  \href {http://dx.doi.org/10.1140/epjc/s10052-011-1680-8}
  {\path{doi:10.1140/epjc/s10052-011-1680-8}}.

\bibitem{Anastasi:2015qla}
A.~Anastasi \textit{et al.},  \textit{Phys. Lett.} \textbf{B750}
  (2015) 633.
  \href {http://arxiv.org/abs/1509.00740} {\path{arXiv:1509.00740}},
  \href {http://dx.doi.org/10.1016/j.physletb.2015.10.003}
  {\path{doi:10.1016/j.physletb.2015.10.003}}.

\bibitem{Babusci:2014sta}
D.~Babusci \textit{et al.},  \textit{Phys.
  Lett.} \textbf{B736} (2014) 459.
  \href {http://arxiv.org/abs/1404.7772} {\path{arXiv:1404.7772}},
  \href {http://dx.doi.org/10.1016/j.physletb.2014.08.005}
  {\path{doi:10.1016/j.physletb.2014.08.005}}.

\bibitem{Abbiendi:2016xup}
G.~Abbiendi \textit{et al.},  \textit{Eur. Phys. J.} \textbf{C77} (2017) 139.
  \href {http://arxiv.org/abs/1609.08987} {\path{arXiv:1609.08987}},
  \href {http://dx.doi.org/10.1140/epjc/s10052-017-4633-z}
  {\path{doi:10.1140/epjc/s10052-017-4633-z}}.

\bibitem{Calame:2015fva}
C.M. Carloni~Calame \textit{et al.},  \textit{Phys. Lett.}
  \textbf{B746} (2015) 325.
  \href {http://arxiv.org/abs/1504.02228} {\path{arXiv:1504.02228}},
  \href {http://dx.doi.org/10.1016/j.physletb.2015.05.020}
  {\path{doi:10.1016/j.physletb.2015.05.020}}.

\bibitem{Eidelman:1995ny}
S.~Eidelman and F.~Jegerlehner,  \textit{Z. Phys.}
  \textbf{C67} (1995) 585.
  \href {http://arxiv.org/abs/hep-ph/9502298}
  {\path{arXiv:hep-ph/9502298}}, \href {http://dx.doi.org/10.1007/BF01553984}
  {\path{doi:10.1007/BF01553984}}.

\bibitem{Jegerlehner:2008rs}
F.~Jegerlehner,  \textit{Nucl. Phys. Proc. Suppl.} \textbf{181--182} (2008) 135.
  \href {http://arxiv.org/abs/0807.4206} {\path{arXiv:0807.4206}},
  \href {http://dx.doi.org/10.1016/j.nuclphysbps.2008.09.010}
  {\path{doi:10.1016/j.nuclphysbps.2008.09.010}}.

\bibitem{lumigg1}
M.~Dam, {Luminosity measurement with diphoton: plan},  FCC-ee Physics Coordination Meeting, (CERN, Geneva, 2018),
  \url{https://indico.cern.ch/event/698497/contributions/2870847/attachments/1589367/2514478/PhysCoord250110.pdf}.
  
  \bibitem{lumigg2}
M.~Dam, {LumiCal for FCC-ee and beam-background impact}, FCC Week,  Amsterdam,
2018,   \url{https://indico.cern.ch/event/656491/contributions/2939126/attachments/1629723/2597664/LumiAmsterdam.pdf}.

\bibitem{Gomez-Ceballos:2013zzn}
M.~Bicer \textit{et al.},  \textit{J. High Energy Phys.} \textbf{01} (2014) 164.
  \href {http://arxiv.org/abs/1308.6176} {\path{arXiv:1308.6176}},
  \href {http://dx.doi.org/10.1007/JHEP01(2014)164}
  {\path{doi:10.1007/JHEP01(2014)164}}.

\bibitem{Jadach:1991by}
S.~Jadach \textit{et al.},  \textit{Comput. Phys. Commun.} \textbf{70} (1992) 305.
  \href {http://dx.doi.org/10.1016/0010-4655(92)90196-6}
  {\path{doi:10.1016/0010-4655(92)90196-6}}.

\bibitem{Jadach:1996is}
S.~Jadach \textit{et al.},   \textit{Comput. Phys. Commun.} \textbf{102} (1997) 229.
  \href {http://dx.doi.org/10.1016/S0010-4655(96)00156-7}
  {\path{doi:10.1016/S0010-4655(96)00156-7}}.

\bibitem{Jadach:1991cg}
S.~Jadach \textit{et al.},   \textit{Phys.
  Lett.} \textbf{B268} (1991) 253.
  \href {http://dx.doi.org/10.1016/0370-2693(91)90813-6}
  {\path{doi:10.1016/0370-2693(91)90813-6}}.

\bibitem{Jadach:1995pd}
S.~Jadach \textit{et al.},  
  \textit{Phys. Lett.} \textbf{B353} (1995) 362 [Erratum: \textit{Phys. Lett.} \textbf{B384} (1996) 488].
  \href {http://dx.doi.org/10.1016/0370-2693(95)00577-8,
  10.1016/0370-2693(96)00923-9} {\path{doi:10.1016/0370-2693(95)00577-8,
  10.1016/0370-2693(96)00923-9}}.

\bibitem{Arbuzov:1996eq}
A.~Arbuzov \textit{et al.},   \textit{Phys. Lett.} \textbf{B383} (1996)
  238.
  \href {http://arxiv.org/abs/hep-ph/9605239}
  {\path{arXiv:hep-ph/9605239}}, \href
  {http://dx.doi.org/10.1016/0370-2693(96)00733-2}
  {\path{doi:10.1016/0370-2693(96)00733-2}}.

\bibitem{Ward:1998ht}
B.F.L. Ward \textit{et al.},   \textit{Phys. Lett.} \textbf{B450} (1999)
  262.
  \href {http://arxiv.org/abs/hep-ph/9811245}
  {\path{arXiv:hep-ph/9811245}}, \href
  {http://dx.doi.org/10.1016/S0370-2693(99)00104-5}
  {\path{doi:10.1016/S0370-2693(99)00104-5}}.

\bibitem{Abbiendi:1999zx}
G.~Abbiendi \textit{et al.},   \textit{Eur. Phys. J.} \textbf{C14} (2000) 373.
  \href {http://arxiv.org/abs/hep-ex/9910066}
  {\path{arXiv:hep-ex/9910066}}, \href
  {http://dx.doi.org/10.1007/s100520000353} {\path{doi:10.1007/s100520000353}}.

\bibitem{Montagna:1998vb}
G.~Montagna \textit{et al.},   \textit{Nucl. Phys.} \textbf{B547} (1999)
  39.
  \href {http://arxiv.org/abs/hep-ph/9811436}
  {\path{arXiv:hep-ph/9811436}}, \href
  {http://dx.doi.org/10.1016/S0550-3213(99)00064-4}
  {\path{doi:10.1016/S0550-3213(99)00064-4}}.

\bibitem{Montagna:1999eu}
G.~Montagna \textit{et al.},   \textit{Phys. Lett.} \textbf{B459} (1999) 649.
  \href {http://arxiv.org/abs/hep-ph/9905235}
  {\path{arXiv:hep-ph/9905235}}, \href
  {http://dx.doi.org/10.1016/S0370-2693(99)00729-7}
  {\path{doi:10.1016/S0370-2693(99)00729-7}}.

\bibitem{CarloniCalame:2015zev}
C.M. Carloni~Calame \textit{et al.},   \textit{Acta Phys.
  Pol.} \textbf{B46} (2015) 2227.
  \href {http://dx.doi.org/10.5506/APhysPolB.46.2227}
  {\path{doi:10.5506/APhysPolB.46.2227}}.

\bibitem{jadach:2006fcal}
S.~Jadach, {\uppercase{QED} calculations for \uppercase{B}habha Luminometer -
  summary of \uppercase{LEP} and lessons for the future,} FCAL  Workshop at IFJ PAN, \url{http://nz42.ifj.edu.pl/\_media/user/jadach/}.

\bibitem{CarloniPisa:2015}
C.~Carloni~Calame, The (theoretical) challenge of precise luminosity
  measurement, FCC-ee
  Physics Workshop (TLEP9), SNS Pisa, \url{https://agenda.infn.it/getFile.py/access?contribId=7&sessionId=7&resId=0&materialId=slides&confId=8830}.

\bibitem{JegerlehnerCERN:2016}
F.~Jegerlehner, $\alpha_{\mathrm{QED}} (M_{\mathrm{Z}})$ and future prospects with low energy e$^{+}$e$^{-}$
  collider data, FCC-ee  Mini-Workshop, `Physics   Behind  Precision',
 \url{https://indico.cern.ch/event/469561/contributions/1977974/attachments/1221704/1786449/SMalphaFCCee16.pdf},
    \url{https://indico.cern.ch/event/469561/}.

\bibitem{Jadach:1999pf}
S.~Jadach \textit{et al.}, \textit{Acta Phys. Pol.} \textbf{B30} (1999) 1745.
  \url{http://www.actaphys.uj.edu.pl/fulltext?series=Reg&vol=30&page=1745}.

\bibitem{Jadach:1996ir}
S.~Jadach and B.F.L. Ward,  \textit{Phys. Lett.} \textbf{B389} (1996) 129.
  \href {http://dx.doi.org/10.1016/S0370-2693(96)01242-7}
  {\path{doi:10.1016/S0370-2693(96)01242-7}}.

\bibitem{Burkhardt:1995tt}
H.~Burkhardt and B.~Pietrzyk,  \textit{Phys. Lett.} \textbf{B356} (1995) 398.
  \href {http://dx.doi.org/10.1016/0370-2693(95)00820-B}
  {\path{doi:10.1016/0370-2693(95)00820-B}}.

\bibitem{Jadach:1996ca}
S.~Jadach \textit{et al.},   \textit{Phys. Rev.} \textbf{D55} (1997)
  1206.
  \href {http://dx.doi.org/10.1103/PhysRevD.55.1206}
  {\path{doi:10.1103/PhysRevD.55.1206}}.

\bibitem{Jadach:1995hv}
S.~Jadach \textit{et al.},   \textit{Phys. Lett. B}  \textbf{353} 
  (1995) 349.
  \href {http://dx.doi.org/10.1016/0370-2693(95)00576-7}
  {\path{doi:10.1016/0370-2693(95)00576-7}}.

\bibitem{Jadach:1990zf}
S.~Jadach \textit{et al.},   \textit{Phys. Lett.} \textbf{B253} (1991)
  469.
  \href {http://dx.doi.org/10.1016/0370-2693(91)91754-J}
  {\path{doi:10.1016/0370-2693(91)91754-J}}.

\bibitem{Jadach:1996gu}
S.~Jadach \textit{et al.}, {Event generators for Bhabha scattering}, {CERN Workshop
  on LEP2 Physics, Geneva,  1995}, (CERN, Geneva,  1996), p. 229.
  \href {http://arxiv.org/abs/hep-ph/9602393}
  {\path{arXiv:hep-ph/9602393}}, \href
  {http://dx.doi.org/10.5170/CERN-1996-001-V-2.229}
  {\path{doi:10.5170/CERN-1996-001-V-2.229}}.

\bibitem{slac-talk}
S.~Jadach, {\uppercase{MC} tools for extracting luminosity spectra,} Seminar
  at SLAC, \url{http://nz42.ifj.edu.pl/\_media/user/jadach/}.

\bibitem{Czakon:2005gi}
M.~Czakon \textit{et al.}, \textit{Acta Phys. Pol.} \textbf{B36} (2005) 3319,
  \url{http://www.actaphys.uj.edu.pl/fulltext?series=Reg&vol=36&page=3319}.
  \href {http://arxiv.org/abs/hep-ph/0511187}
  {\path{arXiv:hep-ph/0511187}}.

\bibitem{Kleiss:1985yh}
R.~Kleiss and W.J. Stirling,  \textit{Nucl. Phys.} \textbf{B262} (1985) 235.
  \href {http://dx.doi.org/10.1016/0550-3213(85)90285-8}
  {\path{doi:10.1016/0550-3213(85)90285-8}}.

\bibitem{Berends:1984qf}
F.A. Berends \textit{et al.}, 
   \textit{Nucl. Phys.} \textbf{B264} (1986) 265.
  \href {http://dx.doi.org/10.1016/0550-3213(86)90482-7}
  {\path{doi:10.1016/0550-3213(86)90482-7}}.

\bibitem{Jadach:1992tf}
S.~Jadach \textit{et al.},  \textit{Phys. Rev.} \textbf{D47} (1993) 2682.
  \href {http://arxiv.org/abs/hep-ph/9211252}
  {\path{arXiv:hep-ph/9211252}}, \href
  {http://dx.doi.org/10.1103/PhysRevD.47.2682}
  {\path{doi:10.1103/PhysRevD.47.2682}}.

\bibitem{Jadach:1995hy}
S.~Jadach \textit{et al.},   \textit{Phys. Lett.} \textbf{B377} (1996) 168.
  \href {http://arxiv.org/abs/hep-ph/9603248}
  {\path{arXiv:hep-ph/9603248}}, \href
  {http://dx.doi.org/10.1016/0370-2693(96)00354-1}
  {\path{doi:10.1016/0370-2693(96)00354-1}}.

\bibitem{Jadach:2006fx}
S.~Jadach \textit{et al.},   \textit{Phys. Rev.} \textbf{D73} (2006) 073001.
  \href {http://arxiv.org/abs/hep-ph/0602197}
  {\path{arXiv:hep-ph/0602197}}, \href
  {http://dx.doi.org/10.1103/PhysRevD.73.073001}
  {\path{doi:10.1103/PhysRevD.73.073001}}.

\bibitem{Actis:2009uq}
S.~Actis \textit{et al.},  \textit{Phys. Lett.} \textbf{B682} (2010) 419.
  \href {http://arxiv.org/abs/0909.1750} {\path{arXiv:0909.1750}},
  \href {http://dx.doi.org/10.1016/j.physletb.2009.11.035}
  {\path{doi:10.1016/j.physletb.2009.11.035}}.

\bibitem{Berends:1987ab}
F.A. Berends \textit{et al.},   \textit{Nucl. Phys.} \textbf{B297} (1988) 429 [Erratum: \textit{Nucl.
  Phys.} \textbf{B304} (1988) 921].
  \href {http://dx.doi.org/10.1016/0550-3213(88)90313-6}
  {\path{doi:10.1016/0550-3213(88)90313-6}}.

\bibitem{Bern:2000ie}
Z.~Bern \textit{et al.},   \textit{Phys. Rev.} \textbf{D63} (2001) 053007.
  \href {http://arxiv.org/abs/hep-ph/0010075}
  {\path{arXiv:hep-ph/0010075}}, \href
  {http://dx.doi.org/10.1103/PhysRevD.63.053007}
  {\path{doi:10.1103/PhysRevD.63.053007}}.

\bibitem{Bonciani:2007eh}
R.~Bonciani \textit{et al.},  \textit{Phys. Rev. Lett.} \textbf{100} (2008) 131601.
  \href {http://arxiv.org/abs/0710.4775} {\path{arXiv:0710.4775}},
  \href {http://dx.doi.org/10.1103/PhysRevLett.100.131601}
  {\path{doi:10.1103/PhysRevLett.100.131601}}.

\bibitem{Jadach:1996bx}
S.~Jadach and B.F.L. Ward,  \textit{Acta Phys. Pol.} \textbf{B28} (1997) 1907,
  \url{http://www.actaphys.uj.edu.pl/fulltext?series=Reg&vol=28&page=1907}.

\bibitem{Jegerlehner:2006ju}
F.~Jegerlehner,  \textit{Nucl. Phys. Proc. Suppl.}
  \textbf{162} (2006) 22.
  \href {http://arxiv.org/abs/hep-ph/0608329}
  {\path{arXiv:hep-ph/0608329}}, \href
  {http://dx.doi.org/10.1016/j.nuclphysbps.2006.09.060}
  {\path{doi:10.1016/j.nuclphysbps.2006.09.060}}.

\bibitem{Jegerlehner:2017gek}
F.~Jegerlehner, \textit{The Anomalous Magnetic Moment of the Muon} (Springer,
Cham, 2017).
  \href {http://dx.doi.org/10.1007/978-3-319-63577-4}
  {\path{doi:10.1007/978-3-319-63577-4}}.

\bibitem{Eidelman:1998vc}
S.~Eidelman \textit{et al.},   \textit{Phys. Lett.} \textbf{B454} (1999)
  369.
  \href {http://arxiv.org/abs/hep-ph/9812521}
  {\path{arXiv:hep-ph/9812521}}, \href
  {http://dx.doi.org/10.1016/S0370-2693(99)00389-5}
  {\path{doi:10.1016/S0370-2693(99)00389-5}}.

\bibitem{Caravaglios:1995cd}
F.~Caravaglios and M.~Moretti,  \textit{Phys. Lett.} \textbf{B358} (1995) 332.
  \href {http://arxiv.org/abs/hep-ph/9507237}
  {\path{arXiv:hep-ph/9507237}}, \href
  {http://dx.doi.org/10.1016/0370-2693(95)00971-M}
  {\path{doi:10.1016/0370-2693(95)00971-M}}.

\bibitem{Barbieri:1972as}
R.~Barbieri \textit{et al.},   \textit{Nuovo Cim.} \textbf{A11} (1972) 824.
  \href {http://dx.doi.org/10.1007/BF02728545}
  {\path{doi:10.1007/BF02728545}}.

\bibitem{Barbieri:1972hn}
R.~Barbieri \textit{et al.},  \textit{Nuovo Cim.} \textbf{A11} (1972) 865.
  \href {http://dx.doi.org/10.1007/BF02728546}
  {\path{doi:10.1007/BF02728546}}.

\bibitem{Arbuzov:1995qd}
A.B. Arbuzov \textit{et al.},   \textit{Nucl. Phys.} \textbf{B485} (1997) 457.
  \href {http://arxiv.org/abs/hep-ph/9512344}
  {\path{arXiv:hep-ph/9512344}}, \href
  {http://dx.doi.org/10.1016/S0550-3213(96)00490-7}
  {\path{doi:10.1016/S0550-3213(96)00490-7}}.

\bibitem{Arbuzov:1995cn}
A.B. Arbuzov \textit{et al.},   \textit{J. Exp. Theor. Phys.} \textbf{81} (1995) 638 [\textit{Zh.
  Ehksp. Teor. Fiz.} \textbf{108} (1995) 1164].
  \href {http://arxiv.org/abs/hep-ph/9509405}
  {\path{arXiv:hep-ph/9509405}}.

\bibitem{Jadach:1992nk}
S.~Jadach \textit{et al.},   \textit{Phys. Rev.} \textbf{D47} (1993) 3733.
  \href {http://dx.doi.org/10.1103/PhysRevD.47.3733}
  {\path{doi:10.1103/PhysRevD.47.3733}}.

\bibitem{Jadach:1993wk}
S.~Jadach \textit{et al.},   \textit{Phys. Rev.} \textbf{D49} (1994) 1178.
  \href {http://dx.doi.org/10.1103/PhysRevD.49.1178}
  {\path{doi:10.1103/PhysRevD.49.1178}}.

\bibitem{Denner:2002cg}
A.~Denner \textit{et al.},  \textit{Comput. Phys.
  Commun.} \textbf{153} (2003) 462.
  \href {http://arxiv.org/abs/hep-ph/0209330}
  {\path{arXiv:hep-ph/0209330}}, \href
  {http://dx.doi.org/10.1016/S0010-4655(03)00205-4}
  {\path{doi:10.1016/S0010-4655(03)00205-4}}.

\bibitem{Berends:1987jm}
F.A. Berends \textit{et al.},  \textit{Nucl. Phys.} \textbf{B304} (1988) 712.
  \href {http://dx.doi.org/10.1016/0550-3213(88)90651-7}
  {\path{doi:10.1016/0550-3213(88)90651-7}}.

\bibitem{Beenakker:1990mb}
W.~Beenakker \textit{et al.},  \textit{Nucl. Phys.} \textbf{B349} (1991) 323.
  \href {http://dx.doi.org/10.1016/0550-3213(91)90328-U}
  {\path{doi:10.1016/0550-3213(91)90328-U}}.

\bibitem{Beenakker:1990es}
W.~Beenakker \textit{et al.},   \textit{Nucl. Phys.} \textbf{B355} (1991) 281.
  \href {http://dx.doi.org/10.1016/0550-3213(91)90114-D}
  {\path{doi:10.1016/0550-3213(91)90114-D}}.

\bibitem{Frixione:2002ik}
S.~Frixione and B.R. Webber,  \textit{J. High Energy Phys.} \textbf{06} (2002) 029.
  \href {http://arxiv.org/abs/hep-ph/0204244}
  {\path{arXiv:hep-ph/0204244}}, \href
  {http://dx.doi.org/10.1088/1126-6708/2002/06/029}
  {\path{doi:10.1088/1126-6708/2002/06/029}}.

\bibitem{Nason:2004rx}
P. Nason,  \textit{J. High Energy Phys.} \textbf{11} (2004) 040.
  \href {http://arxiv.org/abs/hep-ph/0409146}
  {\path{arXiv:hep-ph/0409146}}, \href
  {http://dx.doi.org/10.1088/1126-6708/2004/11/040}
  {\path{doi:10.1088/1126-6708/2004/11/040}}.

\bibitem{Jadach:2015mza}
S. Jadach \textit{et al.},   \textit{J. High Energy Phys.} \textbf{10} (2015)
  052.
  \href {http://arxiv.org/abs/1503.06849} {\path{arXiv:1503.06849}},
  \href {http://dx.doi.org/10.1007/JHEP10(2015)052}
  {\path{doi:10.1007/JHEP10(2015)052}}.

\bibitem{Arbuzov:1995id}
A.~Arbuzov  \textit{et al.},  
  \textit{Comput. Phys. Commun.} \textbf{94} (1996) 128.
  \href {http://arxiv.org/abs/hep-ph/9511434}
  {\path{arXiv:hep-ph/9511434}}, \href
  {http://dx.doi.org/10.1016/0010-4655(96)00005-7}
  {\path{doi:10.1016/0010-4655(96)00005-7}}.

\bibitem{Andonov:2004hi}
A.~Andonov \textit{et al.}, \textit{Comput.
  Phys. Commun.} \textbf{174} (2006) 481 [Erratum: \textit{Comput. Phys.
  Commun.} \textbf{177} (2007) 623].
  \href {http://arxiv.org/abs/hep-ph/0411186}
  {\path{arXiv:hep-ph/0411186}}, \href
  {http://dx.doi.org/10.1016/j.cpc.2005.12.006}
  {\path{doi:10.1016/j.cpc.2005.12.006}}.

\bibitem{Arbuzov:2015yja}
A.~Arbuzov \textit{et al.},   \textit{JETP Lett.} \textbf{103} (2016) 131.
  \href {http://arxiv.org/abs/1509.03052} {\path{arXiv:1509.03052}},
  \href {http://dx.doi.org/10.1134/S0021364016020041}
  {\path{doi:10.1134/S0021364016020041}}.

\bibitem{Richardson:2010gz}
P.~Richardson \textit{et al.},  \textit{J. High Energy Phys.} \textbf{06} (2012) 090.
  \href {http://arxiv.org/abs/1011.5444} {\path{arXiv:1011.5444}},
  \href {http://dx.doi.org/10.1007/JHEP06(2012)090}
  {\path{doi:10.1007/JHEP06(2012)090}}.

\bibitem{MoortgatPick:2005cw}
G.~Moortgat-Pick \textit{et al.},  \textit{Phys. Rep.} \textbf{460}
  (2008) 131.
  \href {http://arxiv.org/abs/hep-ph/0507011}
  {\path{arXiv:hep-ph/0507011}}, \href
  {http://dx.doi.org/10.1016/j.physrep.2007.12.003}
  {\path{doi:10.1016/j.physrep.2007.12.003}}.

\bibitem{Vega:1995cc}
R.~Vega and J.~Wudka, 
  \textit{Phys. Rev.} \textbf{D53} (1996) 5286 [Erratum: \textit{Phys. Rev.} \textbf{D56} (1997) 6037].
  \href {http://arxiv.org/abs/hep-ph/9511318}
  {\path{arXiv:hep-ph/9511318}}, \href
  {http://dx.doi.org/10.1103/PhysRevD.53.5286}
  {\path{doi:10.1103/PhysRevD.53.5286}}.

\bibitem{Ruijl:2017dtg}
B.~Ruijl \textit{et al.},  {FORM version 4.2,} \href
  {http://arxiv.org/abs/1707.06453} {\path{arXiv:1707.06453}}.

\bibitem{Hahn:1999mt}
T.~Hahn,  \textit{Acta Phys.
  Pol.} \textbf{B30} (1999) 3469.
  \href {http://arxiv.org/abs/hep-ph/9910227}
  {\path{arXiv:hep-ph/9910227}}.

\bibitem{Belanger:2003sd}
G.~Belanger \textit{et al.},  \textit{Phys.
  Rep.} \textbf{430} (2006) 117.
  \href {http://arxiv.org/abs/hep-ph/0308080}
  {\path{arXiv:hep-ph/0308080}}, \href
  {http://dx.doi.org/10.1016/j.physrep.2006.02.001}
  {\path{doi:10.1016/j.physrep.2006.02.001}}.

\bibitem{Bardin:2017mdd}
D.~Bardin \textit{et al.},  \textit{Phys. Rev.} \textbf{D98} (2018) 013001.
  \href {http://arxiv.org/abs/1801.00125} {\path{arXiv:1801.00125}},
  \href {http://dx.doi.org/10.1103/PhysRevD.98.013001}
  {\path{doi:10.1103/PhysRevD.98.013001}}.

\bibitem{Hollik:1981bu}
W.~Hollik and A.~Zepeda,  \textit{Z. Phys.} \textbf{C12} (1982) 67.
  \href {http://dx.doi.org/10.1007/BF01475733}
  {\path{doi:10.1007/BF01475733}}.

\bibitem{Hollik:1982wr}
W.~Hollik, \textit{Phys.
  Lett.} \textbf{123B} (1983) 259.
  \href {http://dx.doi.org/10.1016/0370-2693(83)90434-3}
  {\path{doi:10.1016/0370-2693(83)90434-3}}.

\bibitem{Arbuzov:1991pr}
A.~Arbuzov \textit{et al.},  \textit{Mod. Phys. Lett.} \textbf{A7} (1992) 2029
 [Erratum: \textit{Mod. Phys. Lett.} {\bf A9} (1994) 1515].
  \url{http://www--lib.kek.jp/cgi--bin/img\_index?9202059},
    \href {http://dx.doi.org/10.1142/S0217732392001762}
  {\path{doi:10.1142/S0217732392001762}}.

\bibitem{Djouadi:1994wt}
A.~Djouadi \textit{et al.},  \textit{Z. Phys.} \textbf{C67} (1995)
  123.
  \href {http://arxiv.org/abs/hep-ph/9411386}
  {\path{arXiv:hep-ph/9411386}}, \href {http://dx.doi.org/10.1007/BF01564827}
  {\path{doi:10.1007/BF01564827}}.

\bibitem{Bernreuther:2004th}
W.~Bernreuther \textit{et al.},  \textit{Nucl. Phys.} \textbf{B712} (2005) 229.
  \href {http://arxiv.org/abs/hep-ph/0412259}
  {\path{arXiv:hep-ph/0412259}}, \href
  {http://dx.doi.org/10.1016/j.nuclphysb.2005.01.035}
  {\path{doi:10.1016/j.nuclphysb.2005.01.035}}.

\bibitem{Bernreuther:2005rw}
W.~Bernreuther \textit{et al.},  \textit{Nucl. Phys.} \textbf{B723} (2005) 91.
  \href {http://arxiv.org/abs/hep-ph/0504190}
  {\path{arXiv:hep-ph/0504190}}, \href
  {http://dx.doi.org/10.1016/j.nuclphysb.2005.06.025}
  {\path{doi:10.1016/j.nuclphysb.2005.06.025}}.

\bibitem{Bernreuther:2005gw}
W.~Bernreuther \textit{et al.},  \textit{Phys. Rev.}
  \textbf{D72} (2005) 096002.
  \href {http://arxiv.org/abs/hep-ph/0508254}
  {\path{arXiv:hep-ph/0508254}}, \href
  {http://dx.doi.org/10.1103/PhysRevD.72.096002}
  {\path{doi:10.1103/PhysRevD.72.096002}}.

\bibitem{Bernreuther:2004ih}
W.~Bernreuther \textit{et al.},   \textit{Nucl. Phys.} \textbf{B706} (2005) 245.
  \href {http://arxiv.org/abs/hep-ph/0406046}
  {\path{arXiv:hep-ph/0406046}}, \href
  {http://dx.doi.org/10.1016/j.nuclphysb.2004.10.059}
  {\path{doi:10.1016/j.nuclphysb.2004.10.059}}.

\bibitem{Bernreuther:2005gq}
W.~Bernreuther \textit{et al.},   \textit{Phys. Rev. Lett.} \textbf{95} (2005) 261802.
  \href {http://arxiv.org/abs/hep-ph/0509341}
  {\path{arXiv:hep-ph/0509341}}, \href
  {http://dx.doi.org/10.1103/PhysRevLett.95.261802}
  {\path{doi:10.1103/PhysRevLett.95.261802}}.

\bibitem{Gluza:2009yy}
J.~Gluza \textit{et al.},  \textit{J. High Energy Phys.} \textbf{07} (2009) 001.
  \href {http://arxiv.org/abs/0905.1137} {\path{arXiv:0905.1137}},
  \href {http://dx.doi.org/10.1088/1126-6708/2009/07/001}
  {\path{doi:10.1088/1126-6708/2009/07/001}}.

\bibitem{Ablinger:2017hst}
J.~Ablinger \textit{et al.},   \textit{Phys. Rev.}
  \textbf{D97} (2018) 094022.
  \href {http://arxiv.org/abs/1712.09889} {\path{arXiv:1712.09889}},
  \href {http://dx.doi.org/10.1103/PhysRevD.97.094022}
  {\path{doi:10.1103/PhysRevD.97.094022}}.

\bibitem{Lee:2018nxa}
R.N. Lee \textit{et al.},   \textit{J. High Energy Phys.} \textbf{03} (2018) 136.
  \href {http://arxiv.org/abs/1801.08151} {\path{arXiv:1801.08151}},
  \href {http://dx.doi.org/10.1007/JHEP03(2018)136}
  {\path{doi:10.1007/JHEP03(2018)136}}.

\bibitem{Lee:2018rgs}
R.N. Lee \textit{et al.},  \textit{J. High Energy Phys.} \textbf{05} (2018) 187.
  \href {http://arxiv.org/abs/1804.07310} {\path{arXiv:1804.07310}},
  \href {http://dx.doi.org/10.1007/JHEP05(2018)187}
  {\path{doi:10.1007/JHEP05(2018)187}}.

\bibitem{Henn:2016tyf}
J.~Henn \textit{et al.},   \textit{J. High Energy Phys.} \textbf{01} (2017) 074.
  \href {http://arxiv.org/abs/1611.07535} {\path{arXiv:1611.07535}},
  \href {http://dx.doi.org/10.1007/JHEP01(2017)074}
  {\path{doi:10.1007/JHEP01(2017)074}}.

\bibitem{Ablinger:2018yae}
J.~Ablinger \textit{et al.},  \textit{Phys. Lett.} \textbf{B782} (2018)
  528.
  \href {http://arxiv.org/abs/1804.07313} {\path{arXiv:1804.07313}},
  \href {http://dx.doi.org/10.1016/j.physletb.2018.05.077}
  {\path{doi:10.1016/j.physletb.2018.05.077}}.

\bibitem{Henn:2016kjz}
J.M. Henn \textit{et al.},  \textit{J. High Energy Phys.} \textbf{12} (2016) 144.
  \href {http://arxiv.org/abs/1611.06523} {\path{arXiv:1611.06523}},
  \href {http://dx.doi.org/10.1007/JHEP12(2016)144}
  {\path{doi:10.1007/JHEP12(2016)144}}.

\bibitem{Blumlein:2018cms}
J.~Bl\"umlein and C.~Schneider,  \textit{Int. J. Mod. Phys.} \textbf{A33} (2018)
  1830015.
  \href {http://dx.doi.org/10.1142/S0217751X18300156}
  {\path{doi:10.1142/S0217751X18300156}}.

\bibitem{Grozin:2015kna}
A.~Grozin \textit{et al.},   \textit{J. High Energy Phys.} \textbf{01} (2016)
  140.
  \href {http://arxiv.org/abs/1510.07803} {\path{arXiv:1510.07803}},
  \href {http://dx.doi.org/10.1007/JHEP01(2016)140}
  {\path{doi:10.1007/JHEP01(2016)140}}.

\bibitem{Grozin:2014hna}
A.~Grozin \textit{et al.},   \textit{Phys. Rev. Lett.} \textbf{114} (2015) 062006.
  \href {http://arxiv.org/abs/1409.0023} {\path{arXiv:1409.0023}},
  \href {http://dx.doi.org/10.1103/PhysRevLett.114.062006}
  {\path{doi:10.1103/PhysRevLett.114.062006}}.

\bibitem{Mitov:2006xs}
A.~Mitov and S.~Moch,  \textit{J. High Energy Phys.} \textbf{05}
  (2007) 001.
  \href {http://arxiv.org/abs/hep-ph/0612149}
  {\path{arXiv:hep-ph/0612149}}, \href
  {http://dx.doi.org/10.1088/1126-6708/2007/05/001}
  {\path{doi:10.1088/1126-6708/2007/05/001}}.

\bibitem{Becher:2009kw}
T.~Becher and M.~Neubert,  \textit{Phys. Rev.} \textbf{D79} (2009) 125004 [Erratum: \textit{Phys.
  Rev.} \textbf{D80} (2009) 109901].
  \href {http://arxiv.org/abs/0904.1021} {\path{arXiv:0904.1021}},
  \href {http://dx.doi.org/10.1103/PhysRevD.79.125004}
  {\path{doi:10.1103/PhysRevD.79.125004}}.

\bibitem{Kramer:1986sg}
G.~Kramer and B.~Lampe,  \textit{Z. Phys.}
  \textbf{C34} (1987) 497 [Erratum: \textit{Z. Phys.} \textbf{C42} (1989) 504].
  \href {http://dx.doi.org/10.1007/BF01679868}
  {\path{doi:10.1007/BF01679868}}.

\bibitem{Matsuura:1988sm}
T.~Matsuura \textit{et al.}, 
  \textit{Nucl. Phys.} \textbf{B319} (1989) 570.
  \href {http://dx.doi.org/10.1016/0550-3213(89)90620-2}
  {\path{doi:10.1016/0550-3213(89)90620-2}}.

\bibitem{Matsuura:1987wt}
T.~Matsuura and W.L. van Neerven,  \textit{Z. Phys.} \textbf{C38} (1988) 623.
  \href {http://dx.doi.org/10.1007/BF01624369}
  {\path{doi:10.1007/BF01624369}}.

\bibitem{Harlander:2000mg}
R.V. Harlander, \textit{Phys. Lett.} \textbf{B492} (2000) 74.
  \href {http://arxiv.org/abs/hep-ph/0007289}
  {\path{arXiv:hep-ph/0007289}}, \href
  {http://dx.doi.org/10.1016/S0370-2693(00)01042-X}
  {\path{doi:10.1016/S0370-2693(00)01042-X}}.

\bibitem{Ravindran:2004mb}
V.~Ravindran \textit{et al.},  \textit{Nucl. Phys.} \textbf{B704} (2005) 332.
  \href {http://arxiv.org/abs/hep-ph/0408315}
  {\path{arXiv:hep-ph/0408315}}, \href
  {http://dx.doi.org/10.1016/j.nuclphysb.2004.10.039}
  {\path{doi:10.1016/j.nuclphysb.2004.10.039}}.

\bibitem{Gehrmann:2005pd}
T.~Gehrmann \textit{et al.},   \textit{Phys. Lett.} \textbf{B622} (2005) 295.
  \href {http://arxiv.org/abs/hep-ph/0507061}
  {\path{arXiv:hep-ph/0507061}}, \href
  {http://dx.doi.org/10.1016/j.physletb.2005.07.019}
  {\path{doi:10.1016/j.physletb.2005.07.019}}.

\bibitem{Baikov:2009bg}
P.A. Baikov \textit{et al.}, 
  \textit{Phys. Rev. Lett.} \textbf{102} (2009)
  212002.
  \href {http://arxiv.org/abs/0902.3519} {\path{arXiv:0902.3519}},
  \href {http://dx.doi.org/10.1103/PhysRevLett.102.212002}
  {\path{doi:10.1103/PhysRevLett.102.212002}}.

\bibitem{Gehrmann:2010ue}
T.~Gehrmann \textit{et al.},   \textit{J. High Energy Phys.} \textbf{06} (2010)
  094.
  \href {http://arxiv.org/abs/1004.3653} {\path{arXiv:1004.3653}},
  \href {http://dx.doi.org/10.1007/JHEP06(2010)094}
  {\path{doi:10.1007/JHEP06(2010)094}}.

\bibitem{Gehrmann:2010tu}
T.~Gehrmann \textit{et al.},  \textit{J. High Energy Phys.} \textbf{11}
  (2010) 102.
  \href {http://arxiv.org/abs/1010.4478} {\path{arXiv:1010.4478}},
  \href {http://dx.doi.org/10.1007/JHEP11(2010)102}
  {\path{doi:10.1007/JHEP11(2010)102}}.

\bibitem{Heinrich:2009be}
G.~Heinrich \textit{et al.},  \textit{Phys. Lett.} \textbf{B678} (2009)
  359.
  \href {http://arxiv.org/abs/0902.3512} {\path{arXiv:0902.3512}},
  \href {http://dx.doi.org/10.1016/j.physletb.2009.06.038}
  {\path{doi:10.1016/j.physletb.2009.06.038}}.

\bibitem{Gehrmann:2006wg}
T.~Gehrmann \textit{et al.},   \textit{Phys. Lett.} \textbf{B640}
  (2006) 252.
  \href {http://arxiv.org/abs/hep-ph/0607185}
  {\path{arXiv:hep-ph/0607185}}, \href
  {http://dx.doi.org/10.1016/j.physletb.2006.08.008}
  {\path{doi:10.1016/j.physletb.2006.08.008}}.

\bibitem{Heinrich:2007at}
G.~Heinrich \textit{et al.},  \textit{Phys. Lett.} \textbf{B662} (2008) 344.
  \href {http://arxiv.org/abs/0711.3590} {\path{arXiv:0711.3590}},
  \href {http://dx.doi.org/10.1016/j.physletb.2008.03.028}
  {\path{doi:10.1016/j.physletb.2008.03.028}}.

\bibitem{Catani:1998bh}
S.~Catani, 
  \textit{Phys. Lett.} \textbf{B427} (1998) 161.
  \href {http://arxiv.org/abs/hep-ph/9802439}
  {\path{arXiv:hep-ph/9802439}}, \href
  {http://dx.doi.org/10.1016/S0370-2693(98)00332-3}
  {\path{doi:10.1016/S0370-2693(98)00332-3}}.

\bibitem{Sterman:2002qn}
G.F. Sterman and M.E. Tejeda-Yeomans,
  \textit{Phys. Lett.} \textbf{B552} (2003) 48.
  \href {http://arxiv.org/abs/hep-ph/0210130}
  {\path{arXiv:hep-ph/0210130}}, \href
  {http://dx.doi.org/10.1016/S0370-2693(02)03100-3}
  {\path{doi:10.1016/S0370-2693(02)03100-3}}.

\bibitem{Becher:2009cu}
T.~Becher and M.~Neubert,  \textit{Phys. Rev. Lett}. \textbf{102} (2009) 162001 [Erratum: \textit{Phys. Rev.
  Lett.} \textbf{111} (2013) 199905].
  \href {http://arxiv.org/abs/0901.0722} {\path{arXiv:0901.0722}},
  \href {http://dx.doi.org/10.1103/PhysRevLett.102.162001}
  {\path{doi:10.1103/PhysRevLett.102.162001}}.

\bibitem{Gardi:2009qi}
E.~Gardi and L.~Magnea,  \textit{J. High Energy Phys.} \textbf{03} (2009) 079.
  \href {http://arxiv.org/abs/0901.1091} {\path{arXiv:0901.1091}},
  \href {http://dx.doi.org/10.1088/1126-6708/2009/03/079}
  {\path{doi:10.1088/1126-6708/2009/03/079}}.

\bibitem{Henn:2016men}
J.M. Henn \textit{et al.},   \textit{J. High Energy Phys.} \textbf{05} (2016) 066.
  \href {http://arxiv.org/abs/1604.03126} {\path{arXiv:1604.03126}},
  \href {http://dx.doi.org/10.1007/JHEP05(2016)066}
  {\path{doi:10.1007/JHEP05(2016)066}}.

\bibitem{Lee:2016ixa}
J.~Henn \textit{et al.},   \textit{J. High Energy Phys.} \textbf{03} (2017) 139.
  \href {http://arxiv.org/abs/1612.04389} {\path{arXiv:1612.04389}},
  \href {http://dx.doi.org/10.1007/JHEP03(2017)139}
  {\path{doi:10.1007/JHEP03(2017)139}}.

\bibitem{Lee:2017mip}
R.N. Lee \textit{et al.},   \textit{Phys. Rev.} \textbf{D96} (2017)
  014008.
  \href {http://arxiv.org/abs/1705.06862} {\path{arXiv:1705.06862}},
  \href {http://dx.doi.org/10.1103/PhysRevD.96.014008}
  {\path{doi:10.1103/PhysRevD.96.014008}}.

\bibitem{vonManteuffel:2016xki}
A.~von Manteuffel and R.M. Schabinger,  \textit{Phys. Rev.} \textbf{D95} (2017) 034030.
  \href {http://arxiv.org/abs/1611.00795} {\path{arXiv:1611.00795}},
  \href {http://dx.doi.org/10.1103/PhysRevD.95.034030}
  {\path{doi:10.1103/PhysRevD.95.034030}}.

\bibitem{Gnendiger:2017pys}
C.~Gnendiger \textit{et al.}, \textit{Eur. Phys. J.} \textbf{C77} (2017) 471.
  \href {http://arxiv.org/abs/1705.01827} {\path{arXiv:1705.01827}},
  \href {http://dx.doi.org/10.1140/epjc/s10052-017-5023-2}
  {\path{doi:10.1140/epjc/s10052-017-5023-2}}.

\bibitem{tHooft:1972tcz}
G.~'t~Hooft and M.J.G. Veltman,  \textit{Nucl. Phys.} \textbf{B44} (1972) 189.
  \href {http://dx.doi.org/10.1016/0550-3213(72)90279-9}
  {\path{doi:10.1016/0550-3213(72)90279-9}}.

\bibitem{Breitenlohner:1977hr}
P.~Breitenlohner and D.~Maison,  \textit{Commun. Math. Phys.} \textbf{52} (1977) 11.
  \href {http://dx.doi.org/10.1007/BF01609069}
  {\path{doi:10.1007/BF01609069}}.

\bibitem{Larin:1993tq}
S.A. Larin,  \textit{Phys. Lett.} \textbf{B303} (1993) 113.
  \href {http://arxiv.org/abs/hep-ph/9302240}
  {\path{arXiv:hep-ph/9302240}}, \href
  {http://dx.doi.org/10.1016/0370-2693(93)90053-K}
  {\path{doi:10.1016/0370-2693(93)90053-K}}.

\bibitem{Gnendiger:2017rfh}
C.~Gnendiger and A.~Signer, \textit{Phys. Rev.} \textbf{D97} (2018) 096006.
  \href {http://arxiv.org/abs/1710.09231} {\path{arXiv:1710.09231}},
  \href {http://dx.doi.org/10.1103/PhysRevD.97.096006}
  {\path{doi:10.1103/PhysRevD.97.096006}}.

\bibitem{Heinemeyer:2004yq}
S.~Heinemeyer \textit{et al.},  \textit{Nucl. Phys.} \textbf{B699} (2004) 103.
  \href {http://arxiv.org/abs/hep-ph/0405255}
  {\path{arXiv:hep-ph/0405255}}, \href
  {http://dx.doi.org/10.1016/j.nuclphysb.2004.08.014}
  {\path{doi:10.1016/j.nuclphysb.2004.08.014}}.

\bibitem{Freitas:2002ja}
A.~Freitas \textit{et al.},   \textit{Nucl. Phys.} \textbf{B632} (2002) 189 [Erratum: \textit{Nucl. Phys.} \textbf{{B666}} (2003) 305].
  \href{http://arxiv.org/pdf/hep--ph/0202131v4}{http://arxiv.org/pdf/hep--ph/0202131v4}.
  \href {http://dx.doi.org/10.1016/S0550-3213(02)00243-2}
  {\path{doi:10.1016/S0550-3213(02)00243-2}}.

\bibitem{Bednyakov:2015ooa}
A.V. Bednyakov and A.F. Pikelner,  \textit{Phys. Lett.} \textbf{B762} (2016) 151.
  \href {http://arxiv.org/abs/1508.02680} {\path{arXiv:1508.02680}},
  \href {http://dx.doi.org/10.1016/j.physletb.2016.09.007}
  {\path{doi:10.1016/j.physletb.2016.09.007}}.

\bibitem{Zoller:2015tha}
M.F. Zoller,  \textit{J. High Energy Phys.} \textbf{02} (2016) 095.
  \href {http://arxiv.org/abs/1508.03624} {\path{arXiv:1508.03624}},
  \href {http://dx.doi.org/10.1007/JHEP02(2016)095}
  {\path{doi:10.1007/JHEP02(2016)095}}.

\bibitem{Korner:1991sx}
J.G. Korner \textit{et al.},  \textit{Z. Phys.} \textbf{C54} (1992) 503.
  \href {http://dx.doi.org/10.1007/BF01559471}
  {\path{doi:10.1007/BF01559471}}.

\bibitem{Jegerlehner:2000dz}
F.~Jegerlehner,  \textit{Eur. Phys. J.} \textbf{C18} (2001)
  673.
  \href {http://arxiv.org/abs/hep-th/0005255}
  {\path{arXiv:hep-th/0005255}}, \href
  {http://dx.doi.org/10.1007/s100520100573} {\path{doi:10.1007/s100520100573}}.

\bibitem{Bruque:2018bmy}
A.M. Bruque \textit{et al.},   \textit{J. High Energy Phys.} \textbf{08} (2018)
  109.
  \href {http://arxiv.org/abs/1803.09764} {\path{arXiv:1803.09764}},
  \href {http://dx.doi.org/10.1007/JHEP08(2018)109}
  {\path{doi:10.1007/JHEP08(2018)109}}.

\bibitem{Fazio:2014xea}
R.A. Fazio \textit{et al.},  \textit{Eur.
  Phys. J.} \textbf{C74} (2014) 3197.
  \href {http://arxiv.org/abs/1404.4783} {\path{arXiv:1404.4783}},
  \href {http://dx.doi.org/10.1140/epjc/s10052-014-3197-4}
  {\path{doi:10.1140/epjc/s10052-014-3197-4}}.

\bibitem{Martin:1999cc}
C.P. Martin and D.~Sanchez-Ruiz,  \textit{Nucl.
  Phys.} \textbf{B572} (2000) 387.
  \href {http://arxiv.org/abs/hep-th/9905076}
  {\path{arXiv:hep-th/9905076}}, \href
  {http://dx.doi.org/10.1016/S0550-3213(99)00453-8}
  {\path{doi:10.1016/S0550-3213(99)00453-8}}.

\bibitem{Grassi:2001zz}
P.A. Grassi \textit{et al.},  \textit{Nucl. Phys.}
  \textbf{B610} (2001) 215.
  \href {http://arxiv.org/abs/hep-ph/0102005}
  {\path{arXiv:hep-ph/0102005}}, \href
  {http://dx.doi.org/10.1016/S0550-3213(01)00303-0}
  {\path{doi:10.1016/S0550-3213(01)00303-0}}.

\bibitem{Maldacena:1997re}
J.M. Maldacena,  \textit{Int. J. Theor. Phys.} \textbf{38} (1999) 1113 [\textit{Adv. Theor. Math.
  Phys.} \textbf{2} (1998) 231].
  \href {http://arxiv.org/abs/hep-th/9711200}
  {\path{arXiv:hep-th/9711200}}, \href
  {http://dx.doi.org/10.1023/A:1026654312961}
  {\path{doi:10.1023/A:1026654312961}}.

\bibitem{Boels:2017skl}
R.H. Boels \textit{et al.},   \textit{Phys. Rev. Lett.} \textbf{119} (2017)
  201601.
  \href {http://arxiv.org/abs/1705.03444} {\path{arXiv:1705.03444}},
  \href {http://dx.doi.org/10.1103/PhysRevLett.119.201601}
  {\path{doi:10.1103/PhysRevLett.119.201601}}.

\bibitem{Boels:2017ftb}
R.H. Boels \textit{et al.},   \textit{J. High Energy Phys.} \textbf{01} (2018) 153.
  \href {http://arxiv.org/abs/1711.08449} {\path{arXiv:1711.08449}},
  \href {http://dx.doi.org/10.1007/JHEP01(2018)153}
  {\path{doi:10.1007/JHEP01(2018)153}}.

\bibitem{Boels:2012ew}
R.H. Boels \textit{et al.},   \textit{J. High Energy Phys.} \textbf{02} (2013) 063.
  \href {http://arxiv.org/abs/1211.7028} {\path{arXiv:1211.7028}},
  \href {http://dx.doi.org/10.1007/JHEP02(2013)063}
  {\path{doi:10.1007/JHEP02(2013)063}}.

\bibitem{tHooft:1973alw}
G.~'t~Hooft,  \textit{Nucl. Phys.} \textbf{B72}
  (1974) 461.
  \href {http://dx.doi.org/10.1016/0550-3213(74)90154-0}
  {\path{doi:10.1016/0550-3213(74)90154-0}}.

\bibitem{Bern:2005iz}
Z.~Bern \textit{et al.},   \textit{Phys.
  Rev.} \textbf{D72} (2005) 085001.
  \href {http://arxiv.org/abs/hep-th/0505205}
  {\path{arXiv:hep-th/0505205}}, \href
  {http://dx.doi.org/10.1103/PhysRevD.72.085001}
  {\path{doi:10.1103/PhysRevD.72.085001}}.

\bibitem{Mueller:1979ih}
A.H. Mueller,  \textit{Phys.
  Rev.} \textbf{D20} (1979) 2037.
  \href {http://dx.doi.org/10.1103/PhysRevD.20.2037}
  {\path{doi:10.1103/PhysRevD.20.2037}}.

\bibitem{Collins:1980ih}
J.C. Collins, 
  \textit{Phys. Rev.} \textbf{D22} (1980) 1478.
  \href {http://dx.doi.org/10.1103/PhysRevD.22.1478}
  {\path{doi:10.1103/PhysRevD.22.1478}}.

\bibitem{Sen:1981sd}
A.~Sen, \textit{Phys. Rev.} \textbf{D24}
  (1981) 3281.
  \href {http://dx.doi.org/10.1103/PhysRevD.24.3281}
  {\path{doi:10.1103/PhysRevD.24.3281}}.

\bibitem{Magnea:1990zb}
L.~Magnea and G.~Sterman,  \textit{Phys. Rev.} \textbf{D42} (1990) 4222.
  \href {http://dx.doi.org/10.1103/PhysRevD.42.4222}
  {\path{doi:10.1103/PhysRevD.42.4222}}.

\bibitem{Korchemsky:1988si}
G.P. Korchemsky,  \textit{Mod. Phys. Lett.} \textbf{A4} (1989) 1257.
  \href {http://dx.doi.org/10.1142/S0217732389001453}
  {\path{doi:10.1142/S0217732389001453}}.

\bibitem{Kotikov:2000pm}
A.V. Kotikov and L.N. Lipatov,  \textit{Nucl. Phys.} \textbf{B582} (2000) 19.
  \href {http://arxiv.org/abs/hep-ph/0004008}
  {\path{arXiv:hep-ph/0004008}}, \href
  {http://dx.doi.org/10.1016/S0550-3213(00)00329-1}
  {\path{doi:10.1016/S0550-3213(00)00329-1}}.

\bibitem{Gubser:2002tv}
S.S. Gubser \textit{et al.},   \textit{Nucl. Phys.} \textbf{B636} (2002) 99.
  \href {http://arxiv.org/abs/hep-th/0204051}
  {\path{arXiv:hep-th/0204051}}, \href
  {http://dx.doi.org/10.1016/S0550-3213(02)00373-5}
  {\path{doi:10.1016/S0550-3213(02)00373-5}}.

\bibitem{Beisert:2010jr}
N.~Beisert \textit{et al.},  \textit{Lett. Math.
  Phys.} \textbf{99} (2012) 3.
  \href {http://arxiv.org/abs/1012.3982} {\path{arXiv:1012.3982}},
  \href {http://dx.doi.org/10.1007/s11005-011-0529-2}
  {\path{doi:10.1007/s11005-011-0529-2}}.

\bibitem{Bern:2006ew}
Z.~Bern \textit{et al.},   \textit{Phys. Rev.} \textbf{D75} (2007) 085010.
  \href {http://arxiv.org/abs/hep-th/0610248}
  {\path{arXiv:hep-th/0610248}}, \href
  {http://dx.doi.org/10.1103/PhysRevD.75.085010}
  {\path{doi:10.1103/PhysRevD.75.085010}}.

\bibitem{Cachazo:2006az}
F.~Cachazo \textit{et al.},   \textit{Phys. Rev.} \textbf{D75} (2007) 105011.
  \href {http://arxiv.org/abs/hep-th/0612309}
  {\path{arXiv:hep-th/0612309}}, \href
  {http://dx.doi.org/10.1103/PhysRevD.75.105011}
  {\path{doi:10.1103/PhysRevD.75.105011}}.

\bibitem{Henn:2013wfa}
J.M. Henn and T.~Huber,  \textit{J. High Energy Phys.} \textbf{09} (2013) 147.
  \href {http://arxiv.org/abs/1304.6418} {\path{arXiv:1304.6418}},
  \href {http://dx.doi.org/10.1007/JHEP09(2013)147}
  {\path{doi:10.1007/JHEP09(2013)147}}.

\bibitem{Grozin:2017css}
A.~Grozin \textit{et al.},  \textit{J. High Energy Phys.} \textbf{10} (2017) 052.
  \href {http://arxiv.org/abs/1708.01221} {\path{arXiv:1708.01221}},
  \href {http://dx.doi.org/10.1007/JHEP10(2017)052}
  {\path{doi:10.1007/JHEP10(2017)052}}.

\bibitem{Moch:2017uml}
S.~Moch \textit{et al.},   \textit{J. High Energy Phys.} \textbf{10}
  (2017) 041.
  \href {http://arxiv.org/abs/1707.08315} {\path{arXiv:1707.08315}},
  \href {http://dx.doi.org/10.1007/JHEP10(2017)041}
  {\path{doi:10.1007/JHEP10(2017)041}}.

\bibitem{Moch:2018wjh}
S.~Moch \textit{et al.},   \textit{Phys.
  Lett.} \textbf{B782} (2018) 627.
  \href {http://arxiv.org/abs/1805.09638} {\path{arXiv:1805.09638}},
  \href {http://dx.doi.org/10.1016/j.physletb.2018.06.017}
  {\path{doi:10.1016/j.physletb.2018.06.017}}.

\bibitem{Bruser:2018aud}
R.~Bruser \textit{et al.},  \textit{Proc. Sci.}  \textbf{LL2018} (2018) 018.
  \href {http://arxiv.org/abs/1807.05145} {\path{arXiv:1807.05145}},
  \href {http://dx.doi.org/10.22323/1.303.0018}
  {\path{doi:10.22323/1.303.0018}}.

\bibitem{Vogt:2018miu}
A.~Vogt \textit{et al.},  \textit{Proc. Sci.}   \textbf{LL2018} (2018) 050.
  \href {http://arxiv.org/abs/1808.08981} {\path{arXiv:1808.08981}},
  \href {http://dx.doi.org/10.22323/1.303.0050}
  {\path{doi:10.22323/1.303.0050}}.

\bibitem{vanNeerven:1985ja}
W.L. van Neerven,  \textit{Z. Phys.} \textbf{C30} (1986) 595.
  \href {http://dx.doi.org/10.1007/BF01571808}
  {\path{doi:10.1007/BF01571808}}.

\bibitem{Lee:2010cga}
R.N. Lee \textit{et al.},  \textit{J. High Energy Phys.} \textbf{04} (2010) 020.
  \href {http://arxiv.org/abs/1001.2887} {\path{arXiv:1001.2887}},
  \href {http://dx.doi.org/10.1007/JHEP04(2010)020}
  {\path{doi:10.1007/JHEP04(2010)020}}.

\bibitem{vonManteuffel:2015gxa}
A.~von Manteuffel \textit{et al.},   \textit{Phys. Rev.} \textbf{D93} 
  (2016) 125014.
  \href {http://arxiv.org/abs/1510.06758} {\path{arXiv:1510.06758}},
  \href {http://dx.doi.org/10.1103/PhysRevD.93.125014}
  {\path{doi:10.1103/PhysRevD.93.125014}}.

\bibitem{Ahmed:2017gyt}
T.~Ahmed \textit{et al.}, 
  \textit{J. High Energy Phys.} \textbf{06} (2017) 125.
  \href {http://arxiv.org/abs/1704.07846} {\path{arXiv:1704.07846}},
  \href {http://dx.doi.org/10.1007/JHEP06(2017)125}
  {\path{doi:10.1007/JHEP06(2017)125}}.

\bibitem{Davies:2016jie}
J.~Davies \textit{et al.},   \textit{Nucl. Phys.} \textbf{B915}
  (2017) 335.
  \href {http://arxiv.org/abs/1610.07477} {\path{arXiv:1610.07477}},
  \href {http://dx.doi.org/10.1016/j.nuclphysb.2016.12.012}
  {\path{doi:10.1016/j.nuclphysb.2016.12.012}}.

\bibitem{Dixon:2017nat}
L.J. Dixon,  \textit{J. High Energy Phys.} \textbf{01} (2018) 075.
  \href {http://arxiv.org/abs/1712.07274} {\path{arXiv:1712.07274}},
  \href {http://dx.doi.org/10.1007/JHEP01(2018)075}
  {\path{doi:10.1007/JHEP01(2018)075}}.

\bibitem{Grozin:2018vdn}
A.~Grozin,  \textit{J. High Energy Phys.} \textbf{06} (2018) 073.
  \href {http://arxiv.org/abs/1805.05050} {\path{arXiv:1805.05050}},
  \href {http://dx.doi.org/10.1007/JHEP06(2018)073}
  {\path{doi:10.1007/JHEP06(2018)073}}.

\bibitem{Becher:2009qa}
T.~Becher and M.~Neubert,  \textit{J. High Energy Phys.} \textbf{06} (2009) 081 [Erratum: \textit{J. High Energy Phys.} \textbf{11} (2013) 024].
  \href {http://arxiv.org/abs/0903.1126} {\path{arXiv:0903.1126}},
  \href {http://dx.doi.org/10.1088/1126-6708/2009/06/081}
  {\path{doi:10.1088/1126-6708/2009/06/081}}.

\bibitem{Anzai:2010td}
C.~Anzai \textit{et al.},   \textit{Nucl. Phys.} \textbf{B838} (2010) 28 [Erratum:
  \textit{Nucl. Phys.} \textbf{B890} (2015) 569].
  \href {http://arxiv.org/abs/1004.1562} {\path{arXiv:1004.1562}},
  \href {http://dx.doi.org/10.1016/j.nuclphysb.2010.05.012}
  {\path{doi:10.1016/j.nuclphysb.2010.05.012}}.

\bibitem{Dixon:2009gx}
L.J. Dixon,  \textit{Phys. Rev.} \textbf{D79} (2009) 091501.
  \href {http://arxiv.org/abs/0901.3414} {\path{arXiv:0901.3414}},
  \href {http://dx.doi.org/10.1103/PhysRevD.79.091501}
  {\path{doi:10.1103/PhysRevD.79.091501}}.

\bibitem{Dixon:2009ur}
L.J. Dixon \textit{et al.},   \textit{J. High Energy Phys.} \textbf{02} (2010) 081.
  \href {http://arxiv.org/abs/0910.3653} {\path{arXiv:0910.3653}},
  \href {http://dx.doi.org/10.1007/JHEP02(2010)081}
  {\path{doi:10.1007/JHEP02(2010)081}}.

\bibitem{Ahrens:2012qz}
V.~Ahrens \textit{et al.},   \textit{J. High Energy Phys.} \textbf{09} (2012) 138.
  \href {http://arxiv.org/abs/1208.4847} {\path{arXiv:1208.4847}},
  \href {http://dx.doi.org/10.1007/JHEP09(2012)138}
  {\path{doi:10.1007/JHEP09(2012)138}}.

\bibitem{Korchemsky:2017ttd}
G.P. Korchemsky,  \textit{J. High Energy Phys.} \textbf{12} (2017) 093.
  \href {http://arxiv.org/abs/1704.00448} {\path{arXiv:1704.00448}},
  \href {http://dx.doi.org/10.1007/JHEP12(2017)093}
  {\path{doi:10.1007/JHEP12(2017)093}}.

\bibitem{Bern:2008qj}
Z.~Bern \textit{et al.},  \textit{Phys. Rev.} \textbf{D78} (2008) 085011.
  \href {http://arxiv.org/abs/0805.3993} {\path{arXiv:0805.3993}},
  \href {http://dx.doi.org/10.1103/PhysRevD.78.085011}
  {\path{doi:10.1103/PhysRevD.78.085011}}.

\bibitem{Bern:2010ue}
Z.~Bern \textit{et al.},  \textit{Phys. Rev. Lett.} \textbf{105} (2010) 061602.
  \href {http://arxiv.org/abs/1004.0476} {\path{arXiv:1004.0476}},
  \href {http://dx.doi.org/10.1103/PhysRevLett.105.061602}
  {\path{doi:10.1103/PhysRevLett.105.061602}}.

\bibitem{Carrasco:2015iwa}
J.J.M. Carrasco, {Gauge and gravity amplitude relations}, {Proc.   Theoretical Advanced Study Institute in Elementary Particle Physics: Journeys
  Through the Precision Frontier: Amplitudes for Colliders (TASI 2014):
  Boulder, CO, USA,  2014}, (World Scientific, Singapore, 2015), p. 477.
  \href {http://arxiv.org/abs/1506.00974} {\path{arXiv:1506.00974}},
  \href {http://dx.doi.org/10.1142/9789814678766_0011}
  {\path{doi:10.1142/9789814678766_0011}}.

\bibitem{Yang:2016ear}
G.~Yang,  \textit{Phys. Rev. Lett.} \textbf{117}  (2016)
  271602.
  \href {http://arxiv.org/abs/1610.02394} {\path{arXiv:1610.02394}},
  \href {http://dx.doi.org/10.1103/PhysRevLett.117.271602}
  {\path{doi:10.1103/PhysRevLett.117.271602}}.

\bibitem{Chetyrkin:1981qh}
K.~Chetyrkin and F.~Tkachov,  \textit{Nucl. Phys.} \textbf{B192} (1981) 159.
  \href {http://dx.doi.org/10.1016/0550-3213(81)90199-1}
  {\path{doi:10.1016/0550-3213(81)90199-1}}.

\bibitem{Tkachov:1981wb}
F.V. Tkachov,  \textit{Phys. Lett.} \textbf{100B} (1981) 65.
  \href {http://dx.doi.org/10.1016/0370-2693(81)90288-4}
  {\path{doi:10.1016/0370-2693(81)90288-4}}.

\bibitem{Boels:2015yna}
R.~Boels \textit{et al.},   \textit{Nucl. Phys.} \textbf{B902} (2016) 387.
  \href {http://arxiv.org/abs/1508.03717} {\path{arXiv:1508.03717}},
  \href {http://dx.doi.org/10.1016/j.nuclphysb.2015.11.016}
  {\path{doi:10.1016/j.nuclphysb.2015.11.016}}.

\bibitem{vonManteuffel:2012np}
A.~von Manteuffel and C.~Studerus, {Reduze 2 -- distributed Feynman integral
  reduction,} \href {http://arxiv.org/abs/1201.4330} {\path{arXiv:1201.4330}}.

\bibitem{Anastasiou:2004vj}
C.~Anastasiou and A.~Lazopoulos,  \textit{J. High Energy Phys.} \textbf{07} (2004) 046.
  \href {http://arxiv.org/abs/hep-ph/0404258}
  {\path{arXiv:hep-ph/0404258}}, \href
  {http://dx.doi.org/10.1088/1126-6708/2004/07/046}
  {\path{doi:10.1088/1126-6708/2004/07/046}}.

\bibitem{Smirnov:2008iw}
A.V. Smirnov, \textit{J. High Energy Phys.} \textbf{10} (2008)
  107.
  \href {http://arxiv.org/abs/0807.3243} {\path{arXiv:0807.3243}},
  \href {http://dx.doi.org/10.1088/1126-6708/2008/10/107}
  {\path{doi:10.1088/1126-6708/2008/10/107}}.

\bibitem{Smirnov:2013dia}
A.V. Smirnov and V.A. Smirnov,  \textit{Comput. Phys. Commun.} \textbf{184} (2013) 2820.
  \href {http://arxiv.org/abs/1302.5885} {\path{arXiv:1302.5885}},
  \href {http://dx.doi.org/10.1016/j.cpc.2013.06.016}
  {\path{doi:10.1016/j.cpc.2013.06.016}}.

\bibitem{Smirnov:2014hma}
A.V. Smirnov, 
  \textit{Comput. Phys. Commun.} \textbf{189} (2015) 182.
  \href {http://arxiv.org/abs/1408.2372} {\path{arXiv:1408.2372}},
  \href {http://dx.doi.org/10.1016/j.cpc.2014.11.024}
  {\path{doi:10.1016/j.cpc.2014.11.024}}.

\bibitem{2010CoPhC.181.1293S}
C.~{Studerus}, \textit{Comput. Phys.
  Commun.} \textbf{181} (2010) 1293.
  \href {http://arxiv.org/abs/0912.2546} {\path{arXiv:0912.2546}},
  \href {http://dx.doi.org/10.1016/j.cpc.2010.03.012}
  {\path{doi:10.1016/j.cpc.2010.03.012}}.

\bibitem{Laporta:2001dd}
S.~Laporta,  \textit{Int. J. Mod. Phys.} \textbf{A15} (2000) 5087.
  \href {http://arxiv.org/abs/hep-ph/0102033}
  {\path{arXiv:hep-ph/0102033}}, \href
  {http://dx.doi.org/10.1016/S0217-751X(00)00215-7}
  {\path{doi:10.1016/S0217-751X(00)00215-7}}.

\bibitem{Lee:2012cn}
R.N. Lee, {Presenting LiteRed: a tool for the loop integrals reduction,}
\href   {http://arxiv.org/abs/1212.2685} {\path{arXiv:1212.2685}}.

\bibitem{Lee:2013mka}
R.N. Lee, 
  \textit{J. Phys. Conf. Ser.} \textbf{523} (2014) 012059.
  \href {http://arxiv.org/abs/1310.1145} {\path{arXiv:1310.1145}},
  \href {http://dx.doi.org/10.1088/1742-6596/523/1/012059}
  {\path{doi:10.1088/1742-6596/523/1/012059}}.

\bibitem{Blumlein:2009cf}
J. Bl\"umlein \textit{et al.},  \textit{Comput. Phys. Commun.} \textbf{181} (2010) 582.
  \href {http://arxiv.org/abs/0907.2557} {\path{arXiv:0907.2557}},
  \href {http://dx.doi.org/10.1016/j.cpc.2009.11.007}
  {\path{doi:10.1016/j.cpc.2009.11.007}}.

\bibitem{Kotikov:2002ab}
A.V. Kotikov and L.N. Lipatov,  \textit{Nucl. Phys.} \textbf{B661} (2003) 19 [Erratum: \textit{Nucl.
  Phys.} \textbf{B685} (2004) 405].
  \href {http://arxiv.org/abs/hep-ph/0208220}
  {\path{arXiv:hep-ph/0208220}}, \href
  {http://dx.doi.org/10.1016/S0550-3213(03)00264-5}
  {\path{doi:10.1016/S0550-3213(03)00264-5}}.

\bibitem{Kotikov:2004er}
A.V. Kotikov \textit{et al.},   \textit{Phys. Lett.} \textbf{B595} (2004) 521 [Erratum: \textit{Phys.
  Lett.} \textbf{B632} (2006) 754].
  \href {http://arxiv.org/abs/hep-th/0404092}
  {\path{arXiv:hep-th/0404092}}, \href
  {http://dx.doi.org/10.1016/j.physletb.2004.05.078}
  {\path{doi:10.1016/j.physletb.2004.05.078}}.

\bibitem{Gehrmann:2011xn}
T.~Gehrmann \textit{et al.},  \textit{J. High Energy Phys.} \textbf{03} (2012) 101.
  \href {http://arxiv.org/abs/1112.4524} {\path{arXiv:1112.4524}},
  \href {http://dx.doi.org/10.1007/JHEP03(2012)101}
  {\path{doi:10.1007/JHEP03(2012)101}}.

\bibitem{Arkani-Hamed:2014via}
N.~Arkani-Hamed \textit{et al.},  \textit{Phys. Rev. Lett.} \textbf{113} 
  (2014) 261603.
  \href {http://arxiv.org/abs/1410.0354} {\path{arXiv:1410.0354}},
  \href {http://dx.doi.org/10.1103/PhysRevLett.113.261603}
  {\path{doi:10.1103/PhysRevLett.113.261603}}.

\bibitem{Bern:2014kca}
Z.~Bern \textit{et al.},   \textit{J. High Energy Phys.} \textbf{06} (2015) 202.
  \href {http://arxiv.org/abs/1412.8584} {\path{arXiv:1412.8584}},
  \href {http://dx.doi.org/10.1007/JHEP06(2015)202}
  {\path{doi:10.1007/JHEP06(2015)202}}.

\bibitem{Henn:2013pwa}
J.M. Henn, 
  \textit{Phys. Rev. Lett.} \textbf{110} (2013) 251601.
  \href {http://arxiv.org/abs/1304.1806} {\path{arXiv:1304.1806}},
  \href {http://dx.doi.org/10.1103/PhysRevLett.110.251601}
  {\path{doi:10.1103/PhysRevLett.110.251601}}.

\bibitem{Bern:2015ple}
Z.~Bern \textit{et al.},  \textit{J. High Energy Phys.} \textbf{06} (2016) 098.
  \href {http://arxiv.org/abs/1512.08591} {\path{arXiv:1512.08591}},
  \href {http://dx.doi.org/10.1007/JHEP06(2016)098}
  {\path{doi:10.1007/JHEP06(2016)098}}.

\bibitem{Henn:2013nsa}
J.M. Henn \textit{et al.},  \textit{J. High Energy Phys.} \textbf{03} (2014) 088.
  \href {http://arxiv.org/abs/1312.2588} {\path{arXiv:1312.2588}},
  \href {http://dx.doi.org/10.1007/JHEP03(2014)088}
  {\path{doi:10.1007/JHEP03(2014)088}}.

\bibitem{Boels:2016bdu}
R.H. Boels \textit{et al.}, \textit{Proc. Sci.}
  \textbf{LL2016} (2016) 039.
  \href {http://arxiv.org/abs/1607.00172} {\path{arXiv:1607.00172}},
  \href {http://dx.doi.org/10.22323/1.260.0039}
  {\path{doi:10.22323/1.260.0039}}.

\bibitem{Smirnov:1999gc}
V.A. Smirnov,  \textit{Phys. Lett.} \textbf{B460} (1999) 397.
  \href {http://arxiv.org/abs/hep-ph/9905323}
  {\path{arXiv:hep-ph/9905323}}, \href
  {http://dx.doi.org/10.1016/S0370-2693(99)00777-7}
  {\path{doi:10.1016/S0370-2693(99)00777-7}}.

\bibitem{Tausk:1999vh}
J.~Tausk,  \textit{Phys. Lett.} \textbf{B469} (1999) 225.
  \href {http://arxiv.org/abs/hep-ph/9909506}
  {\path{arXiv:hep-ph/9909506}}, \href
  {http://dx.doi.org/10.1016/S0370-2693(99)01277-0}
  {\path{doi:10.1016/S0370-2693(99)01277-0}}.

\bibitem{Anastasiou:2005cb}
C.~Anastasiou and A.~Daleo,  \textit{J. High Energy Phys.} \textbf{0610}
  (2006) 031.
  \href {http://arxiv.org/abs/hep-ph/0511176}
  {\path{arXiv:hep-ph/0511176}}, \href
  {http://dx.doi.org/10.1088/1126-6708/2006/10/031}
  {\path{doi:10.1088/1126-6708/2006/10/031}}.

\bibitem{Binoth:2000ps}
T.~Binoth and G.~Heinrich,  \textit{Nucl. Phys.} \textbf{B585} (2000) 741.
  \href {http://arxiv.org/abs/hep-ph/0004013}
  {\path{arXiv:hep-ph/0004013}}, \href
  {http://dx.doi.org/10.1016/S0550-3213(00)00429-6}
  {\path{doi:10.1016/S0550-3213(00)00429-6}}.

\bibitem{Heinrich:2008si}
G.~Heinrich, \textit{Int. J. Mod. Phys.} \textbf{A23} (2008) 1457.
  \href {http://arxiv.org/abs/0803.4177} {\path{arXiv:0803.4177}},
  \href {http://dx.doi.org/10.1142/S0217751X08040263}
  {\path{doi:10.1142/S0217751X08040263}}.

\bibitem{Smirnov:2008py}
A.V. Smirnov and M.N. Tentyukov,  \textit{Comput. Phys. Commun.} \textbf{180} (2009) 735.
  \href {http://arxiv.org/abs/0807.4129} {\path{arXiv:0807.4129}},
  \href {http://dx.doi.org/10.1016/j.cpc.2008.11.006}
  {\path{doi:10.1016/j.cpc.2008.11.006}}.

\bibitem{Smirnov:2009pb}
A.V. Smirnov \textit{et al.},  \textit{Comput. Phys. Commun.} \textbf{182} (2011) 790.
  \href {http://arxiv.org/abs/0912.0158} {\path{arXiv:0912.0158}},
  \href {http://dx.doi.org/10.1016/j.cpc.2010.11.025}
  {\path{doi:10.1016/j.cpc.2010.11.025}}.

\bibitem{Smirnov:2015mct}
A.V. Smirnov,  \textit{Comput. Phys. Commun.} \textbf{204} (2016) 189.
  \href {http://arxiv.org/abs/1511.03614} {\path{arXiv:1511.03614}},
  \href {http://dx.doi.org/10.1016/j.cpc.2016.03.013}
  {\path{doi:10.1016/j.cpc.2016.03.013}}.

\bibitem{Carter:2010hi}
J.~Carter and G.~Heinrich, 
  \textit{Comput. Phys. Commun.} \textbf{182} (2011) 1566.
  \href {http://arxiv.org/abs/1011.5493} {\path{arXiv:1011.5493}},
  \href {http://dx.doi.org/10.1016/j.cpc.2011.03.026}
  {\path{doi:10.1016/j.cpc.2011.03.026}}.

\bibitem{Borowka:2012yc}
S.~Borowka \textit{et al.},   \textit{Comput. Phys. Commun.}
  \textbf{184} (2013) 396.
  \href {http://arxiv.org/abs/1204.4152} {\path{arXiv:1204.4152}},
  \href {http://dx.doi.org/10.1016/j.cpc.2012.09.020}
  {\path{doi:10.1016/j.cpc.2012.09.020}}.

\bibitem{Smirnov:2004ym}
V.~Smirnov, \textit{Evaluating {Feynman} {Integrals}} (Springer, Berlin,
  2004).
  \href {http://dx.doi.org/10.1007/b95498} {\path{doi:10.1007/b95498}}.

\bibitem{Smirnov:2006ry}
V.~Smirnov, \textit{Feynman Integral Calculus} (Springer, Berlin, 2006).
  \href {http://dx.doi.org/10.1007/3-540-30611-0}
  {\path{doi:10.1007/3-540-30611-0}}.

\bibitem{Czakon:2005rk}
M.~Czakon,   \textit{Comput. Phys. Commun.} \textbf{175} (2006) 559.
  \href {http://arxiv.org/abs/hep-ph/0511200}
  {\path{arXiv:hep-ph/0511200}}, \href
  {http://dx.doi.org/10.1016/j.cpc.2006.07.002}
  {\path{doi:10.1016/j.cpc.2006.07.002}}.

\bibitem{Smirnov:2009up}
A.~Smirnov and V.~Smirnov,  \textit{Eur. Phys. J.} \textbf{C62} (2009) 445.
  \href {http://arxiv.org/abs/0901.0386} {\path{arXiv:0901.0386}},
  \href {http://dx.doi.org/10.1140/epjc/s10052-009-1039-6}
  {\path{doi:10.1140/epjc/s10052-009-1039-6}}.

\bibitem{Hahn:2004fe}
T.~Hahn,  \textit{Comput.
  Phys. Commun}. \textbf{168} (2005) 78.
  \href {http://arxiv.org/abs/hep-ph/0404043}
  {\path{arXiv:hep-ph/0404043}}, \href
  {http://dx.doi.org/10.1016/j.cpc.2005.01.010}
  {\path{doi:10.1016/j.cpc.2005.01.010}}.

\bibitem{Marquard:2016dcn}
P.~Marquard \textit{et al.}, 
   \textit{Phys. Rev.} \textbf{D94} (2016) 074025.
  \href {http://arxiv.org/abs/1606.06754} {\path{arXiv:1606.06754}},
  \href {http://dx.doi.org/10.1103/PhysRevD.94.074025}
  {\path{doi:10.1103/PhysRevD.94.074025}}.

\bibitem{Ferguson:1999:API:307090.307114}
H.R.P. Ferguson \textit{et al.},   \textit{Math. Comput.} \textbf{68} (1999) 351.
  \href {http://dx.doi.org/10.1090/S0025-5718-99-00995-3}
  {\path{doi:10.1090/S0025-5718-99-00995-3}}.

\bibitem{Schabinger:2018dyi}
R.M. Schabinger, {Constructing multi-loop scattering amplitudes with manifest
  singularity structure}, \href {http://arxiv.org/abs/1806.05682}
  {\path{arXiv:1806.05682}}.

\bibitem{Henn:Jan2018}
J.~Henn, Bootstrapping pentagon functions, Mini~Workshop: {Precision EW and QCD
  Calculations for the FCC Studies: Methods and Tools},  2018
  (CERN, Geneva, Switzerland),
  \url{https://indico.cern.ch/event/669224/contributions/2805490/attachments/1582262/2500779/henn_cern_jan_2018.pdf}.

\bibitem{Borwein:2004}
J.M. Borwein \textit{et al.}, \textit{Experimentation in Mathematics:
  Computational Paths to Discovery} (CRC Press, Boca Raton, 2004).
  \href {http://dx.doi.org/10.1.1.145.7158}
  {\path{doi:10.1.1.145.7158}}.

\bibitem{math3020337}
D.H. Bailey and J.M. Borwein,
   \textit{Mathematics} \textbf{3} (2015) 337.
  \href {http://dx.doi.org/10.3390/math3020337}
  {\path{doi:10.3390/math3020337}}.
\url{http://www.mdpi.com/2227-7390/3/2/337}

\bibitem{Ita:Jan2018}
H.~Ita, Expectations from current multi-loop computations with the unitarity
  method, Mini~Workshop: {Precision EW and QCD
  Calculations for the FCC Studies: Methods and Tools},  2018
  (CERN, Geneva, Switzerland),
  \url{https://indico.cern.ch/event/669224/contributions/2805502/attachments/1582263/2500780/Ita_FCC_ee_QCD.pdf}.

\bibitem{Page:2018flm}
B.~Page \textit{et al.},  \textit{Proc.
  Sci.}  \textbf{RADCOR2017}
  (2018) 012.
  \href {http://dx.doi.org/10.22323/1.290.0012}
  {\path{doi:10.22323/1.290.0012}}.

\bibitem{Abreu:2018gii}
S.~Abreu \textit{et al.}, \textit{Proc. Sci.}  \textbf{LL2018} (2018) 016.
  \href {http://arxiv.org/abs/1807.09447} {\path{arXiv:1807.09447}},
  \href {http://dx.doi.org/10.22323/1.303.0016}
  {\path{doi:10.22323/1.303.0016}}.

\bibitem{Smirnov:Jan2018}
V.~Smirnov, A mini review of methods of evaluating Feynman integrals. Solving
  differential equations for Feynman integrals by expansions near singular
  points, Mini~Workshop: {Precision EW and QCD
  Calculations for the FCC Studies: Methods and Tools},  2018
  (CERN, Geneva, Switzerland),
  \url{https://indico.cern.ch/event/669224/contributions/2805433/attachments/1581916/2500083/smirnov.pdf}.

\bibitem{Gituliar:Jan2018}
O.~Gituliar, Fuchsia and differential equations for multi-scale master
  integrals, Mini~Workshop: {Precision EW and QCD
  Calculations for the FCC Studies: Methods and Tools},  2018
  (CERN, Geneva, Switzerland),
  \url{https://indico.cern.ch/event/669224/contributions/2805514/attachments/1582273/2500797/gituliar_fcc.pdf}.

\bibitem{LL2018}
J.~Bl\"umlein and P.~Marquard, {Loops and legs in quantum field theory},  Sankt Goar, Germany, 2018,
  \url{https://indico.desy.de/indico/event/16613/}.

\bibitem{LoopFest2018}
J.~Huston \textit{et al.}, {LoopFest 2018},  \url{https://web.pa.msu.edu/people/huston/LoopFest2018/}.

\bibitem{Bogner:2010kv}
C.~Bogner and S.~Weinzierl,  \textit{Int. J. Mod. Phys.} \textbf{A25}
  (2010) 2585.
  \href {http://arxiv.org/abs/1002.3458} {\path{arXiv:1002.3458}},
  \href {http://dx.doi.org/10.1142/S0217751X10049438}
  {\path{doi:10.1142/S0217751X10049438}}.

\bibitem{Usovitsch:2018shx}
J.~Usovitsch \textit{et al.}, \textit{Proc. Sci.}  \textbf{LL2018} (2018) 046.
  \href {http://arxiv.org/abs/1810.04580} {\path{arXiv:1810.04580}},
  \href {http://dx.doi.org/10.22323/1.303.0046}
  {\path{doi:10.22323/1.303.0046}}.

\bibitem{Usovitsch:PhD2018}
J.~Usovitsch, Ph.D. thesis, Humboldt-Universit{\"a}t zu Berlin, 
 2018.
  \href {http://dx.doi.org/10.3204/PUBDB-2018-05160}
  {\path{doi:10.3204/PUBDB-2018-05160}}.

\bibitem{Heinrich:2004iq}
G.~Heinrich and V.A. Smirnov,  \textit{Phys. Lett.} \textbf{B598} (2004) 55.
  \href {http://arxiv.org/abs/hep-ph/0406053}
  {\path{arXiv:hep-ph/0406053}}, \href
  {http://dx.doi.org/10.1016/j.physletb.2004.07.058}
  {\path{doi:10.1016/j.physletb.2004.07.058}}.

\bibitem{Prausa:2017frh}
M.~Prausa, \textit{Eur. Phys. J.} \textbf{C77} 
  (2017) 594.
  \href {http://arxiv.org/abs/1706.09852} {\path{arXiv:1706.09852}},
  \href {http://dx.doi.org/10.1140/epjc/s10052-017-5150-9}
  {\path{doi:10.1140/epjc/s10052-017-5150-9}}.

\bibitem{Hepp:1966eg}
K.~Hepp,   \textit{Commun. Math. Phys.} \textbf{2} (1966) 301,
  \url{http://www.projecteuclid.org/euclid.cmp/1103815087}.
  \href {http://dx.doi.org/10.1007/BF01773358}
  {\path{doi:10.1007/BF01773358}}.

\bibitem{Binoth:2003ak}
T.~Binoth and G.~Heinrich,  \textit{Nucl. Phys.} \textbf{B680} (2004) 375.
  \href {http://arxiv.org/abs/hep-ph/0305234}
  {\path{arXiv:hep-ph/0305234}}, \href
  {http://dx.doi.org/10.1016/j.nuclphysb.2003.12.023}
  {\path{doi:10.1016/j.nuclphysb.2003.12.023}}.

\bibitem{Binoth:2004jv}
T.~Binoth and G.~Heinrich,  \textit{Nucl. Phys.} \textbf{B693} (2004) 134.
  \href {http://arxiv.org/abs/hep-ph/0402265}
  {\path{arXiv:hep-ph/0402265}}, \href
  {http://dx.doi.org/10.1016/j.nuclphysb.2004.06.005}
  {\path{doi:10.1016/j.nuclphysb.2004.06.005}}.

\bibitem{Denner:2004iz}
A.~Denner and S.~Pozzorini,  \textit{Nucl. Phys.}
  \textbf{B717} (2005) 48.
  \href {http://arxiv.org/abs/hep-ph/0408068}
  {\path{arXiv:hep-ph/0408068}}, \href
  {http://dx.doi.org/10.1016/j.nuclphysb.2005.03.036}
  {\path{doi:10.1016/j.nuclphysb.2005.03.036}}.

\bibitem{ambrewww}
 \url{http://prac.us.edu.pl/~gluza/ambre}, last accessed May 20th 2019.

\bibitem{Gluza:200704v1}
K.~Kajda, {AMBRE} 1.0 (April 2007), a Mathematica package representing Feynman
  integrals by Mellin--Barnes integrals, 
  \url{http://prac.us.edu.pl/~gluza/ambre/}.

\bibitem{Gluza:200704v12}
K.~Kajda, {AMBRE} 1.2 (9 April 2008), a Mathematica package representing
  Feynman integrals by Mellin--Barnes integrals, 
  \url{http://prac.us.edu.pl/~gluza/ambre/}.

\bibitem{Gluza:200704v13}
K.~Kajda, {AMBRE} 1.3 (24 Aug 2011), a Mathematica package representing Feynman
  integrals by Mellin--Barnes integrals, 
  \url{http://prac.us.edu.pl/~gluza/ambre/}.

\bibitem{Gluza:2010v20}
K.~Kajda, {AMBRE} 2.0 (18 June 2010), a Mathematica package representing
  Feynman integrals by Mellin--Barnes integrals, 
  \url{http://prac.us.edu.pl/~gluza/ambre/}.

\bibitem{Gluza:2010v21}
K.~Kajda, {AMBRE} 2.1 (12 Sep 2010), a Mathematica package representing Feynman
  integrals by Mellin--Barnes integrals, 
  \url{http://prac.us.edu.pl/~gluza/ambre/}.

\bibitem{Gluza:2010v22}
K.~Kajda and I.~Dubovyk, {AMBRE} 2.2 (12 Sep 2015), a Mathematica package
  representing Feynman integrals by Mellin--Barnes integrals, 
  \url{http://prac.us.edu.pl/~gluza/ambre/}.

\bibitem{Dubovyk:201509v30x}
I.~Dubovyk, {AMBRE} 3.0 (1 Sep 2015), a Mathematica package representing
  Feynman integrals by Mellin--Barnes integrals, 
  \url{http://prac.us.edu.pl/~gluza/ambre/}.

\bibitem{Gluza:2009mj}
J.~Gluza \textit{et al.}, \textit{Proc. Sci.}  \textbf{ACAT08} (2008) 124.
  \href {http://arxiv.org/abs/0902.4830} {\path{arXiv:0902.4830}},
  \href {http://dx.doi.org/10.22323/1.070.0124}
  {\path{doi:10.22323/1.070.0124}}.

\bibitem{Gluza:2010mz}
J.~Gluza \textit{et al.},  \textit{Nucl.
  Phys. Proc. Suppl.} \textbf{205--206} (2010) 147.
  \href {http://arxiv.org/abs/1006.4728} {\path{arXiv:1006.4728}},
  \href {http://dx.doi.org/10.1016/j.nuclphysbps.2010.08.034}
  {\path{doi:10.1016/j.nuclphysbps.2010.08.034}}.

\bibitem{mbtoolsMBsuite}
 \url{https://mbtools.hepforge.org/}, last accessed May 20th 2019.

\bibitem{Freitas:2010nx}
A.~Freitas and Y.-C. Huang,  \textit{J. High Energy Phys.} \textbf{2010} (2010) 074.
  \href {http://arxiv.org/abs/1001.3243} {\path{arXiv:1001.3243}},
  \href {http://dx.doi.org/10.1007/JHEP04(2010)074}
  {\path{doi:10.1007/JHEP04(2010)074}}.

\bibitem{Gluza:2016fwh}
J.~Gluza \textit{et al.},  \textit{Phys. Rev.} \textbf{D95} (2017) 076016.
  \href {http://arxiv.org/abs/1609.09111} {\path{arXiv:1609.09111}},
  \href {http://dx.doi.org/10.1103/PhysRevD.95.076016}
  {\path{doi:10.1103/PhysRevD.95.076016}}.

\bibitem{Peng:2012zpa}
Z.~Peng,  Ph.D. thesis, IPhT, Saclay, 2012. \url{http://tel.archives-ouvertes.fr/tel-00834200}

\bibitem{Usovitsch:Jan2018}
J.~Usovitsch \textit{et al.}, {MB-suite 2: MBnumerics news}, Mini~Workshop: {Precision EW and QCD
  Calculations for the FCC Studies: Methods and Tools},  2018
  (CERN, Geneva, Switzerland),
  \url{https://indico.cern.ch/event/669224/contributions/2805454/attachments/1581984/2500208/Usovitsch.pdf}.

\bibitem{Bluemlein:2017rbi}
J.~Bl{\"u}mlein \textit{et al.},  \textit{Acta Phys. Pol.} \textbf{B48}
  (2017) 2313.
  \href {http://arxiv.org/abs/1711.05510} {\path{arXiv:1711.05510}},
  \href {http://dx.doi.org/10.5506/APhysPolB.48.2313}
  {\path{doi:10.5506/APhysPolB.48.2313}}.

\bibitem{Borowka:2016ehy}
S.~Borowka \textit{et al.},   \textit{Phys. Rev. Lett.}
  \textbf{117}  (2016) 012001 [Erratum: \textit{Phys. Rev. Lett.} \textbf{117} (2016) 079901].
  \href {http://arxiv.org/abs/1604.06447} {\path{arXiv:1604.06447}},
  \href {http://dx.doi.org/10.1103/PhysRevLett.117.079901}
  {\path{doi:10.1103/PhysRevLett.117.079901}}.

\bibitem{Borowka:2016ypz}
S.~Borowka \textit{et al.},   \textit{J. High Energy Phys.} \textbf{10} (2016) 107.
  \href {http://arxiv.org/abs/1608.04798} {\path{arXiv:1608.04798}},
  \href {http://dx.doi.org/10.1007/JHEP10(2016)107}
  {\path{doi:10.1007/JHEP10(2016)107}}.

\bibitem{Grober:2017uho}
R.~Grober \textit{et al.},   \textit{J. High Energy Phys.} \textbf{03} (2018) 020.
  \href {http://arxiv.org/abs/1709.07799} {\path{arXiv:1709.07799}},
  \href {http://dx.doi.org/10.1007/JHEP03(2018)020}
  {\path{doi:10.1007/JHEP03(2018)020}}.

\bibitem{SABRY1962401}
A.~Sabry,  \textit{Nucl.
  Phys.} \textbf{33} (1962) 401.
  \href {http://dx.doi.org/10.1016/0029-5582(62)90535-7}
  {\path{doi:10.1016/0029-5582(62)90535-7}}.

\bibitem{Aglietti:2007as}
U.~Aglietti \textit{et al.},  \textit{Nucl. Phys.} \textbf{B789} (2008) 45.
  \href {http://arxiv.org/abs/0705.2616} {\path{arXiv:0705.2616}},
  \href {http://dx.doi.org/10.1016/j.nuclphysb.2007.07.019}
  {\path{doi:10.1016/j.nuclphysb.2007.07.019}}.

\bibitem{vonManteuffel:2017hms}
A.~von Manteuffel and L.~Tancredi,  \textit{J. High Energy Phys.} \textbf{06} (2017) 127.
  \href {http://arxiv.org/abs/1701.05905} {\path{arXiv:1701.05905}},
  \href {http://dx.doi.org/10.1007/JHEP06(2017)127}
  {\path{doi:10.1007/JHEP06(2017)127}}.

\bibitem{CaronHuot:2012ab}
S.~Caron-Huot and K.J. Larsen,  \textit{J. High Energy Phys.} \textbf{10}
  (2012) 026.
  \href {http://arxiv.org/abs/1205.0801} {\path{arXiv:1205.0801}},
  \href {http://dx.doi.org/10.1007/JHEP10(2012)026}
  {\path{doi:10.1007/JHEP10(2012)026}}.

\bibitem{Broedel:2017kkb}
J.~Broedel \textit{et al.},  \textit{J. High Energy Phys.} \textbf{05}
  (2018) 093.
  \href {http://arxiv.org/abs/1712.07089} {\path{arXiv:1712.07089}},
  \href {http://dx.doi.org/10.1007/JHEP05(2018)093}
  {\path{doi:10.1007/JHEP05(2018)093}}.

\bibitem{Borowka:2013cma}
S.~Borowka and G.~Heinrich,  \textit{Comput. Phys. Commun.} \textbf{184} (2013) 2552.
  \href {http://arxiv.org/abs/1303.1157} {\path{arXiv:1303.1157}},
  \href {http://dx.doi.org/10.1016/j.cpc.2013.05.022}
  {\path{doi:10.1016/j.cpc.2013.05.022}}.

\bibitem{Borowka:2017idc}
S.~Borowka \textit{et al.}, 
    \textit{Comput. Phys. Commun.} \textbf{222} (2018) 313.
  \href {http://arxiv.org/abs/1703.09692} {\path{arXiv:1703.09692}},
  \href {http://dx.doi.org/10.1016/j.cpc.2017.09.015}
  {\path{doi:10.1016/j.cpc.2017.09.015}}.

\bibitem{Borowka:2018dsa}
S.~Borowka \textit{et al.},   \textit{J. High Energy Phys.} \textbf{08} (2018) 111.
  \href {http://arxiv.org/abs/1804.06824} {\path{arXiv:1804.06824}},
  \href {http://dx.doi.org/10.1007/JHEP08(2018)111}
  {\path{doi:10.1007/JHEP08(2018)111}}.

\bibitem{Bogner:2017xhp}
C.~Bogner \textit{et al.},  \textit{Comput. Phys. Commun.} \textbf{225} (2018) 1.
  \href {http://arxiv.org/abs/1709.01266} {\path{arXiv:1709.01266}},
  \href {http://dx.doi.org/10.1016/j.cpc.2017.12.017}
  {\path{doi:10.1016/j.cpc.2017.12.017}}.

\bibitem{Roth:1996pd}
M.~Roth and A.~Denner, 
  \textit{Nucl. Phys.} \textbf{B479} (1996) 495.
  \href {http://arxiv.org/abs/hep-ph/9605420}
  {\path{arXiv:hep-ph/9605420}}, \href
  {http://dx.doi.org/10.1016/0550-3213(96)00435-X}
  {\path{doi:10.1016/0550-3213(96)00435-X}}.

\bibitem{Cheng:1987ga}
H.~Cheng and T.~Wu, \textit{Expanding Protons: Scattering at High Energies} (MIT
  Press, Cambridge, MA, 1987).

\bibitem{Kaneko:2009qx}
T.~Kaneko and T.~Ueda,  \textit{Comput. Phys.
  Commun.} \textbf{181} (2010) 1352.
  \href {http://arxiv.org/abs/0908.2897} {\path{arXiv:0908.2897}},
  \href {http://dx.doi.org/10.1016/j.cpc.2010.04.001}
  {\path{doi:10.1016/j.cpc.2010.04.001}}.

\bibitem{Kaneko:2010kj}
T.~Kaneko and T.~Ueda, \textit{Proc. Sci.} 
  \textbf{ACAT2010} (2010) 082.
  \href {http://arxiv.org/abs/1004.5490} {\path{arXiv:1004.5490}},
  \href {http://dx.doi.org/10.22323/1.093.0082}
  {\path{doi:10.22323/1.093.0082}}.

\bibitem{Schlenk:2016a}
J.~Schlenk,  Ph.D. thesis, Technische Universit\"at M\"unchen, 2016.

\bibitem{Bogner:2007cr}
C.~Bogner and S.~Weinzierl, \textit{Comput. Phys. Commun.} \textbf{178} (2008) 596.
  \href {http://arxiv.org/abs/0709.4092} {\path{arXiv:0709.4092}},
  \href {http://dx.doi.org/10.1016/j.cpc.2007.11.012}
  {\path{doi:10.1016/j.cpc.2007.11.012}}.

\bibitem{Bogner:2008ry}
C.~Bogner and S.~Weinzierl,  \textit{Nucl. Phys. Proc.
  Suppl.} \textbf{183} (2008) 256.
  \href {http://arxiv.org/abs/0806.4307} {\path{arXiv:0806.4307}},
  \href {http://dx.doi.org/10.1016/j.nuclphysbps.2008.09.113}
  {\path{doi:10.1016/j.nuclphysbps.2008.09.113}}.

\bibitem{Smirnov:2008aw}
A.~Smirnov and V.~Smirnov,  \textit{J. High Energy Phys.} \textbf{0905} (2009) 004.
  \href {http://arxiv.org/abs/0812.4700} {\path{arXiv:0812.4700}},
  \href {http://dx.doi.org/10.1088/1126-6708/2009/05/004}
  {\path{doi:10.1088/1126-6708/2009/05/004}}.

\bibitem{Vermaseren:2000nd}
J.~Vermaseren, {New features of FORM}, \href
  {http://arxiv.org/abs/math-ph/0010025} {\path{arXiv:math-ph/0010025}}.

\bibitem{Kuipers:2013pba}
J.~Kuipers \textit{et al.},   \textit{Comput.
  Phys. Commun.} \textbf{189} (2015) 1.
  \href {http://arxiv.org/abs/1310.7007} {\path{arXiv:1310.7007}},
  \href {http://dx.doi.org/10.1016/j.cpc.2014.08.008}
  {\path{doi:10.1016/j.cpc.2014.08.008}}.

\bibitem{Agrawal:2011tm}
S.~Agrawal \textit{et al.},   \textit{J. Phys. Conf. Ser.} \textbf{368} (2012)
  012054.
  \href {http://arxiv.org/abs/1112.0124} {\path{arXiv:1112.0124}},
  \href {http://dx.doi.org/10.1088/1742-6596/368/1/012054}
  {\path{doi:10.1088/1742-6596/368/1/012054}}.

\bibitem{Hahn:2014fua}
T.~Hahn,  \textit{J. Phys. Conf. Ser.} \textbf{608} (2015) 012066.
  \href {http://arxiv.org/abs/1408.6373} {\path{arXiv:1408.6373}},
  \href {http://dx.doi.org/10.1088/1742-6596/608/1/012066}
  {\path{doi:10.1088/1742-6596/608/1/012066}}.

\bibitem{Gough:2009:GSL:1538674}
M.~Galassi \textit{et al.},  \textit{GNUScientific Library Reference Manual}, 3rd
  ed. (Network Theory Ltd., Boston,  2009).

\bibitem{2013arXiv1301.1493M}
B.D. McKay and A. Piperno, Practical graph isomorphism, II, \href {http://arxiv.org/abs/1301.1493} {\path{arXiv:1301.1493}}.

\bibitem{graphviz}
\url{http://www.graphviz.org}, last accessed May 21st 2019.

\bibitem{Borowka:2017esm}
S.~Borowka \textit{et al.},    \textit{Acta Phys. Pol.   Supp.} \textbf{11} (2018) 375.
  \href {http://arxiv.org/abs/1712.05755} {\path{arXiv:1712.05755}},
  \href {http://dx.doi.org/10.5506/APhysPolBSupp.11.375}
  {\path{doi:10.5506/APhysPolBSupp.11.375}}.

\bibitem{Panzer:2014gra}
E.~Panzer,  \textit{J. High Energy Phys.} \textbf{03} (2014) 071.
  \href {http://arxiv.org/abs/1401.4361} {\path{arXiv:1401.4361}},
  \href {http://dx.doi.org/10.1007/JHEP03(2014)071}
  {\path{doi:10.1007/JHEP03(2014)071}}.

\bibitem{vonManteuffel:2014qoa}
A.~von Manteuffel \textit{et al.}, \textit{J. High Energy Phys.} \textbf{02} (2015) 120.
  \href {http://arxiv.org/abs/1411.7392} {\path{arXiv:1411.7392}},
  \href {http://dx.doi.org/10.1007/JHEP02(2015)120}
  {\path{doi:10.1007/JHEP02(2015)120}}.

\bibitem{Bauer:2000cp}
C.W. Bauer \textit{et al.}, \textit{J. Symb. Comput.}
  \textbf{33} (2000) 1.
  \href {http://arxiv.org/abs/cs/0004015} {\path{arXiv:cs/0004015}},
  \href {http://dx.doi.org/10.1006/jsco.2001.0494}
  {\path{doi:10.1006/jsco.2001.0494}}.

\bibitem{Wolfram}
S.~Wolfram, \textit{The Mathematica Book} (Wolfram  Media, Champaign, IL,
  2003).

\bibitem{Gluza:2007bd}
J.~Gluza \textit{et al.},  \textit{Proc. Sci.}  \textbf{ACAT} (2007) 081.
  \href {http://arxiv.org/abs/0707.3567} {\path{arXiv:0707.3567}},
  \href {http://dx.doi.org/10.22323/1.050.0081}
  {\path{doi:10.22323/1.050.0081}}.

\bibitem{Dubovyk:2016zok}
I.~Dubovyk \textit{et al.},  \textit{Proc. Sci.}  \textbf{LL2016} (2016) 075.
  \href {http://arxiv.org/abs/1610.07059} {\path{arXiv:1610.07059}},
  \href {http://dx.doi.org/10.22323/1.260.0075}
  {\path{doi:10.22323/1.260.0075}}.

\bibitem{Czakon:2007wk}
M.~Czakon \textit{et al.},   \textit{Nucl. Phys.} \textbf{B798} (2008) 210.
  \href {http://arxiv.org/abs/0707.4139} {\path{arXiv:0707.4139}},
  \href {http://dx.doi.org/10.1016/j.nuclphysb.2008.02.001}
  {\path{doi:10.1016/j.nuclphysb.2008.02.001}}.

\bibitem{mbtools-kosower}
D. Kosower, Mathematica program barnesroutines.m version 1.1.1 (July 23, 2009),
    \url{https://mbtools.hepforge.org/}.

\bibitem{mbtools}
 \url{https://mbtools.hepforge.org/}, last accessed May 21st 2019.

\bibitem{Riemann:April2018}
T.~Riemann \textit{et al.},  {Scalar} one-loop Feynman integrals in arbitrary space-time
  dimension,  14th Workshop `Loops and Legs in Quantum Field Theory' LL2018,  Sankt Goar, Germany, 2018.
  \url{
  https://indico.desy.de/indico/event/16613/session/12/contribution/24/material/slides/0.pdf}.

\bibitem{Cvitanovic:1974uf}
P.~Cvitanovic and T.~Kinoshita,  \textit{Phys.
  Rev.} \textbf{D10} (1974) 3978.
  \href {http://dx.doi.org/10.1103/PhysRevD.10.3978}
  {\path{doi:10.1103/PhysRevD.10.3978}}.

\bibitem{Smirnov:2002}
V.~Smirnov, \textit{Applied Asymptotic Expansions in Momenta and Masses} (Springer, Berlin, 2002).
  \href {http://dx.doi.org/10.1007/3-540-44574-9}
  {\path{doi:10.1007/3-540-44574-9}}.

\bibitem{Blumlein:2014maa}
J.~Bl\"umlein \textit{et al.}, \textit{Proc. Sci.}  \textbf{LL2014} (2014) 052.
  \href {http://arxiv.org/abs/1407.7832} {\path{arXiv:1407.7832}},
  \href {http://dx.doi.org/10.22323/1.211.0052}
  {\path{doi:10.22323/1.211.0052}}.

\bibitem{MBasymptotics}
M.~Czakon and A. Smirnov, MBasymptotics.m, \url{https://mbtools.hepforge.org}.

\bibitem{Ochman:2015fho}
M.~Ochman and T.~Riemann,  \textit{Acta Phys. Pol.} \textbf{B46}
  (2015) 2117.
  \href {http://arxiv.org/abs/1511.01323} {\path{arXiv:1511.01323}},
  \href {http://dx.doi.org/10.5506/APhysPolB.46.2117}
  {\path{doi:10.5506/APhysPolB.46.2117}}.

\bibitem{Gonzalez:2007ry}
I.~Gonzalez and I.~Schmidt,  \textit{Nucl. Phys.} \textbf{B769} (2007)
  124.
  \href {http://arxiv.org/abs/hep-th/0702218}
  {\path{arXiv:hep-th/0702218}}, \href
  {http://dx.doi.org/10.1016/j.nuclphysb.2007.01.031}
  {\path{doi:10.1016/j.nuclphysb.2007.01.031}}.

\bibitem{Gonzalez:2010}
I.~Gonzalez and V.H. Moll,  \textit{Adv. Appl. Math.} \textbf{45} (2010) 50.
  \href {http://arxiv.org/abs/0812.3356} {\path{arXiv:0812.3356}},
  \href {http://dx.doi.org/http://dx.doi.org/10.1016/j.aam.2009.11.003}
  {\path{doi:http://dx.doi.org/10.1016/j.aam.2009.11.003}}.

\bibitem{Gonzalez:2010uz}
I.~Gonzalez,  \textit{Nucl. Phys.
  Proc. Suppl.} \textbf{205--206} (2010) 141.
  \href {http://arxiv.org/abs/1008.2148} {\path{arXiv:1008.2148}},
  \href {http://dx.doi.org/10.1016/j.nuclphysbps.2010.08.033}
  {\path{doi:10.1016/j.nuclphysbps.2010.08.033}}.

\bibitem{Fleischer:2003rm}
J.~Fleischer \textit{et al.},   \textit{Nucl. Phys.} \textbf{B672} (2003)
  303.
  \href {http://arxiv.org/abs/hep-ph/0307113}
  {\path{arXiv:hep-ph/0307113}}, \href
  {http://dx.doi.org/10.1016/j.nuclphysb.2003.09.004}
  {\path{doi:10.1016/j.nuclphysb.2003.09.004}}.

\bibitem{Bern:1992em}
Z.~Bern \textit{et al.},   \textit{Phys. Lett.} \textbf{B302} (1993) 299 [Erratum: \textit{Phys.
  Lett.} \textbf{B318} (1993) 649].
  \href {http://arxiv.org/abs/hep-ph/9212308}
  {\path{arXiv:hep-ph/9212308}}, \href
  {http://dx.doi.org/10.1016/0370-2693(93)90469-X}
  {\path{doi:10.1016/0370-2693(93)90469-X}}.

\bibitem{Tarasov:1996br}
O.~Tarasov,  \textit{Phys. Rev.} \textbf{D54} (1996) 6479.
  \href {http://arxiv.org/abs/hep-th/9606018}
  {\path{arXiv:hep-th/9606018}}, \href
  {http://dx.doi.org/10.1103/PhysRevD.54.6479}
  {\path{doi:10.1103/PhysRevD.54.6479}}.

\bibitem{Davydychev:1991va}
A.I. Davydychev,  \textit{Phys. Lett.} \textbf{B263} (1991) 107.
  \href {http://dx.doi.org/10.1016/0370-2693(91)91715-8}
  {\path{doi:10.1016/0370-2693(91)91715-8}}.

\bibitem{Fleischer:1999hq}
J.~Fleischer \textit{et al.},   \textit{Nucl. Phys.} \textbf{B566} (2000) 423.
  \href {http://arxiv.org/abs/hep-ph/9907327}
  {\path{arXiv:hep-ph/9907327}}, \href
  {http://dx.doi.org/10.1016/S0550-3213(99)00678-1}
  {\path{doi:10.1016/S0550-3213(99)00678-1}}.

\bibitem{Maierhoefer:2017hyi}
P.~Maierh{\"o}fer \textit{et al.},  \textit{Comput. Phys. Commun.} \textbf{230} (2018) 99.
  \href {http://arxiv.org/abs/1705.05610} {\path{arXiv:1705.05610}},
  \href {http://dx.doi.org/10.1016/j.cpc.2018.04.012}
  {\path{doi:10.1016/j.cpc.2018.04.012}}.

\bibitem{Melrose:1965kb}
D.B. Melrose,  \textit{Nuovo Cim.} \textbf{40} (1965)
  181.
  \href {http://dx.doi.org/10.1007/BF028329}
  {\path{doi:10.1007/BF028329}}.

\bibitem{Barnes:zbMATH02640947}
E.W. Barnes,  \textit{Proc. London Math. Soc.} \textbf{S2--6} (1908) 141.
  \href {http://dx.doi.org/10.1112/plms/s2-6.1.141}
  {\path{doi:10.1112/plms/s2-6.1.141}}.

\bibitem{Bernshtein1971-from-springer}
I.~Bernshtein, \textit{Funct. Anal. Appl.} \textbf{5} (1971) 89. \href
  {http://dx.doi.org/10.1007/BF01076413} {\path{doi:10.1007/BF01076413}}.

\bibitem{Golubeva:1978}
V.A. Golubeva and V.Z. Enol'skii,  \textit{Math. Notes 
  Acad. Sci. USSR} \textbf{23} (1978) 63.  
  \href{http://link.springer.com/article/10.1007/BF01104888}
  {\path{doi:10.1007/BF01104888}}, 
  \url{http://www.mathnet.ru/php/getFT.phtml?jrnid=mzm&paperid=8124&what=fullt&option_lang=rus}.

\bibitem{Devaraj:1997es}
G.~Devaraj and R.G. Stuart,  \textit{Nucl. Phys.} \textbf{B519} (1998) 483.
  \href {http://arxiv.org/abs/hep-ph/9704308}
  {\path{arXiv:hep-ph/9704308}}, \href
  {http://dx.doi.org/10.1016/S0550-3213(98)00035-2}
  {\path{doi:10.1016/S0550-3213(98)00035-2}}.

\bibitem{Passarino:1978jh}
G.~Passarino and M.~Veltman,  \textit{Nucl. Phys.} \textbf{B160} (1979) 151.
  \href {http://dx.doi.org/10.1016/0550-3213(79)90234-7}
  {\path{doi:10.1016/0550-3213(79)90234-7}}.

\bibitem{Fleischer:2011bi}
J.~Fleischer \textit{et al.},  \textit{J. Phys. Conf. Ser.} \textbf{368} (2012) 012057.
  \href {http://arxiv.org/abs/1112.0500} {\path{arXiv:1112.0500}},
  \href {http://dx.doi.org/10.1088/1742-6596/368/1/012057}
  {\path{doi:10.1088/1742-6596/368/1/012057}}.

\bibitem{Yundin-phd:2012oai}
V. Yundin,  Ph.D thesis,
  Humboldt-Universit{\"a}t zu Berlin, 2012.
  \href{http://edoc.hu-berlin.de/dissertationen/yundin-valery-2012-02-01/PDF/yundin.pdf}{
  http://edoc.hu-berlin.de/dissertationen/yundin-valery-2012-02-01/PDF/yundin.pdf}.

\bibitem{Hahn:1998yk}
T.~Hahn and M.~Perez-Victoria,  \textit{Comput. Phys. Commun.} \textbf{118} (1999) 153.
  \href {http://arxiv.org/abs/hep-ph/9807565}
  {\path{arXiv:hep-ph/9807565}}, \href
  {http://dx.doi.org/10.1016/S0010-4655(98)00173-8}
  {\path{doi:10.1016/S0010-4655(98)00173-8}}.

\bibitem{planarity}
K.~Bielas and I.~Dubovyk, {P}lanarityTest.m, a Mathematica package for testing the
  planarity of Feynman diagrams,
  \url{http://us.edu.pl/~gluza/ambre/planarity/}.

\bibitem{wong2001asymptotic}
R. Wong, \textit{Asymptotic Approximations of Integrals: Classics in Applied
  Mathematics} (Society for Industrial and Applied Mathematics, Philadelphia,
  PA, 2001).
  \href {http://dx.doi.org/10.1137/1.9780898719260}
  {\path{doi:10.1137/1.9780898719260}}.

\bibitem{temme2014asymptotic}
N.M. Temme, \textit{Asymptotic Methods for Integrals} (World Scientific,
Singapore, 2014).
  \href {http://dx.doi.org/10.1142/9195} {\path{doi:10.1142/9195}}.

\bibitem{Sidorov:2017aea}
A.V. Sidorov \textit{et al.},   \textit{Phys. Rev.} \textbf{D97}  (2018) 076009.
  \href {http://arxiv.org/abs/1712.05601} {\path{arXiv:1712.05601}},
  \href {http://dx.doi.org/10.1103/PhysRevD.97.076009}
  {\path{doi:10.1103/PhysRevD.97.076009}}.

\bibitem{axler2001harmonic}
S. Axler \textit{et al.},  \textit{Harmonic Function Theory}
  (Springer, New York, NY, 2001).
  \href {http://dx.doi.org/10.1007/978-1-4757-8137-3}
  {\path{doi:10.1007/978-1-4757-8137-3}}.

\bibitem{Kaminski}
D.~Kaminski,  \textit{Methods  Appl. Anal.} \textbf{01} (1994) 44. \href
  {http://dx.doi.org/10.4310/MAA.1994.v1.n1.a4}
  {\path{doi:10.4310/MAA.1994.v1.n1.a4}}.

\bibitem{Pham1}
F.~Pham,  \textit{Proc.
  Symp. Pure Math.}  \textbf{40} (1983) 319.

\bibitem{Pham2}
F.~Pham, in \textit{Syst\`emes
  Diff\'erentiels et Singularit\'es}, Eds. A. Galligo \textit{et al.},  (Soci\'et\'e math\'ematique de France, Paris, 1985), p. 11. \url{http://www.numdam.org/item/AST_1985__130__11_0/}
 
\bibitem{Delabaere_globalasymptotics}
E.~Delabaere and C.J.  \textit{Duke Math. J.} \textbf{112} (2002) 199.
  \href {http://dx.doi.org/10.1215/S0012-9074-02-11221-6}
  {\path{doi:10.1215/S0012-9074-02-11221-6}}.

\bibitem{Kosower:1997hg}
D.A. Kosower,  \textit{Nucl. Phys.} \textbf{B506} (1997)
  439.
  \href {http://arxiv.org/abs/hep-ph/9706213}
  {\path{arXiv:hep-ph/9706213}}, \href
  {http://dx.doi.org/10.1016/S0550-3213(97)00526-9}
  {\path{doi:10.1016/S0550-3213(97)00526-9}}.

\bibitem{Witten:2010zr}
E.~Witten, {A new look at the path integral of quantum mechanics}, \href
  {http://arxiv.org/abs/1009.6032} {\path{arXiv:1009.6032}}.

\bibitem{Witten:2010cx}
E.~Witten,  \textit{AMS/IP Stud. Adv. Math.} \textbf{50} (2011) 347.
  \href {http://arxiv.org/abs/1001.2933} {\path{arXiv:1001.2933}}.
\url{https://www.ams.org/books/amsip/050/}

\bibitem{Cristoforetti:2012su}
M.~Cristoforetti \textit{et al.}, \textit{Phys.
  Rev.} \textbf{D86} (2012) 074506.
  \href {http://arxiv.org/abs/1205.3996} {\path{arXiv:1205.3996}},
  \href {http://dx.doi.org/10.1103/PhysRevD.86.074506}
  {\path{doi:10.1103/PhysRevD.86.074506}}.

\bibitem{Fujii:2013sra}
H.~Fujii \textit{et al.},   \textit{J. High Energy Phys.} \textbf{10}
  (2013) 147.
  \href {http://arxiv.org/abs/1309.4371} {\path{arXiv:1309.4371}},
  \href {http://dx.doi.org/10.1007/JHEP10(2013)147}
  {\path{doi:10.1007/JHEP10(2013)147}}.

\bibitem{Alexandru:2015xva}
A.~Alexandru \textit{et al.},  \textit{Phys. Rev.} \textbf{D93} (2016) 014504.
  \href {http://arxiv.org/abs/1510.03258} {\path{arXiv:1510.03258}},
  \href {http://dx.doi.org/10.1103/PhysRevD.93.014504}
  {\path{doi:10.1103/PhysRevD.93.014504}}.

\bibitem{Alexandru:2015sua}
A.~Alexandru \textit{et al.},  \textit{J. High Energy Phys.} \textbf{05}
  (2016) 053.
  \href {http://arxiv.org/abs/1512.08764} {\path{arXiv:1512.08764}},
  \href {http://dx.doi.org/10.1007/JHEP05(2016)053}
  {\path{doi:10.1007/JHEP05(2016)053}}.

\bibitem{Tanizaki:2017yow}
Y.~Tanizaki \textit{et al.},   \textit{J. High Energy Phys.} \textbf{10} (2017) 100.
  \href {http://arxiv.org/abs/1706.03822} {\path{arXiv:1706.03822}},
  \href {http://dx.doi.org/10.1007/JHEP10(2017)100}
  {\path{doi:10.1007/JHEP10(2017)100}}.

\bibitem{Alexandru:2017czx}
A.~Alexandru \textit{et al.}, \textit{Phys. Rev.} \textbf{D96} (2017) 094505.
  \href {http://arxiv.org/abs/1709.01971} {\path{arXiv:1709.01971}},
  \href {http://dx.doi.org/10.1103/PhysRevD.96.094505}
  {\path{doi:10.1103/PhysRevD.96.094505}}.

\bibitem{Bluecher:2018sgj}
S.~Bluecher \textit{et al.},   \textit{SciPost
  Phys.} \textbf{5} (2018) 044.
  \href {http://arxiv.org/abs/1803.08418} {\path{arXiv:1803.08418}},
  \href {http://dx.doi.org/10.21468/SciPostPhys.5.5.044}
  {\path{doi:10.21468/SciPostPhys.5.5.044}}.

\bibitem{nicolaescu2007invitation}
L.~Nicolaescu, \textit{An Invitation to Morse Theory} (Springer, New York,
NY, 2007).
  \href {http://dx.doi.org/10.1007/978-1-4614-1105-5}
  {\path{doi:10.1007/978-1-4614-1105-5}}.

\bibitem{Alexandru:2016ejd}
A.~Alexandru \textit{et al.},   \textit{Phys. Rev.} \textbf{D95}
  (2017) 014502.
  \href {http://arxiv.org/abs/1609.01730} {\path{arXiv:1609.01730}},
  \href {http://dx.doi.org/10.1103/PhysRevD.95.014502}
  {\path{doi:10.1103/PhysRevD.95.014502}}.

\bibitem{Kotikov:1990kg}
A.~Kotikov,  \textit{Phys. Lett.} \textbf{B254} (1991) 158.
  \href {http://dx.doi.org/10.1016/0370-2693(91)90413-K}
  {\path{doi:10.1016/0370-2693(91)90413-K}}.

\bibitem{Kotikov1991a}
A.V. Kotikov,  \textit{Phys. Lett.} \textbf{B259} (1991) 314.
  \href {http://dx.doi.org/10.1016/0370-2693(91)90834-D}
  {\path{doi:10.1016/0370-2693(91)90834-D}}.

\bibitem{Kotikov1991b}
A.V. Kotikov,  \textit{Phys. Lett.} \textbf{B267} (1991) 123 [Erratum: \textit{Phys.
  Lett.} \textbf{B295} (1992) 409].
  \href {http://dx.doi.org/10.1016/0370-2693(91)90536-Y}
  {\path{doi:10.1016/0370-2693(91)90536-Y}}.

\bibitem{Remiddi:1997ny}
E.~Remiddi,  \textit{Nuovo Cim.}
  \textbf{A110} (1997) 1435.
  \href {http://arxiv.org/abs/hep-th/9711188}
  {\path{arXiv:hep-th/9711188}}.

\bibitem{Lee2008}
R.N. Lee, \textit{Proc. Sci.} \textbf{ACAT08}
  (2008) 105.
  \href {http://dx.doi.org/10.22323/1.070.0105}
  {\path{doi:10.22323/1.070.0105}}.

\bibitem{SmirPet2010}
A.V. Smirnov and A.V. Petukhov, 
  \textit{Lett. Math. Phys.} \textbf{97} (2011) 37.
  \href {http://arxiv.org/abs/1004.4199} {\path{arXiv:1004.4199}},
  \href {http://dx.doi.org/10.1007/s11005-010-0450-0}
  {\path{doi:10.1007/s11005-010-0450-0}}.

\bibitem{LeePomeransky2013}
R.N. Lee and A.A. Pomeransky, 
  \textit{J. High Energy Phys.} \textbf{2013} (2013) 165.
  \href {http://arxiv.org/abs/1308.6676} {\path{arXiv:1308.6676}},
  \href {http://dx.doi.org/10.1007/JHEP11(2013)165}
  {\path{doi:10.1007/JHEP11(2013)165}}.

\bibitem{Lee2014a}
R.N. Lee, Modern techniques of multiloop calculations,  {Proc. 49th Rencontres de Moriond on QCD and High
  Energy Interactions}, Moriond, Paris, France, 2014, Ed. E.~Auge and 
  J.~Dumarchez, p. 297.
  \href {http://arxiv.org/abs/1405.5616} {\path{arXiv:1405.5616}}.

\bibitem{Larsen:2015ped}
K.J. Larsen and Y.~Zhang,  \textit{Phys. Rev.} \textbf{D93} (2016) 041701.
  \href {http://arxiv.org/abs/1511.01071} {\path{arXiv:1511.01071}},
  \href {http://dx.doi.org/10.1103/PhysRevD.93.041701}
  {\path{doi:10.1103/PhysRevD.93.041701}}.

\bibitem{Bosma:2017hrk}
J.~Bosma \textit{et al.},   \textit{Phys. Rev.} \textbf{D97} (2018) 105014.
  \href {http://arxiv.org/abs/1712.03760} {\path{arXiv:1712.03760}},
  \href {http://dx.doi.org/10.1103/PhysRevD.97.105014}
  {\path{doi:10.1103/PhysRevD.97.105014}}.

\bibitem{Boehm2017}
J.~B{\"o}hm \textit{et al.},   \textit{Phys. Rev.} D98 (2018) 025023.
  \href {http://arxiv.org/abs/1712.09737} {\path{arXiv:1712.09737}},
  \href {http://dx.doi.org/10.1103/PhysRevD.98.025023}
  {\path{doi:10.1103/PhysRevD.98.025023}}.

\bibitem{Bitoun2017}
T.~Bitoun \textit{et al.},  \textit{Lett. Math. Phys.} \textbf{109} (2019)
497. \href {http://arxiv.org/abs/1712.09215}
  {\path{arXiv:1712.09215}}, \href
  {http://dx.doi.org/10.1007/s11005-018-1114-8}
  {\path{doi:10.1007/s11005-018-1114-8}}.

\bibitem{Boehm:2018fpv}
J. B\"ohm \textit{et al.},   \textit{J. High Energy Phys.} \textbf{09} (2018) 024.
  \href {http://arxiv.org/abs/1805.01873} {\path{arXiv:1805.01873}},
  \href {http://dx.doi.org/10.1007/JHEP09(2018)024}
  {\path{doi:10.1007/JHEP09(2018)024}}.

\bibitem{Barkatou1995}
A.~Barkatou, A rational version of Moser's algorithm,  Proc.
  1995 International Symposium on Symbolic and Algebraic Computation, (ACM,
  New York,   1995), p. 297.
  \href {http://dx.doi.org/10.1145/220346.220385}
  {\path{doi:10.1145/220346.220385}}.

\bibitem{BeneSmi1998}
M.~Beneke and V.A. Smirnov, \textit{Nucl.~Phys.~B} \textbf{522} (1998) 321.
  \href {http://arxiv.org/abs/hep-ph/9711391}
  {\path{arXiv:hep-ph/9711391}}, \href
  {http://dx.doi.org/10.1016/S0550-3213(98)00138-2}
  {\path{doi:10.1016/S0550-3213(98)00138-2}}.

\bibitem{HennSmirnov2013}
J.M. Henn and V.A. Smirnov, \textit{J. High Energy Phys.} \textbf{1311} (2013) 041.
  \href {http://arxiv.org/abs/1307.4083} {\path{arXiv:1307.4083}},
  \href {http://dx.doi.org/10.1007/JHEP11(2013)041}
  {\path{doi:10.1007/JHEP11(2013)041}}.

\bibitem{HennSmirnovSmirnov2013}
J.M. Henn \textit{et al.},  \textit{J. High Energy Phys.}
  \textbf{1307} (2013) 128.
  \href {http://arxiv.org/abs/1306.2799} {\path{arXiv:1306.2799}},
  \href {http://dx.doi.org/10.1007/JHEP07(2013)128}
  {\path{doi:10.1007/JHEP07(2013)128}}.

\bibitem{Henn:2014lfa}
J.M. Henn \textit{et al.},  \textit{J. High Energy Phys.} \textbf{1405}
  (2014) 090.
  \href {http://arxiv.org/abs/1402.7078} {\path{arXiv:1402.7078}},
  \href {http://dx.doi.org/10.1007/JHEP05(2014)090}
  {\path{doi:10.1007/JHEP05(2014)090}}.

\bibitem{Goncharov:1998kja}
A.B. Goncharov,   \textit{Math. Res. Lett.} \textbf{5} (1998) 497.
  \href {http://arxiv.org/abs/1105.2076} {\path{arXiv:1105.2076}},
  \href {http://dx.doi.org/10.4310/MRL.1998.v5.n4.a7}
  {\path{doi:10.4310/MRL.1998.v5.n4.a7}}.

\bibitem{Vollinga:2004sn}
J.~Vollinga and S.~Weinzierl, 
  \textit{Comput. Phys. Commun.} \textbf{167} (2005) 177.
  \href {http://arxiv.org/abs/hep-ph/0410259}
  {\path{arXiv:hep-ph/0410259}}, \href
  {http://dx.doi.org/10.1016/j.cpc.2004.12.009}
  {\path{doi:10.1016/j.cpc.2004.12.009}}.

\bibitem{Lee2018a}
R.N. Lee \textit{et al.},  \textit{J. High Energy Phys.} \textbf{03} (2018) 008.
  \href {http://arxiv.org/abs/1709.07525} {\path{arXiv:1709.07525}},
  \href {http://dx.doi.org/10.1007/jhep03(2018)008}
  {\path{doi:10.1007/jhep03(2018)008}}.

\bibitem{Lee:2014ioa}
R.N. Lee, 
  \textit{J. High Energy Phys.} \textbf{04} (2015) 108.
  \href {http://arxiv.org/abs/1411.0911} {\path{arXiv:1411.0911}},
  \href {http://dx.doi.org/10.1007/JHEP04(2015)108}
  {\path{doi:10.1007/JHEP04(2015)108}}.

\bibitem{Gituliar:2017vzm}
O.~Gituliar and V.~Magerya, \textit{Comput. Phys. Commun.} \textbf{219}
  (2017) 329.
  \href {http://arxiv.org/abs/1701.04269} {\path{arXiv:1701.04269}},
  \href {http://dx.doi.org/10.1016/j.cpc.2017.05.004}
  {\path{doi:10.1016/j.cpc.2017.05.004}}.

\bibitem{Prausa:2017ltv}
M.~Prausa,  \textit{Comput. Phys. Commun.} \textbf{219} (2017) 361.
  \href {http://arxiv.org/abs/1701.00725} {\path{arXiv:1701.00725}},
  \href {http://dx.doi.org/10.1016/j.cpc.2017.05.026}
  {\path{doi:10.1016/j.cpc.2017.05.026}}.

\bibitem{Barkatou2007}
M.~Barkatou and E. Pfl{\"u}gel, {Computing super-irreducible forms of systems of
  linear differential equations via Moser-reduction: a new approach},
  Proc. 2007 International Symposium on Symbolic and Algebraic
  Computation (ACM, New York, NY, 2007).
 

\bibitem{barkatou2009moser}
M.~Barkatou and E. Pfl{\"u}gel, \textit{J.
  Symbolic Comput.} \textbf{44} (2009) 1017.
  \href {http://dx.doi.org/10.1016/j.jsc.2009.01.002}
  {\path{doi:10.1016/j.jsc.2009.01.002}}.

\bibitem{Bolibrukh1989}
A.A. Bolibrukh,   Matematicheskie Zametki \textbf{46} (1989) 118.

\bibitem{Moser1959}
J.~Moser,  \textit{Math.
  Z.} \textbf{72} (1959) 379.
  \href {http://dx.doi.org/10.1007/BF01162962}
  {\path{doi:10.1007/BF01162962}}.

\bibitem{Lee:2017oca}
R.N. Lee and A.A. Pomeransky, {Normalized Fuchsian form on Riemann sphere and
  differential equations for multiloop integrals}, \href
  {http://arxiv.org/abs/1707.07856} {\path{arXiv:1707.07856}}.

\bibitem{Argeri:2014qva}
M.~Argeri \textit{et al.},  \textit{J. High Energy Phys.} \textbf{1403} (2014) 082.
  \href {http://arxiv.org/abs/1401.2979} {\path{arXiv:1401.2979}},
  \href {http://dx.doi.org/10.1007/JHEP03(2014)082}
  {\path{doi:10.1007/JHEP03(2014)082}}.

\bibitem{Adams:2018yfj}
L.~Adams and S.~Weinzierl, \textit{Phys. Lett. B} \textbf{781} (2018) 270. \href
  {http://arxiv.org/abs/1802.05020} {\path{arXiv:1802.05020}}, \href
  {http://dx.doi.org/10.1016/j.physletb.2018.04.002}
  {\path{doi:10.1016/j.physletb.2018.04.002}}.

\bibitem{Adams2015a}
L.~Adams \textit{et al.},  \textit{Acta Phys. Pol.}
  \textbf{B46} (2015) 2131.
  \href {http://arxiv.org/abs/1510.02048} {\path{arXiv:1510.02048}},
  \href {http://dx.doi.org/10.5506/APhysPolB.46.2131}
  {\path{doi:10.5506/APhysPolB.46.2131}}.

\bibitem{Adams:2016xah}
L.~Adams \textit{et al.},   \textit{J. Math. Phys.} \textbf{57} (2016)
  122302.
  \href {http://arxiv.org/abs/1607.01571} {\path{arXiv:1607.01571}},
  \href {http://dx.doi.org/10.1063/1.4969060} {\path{doi:10.1063/1.4969060}}.

\bibitem{Remiddi:2016gno}
E.~Remiddi and L.~Tancredi, 
  \textit{Nucl. Phys.} \textbf{B907} (2016) 400.
  \href {http://arxiv.org/abs/1602.01481} {\path{arXiv:1602.01481}},
  \href {http://dx.doi.org/10.1016/j.nuclphysb.2016.04.013}
  {\path{doi:10.1016/j.nuclphysb.2016.04.013}}.

\bibitem{Adams:2017xsu}
L.~Adams \textit{et al.}, \textit{Proc. Sci.} 
  \textbf{RADCOR2017} (2017) 015.
  \href {http://arxiv.org/abs/1712.03532} {\path{arXiv:1712.03532}},
  \href {http://dx.doi.org/10.22323/1.290.0015}
  {\path{doi:10.22323/1.290.0015}}.

\bibitem{Primo:2017ipr}
A.~Primo and L.~Tancredi,  \textit{Nucl.
  Phys.} \textbf{B921} (2017) 316.
  \href {http://arxiv.org/abs/1704.05465} {\path{arXiv:1704.05465}},
  \href {http://dx.doi.org/10.1016/j.nuclphysb.2017.05.018}
  {\path{doi:10.1016/j.nuclphysb.2017.05.018}}.

\bibitem{Primo:2016ebd}
A.~Primo and L.~Tancredi,  \textit{Nucl. Phys.} \textbf{B916} (2017) 94.
  \href {http://arxiv.org/abs/1610.08397} {\path{arXiv:1610.08397}},
  \href {http://dx.doi.org/10.1016/j.nuclphysb.2016.12.021}
  {\path{doi:10.1016/j.nuclphysb.2016.12.021}}.

\bibitem{Remiddi:2017har}
E.~Remiddi and L.~Tancredi,  \textit{Nucl. Phys.} \textbf{B925} (2017) 212.
  \href {http://arxiv.org/abs/1709.03622} {\path{arXiv:1709.03622}},
  \href {http://dx.doi.org/10.1016/j.nuclphysb.2017.10.007}
  {\path{doi:10.1016/j.nuclphysb.2017.10.007}}.

\bibitem{Meyer:2017joq}
C.~Meyer,  \textit{Comput. Phys. Commun.} \textbf{222} (2018) 295.
  \href {http://arxiv.org/abs/1705.06252} {\path{arXiv:1705.06252}},
  \href {http://dx.doi.org/10.1016/j.cpc.2017.09.014}
  {\path{doi:10.1016/j.cpc.2017.09.014}}.

\bibitem{Feynman:1949zx}
R.P. Feynman,  \textit{Phys. Rev.}
  \textbf{76} (1949) 769.
  \href {http://dx.doi.org/10.1103/PhysRev.76.769}
  {\path{doi:10.1103/PhysRev.76.769}}.

\bibitem{Dyson:1949bp}
F.J. Dyson, 
  \textit{Phys. Rev.} \textbf{75} (1949) 486.
  \href {http://dx.doi.org/10.1103/PhysRev.75.486}
  {\path{doi:10.1103/PhysRev.75.486}}.

\bibitem{Dyson:1949ha}
F.J. Dyson,  \textit{Phys. Rev.} \textbf{75} (1949)
  1736.
  \href {http://dx.doi.org/10.1103/PhysRev.75.1736}
  {\path{doi:10.1103/PhysRev.75.1736}}.

\bibitem{Gehrmann:2015bfy}
T.~Gehrmann \textit{et al.},  \textit{Phys. Rev. Lett.} \textbf{116}
  (2016) 062001.
  \href {http://dx.doi.org/10.1103/PhysRevLett.116.189903}
  {\path{doi:10.1103/PhysRevLett.116.189903}}.

\bibitem{Papadopoulos:2015jft}
C.G. Papadopoulos \textit{et al.},  \textit{J. High Energy Phys.} \textbf{04} (2016) 078.
  \href {http://arxiv.org/abs/1511.09404} {\path{arXiv:1511.09404}},
  \href {http://dx.doi.org/10.1007/JHEP04(2016)078}
  {\path{doi:10.1007/JHEP04(2016)078}}.

\bibitem{Andersen:2014efa}
J.R. Andersen \textit{et al.}, {Les Houches 2013: physics at TeV colliders: Standard
  Model Working Group report}, \href {http://arxiv.org/abs/1405.1067}
  {\path{arXiv:1405.1067}}.

\bibitem{Bern:1994cg}
Z.~Bern \textit{et al.}, \textit{Nucl. Phys.} \textbf{B435} (1995) 59.
  \href {http://arxiv.org/abs/hep-ph/9409265}
  {\path{arXiv:hep-ph/9409265}}, \href
  {http://dx.doi.org/10.1016/0550-3213(94)00488-Z}
  {\path{doi:10.1016/0550-3213(94)00488-Z}}.

\bibitem{Bern:1994zx}
Z.~Bern \textit{et al.},  \textit{Nucl. Phys.} \textbf{B425} (1994)
  217.
  \href {http://arxiv.org/abs/hep-ph/9403226}
  {\path{arXiv:hep-ph/9403226}}, \href
  {http://dx.doi.org/10.1016/0550-3213(94)90179-1}
  {\path{doi:10.1016/0550-3213(94)90179-1}}.

\bibitem{Ossola:2006us}
G.~Ossola \textit{et al.}, \textit{Nucl. Phys.} \textbf{B763} (2007) 147.
  \href {http://arxiv.org/abs/hep-ph/0609007}
  {\path{arXiv:hep-ph/0609007}}, \href
  {http://dx.doi.org/10.1016/j.nuclphysb.2006.11.012}
  {\path{doi:10.1016/j.nuclphysb.2006.11.012}}.

\bibitem{Ossola:2008xq}
G.~Ossola \textit{et al.},  \textit{J. High Energy Phys.} \textbf{0805} (2008) 004.
  \href {http://arxiv.org/abs/0802.1876} {\path{arXiv:0802.1876}},
  \href {http://dx.doi.org/10.1088/1126-6708/2008/05/004}
  {\path{doi:10.1088/1126-6708/2008/05/004}}.

\bibitem{Ellis:2011cr}
R.K. Ellis \textit{et al.},  \textit{Phys. Rep.}
  \textbf{518} (2012) 141.
  \href {http://arxiv.org/abs/1105.4319} {\path{arXiv:1105.4319}},
  \href {http://dx.doi.org/10.1016/j.physrep.2012.01.008}
  {\path{doi:10.1016/j.physrep.2012.01.008}}.

\bibitem{vanDeurzen:2015jmn}
H.~van Deurzen \textit{et al.}, \textit{Nucl. Part. Phys. Proc.} \textbf{267--269} (2015) 140.
  \href {http://dx.doi.org/10.1016/j.nuclphysbps.2015.10.094}
  {\path{doi:10.1016/j.nuclphysbps.2015.10.094}}.

\bibitem{Gluza:2010ws}
J.~Gluza \textit{et al.},  \textit{Phys. Rev.} \textbf{D83} (2011) 045012.
  \href {http://arxiv.org/abs/1009.0472} {\path{arXiv:1009.0472}},
  \href {http://dx.doi.org/10.1103/PhysRevD.83.045012}
  {\path{doi:10.1103/PhysRevD.83.045012}}.

\bibitem{Kosower:2011ty}
D.A. Kosower and K.J. Larsen,  \textit{Phys. Rev.} \textbf{D85}
  (2012) 045017.
  \href {http://arxiv.org/abs/1108.1180} {\path{arXiv:1108.1180}},
  \href {http://dx.doi.org/10.1103/PhysRevD.85.045017}
  {\path{doi:10.1103/PhysRevD.85.045017}}.

\bibitem{Johansson:2012zv}
H.~Johansson \textit{et al.},  \textit{Phys. Rev.} \textbf{D87} (2013) 025030.
  \href {http://arxiv.org/abs/1208.1754} {\path{arXiv:1208.1754}},
  \href {http://dx.doi.org/10.1103/PhysRevD.87.025030}
  {\path{doi:10.1103/PhysRevD.87.025030}}.

\bibitem{Johansson:2012sf}
H.~Johansson \textit{et al.}, \textit{Proc. Sci.} \textbf{LL2012} (2012) 066.
  \href {http://arxiv.org/abs/1212.2132} {\path{arXiv:1212.2132}},
  \href {http://dx.doi.org/10.22323/1.151.0066}
  {\path{doi:10.22323/1.151.0066}}.

\bibitem{Johansson:2013sda}
H.~Johansson \textit{et al.},  \textit{Phys. Rev.} \textbf{D89}  (2014) 125010.
  \href {http://arxiv.org/abs/1308.4632} {\path{arXiv:1308.4632}},
  \href {http://dx.doi.org/10.1103/PhysRevD.89.125010}
  {\path{doi:10.1103/PhysRevD.89.125010}}.

\bibitem{Sogaard:2013fpa}
M.~S{\o}gaard and Y.~Zhang,  \textit{J. High Energy Phys.} \textbf{12}
  (2013) 008.
  \href {http://arxiv.org/abs/1310.6006} {\path{arXiv:1310.6006}},
  \href {http://dx.doi.org/10.1007/JHEP12(2013)008}
  {\path{doi:10.1007/JHEP12(2013)008}}.

\bibitem{Ita:2015tya}
H.~Ita,  \textit{Phys. Rev.} \textbf{D94} (2016) 116015.
  \href {http://arxiv.org/abs/1510.05626} {\path{arXiv:1510.05626}},
  \href {http://dx.doi.org/10.1103/PhysRevD.94.116015}
  {\path{doi:10.1103/PhysRevD.94.116015}}.

\bibitem{Johansson:2015ava}
H.~Johansson \textit{et al.},  \textit{Phys. Rev.} \textbf{D92} (2015) 025015.
  \href {http://arxiv.org/abs/1503.06711} {\path{arXiv:1503.06711}},
  \href {http://dx.doi.org/10.1103/PhysRevD.92.025015}
  {\path{doi:10.1103/PhysRevD.92.025015}}.

\bibitem{Mastrolia:2016dhn}
P.~Mastrolia \textit{et al.},  \textit{J. High Energy Phys.} \textbf{08} (2016) 164.
  \href {http://arxiv.org/abs/1605.03157} {\path{arXiv:1605.03157}},
  \href {http://dx.doi.org/10.1007/JHEP08(2016)164}
  {\path{doi:10.1007/JHEP08(2016)164}}.

\bibitem{Mastrolia:2011pr}
P.~Mastrolia and G.~Ossola,  \textit{J. High Energy Phys.} \textbf{1111} (2011) 014.
  \href {http://arxiv.org/abs/1107.6041} {\path{arXiv:1107.6041}},
  \href {http://dx.doi.org/10.1007/JHEP11(2011)014}
  {\path{doi:10.1007/JHEP11(2011)014}}.

\bibitem{Badger:2012dp}
S.~Badger \textit{et al.},  \textit{J. High Energy Phys.} \textbf{1204} (2012) 055.
  \href {http://arxiv.org/abs/1202.2019} {\path{arXiv:1202.2019}},
  \href {http://dx.doi.org/10.1007/JHEP04(2012)055}
  {\path{doi:10.1007/JHEP04(2012)055}}.

\bibitem{Mastrolia:2012wf}
P.~Mastrolia \textit{et al.}, 
  \textit{Phys. Rev.} \textbf{D87}  (2013) 085026.
  \href {http://arxiv.org/abs/1209.4319} {\path{arXiv:1209.4319}},
  \href {http://dx.doi.org/10.1103/PhysRevD.87.085026}
  {\path{doi:10.1103/PhysRevD.87.085026}}.

\bibitem{Badger:2013gxa}
S.~Badger \textit{et al.},  \textit{J. High Energy Phys.} \textbf{1312} (2013) 045.
  \href {http://arxiv.org/abs/1310.1051} {\path{arXiv:1310.1051}},
  \href {http://dx.doi.org/10.1007/JHEP12(2013)045}
  {\path{doi:10.1007/JHEP12(2013)045}}.

\bibitem{Papadopoulos:2013hra}
C.G. Papadopoulos \textit{et al.}, \textit{Proc. Sci.} \textbf{Corfu2012} (2013) 019.
  \href {http://dx.doi.org/10.22323/1.177.0019}
  {\path{doi:10.22323/1.177.0019}}.

\bibitem{Badger:2015lda}
S.~Badger \textit{et al.},  \textit{J. High Energy Phys.} \textbf{10} (2015) 064.
  \href {http://arxiv.org/abs/1507.08797} {\path{arXiv:1507.08797}},
  \href {http://dx.doi.org/10.1007/JHEP10(2015)064}
  {\path{doi:10.1007/JHEP10(2015)064}}.

\bibitem{'tHooft:1978xw}
G. 't Hooft and M. Veltman,  \textit{Nucl. Phys.} \textbf{B153} (1979)
  365.
  \href {http://dx.doi.org/10.1016/0550-3213(79)90605-9}
  {\path{doi:10.1016/0550-3213(79)90605-9}}.

\bibitem{Smirnov:2012gma}
V.A. Smirnov, \textit{Analytic Tools for Feynman Integrals}  (Springer, Berlin, 2012).
  \href {http://dx.doi.org/10.1007/978-3-642-34886-0}
  {\path{doi:10.1007/978-3-642-34886-0}}.

\bibitem{Gehrmann:1999as}
T.~Gehrmann and E.~Remiddi,  \textit{Nucl. Phys.} \textbf{B580} (2000) 485.
  \href {http://arxiv.org/abs/hep-ph/9912329}
  {\path{arXiv:hep-ph/9912329}}, \href
  {http://dx.doi.org/10.1016/S0550-3213(00)00223-6}
  {\path{doi:10.1016/S0550-3213(00)00223-6}}.

\bibitem{Remiddi:1999ew}
E.~Remiddi and J.~Vermaseren,  \textit{Int. J. Mod. Phys.} \textbf{A15}
  (2000) 725.
  \href {http://arxiv.org/abs/hep-ph/9905237}
  {\path{arXiv:hep-ph/9905237}}, \href
  {http://dx.doi.org/10.1142/S0217751X00000367}
  {\path{doi:10.1142/S0217751X00000367}}.

\bibitem{Goncharov:2001iea}
A.B. ~Goncharov, {Multiple polylogarithms and mixed Tate motives}, \href
  {http://arxiv.org/abs/math/0103059} {\path{arXiv:math/0103059}}.

\bibitem{Ablinger:2015tua}
J.~Ablinger \textit{et al.},   \textit{Comput. Phys. Commun.} \textbf{202}
  (2016) 33.
  \href {http://arxiv.org/abs/1509.08324} {\path{arXiv:1509.08324}},
  \href {http://dx.doi.org/10.1016/j.cpc.2016.01.002}
  {\path{doi:10.1016/j.cpc.2016.01.002}}.

\bibitem{Adams:2015gva}
L.~Adams \textit{et al.},   \textit{J. Math. Phys.} \textbf{56} (2015) 072303.
  \href {http://arxiv.org/abs/1504.03255} {\path{arXiv:1504.03255}},
  \href {http://dx.doi.org/10.1063/1.4926985} {\path{doi:10.1063/1.4926985}}.

\bibitem{Bonciani:2016qxi}
R.~Bonciani \textit{et al.},   \textit{J. High Energy Phys.} \textbf{12} (2016) 096.
  \href {http://arxiv.org/abs/1609.06685} {\path{arXiv:1609.06685}},
  \href {http://dx.doi.org/10.1007/JHEP12(2016)096}
  {\path{doi:10.1007/JHEP12(2016)096}}.

\bibitem{Ablinger:2017bjx}
J.~Ablinger \textit{et al.},   \textit{J. Math. Phys.} \textbf{59} (2018) 062305.
  \href {http://arxiv.org/abs/1706.01299} {\path{arXiv:1706.01299}},
  \href {http://dx.doi.org/10.1063/1.4986417} {\path{doi:10.1063/1.4986417}}.

\bibitem{Bourjaily:2017bsb}
J.L. Bourjaily  \textit{et al.}, 
   \textit{Phys. Rev. Lett.} \textbf{120} (2018) 121603.
  \href {http://arxiv.org/abs/1712.02785} {\path{arXiv:1712.02785}},
  \href {http://dx.doi.org/10.1103/PhysRevLett.120.121603}
  {\path{doi:10.1103/PhysRevLett.120.121603}}.

\bibitem{Frellesvig:2017aai}
H.~Frellesvig and C.~G. Papadopoulos,  \textit{J. High Energy Phys.} \textbf{04} (2017) 083.
  \href {http://arxiv.org/abs/1701.07356} {\path{arXiv:1701.07356}},
  \href {http://dx.doi.org/10.1007/JHEP04(2017)083}
  {\path{doi:10.1007/JHEP04(2017)083}}.

\bibitem{Baikov:1996iu}
P.A. Baikov, \textit{Nucl. Instrum. Meth.} \textbf{A389} (1997) 347.
  \href {http://arxiv.org/abs/hep-ph/9611449}
  {\path{arXiv:hep-ph/9611449}}, \href
  {http://dx.doi.org/10.1016/S0168-9002(97)00126-5}
  {\path{doi:10.1016/S0168-9002(97)00126-5}}.

\bibitem{Baikov:1996rk}
P.A. Baikov,  \textit{Phys. Lett.} \textbf{B385} (1996) 404.
  \href {http://arxiv.org/abs/hep-ph/9603267}
  {\path{arXiv:hep-ph/9603267}}, \href
  {http://dx.doi.org/10.1016/0370-2693(96)00835-0}
  {\path{doi:10.1016/0370-2693(96)00835-0}}.

\bibitem{Smirnov:2003kc}
V.A. Smirnov and M.~Steinhauser,  \textit{Nucl. Phys.} \textbf{B672} (2003) 199.
  \href {http://arxiv.org/abs/hep-ph/0307088}
  {\path{arXiv:hep-ph/0307088}}, \href
  {http://dx.doi.org/10.1016/j.nuclphysb.2003.09.003}
  {\path{doi:10.1016/j.nuclphysb.2003.09.003}}.

\bibitem{Lee:2010wea}
R.N. Lee,  \textit{Nucl. Phys. Proc. Suppl.} \textbf{205--206} (2010)
  135.
  \href {http://arxiv.org/abs/1007.2256} {\path{arXiv:1007.2256}},
  \href {http://dx.doi.org/10.1016/j.nuclphysbps.2010.08.032}
  {\path{doi:10.1016/j.nuclphysbps.2010.08.032}}.

\bibitem{Grozin:2011mt}
A.G. Grozin,  \textit{Int. J. Mod. Phys.} \textbf{A26}
  (2011) 2807.
  \href {http://arxiv.org/abs/1104.3993} {\path{arXiv:1104.3993}},
  \href {http://dx.doi.org/10.1142/S0217751X11053687}
  {\path{doi:10.1142/S0217751X11053687}}.

\bibitem{Papadopoulos:2014lla}
C.G. Papadopoulos, \textit{J. High Energy Phys.} \textbf{1407} (2014) 088.
  \href {http://arxiv.org/abs/1401.6057} {\path{arXiv:1401.6057}},
  \href {http://dx.doi.org/10.1007/JHEP07(2014)088}
  {\path{doi:10.1007/JHEP07(2014)088}}.

\bibitem{Papadopoulos:2014hla}
C.G. Papadopoulos \textit{et al.},   \textit{J. High Energy Phys.} \textbf{01} (2015) 072.
  \href {http://arxiv.org/abs/1409.6114} {\path{arXiv:1409.6114}},
  \href {http://dx.doi.org/10.1007/JHEP01(2015)072}
  {\path{doi:10.1007/JHEP01(2015)072}}.

\bibitem{Henn:2014qga}
J.M. Henn,  \textit{J.
  Phys.} \textbf{A48} (2015) 153001.
  \href {http://arxiv.org/abs/1412.2296} {\path{arXiv:1412.2296}},
  \href {http://dx.doi.org/10.1088/1751-8113/48/15/153001}
  {\path{doi:10.1088/1751-8113/48/15/153001}}.

\bibitem{Dick}
\url{https://www.dropbox.com/s/90iiqfcazrhwtso/results.tgz?dl=0}, last accessed
May 16th 2019.

\bibitem{Panzer:2014caa}
E.~Panzer,  \textit{Comput. Phys. Commun.} \textbf{188} (2014)
  148.
  \href {http://arxiv.org/abs/1403.3385} {\path{arXiv:1403.3385}},
  \href {http://dx.doi.org/10.1016/j.cpc.2014.10.019}
  {\path{doi:10.1016/j.cpc.2014.10.019}}.

\bibitem{Gehrmann:2015ora}
T.~Gehrmann \textit{et al.},   \textit{J. High Energy Phys.} \textbf{09} (2015)
  128.
  \href {http://arxiv.org/abs/1503.04812} {\path{arXiv:1503.04812}},
  \href {http://dx.doi.org/10.1007/JHEP09(2015)128}
  {\path{doi:10.1007/JHEP09(2015)128}}.

\bibitem{Veltman:1994wz}
M.J.G. Veltman, \textit{Diagrammatica: The Path to Feynman Rules} (Cambridge
University Press, Cambridge, 1994).
  \href {http://dx.doi.org/10.1017/CBO9780511564079}
  {\path{doi:10.1017/CBO9780511564079}}.

\bibitem{Abreu:2014cla}
S.~Abreu  \textit{et al.},  \textit{J. High Energy Phys.} \textbf{10} (2014) 125.
  \href {http://arxiv.org/abs/1401.3546} {\path{arXiv:1401.3546}},
  \href {http://dx.doi.org/10.1007/JHEP10(2014)125}
  {\path{doi:10.1007/JHEP10(2014)125}}.

\bibitem{Abreu:2015zaa}
S.~Abreu \textit{et al.},   \textit{J. High Energy Phys.} \textbf{07} (2015) 111.
  \href {http://arxiv.org/abs/1504.00206} {\path{arXiv:1504.00206}},
  \href {http://dx.doi.org/10.1007/JHEP07(2015)111}
  {\path{doi:10.1007/JHEP07(2015)111}}.

\bibitem{Anastasiou:2002yz}
C.~Anastasiou and K.~Melnikov,  \textit{Nucl. Phys.} \textbf{B646} (2002) 220.
  \href {http://arxiv.org/abs/hep-ph/0207004}
  {\path{arXiv:hep-ph/0207004}}, \href
  {http://dx.doi.org/10.1016/S0550-3213(02)00837-4}
  {\path{doi:10.1016/S0550-3213(02)00837-4}}.

\bibitem{Lee:2012te}
R.N. Lee and V.A. Smirnov,  \textit{J. High Energy Phys.} \textbf{12} (2012) 104.
  \href {http://arxiv.org/abs/1209.0339} {\path{arXiv:1209.0339}},
  \href {http://dx.doi.org/10.1007/JHEP12(2012)104}
  {\path{doi:10.1007/JHEP12(2012)104}}.

\bibitem{Bosma:2017ens}
J.~Bosma \textit{et al.},   \textit{J. High Energy Phys.}
  \textbf{08} (2017) 051.
  \href {http://arxiv.org/abs/1704.04255} {\path{arXiv:1704.04255}},
  \href {http://dx.doi.org/10.1007/JHEP08(2017)051}
  {\path{doi:10.1007/JHEP08(2017)051}}.

\bibitem{Harley:2017qut}
M.~Harley \textit{et al.},   \textit{J. High Energy Phys.} \textbf{06} (2017) 049.
  \href {http://arxiv.org/abs/1705.03478} {\path{arXiv:1705.03478}},
  \href {http://dx.doi.org/10.1007/JHEP06(2017)049}
  {\path{doi:10.1007/JHEP06(2017)049}}.

\bibitem{Argeri:2007up}
M.~Argeri and P.~Mastrolia,  \textit{Int. J.
  Mod. Phys.} \textbf{A22} (2007) 4375.
  \href {http://arxiv.org/abs/0707.4037} {\path{arXiv:0707.4037}},
  \href {http://dx.doi.org/10.1142/S0217751X07037147}
  {\path{doi:10.1142/S0217751X07037147}}.

\bibitem{MullerStach:2012mp}
S.~M{\"u}ller-Stach \textit{et al.}, \textit{Commun. Math. Phys.} \textbf{326} (2014) 237.
  \href {http://arxiv.org/abs/1212.4389} {\path{arXiv:1212.4389}},
  \href {http://dx.doi.org/10.1007/s00220-013-1838-3}
  {\path{doi:10.1007/s00220-013-1838-3}}.

\bibitem{Borwein}
J.M. Borwein \textit{et al.},   \textit{Trans. Amer. Math. Soc.} \textbf{353} (2001) 907.
  \href {http://arxiv.org/abs/math.CA/9910045}
  {\path{arXiv:math.CA/9910045}}, \href
  {http://dx.doi.org/10.1090/S0002-9947-00-02616-7}
  {\path{doi:10.1090/S0002-9947-00-02616-7}}.

\bibitem{Moch:2001zr}
S.~Moch \textit{et al.},  \textit{J. Math. Phys.} \textbf{43} (2002)
  3363.
  \href {http://arxiv.org/abs/hep-ph/0110083}
  {\path{arXiv:hep-ph/0110083}}, \href {http://dx.doi.org/10.1063/1.1471366}
  {\path{doi:10.1063/1.1471366}}.

\bibitem{Gehrmann:2014bfa}
T.~Gehrmann \textit{et al.},   \textit{J. High Energy Phys.} \textbf{1406} (2014) 032.
  \href {http://arxiv.org/abs/1404.4853} {\path{arXiv:1404.4853}},
  \href {http://dx.doi.org/10.1007/JHEP06(2014)032}
  {\path{doi:10.1007/JHEP06(2014)032}}.

\bibitem{Meyer:2016slj}
C.~Meyer,  \textit{J. High Energy Phys.} \textbf{04} (2017) 006.
  \href {http://arxiv.org/abs/1611.01087} {\path{arXiv:1611.01087}},
  \href {http://dx.doi.org/10.1007/JHEP04(2017)006}
  {\path{doi:10.1007/JHEP04(2017)006}}.

\bibitem{Adams:2017tga}
L.~Adams \textit{et al.},   \textit{Phys. Rev.
  Lett.} \textbf{118} (2017) 141602.
  \href {http://arxiv.org/abs/1702.04279} {\path{arXiv:1702.04279}},
  \href {http://dx.doi.org/10.1103/PhysRevLett.118.141602}
  {\path{doi:10.1103/PhysRevLett.118.141602}}.

\bibitem{Becchetti:2017abb}
M.~Becchetti and R.~Bonciani,  \textit{J. High Energy Phys.} \textbf{01} (2018)
  048.
  \href {http://arxiv.org/abs/1712.02537} {\path{arXiv:1712.02537}},
  \href {http://dx.doi.org/10.1007/JHEP01(2018)048}
  {\path{doi:10.1007/JHEP01(2018)048}}.

\bibitem{Broadhurst:1993mw}
D.J. Broadhurst \textit{et al.},   \textit{Z.
  Phys.} \textbf{C60} (1993) 287.
  \href {http://arxiv.org/abs/hep-ph/9304303}
  {\path{arXiv:hep-ph/9304303}}, \href {http://dx.doi.org/10.1007/BF01474625}
  {\path{doi:10.1007/BF01474625}}.

\bibitem{Berends:1993ee}
F.A. Berends \textit{et al.},  \textit{Z.Phys.} \textbf{C63} (1994)
  227.
  \href {http://dx.doi.org/10.1007/BF01411014}
  {\path{doi:10.1007/BF01411014}}.

\bibitem{Bauberger:1994nk}
S.~Bauberger \textit{et al.},   \textit{Nucl. Phys. Proc.
  Suppl.} \textbf{37B} (1994) 95.
  \href {http://arxiv.org/abs/hep-ph/9406404}
  {\path{arXiv:hep-ph/9406404}}, \href
  {http://dx.doi.org/10.1016/0920-5632(94)90665-3}
  {\path{doi:10.1016/0920-5632(94)90665-3}}.

\bibitem{Bauberger:1994by}
S.~Bauberger \textit{et al.},   \textit{Nucl. Phys.} \textbf{B434} (1995)
  383.
  \href {http://arxiv.org/abs/hep-ph/9409388}
  {\path{arXiv:hep-ph/9409388}}, \href
  {http://dx.doi.org/10.1016/0550-3213(94)00475-T}
  {\path{doi:10.1016/0550-3213(94)00475-T}}.

\bibitem{Bauberger:1994hx}
S.~Bauberger and M.~B{\"o}hm,  \textit{Nucl. Phys.} \textbf{B445} (1995) 25.
  \href {http://arxiv.org/abs/hep-ph/9501201}
  {\path{arXiv:hep-ph/9501201}}, \href
  {http://dx.doi.org/10.1016/0550-3213(95)00199-3}
  {\path{doi:10.1016/0550-3213(95)00199-3}}.

\bibitem{Caffo:1998du}
M.~Caffo \textit{et al.},   \textit{Nuovo Cim.} \textbf{A111} (1998) 365,
   \href {http://arxiv.org/abs/hep-th/9805118}
  {\path{arXiv:hep-th/9805118}}.

\bibitem{Laporta:2004rb}
S.~Laporta and E.~Remiddi,  \textit{Nucl. Phys.} \textbf{B704} (2005) 349.
  \href {http://arxiv.org/abs/hep-ph/0406160}
  {\path{arXiv:hep-ph/0406160}}, \href
  {http://dx.doi.org/10.1016/j.nuclphysb.2004.10.044}
  {\path{doi:10.1016/j.nuclphysb.2004.10.044}}.

\bibitem{Kniehl:2005bc}
B.A. Kniehl \textit{et al.}, \textit{Nucl. Phys.} \textbf{B738} (2006) 306.
  \href {http://arxiv.org/abs/hep-ph/0510235}
  {\path{arXiv:hep-ph/0510235}}, \href
  {http://dx.doi.org/10.1016/j.nuclphysb.2006.01.013}
  {\path{doi:10.1016/j.nuclphysb.2006.01.013}}.

\bibitem{Groote:2005ay}
S.~Groote \textit{et al.}, \textit{Ann. Phys.} \textbf{322} (2007) 2374.
  \href {http://arxiv.org/abs/hep-ph/0506286}
  {\path{arXiv:hep-ph/0506286}}, \href
  {http://dx.doi.org/10.1016/j.aop.2006.11.001}
  {\path{doi:10.1016/j.aop.2006.11.001}}.

\bibitem{Groote:2012pa}
S.~Groote \textit{et al.},  \textit{Eur. Phys. J.} \textbf{C72} (2012) 2085.
  \href {http://arxiv.org/abs/1204.0694} {\path{arXiv:1204.0694}},
  \href {http://dx.doi.org/10.1140/epjc/s10052-012-2085-z}
  {\path{doi:10.1140/epjc/s10052-012-2085-z}}.

\bibitem{Bailey:2008ib}
D.H. Bailey \textit{et al.},   \textit{J. Phys.} \textbf{A41} (2008) 205203.
  \href {http://arxiv.org/abs/0801.0891} {\path{arXiv:0801.0891}},
  \href {http://dx.doi.org/10.1088/1751-8113/41/20/205203}
  {\path{doi:10.1088/1751-8113/41/20/205203}}.

\bibitem{MullerStach:2011ru}
S.~M\"uller-Stach \textit{et al.},   \textit{Commun. Num.
  Theor. Phys.} \textbf{6} (2012) 203.
  \href {http://arxiv.org/abs/1112.4360} {\path{arXiv:1112.4360}},
  \href {http://dx.doi.org/10.4310/CNTP.2012.v6.n1.a5}
  {\path{doi:10.4310/CNTP.2012.v6.n1.a5}}.

\bibitem{Adams:2013nia}
L.~Adams \textit{et al.},   \textit{J. Math. Phys.} \textbf{54} (2013) 052303.
  \href {http://arxiv.org/abs/1302.7004} {\path{arXiv:1302.7004}},
  \href {http://dx.doi.org/10.1063/1.4804996} {\path{doi:10.1063/1.4804996}}.

\bibitem{Bloch:2013tra}
S.~Bloch and P.~Vanhove,  \textit{J.
  Number Theor.} \textbf{148} (2015) 328.
  \href {http://arxiv.org/abs/1309.5865} {\path{arXiv:1309.5865}},
  \href {http://dx.doi.org/10.1016/j.jnt.2014.09.032}
  {\path{doi:10.1016/j.jnt.2014.09.032}}.

\bibitem{Adams:2014vja}
L.~Adams \textit{et al.},  \textit{J. Math. Phys.} \textbf{55} (2014) 102301.
  \href {http://arxiv.org/abs/1405.5640} {\path{arXiv:1405.5640}},
  \href {http://dx.doi.org/10.1063/1.4896563} {\path{doi:10.1063/1.4896563}}.

\bibitem{Adams:2015ydq}
L.~Adams \textit{et al.},  \textit{J. Math. Phys.} \textbf{57} (2016)
  032304.
  \href {http://arxiv.org/abs/1512.05630} {\path{arXiv:1512.05630}},
  \href {http://dx.doi.org/10.1063/1.4944722} {\path{doi:10.1063/1.4944722}}.

\bibitem{Remiddi:2013joa}
E.~Remiddi and L.~Tancredi,  \textit{Nucl. Phys.} \textbf{B880}
  (2014) 343.
  \href {http://arxiv.org/abs/1311.3342} {\path{arXiv:1311.3342}},
  \href {http://dx.doi.org/10.1016/j.nuclphysb.2014.01.009}
  {\path{doi:10.1016/j.nuclphysb.2014.01.009}}.

\bibitem{Bloch:2016izu}
S.~Bloch \textit{et al.}, \textit{Adv. Theor. Math. Phys.} \textbf{21} (2017) 1373.
  \href {http://arxiv.org/abs/1601.08181} {\path{arXiv:1601.08181}},
  \href {http://dx.doi.org/10.4310/ATMP.2017.v21.n6.a1}
  {\path{doi:10.4310/ATMP.2017.v21.n6.a1}}.

\bibitem{Adams:2017ejb}
L.~Adams and S.~Weinzierl,  \textit{Commun. Num. Theor. Phys.} \textbf{12} (2018) 193.
  \href {http://arxiv.org/abs/1704.08895} {\path{arXiv:1704.08895}},
  \href {http://dx.doi.org/10.4310/CNTP.2018.v12.n2.a1}
  {\path{doi:10.4310/CNTP.2018.v12.n2.a1}}.

\bibitem{Bogner:2017vim}
C.~Bogner \textit{et al.},  \textit{Nucl.
  Phys.} \textbf{B922} (2017) 528.
  \href {http://arxiv.org/abs/1705.08952} {\path{arXiv:1705.08952}},
  \href {http://dx.doi.org/10.1016/j.nuclphysb.2017.07.008}
  {\path{doi:10.1016/j.nuclphysb.2017.07.008}}.

\bibitem{Lee:2009dh}
R.N. Lee, \textit{Nucl. Phys.} \textbf{B830} (2010) 474.
  \href {http://arxiv.org/abs/0911.0252} {\path{arXiv:0911.0252}},
  \href {http://dx.doi.org/10.1016/j.nuclphysb.2009.12.025}
  {\path{doi:10.1016/j.nuclphysb.2009.12.025}}.

\bibitem{Beilinson:1994}
A.~Beilinson and A.~Levin,  in {\it Motives}, Eds. U.
  Jannsen \textit{et al.} (AMS, New York, 1994), Vol.  55, part 2, p. 97.

\bibitem{Levin:1997}
A.~Levin, \textit{Comp. Math.} \textbf{106} (1997)
  267.
  \href {http://dx.doi.org/10.1023/A:1000193320513}
  {\path{doi:10.1023/A:1000193320513}}.

\bibitem{Levin:2007}
A.~Levin and G.~Racinet, Towards multiple elliptic polylogarithms, \href
  {http://arxiv.org/abs/math/0703237} {\path{arXiv:math/0703237}}.

\bibitem{Enriquez:2010}
B.~{Enriquez},  \textit{Selecta Math.} \textbf{20} (2014) 491.
  \href {http://arxiv.org/abs/1003.1012} {\path{arXiv:1003.1012}},
  \href {http://dx.doi.org/10.1007/s00029-013-0137-3}
  {\path{doi:10.1007/s00029-013-0137-3}}.

\bibitem{Brown:2011}
F.~Brown and  A.~Levin, Multiple elliptic polylogarithms, \href
  {http://arxiv.org/abs/1110.6917} {\path{arXiv:1110.6917}}.

\bibitem{Wildeshaus}
J.~Wildeshaus, \textit{Realizations of Polylogarithms} (Springer, Berlin,
1997).
  \href {http://dx.doi.org/10.1007/BFb0093051}
  {\path{doi:10.1007/BFb0093051}}.

\bibitem{Bloch:2014qca}
S.~Bloch \textit{et al.},  \textit{Compos. Math.} \textbf{151} (2015) 2329.
  \href {http://arxiv.org/abs/1406.2664} {\path{arXiv:1406.2664}},
  \href {http://dx.doi.org/10.1112/S0010437X15007472}
  {\path{doi:10.1112/S0010437X15007472}}.

\bibitem{Broedel:2017siw}
J.~Broedel \textit{et al.},   \textit{Phys. Rev.} \textbf{D97}  (2018) 116009.
  \href {http://arxiv.org/abs/1712.07095} {\path{arXiv:1712.07095}},
  \href {http://dx.doi.org/10.1103/PhysRevD.97.116009}
  {\path{doi:10.1103/PhysRevD.97.116009}}.

\bibitem{Broedel:2018iwv}
J.~Broedel \textit{et al.},   \textit{J. High Energy Phys.} \textbf{08} (2018) 014.
  \href {http://arxiv.org/abs/1803.10256} {\path{arXiv:1803.10256}},
  \href {http://dx.doi.org/10.1007/JHEP08(2018)014}
  {\path{doi:10.1007/JHEP08(2018)014}}.

\bibitem{Passarino:2017EPJC}
G.~{Passarino}, 
  \textit{Eur. Phys. J. C} \textbf{77} (2017) 77.
  \href {http://arxiv.org/abs/1610.06207} {\path{arXiv:1610.06207}},
  \href {http://dx.doi.org/10.1140/epjc/s10052-017-4623-1}
  {\path{doi:10.1140/epjc/s10052-017-4623-1}}.

\bibitem{Sogaard:2014jla}
M.~S{\o}gaard and Y.~Zhang,  \textit{Phys. Rev.}
  \textbf{D91}  (2015) 081701.
  \href {http://arxiv.org/abs/1412.5577} {\path{arXiv:1412.5577}},
  \href {http://dx.doi.org/10.1103/PhysRevD.91.081701}
  {\path{doi:10.1103/PhysRevD.91.081701}}.

\bibitem{Hidding:2017jkk}
M.~Hidding and F.~Moriello, {All orders structure and efficient computation of
  linearly reducible elliptic Feynman integrals}, \href
  {http://arxiv.org/abs/1712.04441} {\path{arXiv:1712.04441}}.

\bibitem{vanHoeij:1997}
M.~van Hoeij,  \textit{J. Symbolic Comput.} \textbf{24} (1997) 537.
  \href {http://dx.doi.org/10.1006/jsco.1997.0151}
  {\path{doi:10.1006/jsco.1997.0151}}.

\bibitem{Pittau:2012zd}
R.~Pittau,  \textit{J. High Energy Phys.} \textbf{11}
  (2012) 151.
  \href {http://arxiv.org/abs/1208.5457} {\path{arXiv:1208.5457}},
  \href {http://dx.doi.org/10.1007/JHEP11(2012)151}
  {\path{doi:10.1007/JHEP11(2012)151}}.

\bibitem{Sborlini:2016gbr}
G.F.R. Sborlini \textit{et al.}, 
   \textit{J. High Energy Phys.} \textbf{08} (2016)
  160.
  \href {http://arxiv.org/abs/1604.06699} {\path{arXiv:1604.06699}},
  \href {http://dx.doi.org/10.1007/JHEP08(2016)160}
  {\path{doi:10.1007/JHEP08(2016)160}}.

\bibitem{Seth:2016hmv}
S.~Seth and S.~Weinzierl,  \textit{Phys. Rev.}
  \textbf{D93}  (2016) 114031.
  \href {http://arxiv.org/abs/1605.06646} {\path{arXiv:1605.06646}},
  \href {http://dx.doi.org/10.1103/PhysRevD.93.114031}
  {\path{doi:10.1103/PhysRevD.93.114031}}.

\bibitem{Gastmans:2011wh}
R.~Gastmans \textit{et al.},  {Higgs decay into two photons, revisited},
\href
  {http://arxiv.org/abs/1108.5872} {\path{arXiv:1108.5872}}.

\bibitem{Melnikov:2016nvo}
K.~Melnikov and A.~Vainshtein, \textit{Phys. Rev.} \textbf{D93} (2016) 053015.
  \href {http://arxiv.org/abs/1601.00406} {\path{arXiv:1601.00406}},
  \href {http://dx.doi.org/10.1103/PhysRevD.93.053015}
  {\path{doi:10.1103/PhysRevD.93.053015}}.

\bibitem{Donati:2013voa}
A.M. Donati and R.~Pittau,  \textit{Eur. Phys. J.} \textbf{C74} (2014) 2864.
  \href {http://arxiv.org/abs/1311.3551} {\path{arXiv:1311.3551}},
  \href {http://dx.doi.org/10.1140/epjc/s10052-014-2864-9}
  {\path{doi:10.1140/epjc/s10052-014-2864-9}}.

\bibitem{Pittau:2014tva}
R.~Pittau,  \textit{Fortschr. Phys.} \textbf{63} (2015)
  601.
  \href {http://arxiv.org/abs/1408.5345} {\path{arXiv:1408.5345}},
  \href {http://dx.doi.org/10.1002/prop.201500040}
  {\path{doi:10.1002/prop.201500040}}.

\bibitem{Page:2015zca}
B.~Page and R.~Pittau,  \textit{J. High Energy Phys.} \textbf{11} (2015)
  183.
  \href {http://arxiv.org/abs/1506.09093} {\path{arXiv:1506.09093}},
  \href {http://dx.doi.org/10.1007/JHEP11(2015)183}
  {\path{doi:10.1007/JHEP11(2015)183}}.

\bibitem{Pittau:2013qla}
R.~Pittau, \textit{Eur. Phys. J.} \textbf{C74} (2014)
  2686.
  \href {http://arxiv.org/abs/1307.0705} {\path{arXiv:1307.0705}},
  \href {http://dx.doi.org/10.1140/epjc/s10052-013-2686-1}
  {\path{doi:10.1140/epjc/s10052-013-2686-1}}.

\bibitem{Bogoliubov:1957gp}
N.N. Bogoliubov and O.S. Parasiuk,  \textit{Acta Math.} \textbf{97} (1957) 227.
  \href {http://dx.doi.org/10.1007/BF02392399}
  {\path{doi:10.1007/BF02392399}}.

\bibitem{Zimmermann:1969jj}
W.~Zimmermann,  \textit{Commun. Math. Phys.} \textbf{15} (1969) 208.
  \href {http://dx.doi.org/10.1007/BF01645676}
  {\path{doi:10.1007/BF01645676}}.

\bibitem{Pittau:2013ica}
R.~Pittau,  \textit{Fortschr. Phys.} \textbf{63} (2015) 132.
  \href {http://arxiv.org/abs/1305.0419} {\path{arXiv:1305.0419}},
  \href {http://dx.doi.org/10.1002/prop.201400079}
  {\path{doi:10.1002/prop.201400079}}.

\bibitem{Zirke:2015spg}
T.J.E. Zirke, \textit{J. High Energy Phys.} \textbf{02}
  (2016) 029.
  \href {http://arxiv.org/abs/1512.04920} {\path{arXiv:1512.04920}},
  \href {http://dx.doi.org/10.1007/JHEP02(2016)029}
  {\path{doi:10.1007/JHEP02(2016)029}}.

\bibitem{Alwall:2014hca}
J.~Alwall \textit{et al.},  \textit{J. High Energy Phys.} \textbf{07} (2014) 079.
  \href {http://arxiv.org/abs/1405.0301} {\path{arXiv:1405.0301}},
  \href {http://dx.doi.org/10.1007/JHEP07(2014)079}
  {\path{doi:10.1007/JHEP07(2014)079}}.

\bibitem{Cacciari:2011ma}
M.~Cacciari \textit{et al.},   \textit{Eur. Phys. J.} \textbf{C72}
  (2012) 1896.
  \href {http://arxiv.org/abs/1111.6097} {\path{arXiv:1111.6097}},
  \href {http://dx.doi.org/10.1140/epjc/s10052-012-1896-2}
  {\path{doi:10.1140/epjc/s10052-012-1896-2}}.

\bibitem{Bollini:1972ui}
C.G. Bollini and J.J. Giambiagi,  \textit{Nuovo Cim.} \textbf{B12} (1972) 20.
  \href {http://dx.doi.org/10.1007/BF02895558}
  {\path{doi:10.1007/BF02895558}}.

\bibitem{Cicuta:1972jf}
G.M. Cicuta and E.~Montaldi,  \textit{Lett. Nuovo Cim.} \textbf{4} (1972) 329.
  \href {http://dx.doi.org/10.1007/BF02756527}
  {\path{doi:10.1007/BF02756527}}.

\bibitem{Ashmore:1972uj}
J.F. Ashmore,  \textit{Lett. Nuovo Cim.} \textbf{4}
  (1972) 289.
  \href {http://dx.doi.org/10.1007/BF02824407}
  {\path{doi:10.1007/BF02824407}}.

\bibitem{Wilson:1972cf}
K.G. Wilson,  \textit{Phys.
  Rev.} \textbf{D7} (1973) 2911.
  \href {http://dx.doi.org/10.1103/PhysRevD.7.2911}
  {\path{doi:10.1103/PhysRevD.7.2911}}.

\bibitem{Kunszt:1992tn}
Z.~Kunszt and D.E. Soper,  \textit{Phys. Rev.} \textbf{D46} (1992) 192.
  \href {http://dx.doi.org/10.1103/PhysRevD.46.192}
  {\path{doi:10.1103/PhysRevD.46.192}}.

\bibitem{Frixione:1995ms}
S.~Frixione \textit{et al.},  \textit{Nucl. Phys.} \textbf{B467} (1996) 399.
  \href {http://arxiv.org/abs/hep-ph/9512328}
  {\path{arXiv:hep-ph/9512328}}, \href
  {http://dx.doi.org/10.1016/0550-3213(96)00110-1}
  {\path{doi:10.1016/0550-3213(96)00110-1}}.

\bibitem{Catani:1996jh}
S.~Catani and M.H. Seymour,  \textit{Phys. Lett.} \textbf{B378} (1996) 287.
  \href {http://arxiv.org/abs/hep-ph/9602277}
  {\path{arXiv:hep-ph/9602277}}, \href
  {http://dx.doi.org/10.1016/0370-2693(96)00425-X}
  {\path{doi:10.1016/0370-2693(96)00425-X}}.

\bibitem{Catani:1996vz}
S.~Catani and M.H. Seymour,  \textit{Nucl. Phys.} \textbf{B485} (1997) 291 [Erratum: \textit{Nucl.
  Phys.} \textbf{B510} (1998) 503].
  \href {http://arxiv.org/abs/hep-ph/9605323}
  {\path{arXiv:hep-ph/9605323}}, \href
  {http://dx.doi.org/10.1016/S0550-3213(96)00589-5}
  {\path{doi:10.1016/S0550-3213(96)00589-5}}.

\bibitem{GehrmannDeRidder:2005cm}
A.~Gehrmann-De~Ridder \textit{et al.},   \textit{J. High Energy Phys.} \textbf{09} (2005) 056.
  \href {http://arxiv.org/abs/hep-ph/0505111}
  {\path{arXiv:hep-ph/0505111}}, \href
  {http://dx.doi.org/10.1088/1126-6708/2005/09/056}
  {\path{doi:10.1088/1126-6708/2005/09/056}}.

\bibitem{Catani:2007vq}
S.~Catani and M.~Grazzini,  \textit{Phys. Rev. Lett.} \textbf{98}
  (2007) 222002.
  \href {http://arxiv.org/abs/hep-ph/0703012}
  {\path{arXiv:hep-ph/0703012}}, \href
  {http://dx.doi.org/10.1103/PhysRevLett.98.222002}
  {\path{doi:10.1103/PhysRevLett.98.222002}}.

\bibitem{Czakon:2010td}
M.~Czakon, 
  \textit{Phys. Lett.} \textbf{B693} (2010) 259.
  \href {http://arxiv.org/abs/1005.0274} {\path{arXiv:1005.0274}},
  \href {http://dx.doi.org/10.1016/j.physletb.2010.08.036}
  {\path{doi:10.1016/j.physletb.2010.08.036}}.

\bibitem{Bolzoni:2010bt}
P.~Bolzoni \textit{et al.},   \textit{J. High Energy Phys.} \textbf{01} (2011) 059.
  \href {http://arxiv.org/abs/1011.1909} {\path{arXiv:1011.1909}},
  \href {http://dx.doi.org/10.1007/JHEP01(2011)059}
  {\path{doi:10.1007/JHEP01(2011)059}}.

\bibitem{Boughezal:2015dva}
R.~Boughezal \textit{et al.},  
  \textit{Phys. Rev. Lett.} \textbf{115} (2015) 062002.
  \href {http://arxiv.org/abs/1504.02131} {\path{arXiv:1504.02131}},
  \href {http://dx.doi.org/10.1103/PhysRevLett.115.062002}
  {\path{doi:10.1103/PhysRevLett.115.062002}}.

\bibitem{Gaunt:2015pea}
J.~Gaunt \textit{et al.},   \textit{J. High Energy Phys.} \textbf{09} (2015) 058.
  \href {http://arxiv.org/abs/1505.04794} {\path{arXiv:1505.04794}},
  \href {http://dx.doi.org/10.1007/JHEP09(2015)058}
  {\path{doi:10.1007/JHEP09(2015)058}}.

\bibitem{DelDuca:2016ily}
V.~Del~Duca \textit{et al.},   \textit{Phys. Rev.} \textbf{D94} (2016) 074019.
  \href {http://arxiv.org/abs/1606.03453} {\path{arXiv:1606.03453}},
  \href {http://dx.doi.org/10.1103/PhysRevD.94.074019}
  {\path{doi:10.1103/PhysRevD.94.074019}}.

\bibitem{Caola:2017dug}
F.~Caola \textit{et al.},  \textit{Eur. Phys. J.} \textbf{C77} (2017) 248.
  \href {http://arxiv.org/abs/1702.01352} {\path{arXiv:1702.01352}},
  \href {http://dx.doi.org/10.1140/epjc/s10052-017-4774-0}
  {\path{doi:10.1140/epjc/s10052-017-4774-0}}.

\bibitem{Magnea:2018hab}
L.~Magnea \textit{et al.},   \textit{J. High Energy Phys.} \textbf{12} (2018) 107.
  \href {http://arxiv.org/abs/1806.09570} {\path{arXiv:1806.09570}},
  \href {http://dx.doi.org/10.1007/JHEP12(2018)107}
  {\path{doi:10.1007/JHEP12(2018)107}}.

\bibitem{Hernandez-Pinto:2015ysa}
R.J. Hernandez-Pinto \textit{et al.}, \textit{J. High Energy Phys.} \textbf{02} (2016) 044.
  \href {http://arxiv.org/abs/1506.04617} {\path{arXiv:1506.04617}},
  \href {http://dx.doi.org/10.1007/JHEP02(2016)044}
  {\path{doi:10.1007/JHEP02(2016)044}}.

\bibitem{Sborlini:2016hat}
G.F.R. Sborlini \textit{et al.},   \textit{J. High Energy Phys.} \textbf{10} (2016) 162.
  \href {http://arxiv.org/abs/1608.01584} {\path{arXiv:1608.01584}},
  \href {http://dx.doi.org/10.1007/JHEP10(2016)162}
  {\path{doi:10.1007/JHEP10(2016)162}}.

\bibitem{Driencourt-Mangin:2017gop}
F.~Driencourt-Mangin \textit{et al.},  \textit{Eur. Phys. J.} \textbf{C78} (2018) 231.
  \href {http://arxiv.org/abs/1702.07581} {\path{arXiv:1702.07581}},
  \href {http://dx.doi.org/10.1140/epjc/s10052-018-5692-5}
  {\path{doi:10.1140/epjc/s10052-018-5692-5}}.

\bibitem{Ramirez-Uribe:2017gbf}
N.~Selomit Ramirez-Uribe \textit{et al.},   \textit{J. Phys. Conf. Ser.}
  \textbf{912} (2017) 012013.
  \href {http://arxiv.org/abs/1709.07802} {\path{arXiv:1709.07802}},
  \href {http://dx.doi.org/10.1088/1742-6596/912/1/012013}
  {\path{doi:10.1088/1742-6596/912/1/012013}}.

\bibitem{Catani:2008xa}
S.~Catani \textit{et al.},   \textit{J. High Energy Phys.} \textbf{09} (2008) 065.
  \href {http://arxiv.org/abs/0804.3170} {\path{arXiv:0804.3170}},
  \href {http://dx.doi.org/10.1088/1126-6708/2008/09/065}
  {\path{doi:10.1088/1126-6708/2008/09/065}}.

\bibitem{Bierenbaum:2010cy}
I.~Bierenbaum \textit{et al.},  \textit{J. High Energy Phys.} \textbf{10} (2010) 073.
  \href {http://arxiv.org/abs/1007.0194} {\path{arXiv:1007.0194}},
  \href {http://dx.doi.org/10.1007/JHEP10(2010)073}
  {\path{doi:10.1007/JHEP10(2010)073}}.

\bibitem{Bierenbaum:2012th}
I.~Bierenbaum \textit{et al.},   \textit{J. High Energy Phys.} \textbf{03} (2013) 025.
  \href {http://arxiv.org/abs/1211.5048} {\path{arXiv:1211.5048}},
  \href {http://dx.doi.org/10.1007/JHEP03(2013)025}
  {\path{doi:10.1007/JHEP03(2013)025}}.

\bibitem{Buchta:2014dfa}
S.~Buchta \textit{et al.},  \textit{J. High Energy Phys.} \textbf{11}
  (2014) 014.
  \href {http://arxiv.org/abs/1405.7850} {\path{arXiv:1405.7850}},
  \href {http://dx.doi.org/10.1007/JHEP11(2014)014}
  {\path{doi:10.1007/JHEP11(2014)014}}.

\bibitem{Buchta:2015wna}
S.~Buchta \textit{et al.},   \textit{Eur. Phys. J.} \textbf{C77}  (2017) 274.
  \href {http://arxiv.org/abs/1510.00187} {\path{arXiv:1510.00187}},
  \href {http://dx.doi.org/10.1140/epjc/s10052-017-4833-6}
  {\path{doi:10.1140/epjc/s10052-017-4833-6}}.

\bibitem{Tomboulis:2017rvd}
E.T. Tomboulis, 
  \textit{J. High Energy Phys.} \textbf{05} (2017) 148.
  \href {http://arxiv.org/abs/1701.07052} {\path{arXiv:1701.07052}},
  \href {http://dx.doi.org/10.1007/JHEP05(2017)148}
  {\path{doi:10.1007/JHEP05(2017)148}}.

\bibitem{CaronHuot:2010zt}
S.~Caron-Huot,  \textit{J. High Energy Phys.} \textbf{05} (2011) 080.
  \href {http://arxiv.org/abs/1007.3224} {\path{arXiv:1007.3224}},
  \href {http://dx.doi.org/10.1007/JHEP05(2011)080}
  {\path{doi:10.1007/JHEP05(2011)080}}.

\bibitem{Jurado:2017xut}
J.L. Jurado \textit{et al.},   \textit{J. High Energy Phys.} \textbf{12} (2017) 122.
  \href {http://arxiv.org/abs/1710.11010} {\path{arXiv:1710.11010}},
  \href {http://dx.doi.org/10.1007/JHEP12(2017)122}
  {\path{doi:10.1007/JHEP12(2017)122}}.

\bibitem{Bern:2002zk}
Z.~Bern \textit{et al.}, \textit{Phys. Rev.} \textbf{D66}
  (2002) 085002.
  \href {http://arxiv.org/abs/hep-ph/0202271}
  {\path{arXiv:hep-ph/0202271}}, \href
  {http://dx.doi.org/10.1103/PhysRevD.66.085002}
  {\path{doi:10.1103/PhysRevD.66.085002}}.

\bibitem{Battistel:1998sz}
O.A. Battistel \textit{et al.},   \textit{Mod. Phys. Lett.} \textbf{A13} (1998) 1597.
  \href {http://dx.doi.org/10.1142/S0217732398001686}
  {\path{doi:10.1142/S0217732398001686}}.

\bibitem{Feynman:1963ax}
R.P. Feynman, \textit{Acta Phys. Pol.} \textbf{24} (1963)
  697.

\bibitem{MapleCuba}
Maplesoft,
 Maple
  18+, 
\url{{https://www.maplesoft.com/support/help/AddOns/view.aspx?path=evalf/Int/cuba}}.

\bibitem{R2Cuba}
A.~Bouvier, K.~Ki\^eu,
 R2Cuba,
\url{{https://cran.r-project.org/package=R2Cuba}}.

\bibitem{REDUCECuba}
K.~Oikonomou, Cuba
  library for multidimensional numerical integration Interface to REDUCE,
\url{{https://sourceforge.net/projects/reduce-algebra}}.

\bibitem{PythonCuba}
J.~Buchner,
  pyCuba
  library for multidimensional numerical integration,
\url{{https://johannesbuchner.github.io/PyMultiNest/pycuba.html}}.

\bibitem{JuliaCuba}
M.~Giordano, https://github.com/giordano/Cuba.jlCuba.jl library for
  multidimensional numerical integration,
\url{{https://github.com/giordano/Cuba.jl}}.

\bibitem{Genz:1983}
A.~Genz and A.~Malik,  \textit{SIAM J. Numer. Anal.} \textbf{20}
  (1983) 580.
  \href {http://dx.doi.org/10.1137/0720038}
  {\path{doi:10.1137/0720038}}.
\url{https://doi.org/10.1137/0720038}

\bibitem{Press:1992zz}
W.H. Press \textit{et al.}, \textit{Numerical
  Recipes in FORTRAN: The Art of Scientific Computing} (Cambridge University Press, Cambridge, 1992).

\bibitem{Bratley:1988:AIS:42288.214372}
P.~Bratley and B.L. Fox,  \textit{ACM Trans. Math.
  Softw.} \textbf{14} (1988) 88.
  \href {http://dx.doi.org/10.1145/42288.214372}
  {\path{doi:10.1145/42288.214372}}.
\url{http://doi.acm.org/10.1145/42288.214372}

\bibitem{Matsumoto:1998:MTE:272991.272995}
M.~Matsumoto and T.~Nishimura,
   \textit{ACM Trans. Model. Comput. Simul.} \textbf{8} (1998) 3.
  \href {http://dx.doi.org/10.1145/272991.272995}
  {\path{doi:10.1145/272991.272995}}.
\url{http://doi.acm.org/10.1145/272991.272995}

\bibitem{Luscher:1993dy}
M.~Luscher,  \textit{Comput. Phys. Commun.} \textbf{79} (1994) 100.
  \href {http://arxiv.org/abs/hep-lat/9309020}
  {\path{arXiv:hep-lat/9309020}}, \href
  {http://dx.doi.org/10.1016/0010-4655(94)90232-1}
  {\path{doi:10.1016/0010-4655(94)90232-1}}.

\bibitem{James:1993np}
F.~James,  \textit{Comput. Phys. Commun.} \textbf{79} (1994) 111
  [Erratum: \textit{Comput. Phys. Commun.} \textbf{97}  (1996) 357].
  \href {http://dx.doi.org/10.1016/0010-4655(94)90233-X}
  {\path{doi:10.1016/0010-4655(94)90233-X}}.

\bibitem{Keng:1981}
H.~Keng and W.~Yuan, \textit{Applications of Number Theory to Numerical Analysis},
  (Springer, Berlin, 1981). \href
  {http://dx.doi.org/10.1007/978-3-642-67829-5}
  {\path{doi:10.1007/978-3-642-67829-5}}.

\bibitem{Lepage:1980dq}
G.P. Lepage,
 VEGAS:
  an adaptive multidimensional integration program.
\url{{https://lib-extopc.kek.jp/preprints/PDF/1980/8006/8006210.pdf}}

\bibitem{Lepage:1977sw}
G.P. Lepage,  \textit{J.
  Comput. Phys.} \textbf{27} (1978) 192.
  \href {http://dx.doi.org/10.1016/0021-9991(78)90004-9}
  {\path{doi:10.1016/0021-9991(78)90004-9}}.

\bibitem{Press:1989vk}
W.H. Press and G.R. Farrar, \textit{Comp. Phys.}  190 (1990) 190. \href
  {http://dx.doi.org/10.1063/1.4822899} {\path{doi:10.1063/1.4822899}}.

\bibitem{Friedman:1981ak}
J.H. Friedman and M.H. Wright, {DIVONNE4}: a program for multiple integration
  and adaptive importance sampling (1981), CGTM-193-REV.

\bibitem{Friedman:1981:NPP:355934.355939}
J.H. Friedman and M.H. Wright,  \textit{ACM Trans. Math. Softw.} \textbf{7} (1981) 76.
  \href {http://dx.doi.org/10.1145/355934.355939}
  {\path{doi:10.1145/355934.355939}}.

\bibitem{Berntsen:1991:ADA:210232.210234}
J.~Berntsen \textit{et al.},  \textit{ACM Trans.
  Math. Softw.} \textbf{17} (1991) 452.
  \href {http://dx.doi.org/10.1145/210232.210234}
  {\path{doi:10.1145/210232.210234}}.


\end{thebibliography}
\clearpage  \pagestyle{empty}
\chapter*{Bibliography}
\phantomsection       
\addcontentsline{toc}{chapter}{Bibliography}
\renewcommand{\bibname}{} 
\pagestyle{plain}
\fancyhead[LO]{}
\fancyhead[RO]{}
\fancyhead[CO]{Bibliography}
\fancyhead[LE]{}
\fancyhead[CE]{Bibliography}
\fancyhead[RE]{}

\end{flushleft}


\end{document}